\documentclass[fleqn,12pt]{article}
 \usepackage{epsfig}
\gdef\Feynmanlength{\setlength{\unitlength}{0.01pt}}  % Say \Feynmanlength

\global\newcount\LINETYPE                     
\global\newcount\LINEDIRECTION
\global\newcount\LINECONFIGURATION
\newcommand{\LTYPE}{\LINETYPE}
\newcommand{\LDIR}{\LINEDIRECTION}

\global\LINETYPE=1  \global\LINEDIRECTION=0  \global\LINECONFIGURATION=0
\global\newcount\fermion    \fermion=1
\global\newcount\scalar     \scalar=2
\global\newcount\photon     \photon=3
\global\newcount\gluon      \gluon=4
\global\newcount\especial   \especial=5
\gdef\N{0}  \gdef\NE{1}  \gdef\E{2}   \gdef\SE{3}
\gdef\S{4}  \gdef\SW{5}  \gdef\W{6}   \gdef\NW{7}
\global\newcount\REG            \global\REG=0
\global\newcount\FLIPPED        \global\FLIPPED=1
\global\newcount\CURLY          \global\CURLY=2
\global\newcount\FLIPPEDCURLY   \global\FLIPPEDCURLY=3
\global\newcount\FLAT           \global\FLAT=4
\global\newcount\FLIPPEDFLAT    \global\FLIPPEDFLAT=5
\global\newcount\CENTRAL        \global\CENTRAL=6
\global\newcount\FLIPPEDCENTRAL \global\FLIPPEDCENTRAL=7
             
\global\newcount\SQUASHEDGLUON  \global\SQUASHEDGLUON=8

\newcount\adjx \adjx=0
\newcount\adjy \adjy=0
\global\newdimen\BIGPHOTONS     \BIGPHOTONS=0pt  %  DEFAULT:  10 & 11-PT PHOTONS
\global\newdimen\THICKPHOTONS     \THICKPHOTONS=0pt  %  FOR E-W PHOTONS 
\global\newdimen\THICKPHOTONSWITCH    \THICKPHOTONSWITCH=0pt
\gdef\THICKPHOTONTEST{
\THICKPHOTONSWITCH=0pt
\ifdim\THICKPHOTONS=0pt \relax
  \else \ifnum\LTYPE=3
	   \ifnum\LDIR=2 \THICKPHOTONSWITCH=1pt \fi % THICK \E PHOTON
	   \ifnum\LDIR=6 \THICKPHOTONSWITCH=1pt \fi % THICK \W PHOTON
	\fi
\fi
}  % end of THICKPHOTONTEST
\gdef\THICKLINES{\thicklines  \THICKPHOTONS=1pt}
\gdef\THINLINES{\thinlines  \THICKPHOTONS=0pt}
\global\newcount\phantomswitch   \global\phantomswitch=0
\global\newcount\stemlength   \global\stemlength=275   % Default STEM length.
\global\newcount\absstemlength        % A copy of STEM length.
\global\newcount\stemlengthx          % FOR STEMS on particle lines
\global\newcount\stemlengthy          % FOR STEMS on particle lines
\newdimen\FRONTSTEM  \FRONTSTEM=0pt   % FOR STEMS
\newdimen\BACKSTEM   \BACKSTEM=0pt    % FOR STEMS
\newdimen\EITHERSTEM \EITHERSTEM=0pt  % FOR STEMS
\global\newcount\arrowlength                % FOR ARROWS
\global\newdimen\ATTIP   \global\ATTIP=0pt  % FOR ARROWS
\global\newdimen\ATBASE  \global\ATBASE=1pt % FOR ARROWS
\global\newcount\unitboxnumber  % SHOWS THE NUMBER OF `UNIT BOXES' IN LINE
\global\newcount\unitboxnumberpo  % One more than \unitboxnumber (in GLUONSETUP)
\global\newcount\particlelengthx  % THE X-LENGTH OF THE PARTICLE LINE
\gdef\plengthx{\particlelengthx}
\global\newcount\particlelengthy  % THE Y-LENGTH OF THE PARTICLE LINE
\gdef\plengthy{\particlelengthy}  
\global\newcount\boxlengthx  % THE X-LENGTH OF THE BOX:  abs(plengthx) usually
\global\newcount\boxlengthy  % THE y-LENGTH OF THE box:  abs(plengthy) usually
\global\newcount\particleadjustx  % Replaces \gluonadjustx, \scalaradjustx etc.
\global\newcount\particleadjusty  % Replaces \gluonadjusty, \scalaradjusty etc.
\global\newcount\particlelength   % The LENGTH of a particle line BOX (x)
\global\newcount\particlefrontx
\gdef\pfrontx{\particlefrontx}
\global\newcount\PFRONTx
\global\newcount\particlefronty
\gdef\pfronty{\particlefronty}
\global\newcount\PFRONTy
\global\newcount\particlebackx
\gdef\pbackx{\particlebackx}
\global\newcount\particlebacky
\gdef\pbacky{\particlebacky}
\global\newcount\particlemidx
\gdef\pmidx{\particlemidx}
\global\newcount\particlemidy
\gdef\pmidy{\particlemidy}
\global\newcount\seglength  \global\newcount\gaplength
\global\gaplength=850  %default
\global\seglength=1416  % Length of each seg not including `ends' for attachment
\global\newcount\Xone    \global\newcount\Yone    % user co-ords (\Xone,\Yone)
\global\newcount\Xtwo    \global\newcount\Ytwo    % user co-ords (\Xtwo,\Ytwo)
\global\newcount\Xthree  \global\newcount\Ythree  % user's (\Xthree,\Ythree)
\global\newcount\Xfour   \global\newcount\Yfour   % user co-ords (\Xfour,\Yfour)
\global\newcount\Xfive   \global\newcount\Yfive   % user co-ords (\Xfive,\Yfive)
\global\newcount\Xsix    \global\newcount\Ysix    % user co-ords (\Xsix,\Ysix)
\global\newcount\Xseven  \global\newcount\Yseven  % user's (\Xseven,\Yseven)
\global\newcount\Xeight  \global\newcount\Yeight  % user's (\Xeight,\Yeight)
\newsavebox{\lastline}  %  Default name for an unnamed particle line.
\global\newcount\numlineparts   % Num of pieces each unitbox of the line needs
\global\newcount\upperlineadjx  \upperlineadjx=0  %Default
\global\newcount\upperlineadjy  \upperlineadjy=0  %Default
\global\newcount\lowerlineadjx  \lowerlineadjx=0  %Default
\global\newcount\lowerlineadjy  \lowerlineadjy=0  %Default
\global\newcount\thirdlineadjx  \thirdlineadjx=0  %Default
\global\newcount\thirdlineadjy  \thirdlineadjy=0  %Default
\global\newcount\fourthlineadjx \fourthlineadjx=0  %Default
\global\newcount\fourthlineadjy \fourthlineadjy=0  %Default
\global\newcount\unitboxwidth   \unitboxwidth=1000%Default
\global\newcount\unitboxheight  \unitboxheight=0  %Default
\global\newcount\numupperunits  \numupperunits=8  %Default
\global\newcount\numlowerunits  \numlowerunits=8  %Default
\global\newcount\numthirdunits  \numthirdunits=8  %Default
\global\newcount\numfourthunits \numfourthunits=8  %Default
\global\newcount\fermioncount   \global\fermioncount=0    
\global\newcount\scalarcount    \global\scalarcount=0    
\global\newcount\photoncount    \global\photoncount=0    
\global\newcount\gluoncount     \global\gluoncount=0    
\global\newcount\especialcount  \global\especialcount=0    
\global\newcount\vertexcount    \global\vertexcount=-1
\global\newcount\XDIR
\global\newcount\YDIR
\gdef\SETDIR{  % SETS THE DIRECTIONS
\ifcase\LDIR 
     \global\XDIR=0  \global\YDIR=1   %\N  case.
\or  \global\XDIR=1  \global\YDIR=1   %\NE case.
\or  \global\XDIR=1  \global\YDIR=0   %\E  case.
\or  \global\XDIR=1  \global\YDIR=-1  %\SE case.
\or  \global\XDIR=0  \global\YDIR=-1  %\S  case.
\or  \global\XDIR=-1 \global\YDIR=-1  %\SW case.
\or  \global\XDIR=-1 \global\YDIR=0   %\W  case.
\or  \global\XDIR=-1 \global\YDIR=1   %\NW case.
\else\DIRECTERROR 
\fi}  % END OF \SETDIR
\gdef\moduloeight#1{
\ifnum#1>7 \global\advance #1 by -8 
\relax
\moduloeight#1 
\relax
\else \relax  
\fi}
\gdef\multroothalf#1{\global\multiply #1 by 7071 \global\divide #1 by 10000}
\gdef\negate#1{\global\multiply #1 by -1}

\gdef\slanttest(#1,#2){ 
\ifodd\LDIR
\multiply #1 by 7071  \divide #1 by 10000
\multiply #2 by 7071  \divide #2 by 10000
\fi
}
\gdef\gslanttest(#1,#2){
\ifodd\LDIR
\multroothalf#1
\multroothalf#2
\fi
}
\gdef\setplength{ % calcs length of particle line
\global\particlelengthx=\unitboxwidth
\global\particlelengthy=\unitboxheight
\global\multiply \particlelengthx by \unitboxnumber
\global\multiply \particlelengthy by \unitboxnumber
\global\advance \particlelengthx by \particleadjustx
\global\advance \particlelengthy by \particleadjusty
}
\gdef\boxlengthdefault{  % DEFAULT FOR BOX SIZES IN \drawas
\global\boxlengthx=\plengthx
\global\boxlengthy=\plengthy
\ifnum\plengthx<0 \global\multiply\boxlengthx by -1 \fi
\ifnum\plengthy<0 \global\multiply\boxlengthy by -1 \fi
}
\gdef\rearcoords{  %  CALCULATES THE CO-ORDINATES OF THE BACK OF PARTICLE LINE
\global\particlebacky=\particlefronty 
\global\particlebackx=\particlefrontx 
\global\advance \particlebackx by \particlelengthx
\global\advance \particlebacky by \particlelengthy
}
\gdef\midcoords{  %  CALCULATES THE CO-ORDINATES OF THE MID OF PARTICLE LINE
\global\particlemidy=\particlefronty
\global\particlemidx=\particlefrontx
\global\stemlengthx=\particlelengthx  % Convenient variables not being used
\global\stemlengthy=\particlelengthy  
\global\divide\stemlengthx by 2
\global\divide\stemlengthy by 2
\global\advance \particlemidx by \stemlengthx
\global\advance \particlemidy by \stemlengthy
}
\gdef\setcoords(#1,#2,#3)(#4,#5,#6)[#7,#8]{  
\global\upperlineadjx=#1
\global\lowerlineadjx=#2
\global\thirdlineadjx=#3
\global\upperlineadjy=#4
\global\lowerlineadjy=#5
\global\thirdlineadjy=#6
\global\unitboxwidth=#7
\global\unitboxheight=#8
}
\gdef\drawoldpic#1(#2,#3){  % DRAWS PRE-SAVED PICTURE
\global\particlefrontx=#2
\global\particlefronty=#3
\rearcoords  
\midcoords
\put(#2,#3){\usebox{#1}}
}
\gdef\drawsavedline`#1' as #2[#3#4](#5,#6)[#7]{
\global\LINETYPE=#2
\global\LINEDIRECTION=#3
\global\LINECONFIGURATION=#4
\global\particlefrontx=#5
\global\particlefronty=#6
\global\unitboxnumber=#7  
\selectcase
\rearcoords% moved from before selectcase.
\midcoords
\ifnum\phantomswitch=0 \drawas{#1}\fi
}

\gdef\drawas#1{
\global\savebox{#1}(\boxlengthx,\boxlengthy){
\setlength{\unitlength}{0.01pt}
\begin{picture}(\boxlengthx,\boxlengthy)
\multiput(\upperlineadjx,\upperlineadjy)(\unitboxwidth,\unitboxheight)
{\numupperunits}{\upperunitbox}
\ifnum\numlineparts > 1  %  If the line needs 2 parts per unit or more
\multiput(\lowerlineadjx,\lowerlineadjy)(\unitboxwidth,\unitboxheight)
{\numlowerunits}{\lowerunitbox}  
\fi
\ifnum\numlineparts > 2  %  If the line needs 3 parts per unit or more
\multiput(\thirdlineadjx,\thirdlineadjy)(\unitboxwidth,\unitboxheight)
{\numthirdunits}{\thirdunitbox}  
\fi
\ifnum\numlineparts > 3  %  If the line needs 4 parts per unit or more
\multiput(\fourthlineadjx,\fourthlineadjy)(\unitboxwidth,\unitboxheight)
{\numfourthunits}{\lowerunitbox}  
\fi
\end{picture} }
\global\PFRONTx=\pfrontx  \global\PFRONTy=\pfronty   %save this value
\SETFRONTSTEM
\THICKPHOTONTEST
\ifdim\THICKPHOTONSWITCH=1pt\global\advance\PFRONTy by 20  \fi
\put(\PFRONTx,\PFRONTy) {\usebox{#1}}   %\pfrontX,Y=\particlefrontx,y
\ifdim\THICKPHOTONSWITCH=1pt
\global\advance\PFRONTy by -40
\put(\PFRONTx,\PFRONTy) {\usebox{#1}}   % The second \E or \W photon ->thicker
\global\advance \PFRONTy by 20  %re-adjust:  advanced by -20 in total above.
\fi  %End of \ifdim\THICKPHOTONSWITCH=1
\SETBACKSTEM
\seglength=1416   \gaplength=850   % Re-set \SCALR defaults.
}
\gdef\drawandsaveline`#1' as #2[#3#4](#5,#6)[#7]{
\global\newsavebox{#1}
\drawsavedline`#1' as #2[#3#4](#5,#6)[#7]
}

\gdef\drawline#1[#2#3](#4,#5)[#6]{   % Draw line but don't name it.
\drawsavedline`\lastline' as #1[#2#3](#4,#5)[#6]}

\gdef\TYPEERROR{\message{*** ERROR IN PARTICLE TYPE SELECTION ***}
\message{+++ Try with line type \fermion,\scalar,\photon,\gluon 
(see manual) +++}\SETERR}
\gdef\DIRECTERROR{\SETERR\message{*** ERROR IN PARTICLE DIRECTION SELECTION ***}
\message{+++ Try again with direction N, NE, E, SE  etc. or see manual +++}}
\gdef\UNIMPERROR{\message{*** ERROR IN PARTICLE OPTIONS SELECTION ***}
\message{
+++ The requested options combination has not yet been implemented +++}\SETERR}
\gdef\SETERR{\gdef\upperunitbox{{\tiny Error}}  % PRINTS `error' in diagram.
\gdef\lowerunitbox{\relax}
\gdef\thirdunitbox{\relax}
}
\gdef\neglengthcheck{\ifnum\unitboxnumber < 1 
\message{   *** ERROR:  PARTICLE OF NEGATIVE OR ZERO LENGTH REQUESTED. ***   }
\message{   ***         TAKING ABSOLUTE VALUE. ***   }\negate\unitboxnumber \fi}
\gdef\selectcase{  
\neglengthcheck   %  check for particles of negative length.
\SETDIR  
\ifcase\LINETYPE
\TYPEERROR  % \LINETYPE=0 case.
\or \selectfermion  % \LINETYPE=1 case.
\or \selectscalar   % \LINETYPE=2 case.
\or \selectphoton   % \LINETYPE=3 case.
\or \selectgluon    % \LINETYPE=4 case.
\or \selectespecial % \LINETYPE=5 case.
\else \TYPEERROR \fi  }
\gdef\selectfermion{
\ifnum\fermioncount=0 %Subj:   fermionsetup.tex

\global\newcount\fermionlength  %  THE TOTAL FERMION LINE LENGTH.
\global\newcount\fermionlengthx
\global\newcount\fermionlengthy
\global\newcount\fermionfrontx  %}(x,y) co-ord of left of fermion
\global\newcount\fermionfronty  %}
\global\newcount\fermionbackx
\global\newcount\fermionbacky
\gdef\ALLfermion{  % READ IN FROM FEYNMAN \selectfermion
\global\fermionfrontx=\particlefrontx \global\fermionfronty=\particlefronty
\ifnum\unitboxnumber > 50000
\message{   *** WARNING *** Fermion of length
\the\unitboxnumber\space requested ***   }
\ifnum\unitboxnumber > 80000
\message{   *** Reducing fermion length to 30000 (max 80000) ***   }
\global\unitboxnumber=30000 \fi \fi  % end of length error
\global\fermionlength=\unitboxnumber % The TOTAL line length
\global\particleadjustx=0   \global\particleadjusty=0 %Default
\global\numlineparts = 1    \global\numupperunits=1
\global\upperlineadjx=-200  \global\upperlineadjy=0
\global\fermionlengthx=\fermionlength    \global\fermionlengthy=\fermionlength
\gslanttest(\fermionlengthx,\fermionlengthy)  % See FEYNMAN22.TEX (FOR \XDIR).
\global\multiply\fermionlengthx by \XDIR  %  In keeping with photons and gluons.
\global\multiply\fermionlengthy by \YDIR  %  In keeping with photons and gluons.
\global\unitboxheight=\fermionlengthy   \global\unitboxwidth=\fermionlengthx   
\global\advance \fermionlengthx by \particleadjustx
\global\advance \fermionlengthy by \particleadjusty
\global\particlelengthx=\fermionlengthx
\global\particlelengthy=\fermionlengthy  
\boxlengthdefault    \rearcoords    \midcoords
\global\fermionbackx=\particlebackx     \global\fermionbacky=\particlebacky
\ifcase\LINECONFIGURATION  %\REG case
\ifnum\XDIR=0 
\gdef\upperunitbox{\line(\XDIR,\YDIR){\boxlengthy}} %\N or \S
\else
\gdef\upperunitbox{\line(\XDIR,\YDIR){\boxlengthx}}
\fi
\else \UNIMPERROR
\fi
}
 \fi   
\global\advance\fermioncount by 1  % Counts number of fermions drawn. 
\ALLfermion   
}
\gdef\selectscalar{
\ifnum\scalarcount=0 \input SCALARS.tex \fi   
\global\advance\scalarcount by 1  % Counts number of scalars drawn. 
\ALLscalar
}
\gdef\selectphoton{   % RECURSIVELY RE-DEFINED IN PHOTONSETUP(23+).TEX.
\ifnum\photoncount=0 %Subj:   photonsetup.tex

\newcount\numwiggles    \newcount\numwigglespo
\global\newcount\photonlengthx
\global\newcount\photonlengthy
\global\newcount\photonfrontx  %}(x,y) co-ord of left of photon
\global\newcount\photonfronty  %}
\global\newcount\photonbackx
\global\newcount\photonbacky
\newcount\halfwigglelength
\global\font\Twelverom=cmr12
\global\font\Tenrom=cmr10
\gdef\Lbr{{\Twelverom(}}   \gdef\Rbr{{\Twelverom)}}
\gdef\SLbr{{\Tenrom(}}     \gdef\SRbr{{\Tenrom)}}
\gdef\Smile{{\large$\smile$}}  % Default for 10 and 11-point documents.
\gdef\Frown{{\large$\frown$}}  % Default for 10 and 11-point documents.
\ifdim\BIGPHOTONS>0pt  \gdef\Smile{$\smile$} \gdef\Frown{$\frown$} \fi
\gdef\selectphoton{   % RECURSIVELY RE-DEFINED HERE.  Define in FEYNMAN.
\global\advance\photoncount by 1  % Counts number of photons drawn. 
\global\photonfrontx=\particlefrontx   % READ IN FROM FEYNMAN \selectphoton
\global\photonfronty=\particlefronty   % READ IN FROM FEYNMAN \selectphoton
\ifnum\unitboxnumber > 50
\message{   *** WARNING *** Photon with 
\the\unitboxnumber\space half-wiggles requested ***   }
\ifnum\unitboxnumber > 150
\message{   *** Reducing photon length to 10 half-wiggles (max 150) ***   }
\ifnum\unitboxnumber > 1000
\message{   *** Probable Cause:  Photon selected instead of Fermion ***   }
\fi \global\unitboxnumber=10 \fi \fi  % end of length error
\numwiggles=\unitboxnumber
\divide\numwiggles by 2
\global\unitboxnumberpo=\numwiggles % here \unitboxnumberpo is an unused counter
\global\multiply \unitboxnumberpo by -1
\numwigglespo=\unitboxnumber
\advance\numwigglespo by \unitboxnumberpo %\numwigglespo is one greater than 
\global\numlineparts = 2  % DEFAULT                %\numwiggles in this case.
\global\numupperunits=\numwigglespo  % DEFAULT
\global\numlowerunits=\numwiggles  % DEFAULT
\particleadjustx=0  %DEFAULT
\particleadjusty=0  %DEFAULT
\ifcase\LINEDIRECTION
     \Nphoton    %\LINEDIRECTION=0 (NORTH) CASE
\or  \NEphoton   % 1 case
\or  \Ephoton    % 2 case...horizontal photon.
\or  \SEphoton   % .
\or  \Sphoton    % .
\or  \SWphoton   % .
\or  \Wphoton    % .
\or  \NWphoton   % 7 case
\else\DIRECTERROR \fi
\setplength
\global\divide\plengthx by 2  \global\divide\plengthy by 2
\rearcoords  \boxlengthdefault   \midcoords
\global\photonbackx=\pbackx  %PHOTONSETUP26
\global\photonbacky=\pbacky  %PHOTONSETUP26
\global\photonlengthx=\plengthx  %PHOTONSETUP26
\global\photonlengthy=\plengthy  %PHOTONSETUP26
}
\gdef\SETUNITBOX(#1)[#2][#3]{ % For slanted photons only.
\gdef\upperunitbox{\oval(#1,#1)[#2]}
\gdef\lowerunitbox{\oval(#1,#1)[#3]}
}
\gdef\Nphoton{  % VERTICAL PHOTONS
\ifcase\LINECONFIGURATION  %\REG case
\setcoords(-490,-250,0)(260,1250,0)[0,2000]
\gdef\upperunitbox{\SLbr}   \gdef\lowerunitbox{\SRbr}
\particleadjusty=10
\or % \FLIPPED case
\setcoords(-271,-501,0)(250,1250,0)[0,2000]   
\gdef\upperunitbox{\SRbr}   \gdef\lowerunitbox{\SLbr}
\or %\CURLY case (a bit shorter).
\particleadjusty=0
\setcoords(-501,-351,0)(300,1400,0)[0,2200]
\gdef\upperunitbox{\Lbr}   \gdef\lowerunitbox{\Rbr}
\or %\FLIPPEDCURLY case.
\setcoords(-353,-499,0)(300,1400,0)[0,2200]
\gdef\upperunitbox{\Rbr}   \gdef\lowerunitbox{\Lbr}
\or % \FLAT case.  Flatter and shorter than \CURLY.
\setcoords(-481,-371,0)(280,1300,0)[0,2000]
\gdef\upperunitbox{\Lbr}   \gdef\lowerunitbox{\Rbr}
\particleadjusty=150
\ifnum\numwiggles=\number\numwigglespo \particleadjustx=-50 \fi
\or %\FLIPPEDFLAT case.  \LINECONFIGURATION=5.
\setcoords(-321,-391,0)(280,1300,0)[0,2000]
\gdef\upperunitbox{\Rbr}   \gdef\lowerunitbox{\Lbr}
\particleadjusty=150
\ifnum\numwiggles=\number\numwigglespo \particleadjustx=80 \fi
\or % \LONGPHOTON
\setcoords(-490,-260,0)(300,1500,0)[0,2400]
\gdef\upperunitbox{\Lbr}   \gdef\lowerunitbox{\Rbr}
\or % \FLIPPEDLONGPHOTON
\setcoords(-301,-531,0)(300,1500,0)[0,2400]
\gdef\upperunitbox{\Rbr}   \gdef\lowerunitbox{\Lbr}
\else \UNIMPERROR
\fi
}
\gdef\NEphoton{    % NE   SLANTED PHOTONS:  RE-ORDERED IN PHOTONSETUP27
\ifcase\LINECONFIGURATION  %\REG case
\setcoords(425,425,0)(1250,0,0)[1250,1250]       \SETUNITBOX(1250)[br][tl]  
\ifnum\numwigglespo > \number \numwiggles \particleadjustx=15 \fi
\or % \FLIPPED case
\setcoords(1050,-200,0)(625,625,0)[1250,1250]    \SETUNITBOX(1250)[tl][br]
\ifnum\numwigglespo > \number \numwiggles \particleadjustx=25 \fi
\or % \CURLY case.
\setcoords(500,500,0)(1400,0,0)[1400,1400]       \SETUNITBOX(1400)[br][tl]
\or % \FLIPPEDCURLY case
\setcoords(1200,-200,0)(700,700,0)[1400,1400]    \SETUNITBOX(1400)[tl][br]  
\or % \FLAT case
\setcoords(400,400,0)(1200,0,0)[1200,1200]       \SETUNITBOX(1200)[br][tl]  
\or % \FLIPPEDFLAT case
\setcoords(1000,-200,0)(600,600,0)[1200,1200]    \SETUNITBOX(1200)[tl][br]
\else \UNIMPERROR
\fi
\numupperunits=\numwiggles   \numlowerunits=\numwigglespo
}
\gdef\Ephoton{    %  EASTWARD  HORIZONTAL PHOTONS
\ifcase\LINECONFIGURATION  % REG case
\setcoords(-285,715,0)(-150,-400,0)[2005,0]
\gdef\upperunitbox{\Frown}   \gdef\lowerunitbox{\Smile}
\or  % \FLIPPED case
\setcoords(-285,715,0)(-420,-170,0)[2005,0]
\gdef\upperunitbox{\Smile}   \gdef\lowerunitbox{\Frown}
\else \UNIMPERROR
\fi
\particleadjustx=-15 % Lengths are in centipoints.
}
\gdef\SEphoton{   % SE   SLANTED PHOTONS:  RE-ORDERED IN PHOTONSETUP27
\ifcase\LINECONFIGURATION  %\REG case
\setcoords(-200,1050,0)(-625,-625,0)[1250,-1250] \SETUNITBOX(1250)[tr][bl]
\ifnum\numwigglespo > \number \numwiggles \particleadjustx=25 \fi
\or % \FLIPPED case
\setcoords(425,425,0)(0,-1250,0)[1250,-1250]     \SETUNITBOX(1250)[bl][tr]
\ifnum\numwigglespo > \number \numwiggles \particleadjustx=15 \fi
\or % \CURLY case.
\setcoords(-200,1200,0)(-700,-700,0)[1400,-1400] \SETUNITBOX(1400)[tr][bl]  
\or % \FLIPPEDCURLY case
\setcoords(500,500,0)(0,-1400,0)[1400,-1400]     \SETUNITBOX(1400)[bl][tr]  
\or % \FLAT case
\setcoords(-200,1000,0)(-600,-600,0)[1200,-1200] \SETUNITBOX(1200)[tr][bl]
\particleadjustx=-20
\or % \FLIPPEDFLAT case
\setcoords(420,420,0)(0,-1200,0)[1200,-1200]     \SETUNITBOX(1200)[bl][tr]
\particleadjustx=40
\else \UNIMPERROR
\fi
}
\gdef\Sphoton{  % DOWN, DOWN VERTICAL PHOTONS
\ifcase\LINECONFIGURATION  %\REG case
\setcoords(-252,-490,0)(-740,-1740,0)[0,-2000]
\gdef\upperunitbox{\SRbr}   \gdef\lowerunitbox{\SLbr}
\or % \FLIPPED case
\setcoords(-490,-260,0)(-740,-1740,0)[0,-2002]
\gdef\upperunitbox{\SLbr}   \gdef\lowerunitbox{\SRbr}
\or %\CURLY case (a bit shorter).
\setcoords(-299,-449,0)(-870,-1970,0)[0,-2200]
\gdef\upperunitbox{\Rbr}    \gdef\lowerunitbox{\Lbr}
\particleadjusty=-95
\or %\FLIPPEDCURLY case.
\setcoords(-517,-371,0)(-900,-2000,0)[0,-2200]
\gdef\upperunitbox{\Lbr}    \gdef\lowerunitbox{\Rbr}
\particleadjusty=-165
\or % \FLAT case.  Flatter and shorter than \CURLY.  \LINECONFIGURATION=4.
\setcoords(-299,-409,0)(-885,-1905,0)[0,-2000]
\gdef\upperunitbox{\Rbr}   \gdef\lowerunitbox{\Lbr}
\particleadjustx=50     \particleadjusty=-380
\ifodd\unitboxnumber\relax\else\particleadjustx=250 \particleadjusty=-400 \fi
\or %\FLIPPEDFLAT case.  \LINECONFIGURATION=5.
\setcoords(-519,-449,0)(-900,-1920,0)[0,-2000]
\gdef\upperunitbox{\Lbr}   \gdef\lowerunitbox{\Rbr}
\particleadjusty=-370
\ifodd\unitboxnumber\relax\else\particleadjustx=-240 \particleadjusty=-400 \fi
\or % \LONGPHOTON
\gdef\upperunitbox{\Rbr}   \gdef\lowerunitbox{\Lbr}
\setcoords(-325,-555,0)(-900,-2100,0)[0,-2400]
\particleadjusty=-40
\or % \FLIPPEDLONG
\setcoords(-505,-275,0)(-900,-2100,0)[0,-2400]
\gdef\upperunitbox{\Lbr}   \gdef\lowerunitbox{\Rbr}
\particleadjusty=-30  % Lengths are in centipoints.
\else \UNIMPERROR
\fi
}
\gdef\SWphoton{  % SW SLANTED PHOTONS:  RE-ORDERED IN PHOTONSETUP27
\ifcase\LINECONFIGURATION  %\REG case
\setcoords(-825,-825,0)(0,-1250,0)[-1250,-1250]     \SETUNITBOX(1250)[br][tl]  
\or % \FLIPPED case
\setcoords(-175,-1425,0)(-625,-625,0)[-1250,-1250]  \SETUNITBOX(1250)[tl][br]  
\or % \CURLY case.
\setcoords(-900,-900,0)(0,-1410,0)[-1400,-1400]     \SETUNITBOX(1400)[br][tl]  
\or % \FLIPPEDCURLY case
\setcoords(-200,-1600,0)(-700,-700,0)[-1400,-1400]  \SETUNITBOX(1400)[tl][br]  
\or % \FLAT case
\setcoords(-800,-800,0)(0,-1200,0)[-1200,-1200]     \SETUNITBOX(1200)[br][tl]  
\or % \FLIPPEDFLAT case
\setcoords(-200,-1400,0)(-600,-600,0)[-1200,-1200]  \SETUNITBOX(1200)[tl][br]  
\else \UNIMPERROR
\fi
}
\gdef\Wphoton{
\ifcase\LINECONFIGURATION %\REG case
\setcoords(-2245,-1245,0)(-150,-400,0)[-2005,0]
\gdef\upperunitbox{\Frown}   \gdef\lowerunitbox{\Smile}
\or % \FLIPPED case
\setcoords(-2245,-1245,0)(-400,-150,0)[-2005,0]
\gdef\upperunitbox{\Smile}   \gdef\lowerunitbox{\Frown}
\else \UNIMPERROR
\fi
\particleadjustx=57 % Lengths are in centipoints.
\ifnum\numwigglespo=\number\numwiggles \particleadjustx=0  \fi
\numlowerunits=\numwigglespo   \numupperunits=\numwiggles
}
\gdef\NWphoton{  % NW   SLANTED PHOTONS:  RE-ORDERED IN PHOTONSETUP27
\ifcase\LINECONFIGURATION  %\REG case
\setcoords(-200,-1425,0)(625,625,0)[-1250,1250]   \SETUNITBOX(1250)[bl][tr]
\or % \FLIPPED case
\setcoords(-825,-825,0)(0,1250,0)[-1250,1250]     \SETUNITBOX(1250)[tr][bl]
\ifnum\numwigglespo > \number \numwiggles \particleadjusty=-15 \fi
\or % \CURLY case.
\setcoords(-200,-1600,0)(700,700,0)[-1400,1400]   \SETUNITBOX(1400)[bl][tr]
\or % \FLIPPEDCURLY case
\setcoords(-900,-900,0)(0,1400,0)[-1400,1400]     \SETUNITBOX(1400)[tr][bl]
\or % \FLAT case.
\setcoords(-200,-1400,0)(600,600,0)[-1200,1200]   \SETUNITBOX(1200)[bl][tr]  
\or % \FLIPPEDFLAT case
\setcoords(-800,-800,0)(0,1200,0)[-1200,1200]     \SETUNITBOX(1200)[tr][bl]  
\else \UNIMPERROR
\fi
}
  \fi
\selectphoton
}
\gdef\selectgluon{   % RECURSIVELY RE-DEFINED IN GLUONSETUP(25+).TEX.
\ifnum\gluoncount=0 \input GLUONS.tex  \fi
\selectgluon
}
\gdef\selectespecial{\UNIMPERROR}
\gdef\checkvertex{ %immediately re-defines \drawvertex,\vertexcap,\linkvertex...
\ifnum\vertexcount=-1   \input VERTEX  \fi}
\gdef\drawvertex#1[#2#3](#4,#5)[#6]{\checkvertex\drawvertex#1[#2#3](#4,#5)[#6]}
\gdef\vertexcap#1{\checkvertex\vertexcap#1}
\gdef\vertexcaps{\checkvertex\vertexcaps}
\gdef\vertexlink#1{\checkvertex\vertexlink#1}
\gdef\vertexlinks{\checkvertex\vertexlinks}
\gdef\stemvertex#1{\checkvertex\stemvertex#1}
\gdef\stemvertices{\checkvertex\stemvertices}
\gdef\flipvertex{\checkvertex\flipvertex}
\global\arrowlength=349  % Length of arrow
\gdef\drawarrow[#1#2](#3,#4){
\global\LDIR=#1
\SETDIR
\global\boxlengthx=#3  % Just a convenient variable name.  No significance.
\global\boxlengthy=#4  % The Arrow co-ordinates.
\ifdim#2=1pt  % CASE \ATBASE WHERE THE CO-ORDS ARE AT THE ARROWS BASE.
\adjx=\arrowlength      \adjy=\arrowlength
\multiply\adjx by \XDIR \multiply\adjy by \YDIR  % Set in \SETDIR
\slanttest(\adjx,\adjy)
\global\advance\boxlengthx by \adjx    \global\advance\boxlengthy by \adjy
\fi
\ifnum\phantomswitch=0\put(\boxlengthx,\boxlengthy){\vector(\XDIR,\YDIR){0}}\fi
}  % END OF \drawarrow.
\gdef\SETFRONTSTEM{
\EITHERSTEM=\FRONTSTEM   \advance\EITHERSTEM by \BACKSTEM
\ifdim\EITHERSTEM>0pt
\global\stemlengthx=\stemlength   \global\stemlengthy=\stemlength   
\global\absstemlength=\stemlength   
\SETDIR
\gslanttest(\stemlengthx,\stemlengthy)
\gslanttest(\absstemlength,\REG)  % the \REG is to use up the parameter space.
\ifnum\XDIR=0 \stemlengthx=0 \fi
\ifnum\YDIR=0 \stemlengthy=0 \fi
\global\multiply\stemlengthx by \XDIR
\global\multiply\stemlengthy by \YDIR
\ifdim\FRONTSTEM=1pt 
\ifnum\phantomswitch=0
	  \put(\pfrontx,\pfronty){\line(\XDIR,\YDIR){\absstemlength}}\fi
\global\advance\plengthx by \stemlengthx
\global\advance\plengthy by \stemlengthy
\global\advance\PFRONTx by \stemlengthx   
\global\advance\PFRONTy by \stemlengthy
\global\advance\pmidx by \stemlengthx
\global\advance\pmidy by \stemlengthy
\global\advance\pbackx by \stemlengthx
\global\advance\pbacky by \stemlengthy
\ifnum\LTYPE=3
\global\photonfrontx=\PFRONTx  \global\photonfronty=\PFRONTy
\global\photonbackx=\pbackx    \global\photonbacky=\pbacky
\fi  % END LTYPE
\ifnum\LTYPE=4
\global\gluonfrontx=\PFRONTx  \global\gluonfronty=\PFRONTy
\global\gluonbackx=\pbackx    \global\gluonbacky=\pbacky
\fi  % END LTYPE
\fi  % END FRONTSTEM
\fi  % END EITHERSTEM
}    % end of \SETFRONTSTEM
\gdef\SETBACKSTEM{
\ifdim\BACKSTEM=1pt 
\ifnum\phantomswitch=0
       \put(\pbackx,\pbacky){\line(\XDIR,\YDIR){\absstemlength}}\fi
\global\advance\plengthx by \stemlengthx
\global\advance\plengthy by \stemlengthy
\global\advance\pbackx by \stemlengthx
\global\advance\pbacky by \stemlengthy
\fi  % END BACKSTEM
\global\stemlength=275  \FRONTSTEM=0pt  \BACKSTEM=0pt % Reset default switches.
}    % END OF \SETBACKSTEM 
\gdef\drawloop#1[#2#3](#4,#5){  %RECURSIVE.  
\input LOOPS  % contains loops definitions
\drawloop#1[#2#3](#4,#5)}
\Feynmanlength  % Set length scale to centipoints.

\def\cal{\fam2 }
\newcommand{\ba}{\begin{eqnarray}}
\newcommand{\re}{{\rm Re\,}}
\newcommand{\im}{{\rm Im\,}}
\newcommand{\ea}{\end{eqnarray}}
\newcommand{\be}{\begin{equation}}
\newcommand{\ee}{\end{equation}}
\newcommand{\ie}{{i.e.,\  }}
\newcommand{\eq}[1]{Eq.\,(\ref{#1})}
\newcommand{\fig}[1]{Fig.\,\ref{#1}}
\newcommand{\chii}{\chi_{{}_{\rm I}}}

\newcommand{\pbar}{\bar p}

\newcommand{\alphabold}{\mbox{\small\boldmath $\alpha$}}
\newcommand{\xbold}{\mbox{\boldmath $x$}}

\newcommand{\delchisq}{\Delta \chi^2_i(x_i;\alphabold)}
\newcommand{\delchi}{\Delta \chi^2_i}
\newcommand{\delchisqmax}{{\Delta \chi^2_i(x_i;\alphabold)}_{\rm max}}
\newcommand{\delchimax}{{\delchi}_{\rm max}}
\newcommand{\x}{(\nu/m)}
\newcommand{\y}{(\nu_0/m)}
\newcommand{\sigtot}{\sigma_{\rm tot}}
\newcommand{\F}{{\cal F}_1}
\newcommand{\pbarp}{\mbox{$\bar p p$\ }}
\newcommand{\calF}{\mbox{${\cal F}$}}
\newcommand{\calG}{\mbox{${\cal G}$}}
\newcommand{\ppbar}{\mbox{$p\bar p$\ }}
\newcommand{\betaP}{\beta_{\cal P'}}

\newcommand{\rnd}{{\rm RND}}
\def\stot{\sigma_{\rm tot}}
\newcommand{\Em}[1]{{\em {#1}}} 
\def\bea{\begin{eqnarray}} 
\def\eea{\end{eqnarray}} 
\def\rd{{\mathrm d}} 
 
\def\intd4x{\int{\rd}^4x}

\def\m32{{m_{3/2}}}

\newcommand{\la}{\raisebox{-.8ex}{\,$\stackrel{\textstyle <}{\sim}$}\,} 
 
\newcommand{\sumovern}{\raisebox{-1.8ex}{$\stackrel{\displaystyle \sum}{\scriptstyle n}$}}

\def\signn{\sigma_{\rm nn}}
\def\siggp{\sigma_{\gamma\rm p}}
\def\siggg{\sigma_{\gamma\gamma}}
\def\rhonn{\rho_{\rm nn}}
\def\rhogp{\rho_{\gamma\rm p}}
\def\rhogg{\rho_{\gamma\gamma}}
\def\Bnn{B_{\rm nn}}
\def\Bgp{B_{\gamma\rm p}}
\def\Bgg{B_{\gamma\gamma}}       
\def\lpa{\lambda_{p{-}\rm air}}
\def\spa{\sigma_{p{-}\rm air}}
\def\spai{\sigma_{p{-}\rm air}^{\rm prod}}

\def\spae{\sigma_{p{-}\rm air}^{\rm el}}
\def\spaqe{\sigma_{p{-}\rm air}^{q{-}\rm el}}
\begin{document}
\renewcommand\thepage{\ }
\begin{titlepage} % This environment allows me to set up an "empty"
          % page. I must fill it with \title, \abstract and \date.
     % Make the NU logo here.
     % We make a box of the perfect size for the current report number.
%
     % We must ENTER Report Number and DATE below.
\newcommand\mydate{\today} %Set date up in logo.
\newlength{\nulogo} % define a new length "nulogo".
     % Use  "\settowidth" to get correct width of box, called "nulogo".
\settowidth{\nulogo}{\small\sf{NUHEP Report XXXX}}
\title{
\vspace{-.8in} % Puts everything toward top of page
\hfill
         {\small\sf \mydate}
\vspace{0.5in} \\
          % The "\fbox" puts a BOX around text and the
          % "\small\sf" sets it in small, sans serif type,
          % the "\parbox{nulogo}{} makes a paragraph box, of correct width,
          % the "\fbox makes a box around the "parbox", and finally,
          % the "\hfill" pushes it to right.
          % Finally, the "\vspace{1in}" sets the NEXT line down by one inch.
% We now continue to put in MAIN title.
{
Hadronic forward scattering: Predictions for the Large Hadron Collider and cosmic rays 
}}

\author{
M.~M.~Block\\
{\small\em Department of Physics and Astronomy,} \vspace{-5pt} \\ %Make
    %smaller separation between lines.
{\small\em Northwestern University, Evanston, IL 60208}\\
\vspace{-5pt}
\  \\
\vspace{-5pt}\\
   % The negative \vspace decreases the space between last line.
%
\vspace{-5pt}\\
   % The negative \vspace decreases the space between last line.
%
}    %    End of title section.
\vspace{.5in}
\vfill
     %    End of title section.
     % Command to print title.
%\date{} % This stops printing of date.
\date {}%{April 23, 2000}
\maketitle
\vspace{-.2in}
\begin{abstract}
%\vskip-5ex %% reduce space between "Abstract" and it's text
The  status of hadron-hadron interactions is reviewed, with emphasis on the forward and near-forward scattering regions. Using unitarity, the optical theorem is derived. Analyticity and crossing symmetry, along with integral dispersion relations, are used to connect particle-particle and antiparticle-particle total cross sections and $\rho$-values, e.g.,  $\sigma_{pp}$, $\sigma_{\pbar p}$, $\rho_{pp}$ and $\rho_{\bar pp}$, where $\rho$ is the ratio of the real to the imaginary portion of the forward scattering amplitude.  Real analytic amplitudes are then introduced to exploit analyticity and crossing symmetry.  Again, from analyticity,  Finite Energy Sum Rules (FESRs) are introduced from which new analyticity constraints are derived.   These new analyticity conditions exploit the many very accurate low energy experimental cross sections, i.e., they constrain the values of the asymptotic cross sections and their derivatives at low energies just above the resonance regions, allowing us new insights into duality. A robust fitting technique---using a minimization of the Lorentzian squared followed by the ``Sieve'' algorithm---is introduced in order to `clean up' large data samples that are contaminated by outliers, allowing us to make much better fits to hadron-hadron scattering data over very  large regions of energy. Experimental evidence for factorization theorems for $\gamma\gamma$, $\gamma p$ and nucleon-nucleon collisions is presented.  The Froissart bound is discussed---what do we mean here by the saturation of the Froissart bound? Using our analyticity constraints, new methods of fitting high energy hadronic data are introduced which result in much more precise estimates of the fit parameters, allowing accurate extrapolations to much higher energies.   It's shown that the $\gamma p$, $\pi^\pm p$ and nucleon-nucleon cross sections {\em all} go asymptotically as $\ln^2s$,  saturating the bound, while conclusively ruling out $\ln s$ and $s^{\alpha}$ ($\alpha\sim0.08$) behavior.  Implications of this saturation for predictions of $\sigma_{pp}$ and $\rho_{pp}$ at the LHC and for cosmic rays are given. We discuss present day cosmic ray measurements, what they measure and how they deduce  $p$-air cross sections. Connections are made between very high energy measurements of $\sigma_{\rm p-air}^{\rm prod}$---which have rather poor energy determination---and  predictions of $\spai$ obtained, using a Glauber model, from values of $\sigma_{pp}$  that are extrapolated from fits of  accelerator data at very precisely known, albeit lower, energies.  
\end{abstract}
\end{titlepage} % End of titlepage environment.
% main text
%\pagenumbering{roman}
\renewcommand{\thepage}{\arabic{page}}  
{\small \tableofcontents} \newpage
{\small \listoffigures} \newpage
{\small \listoftables} \newpage
%\pagenumbering{arabic}
\setcounter{tocdepth}{4}
%\setcounter{tocdepth}{3}
%\setcounter{secnumdepth}{3}
%    Text starts here.
\section{Introduction}
In the last 20 years, high energy $\pbar p$ colliders have extended the maximum $\pbar p$ c.m. (center-of-mass) energy  from $\sqrt s\sim 20$ GeV to $\sqrt s\sim2000$ GeV. Further, during this period, the maximum c.m. energy for $\pi p$ scattering has gone to $\sim 35$ GeV, whereas the top c.m. energy for $\gamma p$ collisions is now $\sim 200$ GeV and about $\sim 180$ GeV for $\gamma \gamma$ collisions.  All of these total cross sections rise with energy.  Up until recently, it has not been clear whether they rose as $\ln s$ or as $\ln^2 s$ as $s \rightarrow\infty$. The latter would  saturate the Froissart bound, which tells us  that hadron-hadron cross sections should be bounded by $\sigma\sim \ln^2 s$. This fundamental result is derived from unitarity and analyticity by Froissart\cite{froissart}, who states: 
\begin{itemize}
\item[]``At forward or backward angles, the modulus of the amplitude behaves at most like $s\ln^2s$, as $s$ goes to infinity.  We can use the optical theorem to derive that the total cross sections behave at most like $\ln^2s$, as $s$ goes to infinity".
\end{itemize}  
In this context, saturating the Froissart bound refers to an energy dependence of the total  cross section rising asymptotically as $\ln^2s$. 

If the Froissart bound is saturated, we know the high energy dependence of hadron-hadron interactions---it gives us a important tool to use in constraining the extrapolation of present day accelerator data to much higher energies. In a few years, the CERN Large Hadron Collider (LHC) will take $pp$ collisions up to $\sqrt s=14$ TeV. Cosmic ray experiments now under way  will extend the energy frontiers enormously. The HiRes experiment is currently exploring $p$-air collisions up to $\sqrt s\approx 80$ TeV, and the Pierre Auger collaboration is also planning to  measure $p$-air cross sections in this energy range. This is  indeed an exciting era in high energy hadron-hadron collisions.

Often following the path of the 1985 review of Block and Cahn\cite{bc} and almost always using their notation, we will make a thorough review of some of the fundamental tenets of modern physics, including unitarity, analyticity and crossing symmetry, in order to derive the necessary tools to understanding dispersion relations, finite energy sum rules and real analytic amplitudes. These theorems are needed in fitting high energy hadron-hadron scattering. Building on them, we will use these tools to derive new analyticity constraints for hadron-hadron scattering, constraints that exploit the large amount of accurate low energy hadron-hadron experimental cross sections by anchoring high energy cross section fits to their values. These analyticity constraints will allow us to understand duality in a new way.

En route, we will make a brief discussion of phase space, going from Fermi's ``Golden Rule'' to modern Lorentz invariant phase space, reserving details for the Appendices, where we will discuss multi-body phase space and some computing techniques needed to evaluate them.

Next, our attention  will be turned to actual data fitting techniques.   After reviewing maximum likelihood techniques, the concept of robust fitting will be introduced. The ``Sieve'' algorithm  will be introduced, to rid ourselves of annoying `outliers' which skew $\chi^2$ fitting techniques and give huge total $\chi^2$, making  error assignments and goodness-of-fit problematical.  We will show how to make a `sifted' set of data where  outliers have been eliminated, as well as how to modify fitting algorithms in order to make a robust fit to the original data, including goodness-of-fit and error estimates.

A QCD-inspired eikonal model called the Aspen model will be introduced, whose parameters will be determined using the new analyticity constraints we have derived.  We will then exploit the richness of the eikonal to allow us to predict $\sigtot,\ \sigma_{\rm el}$, the $\rho$-value (the ratio of the real to the imaginary portion of the forward scattering amplitude), the nuclear slope parameter $B$, the survival probability of large rapidity gaps, as well as the differential elastic scattering $d\sigma/dt$ as a function of $|t|$, at the Tevatron ($\sqrt s=1.8$ TeV), the LHC ($\sqrt s=14$ TeV) and at cosmic ray energies ($\sqrt s\approx 80$ TeV).  Using our new analyticity constraints, the Aspen model parameters are obtained by fitting accelerator $pp$ and $\bar pp$ data for $\sigtot, \ \rho$ and $B$.

A detailed discussion of the factorization properties of the Aspen model eikonal is made, allowing numerical comparisons of nucleon-nucleon, $\gamma p$ and $\gamma\gamma$ scattering, by using the additive quark model and vector dominance as input. 

Methods  will then be introduced for fitting high energy cross section data using real analytic amplitudes, where the fits are anchored at low energy by our new analyticity constraints.  These take advantage of the  prolific amount of very accurate low energy experimental cross section data in  constraining  high energy parametrizations at energies slightly above the resonance regions.  Using analyticity constraints, the $\gamma p$,  $\pi^+p$ and $\pi^-p$, and the $pp$ and $\pbarp$\ systems will be fit. These new techniques---a sifted data set and the imposition of the new analytic constraints---will be shown to produce much smaller errors of the fit parameters, and consequently, much more accurate cross section and $\rho$-value predictions when extrapolated to ultra-high energies.

Further, these techniques  completely rule out $\ln s$ fits statistically, for the first time. Also, they give new and very restrictive limits on `odderons'---unconventional odd amplitudes that do not vanish with increasing energy. Further, popular high energy fits of the form $s^{\alpha}$, where $\alpha\approx0.08$ are shown to be deficient when the new analyticity requirements are satisfied.

Using a $\ln^2s$ fit for $pp$ scattering, we will make the predictions  that $\rho=0.132\pm0.001$ and $\sigma_{pp}=107.3\pm1.2$ mb at the LHC collider. 

Finally, a detailed discussion of the cosmic ray measurements of the $p$-air cross section at ultra-high energies will be made, including the very recent HiRes measurement. Since the Froissart bound has been shown to be saturated, we will make a $\ln^2s$ fit to accelerator data to produce accurate $pp$ cross section predictions at ultra-high energies, $\sqrt s\approx 80$ TeV.  Using a Glauber calculation requiring a knowledge of the nuclear slope parameter $B(s)$ and  $\sigma_{pp}(s)$ at these  energies, we will convert our extrapolated $pp$ total cross sections at cosmic ray energies into $p$-air particle production cross sections, making possible comparisons with cosmic ray experiments, tying together measurements from colliders to cosmic rays (C2CR).

%%%%%%%%%%%%%%%%%%%%%%%%%%%%%%%%%%%%%%%%%%%%%%%%%%%%%%%%%%%%%%%%%%%%%
\section{Scattering amplitude and kinematics}
We will consider here elastic scattering of $a+b\rightarrow a+b$ and $\bar a +b\rightarrow \bar a+b$, where the initial state projectile 4-momentum of $a$  ($\bar a$) is  $p_1$ and  the initial state target 4-momentum of $b$ is $p_2$, and where the final state   4-momentum  of $a$  ($\bar a$) is $p_3$ and of $b$ is $p_4$. 
%%%%%%%%%%%%%%%%%%%%%%%%%%%%%%%%%%%%%%%%%%%%%
\subsection{Kinematics}\label{sec:kinematics}
The Mandelstam invariant $s$, the square of the c.m. (center of mass system) energy, is given by
\be
s\equiv(p_1+p_2)^2=m_1^2+m_2^2+2\left(\sqrt{k^2+m_1^2}\sqrt{k^2+m_2^2}+k^2\right), \label{scms}
\ee
where $k$ is the magnitude of the c.m. 3-momentum $\vec k$. For $pp$ ($\pbar p$) scattering, $m_1=m_2=m$, the proton mass, and for $\pi p$ scattering, $m_1=m_\pi$, $m_2=m$.  We find, using c.m. variables, that
\ba
s_{pp}&=&4(k^2+m^2)\label{sppcms}\\
s_{\pi p}&=& m^2+m_\pi^2+2\left(\sqrt{k^2+m^2}\sqrt{k^2+m_{\pi}^2}+k^2\right)\label{spipcms}
\ea
and introducing the laboratory momentum $p$ and laboratory energy $E=\left(p^2+m_1^2\right)^{1/2}$, we find
\ba
s_{pp}&=&2\left(m^2+mE)\right)\label{spplab}\\
s_{\pi p}&=&m^2+m_{\pi}^2+2mE \label{spiplab}.
\ea
The invariant 4-momentum transfer squared $t$ is given by
\be
t\equiv(p_1-p_3)^2=-4k^2\sin^2\frac{\theta}{2},\label{t}
\ee
where $\theta$ is the c.m. scattering angle. 
The third Mandelstam invariant $u$ is given by
\be
u=(p_1-p_4)^2,\label{u}
\ee
and we have
\ba
s_{pp}+t_{pp}+u_{pp}&=&4m^2\label{upp}\\
s_{\pi p}+t_{\pi p}+u_{\pi p}&=&2m_\pi^2+2m^2\label{upip}.
\ea
\subsection{Scattering amplitude conventions}

We will use units where $\hbar=c=1$, throughout this work. We now introduce  elastic scattering amplitudes with various normalizations. 

The c.m. amplitude $f_{\rm c.m.}$ is given by
\ba
\frac{d\sigma}{d\Omega_{\rm c.m.}}&=&\left |f_{\rm c.m.}\right |^2,\label{cms1}\\
\frac{d\sigma}{dt}&=&\frac{\pi}{k^2}\,\left |f_{\rm c.m.}\right |^2,\label{cms2}\\
\sigma_{\rm tot}&=&\frac{4\pi}{k}\,\im f_{\rm c.m.}(\theta=0)\label{cmsampl}.
\ea

The laboratory scattering amplitude, $f$, is given by
\ba
\frac{d\sigma}{d\Omega_{\rm lab}}&=&\left |f\right |^2,\\
\frac{d\sigma}{dt}&=&\frac{\pi}{p^2}\,\left |f\right |^2,\\
\sigma_{\rm tot}&=&\frac{4\pi}{p}\,\im f(\theta_{\rm L}=0)\label{labampl},
\ea
where $\theta_{\rm L}$ is the laboratory scattering angle.

The Lorentz-invariant amplitude ${\cal M}$ is related to the laboratory scattering amplitude $f$ for the nucleon-nucleon system by
\ba
\cal M&=& -8\pi\sqrt s(k/p) f\\
&=&-8\pi mf.
\ea
Thus, we find 
\ba
\sigma_{\rm tot}&=& -\frac{1}{2pm}\im {\cal M}(t=0)\\
&=&-\frac{1}{2k\sqrt s }\im{\cal M}(t=0).
\ea

Lastly, we introduce the amplitude $F$, with the properties
\ba
\frac {d\sigma}{dt}&=& \left|F\right|^2,\label{F1}\\
\sigtot&=&4\sqrt \pi\, \im F(t=0)\label{F2}.
\ea

The elastic scattering amplitudes are related by 
\be
f=\frac{p}{k}f_{\rm c.m.}=\frac{p}{\sqrt \pi}F=-\frac{1}{8\pi m}{\cal M},
\ee
and they are interchangeably introduced whenever convenient to the discussion.
%%%%%%%%%%%%%%%%%%%%%%%%%%%%%%%%%%%%%%%%
%%%%%%%%%%%%%%%%%%%%%%%%%%%%%%%%%%%%%%%%%%%%5
\section{Theory of $pp$ and $\pbar p$ elastic hadronic  scattering in the presence of the Coulomb field} 

The interference at small $|t|$ of the Coulomb scattering amplitude
$f_c$
and the nuclear amplitude $f_n$ is used to measure the phase of the nuclear
scattering 
amplitude, and hence the $\rho$-value, where $\rho\equiv \left(\re 
f_n/\im f_n\right)_{t=0}$. 
The ``normal'' analysis of $\pbar p$ and $pp$ elastic scattering uses 
a `spinless'  
Coulomb amplitude, \ie a Rutherford amplitude---$2\sqrt{\pi}\alpha/t$---multiplied by a Coulomb  
form factor $G^2(t)$. This   conventional {\em ansatz}  that neglects  any magnetic  
scattering and spin effects is used by all experimenters  .

We will only calculate electromagnetic amplitudes accurate to order $\alpha$, 
\ie the one-photon exchange diagram shown in Fig. \ref{withkappa}. Further, we will consider 
only high energy 
scattering ($E_{\rm lab} \gg m$, where  $m$ is the nucleon mass) in the region 
of small $|t|$, 
where $t$ is the squared 4-momentum transfer. We will measure $m$ and  $E_{\rm 
lab}$ in GeV and 
$t$ in (GeV)$^2$, and will use $\hbar=c=1$.  
\subsection{`Spinless' Coulomb scattering}  
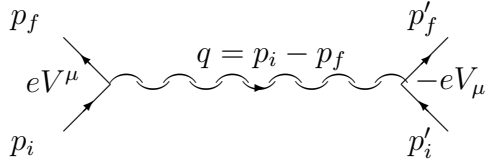
\begin{figure}[htb] 
\begin{picture}(30000,10000) 
\THINLINES 
\drawline\fermion[\NW\REG](15000,5000)[2500] 
\drawarrow[\NW\ATBASE](\pmidx,\pmidy)%
\advance\pbacky by 500  
\advance\pbackx by -2000 
\put(\pbackx,\pbacky){$p_f$}%
\drawline\fermion[\SW\REG](15000,5000)[2500] 
\advance\pbackx by -2000 
\put(\pbackx,2500){$p_i$}%
\drawarrow[\NE\ATBASE](\pmidx,\pmidy)%
\drawline\photon[\E\REG](15000,5000)[11] 
\advance\pfrontx by -3200 %
\advance\pfronty by -300 
\put(\pfrontx,\pfronty){$eV^\mu$} 
\advance\pbackx by 500 
\advance\pbacky by -200 
\put(\pbackx,\pbacky){$-eV_\mu$} 
\advance\pmidy by -250 
\advance\pmidy by 100 
\drawarrow[\E\ATBASE](\pmidx,\pmidy)%
\advance\pmidx by -2250 
\advance\pmidy by 1000 
\put(\pmidx,\pmidy){$q=p_i-p_f$} 
\drawline\fermion[\NE\REG](\photonbackx,\photonbacky)[2500] 
\advance\pbacky by 500  
\advance\pbackx by -1500 
\put(\pbackx,\pbacky){$p'_f$} 
\drawarrow[\NE\ATBASE](\pmidx,\pmidy)%
\drawline\fermion[\SE\REG](\photonbackx,\photonbacky)[2500] 
\advance\pbackx by -1500 
\put(\pbackx,2500){$p'_{i}$}%
\drawarrow[\NW\ATBASE](\pmidx,\pmidy)%
\end{picture} 
\caption [Feynman diagram for $p\bar p$ Coulomb scattering] {\protect{\footnotesize {One-Photon Feynman diagram for $p\bar p$ Coulomb scattering,  
$p_i +p'_i\rightarrow p_f+p'_f$, with couplings 
$eV^\mu$ and $-eV_\mu$.}}}\label{withkappa} 
\end{figure}

If we consider `spinless' proton-antiproton Coulomb scattering, the relevant 
Feynman diagram  
is shown in Fig. \ref{withkappa}, with $V^\mu=G\left(p_i+p_f\right)^\mu$ and $G(t)$ is the 
electromagnetic charge form factor of the nucleon.
The electromagnetic differential cross section is readily evaluated as
\begin{eqnarray} 
\frac{d\sigma}{dt}&=& 
\pi 
\left| 
\frac{\mp 2G^2(t)\alpha}{\beta_{\rm lab}|t|}\times 
\left(1-\frac{|t|}{4mE_{\rm lab}}\right)\right|^2, 
\label{spin0answer} 
\end{eqnarray} 
where the upper (lower) sign is for like (unlike) charges, $t$ is the 
(negative) 4-momentum  
transfer squared, and $m$ is the nucleon mass.

For small angle scattering and at high energies, the correction term  
$\frac{|t|}{2mE_{\rm lab}}$ becomes  negligible and $\beta_{\rm lab} 
\rightarrow 1$, so 
\eq{spin0answer}  goes over into the well-known Rutherford  
scattering formula, 
\be 
\frac{d\sigma}{dt} 
=\pi\left| 
\frac{\mp 2\alpha G^2(t)} 
{|t|}\right|^2 
,\label{rutherford} 
\ee 
where the electromagnetic charge form factor $G(t)$ is commonly parametrized 
by the dipole form 
\be 
G(t)=\frac{1}{\left(1-\frac{t}{\Lambda^2}\right)^2},\label{G} 
\ee 
where $\Lambda^2=0.71$ (GeV)$^2$ when $t$ is measured in (GeV)$^2$. 
We note that this is the Coulomb amplitude that is normally used in experimental analyses of $\pbar p$ and $pp$ elastic scattering, \ie  the `spinless' analysis\cite{bc}.
Thus, the `spinless' Coulomb amplitude $F_c$ is given by
\be
F_c=\left(\mp\right)\frac{2\alpha G^2(t)\sqrt\pi}{|t|}\label{spinless}
\ee
%%%%%%%%%%%%%%%%%%%%%%%%%%%%%%%%%
\subsection{$\pbar p$ Coulombic scattering, including magnetic scattering} 
The relevant Feynman diagram is shown in Fig. 1, where magnetic 
scattering  
is explicitly taken into account via the anomalous 
magnetic moment  $\kappa$ ($ \approx 1.79$).  

The fundamental electromagnetic interaction is 
\be 
eV^\mu=e\left(F_1\gamma^\mu+i\frac{\kappa}{2m}F_2\sigma^{\nu 
\mu}q_\nu\right),\qquad q=p_f-p_i 
\label{interaction} 
\ee  
which has two form factors  
$F_1(q^2)$ and $F_2(q^2)$, both normalized to 1 at $q^2=0$.  The anomalous 
magnetic moment of  
the nucleons is $\kappa$ and $m$ is the nucleon mass. Because of the rapid 
form factor  
dependence on $t$, the annihilation  diagram  for $\pbar p$ scattering (or the 
exchange  
diagram for $pp$ scattering) is negligible in the small $|t|$ region of 
interest and has been  
ignored. 
The interaction of \eq{interaction} is most simply treated by using Gordon 
decomposition and  
can be rewritten as 
$eV^\mu=e\left[(F_1+\kappa F_2)\gamma^\mu -\kappa 
F_2\left(\frac{p_f+p_i}{2m}\right)^\mu\right].$ Thus, using this form, the 
matrix element for the scattering is given by
\begin{eqnarray} 
M&=&\mp e\bar u(p_f)\left[-\kappa F_2\left(\frac{p_f+p_i}{2m}\right)^\mu 
+(F_1+\kappa F_2) 
\gamma^\mu\right]u(p_i)\times \frac{1}{t}\times\nonumber\\ 
&& e\bar u(p'_f)\left[-\kappa F_2\left(\frac{p'_f+p'_i}{2m'}\right)_\mu + 
(F_1+\kappa F_2) 
\gamma_\mu\right]u(p'_i),\label{Mkappa} 
\end{eqnarray} 
where the upper (lower) sign is for $\pbar p$ ($pp$) scattering. 
A straightforward, albeit laborious calculation, gives a differential 
scattering cross section 
\begin{eqnarray} 
\frac{d\sigma}{dt} 
&=&4\pi \frac{\alpha^2}{\beta_{\rm lab}^2t^2}\times \nonumber\\ 
&&\left\{(F_1+\kappa F_2)^4\left[1+\frac{t}{2}\left(\frac{1}{mE_{\rm lab}} 
+\frac{1}{E_{\rm lab}^2}\right)+\frac{t^2}{8m^2E_{\rm 
lab}^2}\right]\right.\nonumber\\ 
&&-\,2(F_1+\kappa F_2)^2 
\left[\kappa^2F_2^2\left(1+\frac{t}{4m^2}\right)+2\kappa 
F_1F_2\right]\times\nonumber\\ 
&&\qquad\quad\left[1+\frac{t}{2}\left(\frac{1}{mE_{\rm lab}} 
+\frac{1}{2E_{\rm lab}^2}\right)\right]\nonumber\\ 
&&+\,\left. 
\left[\kappa^2F_2^2\left(1+\frac{t}{4m^2}\right)+2\kappa F_1F_2\right]^2 
\left[1+\frac{t}{2mE_{\rm lab}}+\frac{t^2}{16m^2E_{\rm lab}^2}\right]\right\} 
.\label{sigmaF1F2} 
\end{eqnarray} 
We now introduce the \Em{electric} and \Em{magnetic} form factors, $G_E(t)$
and 
$G_M(t)$, 
defined as 
\begin{eqnarray} 
G_E(t)&\equiv&F_1(t)+\frac{\kappa t}{4m^2}F_2(t)%
\qquad{\rm and}\quad G_M(t)\equiv F_1(t)+\kappa F_2(t),\label{GEGM} 
\end{eqnarray} 
and rewrite the differential cross section of \eq{sigmaF1F2} as 
\begin{eqnarray}
\frac{d\sigma}{dt}
&=&4\pi \frac{\alpha^2}{\beta_{\rm lab}^2t^2}\times\nonumber\\
&&\left\{\left(\frac{G_E^2(t)-\frac{t}{4m^2}G_M^2(t)}
{1-\frac{t}{4m^2}}
\right)^2
\left(1+\frac{t}
{2mE_{\rm lab}}\right)
+G_M^2(t)\frac{G_E^2(t)-\frac{t}{4m^2}G_M^2(t)}{1-\frac{t}{4m^2}}
\frac{t}{2E_{\rm lab}^2}\right.\nonumber\\
&&%
\quad +\,\left.\left[G_M^4(t)
+\frac{1}{2}\left(\frac{G_E^2(t)-G_M^2(t)}{1-\frac{t}{4m^2}}
\right)^2
\right]\frac{t^2}{8m^2E_{\rm lab}^2}\right\}
.\label{dsigmadtpp}
\end{eqnarray}
We can parametrize these new form factors with 
\be 
\begin{array}{rclcl} 
G_E(t)&=&G(t)&=&\displaystyle \frac{1}{\left(1-\frac{t}{\Lambda^2}\right)^2}, 
\qquad{\rm where } \quad \Lambda^2=0.71 {\rm \ (GeV/c)}^2,\\[7mm] 
G_M(t)&=&(1+\kappa)G(t)&=&\displaystyle \frac{1+\kappa}{\left(1-
\frac{t}{\Lambda^2}\right)^2}, 
\end{array} 
\label{GEGMparameter} 
\ee 
with $t$ in (GeV/c)$^2$, and where $G(t)$ is the dipole form factor already 
defined in  
\eq{G}, \ie the form factor that is traditionally used in experimental 
analyses\cite{bc}. 
   
We now expand \eq{dsigmadtpp} for \Em{very small} $|t|$, and find that 
\begin{eqnarray} 
\frac{d\sigma}{dt} 
&\approx &4\pi \frac{\alpha^2}{\beta_{\rm lab}^2t^2}G^4(t)%
\left\{1-\kappa(\kappa +2)\frac{t}{2m^2}+\frac{t}{2mE_{\rm lab}} 
+(\kappa +1)^2\frac{t}{2E_{\rm lab}^2}\right\} 
,\label{dsigmadtppsmallt} 
\end{eqnarray} 
where the \Em{new term} in $t$, compared to \eq{spin0answer}, is  
$-\frac{\kappa(2+\kappa)}{2m^2}t +\kappa(\kappa +1)\frac{t}{2E_{\rm lab}^2} 
\approx 1+3.86|t|-\frac{3.39}{E_{\rm lab}^2}|t|$,  
and is due to the anomalous magnetic moment of the proton (antiproton). To get 
an estimate of 
its effect, we note that $G^4(t)\approx 1-11.26|t|$, in our units where $t$ is 
in (GeV/c)$^2$.  
We note that  the new term 
is \Em{not negligible} in comparison to the squared form factor, reducing  
the form factor effect by about 35\%
if the  
energy $E_{\rm lab}$ 
is large compared to $m$. In this limit, we find now a t-dependent term,
independent of the energy 
$E_{\rm lab}$, i.e., 
\be 
\frac{d\sigma}{dt} 
\approx 4\pi \frac{\alpha^2}{t^2}G^4(t)\left\{ 
1+3.86|t|\right\} 
,\label{epsmallt2} 
\ee 
which is to be compared with the `spinless' Rutherford formula of 
\eq{rutherford}. However, we will use the `spinless' ansatz of \eq{spinless}, since this is what experimenters typically use, neglecting magnetic effects.
%%%%%%%%%%%%%%%%%%%%%%%%%%%%%%%%%%%%%%%%
\section{$\rho$-value analysis} 
The $\rho$-value, where $\rho\equiv \re 
f_{\rm c.m.}(0)/\im f_{\rm c.m.}(0)$,
 is found by measuring the interference term between the Coulomb and nuclear scattering.  In the following sections, we will give a theoretical formulation of elastic hadronic scattering in the presence of a Coulomb field.

%%%%%%%%%%%%%%%%%%%%%%%%%%%%%%%%%%%%%
\subsection{Spinless analysis neglecting magnetic 
scattering}\label{sec:spinless} 
 
For small $|t|$ values, it is found from experiment that  the hadronic portion of the  elastic nuclear cross section can be adequately parametrized  as
\be
\frac{d\sigma_n}{dt}=\left[\frac{d\sigma_n}{dt}\right]_{t=0}e^{-B|t|}.\label{dsdt}
\ee
Hence, if we were to plot $\ln({d\sigma_n}/{dt})$ against $|t|$ for small $|t|$, we would get a straight line whose slope is $B$, the nuclear slope parameter.  
Using \eq{cms1} and \eq{cms2}, we write  \eq{dsdt} at $t=0$ as
\ba
\left[\frac{d\sigma_n}{dt}\right]_{t=0}&=&\frac{\pi}{k^2}\left[ \frac{d\sigma_n\ \ }{d\Omega_{\rm c.m.}}\right]_{\theta=0}\nonumber\\
&=&\frac{\pi}{k^2}\left|\re f_{\rm c.m.}(0)+i\im f_{\rm c.m.}(0)\right|^2\nonumber\\
&=&\pi\left|\frac{(\rho+i)\im f_{\rm c.m.}(0)}{k} \right|^2\nonumber\\
&=&\pi\left|\frac{(\rho+i)\sigtot)}{4\pi} \right|^2.\label{sigtot}
\ea
For the last step, we used the optical theorem of \eq{cmsampl}. We now rewrite the hadronic elastic scattering cross section at small $|t|$, \eq{dsdt}, as
\be
\frac{d\sigma_n}{dt}=\pi\left|\frac{\sigma_{\rm tot}(\rho + i)}{4\sqrt{\pi}}e^{-B|t|/2}\right|^2.
\ee 
Introducing the notation of \eq{F1}, we now write 
\be 
F_n(|t|)=\frac{\sigma_{\rm tot}(\rho + i)}{4\sqrt{\pi}}e^{-B|t|/2}, 
\label{fn} 
\ee 
so that 
\be
\frac{d\sigma_n}{dt}=\left|F_n\right|^2.
\ee

For the Coulomb amplitude,  the `spinless' Rutherford amplitude of \eq{spinless}
\be 
F_c(t)= \frac{2\sqrt{\pi}\alpha}{|t|}G^2(t)
\ee 
is conventionally used, so that
\be
\frac{d\sigma_c}{dt}=\left|F_c\right|^2.
\ee
\subsection{Addition of Coulomb and nuclear amplitudes}
The preceding work has considered only one amplitude at a time. When { both the nuclear and the Coulomb amplitudes are {\em simultaneously} present,   one can {\em not} simply add up the amplitudes and square them. Rather,  a phase factor $\alpha\phi(t)$ must be introduced into the Coulomb amplitude so that the elastic scattering cross section is now given by
\be
\frac{d\sigma}{dt}=\left|F_ce^{i\alpha \phi(t)}+F_n\right|^2.
\ee
We can understand this most simply by using the language of Feynman diagrams, where $F_n$ might correspond to summing over all Feynman diagrams with {\em only} pions present and $F_c$ might correspond to summing over  all of those diagrams with {\em only} photons present. Simply summing $F_c$ and $F_n$ and squaring would miss all of those mixed diagrams that had {\em both } pions and photons present.  The phase $\phi(t)$ takes care of this problem.

This phase was first investigated by Bethe\cite{bethephase} and later by West and Yennie\cite{westyennie} who used QED calculations of Feynman diagrams. The approach of Cahn\cite{cahnphase} was to evaluate $ \phi(t)$ using an eikonal formulation, and this is the phase that will be used here, given by
\be 
\phi(t)=\mp \left\{\gamma +\ln \left(\frac{B|t|}{2}\right) 
     +\ln\left(1+\frac{8}{B\Lambda^2}\right)+\left(\frac{4|t|}{\Lambda^2}
\right ) 
     \ln\left(\frac{4|t|}{\Lambda^2}\right)+\frac{2|t|}{\Lambda^2}\right\},
\label{cahn} 
\ee 
where $\gamma=0.577\ldots$ is Euler's constant, $B$ is the slope parameter, and $\Lambda^2=0.71$ GeV$^2$ appears in the dipole fit to the proton's electromagnetic form factor, $G(t)$. The upper sign is for $pp$ and the lower sign for $\bar p p$.   

Using these `standard' parametrizations\cite{bc}, the differential elastic
scattering cross 
section is  
\begin{eqnarray} 
\frac{d\sigma}{d|t|}&=&\left|F_ce^{i\alpha \phi(t)}+F_n\right|^2\\%
&=&\pi\left|\frac{\left(\mp\right) 2\alpha G^2(t)} 
{|t|}e^{i\alpha \phi(t)}+ (\rho + i)\frac{\sigtot}{4\pi}e^{-B|t|/2}   \right|^2\\
&=&\frac{\sigma_{\rm 
tot}^2}{16\pi}\left\{G^4(t)\frac{t_0^2}{t^2} 
+2\frac{t_0}{|t|}(\rho +\alpha\phi(t))\,G^2(t)e^{-B|t|/2}+(1+\rho^2)e^{-
B|t|}\right\}.
\label{sumamp} 
\end{eqnarray} 
In \eq{sumamp}, we have  introduced the parameter $t_0$, defined as the absolute value of $t$ 
where the nuclear and 
Coulomb amplitudes have the same magnitude, i.e.,  
\ba
t_0&=&\frac{8\pi\alpha}{\sigma_{\rm tot}}\\
&=&\frac{1}{14.00\sigma_{\rm tot}}, 
\ea
when $\sigma_{\rm tot}$ is in mb, and $t_0$ is in (GeV/c)$^2$. The importance of $t_0$ is that, at momentum transfers $|t|\sim t_0$,  the interference term is maximum and thus, the experiment has the most sensitivity to $\rho$. 

Table \ref{table:tmin} shows some values of $t_0$ as a function of some typical collider c.m. energies, $\sqrt s$, in GeV.  Also shown is the scattering angle $\theta_0=\sqrt {t_0/p^2}$ in mr, where the collider  beam momentum is given by $p=\sqrt{ (s/4)-m^2}$ in GeV. In Table \ref{table:tmin} we also show  the equivalent accelerator straight section length $L_{\rm eff}$ (in m) that is needed to get  $\theta_0$, assuming that the minimum distance from the center of beam that the detector is placed is 2 mm, i.e. $L\theta_0=2$ mm. Here , the beam and its halo has to be smaller than 2mm, in order to place a detector such as a scintillation counter at 2mm from the beam center and not have it swamped by background counts.  The extreme difficulty of achieving this at the LHC is apparent, where an equivalent straight section of $\sim 0.5 $ km would be required to get to $t_0$.  
%%%%%%%%%%%%%%%%%%%%%%%%%%%%%%%%%%%%%%
\subsection{Example: $pp$ scattering at 23.5 GeV at the ISR}
An elegant experimental example of Coulomb normalization of the elastic scattering, and hence an absolute determination of the total cross section, was made by the Northwestern-Louvain group\cite{nulouvain} at the CERN ISR accelerator, using $pp$ collisions at $\sqrt s=23.5$ GeV. Using \eq{sumamp}, they were able to determine $\sigtot,\ \rho,\  B$ and by integrating $d\sigma_{\rm el}/dt$, the elastic cross section, $\sigma_{\rm el}$. Figure \ref{fig:rhoISR} shows the data  needed for normalization, which are deep inside the Coulomb region.
\begin{figure}[h,t,b]
\begin{center}
\mbox{\epsfig{file=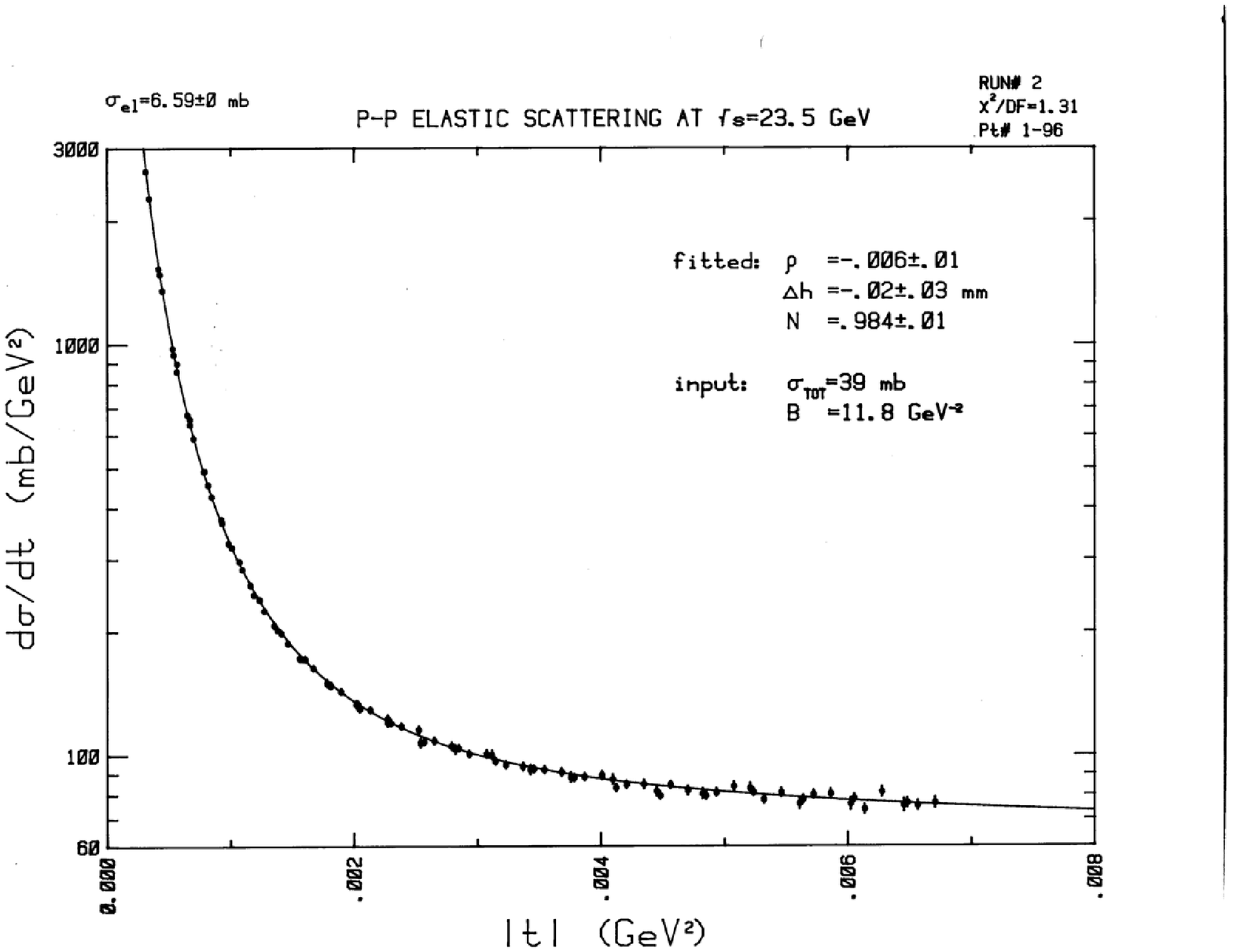%
,width=4in,bbllx=0pt,bblly=0pt,bburx=537pt,bbury=425pt,clip=%
}}
\end{center}
\caption[$d\sigma_{pp}/dt$ at $\sqrt s=23.5$ GeV]
{ \footnotesize 
$d\sigma_{\rm el}/dt$, the differential elastic cross section for $pp$ scattering at $\sqrt s=23.5$ GeV, in mb/GeV$^{2}$ vs. $|t|$, the 4-momentum transfer squared, in GeV$^2$. 
}
\label{fig:rhoISR}
\end{figure}
This experiment probed directly into the Coulomb region, getting into a minimum $t$ value, $|t|_{\rm min}\sim0.0004$ (Gev/c)$^2$, whereas the interference region is centered at $t_0=0.0017$ (GeV/c)$^2$ (see Table \ref{table:tmin}). For larger $|t|\sim0.005$ (GeV/c)$^2$,  the pure nuclear cross section takes over and on a semi-log plot, the cross section  approaches a straight line with nuclear slope $B=11.8 $  GeV$^{-2}$. 

In the same experiment, one sees the Coulomb region, the Coulomb-nuclear interference region and the pure nuclear region, a real experimental tour-de-force. 
\subsection{Example: the UA4/2 $\rho$-value measurement}
At the S$\pbar p$S, at $\sqrt s=541$ GeV, the UA4/2 group measured the interference term of \eq{sumamp} in a dedicated experiment.  
UA4/2 has  made a precision measurement\cite{ua42} of $\bar p$-$p$ 
scattering at  
$\sqrt s=541$ GeV, at the S$\bar pp$S at CERN, in order to extract the 
$\rho$ value for elastic scattering.  
They measured the nuclear slope parameter $B=15.5\pm 0.2$ GeV$^{-2}$.  
They
constrained the total cross section by an 
independent  
measurement\cite{ua4old} of $(1+\rho^2)\sigtot=63.3\pm 1.5$ mb. For their 
published 
$\rho$-value of $0.135\pm0.015$, this implies that they fixed the total cross 
section  at  
$\sigtot= 62.17\pm 1.5$ mb. Using  $\sigma_{\rm 
tot}= 62.17$ mb,   
they  substituted $t_0=0.00115$ GeV$^2$ into \eq{sumamp}. They then
fit the $\pbar p$  
elastic scattering data over the  t-interval  
$0.00075\le |t| \le 0.12$ GeV$^{2}$, reaching values of $|t|\la t_0$,
which gave them maximum sensitivity to $\rho$.

From their measurement of the interference term, they deduced  
the value $\rho=0.135\pm 0.015$---the most precise high energy $\rho$-value ever measured. 
%%%%%%%%%%%%%%%%%%%%%%%%%%%%%%
\section{Measurements of $\sigtot$ and $B$ from elastic scattering}\label{sec:measurements}
Counting rates are the experimentally measured quantities, and not cross sections. In an elastic scattering experiment to measure the differential cross section, what is measured is the differential counting rate $\Delta  N(|t|)$ at $|t|$, i.e.,  the number of counts per second per $\Delta t$ in a small interval $\Delta t$ around $|t|$, after corrections for background and inefficiencies such as  dead time, azimuthal coverage,  etc. In order to obtain the differential elastic nuclear scattering $d\sigma_{\rm el}/dt$, we must normalize this rate.  Thus, we write
\be
\Delta N(t)={\cal L}\left(\frac{d\sigma_{\rm el}}{dt}\right),\label{lum}
\ee
where $\cal L$ is a normalization factor with units of ({\mbox area})$^{-1}\times ${\mbox (time)}$^{-1}$.  For colliding beam experiments, $\cal L$ is the luminosity.

If the experiment can get into the Coulombic region $|t|<<t_0$, then $d\sigma_{\rm el}/dt$ is, for intents and purposes, given by the differential Coulomb cross section $d\sigma_c/dt\approx 4\pi(\alpha/t)^2$. This self-normalization of the experiment allows one to obtain $\cal L$ directly from \eq{lum}. Unfortunately, we see from Table \ref{table:tmin} that this is only possible for energies in the IRS region. Even at the $S\pbarp S$, where UA4/2 got down to $|t_{\rm min}|=0.0075$ (GeV/c)$^2$ (from Table \ref{table:tmin}, we find $t_0=0.0010$ (GeV/c)$^2$, so that they were unable to penetrate sufficiently into the Coulomb region---where $|t|\la t_0/2$---to normalize using the known Coulomb cross section.   Other techniques such as the van der Meer\cite{vandermeer} technique of sweeping colliding beams through each other, etc., also give direct measures of $\cal L$.  At the Tevatron, the experimenters got to $|t_{\rm min}|=0.0014$ (GeV/c)$^2$ (from Table \ref{table:tmin}, $t_0= 0.0010$ (GeV/c)$^2$, just a bit hit larger than $t_0$, but small enough to do a $\rho$-value measurement\cite{amos2} of $\rho=0.14\pm0.07$.

%%%%%%%%%%%%%%%%%%%%%%%%%%%%%%%%Table #1
\begin{table}[h,t]                   % Use "table" environment, but also
				 % use  "tabular" environment below.
%
\def\arraystretch{1.5}            % Make the space between rows in the Table,
				  % 1.5 x bigger than the default spacing.

\begin{center}
\caption[$t_0$ and $\theta_0$ for the Coulomb interference region for $\pbarp$ ($pp$) scattering]{Values of $t_0$ and $\theta_0$ for the Coulomb interference region for $\pbarp$ ($pp$) scattering, assuming $L_{\rm eff}$(m)=2mm/$\theta_0$(mr). 
\label{table:tmin}
}
\vspace{.2in}
\begin{tabular}[b]{ccccc}
\hline\hline
$\sqrt s$&&$t_0$&$\theta_0$&$L_{\rm eff}$\\
(GeV)&Accelerator&(GeV/c)$^2$&(mrad)&(m)\\
\hline
23.5&ISR&0.0017&3.6&0.56\\
62.5&ISR&0.0016&1.5&1.5\\
540&$S\pbarp S$&0.0011&0.12&16.0\\
1800&Tevatron&0.0010&0.035&57.4\\
14000&LHC&0.00067&0.0037&544\\
\hline\hline
\end{tabular}
     %\vspace{1in} \\
\end{center}
\end{table}
\def\arraystretch{1}  %Restore the default row spacing in the Table.
%%%%%%%%%%%%%%%%%%%%%%%%%%%%%%%%%%%%%%%%%%%%%%
%%%%%%%%%%%%%%%%%%%%%%%%%%%%%%%%Table #1

In any event, if $\cal L$ is known, one goes  to the nuclear region where $|t|>>t_0$, and  plots the logarithms of the counting rates against $|t|$. After fitting with a straight line, the line is extrapolated to $t=0$, to obtain the hadronic counting rate $\Delta N(0)$. 

When  $\cal L$ is known, by using \eq{dsdt}, \eq{sigtot} and \eq{lum}, we can write
\ba
\sigtot(1+\rho^2)^{1/2}&=& 4\left[\pi\left(\frac{d\sigma_n}{dt}\right)_{t=0}\right]^{1/2}\\
&=&4\left[\pi\left(\frac{\Delta N(0)}{\cal L}\right)\right]^{1/2}.\label{siglum}
\ea

Thus, a direct method of determining $\cal L$ allows one to measure the combination $\sigtot(1+\rho^2)^{1/2}$.  Often, at high energy, the real portion is sufficiently small that $\rho << 1$, and hence, $\sigtot\approx \sigtot(1+\rho^2)^{1/2}$, obviating the necessity of a separate determination of $\rho$.

A very important technique for determining the total cross section is the ``Luminosity-free'' method, where one simultaneously measures $N_{\rm tot}$, the total counting rate due to {\em any} interaction, together with $\Delta N(0)$,  the differential elastic scattering rate at $t=0$.

We now write 
\ba
\Delta N(0)&=&{\cal L}\left[\frac{d\sigma_n}{dt}\right]_{t=0}\label{N0}\\
N_{\rm tot}&=&{\cal L}\,\sigtot.\label{Ntot}
\ea
From \eq{siglum} and \eq{N0}, we find $\cal L$ and substitute into \eq{Ntot} to get
\be
\sigtot(1+\rho^2)=16\pi\frac{\Delta N(0)}{N_{\rm tot}}.\label{sigtot2}
\ee

From \eq{sigtot2}, we see that the ``luminosity-free'' method measures 
$\sigtot(1+\rho^2)$, whereas the direct luminosity determination method measures $\sigtot(1+\rho^2)^{1/2}$. As mentioned earlier, both of these techniques only weakly depend on $\rho$ when $\rho$ is small---a very good approximation for high energies---so a very inaccurate measurement of $\rho$  can still yield a highly accurate measurement of $\sigtot$.

Using the parametrization of \eq{dsdt}, the total elastic cross section $\sigma_{\rm el}$ is given by
\ba
\sigma_{\rm el}&\equiv &\int_{-\infty}^0\frac{d\sigma_n}{dt}\,dt\nonumber\\
&=&\int_0^\infty \left[\frac{d\sigma_n}{dt}\right]_{t=0}e^{-B|t|}\,d|t|\nonumber\\
&=&\frac{1}{B}\left[\frac{d\sigma_n}{dt}\right]_{t=0}\nonumber\\
&=&\frac{\sigtot^2(1+\rho^2)}{16\pi B}\label{Siggreek}.
\ea

We will give this value a special name, $\Sigma_{\rm el}$ and rewrite \eq{Siggreek} as 
\be
\Sigma_{\rm el}=\frac{\sigtot^2(1+\rho^2)}{16\pi B}.\label{Siggreek2}
\ee

If the parametrization of \eq{dsdt} used above were valid over the full $t$ range, then $\Sigma_{\rm el}$ would be equal to $\sigma_{\rm el}$. It should be noted that very often, experimenters quote $\Sigma_{\rm el}$ as the experimental value of $\sigma_{\rm el}$.  

We rewrite \eq{Siggreek2} in a very useful form as
\be
\frac{\Sigma_{\rm el}}{\sigtot}=\frac{\sigtot(1+\rho^2)}{16\pi B}.\label{sigtotoverB}
\ee
At high energies, where $\rho$ is small, \eq{sigtotoverB} essentially says that the ratio of the elastic to  total cross section, $\Sigma_{\rm el}/\sigtot$, varies as the ratio of the total cross section to the nuclear slope parameter $B$, $\sigtot/B$.  The ratio $\Sigma_{\rm el}/\sigtot$ is bounded by unity as  $s\rightarrow\infty$. Thus \eq{sigtotoverB} tells us that the ratio  $\sigtot/B$ also approaches a constant, i.e.,  $\sigtot$ and $B$ have the {\em same}  dependence on $s$ as $s\rightarrow \infty$.

\section{Unitarity}
We next will discuss unitarity, first in reactions with only elastic scattering and then in reactions involving both elastic and inelastic scattering. It is convenient to  work in the c.m. frame, where we will show that  unitarity is implied by the optical theorem,
 
\be\sigma_{\rm tot}=\frac{4\pi}{k}\,\im f_{\rm c.m.}(t=0)\label{cmsampl2},
\ee
and vice versa.

\subsection{Unitarity in purely elastic scattering}\label{sect:unitarity}
For elastic scattering, we expand the c.m. amplitude in \eq{cmsampl2} in terms of   a standard partial-wave Legendre polynomial expansion.  Ignoring spin, in the c.m. system we have
\be
f_{\rm c.m.}(s,t)=\frac{1}{k}\sum_{\ell=0}^\infty (2\ell+1)P_\ell\cos\theta)a_\ell(k),
\ee
where $a_\ell$ is the $\ell$th partial wave scattering amplitude.

For purely elastic scattering, since $\sigtot=\sigma_{\rm el}$, we have 
\ba
\sigtot&=&\int\frac{d\sigma_{\rm el}}{dt}\,dt\nonumber\\
&=&\int\frac{\pi}{k^2}\left|f_{\rm c.m.}\right|^2\,dt\label{fsq}\\
&=&\frac{4\pi}{k}\im f_{\rm c.m.}(t=0)\label{imf}.
\ea
Comparing coefficients in \eq{fsq} and \eq{imf}, we see that unitarity is expressed in the relation
\be
\im a_\ell=\re a_\ell^2+\im a_\ell^2.\label{argand}
\ee
Therefore, the amplitude for each partial wave $a_\ell$ lies on the Argand circle, shown in Fig. \ref{fig:argand}.
%%%%%%%%%%%%%%%%%%%%%%%%%%%%%%%%%%%%%%%%%%%%%
\begin{figure}[h,t,b] %Fig. 1
\begin{center}
\mbox{\epsfig{file=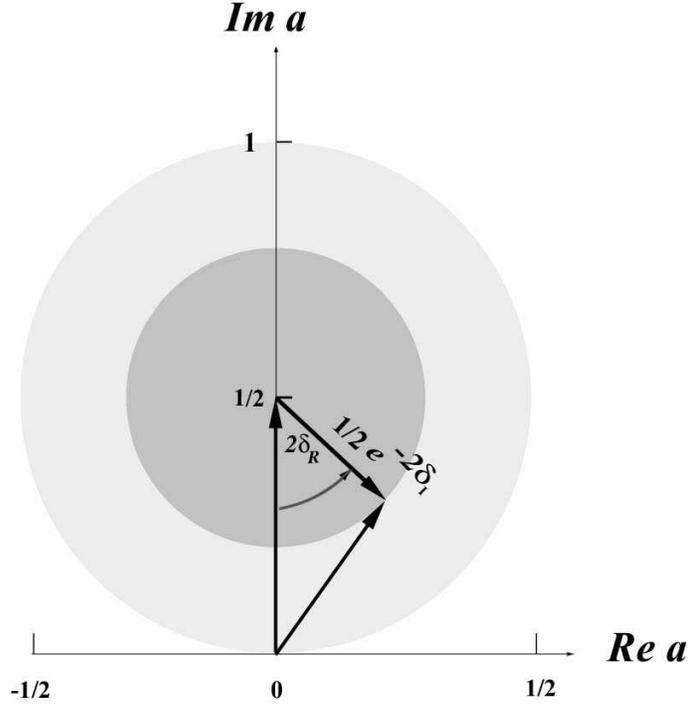,		width=4in,bbllx=96pt,bblly=178pt,bburx=530pt,bbury=600pt,clip=%
}}
\end{center}
\caption[Argand circle]{ \footnotesize The Argand circle. The partial wave scattering amplitude $a_\ell$ (or $a(b,s)$) is given graphically by the Argand circle.   For  {\em pure elastic} scattering, the imaginary portion of $\delta_\ell$ (called $\delta_{I}$) is zero and the amplitude lies on the outer circle of radius $\frac{1}{2}$---otherwise, if there is absorption, the amplitude $a_\ell$ lies on the circle of radius $\frac{1}{2}e^{-2\delta_I}$. Note the discontinuity at $a=i/2$; for a {\em pure imaginary } amplitude ($\re a=0$), the real portion of $a_\ell$  is given by $\delta_R=0$ if $0\le a<\frac{i}{2}$; however, $\delta_R=\pi$ if $\frac{i}{2}<a \le i$. 
}
\label{fig:argand}
\end{figure}
%%%%%%%%%%%%%%%%%%%%%%%%%%%%%%%%%%%%%%%%%%%%%%%%%%%
If there is inelasticity, the amplitude lies {\em inside} the Argand circle of Fig. \ref{fig:argand}, and the amplitude can be represented as
\be
a_\ell=\frac{e^{2i\delta_\ell}-1}{2i},\label{inelastic}
\ee
where $\im \delta_\ell>0$ if there is inelastic scattering, whereas $\delta_\ell$ is pure real if there is only elastic scattering. In Fig. \ref{fig:argand}, $\delta_\ell$ is called $\delta$, with $\delta=\delta_R+i\delta_I$ and $a_\ell$ is called $a$, with $a=\re a+i\,\im a$. 

\subsection{Unitarity in inelastic scattering}
For inelastic scattering, the situation is much more complicated.  It is convenient to introduce here the Lorentz-invariant amplitude $\cal M$, given by the $S$ matrix for the production of $n$ particles in the reaction $p_1+p_2\rightarrow p_1'+p_2'+\cdots p_n'$ by 
\ba
S&=& I -i(2\pi)^4\,\delta^4\!\left( p_1+p_2-\sum_{i=1}^n p_i'\right)\nonumber\\
&&\quad \times\frac{(p_1'p_2'\cdots p_n'|{\cal M}|p_1p_2)}{(2E_1)(2E_2)\stackrel{n}{\raisebox{-1.8ex}{$\stackrel{\displaystyle  \prod }{\scriptstyle i=1}$}}(2E'_i)},\label{Smatrix}
\ea
where $p_1,\ (E_1)$ and $p_2,\ (E_2)$ are the initial 4-momenta (energies) and the primes indicate final state 4-momenta and energies, with $I$ being the  unit matrix. The normalization for the states is
\be
<p'|\,p>=(2\pi)^3\delta^3(p-p'),
\ee
so the completeness relation is given by
\be
I=\sumovern\stackrel{n}{\raisebox{-1.8ex}{$\stackrel{\displaystyle  \prod }{\scriptstyle i=1}$}}\left( \int\,\frac{d^3\vec p_i'}{(2\pi)^3}   \right)
|\,p_1'p_2'\cdots p'_n><p_1'p_2'\cdots p'_n|. \label{completeness1}
\ee
The completeness relation of \eq{completeness1} is more readily envisioned when we rewrite it symbolically as 
\be
I=\sumovern|n><n|.\label{completeness2}
\ee

Unitarity is expressed by the fact that the $S$ matrix is unitary, i.e., 
\be
S^\dagger S=1.\label{Sunitarity}
\ee
Evaluating \eq{Sunitarity} between 2-body states $|p_1p_2>$ and $|p_3p_4>$, we find
\ba
<p_3p_4|-i{\cal M}+i{\cal M}^\dagger|p_1p_2>&=&2\,\im\!\!<p_3p_4|{\cal M}|p_1p_2>\nonumber\\
&\!\!\!\!\!\!\!\!\!\!\!\!\!\!\!\!\!\!\!\!\!\!\!\!\!\!\!\!\!\!\!\!\!\!\!\!\!\!\!\!\!\!\!\!\!\!\!\!\!\!\!\!\!\!\!\!\!\!\!\!\!\!\!\!\!\!\!\!\!\!\!\!=&\!\!\!\!\!\!\!\!\!\!\!\!\!\!\!\!\!\!\!\!\!\!\!\!\!\!\!\!\!\!\!\!\!\!\!\!-(2\pi)^4\sumovern \stackrel{n}{\raisebox{-1.8ex}{$\stackrel{\displaystyle  \prod }{\scriptstyle i=1}$}}\left( \int\frac{d^3\vec p_i'}{(2\pi)^32E_i'} \right)\delta^4\left( p_1+p_2-\sumovern\stackrel{n}{\raisebox{-1.8ex}{$\stackrel{\displaystyle  \prod }{\scriptstyle j=1}$}}\right)\nonumber\\
&&\!\!\!\!\!\!\!\!\!\!\!\!\!\!\!\!\!\!\!\!\!\!\!\!\!\!\!\!\!\!\!\!\times <p_3p_4|{\cal M}^\dagger| p_1'p_2'\cdots p_n'><p_1'p_2'\cdots p_n'|{\cal M}|p_1p_2>.
\ea
The $n$-body invariant phase space $d\Phi_n$\footnote{The factor $2E_i'$ in the denominator is for  bosons. For fermions, $2E_i'$ is replaced by $E'_i$.} which we will discuss in detail later in Section \ref{section:Lorentz}, is defined in  \eq{phspace} as
\be
d\Phi_n=\sumovern \stackrel{n}{\raisebox{-1.8ex}{$\stackrel{\displaystyle  \prod }{\scriptstyle i=1}$}}\left( \frac{d^3\vec p_i'}{(2\pi)^32E_i'} \right)\delta^4\left( p_1+p_2-\sumovern\stackrel{n}{\raisebox{-1.8ex}{$\stackrel{\displaystyle  \prod }{\scriptstyle j=1}$}}\right).
\ee
The cross section $d\sigma_n$ for the production of $n$ particles is given by 
\be
d\sigma_n=\frac{(2\pi)^4}{2E_12E_2}\frac{|{\cal M}|^2d\Phi_n}{ |\vec v_1-\vec v_3|},\label{dsigman}
\ee
where $ v_{\rm rel}\equiv  |\vec v_1-\vec v_3|$ is the flux factor (the `relative velocity' of the colliding particles). The invariant $\cal F$ associated with the flux factor is given by
\ba
{\cal F}&=&\sqrt{(p_1\cdot p_2)^2-m_1^2m_2^2}\nonumber\\
&=&\sqrt{ \left|\vec p_1E_2-\vec p_2E_1\right|^2-\left| \vec p_1 \times \vec p_2\right|^2 }.\label{fluxfactor}
\ea
It is easy to show that the invariant ${\cal F}$ is given by $ E_1E_2|\vec v_1-\vec v_3|=E_1E_2v_{\rm rel}$ in {\em both} the c.m. and the laboratory systems. 
In the laboratory system, we find that
\ba 
{\cal F}&= &m_2p_{\rm lab}\\
&=&\left[E_1E_2v_{\rm rel}\right]_{\rm lab}
\label{E1E2vrellab},
\ea
since $E_2=m_2$ and $v_{\rm rel}=p_1/E_1$, 
evaluating $E_1,\  E_2$ and $v_{\rm rel}$ in the laboratory frame. In the c.m. frame, we find
\be
{\cal F}= k(E^*_1+E^*_2)=k\sqrt s\quad\mbox{ from direct evaluation of \ }{\cal F},\label{direct}
\ee
whereas an evaluation of $E^*_1E^*_2v^*_{\rm rel}$ gives
\ba
E^*_1E^*_2v^*_{\rm rel}&=&E_1^*E_2^*(k/E_1^*+k/E_2^*)\nonumber\\
&=&k(E_1^*+E_2^*)=k\sqrt s={\cal F}.\label{E1E2vrelcms}
\ea
Thus we find that  $E^*_1E^*_2v^*_{\rm rel}={\cal F}=k\sqrt s$.

Using \eq{E1E2vrelcms}, we see that in \eq{dsigman} that the factor
$2E_12E_2|\vec v_1-\vec v_2|= 4k\sqrt s.$ Therefore, we can now rewrite the cross section $d\sigma_n$ as
\ba
d\sigma_n&=&(2\pi)^4\frac{|{\cal M}|^2d\Phi_n}{4E_1E_2v_{\rm rel}}\nonumber\\
&=&(2\pi)^4\frac{|{\cal M}|^2d\Phi_n}{4{\cal F}}\label{siginvariant}.
\ea
We will later show that $d\Phi_n$ is a Lorentz invariant. Thus, the partial cross section $d\sigma_n$ is {\em manifestly} Lorentz invariant, since
${\cal M},\ \Phi_n$ and ${\cal F}$ are all  Lorentz invariants.
Thus, the cross section for production of $n$ particles is given by
\be
\sigma_n=(2\pi)^4\int \frac{|{\cal M}|^2d\Phi_n}{4{\cal F}}.
\ee
%%%%%%%%%%%%%%%%%%%%%%%%%%%%%%%%%%%%%%%%%%%%%%%%%%%%
\subsection{The optical theorem}\label{section:opticaltheorem}
For forward scattering, where $p_1=p_2$, $p_3=p_4$ and ${\cal M}(t=0)=<p_1p_2|{\cal M}|p_1p_2>$, we find
\ba
2\,\im{\cal M}(t=0)&=&-4k\sqrt s\,\,\sumovern\,\sigma_n
=-4k\sqrt s \,\sigtot.\label{ImM}
\ea
Rewriting \eq{ImM}, we have finally obtained the optical theorem, true for either elastic or inelastic scattering, 
\ba
\sigtot&=&-\frac{1}{2k\sqrt s}\,\im {\cal M}(t=0)\label{opticalImM}\\
&=&\frac{4\pi}{k}\,\im f_{\rm c.m.}(\theta=0)\label{opticalcm}\\
&=&\frac{4\pi}{p}\,\im f(\theta_L=0),\label{opticallab}
\ea
which are consequences of unitarity.
%%%%%%%%%%%%%%%%%%%%%%%%%%
\section{Phase space}
\subsection{Fermi's ``Golden Rule''}
The second `Golden Rule' of Fermi\cite{goldenrule} asserts that the transition 
probability of any physical process is proportional to the squared
modulus of the matrix element times {\em the number of final states per
unit energy} that are realizable with energy and momentum conservation,
i.e., 
\begin{equation}
d\Gamma\propto |H|^2 \frac{dN}{dE}\,,\label{Fermi}
\end{equation}
where $\frac{dN}{dE}$ is the number of final states per unit energy,
which is commonly called the ``phase space''.  In \eq{Fermi}, neither the phase space nor the matrix element $H$ is Lorentz invariant, whereas their product is.
%%%%%%%%%%%%%%%%%%%%%
\subsection{Modern form of Fermi's ``Golden Rule''}
The original version of Fermi's Golden Rule\cite{Fermi} used a non-invariant form, 
whereas the 
more modern version substitutes for $|H|^2$ the Lorentz invariant squared matrix element
$|{\cal M}|^2$, and for $\frac{dN}{dE}$ (see refs. \cite{Block,Kretzschmar,Hagedorn}) 
the Lorentz invariant phase space
$d\Phi_n$. In the next section, we will prove the invariance of $\Phi_n$,  defined in \eq{phspace}. We now write a modern (invariant) form\cite{sudarshan} of Fermi's second ``Golden Rule'' as 
\begin{equation}
d\Gamma\propto |{\cal M}|^2 d\Phi_n\,.\label{modern}
\end{equation}
In terms of the cross section $d\sigma_n$, we see from \eq{siginvariant} that
\ba
d\sigma_n&=&(2\pi)^4\frac{|{\cal M}|^2d\Phi_n}{4{\cal F}}\nonumber\\
&=&(2\pi)^4\frac{|{\cal M}|^2d\Phi_n}{4E_1E_2v_{\rm rel}}\label{siginvariant2}.
\ea
Since a cross section is an area perpendicular to the direction of
motion of the incident particle, it must of course be Lorentz invariant. 
 
We note that if the invariant matrix element ${\cal M}$ in eq.~(\ref{siginvariant2})  is a 
function that
has little variability compared to the phase space, i.e.,  $|{\cal M}|^2$
is approximately constant, then the distribution in momentum space, angle, etc., of the final state particles is given by \eq{siginvariant2}, which  
depends 
only on the Lorentz invariant phase space $d\Phi_n$.

We quote from a 1980 paper by Block and Jackson\cite{BlockJackson}:
\begin{quote}
``Phase space considerations have a long and honorable history, 
from Dalitz plots for three particles to statistical models of particle 
production
for large numbers of particles\cite{Fermi,Block,Kretzschmar}. In attempts
to unravel interaction dynamics or hunt for the production of new particles, 
the experimenter uses phase space to estimate the shapes of backgrounds in
various mass or other distributions. High-speed computers have led to an 
increasing reliance on Monte Carlo methods to generate the phase-space 
plots\cite{Hagedorn}.''
\end{quote}
For these reasons, the function $d\Phi_n$ plays an 
particularly important
role in both the strong and weak interactions of elementary particles, 
where 
we 
{\em almost never} know the detailed structure of the matrix elements. 
As an example, for either a decay that produces $n$ particles, 
$$M_{n}\rightarrow m_0+m_1+\cdots+m_{n-1},\qquad i=0,1,\cdots,n-1$$ or 
an inelastic reaction producing $n$ particles,
$$a+b\rightarrow m_0+m_1+\cdots+m_{n-1},\qquad  i=0,1,\cdots,n-1,$$ the 
phase 
space of the $n$ particles with masses $m_0, m_1,\ldots m_{n-1}$ plays
a dominant role. The final state particles have a enormous variety of
possible momenta, limited only by conservation of energy and momentum.
The 
phase space (for a {\em constant} matrix element) determines the 
probability
distribution for the momentum of each of the final-state particles 
which is a function the kinematics of the process, i.e.,  the total 
c.m. energy of the system and the $n$ masses, $m_0, m_1,\ldots, m_{n-1}$.
%%%%%%%%%%%%%%%%%%%%%%%%%%%%%%%%%%%
\subsection{Lorentz invariant phase space}\label{section:Lorentz}
For our discussion of invariant phase space, we introduce the notation that the $n$ final state particles have the masses $m_0, m_1,\ldots,m_{n-1}$ and
define $p_{i}$ as the 4-vector $(E_{i},\vec{p_{i}})$ of 
particle
of mass $m_i$, $i=0, 1,\ldots, n-1$, where
we use the metric $m_i^2=E_i^2-\vec p_i^{\,\,2}$. We  define $P_{n}$ 
as
the 4-vector of the whole system, so that energy-momentum 
conservation
leads to $P_n=\sum_{i=0}^{n-1}P_i$. We  further define the invariant $M_n$ as 
$M_{n}^{\,\,2}=P_{n}^{\,\,2}$ and note that 
$p_{i}^{\,\,2}=m_{i}^{2}$.

For the $n$ particle system, we can write
$
\Phi_{n}(M_{n}^{2};m_{n-1}^{2},....,m_{1}^{2},m_{0}^{2}),
$
the integral
of the Lorentz invariant phase space of $n$ bodies (using units of
$\hbar=c=1$), as 
\begin{equation}
\Phi_{n}(M_{n}^{2};m_{n-1}^{2},....,m_{1}^{2},m_{0}^{2})  = 
\prod_{i=0}^{n-1} \int_{\vec{p}_{i_{min}}}^{\vec{p}_{i_{max}}}
  \frac{d^{3}\vec{p_{i}}}{(2 \pi)^{3} (2 E_{i})} \delta^{4}
(P_{n} - \sum_{i=0}^{n-1} p_{i})\label{phspace}.
\end{equation}
The factor $(2 \pi)^3$ arises because we must divide the phase space by 
$h^3$ to get the number of quantum states% 
%%%%%%%%%%%%%%%%%%%%%%%%%%%%%%%%%%%%%%%%%%%%%
\footnote{Statistical mechanics states that  the number of quantum mechanical states is given by the  true phase space for a particle divided by $h^3$, i.e., is given by $d^3\vec pd^3\vec x/h^3$.  When wave functions are normalized in  a space volume $V$,  the number of quantum states is $d^3\vec p\, V/h^3$.   It can be shown that all of the factors of $V^n$ due to the phase space cancel out in \eq{dsigman} and that $d\Phi_n$ in \eq{siginvariant2} is independent of the normalization volume $V$, depending only on the ``invariant'' phase space (more correctly, the invariant momentum space) of \eq{phspace}.},
%%%%%%%%%%%%%%%%%%%%%%%%%%%%%%%%%%%%%%%%%%%%%%%%%%%%%%%
since  
we are using units where $\hbar=1$.  The factor 2 appearing in the
denominator, in the term $(2 E_i)$, is the appropriate 
normalization
if the particles are all  bosons (it is simply $E_i$ for 
 fermions).  The four-dimensional delta function 
$\delta^{4}(P_{n} - \sum_{i=0}^{n-1} p_{i})$ which is manifestly 
Lorentz-invariant
insures the conservation of energy and momentum in the process. 

We will now
prove that each factor $d^3\vec{p_i}/E_i$ is {\em also} 
Lorentz-invariant. 
Consider two different frames of reference, systems
$O$ and $O^*$, having four-vectors $(E,p_x,p_y,p_z)$ and 
$(E^*,p^*_x,p^*_y,p^*_z)$, with the two  systems 
being connected by a Lorentz transformation along the $z$ axis, so that
\begin{eqnarray}
p_z&=&\beta\gamma E^* + \gamma p^*_z \label{lor1}\\
E  &=&\gamma E^* + \beta\gamma p^*_z \label{lor2}\,.
\end{eqnarray}
Differentiating eq.~(\ref{lor1}) with respect to $p^*_z$, we immediately
obtain
\begin{equation}
\frac{\partial{p_z}}{\partial{p^*_z}}=\beta\gamma
\frac{\partial{E^*}}{\partial{p^*_z}} +\gamma \label{lor3}\,.
\end{equation}
Invoking the relation ${E^*}^2={p^*}^2+m^2=p_x^2+p_y^2+p_z^2+m^2$, we have
\begin{equation}
\frac{\partial{p_z}}{\partial{p^*_z}}=\beta\gamma\frac{p^*}{E^*}
+\gamma
=\frac{\beta\gamma p^* + \gamma E^*}{E^*}
\end{equation}
which, using eq.~(\ref{lor2}), becomes
\begin{equation}
\frac{\partial{p_z}}{\partial{p^*_z}}=\frac{E}{E^*},
\mbox{    hence   } \frac{dp_z}{E}=\frac{dp^*_z}{E^*}\,.
\end{equation}
Since $dp_x=dp^*_x$ and $dp_y=dp^*_y$, it becomes evident that
\begin{equation}
\frac{dp_x \, dp_y \, dp_z}{E} = \frac {dp^*_x \, dp^*_y \, dp^*_z} {E^*} \,,
\end{equation}
 i.e., we have shown that  $d^3\vec p/E$ is a Lorentz invariant.

Thus,  the entire {phase space integral
\begin{equation}
\Phi_{n}(M_{n}^{2};m_{n-1}^{2},....,m_{1}^{2},m_{0}^{2})  = 
\prod_{i=0}^{n-1} \int_{\vec{p}_{min}}^{\vec{p}_{max}}
  \frac{d^{3}\vec{p_{i}}}{(2 \pi)^{3} (2 E_{i})} \delta^{4}
(P_{n} - \sum_{i=0}^{n-1} p_{i})\label{phspace2}
\end{equation}
is  now manifestly Lorentz-invariant, since each portion of it has been shown to be invariant. The flux factor $\cal F$ in \eq{siginvariant} was already shown to be a Lorentz invariant (see \eq{fluxfactor}). Therefore the cross section $d\sigma_n$ of \eq{siginvariant2} is also now manifestly Lorentz invariant, since $|{\cal M}|^2$, $d\Phi_n$ and $\cal F$  are  each  separately Lorentz invariant. 

In  Appendix \ref{sec:phspace} we derive the necessary theorems for the evaluation of Lorentz-invariant phase space for 2-bodies, 3-bodies, up to n-bodies. In Appendix \ref{sec:MCtechniques}, we discuss the Monte Carlo techniques necessary for a fast computer implementation of $n-$body phase space, allowing us to make distributions  with `events' (of {\em unit} weight, rather than weighted `events', as discussed there. Finally, in Appendix \ref{sec:MCphasespace}, we develop a very fast computer algorithm for the evaluation of $n$-body phase space, even when $n$ is very large. 
%
% 
%
%%%%%%%%%%%%%%%%%%%%%%%%%%%%%%%%%%%%%%%%
\section{Impact parameter representation}
In Section \ref{sect:unitarity} we found that the c.m. amplitude for spinless particles could be written as 
\be
f_{\rm c.m.}(s,t)=\frac{1}{k}\sum_{\ell=0}^\infty (2\ell+1)P_\ell(\cos\theta)a_\ell(k),\label{discretesum}
\ee
with $a_\ell(k)$,  the $\ell$th partial wave scattering amplitude, given by
\be 
a_\ell(k)=\frac{e^{2i\delta_\ell}-1}{2i},
\ee
where $\delta_\ell$ is the (complex) phase shift of the $\ell$th partial wave. For purely elastic scattering, $\delta_\ell$ is real.  If there is inelasticity, $\im \delta_\ell>0$. From \eq{opticalcm}, we find that the contribution of the $\ell$th partial wave to the cross section is bounded, i.e., 
\be
\sigma_\ell \le\frac{4\pi(2\ell +1)}{k^2}.\label{partialwavesigma}
\ee
Since the upper bound $\frac{4\pi(2\ell +1)}{k^2}$ decreases with energy,  the high energy amplitude must contain a very large number of partial waves.  Thus it is reasonable to change the discrete sum of \eq{discretesum} into an integral.  

Let us now introduce the impact parameter $b$. A classical description of the scattering would relate the angular momentum to $\ell$ by $kb=\ell +1/2$,
with the  extra 1/2  put in for convenience.  We then convert the discrete \eq{partialwavesigma} into an integral equation via the substitutions $\sum_\ell\rightarrow\int dl\rightarrow \int k\,db$ and $a_\ell(k)\rightarrow a(b,s)$. Defining $q^2=-t=4k^2\sin^2(\theta/2)$, we will reexpress $P_\ell(\cos \theta)$ in terms of the new variables $b$ and $q^2$. Since we have many partial waves, we note that\cite{erdelyi}
\be
P_\ell(\cos\theta)\rightarrow J_0\left[ (2\ell +1) \sin(\theta/2)\right]\quad\mbox{as\ } \ell\rightarrow\infty.
\ee
Rewriting \eq{discretesum} as a continuous integral over the 2-dimensional impact parameter space $b$,  we find
\ba
f_{\rm c.m.}(s,t)&=&2k\int^\infty_0 b\,db J_0(qb)a(b,s)\label{J0}\\
&=&\frac{k}{\pi}\int a(b,s)\,e^{i\vec q\cdot\vec b}\,d^2\vec b\label{fofb},
\ea
where $b=|\vec b|$, $q=|\vec q|$, $\vec q\cdot \vec b=qb\cos\phi$ and $d^2\vec b=b\,db\, d\phi$.
To get  \eq{fofb}  we substituted the integral representation\cite{Stegun} of $J_0$,
\be J_0(z)=\frac{1}{2\pi}\int^{2\pi}_{0}e^{iz\cos\phi}\,d\phi
\ee
into \eq{J0} .

Inverting  the Fourier transform of $a(b,s)$,  we find
\be
a(b,s)=\frac{1}{4\pi}\int f_{\rm c.m.}(s,t)\, e^{-i\vec q\cdot\vec b}\,d^2\vec q .\label{aofb}
\ee
%%%%%%%%%%%%%%%%%%
\subsection{$d\sigma_{\rm el}/dt$, \ $\sigma_{\rm el}$ and $\sigtot$ in impact parameter space}
From \eq{cms2} we find that
\be
\frac{d\sigma_{\rm el}}{dt}=\frac{\pi}{k^2} |f_{\rm c.m.}|^2=4\pi\left|\int a(b,s)J_0(qb)b\,db\right|^2. \label{dsdtb}
\ee
Integrating \eq{dsdtb} over all $t$, we see that $\sigma_{\rm el}$, the total elastic scattering cross section, is given by
\ba
\sigma_{\rm el}&=&\frac{\pi}{k^2}\int |f_{\rm c.m.}|^2 \,dt=\frac{1}{k^2}\int |f_{\rm c.m.}|^2 \,d^2\vec q \nonumber\\
&=&4\int |a(b,s)|^2\,d^2\vec b.\label{sigelofb}
\ea
From the optical theorem of \eq{opticalcm},  the total cross section $\sigtot$ is given by
\ba
\sigtot&=&\frac{4\pi}{k}\im f_{\rm c.m.}(s,0)=4\int \im a(b,s) \,d^2\vec b\label{sigtotofb1}.
\ea

Since the impact parameter vector $\vec b$ is a two-dimensional vector perpendicular to the direction of the projectile, the amplitude a(b,s) is an Lorentz invariant, being the same in the laboratory frame as in the c.m. frame.
This amplitude  lies on the Argand plot of Fig. \ref{fig:argand} and again, we can write it as
\be
a(b,s)=\frac{e^{2i\delta(b,s)}-1}{2i}\label{aofb&s},
\ee 
where the phase shift is now a function of $b$, as well as $s$. 
The total cross section can now be written as
\be
\sigtot=2\int\im \left[i(1-e^{2i\delta(b,s)}) \right]\,d^2\vec b.
\ee
It is important to note that the impact-parameter formulation of $\sigtot$ in \eq{sigtotofb1} satisfies unitarity.

Once again, elastic scattering corresponds to the phase shift $\delta(b,s)$ being real. For inelastic scattering, $\im \delta>0$, and consequently, $a(b,s)$ lies {\em within} the Argand circle of Fig. \ref{fig:argand}.
%%%%%%%%%%%%%%%%%%%%%%%%%%%%%%%%%%%%%%
\subsection{The nuclear slope parameter $B$ in impact parameter space}
The nuclear slope parameter 
\be
B(s,t)\equiv \frac{d}{dt}\left(\ln\frac{d\sigma_{\rm el}}{dt}\right)
\ee
is most often evaluated at $t=0$, where it has the special name $B$.  Thus, 
\be
B=B(s)\equiv B(s, t=0)=\left.\frac{d}{dt}\left(\ln\frac{d\sigma_{\rm el}}{dt}\right)\right|_{t=0}.\label{Bof0}
\ee
Since 
\be
f_{\rm c.m.}\propto\int a(b,s) e^{i\vec q\cdot\vec b} \,d^2\vec b \quad\mbox{and \ }\frac{d\sigma}{dt}\propto |f_{\rm c.m.}|^2,\label{foft}
\ee
we expand $f_{\rm c.m.}$ about $q=0$ to obtain 
\be
f_{\rm c.m.}\propto\int a(b,s)\, d^2\vec b\left[1+i\vec q\cdot\vec b-\frac{1}{2}(\vec q\cdot\vec b)^2\cdots \right], \label{fexpansion}
\ee
so that substituting \eq{fexpansion} into \eq{Bof0}, where we take the logarithmic derivative of $d\sigma/dt$ at $t=0$, we find 
\be
B=\frac{\re [\int a(b,s)\,b\,db\times\int a^*(b,s)\, b^3\,db ]}{2\left|\int a(b,s)\, b \,db\right|^2}. \label{Bcomplicated}
\ee
A physical interpretation of \eq{Bcomplicated} in nucleon-nucleon scattering is that $B(s)$ measures the size of the proton at $s$, or more accurately, $B$ is one-half the average value of  $b^2$, weighted by $a(b,s)$.

When the phase of $a(b,s)$ is independent of $b$, so that $\rho=\re a(b,s)/\im a(b,s)$ is a function only of $s$---a useful example is when $a(b,s)$ factorizes into $a(b,s)=\alpha(s)\beta(b)$---we can write
\be
a(b,s)=\frac{\rho +i}{(1+\rho^2)^{1/2}}|a(b,s)|\label{a}
\ee
and  \eq{Bcomplicated} simplifies to
\be
B=\frac{1}{2}\frac{\int |a(b,s)|\, b^2\,d^2 \vec b}{\int| a(b,s)|\,d^2\vec b}.\label{Bsimple}
\ee
We find, when inserting \eq{a} into \eq{sigelofb} and \eq{sigtotofb1}, that 
\ba
\frac{\sigma_{\rm el}}{\sigtot}&=&(1+\rho^2)^{1/2}\frac{\int |a(b,s)|^2\,b\,db}{\int |a(b,s)|\,\,b\,db},\label{seloverstot}\\
\frac{\Sigma_{\rm el}}{\sigtot}&=&\frac{\sigtot (1+\rho^2)}{16\pi B},\quad\mbox{using \eq{sigtotoverB}},\nonumber\\
&=&(1+\rho^2)^{1/2}\frac{\left[\int |a(b,s)|\,\,b\,db\right]^2 }{\int |a(b,s)|\,\,b^3\,db }.\label{Seloverstot}
\ea

%%%%%%%%%%%%%%%%%%%%%%%%%%%%%%%%%%%%%%%%%%5
\subsubsection{$d\sigma_{\rm el}/dt$,\ $\sigma_{\rm el}$,\  $\sigtot$ and $B$ for a disk}

An important example is that of a disk with  a purely imaginary amplitude $a(b,s)=iA/2$ for $b\le R$, and $a(b,s)=0$ for $b>R$, where $0\le A\le2$, where a perfectly black disk has $A=1$. We get 
\ba
\frac{d\sigma_{\rm el}}{dt}&=&\pi A^2\left[ \int^R_0J_0(qb)b\,db \right]^2=\pi R^4A^2\left[ \frac{J_1(qR)}{ qR } \right]^2,\\
\sigma_{\rm el}&=&\pi R^2A^2,\\
\sigtot&=&2\pi R^2A,\\
\frac{\sigma_{\rm el}}{\sigtot}&=&\frac{\Sigma_{\rm el}}{\sigtot}=\frac{A}{2},\\
B&=&\frac{R^2}{4}.\\
\label{diskR}
\ea

For a black disk, $A=1$, and $\sigma_{\rm el}/\sigtot=\Sigma_{\rm el}/\sigtot=1/2$. When $s\rightarrow\infty$, many high energy scattering models have $\sigma_{\rm el}/\sigtot\rightarrow 1/2$, i.e., they approach the black disk ratio.
%%%%%%%%%%%%%%%%%%%%%%%%%%
\subsubsection{$d\sigma_{\rm el}/dt$,\ $\sigma_{\rm el}$,\  $\sigtot$ and $B$ for a Gaussian distribution}
Another important example is when the $b$ profile is Gaussian, with
\be
 a(b,s)=iA\exp[-2(b/R)^2].
\ee 
For the Gaussian, we find
\ba
\frac{d\sigma_{\rm el}}{dt}&=&4\pi A^2\left[ \int^\infty_0 bJ_0(qb)\exp[-2(b/R)^2] \,db \right]^2\nonumber\\
&=&\pi R^4 A^2 \exp[-(qR)^2/4]=\left(\frac{d\sigma_{\rm el}}{dt}\right)_{t=0}e^{-B|t|},\label{dsdtg}\\
\sigma_{\rm el}&=&\pi R^2A^2,\\
\sigtot&=&2\pi R^2A,\\
\frac{\sigma_{\rm el}}{\sigtot}&=&\frac{\Sigma_{\rm el}}{\sigtot}=\frac{A}{2}\\
B&=&\frac{R^2}{4},\label{gaussR}
\ea
where in the right-hand side of \eq{dsdtg}, we have reexpressed $d\sigma_{\rm el}/dt$  in terms of $B$ and $|t|=q^2$.   

We see that $\sigma_{\rm el}$, $\sigtot$, $\sigma_{\rm el}/\sigtot$, $B$ and $d\sigma_{\rm el}/dt)_{t=0}$ are all the same as for the case of a disk having the same $A$. The only difference is that for the Gaussian, the differential elastic scattering $d\sigma_{\rm el}/dt$ is a featureless exponential in $|t|$, whereas  for the disk, it  is the well-known diffraction pattern. 
 
We note from \eq{dsdtg} that  the logarithm of $d\sigma_{\rm el}/dt$ is a straight line in $|t|$,  whose slope is $B$. Further, after integration over $|t|$, we also find that $\sigma_{\rm el}$, the total elastic scattering cross section is given by $\Sigma_{\rm el}=(d\sigma_{\rm el}/dt)_{t=0}/B$. Indeed, this is the method experimenters use to  measure $B$ and $\Sigma_{\rm el}$. Thus, we see that the Gaussian profile in impact parameter space  gives rise to the exponential parametrization in $|t|$ that was used earlier in \eq{dsdt}, which in turn led to the definition of $\Sigma_{\rm el}$ in \eq{Siggreek2}.

Again, as we have pointed out, $\sigma_{\rm el}=\Sigma_{\rm el}$, as was true for the disk. Indeed, for most reasonable shapes of $a$, the ratio $\sigma_{\rm el}/\Sigma_{\rm el}$ is very close to unity.  In fact, the MacDowell-Martin bound\cite{macdowell} states that 
\be
\frac{\sigma_{\rm el}}{\Sigma_{\rm el}}\ge\frac{8}{9}. 
\ee
For a proof of this bound, using an impact-parameter representation, see Block and Cahn\cite{bc}, p. 573. 

Later, in Fig. \ref{fig:sigoverSIG} of Section \ref{section:seloverSel}, we will  plot the ratio $\sigma_{\rm el}/\Sigma_{\rm el}$ as a function of the c.m. energy $\sqrt s$, obtained from analyzing $\sigtot$, $\rho$ and $B$ data for $pp$ and $\bar pp$ collisions using the Aspen model, a QCD-inspired eikonal model, where the correction from unity is typically smaller than 1\%  for energies $\sqrt s >200$ GeV. 

%%%%%%%%%%%%%%%%%%%%%%%%%%%%%%%%%%%%%
\section{Eikonal amplitudes}\label{sec:eikonal}
The complex eikonal $\chi(b,s)=\chi_R+i\chi_I$ is conventionally defined as $\chi\equiv -2i\delta(b,s)$.  Hence, $\chi_R=2\delta_I$ and $\chi_I=-2\delta_R$.  We rewrite the amplitude $a(b,s)$ of \eq{aofb&s}
as 
\be
a(b,s)=\frac{e^{-\chi(b,s)}-1}{2i}=-\frac{e^{-\chi_R}\sin(\chi_I)}{2}+i\frac{1-e^{-\chi_R}\cos(\chi_I)}{2}. \label{aofchi}
\ee
Using the amplitude of \eq{aofchi} in \eq{sigelofb} and in \eq{sigtotofb1}, we find
\ba
\frac{d\sigma_{\rm el}}{dt}&=&\pi\left|\int J_0(qb)\left[ e^{-\chi(b,s)}-1\right]b\,db\right|^2,\\
\sigma_{\rm el}(s)&=&\int\left|e^{-\chi(b,s)}-1\right|^2\,d^2\vec b,\label{sigelofb2},\\
\sigtot(s)&=&2\int \left[1-e^{-\chi_R}\cos\left(\chi_I\right)\right]\, d^2\vec b,\label{sigtotofb2}
\ea 
respectively.
Again, using \eq{aofchi}, we find
\ba
\rho(s)&=&-\frac{\int e^{-\chi_R}\sin(\chi_I)\,d^2\vec b}{\int\left[ 1-e^{-\chi_R}\cos(\chi_I)\right]\,d^2\vec b}\quad, \label{rhoofb}\\
B(s)&=&\frac{1}{2}\frac{\int |e^{-\chi(b,s)}-1|b^2\,d^2\vec b}{\int |e^{-\chi(b,s)}-1|\,d^2\vec b}.
\ea

Since  eikonals are unitary, we have the important result that the cross sections of \eq{sigelofb2} and \eq{sigtotofb2} are guaranteed to satisfy unitarity,  a major reason for using an eikonal formulation. 

For nucleon-nucleon scattering , let us introduce a factorizable eikonal, 
\be 
\chi(b,s)=A(s)\times W(b),\label{chifactorize}
\ee
with $W(b)$ normalized so that $\int W(b)\,d^2\vec b=1$.

If we assume that the matter distribution in a proton is the same as the electric charge distribution\cite{durand} and is given by a dipole form factor
\be
G(q^2)=\left[\frac{\mu^2}{q^2+\mu^2}\right]^2,
\ee 
with $\mu^2=0.71$ (GeV/c)$^2$, then 
\ba
W(\mu b)&\propto&\frac{1}{(2\pi)^2}\int  G^2(q^2)e^{i\vec q\cdot \vec b}d^2\vec b
=\int \frac{1}{(2\pi)^2}\left[\frac{\mu^2}{q^2+\mu^2}\right]^2e^{i\vec q\cdot \vec b}d^2\vec b\nonumber\\
&\propto&(\mu b)^3K_3(\mu b).
\ea
When we require normalization, we find 
\be
W(\mu b)=\frac{\mu^2}{96\pi}(\mu b)^3K_3(\mu b).
\ee
\section{Analyticity}
We will limit ourselves here to a discussion of $pp$ elastic scattering, where $f(s,t)$, the physical amplitude for scattering, is defined only for Mandelstam values $s\ge4m^2$ and $-4m^2\ge t\ge 0$. It can be shown that $f(s,t)$ is the limit of a more general function ${\cal F}(s,t)$ where $s$ and $t$ can take on {\em complex} values. For details, see ref. \cite{jackson1, eden, martinandcheung, jackson2}.

\subsection{Mandelstam variables and crossed channels}
Consider the following three  reactions named  $s$-channel, $t$-channel and $u$-channel, of particles 1,2,3,4 having  masses $m_1, m_2,m_3$,$m_4$, :
\ba
1+2\rightarrow3+4,&&\qquad\mbox{$s$-channel}\label{schannel}\\
1+\bar 3\rightarrow\bar 2+4,&&\qquad\mbox{$t$-channel}\label{tchannel}\\
1+\bar 4\rightarrow\bar 3+\bar 2,&&\qquad\mbox{$u$-channel}.\label{uchannel}\nonumber
\ea
We then relate the properties for reactions in the following channels: 
\begin{description}  
\item[s-channel] the $t$-channel and the $u$-channel are the {\em crossed} channels.
\item [$t$-channel]the $u$-channel and the $s$-channel
  are the {\em crossed} channels.
\item[$u$-channel]the $s$-channel and the $t$-channel are the {\em crossed} channels. 
\end{description}
We have three mutually crossed channels,  where we associate each   particle with a 4-momentum $p_j$ as follows:
\begin{itemize}
\item  for the particle $j=1,2,3,4$, the $j$th particle has 4-momentum $p_j$.
\item for the anti-particle $\bar j$, for $\bar j=\bar 1,\bar 2,\bar 3,\bar 4$, the anti-particle has 4-momentum $-p_j$.  
\end{itemize}
Thus, in the $s$-channel and $u$-channel,  $p_3$ is the 4-momentum of particle 3, whereas in the $t$-channel, the 4-momentum of anti-particle 3 is $-p_3$.

We summarize:
\begin{itemize}
\item For the $s$-channel reaction $1+2\rightarrow3+4$, the c.m. energy squared is given by the squared sum of the initial state 4-momenta,  $(p_1+p_2)^2=s\ge (m_1+m_2)^2\ge 0$. The crossed channels, the $t$-channel and $u$-channel, have negative 4-momentum transfer squared, i.e., $t,u\le0$.

\item For the $t$-channel reaction $1+\bar 3\rightarrow\bar2+4$, the c.m. energy squared is given by the squared sum of the initial state 4-momenta, $(p_1-p_3)^2=t\ge (m_1+m_3)^2\ge 0$.  The crossed channels, the $u$-channel and $s$-channel,  have negative 4-momentum transfer squared, i.e., $u,s\le0$.

\item For the $u$-channel reaction  $1+\bar 4\rightarrow\bar3+4$, the c.m. energy squared is given by the squared sum of the initial state 4-momenta, $(p_1-p_4)^2=u\ge (m_1+m_4)^2\ge 0$. The  crossed channels, the $s$-channel and $t$-channel, have  negative 4-momentum transfer squared, i.e.,  $s,t\le0$.
\end{itemize}

The three Mandelstam variables, earlier introduced in Section \ref{sec:kinematics}, are given by $s=(p_1+p_2)^2$, $t=(p_1-p_3)^2$ and $u=(p_1-p_4)^2$, with the constraint that
\be 
s+t+u=m_1^2+m_2^2+m_3^3+m_4^2\label{sumstu},
\ee 
so that only 2 of the 3 Mandelstam variables are independent.

\subsection{Crossing symmetry}
Let particles 1,2,3,4 in the $s$-channel reaction of \eq{schannel} all be protons and consider the elastic $pp$ scattering reaction 
\be
p_1+p_2\rightarrow p_3+p_4,\label{pp}
\ee
which has the amplitude $f_{pp}(s,t)$.  In particular, we will only consider the amplitude for forward scattering, where $t=0$, i.e., $f_{pp}(s)$. 

When we consider the {\em crossed} reaction, the $u$-channel reaction of \eq{uchannel}, again for $t=0$, we study elastic $\ppbar$\ scattering reaction in the forward direction, i.e.,  
\be
p_1+\bar p_4\rightarrow p_3 +\bar p_2,\label{pbarp}
\ee
which has the amplitude $f_{p\bar p}(s, t=0)=f_{p\bar p}(s).$

The principle of crossing symmetry states that$f_{p\bar p}(s)$, the scattering amplitude for forward  elastic $p\pbar$ scattering,  is given  by $f_{pp}(u)$.  To get the reaction in \eq{pbarp}, we need to make the substitutions $2\rightarrow\bar 4$ and $4\rightarrow\bar2$, with $p_2\rightarrow -p_4$ and $p_4\rightarrow -p_2$, i.e., $s\rightarrow u$ and $u\rightarrow  s$, for $t=0$.  In other words,  if we know  the $pp$ scattering amplitude $f_{pp}(s)$, we know  $f_{p\bar p}(s)$, which is given by $f_{pp}(u)$.

Substituting  $t=0$ in \eq{sumstu}, we find 
\be
u=-s+4m^2.\label{uofs} 
\ee
Since $f_{pp}(s)=f_{p\bar p}(u)$, using \eq{uofs}, we find that
\be 
f_{p\bar p}=f_{pp}(-s+4m^2).\label{fpbarp1}
\ee
After evaluating $s$ in the laboratory frame as a function of the projectile energy $E$, and then using \eq{uofs},  for $t=0$ we find
\ba
s=&&2mE+2m^2\label{sofE},\\
u=&-&2mE+2m^2\label{uofE}.
\ea
From inspection of \eq{sofE} and \eq{uofE}, we see that when $E\rightarrow -E,\quad s\rightarrow u $ and $u\rightarrow s$. 

From here on in, to clarify the crossing symmetry, we will write  $f$ as a function of the variable $E$ rather than $s$, i.e.,  
\be
f_{p\bar p}(E)=f_{pp}(-E).
\ee
Simply put, crossing symmetry for forward elastic scattering states that the $\ppbar$\ scattering amplitude is obtained from the $pp$ scattering amplitude by the substitution $E\rightarrow -E$ and vice versa. 
%%%%%%%%%%%%%%%%%%%%%%%%%%%%%%%%%%%%%%%%%%%%
\subsection{Real analytic amplitudes}
As discussed earlier, $f_{pp}(s,t)$ is the limit of a more general analytic function ${\cal F}(s,t)$ when $s$ and $t$ take on complex values. Fixing $t=0$ and writing $f(s,t)$ as a function of the laboratory energy $E$, we have
\be
f_{pp}(E, t=0)=f_{pp}(E)=\lim_{\epsilon\rightarrow 0}{\cal F}(E+i\epsilon, t=0),
\ee
where $\epsilon>0$ (see Fig. \ref{fig:complexe}). The $p\bar p$ forward amplitude is found from crossing symmetry to be
\be
f_{p \bar p}(E, t=0)=f_{p\bar p}(E)=\lim_{\epsilon\rightarrow 0}{\cal F}(-E-i\epsilon, t=0),
\ee
again for  $\epsilon>0$, where we see from Fig. \ref{fig:complexe} that $\cal F$ approaches $f_{p\bar p}$ from below. 
%%%%%%%%%%%%%%%%%%%%%%%%%%\begin{figure}[h,t,b] %Fig. 2
\begin{figure}[h,t,b]
\begin{center}
\mbox{\epsfig{file=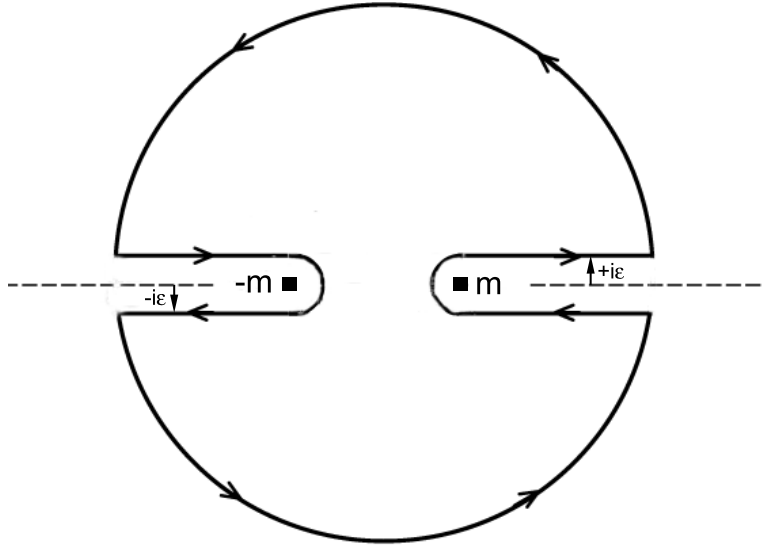%
            ,%width=6in,bbllx=122pt,bblly=403pt,bburx=494pt,bbury=660pt,clip=%
		width=4in,bbllx=24pt,bblly=174pt,bburx=590pt,bbury=580pt,clip=%
}}
\end{center}
\caption[Complex $E$ plane for $\pbarp$ and $pp$ scattering]
{ \footnotesize
Complex $E$ plane for $\pbarp$ and $pp$ scattering.   The physical cuts shown are from $m\le E\le\infty$ and $-\infty\le E\le -m$, where $m$ is the proton mass.  The unphysical cut and pion poles are not shown. Integration is  along the indicated contour and ignores the unphysical region. Note that the contour is really closed  by infinite semicircles.
}
\label{fig:complexe}
\end{figure}
%%%%%%%%%%%%%%%%%%%%%%%%%%%%%%%%%%%%%%%%%%%%%%%%%%%%%%%

To be more complete, the physical amplitudes $f_{pp}(E)$ and $f_{p\bar p}(E)$ are the limits of an analytic function ${\cal F}(E)$, for complex $E$, with physical cuts along the real axis from $m$ to $+\infty$ and $-\infty$ to $-m$, illustrated in Fig. \ref{fig:complexe}. At the very high energies we will be considering,  we will be very far from the ``unphysical'' two-pion cut and pion pole between $-m$ and $m$ that communicate with the $\ppbar$\ system and  we will ignore these singularities, since they will have negligible effect.   Then,  $\calF$ is real for $-m\ge E\ge m$, since there is no physical scattering for these energies, i.e., $\im {\calF}=0.$ An analytic function such as $\calF$ that is real on some segment of the real axis is called a real analytic function. 
%%%%%%%%%%%%%%%%%%%%%%%%%%%%%%%%%%%%%%%%%%%%%%%%%%%%%%%%  
\subsubsection{Schwarz reflection principle}
If $\calF (z)$ is real analytic, then the Schwarz reflection principle states that:
\be
\calF (z^*)=\calF ^*(z).\label{schwarz}
\ee

\begin{description}
\item[Proof:] Let $\calF (z)$ be analytic in some region, where this region includes a finite segment, however small, of the real axis. Define 
\calG\ to be $\calG(z)\equiv \calF^*(z^*)$. We  expand \calF\ in a power series, i.e.,   $\calF(z)=\sum_{i=0}^\infty a_{i}z^i$. Then, $\calG(z)=\sum_{i=0}^\infty a_{i}^*z^i$. Since the two series have the same radius of convergence, \calG\ is analytic. By construction, \calF\ and \calG\ have the same values where they coincide on the real axis. The principal of analytic continuation states that a function is uniquely determined by its values on a segment.  Hence, \calG\ and \calF\  are the {\em same} function. Thus, \calG$^*(z)=\calF^*(z)$ and \calG$^*(z)=\calF(z^*)$, so \calF$^*(z)=\calF(z^*)$. {\bf Q.E.D.}
\end{description}

If $\calF$\ has a cut along the real axis, the real part is the same on both sides of the cut, but the imaginary part changes sign, i.e., the discontinuity across the cut is imaginary, a result we will use in Section \ref{sec:dispersion}.

%%%%%%%%%%%%%%%%%%%%%%%%%%%%%%%%%%%%%%%%%%%%%%%%%%%%
\subsubsection{Construction of real analytic amplitudes}
Define the linear combinations  of $pp$ and \ppbar\ amplitudes in terms of even and odd amplitudes 
\be
f_{\pm}\equiv\frac{f_{pp} \pm f_{p\bar p}}{2},\label{evenodd}
\ee
such that $f_+(-E)=f_+(E)$ and $f_-(-E)=-f_-(E)$, i.e., $f_+$ is even and $f_-$ is odd under the exchange $E\rightarrow -E$. 

The analytic function 
\be
\calG_-(E)\equiv (m+E)^\alpha-(m-E)^\alpha\label{m+E}
\ee 
has the properties of a forward elastic scattering amplitude that has branch points at $\pm \,m$, with cuts from $-\infty$ to $-m$ and from $m$ to $+\infty$,  and is patently odd. We have  defined the function to be real on the real axis.  Hence, it is a real analytic amplitude and thus is a candidate for an odd elastic scattering amplitude. 

The phase structure of this amplitude is clarified in Fig. \ref{fig:cuts}. Just above the right-hand cut of Fig. \ref{fig:complexe},
\be
\calG_-(E)=(E+m)^\alpha-(E-m)^\alpha e^{-i\pi\alpha},\label{ppeverywhere}
\ee
and  for $E\gg m$,
\be
\calG_-(E)\approx 2i\sin(\pi\alpha/2)|E|^\alpha e^{-i\pi\alpha/2},\label{pphighE}
\ee%%%%%%%%%%%%%%%%%%%%%%%%%%\begin{figure}[h,t,b] %Fig. 1
\begin{figure}[h,t,b]
\begin{center}
\mbox{\epsfig{file=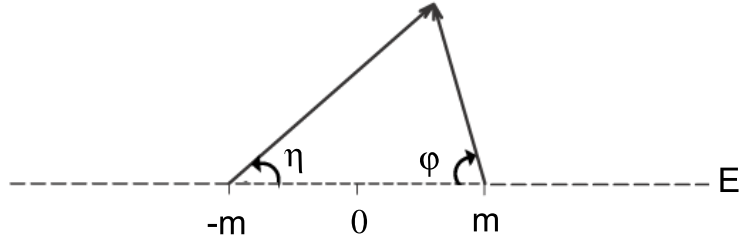%
            ,%width=6in,bbllx=122pt,bblly=403pt,bburx=494pt,bbury=660pt,clip=%
		width=4in,bbllx=35pt,bblly=345pt,bburx=575pt,bbury=530pt,clip=%
}}
\end{center}
\caption[The cut structure for ${\cal G}_-$ of \eq{m+E}]
{ \footnotesize
The cut structure for ${\cal G}_-$ of \eq{m+E}. ${\cal G}_-$ becomes well-defined by specifying that ${\cal G}_-=|m+E|^\alpha e^{i\eta\alpha}-|m-E|^\alpha e^{-i\phi\alpha}$.  For the $\pbarp$ amplitude, $\eta\rightarrow-\pi,\quad\phi\rightarrow0$. For the $pp$ amplitude, $\eta\rightarrow 0,\quad\phi\rightarrow \pi$. 
}
\label{fig:cuts}
\end{figure}
%%%%%%%%%%%%%%%%%%%%%%%%%%%%%%%%%%%%%%%%%%%%%%%%%%%
whereas just below the left-hand cut,
\be
\calG_-(E)=(-E-m)^\alpha e^{-i\pi\alpha}-(-E+m)^\alpha ,\label{pbarpeverywhere}
\ee
and  for $-E\gg m$,
\be
\calG_-(E)\approx -2i\sin(\pi\alpha/2)|E|^\alpha e^{-i\pi\alpha/2}.\label{pbarphighE}
\ee
If $\calG$ is the analytic continuation of the $pp$ amplitude, then the elastic $pp$ amplitude  $f_{pp}$ is given by \eq{ppeverywhere} and \eq{pphighE}, whereas the elastic \ppbar\  amplitude $f_{p\bar p}$ is given  by \eq{pbarpeverywhere} and \eq{pbarphighE}, since $f_{p\bar p}(E)=f_{pp}(-E)$.

For  $E\gg m$, the odd power-law scattering amplitude is given by
\be f_-(E)=2i\sin(\pi\alpha/2)E^\alpha e^{-i\pi\alpha/2},\label{f-odd}
\ee
which has the phase $ie^{-i\pi\alpha/2}$. 

A similar analysis for an even power law amplitude indicates that
\be
\calG_+(E)=(E+m)^\alpha+(E-m)^\alpha e^{-i\pi\alpha},\label{ppevenanywhere}
\ee
and  for $E\gg m$,
\be
\calG_+(E)\approx 2\sin(\pi\alpha/2)|E|^\alpha e^{-i\pi\alpha/2}.\label{ppevenhigh}
\ee
For  $E\gg m$, the even power-law scattering amplitude is given by
\be 
f_+(E)=2\cos(\pi\alpha/2)E^\alpha e^{-i\pi\alpha/2},\label{f+odd}
\ee
which has  the phase $e^{-i\pi\alpha/2}$. 

Other useful amplitudes are given in Table \ref{table:amps}.
%%%%%%%%%%%%%%%%%%%%%%%%%%
%  Table 2 
\begin{table}[h,t]                   % Use "table" environment, but also
				 % use  "tabular" environment below.
     \caption[A list of commonly used even and odd  real analytic amplitudes]{\protect\small A list of commonly used even and odd  analytic amplitudes. The amplitudes $\calF_{\pm}(E)$\  and their corresponding even and odd physical amplitudes $f_{\pm}(E)$, along with their corresponding even and odd physical amplitudes $\tilde f_{\pm}(E)$, where $\tilde f_{\pm}(E)$ is $f_{\pm}(E)$ when $E\gg m$, are shown. 
 $E_0$ is a scale factor. \label{table:amps}}

\def\arraystretch{1.5}            % Make the space between rows in the Table 1.5 x bigger than the default spacing.
\begin{center}	
\begin{tabular}[b]{l|l|l}
\hline\hline
\multicolumn{3}{c}{Even Amplitudes}\\
\hline
\multicolumn{1}{c|}{$\calF_+(E)$}&\multicolumn{1}{c|}{$f_+(E)$}&\multicolumn{1}{c}{$\tilde f_+(E)$}\\
\hline
$(E+m)^\alpha+(E-m)^\alpha$&$(E+m)^\alpha+(E-m)^\alpha e^{-i\pi\alpha}$&$E^\alpha e^{i\pi(1-\alpha)/2}$\\
$\sqrt{(m+E)(m-E)}$&$-ip$&$-iE$\\
$\frac{1}{2}\left\{\ln\left[(m-E)/E_0\right]\right.$&$\ln\left(p/E_0\right)-i\pi/2$&$\ln\left(E/E_0\right)-i\pi/2$\\
$\left.\qquad +\ln\left[(m+E)/E_0\right]\right\}$&&\\
$\left(\frac{1}{2}\left\{\ln\left[(m-E)/E_0\right]\right.\right.$&$\left[\ln\left(p/E_0\right)-i\pi/2\right]^2$&$\left[\ln\left(E/E_0\right)-i\pi/2\right]^2$\\
$\left.\left.\qquad +\ln\left[(m+E)/E_0\right]\right\}\right)^2$&&\\
\hline
\hline
\multicolumn{3}{c}{Power Law Odd Amplitude}\\
\hline
\multicolumn{1}{c|}{$\calF_-(E)$}&\multicolumn{1}{c|}{$f_-(E)$}&\multicolumn{1}{c}{$\tilde f_-(E)$}\\
\hline
$(m+E)^\alpha-(m-E)^\alpha$&$(E+m)^\alpha-(E-m)^\alpha e^{-i\pi\alpha}$&$iE^\alpha e^{i\pi(1-\alpha)/2}$   \\
\hline\hline
\end{tabular}
     %\vspace{1in} \\
\end{center}
\end{table}
\def\arraystretch{1}  %Restore the default row spacing in the Table.
%%%%%%%%%%%%%%%%%%%%%%%%%%%%%%%%%%%%%
%%%%%%%%%%%%%%%%%%%%%%%%%%%%%%%%%%%%%%%%%%
\subsubsection{Application of the Phragm\`en-Lindel\"of theorem to amplitude building}\label{phragmen}
An important generalization is possible using the Phragm\`en-Lindel\"of theorem\cite{titchmarsh}. Let us consider the amplitude as a function of $s$, rather than $E$.  For high energies, $s\rightarrow 2mE$. We rewrite the power-law odd and even amplitudes \eq{f-odd} and \eq{f+odd} as
\ba
f_-(s)&=&is^\alpha e^{-i\pi\alpha/2},\label{f-ofs}\\
f_+(s)&=&\ s^\alpha e^{-i\pi\alpha/2}.\label{f+ofs}
\ea
Using  the Phragm\`en-Lindel\"of theorem\cite{titchmarsh},   {\em any function} of $s$ can be made to have the proper phase by the substitution 
\ba s&\rightarrow& se^{-i\pi/2}\qquad\mbox{\ (for even functions),}\label{odd}\\
s&\rightarrow& ise^{-i\pi/2}\qquad\mbox{(for odd functions)},\quad\mbox{ when $s>>4m^2$}.\label{even}
\ea
More precisely, what we mean by the above substitution scheme is that first, one fashions the amplitude $f_1(s)$ that is desired as a function of $s$ only, {\em ignoring its phase}. Assuming that the amplitude $f_1(s)$ is to be transformed into an even amplitude $f_+(s)$, it is given by $f_+(s)=f_1(se^{-i\pi/2})$. To transform it into the  odd amplitude $f_-(s)$,  $f_-(s)=f_1(ise^{-i\pi/2})$. 

Obviously, \eq{f-ofs} and \eq{f+ofs} satisfy these substitution rules. If we rewrite  the amplitudes of Table \ref{table:amps} in terms of the variable $s$ (for large $s$, both $E, p\rightarrow s/2m$), we see that they also satisfy these rules, i.e, $p$ is replaced by $s\rightarrow sie^{-i\pi}=-is$. Replacing  $\ln(E/E_0)$ by $\ln(s/s_0)$,  we see that $\ln(s/s_0)\rightarrow\ln(-is/s_0)=\ln(se^{-i\pi/2}/s_0)=\ln(s/s_0)-i\pi/2$. Here, $s_0$ is a scale factor which makes the argument of the logarithm  dimensionless.

Making these substitutions are an easy way of guaranteeing  the analyticity of your high energy amplitudes, however complicated they may be. 
%%%%%%%%%%%%%%%%%%%%%%%%%%%%%%%%%%%%%%
\subsection{High energy real analytic amplitudes}\label{sec:analyticamplitudes}
We will divide the high energy real analytic amplitudes up into two groups, conventional amplitudes and ``odderons''.
%%%%%%%%%%%%%%%%%%%%%%%%%%
\subsubsection{Conventional high energy amplitudes}
  High energy even and odd forward scattering amplitudes---constructed of real analytic amplitudes from Table \ref{table:amps}---used by Block and Cahn\cite{bc} to fit high energy $pp$ and $\ppbar$\ forward scattering---are given by\footnote{We note that from here on in, the laboratory energy will be denoted interchangeably by $\nu$ or $E$, depending on context. Its usage will be clarified where necessary.}
\begin{equation}
\frac{4\pi}{p}f_+(s)=i\left\{A+\beta[\ln (s/s_0) -i\pi/2]^2+cs^{\mu-1}e^{i\pi(1-\mu)/2}\right\}\label{evenamplitude_gp}
\end{equation}
for the  crossing-even real analytic amplitude, where we have ignored the real subtraction constant $f_+(0)$, and
\be
\frac{4\pi}{p}f_-(s)=-Ds^{\alpha -1}e^{i\pi(1-\alpha)/2}\label{oddamplitude2}
\ee
for the crossing-odd real analytic amplitude. Here $\alpha < 1$ parameterizes the Regge behavior of the crossing-odd amplitude which vanishes at high energies and $A$, $\alpha$, $\beta$, $c$, $D$, $s_0$ and $\mu$ are real constants. The variable $s$ is the square of the c.m. energy and we now introduce $\nu$ as the laboratory energy. In \eq{evenamplitude_gp}} we have neglected the real constant $f_+(0)$, the subtraction constant\cite{bc} required at $\nu=0$.  In the high energy limit where \eq{evenamplitude_gp} and \eq{oddamplitude2} are valid, $s\rightarrow2m\nu$ where $m$ is the proton mass.

 From the optical theorem we obtain the total cross section
\be
\sigma^\pm= A+\beta\left[\ln^2 s/s_0-\frac{\pi^2}{4}\right]+c\,\sin(\pi\mu/2)s^{\mu-1}\pm D\cos(\pi\alpha/2)s^{\alpha -1}  \label{sigmatot}
\ee
with  $\rho$, the ratio of the real to the imaginary part of the forward scattering amplitude,  given by
\be
\rho^\pm={1\over\sigma_{\rm tot}}\left\{\beta\,\pi\ln s/s_0-c\,\cos(\pi\mu/2)s^{\mu-1}\pm D\sin(\pi\alpha/2)s^{\alpha -1}\right\},\label{rhogeneral}
\ee 
where the upper (lower) sign refer to $pp$ ($\bar p p$) scattering. The even amplitude describes the even cross section $(\sigma_{pp}+\sigma_{p\bar p})/2$. Later, we will invoke Regge behavior and fix $\mu=0.5$. 

Let us introduce the relations 
\ba
A &=& c_0 + \frac{\pi^2}{4}c_2 - \frac{c_1 ^ 2}{ 4c_2}\label{transform}\\ 
s_0 &=& 2m ^ 2 e^{-c_1 / (2c_2)}\label{s0}\\
\beta&=&c_2\\
c &=& \frac{(2m^2)^{1 - \mu} } {\sin(\pi\mu/ 2)}\beta_{\cal P'}\label{evenchanges} 
\ea
for the even amplitude, and
\ba
D&=&\frac{(2m^2)^{1-\alpha}}{\cos(\pi\alpha/2)}\delta\label{oddchanges}
\ea
for the odd amplitude, where $m$ is the proton mass.

For fixed $\mu$ and $\alpha$, these transformations make  \eq{sigmapm} linear in the real  coefficients $c_0,\  c_1,\  c_2, \ \beta_{\cal P'}$ and $\delta$, a useful property  in minimizing  a $\chi^2$ fit to the experimental total cross sections and $\rho$-values. The real coefficients $c_0,\  c_1,\  c_2,\  \beta_{\cal P'}$ and $\delta$ have dimensions of a cross section, which we will later take to be mb when fitting data. 
\eq{sigmatot} and \eq{rhogeneral} can now be rewritten as
 \begin{eqnarray}
\sigma^0&=&c_0+c_1\ln\left(\frac{\nu}{m}\right)+c_2\ln^2\left(\frac{\nu}{m}\right)+\beta_{\cal P'}\left(\frac{\nu}{m}\right)^{\mu -1},\label{sig0}\\
\rho^0&=&{1\over\sigma^0}\left\{\frac{\pi}{2}c_1+c_2\pi \ln\left(\frac{\nu}{m}\right)-\beta_{\cal P'}\cot\left({\pi\mu\over 2}\right)\left(\frac{\nu}{m}\right)^{\mu -1}+\frac{4\pi}{\nu}f_+(0)\right\}\label{rho0}\\
\sigma^\pm&=&\sigma^0\pm\  \delta\left({\nu\over m}\right)^{\alpha -1},\label{sigmapm}\\
\rho^\pm&=&{1\over\sigma^\pm}\left\{\frac{\pi}{2}c_1+c_2\pi \ln\left(\frac{\nu}{m}\right)-\beta_{\cal P'}\cot\left({\pi\mu\over 2}\right)\left(\frac{\nu}{m}\right)^{\mu -1}+\frac{4\pi}{\nu}f_+(0)\right.\nonumber\\
&&\left.\qquad\qquad\qquad\pm \delta\tan({\pi\alpha\over 2})\left({\nu\over m}\right)^{\alpha -1} \right\}\label{rhopm},\\
\frac{d\sigma^{\pm}}{d(\x}&=&c_1\left\{\frac{1}{(\x}\right\} +c_2\left\{ \frac{2\ln(\x}{(\x}\right\}+\beta_{\cal P'}\left\{(\mu-1)\x^{\mu-2}\right\}\nonumber\\
&&\ \ \ \ \ \ \ \ \ \ \ \ \ \ \ \ \ \ \ \ \  \pm \ \delta\left\{(\alpha -1)\x^{\alpha - 2}\right\},\label{derivpm}
\end{eqnarray} 
with $\sigma^0$ being the even cross section  and $\rho^0$  the even $\rho$-value (needed for an analysis of  $\gamma p$ scattering, which only has an even amplitude), and 
where the upper sign is for $pp$  and the lower sign is for $\bar p p$  scattering. For later use, we have included the first derivatives in the last line, in \eq{derivpm}. When applied to $\pi^\pm p$ scattering, one uses Eqns. (\ref{sig0})--(\ref{derivpm}) with $m$ being the pion mass, along with slight  modifications of Eqns. (\ref{transform})--(\ref{oddchanges}). The upper sign is for $\pi^+ p$ and the lower sign is for 
$\pi^- p$.

In the high energy limit, after using the transformations of \eq{transform} to \eq{oddchanges}, we can write $f_+(\nu)$ and $f_-(\nu)$, the conventional even and odd amplitudes of \eq{evenamplitude_gp} and \eq{oddamplitude2}, as
\ba
\frac{4\pi}{\nu}\im f_+(\nu)&=&c_0+c_1\ln\left(\frac{\nu}{m}\right)+c_2\ln^2\left(\frac{\nu}{m}\right)+\beta_{\cal P'}\left(\frac{\nu}{m}\right)^{\mu -1},\nonumber\\
\frac{4\pi}{\nu}\re f_+(\nu)&=&\left\{\frac{\pi}{2}c_1+c_2\pi \ln\left(\frac{\nu}{m}\right)-\beta_{\cal P'}\cot\left({\pi\mu\over 2}\right)\left(\frac{\nu}{m}\right)^{\mu -1}\right\},\label{imref+}
\ea
and
\ba
\frac{4\pi}{\nu}\im f_-(\nu)&=&- \delta\left({\nu\over m}\right)^{\alpha -1},\nonumber\\
\frac{4\pi}{\nu}\re f_-(\nu)&=&-\delta\tan\left({\pi\alpha\over 2}\right)\left({\nu\over m}\right)^{\alpha -1},\label{imref-}
\ea
results that we will utilize quite often later in fitting experimental data.
\subsubsection{Odderon amplitudes}\label{sec:odderon}
Using  $E$ as the laboratory energy, with  $\epsilon^{(0)},\epsilon^{(1)}$ and $\epsilon^{(2)}$ being real  constants, and introducing $s_0$ as a scale fctor, Block and Cahn\cite{bc} introduced three new ``odderon'' amplitudes, $f_-^{(0)}$, $f_-^{(1)}$, and $f_-^{(2)}$, constructed from odd amplitudes of Table \ref{table:odderons}, i.e., 
\ba
4\pi f_-^{(0)}&=&-\epsilon^{(0)}E \label{odd0}\\
4\pi f_-^{(1)}&=&-\epsilon^{(1)}E\left[\ln\frac{s}{s_o}-i\frac{\pi}{2}\right] \label{odd1}\\
4\pi f_-^{(2)}&=&-\epsilon^{(2)}E\left[\ln\frac{s}{s_o}-i\frac{\pi}{2}\right]^2 \label{odd2}.
\ea
They then fit data  by combining each  of the three ``odderon'' amplitudes  with the conventional odd amplitude $f_-(s)$ of \eq{oddamplitude2}
\be
\frac{4\pi}{p}f_-(s)=-Ds^{\alpha -1}e^{i\pi(1-\alpha)/2},\label{oddamplitude3}
\ee
to make a new total odd amplitude. Odderon 0 changes the $\rho$-values, but not the cross sections, since it is pure real.  Odderon 1 gives a constant cross section difference and odderon 2 gives a cross section difference growing as $\ln s$, as  well as  $\rho$-values that are not equal as $s\rightarrow\infty$, which is the maximal behavior allowed, from analyticity and the Froissart bound. Odderon 2 is often called the ``maximal'' odderon.

Odderon amplitudes were first introduced by Lukaszuk and Nicolescu\cite{nicolescu1} and later used by Kang and Nicolescu\cite{nicolescu2} and Joynson et al.\cite{nicolescu3}.  We will put stringent limits on the size of  the three ``odderon'' amplitudes in Eqns. ({\ref{odd0})--(\ref{odd1}) later when we will discuss fits to  experimental data in Section \ref{section:oddlimits}.

%%%%%%%%%%%%%%%%%%%%%%%%%%%%%%%%%%%%%%%%%%%%%%
%  Table  
\begin{table}[h,t]                   
    \caption[A list of ``Odderon'' amplitudes]{\protect\small A list of ``Odderon'' amplitudes. A special class of odd  analytic amplitudes $\calF_{-}(E)$\  and their corresponding odd physical amplitudes $f_{-}(E)$,  along with  their corresponding high energy odd physical amplitudes $\tilde f_{-}(E)$, where $\tilde f_{-}(E)$ is $f_{-}(E\gg m)$, are shown.  $E$ is the laboratory projectile energy and $E_0$ and $s_0$ are  scale factors. \label{table:odderons}}
\def\arraystretch{1.5}            % Make the space between rows in the Table 1.5 x bigger than the default spacing.
\begin{center}	
\begin{tabular}[b]{l|l|l}
\hline\hline
\multicolumn{3}{c}{Three Odderon Amplitudes}\\
\hline
\multicolumn{1}{c|}{$\calF_-(E)$}&\multicolumn{1}{c}{$f_-(E)$}&\multicolumn{1}{|c}{$\tilde f_-(E)$}\\
\hline
$(0)\quad E$&$E$&$E$\\
$(1)\quad E\left\{\frac{1}{2}\ln\left[(m-E)/E_0\right] \right.$&
$E\left[\ln\left(p/E_0\right)-i\pi/2\right]$&$E\left[\ln\left(s/s_0\right)-i\pi/2\right]$\\
$\quad\qquad\left.+\ln\left[(m+E)/E_0\right]\right\}$&&\\
$(2)\quad E\left(\frac{1}{2}\left\{\ln\left[(m-E)/E_0\right] \right.\right.$&$E\left[\ln\left(p/E_0\right)-i\pi/2\right]^2$&$E\left[\ln\left(s/s_0\right)-i\pi/2\right)]^2$\\
$\quad\qquad\left.\left.+\ln\left[(m+E)/E_0\right]\right\}\right)^2$&&\\
\hline\hline
\end{tabular}
\end{center}
\end{table}
\def\arraystretch{1}  %Restore the default row spacing in the Table
%%%%%%%%%%%%%%%%%%%%%%%%%%%%%%%%%%%%%
%%%%%%%%%%%%%%%%%%%%%%%%%%%%%%%%%%%%
%%%%%%%%%%%%%%%%%%%%%%%%%%%%%%%%%%%%%%%%%%%%%%%%
%%%%%%%%%%%%%%%%%%%%%%%%%%%%%%%%%%%%%%%%%%%%%%%%%
\subsection{Integral dispersion relations}\label{sec:dispersion}
Again, we restrict our discussion to  $pp$ and \ppbar\  scattering. We will use Cauchy's theorem to derive integral dispersion relations that   are relations between the real and the imaginary portions of the forward scattering amplitudes.  Let  $\calF(E)$, where $E$ is the laboratory energy,  be the analytic continuation of $f(E, t=0)$ and therefore is analytic in the region shown in Fig. \ref{fig:complexe}. Using Cauchy's theorem, we write
\be
\calF(E)=\frac{1}{2\pi i}\oint \frac{\calF(E\,')}{E\,'-E}\,dE\,',
\ee
where we use the counterclockwise contour shown, since it doesn't cross any of the cuts or encircle any poles. As mentioned earlier, we are neglecting the ``unphysical''  two-pion cut and pion pole, since we will be interested only in high energies, where their influence is very small. In the complex plane, the contours pass just above and below the cuts, at $\pm i\epsilon$, as seen in Fig. \ref{fig:complexe}. The semi-circles are really of infinite radius and we assume that  the contributions of these semicircles at $\infty$ vanish. Replacing $E$ by $E+i\epsilon'$ and evaluating $E\,'$ at the appropriate contour value of $E\,'\pm i\epsilon$ as shown in Fig. \ref{fig:complexe}, with $\epsilon'>\epsilon>0$,  we must evaluate the  four integrals
\ba
\calF(E+i\epsilon')&=&\frac{1}{2\pi i}\left[ \int_m^\infty \frac{\calF(E\,'+i\epsilon)}{E\,'-E-i(\epsilon'-\epsilon)}\,dE\,' \right]\ \ \qquad I\nonumber\\
&+&\frac{1}{2\pi i}\left[ \int^{-m}_{-\infty} \frac{\calF(E\,'+i\epsilon)}{E\,'-E-i(\epsilon'-\epsilon)}\,dE\,' \right]\qquad II\nonumber\\
&+&\frac{1}{2\pi i}\left[ \int_{-m}^{-\infty} \frac{\calF(E\,'-i\epsilon)}{E\,'-E-i(\epsilon'+\epsilon)}\,dE\,' \right]\qquad III\nonumber\\
&+&\frac{1}{2\pi i}\left[ \int^{m}_{\infty} \frac{\calF(E\,'-i\epsilon)}{E\,'-E-i(\epsilon'+\epsilon)}\,dE\,' \right]\quad\qquad IV,\label{FofE}\nonumber
\ea
where $\calF(E)$\  is found by first taking the limit of \eq{FofE} when $\epsilon\rightarrow 0$,  followed by taking the limit when $\epsilon'\rightarrow 0$.

Using Schwarz reflection for a real analytical amplitude yields $\calF(E\,'-i\epsilon)=\re \calF(E\,'+i\epsilon)-i\,\im \calF(E\,'+i\epsilon)$. Substituting this into integral $IV$ and interchanging the integration limits,  we can combine integral $I$  with integral $IV$ to give:
\ba
I+IV&=&\frac{1}{\pi }\left[\int_m^\infty\frac{\im \calF(E\,'+i\epsilon)}{E\,'-E-i(\epsilon'-\epsilon)}\,dE\,'\right].
\ea
We note that \[\lim_{\epsilon\rightarrow 0}\calF(E\,'+i\epsilon)= f(E\,'),\] so that letting $\epsilon\rightarrow 0$, we find
\ba
\lim_{\epsilon\rightarrow 0}(I+IV)&=&\frac{1}{\pi }\left[\int_m^\infty\frac{\im f(E\,')}{E\,'-E-i\epsilon'}\,dE\,'\right]\nonumber\\
&=&\frac{1}{\pi}\left[\int_m^\infty\frac{\im f(E\,')(E\,'-E+i\epsilon')}{(E\,'-E)^2+\epsilon'^2}\,dE\,'\right]\nonumber\\
&=&\frac{1}{\pi}\left[\int_m^\infty\frac{\im f(E\,')(E\,'-E)}{(E\,'-E)^2+\epsilon^2}\,dE\,' +i\epsilon'\int_m^\infty\frac{\im f(E\,')}{(E\,'-E)^2+\epsilon'^2}\,dE\,'\right].\label{I+IV}
\ea
In the limit $\epsilon'\rightarrow 0$,  the last term in \eq{I+IV} can be rewritten in  terms of the Dirac $\delta $ function  as
\be
\lim_{\epsilon'\rightarrow 0}\frac{i}{\pi}\int_m^\infty\frac{\epsilon'f(E\,')}{(E\,'-E)^2+\epsilon'^2}\,dE\,'=\frac{i}{2}\int_m^\infty\im f(E\,')\,\delta(E\,'-E)\,dE\,'.
\ee
We now let $\epsilon' \rightarrow 0$ in \eq{I+IV} to obtain
\ba
 \lim_{\epsilon',\epsilon\rightarrow 0}\,(I+IV)&= &\frac{1}{\pi}\int_m^\infty \frac{\im f(E\,')}{E\,'-E}\,dE\,'
+\frac{i}{2}\int_m^\infty\im f(E\,')\,\delta(E\,'-E)\,dE\,'\nonumber\\
&=&\frac{1}{\pi}\int_m^\infty \frac{\im f(E\,')}{E\,'-E}\,dE\,'
+\frac{i}{2}\im f(E).\label{fE'-E}
\ea
In a similar fashion, integrals $II$ and $III$ combine to give
\ba
 \lim_{\epsilon,\epsilon'\rightarrow 0}\,(II+III)
&=&\frac{1}{\pi}\int_m^\infty \frac{\im f(-E\,')}{E\,'+E}\,dE\,'
+\frac{i}{2}\int_m^\infty\im f(-E\,')\,\delta(E\,'+E)\,dE\,'\nonumber\\
&=&\frac{1}{\pi}\int_m^\infty \frac{\im f(-E\,')}{E\,'+E}\,dE\,'
+\frac{i}{2}\im f(E).\label{fE'+E}
\ea
Thus, summing \eq{fE'-E} and \eq{fE'+E}, we find
\ba
f(E)&=&\re f(E)+i\,\im f(E)\nonumber\\
&=&\frac{1}{\pi}\int_m^\infty \left[\frac{\im f(E\,')}{E\,'-E}+\frac{\im f(-E\,')}{E\,'+E}\right]\,dE\,'
+i\,\im f(E)\label{fofE},
\ea
resulting in the identity $\im f(E)=\im f(E)$ for the imaginary part, together with the principal value integral for the real part,
\be
\re f(E)=P\frac{1}{\pi}\int_m^\infty \left[\frac{\im f(E\,')}{E\,'-E}+\frac{\im f(-E\,')}{E\,'+E}\right]\,dE.\,'
\ee

If $f$ is an even function ($f_+(-E)=f_+(E)$), the real part of $f_+(E)$ is given by the principal value integral
\ba
\re f_+(E)&=&P\frac{1}{\pi}\int_m^\infty \im f_+(E\,')\left[
\frac{1}{E\,'-E}+\frac{1}{E\,'+E}\right]\,dE\,'\nonumber\\
&=& P\frac{1}{\pi}\int_m^\infty \im f_+(E\,')\left[
\frac{2E'}{E\,'^2-E^2}\right]\,dE\,'.\label{Peven}
\ea

If $f$ is an odd function ($f_-(-E)=-f_-(E)$), the real part of $f_-(E)$ is given by the principal value integral
\ba
\re f_-(E)&=&P\frac{1}{\pi}\int_m^\infty \im f_-(E\,')\left[
\frac{1}{E\,'-E}-\frac{1}{E\,'+E}\right]\,dE\,'\nonumber\\
&=& P\frac{1}{\pi}\int_m^\infty \im f_-(E\,')\left[
\frac{2E}{E\,'^2-E^2}\right]\,dE\,'\label{Podd}.
\ea

Since $f_{pp}\equiv(f_+-f_-)/2$ and $f_{\ppbar}\equiv(f_++f_-)/2$, after using the optical theorem of \eq{opticallab} we find that
\ba
\re f_{pp}(E)&=&P\frac{1}{4\pi^2}\int_m^\infty \left[\frac{\sigma_{pp}(E\,')}{E\,'-E}-\frac{\sigma_{p \bar p}(E\,')}{E\,'+E}\right]p'\,dE\,'.  \\
\re f_{p \bar p}(E)&=&P\frac{1}{4\pi^2}\int_m^\infty \left[\frac{\sigma_{p \bar p}(E\,')}{E\,'-E}-\frac{\sigma_{pp}(E\,')}{E\,'+E}\right]p'\,dE\,'.  \\
\ea

This is a rather slowly converging integral at $\pm\infty$ and isn't terribly useful, except for cross sections that are approaching zero more rapidly than $1/E$ as $E\rightarrow \infty$.  To achieve better convergence, we introduce  the {\em odd} scattering amplitude $g_-(E)\equiv f_+(E)/E$. After substituting it for $f_-$ in \eq{Podd}, taking into account the pole ($1/E\,'$) that was introduced  and finally,  multiplying both sides by $E$, we find the singly-subtracted dispersion relations given by  the principal value integrals
\be
\re f_+(E)=\re f_+(0)+P\frac{1}{\pi}\int_m^\infty \im f_+(E\,')\frac{E}{E\,'}\left[
\frac{1}{E\,'-E}-\frac{1}{E\,'+E}\right]\,dE\,', \label{f+single}
\ee
\be
\re f_-(E)=\re f_-(0)+P\frac{1}{\pi}\int_m^\infty \im f_-(E\,')\frac{E}{E\,'}\left[
\frac{1}{E\,'-E}+\frac{1}{E\,'+E}\right]\,dE\,'.\label{f-single}
\ee
 Both $f_+(E)$ and $f_-(E)$ are  real on the real axis between $-m$ and $m$. We note that  $\re f_-(0)=f_-(0)=0$, because $f_-(0)=-f_-(0)=0$. Therefore we see that \eq{f-single} collapses into \eq{Podd}, the unsubtracted odd amplitude,  thus giving us nothing new.  

From \eq{f+single} and \eq{f-single}, remembering that $f_-(0)=0$, the singly-subtracted dispersion relations for $pp$ and $\ppbar$ are given by the principal value integrals
\ba
\re f_{pp}(E)&=&  f_{+}(0)+P\frac{E}{4\pi^2}\int_m^\infty \left[\frac{\sigma_{pp}(E\,')}{E\,'-E}-\frac{\sigma_{p \bar p}(E\,')}{E\,'+E}\right]\frac{p'}{E\,'}\,dE\,',\label{ppsingle} \\
\re f_{p \bar p}(E)&=&f_{+}(0)+P\frac{E}{4\pi^2}\int_m^\infty \left[\frac{\sigma_{p \bar p}(E\,')}{E\,'-E}-\frac{\sigma_{pp}(E\,')}{E\,'+E}\right]\frac{p'}{E\,'}\,dE\,', \label{pbarpsingle}
\ea 
 where we have introduced the subtraction constant $f_+(0)$ and the laboratory momentum $p'=\sqrt{E\,'^2-m^2}$. The singly-subtracted dispersion relations converge more rapidly, but the price you pay is the evaluation of one additional parameter, the subtraction constant $f_+(0)$.

At high energies , where  $s\rightarrow 2mE$, let us replace the variable $E$ by the  the invariant variable $s$.  If $\sigma_{p \bar p}\rightarrow\sigma_{pp}$ at large $s$, then the singly-subtracted dispersion relations of \eq{ppsingle} and \eq{pbarpsingle} converge for cross sections that asymptotically grow as fast as $s^\alpha$, if $\alpha < 1$.

To get even higher convergence, we can evaluate doubly-subtracted dispersion relations by introducing the  odd amplitude $g_-(E)=f_-(E)/E^2$ into in \eq{Podd}, carefully taking into account the double pole ($1/E\,'^2$) we have introduced and finally, multiplying both sides by $E^2$.  We write them as the principal value integrals
\ba
\re f_{pp}(E)&=&f_{+}(0)+E\left.\frac{df_{pp}}{dE}\right|_{E=0}+P\frac{E^2}{4\pi^2}\int_m^\infty \left[\frac{\sigma_{pp}(E\,')}{E\,'-E}-\frac{\sigma_{p\bar p}(E\,')}{E\,'+E}\right]\frac{p'}{E\,'^2}\,dE\,',\label{ppdouble} \\
\re f_{p \bar p}(E)&=& f_{+}(0)-E\left.\frac{df_{pp}}{dE}\right|_{E=0}+P\frac{E^2}{4\pi^2}\int_m^\infty \left[\frac{\sigma_{p \bar p}(E\,')}{E\,'-E}-\frac{\sigma_{pp}(E\,')}{E\,'+E}\right]\frac{p'}{E\,'^2}\,dE\,'.\label{pbarpdouble}
\ea 

We have two real subtraction constants, $f_{+}(0)$ and $(df_{pp}/dE)_{E=0}=-(df_{p\bar p}/dE)_{E=0}$, to evaluate in \eq{ppdouble} and \eq{pbarpdouble}, so here the price one pays for faster convergence is the evaluation of two additional constants.

We have been a bit cavalier about always assuming that the integrals along the infinite semicircles vanish, and sometimes, care must be taken to assure this. For example, if we use the obviously odd function $f_-(E)=E$ in the unsubtracted dispersion relation of \eq{Podd}, we get the nonsense answer $\re E=E=0$, since the imaginary part of $E$ is zero---the principal value integral clearly converges since it vanishes everywhere. Since the singly-subtracted odd dispersion relation of \eq{f-single} collapses into \eq{Podd}, we must use   one-half the difference of \eq{ppdouble} and \eq{pbarpdouble} (the doubly-subtracted dispersion integrals) for the odd dispersion relation.  The principal value integrals vanish identically because the imaginary portion of $E$ is zero. Since $(df_{pp}/dE)_{E=0}=dE/dE=1$, we find that $E=E$, a comforting tautology.  In this case, the contribution of the infinite semi-circular contours does vanish and we now get the right answer.

A brief history of applications of dispersion relations to $pp$ and $\ppbar$\ scattering is in order. By the early 1960's,  the experimental cross sections for $pp$ and $\ppbar$\ cross sections and $\rho$-values were known up to $\sqrt s<6$ GeV. S\"oding\cite{soding}, in 1964, was the first to use dispersion relations to analyze $pp$ and $\ppbar$ interactions, using a singly-subtracted dispersion relation that took  into account the unphysical regions by a sum over poles.  For c.m. energies $\sqrt s<4.7$ GeV, experimental cross sections were used that were numerically inserted into the evaluation of the principal value integrals. For higher energies, the cross sections were parametrized by asymptotic power laws, under the assumption, then widely held, that the cross section was approaching a constant value. The $\rho$-values for both $pp$ and $\ppbar$ scattering were calculated from his fit.  

The experimental situation  had  markedly changed a decade later. Perhaps the most important  physics contribution of the CERN ISR in the early 1970's  was the discovery that the $pp$ cross section was rising at c.m. energies above $\approx 20$ GeV. By the mid-1970's,  data were available for $pp$ cross sections and $\rho$ values for  interactions up to $\sqrt s=62$ GeV and for $\ppbar$\ interactions up to $\sqrt s=15$ GeV.  In 1977, Amaldi et. al.\cite{amaldi1} used a singly-subtracted dispersion relation to predict $\rho_{pp}$-values, but did not use any of the lower energy $\rho_{p\bar p}$ values. They employed a different strategy from S\"oding---they (a) neglected the unphysical region, (b) did not use experimental cross sections, but rather parametrized them by 
\ba
\sigma_{pp}&=&B_1+C_1E^{-\nu_1}+B_2\ln^\gamma s-C_2E^{-\nu_2},\label{amaldi1}\\
\mbox{and}\nonumber\\
\sigma_{p\bar p}&=&B_1+C_1E^{-\nu_1}+B_2\ln^\gamma s+C_2E^{-\nu_2},\label{amaldi2}
\ea 
with $E$ in GeV and $s$ in (Gev)$^2$, inserting these analytic forms into the dispersion relation. 

They made a $\chi^2$ fit simultaneously to the data for $\sigma_{pp}$, $\sigma_{p\bar p}$ and $\rho_{pp}$, using data from $5<\sqrt s<62$ GeV, fitting 8 real parameters: the even parameters $B_1,\ C_1,\ \nu_1,\ B_2,\ \gamma,$ as well as the odd parameters $\ C_2,\ \nu_2$,  along with the subtraction constant associated with a singly-subtracted dispersion relation. They arbitrarily chose 1 GeV$^2$ for the scale of $s$, rather than fitting $s/s_0$, which would have been a more proper procedure, since it would allow the experimental data to determine the scale of $s$. In spite of this, their fit gave  reasonable agreement with the newly-measured high energy $pp$ $\rho$-values. Since they did {\em not} use any experimental data in their dispersion relation, they could have achieved their goal in a much more simple and elegant form through the use of real analytic functions
 which obviate the computational need to evaluate numerically  a principal value integral for each of the multitude of times  a $\rho$-value is called for in a $\chi^2$ minimization program. 

Indeed, \eq{evenamplitude_gp} with the term $\beta[\ln(s/s_0-i\pi/2)]^2$ replaced by the term  $\beta [\ln(s/s_0-i\pi/2)]^\gamma$, and \eq{oddamplitude2} are examples of real analytic amplitudes which reproduce the cross section energy dependence of \eq{amaldi1} and \eq{amaldi2}. The appropriate linear combination of the  imaginary parts of \eq{evenamplitude_gp} and \eq{oddamplitude2} give the cross sections, \eq{amaldi1} and \eq{amaldi2}. The $\rho$ value for $pp$ and $\ppbar$\ interactions is {\em immediately} found in an analytic form  by taking the ratio of the real to the imaginary parts of these linear combinations, eliminating the numerical integration of an enormous number of principal value integrals, a great computational advantage.  Thus, the same results can be achieved using real analytic amplitudes in a much simpler calculation.  

In 1983, Del Prete\cite{delprete} hypothesized that the {\em difference} of the $pp$ and the $\ppbar$\ cross sections grew asymptotically as $\ln s$. In this case, as we have seen earlier, the singly-subtracted dispersion relations given in \eq{ppsingle} and \eq{pbarpsingle}  {\em do not converge} and the doubly-subtracted relations of \eq{ppdouble} and \eq{pbarpdouble} are required for convergence.  Since Del Prete claimed that he used the singly-subtracted relation of S\"oding, the analysis cannot be correct. The reported results are presumably the artifacts of the numerical integration routines that were used.
%%%%%%%%%%%%%%%%%%%%%%%%%%%%%%%%%%%%%%%%%%%%%%%%%%%%
\subsection{Finite energy sum rules}
We restrict ourselves to forward scattering, where $t=0$. 
The finite energy sum rules (FESR)  given by
\be
S_n(N)\equiv \int^N_0\nu^n\im f(\nu)\,d\nu=\sum_j \beta_j\frac{ N^{\alpha_j+n+1}}{\alpha_j+n+1}\label{sumrule}
\ee
were derived by Dolen, Horn and Schmid\cite{dolen-horn-schmid} in 1968.  In \eq{sumrule},  $\nu$ is  the laboratory energy and $n$ is integer, so that then $n^{\rm th}$ moment is given by $\nu^n\im f(E)$. $N$ is a finite, but high energy, cutoff (hence, the name Finite Energy Sum Rule).  They used a Regge amplitude normalized so that 
\be
R_j(\nu)=\beta_j \frac{\pm 1-e^{-i\pi\alpha_j}}{\sin\pi\alpha_j}\nu^\alpha_j,
\ee 
with $\alpha_j=\alpha_j(t=0)=\alpha_{0_j}$ and $\beta_j=\beta_j(t=0)$, and the upper (lower) sign for odd (even) amplitudes. \eq{sumrule} is useful if the high energy behavior ($\nu\ge N$) can be expressed as a sum of a few Regge poles. 

We sketch below their derivation in which they only considered functions that can be expanded at high energies $\nu>N$ as a sum of Regge poles. The finite-energy sum rules are the consistency  conditions that analyticity imposes on these functions.

Consider an odd amplitude $f_-(\nu)$  that obeys the unsubtracted dispersion relation 
\be
\re f_-(\nu)=\frac{2\nu}{\pi}\int_0^\infty\frac{\im f_-(\nu')}{\nu'^2-\nu^2}\,d\nu,\label{fodd2}
\ee
which is the dispersion relation which we would have found in deriving \eq{Podd} had we let the contour in Fig. \ref{fig:complexe} approach arbitrarily close to zero energy, i.e., we replace the lower limit $m$ in the integral by 0. If the leading Regge term of the expansion of $f_-(\nu)$ has $\alpha<-1$, the super-convergence relation
\be
\int^\infty_0\im f_-(\nu)\,d\nu=0,  \qquad\mbox{for  $\alpha<-1$}\label{super}
\ee 
is obeyed.  However, if the leading term $R(\nu)$ has $1>\alpha>-1$, subtract it out of $f_-(\nu)$.  Now $R(\nu)$ satisfies the unsubtracted dispersion relation, \eq{fodd2}, so we now write
\be
 R(\nu)=\frac{2\nu}{\pi}\int_0^\infty\frac{\im R(\nu')}{\nu'^2-\nu^2}\,d\nu \qquad\mbox{for all $-1<\alpha<1$}.\label{R}
\ee
Therefore, we can now write
\be
 f_-(\nu)-R(\nu)=\frac{2\nu}{\pi}\int_0^\infty\frac{\im f_-(\nu')-\im R(\nu')}{\nu'^2-\nu^2}\,d\nu,  \qquad\mbox{for all $-1<\alpha<1$},\label{difference}
\ee
and hence, the difference of amplitudes $f_-(\nu)-R(\nu)$ satisfies the super-convergence relation
\be\int_0^\infty\left[\im f_-(\nu)-\im R(\nu)\right]\,d\nu=0, \qquad\mbox{for all $\alpha<1$},\label{R-f}
\ee
even if neither of them satisfy it alone. 

Using the notation that $\beta_{-1}$ corresponds to the pole at $\alpha=-1$ and replacing $R$ by a sum of poles with $\alpha_j>-1$, we see that
\be
\int_0^\infty\left[\im f_-(\nu)-\sum_{{\alpha_{j}>-1}}\beta_j\nu^{\alpha_j}\right]\,d\nu=\beta_{-1}.\label{intf-R}
\ee
Although neither integrand converges, their difference is convergent. 
In order to demonstrate \eq{intf-R} in a manifestly convergent form, we will cut off the integration at some maximum energy $\nu_{\rm max}=N$ and use those Regge terms with $\alpha_j<-1$ for the high energy behavior. We now rewrite \eq{intf-R} as
\be
\int_0^N\left[\im f_-(\nu)-\sum_{{\alpha_{j}>-1}}\beta_j\nu^{\alpha_j}\right]\,d\nu +\int_N^\infty\sum_{{\alpha_{j}<-1}}\beta_j\nu^{\alpha_j}\,d\nu=\beta_{-1},\label{intf-R2}
\ee 
after splitting off the high energy contribution of those Regge poles with $\alpha_j<-1$.  After evaluating the integrals of the Regge terms in  \eq{intf-R2}, we obtain the finite energy sum rules
\ba
S(N)&=&\int_0^N\im f_-(\nu)\,d\nu\nonumber\\ 
&=&\sum_{{\alpha_{k}>-1}}\beta_k\frac{N^{\alpha_k+1}}{\alpha_k+1}+\sum_{{\alpha_{j}<-1}}\beta_j\frac{N^{\alpha_j+1}}{\alpha_j+1}+\beta_{-1}\nonumber\\
&=&\sum_{{{\rm all\ }\alpha_{k}}}\beta_k\frac{N^{\alpha_k+1}}{\alpha_k+1},\label{SofN}
\ea
which we see is the FESR of \eq{sumrule} for $n=0$.  The generalization of \eq{SofN} to \eq{sumrule}, for all even integer $n$---where we use odd amplitudes $f_-(\nu)$---is straightforward.  The extension to odd integer $n$, using  even amplitudes $f_+(\nu)$, is also straightforward.
It should be  emphasized that for all moments, the relative importance of successive terms in the FESR is the same as that for the usual Regge expansion at high energies.  If a secondary pole or cut is unimportant in the high energy expansion, it is unimportant to exactly the same extent in the FESR. 

Further, in \eq{sumrule}, for $0\le\nu\le N$, the Regge representation has {\em not} been used in $\im f(\nu)$ which appears in the integrand. The integral that appears in \eq{sumrule} can be broken into two parts, an integration over the `unphysical' region ($0\le \nu < m$) and an integration over the physical region $m< \nu \le N$. Using the optical theorem in the integral over the physical region, we can rewrite \eq{sumrule} as
\be
 \int^m_0\nu^n\im f(\nu)\,d\nu+\frac{1}{4\pi}\int^N_m\nu^np\sigtot(\nu)\,d\nu=\sum_j \beta_j\frac{ N^{\alpha_j+n+1}}{\alpha_j+n+1}\label{sumrule2},
\ee 
where $p$ is the laboratory momentum and $n$ is integer. The practical importance of \eq{sumrule2} is that one can now use the  rich amount of very accurate {\em experimental} total cross section data, substituting it for $\sigtot(\nu)$ in the integral over the physical region and then  evaluating the integral numerically. 

The above Finite Energy Sum Rules of \eq{sumrule2}---using moments of integer $n$---were later extended to {\em continuous} moments (effectively by making $n$ continuous) by Barger and Phillips\cite{bargerandphillips} and used successfully in investigations of hadron-hadron scattering.
%%%%%%%%%%%%%%%%%%%%%%%%%%%%%%%%%%%%%%%%%%
\subsubsection{FESR(2), an even amplitude FESR for nucleon-nucleon scattering }\label{section:igi}
Recently, Igi and Ishida developed finite energy sum rules for pion-proton scattering\cite{igiandishidapip}  and for $pp$ and $\ppbar$\ scattering\cite{igiandishidapp}. At {\em high energies} they  fit  the even cross section $\sigtot^+(\nu)$  for $pp$ and $\ppbar$\ with real analytic amplitudes, constraining the  fit coefficients by using a FESR which exploited the very  precise experimental cross section information, $\sigtot(pp)$ and $\sigtot(\ppbar)$, available for {\em low energy} scattering. 

Their derivation of their FESR used a slightly different philosophy from that of Dolen, Horn and Schmid, in that Igi and Ishida used terms for the high energy behavior that involved  non-Regge amplitudes, in addition to Regge poles. We here outline their derivation\footnote{We have changed their notation slightly. In what follows, $m$ is the proton mass, $p$ is the laboratory momentum, $\nu$ is the laboratory energy, $\mu$ is the Regge intercept and the transition energy $N$ is replaced by $\nu_0$.}
 of the rule that they called FESR(2)\cite{igiandishidapp}. For the high energy behavior for $pp$ and $\ppbar$, they used a cross section that   corresponds to  multiplying a factor of $4\pi/m^2$ times the even real analytic amplitude of \eq{imref+} that we discussed earlier, i.e., they used 
\be
\tilde\sigtot^+(\nu)=\frac{4\pi}{m^2} \left[c_0+c_1\ln(\nu/m)+c_2\ln^2(\nu/m)+\betaP(\nu/m)^{\mu-1}\right],\label{sig+}
\ee
valid in the high energy region  $\nu > \nu_0$. From \eq{imref+} (after dividing the amplitude by $4\pi/m^2$),  we see that the imaginary and real portions of $\tilde f_+(\nu)$ are given by
\ba
\im \tilde f_+(\nu)&=&\frac{\nu}{m^2}\left[c_0+c_1\ln\left(\frac{\nu}{m}\right)+c_2\ln^2\left(\frac{\nu}{m}\right)+\betaP\left(\frac{\nu}{m}\right)^{\mu-1}\right],\label{imagf+}\\
\re \tilde f_+(\nu)&=&\frac{\nu}{m^2}\left[\frac{\pi}{2}c_1+c_2\pi \ln\left(\frac{\nu}{m}\right)-\beta_{\cal P'}\cot\left({\pi\mu\over 2}\right)\left(\frac{\nu}{m}\right)^{\mu -1}\right],\label{realf+}
\ea
making the real coefficients $c_0,\ c_1,\ c_2$ and $\betaP$ in \eq{imagf+} and \eq{realf+}  all dimensionless.
Igi and Ishida used a  Regge trajectory with intercept $\mu=0.5$. We note that the non-Reggeon portions of their asymptotic amplitude are given by $\frac{\nu}{m^2}[c_0+c_1\ln(\nu/m)+c_2\ln^2(\nu/m)]$. 

Let us define $f_+(\nu)$ as the {\em true} even forward scattering amplitude, valid for all $\nu$.  In terms of the forward scattering amplitudes for $pp$ and $p\bar p$ collisions, we define
\be
f_+(\nu)\equiv\frac{f_{pp}(\nu)+f_{p\bar p}(\nu)}{2}.
\ee
Using the optical theorem, the imaginary portion of the even amplitude is related to the physical even cross section $\sigtot^+ (\nu)$ by
\be
\im f_+(\nu)\equiv\frac{p}{4\pi}\sigtot^+ (\nu),\qquad\mbox{for $\nu\ge m$},\label{sig+1}
\ee
with the laboratory momentum  given by $p=\sqrt{\nu^2-m^2}$, where $m$ is the proton mass. Of course, the problem is that we do not really know the true  amplitude $f_+(\nu)$ for all energies, but rather are attempting to parametrize it adequately at high energies.

They then define the super-convergent odd amplitude $\nu\hat f_+(\nu)$ as
\ba
\nu\hat f_+(\nu)&\equiv& \nu\left[f_+(\nu)-\tilde f_+(\nu)\right]
,\label{fhat}
\ea
for {\em all} $\nu$. In analogy to the $n=1$ FESR of \eq{sumrule} which requires the odd amplitude $\nu f_+(\nu)$, we now insert the odd amplitude $\nu\hat f_+(\nu)$ into the equivalent of \eq{super}, i.e., we write the super-convergence integral as 
\be
\int_0^{\nu_0}\nu\, \im \nu\left[f_+(\nu)-\tilde f_+(\nu)\right]\,d\nu=0,\label{super2}
\ee 
or,
\be
\int_0^{\nu_0}\nu\, \im f_+(\nu)\,d\nu=\int_0^{\nu_0}\nu\, \im \tilde f_+(\nu)\,d\nu.\label{f=ftilde}
\ee
We break up the left-hand integral of \eq{f=ftilde} into two parts, the integral from $0$ to $m$ (the `unphysical' region) and the integral from $m$ to ${\nu_0}$, the physical region. Using the optical theorem, after changing variables in the physical integral from $\nu$ to $p$ and inserting  \eq{sig+1} into its integrand, we find
\ba
\int_0^m\nu\, \im f_+(\nu)\,d\nu+\frac{1}{4\pi}\int_m^{\tilde {\nu_0}} p^2\sigtot^+(p)\,dp&=&\int_0^{\nu_0}\nu\,\im \tilde f_+(\nu)\, d\nu,\nonumber\\\label{intftilde}
\ea
where $\tilde {\nu_0}\equiv\sqrt{{\nu_0}^2-m^2}$.
Substituting the high energy amplitude (\eq{imagf+}) into the right-hand integral of \eq{intftilde} and then evaluating the high-energy integral, we finally have the sum rule that Igi and Ishida called FESR(2):
\ba
\int_0^m\nu \,\im f_+(\nu)\,d\nu+\frac{1}{4\pi}\int_m^{\tilde {\nu_0}} p^2\sigtot^+(p)\,dp&=&\int_0^{\nu_0} \left(\frac{\nu}{m}\right)^2\left[c_0+c_1\ln\left(\frac{\nu}{m}\right)+c_2\ln^2\left(\frac{\nu}{m}\right)\right.\nonumber \\
&&\left.\qquad\qquad +\beta_{\cal P'}\left( \frac{\nu}{m}\right)^{\mu -1}\right]\,d\nu\nonumber
\ea
\ba 
=\frac{m}{3}\left(\frac{{\nu_0}}{m}\right)^3\left\{
\vphantom{\left.+\frac{6}{5}\left(\frac{{\nu_0}}{m}\right)^{\mu -1}\betaP\right]}
c_0+\left[-\frac{1}{3}+\ln\left(\frac{{\nu_0}}{m}\right)\right]c_1+\left[\frac{2}{9}-\frac{2}{3}\ln\left(\frac{{\nu_0}}{m}\right)+\ln^2\left(\frac{{\nu_0}}{m}\right)\right]c_2\right.&&\nonumber\\
\left.+\frac{6}{5}\left(\frac{{\nu_0}}{m}\right)^{\mu -1}\betaP\right\}.\qquad\qquad\mbox{FESR(2)}&&\label{ksqintegralfinal}
\ea
The authors used experimental cross sections to numerically evaluate the second integral on the left-hand side of \eq{ksqintegralfinal}, obtaining 
\be
\frac{1}{4\pi}\int_m^{\tilde {\nu_0}} p^2\sigtot^+(p)\,dp=3404\pm 20\ {\rm GeV}.
\ee
They also numerically estimated  the first integral (the `unphysical' integral) on the left-hand side to be   $\int_0^m\nu f_+(\nu)\,d\nu\simeq 3.2$ GeV, negligible compared to the second term, $3404\pm 20$ GeV. Neglecting  the contribution of the integral $\int_0^m\nu \im f_+(\nu)\,d\nu$, the final form of \eq{ksqintegralfinal}  is
\ba
c_0+\left[-\frac{1}{3}+\ln\left(\frac{{\nu_0}}{m}\right)\right]c_1+\left[\frac{2}{9}-\frac{2}{3}\ln\left(\frac{{\nu_0}}{m}\right)+\ln^2\left(\frac{{\nu_0}}{m}\right)\right]c_2+\frac{6}{5}\left(\frac{{\nu_0}}{m}\right)^{\mu -1}&&\nonumber \\
=\frac{3}{m}\left(\frac{m}{{\nu_0}}\right)^3\times\frac{1}{4\pi}\int^{\bar {\nu_0}}_0 k^2\sigma_{\rm even}(k)\,dk.
\hphantom{\frac{3}{m}\left(\frac{m}{{\nu_0}}\right)^3\times\frac{1}{4\pi}\int^{\bar {\nu_0}}_0 k^2}
&&\label{finalform}
\ea
They chose  $\tilde {\nu_0}=10$ GeV as the upper limit, so that  ${\nu_0}=10.043$ GeV.  Clearly, their FESR(2) result should be essentially independent of their choice of $\nu_0$, an energy that should be above the resonance region. Numerically, \eq{finalform} reduces to
\be
c_0+2.04 c_1+4.26c_2+0.367\beta_{\cal P'}=8.87,\label{constraint2}
\ee
where the parameters $c_0,\ c_1,\ c_2$ and $\betaP$\  are dimensionless. 
Later, we will use a fit where we parametrize the high energy cross section as
\be
\sigtot^+(\nu)=c_0+c_1\ln\left(\frac{\nu}{m}\right)+c_2\ln^2\left(\frac{\nu}{m}\right)+\betaP\left(\frac{\nu}{m}\right)^{\mu-1}.\label{sigblock}\hphantom{\left[{\nu_0}\sigtot^+({\nu_0})-m\sigtot^+(m)\right]   }
\ee
These coefficients $c_0,\ c_1,\ c_2$ and $\betaP$ in \eq{sigblock} are in mb. The factor 8.87 in \eq{constraint2} has to be multiplied by $0.389\times 4\pi/m_p^2=5.56$, if the constraint of \eq{constraint2} is to be used in conjunction with the coefficients appearing in \eq{sigblock}, which have units of mb. Thus, the appropriate constraint equation FESR(2) to be used  with \eq{sigblock}, where the coefficients are in mb, is
\be
c_0+2.04 c_1+4.26c_2+0.367\beta_{\cal P'}=49.3 {\rm \ mb,}\label{final2}
\ee
a result we will use later in Section \ref{section:compare} when we compare results using FESR(2) to  those using the new analyticity constraints derived in the next Section.
%%%%%%%%%%%%%%%%%%%%%%%%%%%%%%%%%%%%%%%%%%%%%%%%
\subsubsection{ New analyticity constraints for even amplitudes: extensions of FESR(2)}\label{sec:analyticityextensions}
In this Section, we make some important extensions---very recently published by Block\cite{blockalone}---to the FESR(2) sum rule of Igi and Ishida\cite{igiandishidapp}. These new extensions will  have a major influence on the techniques we will adopt later  for fitting high energy hadron-hadron cross sections.

Clearly, as noted earlier, the FESR rule of \eq{intftilde} is only valid if it is essentially independent of ${\nu_0}$,  the upper energy cut-off choice, where valid values of ${\nu_0}$ are those low  energies  above where resonant behavior stops and  smooth high energy behavior takes over. For simplicity, we now call   ${\nu_0}$  a low energy cut-off.

Our starting point is \eq{ksqintegralfinal}, which we rewrite in the form 
\ba
\int_0^m\nu\,\im f_+(\nu)\,d\nu+\frac{1}{4\pi}\int_m^{ {\nu_0}} \nu p\,\sigtot^+(\nu)\,d\nu&=&\frac{1}{4\pi}\int_m^{ {\nu_0}} \nu^2\,\tilde\sigtot^+(\nu)\,d\nu\label{FESR(2)_2},
\ea
where now the right-hand side is expressed in terms of our high energy parametrization to the total even cross section. 
We note that if \eq{FESR(2)_2} is valid at  upper limit ${\nu_0}$, it certainly is also valid at upper limit ${\nu_0}+\Delta {\nu_0}$, where $0<\Delta {\nu_0}\ll {\nu_0}$, i.e., $\Delta {\nu_0}$ is a very small positive energy interval.  Evaluating \eq{FESR(2)_2} at its new upper limit ${\nu_0}+\Delta {\nu_0}$,  we find  
\ba
\int_0^m\nu\, \im f_+(\nu)\,d\nu+\frac{1}{4\pi}\int_m^{ {\nu_0}+\Delta \nu_0} \nu p\,\sigtot^+(\nu)\,d\nu
&=&\frac{1}{4\pi}\int_{\nu_0}^{\nu_0+\Delta {\nu_0}} \nu^2\,\tilde\sigtot^+(\nu)\,d\nu,
\label{FESR(2)_3}
\ea
which, after subtracting \eq{FESR(2)_2},  reduces to 
\ba
\frac{1}{4\pi}\int^{\nu_0+\Delta \nu_0}_{{\nu_0}} \nu p\,\sigtot^+(\nu)\,d\nu&=&\frac{1}{4\pi}\int^{\nu_0+\Delta \nu_0}_{{\nu_0}} \nu^2 \,\tilde\sigtot^+(\nu)\,d\nu. \label{FESR(2)_4}
\ea
We emphasize  two very important results\cite{blockalone} from \eq{FESR(2)_4}: 
\begin{enumerate}
\item There no longer is any reference to the unphysical region $0\le \nu <m$.
\item The left-hand integrand {\em only} contains reference to $\sigtot^+(\nu)$, the {\em true} even cross section,  which can now be replaced by the  physical {\em experimental} even cross section $\sigtot^+(\nu)\equiv[\sigtot^{pp}(\nu)+\sigtot^{\bar pp}(\nu)]/2$. 
\end{enumerate}

After taking the limit of $\Delta {\nu_0}\rightarrow 0$ and some minor rearranging, \eq{FESR(2)_4}  goes into
\ba
\sigtot^+({\nu_0})&=&\left(\frac{\nu_0}{\tilde\nu_0}\right) \tilde\sigtot^+({\nu_0})\qquad\approx\qquad \tilde\sigtot^+({\nu_0}).\label{sig+=sighigh}
\ea
A value of $\tilde {\nu_0}=10$ GeV was used, corresponding to ${\nu_0}=10.04$ GeV. Noting that the ratio of ${\nu_0}/\tilde {\nu_0}=1.004$, we see that \eq{sig+=sighigh} is numerically accurate to a precision of $\approx 0.04$\%. 

In summary, due to the imposition of analyticity, we find a new constraint equation\cite{blockalone}
\be
\sigtot^+({\nu_0})=\tilde\sigtot^+({\nu_0}),\label{sig+=sighigh2}
\ee
whose right-hand side  is $\tilde \sigtot^+({\nu_0})$, the {\em high energy phenomenological parametrization} to the cross section at energy ${\nu_0}$, the (low) transition  energy, whereas  its left-hand side of  \eq{sig+=sighigh} is $ \sigtot^+({\nu_0})$, the {\em low energy experimental cross section} at energy ${\nu_0}$. In summary, \eq{sig+=sighigh} forces the equality $ \sigtot^+({\nu_0})=\tilde \sigtot^+({\nu_0})$, tying the two together by analyticity. Clearly, since $\nu_0$ is not unique, it means that they must be the same over a large region of energy, i.e., the constraint of \eq{sig+=sighigh2} must be essentially independent of the exact value of the transition energy $\nu_0$.

The forced equating  of the high energy cross section $ \tilde \sigtot^+({\nu_0})$ to $\sigtot^+({\nu_0})$, the low energy experimental cross section, produces essentially {\em identical} fit parameters as those obtained making a fit using FESR(2) of Igi and Ishida, \eq{constraint2}, as we will show later in Section \ref{section:compare}.   Thus, one can avoid the tedious numerical evaluations needed to evaluate the integrals of FESR(2) and simply replace it completely by evaluating the $pp$ and $\ppbar$ experimental cross sections at energy $\nu_0$, a far simpler---and perhaps more accurate---task. This is our first important extension---our first new analyticity constraint---which we will return to later in some detail.

The next extension\cite{blockalone} is to note that if \eq{sig+=sighigh2} holds for $\nu_0=\nu_1$, it obviously also holds for $\nu_0=\nu_2>\nu_1$, i.e.,
\ba
\sigtot^+(\nu_1)&=&\tilde\sigtot^+(\nu_1),\nonumber\\
\sigtot^+(\nu_2)&=&\tilde\sigtot^+(\nu_2),\label{sig+=sighigh2twice}
\ea
providing that {\em both} $\nu_1$ and $\nu_2$  occur {\em after} resonance behavior is finished and {\em before} we start fitting the high energy data.

Clearly, if \eq{sig+=sighigh2} holds for any $\nu_0$, then the constraints can be generalized\cite{blockalone} to 
\be
\frac{d^n}{d\nu^n}\,\sigtot({\nu_0})=\frac{d^n}{d\nu^n}\,\tilde\sigtot({\nu_0}),\qquad \mbox{for integer $n\ge 0$.}\label{allderiv0}
\ee

%%%%%%%%%%%%%%%%%%%%%%%%%%%%%%%%%%%
\subsubsection{New analyticity constraints for odd amplitudes}
Block\cite{blockalone} next extended his arguments to analyticity constraints for odd forward scattering amplitudes $f_-(\nu)$, where
\be
f_-(\nu)= \frac{f_{pp}-f_{p\bar p}}{2}.
\ee 
Using the optical theorem, the imaginary portion of the odd amplitude is related to the physical odd cross section $\sigtot^- (\nu)$ by
\be
\im f_-(\nu)\equiv\frac{p}{4\pi}\sigtot^- (\nu),\qquad\mbox{for $\nu\ge m$}\label{sig-1}.
\ee

He introduced $\hat f_-(\nu)$, a super-convergent odd amplitude, defined as  
\ba
\hat f_-(\nu)&\equiv& f_-(\nu)-\tilde f_-(\nu)
\label{fhatodd}
\ea
that satisfies the super-convergent dispersion relation
\be
\int_0^\infty {\rm Im}\,\left[f_-(\nu)-\tilde f_-(\nu)\right]\, d\nu=0,\label{super3}
\ee 
even if neither term separately satisfies it. This is in analogy to the $n=0$ FESR of \eq{sumrule} which required the odd amplitude $\nu \hat f_+(\nu)$, whereas here we inserted the odd amplitude $\hat f_-(\nu)$. 

Cutting  off the integration at $\nu_0$, Block\cite{blockalone} wrote \eq{super3} as 
\be
\int_0^{\nu_0}{\rm Im}\,f_-(\nu)\, d\nu=be
\int_0^{\nu_0} {\rm Im}\,\tilde f_-(\nu)\, d\nu.
\ee

For the left-hand side he used 
\be
\int_0^{\nu_0} \im  f_-(\nu)\,d\nu\label{truefodd}=\int_0^m\im f_-(\nu)\,d\nu+\frac{1}{4\pi}\int_m^{ {\nu_0}} p\,\sigtot^-(\nu)\,d\nu
\ee 
and for the right-hand side he substituted 
\be
\int_0^{\nu_0}\im \tilde f_-(\nu)\, d\nu=\frac{1}{4\pi}\int_m^{ {\nu_0}}  \nu\tilde\sigma_{\rm tot}^-(\nu)\,d\nu,
\ee 
obtaining
\ba
\int_0^m\im f_-(\nu)\,d\nu+\frac{1}{4\pi}\int_m^{ {\nu_0}} p\,\sigtot^-(\nu)\,d\nu&=&\frac{1}{4\pi}\int_m^{ {\nu_0}}  \nu\tilde\sigma_{\rm tot}^-(\nu)\,d\nu.\label{Fodd}
\ea
Again, since ${\nu_0}$ is arbitrary,  Block\cite{blockalone} found
\ba
\sigtot^-({\nu_0})&\approx&\tilde\sigma_{\rm tot}^-({\nu_0})\label{sigodd}\\
\frac{d\sigtot^-}{d\nu}({\nu_0})&\approx&\frac{d\tilde\sigma_{\rm tot}^-}{d\nu}({\nu_0}), \qquad{\rm or}\label{dsigodd}\\
\sigtot^-{(\nu_1)}&\approx&\tilde\sigma_{\rm tot}^-(\nu_1)\nonumber\\
\sigtot^-(\nu_2)&\approx&\tilde\sigma_{\rm tot}^-(\nu_2),\qquad \nu_2>\nu_1,\qquad{\rm etc.}\label{sigodd2}
\ea

%%%%%%%%%%%%%%%%%%%%%%%%
\subsubsection{New analyticity constraints: summary}\label{sec:4constraints}
Thus, we have now derived new analyticity constraints for both even and odd amplitudes, and therefore, for {\em all} hadronic reactions of the type
\ba
a+b&\rightarrow& a+b\nonumber\\
\bar a+b&\rightarrow& \bar a+ b.
\ea

The even constraints of  \eq{sig+=sighigh2} and its companions, \eq{sig+=sighigh2twice} and \eq{allderiv0}, together with the odd constraints, \eq{sigodd}, \eq{dsigodd}  and \eq{sigodd2}, are consequences of imposing analyticity. They imply several important conditions:
\begin{itemize}
\item 
On the left-hand side of \eq{sig+=sighigh2}  the cross section $\sigtot^+({\nu_0})$ that appears is the {\em experimental} even  cross section $\sigtot^+({\nu_0})=[\sigtot(pp)+\sigtot(\ppbar)]/2$, whereas in \eq{sigodd} the cross section that appears on the left-hand side is the experimental odd cross section  $\sigtot^-({\nu_0})=[\sigtot(pp)-\sigtot(\ppbar)]/2$), all  evaluated at energy ${\nu_0}$; similar remarks are true about the  derivatives of the {\em experimental} even and odd cross section.  Therefore, the new  constraints that were derived above---extensions of Finite Energy Sum Rules---tie together {\em both} the $pp$ and $\ppbar$\ experimental cross sections and their derivatives with the high energy approximation that is used to fit data at energies high above the resonance region. Analyticity then requires that there  should be a {\em good} fit to the high energy data after using these new constraints, i.e., the $\chi^2$ per degree of freedom should be $\sim 1$, {\em if, and only if,} the high energy amplitude  provides  a {\em good} approximation to the high energy data.
%%%%%%%%%%%%%%%%%%%%%%%%%%  
\item The results should be independent of ${\nu_0}$ when the energy ${\nu_0}$ is in the region where there is smooth energy variation, just above the resonance region,   Thus, if the phenomenologist has chosen a reasonably valid high energy parametrization,  consistency with analyticity  require that the fitted parameters be essentially independent of ${\nu_0}$, a condition which was explicitly indicated in \eq{sig+=sighigh2twice}. 
\item The new  constraints, \eq{sig+=sighigh2}, \eq{sig+=sighigh2twice}, \eq{allderiv0},  {\em do not depend} on values of the non-physical integral $\int_0^m\nu\, \im f_+(\nu)\,d\nu$ of the type used in \eq{FESR(2)_2}, as long as they are finite.  Therefore, no evaluation of the unphysical region is needed for our new analyticity constraints---it's exact value doesn't matter, even had it been comparable to  the main integral $\int_m^{\nu_0}\nu\, \im f_+(\nu)\,d\nu$.
\end{itemize}
%%%%%%%%%%%%%%%%%%%%%%%%%%%%%%%%%

Block and Halzen\cite{bhfroissart,bhfroissartnew} recently used these analyticity constraints, forming linear combinations of cross sections and derivatives to anchor their high energy cross section fits to an even low energy experimental cross sections\cite{bhfroissart} and its first derivative for $\gamma p$ scattering, and  to {\em both} even and odd cross sections\cite{bhfroissartnew} and their first derivatives  for $\pi^-p$ and $\pi^+p$, and  $\pbar p$ and $pp$ scattering. We will discuss this new  method of fitting and their results in detail later. 

They used four constraints in their successful high energy $\ln^2s$ fits\cite{bhfroissartnew} to $pp\ (\pi^+p)$ and $\pbar p\ (\pi^-p)$  cross sections and $\rho$-values. They first did a local fit to $pp\ (\pi^+p)$ and $\pbar p\ (\pi^-p)$  cross sections---in the neighborhood of ${\nu_0}$---to determine the cross sections and the slopes of the $pp\ (\pi^+p)$ and $\ppbar\ (\pi^-p)$ cross sections at $\sqrt s=4$ GeV for nucleon-nucleon ($\sqrt s=2.6$ GeV for pion-nucleon) scattering, where they anchored their data. Their actual fitted data were to cross sections and $\rho$-values with much higher energies, $\sqrt s\ge 6$ GeV, for both nucleon-nucleon and pion-nucleon scattering.

Because it was relatively easy for them to make an accurate local fit to the experimental cross sections and their derivatives,  whereas determining accurate values of 2nd derivatives and higher was difficult, they stopped with the 4 constraints, $\tilde\sigtot(pp)$, $\tilde\sigtot(\ppbar)$, $d\tilde\sigtot(pp)/d\nu$ and $d\tilde\sigtot(\ppbar)/d\nu$, which they evaluated at $\sqrt s=4$ GeV. Similarly, they evaluated $\tilde\sigtot(\pi^+p)$, $\tilde\sigtot(\pi^-p),$ $d\tilde\sigtot(\pi^+p)/d\nu$ and $d\tilde\sigtot(\pi^-p)/d\nu$ at $\sqrt s=2.6$ GeV. In both cases, they made their fits using only data having an energy $\sqrt s\ge 6$ GeV. 

The advantage of having these 4 analyticity constraints in a high energy fit is multi-fold:
\begin{enumerate}
\item The number of parameters needed to be evaluated in a $\chi^2$ fit is reduced by the number of new constraints, i.e.,  4 in the case of nucleon-nucleon and pion-nucleon scattering, therefore reducing the number of parameters to be determined from 7 to 3. 
\item The statistical errors of the remaining coefficients are markedly reduced, an important result needed for accurate high energy extrapolations.
\item If the $\chi^2$ per degree of freedom---corresponding to the newly reduced number of degrees of freedom---is $\sim1$, the goodness-of-fit of the high energy data is  quite satisfactory, i.e., a good fit was obtained using the constraints.  Because the fit is anchored  at low energies, this satisfactory goodness-of-fit in addition signifies that the high energy amplitude employed by the phenomenologist {\em also satisfies the new analyticity constraints}, giving a very important additional  validation of the choice of the high energy amplitude.
\item Conversely, let us assume that the phenomenologist's model of high energy behavior was not very good.   Because of the low energy constraints, the effects of a poorer model are magnified enormously, and yield a $\chi^2$ per degree of freedom $\gg 1$. This leverage allows the model builder to make  sharp distinctions between models that otherwise might not be distinguishable by a goodness-of-fit criterion.
\end{enumerate}
%%%%%%%%%%%%%%%%%%%%%%%%%%%%%%%%%%%%%
\subsubsection{A new interpretation of duality}
Duality has been previously used to state that the average value of the energy moments of the imaginary portion of the  true amplitude over the energy interval 
0 to ${\nu_0}$ are the same as the average value of the energy moments of the imaginary portion of the  high energy approximation amplitude over the {\em same} interval\cite{dolen-horn-schmid}. The extensions made here show that the imaginary portion of the amplitude itself at energy ${\nu_0}$ is equal to the imaginary portion of the high energy amplitude, when it is also evaluated at ${\nu_0}$. Conversely, if the high energy amplitude is a faithful reproduction of the high energy data, analyticity forces 
the high energy cross sections---with {\em all} of their derivatives---deduced from the high energy amplitude at the low energy ${\nu_0}$ be approximately equal to those deduced from the low energy experimental cross sections at energy ${\nu_0}$, together with all of their derivatives, i.e.,
\be
\frac{d^n}{d\nu^n}\,\sigtot({\nu_0})=\frac{d^n}{d\nu^n}\,\tilde\sigtot({\nu_0}),\qquad \mbox{for integer $n\ge 0$,}\label{allderiv}
\ee
true for both $pp$ and $\ppbar$\ cross sections, providing us with a new interpretation of duality. 

%%%%%%%%%%%%%%%%%%%%%%%%%%%%%
\subsection{Differential dispersion relations}
For completeness, we include  differential dispersion relations. They have been derived in Ref. \cite{bc} and a complete list of references can be found there. They are valid for high energies and are:
\ba
\re f_-(\zeta)&=&\quad\left[\tan\frac{\pi}{2}\frac{\partial}{\partial\zeta}\right]\im f_-(\zeta),\label{dispersion-}\\
\im f_+(\zeta)&=&-\left[\tan\frac{\pi}{2}\frac{\partial}{\partial\zeta}\right]\re f_+(\zeta). \label{dispersion+}
\ea

These relations are quite intractable unless the amplitudes are simple functions of the variable $\zeta$, which can be the laboratory energy $E$ (or $s\approx 2mE$). In that case, we quote Block and Cahn\cite{bc}:

\begin{itemize}
\item []``This prompts the following question: Why not just use the simple analytic forms themselves and bypass the differential dispersion relations?''
\end{itemize}
\noindent We will follow their advice.
%%%%%%%%%%%%%%%%%%%%%%%%%%%%%%%%%%%%%%%%%%%%%%%%%%%%
\section{Applications of Unitarity}
In Section \ref{sec:eikonal}, we saw that the scattering amplitude  corresponding to the  eikonal $\chi(s,b)$ was given by
\be
a(b,s)=\frac{e^{-\chi(b,s)}-1}{2i}=-\frac{e^{-\chi_R}\sin(\chi_I)}{2}+i\frac{1-e^{-\chi_R}\cos(\chi_I)}{2}, \label{aofchi1}
\ee
where $a(b,s)$ satisfied unitarity by being in the Argand circle of Fig. \ref{fig:argand}.

After inserting  this amplitude in \eq{sigtotofb1}, 
\ba
\sigtot(s)&=&2\int \left[1-e^{-\chi_R}\cos\left(\chi_I\right)\right]\, d^2\vec b,\label{sigtotofb3}
\ea 
we see that  cross sections derived from an eikonal satisfy unitarity. In this Section, we will illustrate some applications of unitarity and analyticity, giving heuristic derivations of the Froissart bound and a revised Pomeranchuk theorem.
%%%%%%%%%%%%%%%%%%%%%%%%%%%%%%%%%%%%%%%%%%%
\subsection{The Froissart bound}\label{sec:froissart}
For simplicity, we will assume a factorizable amplitude in \eq{chifactorize}, i.e., 
\be 
\chi(b,s)=A(s)\times W(b),\label{chifactorize2}
\ee
with $W(b)$ normalized so that $\int W(b)\,d^2\vec b=1$. Also assuming  the matter distribution in a proton is the same as the electric charge distribution\cite{durand} and is given by a dipole form factor
$G(q^2)=\left[\mu^2/(q^2+\mu^2)\right]^2,$
with $\mu^2=0.71$ (GeV/c)$^2$, we find the impact parameter distribution $W(\mu b)=\frac{\mu^2}{96\pi}(\mu b)^3K_3(\mu b).$ 

We will first consider the case where the eikonal of \eq{chifactorize2} is pure real (corresponding to a purely imaginary phase shift), factorizable,  and  is also very small.  The total cross section in \eq{sigtotofb3} becomes
\ba
\sigtot(s)&=&2\int \left[1-e^{-\chi(s,b)}\right]\, d^2\vec b\label{realchi}
\ea
and for very small $\chi$
\ba
\sigtot(s)&\approx&2\int \chi(b,s)\, d^2\vec b\nonumber\\
&=&2 A(s)\int W(b)\, d^2\vec b\nonumber\\
&=&2A(s),
\label{sigtotofb4}
\ea 
since $W(b)$ is normalized so its integral is unity.
In other words, for small $\chi(s,b)$, the forward scattering amplitude is given by 
\be
f(s)=i\frac{p}{2\pi}A(s)\label{fsmallchi},
\ee
corresponding to {\em small} amplitudes.  We  take as a given that the forward scattering amplitude can rise no faster than $s^2$, so we can write $A(s)={\rm const}\times s^{\epsilon}$, where $\epsilon\le 1$.  However, we  soon bump into the unitarity boundary for large $s$. 

Since we  showed in Section \ref{sec:eikonal} that cross sections from  eikonals satisfy unitarity,  we will use an eikonal to ``unitarize'' our amplitude, which is proportional to $s^\epsilon$.  Using the  eikonal 
\be
\chi(b,s)=c_1s^\epsilon\frac{\mu^2}{96\pi}(\mu b)^3K_3(\mu b),\label{chifactorize3}
\ee
where $c_1$ is a real constant,   the cross section given by \eq{sigtotofb3} satisfies unitarity.

We observe that the integrand of \eq{realchi} is $O(1)$ if $\chi(s,b)$ is large, whereas it is $O(0)$ if $\chi(b,s)\approx 1$.  In other words, there is a impact parameter space cutoff  value which we call $b_{\rm max}$ (for notational simplicity, we have dropped its explicit dependence on $s$), with the property that the eikonal cross section integrand
\ba
\left[1-e^{-\chi(s,b)}\right]&\approx 1,\qquad b\le b_{\rm max}\nonumber\\
\left[1-e^{-\chi(s,b)}\right]&\approx 0, \qquad b > b_{\rm max},\label{integrand}
\ea
so that  $\sigtot(s)=2\pi b_{\rm max}^2$. For large $b$, 
$W(b)\rightarrow c_2(\mu b)^{5/2}e^{-\mu b}$, where $c_2$ is a constant.  Thus, the cutoff condition on the exponential (yielding $\exp(-\chi)\approx 0$) is given by
\be
\chi(b_{\rm max},s)=c_1s^\epsilon c_2(\mu b_{\rm max})^{5/2}e^{-\mu b_{\rm max}}\sim 1
\ee 
which implies 
\be
b_{\rm max}=\left(\frac{\epsilon}{\mu}\right)\ln s/s_0 +\left(\frac{5}{2}\right)\ln \mu b_{\rm max} ,
\ee
with $s_0$ is a scale constant. 
Hence,  
\be
b_{\rm max}=\left(\frac{\epsilon}{\mu}\right)\ln s/s_0 + O(\ln\ln s/s_0)\label{bmax},
\ee
where $0 < \epsilon\le 1$.  Thus, we find that the total cross section is asymptotically  given by
\be
\sigtot(s)=2\pi\left(\frac{\epsilon}{\mu}\right)^2\ln^2 s/s_0,
\ee 
whose energy dependence is bounded by $\ln^2 s/s_0$, which summarizes  our heuristic derivation of the Froissart bound. We find that the amplitude that saturates the Froissart bound  is 
\ba 
a(s,b)&=&i/2,\qquad b\le \left(\frac{\epsilon}{\mu}\right)\ln s/s_0,\nonumber\\
a(s,b)&=&0,\quad\qquad b>\left(\frac{\epsilon}{\mu}\right)\ln s/s_0,\label{froissartdisk}
\ea
which is the amplitude of  a black disk whose radius grows as $\ln s$. We note that our particular choice of $W(\mu b)$ does not influence this result, since any impact parameter  distribution that falls off as rapidly as $e^{-\mu b}$ as $b\rightarrow\infty$ will give the same result.

The original statement of Froissart\cite{froissart}, who derived 
his fundamental result  from unitarity and analyticity is repeated here:
\begin{itemize}
 \item []``At forward or backward angles, the modulus of the amplitude behaves at most like $s\ln^2s$, as $s$ goes to infinity.  We can use the optical theorem to derive that the total cross sections behave at most like $\ln^2s$, as $s$ goes to infinity.''  
\end{itemize}
According to Froissart, saturating the Froissart bound refers to an energy dependence of all hadron-hadron total  cross sections rising asymptotically as $\ln^2s$, which is the interpretation we will use in this work.

%%%%%%%%%%%%%%%%%%%%%%%%%%%%%%%%%%%%%%%%%%%%%%%%%%%%%%%%%%%%%%%%
\subsection{The Pomeranchuk theorem}\label{sec:pomeranchuk}
The original Pomeranchuk theorem stated that if the $pp$ and $\ppbar$\ cross sections\footnote{More generally, the total cross sections for $ab$ and $a \bar b$.\label{foot:ab}} asymptotically went to a constant and if $\rho$ increased less rapidly than $\ln s$, then the two cross sections became equal asymptotically.

Since we know now that cross sections rise with energy, bounded by a $\ln^2 s$ behavior, the original theorem is inapplicable. Eden\cite{eden2} and Kinoshita\cite{kinoshita} showed that if the $pp$ and $\ppbar$\ cross sections (see footnote \ref{foot:ab}) grow as $\ln^\gamma s$, then the difference cross sections can't grow faster than $\ln^{\gamma/2}$, with $\gamma\le2$,  a relation which is now called the revised Pomeranchuk theorem.

A short heuristic proof follows. For notational simplicity, we again drop the amplitude's explicit  dependence on $s$ and rewrite it as $a(b)$. 

As before, unitarity requires that $a(b)$ must lie inside or on the Argand circle of Fig. \ref{fig:argand}, so that
\be
\left|\re a(b)\right|^2\le\im a(b).\label{unitary1}
\ee

As in our derivation of the Froissart bound in Section \ref{sec:froissart}, the impact parameters that contribute substantially to the scattering must lie below some maximum impact parameter $b_{\rm max}$ which grows with energy as  $\ln s$.  Thus, we approximate the scattering amplitude as
\ba
f(q=0)&=&\frac{p}{\pi}\int a(b)\,d^2\vec b\nonumber\\
&\approx&2p\int_0^{b{_{\rm max}}}a(b)\, b\, db.\label{f1}
\ea
Hence,  
\ba
\left| \re f(0)\right|&\approx& 2p \left|\int_0^{b{_{\rm max}}}\re a(b)\, b\, db \right|\nonumber\\
&\le& 2p\int_0^{b{_{\rm max}}} \left|\re a(b)\right|\, b\, db \nonumber\\
&\le&2p\int_0^{b{_{\rm max}}} \left[\im a(b)\right]^{1/2}\, b\, db,\label{inequality} 
\ea
where in the last step we used the unitary condition of \eq{unitary1}.

The Schwarz inequality states that 
\be
\left[\int_{c_1}^{c_2}f(x)g(x)dx\right]^2\le\int_{c_1}^{c_2}f(x)dx\times\int_{c_1}^{c_2}g(x)dx.\label{schwarz}
\ee  
Using $c_1=0$, $c_2=b_{\rm max}$, $f(x)=\left[\im a(b)\right]^{1/2}$, $g(x)=1$, $x=b^2/2$ and $dx=b\,db$, we can rewrite the preceeding  inequality in terms of $\left|\re f(0)\right|$ and $\left|\im f(0)\right|$ as
\ba
\left|\re f(0)\right|&\le&2p\left[\int_0^{b{_{\rm max}}} \im a(b)\, b\, db\right]^{1/2}\left[\int_0^{b{_{\rm max}}}\, b\, db\right]^{1/2}\nonumber\\
&\le&2p\left[\frac{\sigtot}{8\pi}\right]^{1/2}\left(\frac{1}{2}b_{\rm max}^2\right)^{1/2}\nonumber\\
&\le&{\rm const}\times s\left[\ln\left(s/s_0\right)\right]^{\frac{\gamma}{2}+1},\label{realinequality}
\ea
where we used the optical theorem in step 2. 
It is the odd amplitude that gives rise to a cross section difference.
A generic high energy form for the odd amplitude (see the high energy behavior of the odderons of Section \ref{sec:odderon}, when they are expressed as a function  of the variable $s$ rather than the laboratory energy $E$) which  gives rise to a non-zero $\Delta\sigtot$ is given by 
\be
f_-\sim s\left[\ln\left(\frac{s}{s_0}\right)-i\frac{\pi}{2}\right]^{\gamma'},\label{fodderon1}
\ee
with $\gamma'>0$.
Comparison of \eq{fodderon1} and \eq{realinequality} shows that
\be
\gamma'\le\frac{\gamma}{2}+1.
\ee
Thus, we have demonstrated that the energy dependence of the difference of the cross sections, when combined with the Froissart bound, is limited by
\ba
\Delta \sigtot&\sim&\left[\ln\left(\frac{s}{s_0}\right)\right]^{\gamma'-1}\le{\rm const}\times\left[\ln\left(\frac{s}{s_0}\right)\right]^{\gamma/2},\qquad \gamma\le 2,
\ea
which is the revised Pomeranchuk theorem .
%%%%%%%%%%%%%%%%%%%%%%%%%%%%%%%%%%%%%
\section{The ``Sieve'' algorithm, a new robust data fitting technique}\label{section:sievealgorithm}
Typically, in the past,  eikonal models or real analytic amplitude models have been fit to high energy hadron scattering data---such as cross sections and $\rho$-values  as a function of the c.m. energy $\sqrt s$---by using standard  $\chi^2$ data fitting routines.  Almost always, the source of these multitude of  datum points was the archives of the Particle Data Group (PDG), the latest version of which is ref. \cite{pdg}. This is an unedited compendium of almost all  scattering data published since the early 1950's.  None of the data have been screened for compatibility, accuracy,  etc. For example, if one plots  the $pp$ and $\ppbar$\ total cross sections as a function of energy, along with their quoted error bars, from the lowest energies around $\sqrt s=2$ GeV to the highest energies at the Tevatron Collider ($\sqrt s=1800$ GeV for $\ppbar$) and then examines them carefully (almost microscopically!) in  local (small) energy subregions, it becomes clear that some points, when weighted by their error bars, lie well outside the local averages.  These points are called ``outliers'' by statisticians.  If numerous enough and bad enough when used in a $\chi^2$ fit, outliers will:
\begin{enumerate}
\item  seriously distort the parameters found in a $\chi^2$ fit, giving sufficiently unreliable answers that the statistical errors found from the fitting routine become meaningless.
\item markedly increase the value of the minimum $\chi^2$ found, so that if we here use the symbol $\nu$ to be the number of degrees of freedom of our system, then $\chi^2_{\rm min}/\nu \gg 1$, even if the model was a good one.%
\footnote{The $\chi^2$ probability density distribution has $\nu$, the number of degrees of freedom, as its mean value and  has a variance equal to  $ {2\nu}$. To have an intuitive feeling for the goodness-of-fit, i.e., the probability that $\chi^2>\chi^2_{\rm min}$, we note that for the large number of degrees of freedom $\nu$ that we are considering here,  the probability density distribution for $\chi^2$ is well approximated by a  Gaussian, with a mean at $\nu$ and a width of $\sqrt {2 \nu}$, where $0<\chi^2<\infty$ ({\em n.b.,} the usual lower limit of $-\infty$ is truncated here to 0, since by definition $\chi^2\ge 0$).  In this approximation, we have the {\em most probable} situation if   $\chi^2_{\rm min}/\nu=1$, which corresponds to a goodness-of-fit probability of 0.5.  The chance of having  small $\chi^2_{\rm min}\sim 0$,  corresponding to a goodness-of-fit probability $\sim 1$, is exceedingly small.\label{footnote:probability}}%
  \ When the model is a good one and the data are reasonably Gaussianly distributed about it with correct  error assignments, one should find, on average, $\chi^2_{\rm min}/\nu\sim 1\pm\sqrt{2/\nu}$.  Here, it is a case of {\em  bad data} causing $\chi^2_{\rm min}/\nu\gg1$, and {\em not a bad model} doing it. 
\item not allow the model builder to distinguish between models, since $\chi^2_{\rm min}$, the goodness-of-fit criterion,  is already rendered almost completely useless, even for the correct model.
\end{enumerate}

The obvious moral to this story is to use only good data. But how do we make an objective separation of  the good from the bad? 

An answer to this question is to get rid of the outliers.  Hopefully, the data set is not sufficiently contaminated that the outliers outnumber the good datum points. If they do, one can go no further---the situation is hopeless. However, if the contamination is not too severe, the ``Sieve'' algorithm, recently introduced by Block\cite{sieve} solves our problem by eliminating the outliers. The price one pays for eliminating the outliers is that you simultaneously lose a (hopefully) small amount of good data.

The essence of the ``Sieve'' algorithm is its first step, making a ``robust'' fit to the entire data set, warts and all. The idea of ``robustness''%
\footnote{The terminology is attributed in ``Numerical Recipes''\cite{nr} to G. E. P. Box in 1953.\label{foot:box}} % 
 is a statistical concept indicating that a statistical procedure is relatively insensitive to the presence of outliers, and robust fitting  techniques have been invented to take care of some of these problems\cite{nr,huber, hampel}.  

A very simple, but important, example of a robust estimator is to use the median of a discrete distribution rather than the mean in order to characterize a typical characteristic of the distribution. For example, I live in Aspen, Colorado, a premier ski resort (however, it is also the home of the Aspen Center for Physics!) where the mean selling price of a home is seriously distorted because of several forty to sixty million dollar homes (really mansions)---outliers  in the housing market distribution. Thus, the mean selling price is essentially useless (to most people!).  If you are shopping for a house here, most physicists would be more interested in the median selling price, which is a statistic that is  scarcely  affected by the outliers.

The terminology, ``robust'' statistical estimators (see footnote \ref{foot:box}), was first introduced around 1953 to deal with small numbers of data points which have a large departure from their model predictions,  i.e., outlier points, due to a myriad of reasons, none of which are really known in advance.  Later, in the late 1980's, research on robust estimation\cite{huber,hampel} based on influence functions, which are discussed later in Section \ref{influence}, was carried out. More recently, robust estimations using  linear regression models\cite{regression} were made---these are not useful in fitting the non-linear models often  needed in practical  applications.  For example, the fit needed for  \eq{rhopm} is a {\em non-linear} function of the coefficients $c_0,\ c_1,\ c_2,\ \ldots$, since it is the ratio of two linear functions.
 We will discuss a technique for handling outlier points in a non-linear fit when we introduce the Lorentz probability density function in Section \ref{section:lorentz}.

Up until very recently, no technique that fits non-linear functions of fit parameters has been invented to get rid of outliers. Today, the ``Sieve'' algorithm\cite{sieve} solves this problem. 
%%%%%%%%%%%%%%%%%%%%%%%%%%%%%%%%%%%%%%%%%%%%%%%%%%%%
\subsection{$\chi^2$ and robust fitting routines}
We digress at this point to discuss mathematical details about the $\chi^2$ fitting technique, which is {\em not} robust 
 and compare it to a Lorentzian squared fit, which {\em is} robust, showing why outliers severely influence the standard $\chi^2$ fit by distorting the fit parameters. We will use a generalization of the maximum likelihood function for our discussion.
%%%%%%%%%%%%%%%%%%%%%%%%%%%%%%%%%%%%%%%%%%%%%%%%%
\subsubsection{Maximum likelihood estimates}
We start by introducing the maximum likelihood technique for fitting $N$ measurements at position $x_i$ of physical data which we will assume are Gaussianly distributed about their measured values $y_i$, with error estimates $\sigma_i$, which are standard deviations of the fluctuations, for  $i=1,2,\ldots N$.

Let $P_i$ be the probability density of the $i$th individual measurement, $i=1,\ldots,N$, in the interval $\Delta y$. Then the probability of the total data set is 
\be
{\cal P}=\prod_{i=1}^{N} P_i\Delta y.
\ee 
Let us define the quantity
\be
\Delta \chi^2_i(x_i;\alphabold)\equiv \left(\frac{y_i-y(x_i;\alphabold)}{\sigma_i}\right)^2\label{delchisq},
\ee
where $y_i$ is the measured value at $x_i$, $y(x_i;\alphabold)$ is the expected (theoretical) value from the model under consideration, and $\sigma_i$ is the experimental error of the $i$th measurement. The $M$ model parameters $\alpha_k$ are given by the  $M$-dimensional vector
$
\alphabold =\{\alpha_1,\ldots,\alpha_M\}.
$

$\cal P$ is identified as the likelihood function, which we shall maximize as a function of the parameters $\alphabold=\{\alpha_1,\ldots,\alpha_M\}$.

For the special case where the data are normally  distributed (Gaussian distribution), we have the likelihood function $\cal P$ given as
\be
{\cal P}=\prod_{i=1}^{N}\left\{\exp\left [-\frac{1}{2}\left(\frac{y_i-y(x_i;\alphabold)}{\sigma_i}\right )^2\right ]\frac{\Delta y}{\sqrt{2\pi}\sigma_i}\right\}=\prod_{i=1}^{N}\left\{\exp\left [-\frac{1}{2}
\Delta \chi^2_i\right ]\frac{\Delta y}{\sqrt{2\pi}\sigma_i}\right\},\label{likelihood}
\ee
Maximizing the likelihood function $\cal P$ in \eq{likelihood} is the same as minimizing the negative logarithm of $\cal P$, namely,
\be
\sum_{i=1}^N  \frac{1}{2}\left(\frac{y_i-y(xi;\alphabold)}{\sigma_i}\right)^2-N\ln\frac{\Delta y}{\sqrt{2\pi}\sigma_i}.\label{minchi}
\ee
Using \eq{delchisq} to introduce $\delchisq$, since $N$, $\Delta y$ and $\sigma_i$ are constants, minimizing \eq{minchi} is equivalent to minimizing the quantity
\be
\frac{1}{2}\sum_{i=1}^N\delchisq.\label{2chisq}
\ee

We now define $\chi^2(\alphabold;\xbold)$ as
\be
\chi^2(\alphabold;\xbold)=\sum_{i=1}^N\delchisq,
\ee
where $\xbold\equiv\{x_1,\dots,x_i,\ldots,x_N\}$.

As expected, the maximum likelihood minimization problem appropriate to a Gaussian distribution reduces to finding $\chi^2_{\rm min}$, i.e., 
\be
{\rm minimize\ over\ }\alphabold,\quad\quad\chi^2(\alphabold;\xbold)=\sum_{i=1}^N\delchisq
\ee
   for the set of $N$ experimental points at $x_i$ having the value $y_i$ and error $\sigma_i$.
%%%%%%%%%%%%%%%%%%%%%%%%%%%%%%%%
\subsubsection{Gaussian distribution}\label{appendix:chisq}
To minimize $\chi^2$, we must solve the (in general, non-linear) set of $M$ equations
\be
\sum_{i=1}^N\frac{1}{\sigma_i}\left(\frac{y_i-y(x_i;\alphabold)}{\sigma_i}\right)\left(\frac{\partial y(x_i;\ldots\alpha_j\ldots)}{\partial\alpha_j}\right)=0,\quad j=1,\ldots,M.\label{chieqns}
\ee
The Gaussian distribution allows a $\chi^2$ minimization routine to return several exceedingly useful statistical quantities: 
\begin{enumerate}
\item
It returns the best-fit parameter space $\alphabold_{\rm min}$.
 \item The value of $\chi^2_{\rm min}$, when compared to the number of degrees of freedom   ( d.f.$\equiv\nu= N-M$, the number of data points minus the number of fitted parameters), allows one to make  standard estimates of the goodness-of-fit of the data set to the model, using the $\chi^2$ probability distribution function given in standard texts  for $\nu$ degrees of freedom\cite{nr}.  
\item We can compute the standard covariance matrix $C$ for the individual parameters $\alpha_i$, as well as the correlations between $\alpha_j$ and $\alpha_k$, using $C^{-1}$, the $M\times M$ matrix of the partial derivatives at the minimum\cite{nr}, given by 
\be
\left[C^{-1}\right]_{jk}=\frac{1}{2}\left(\frac{\partial^2\chi^2}{\partial\alpha_j\partial\alpha_k}\right)_{\alphabold=\alphabold_{\rm min}}.
\ee

Thus, we have complete knowledge of the error ellipse in $\alphabold$ space. 
\end{enumerate}

In summary, when the errors are distributed normally, the standard $\chi^2$ technique not only gives us the desired parameters $\alphabold_{\rm min}$, but also furnishes  us with  statistically meaningful error estimates of the fitted parameters, along with goodness-of-fit information of the data to the chosen model---very valuable quantities for any model under consideration.
%%%%%%%%%%%%%%%%%%%%%%%%%%%%%%%%%%%%%5 
\subsubsection{$\psi(z)$, the influence function}\label{influence}
We can generalize the maximum likelihood function of \eq{likelihood}, which is  a function of the variable $z=(y_i-y(x_i; \alphabold))/\sigma_i$, as
\be
{\cal P}=\prod_{i=1}^{N}\left\{\exp\left [-\rho\left( \frac{y_i-y(x_i; \alphabold)}{\sigma_i}\right)
\right ]\Delta y\right\},\label{genP}
\ee
where the function $\rho\left( \frac{y_i-y(x_i; \alphabold)}{\sigma_i}\right)$ is the negative logarithm of the probability density. Note that the statistical function $\rho(z)$ used here has nothing to do with the $\rho$-value used in \eq{rhopm}. 
Thus, we now have to minimize the generalization of \eq{2chisq},  i.e.,
\be
{\rm minimize\ over\ }\alphabold,\quad\quad \sum_{i=1}^N\rho\left( \frac{y_i-y(x_i; \alphabold)}{\sigma_i}\right),\label{newmin}
\ee
for the $N$ measurements $y_1$ to $y_N$.

Introducing the influence function $\psi(z)$, minimization yields the more general set of $M$ equations
\be
\sum_{i=1}^N\frac{1}{\sigma_i}\psi\left(\frac{y_i-y(x_i;\alphabold)}{\sigma_i}\right)\left(\frac{\partial y(x_i;\ldots\alpha_j\ldots)}{\partial\alpha_j}\right)=0,\quad j=1,\ldots,M,\label{robusteqns}
\ee
where the influence function $\psi(z)$ in \eq{robusteqns} is defined as 
\be
\psi(z)\equiv\frac{d\rho(z)}{dz},\quad z\equiv\frac{y_i-y(x_i;\alphabold)}{\sigma_i}={\rm sign}(y_i-y(x_i;\alphabold))\times\sqrt{\delchisq}.\label{z}
\ee
Comparison of \eq{robusteqns} with the Gaussian equivalent of \eq{chieqns} shows that
\be
\rho(z)=\frac{1}{2}z^2,\quad \psi(z)=z \quad{\rm (for\ a \ Gaussian \ distribution)}.
\ee
We note that for a Gaussian distribution,
the influence function $\psi(z)$ for each experimental point $i$ is proportional to $\sqrt{\Delta \chi^2_i}$, the normalized departure of the point from the theoretical value. Thus, the more the departure from the theoretical value, the more ``influence'' the point has in minimizing $\chi^2$. This gives outliers (points with large departures from their theoretical values) unduly large ``influence'' in computing the best vector \mbox{\boldmath $\alpha$}, easily skewing the answer due to the inclusion of these outliers---a major disadvantage of using $\chi^2$ minimization when outliers are present.
%%%%%%%%%%%%%%%%%%%%%%%%%%%%%%%
\subsubsection{Lorentz distribution}\label{section:lorentz}
Let us now consider the normalized Lorentz probability density distribution (also known as the Cauchy distribution or the Breit-Wigner line width distribution), given by 
\begin{eqnarray}
P(z)&=&\frac{\sqrt{\gamma}}{\pi}\frac{1}{1+\gamma z^2},\label{cauchy1}
\end{eqnarray}
where $\gamma$ is a normalization constant whose significance  will be discussed later. 
Using \eq{delchisq},  we first rewrite \eq{cauchy1} in terms of the measurement errors $\sigma_i$ and the  experimental measurements  $y_i$ at $x_i$. Next, we introduce the quantity $\delchisq$ from \eq{z},  and write $P$ as 
\be
P\left(\frac{y_i-y(x_i;\alphabold)}{\sigma_i}\right)=\frac{\sqrt{\gamma}}{\pi}\frac{1}{1+\gamma\left(\frac{y_i-y(x_i;\alphabold)}{\sigma_i}\right)^2}=\frac{\sqrt{\gamma}}{\pi}\frac{1}{1+\gamma\delchisq}.\label{cauchy}
\ee
It has long tails and therefore is more suitable for robust fits than is the Gaussian distribution, which is exceedingly compact. 
Taking the negative logarithm of \eq{cauchy} and using it in \eq{newmin}, we see that 
\begin{eqnarray}
\rho(z)&=&\ln\left(1+\gamma z^2\right)=\ln\left\{1+\gamma\delchisq\right\},\nonumber\\
\psi(z)&=&\frac{z}{1+\gamma z^2}
=\frac{{\rm sign}(y_i-y(x_i;\alphabold))\times\sqrt{\delchisq}}{1+\gamma\delchisq}.\label{lorentz}
\end{eqnarray}

In analogy to  $\chi^2$ minimization, we must now minimize $\Lambda^2(\alphabold;\xbold)$, the Lorentzian squared, 
with respect to the parameters $\alphabold$, for a given set of experimental points $\xbold$, i.e.,
\be
{\rm minimize\ over\ }\alphabold,\quad\quad\Lambda^2(\alphabold;\xbold)\equiv\sum_{i=1}^N\ln\left\{1+\gamma\delchisq\right\}\label{lormin},
\ee
 for the set of $N$ experimental points at $x_i$ having the value $y_i$ and error $\sigma_i$.

We note from \eq{lorentz} that the influence function for a point $i$ for small $\sqrt{\Delta \chi^2_i}$ has $\psi(z_i)\propto \sqrt{\Delta \chi^2_i}$ (just like the Gaussian distribution does), whereas for large $\sqrt{\Delta \chi^2_i}$, $\psi(z_i)\propto 1/\sqrt{\Delta \chi^2_i}$, i.e., it  {\em decreases} with large  $\chi^2_i$, so large outliers have  {\em much less} ``influence'' on the fit than do points close to the model curve. This is the feature that gives $\Lambda^2$ minimization its robust character. It is important to note that outliers have  little influence on the choice of the parameters $\alphabold_{\rm min}$ resulting from the minimization of $\Lambda^2_0$, a major consideration for a robust minimization method.

Extensive computer simulations\cite{sieve} have been made using Gaussianly generated data (constant and straight line models) which  showed empirically that the choice $\gamma=0.18$ minimized the rms (root mean square) parameter widths found in $\Lambda^2$ minimization.  Further, it gave rms widths that were almost as narrow as those found in $\chi^2$ minimization on the same Gaussianly distributed data.   We will adopt this value of $\gamma$, since it effectively minimizes the width for the  $\Lambda^2$ routine, which we now call $\Lambda^2_0(\alphabold;\xbold)$. 
Therefore we select for our robust algorithm, 
\be
{\rm minimize\ over\ }\alphabold,\quad\quad\Lambda^2_0(\alphabold;\xbold)\equiv\sum_{i=1}^N\ln\left\{1+0.18\delchisq\right\}\label{lormin0}.
\ee

Summarizing, if there were no outliers and the data were approximately Gaussianly distributed with proper errors, we would get the same results from the $\Lambda^2_0(\alphabold;\xbold)$ minimization of \eq{lormin} as we would from a conventional $\chi^2$ minimization, thus {\em doing no harm}. Therefore, it is always safe to first minimize  $\Lambda^2_0$.  If the $\chi^2_{\rm min}$ you get from it is satisfactorily small, then you are finished.  If not, you have a robust estimate of the parameters $\alphabold$.  It is a fail-safe procedure.

Clearly there are other possible long-tailed distributions that are also suitable for robust fitting.  However, it seems that the Lorentz distribution, and in particular, the minimization of $\Lambda^2_0(\alphabold;\xbold)$, satisfies our needs.

A very useful and simple program for minimization of $\Lambda^2_0(\alphabold;\xbold)$ is the AMOEBA program described in ``Numerical Recipes''\cite{nr}. 

%%%%%%%%%%%%%%%%%%%%%%%%
\subsection{The advantages of a $\chi^2$ fit in an idealized world}
In an idealized world where all of the data follow a normal (Gaussian) distribution, the use of the $\chi^2$ likelihood technique, through minimization of $\chi^2$, described in detail in \ref{appendix:chisq}, offers a powerful statistical analysis tool when fitting  models to a data sample.
It allows the phenomenologist to conclude either:
\begin{itemize}
\item The model is accepted, based on the value of its $\chi^2_{\rm min}$. It certainly fits well when $\chi^2_{\rm min}$ for a given $\nu$, the numbers of degrees of freedom, has a reasonably high probability ($\chi^2_{\rm min}\sim \nu$ for $\nu\gg 1$). 
On the other hand, it might be accepted with a much poorer $\chi^2_{\rm min}$, depending on the phenomenologist's judgment. In any event, the goodness-of-fit of the data to the model is known and an informed judgment can be made.
\item  Its parameter errors are such that  a change of $\Delta \chi^2=1$ from  $\chi^2_{\rm min}$ corresponds to changing a parameter by its standard error $\sigma$. These errors and their correlations are summarized in the standard covariance matrix $C$ discussed in Section \ref{appendix:chisq}.
\end{itemize}
or
\begin {itemize}
\item  The model is rejected, because  the probability that the data set fits the model  is too low,  i.e., $\chi^2_{\rm min}>>\nu$.
\end{itemize}
This decision-making capability (of accepting or rejecting the model) is of primary importance, as is the ability to estimate the parameter errors and their correlations.

Unfortunately, in the real world, experimental data sets are at best only approximately Gaussian and often are riddled with outliers---points far off from a best fit curve to the data, being many standard deviations away. This can be due to many sources, copying errors, bad measurements, wrong calibrations, misassignment of experimental errors,  etc. It is this world that this technique wishes to address---a world with many data points, and perhaps, many different experiments from many different experimenters, with possibly a significant number of  outliers. 

In Section \ref{section:sieve} we will propose the ``Sieve'' algorithm, an adaptive technique for discarding outliers while retaining the vast majority of the good data.
This then allows us to estimate the goodness-of-fit and make a  robust determination of both the parameters and their errors. In essence, we then retain all of the statistical benefits of the conventional $\chi^2$ technique.   
%%%%%%%%%%%%%%%%%%%%%%%%%%%%%%%%%%%%%%%%%%%%%%%%%%%%%%%%%%%%%
\subsection{The Adaptive Sieve Algorithm}\label{section:sieve} 
%%%%%%%%%%%%%%%%%%%%%%%%%%%%%%%%%%%%%%%%%%%%%%%%%%%%%%%%%
\subsubsection{Major assumptions}
Our 4 major assumptions about the experimental data are: 
\begin{enumerate}
\item The experimental data can be  fitted by a model which successfully describes the data.
\item The signal data are Gaussianly distributed, with Gaussian errors.
\item That we have ``outliers'' only, so that the background consists only of points ``far away'' from the true signal. \label{outlier}
\item The noise data,  i.e., the outliers, do not completely swamp the signal data. 
\end{enumerate} 
%%%%%%%%%%%%%%%%%%%%%%%%%%%%%%%%%%%%%%%%%%%%%%%%%%%%%%%%%%
\subsubsection{Algorithmic steps}
We now outline our adaptive Sieve  algorithm, consisting of several steps:
\begin{enumerate}
\item{Make a robust fit (see Section \ref{section:lorentz}) of {\em all} of the data (presumed outliers and all)\ by minimizing $\Lambda^2_0$, the Lorentzian squared, defined as
\be
\Lambda^2_0(\alphabold;\xbold)\equiv\sum_{i=1}^N\ln\left\{1+0.18\delchisq\right\}.\label{lambda0}
\ee 
 The $M$-dimensional parameter space of the fit is given by  $\alphabold=\{\alpha_1,\ldots,\alpha_M\}$, $\xbold=\{{x_1,\ldots,x_N}\}$ represents the abscissa of the $N$ experimental measurements $\mbox{\boldmath $y$}=\{y_1,\ldots,y_N\}$ that are  being fit and $\delchisq\equiv [(y_i-y(x_i;\alphabold))/\sigma_i]^2,$
  where $y(x_i;\alphabold)$ is the theoretical value at $x_i$ and $\sigma_i$ is the experimental error.  As discussed in Section {\ref{section:lorentz}}, minimizing $\Lambda^2_0$ gives  the same total $\chi^2_{\rm min}\equiv\sum_{i=1}^N \delchisq$ from \eq{lambda0} as that found in a $\chi^2$ fit,  as well as  rms widths (errors) for the parameters---for Gaussianly distributed data---that are almost the same as those found in a $\chi^2$ fit}. The quantitative measure of ``far away'' from the true signal,  i.e., point $i$ is an  outlier corresponding to Assumption (\ref{outlier}),  is the magnitude of its $\delchisq=[(y_i-y(x_i;\alphabold))/\sigma_i]^2$. 

If $\chi^2_{\rm min}$ is satisfactory, make a conventional $\chi^2$ fit to get the errors and you are finished.   If $\chi^2_{\rm min}$ is not satisfactory, proceed to step 
 \ref{nextstep}.
\item {Using the above robust $\Lambda^2_0$ fit as the initial estimator for the theoretical curve, evaluate $\delchisq$, for each of the $N$ experimental points.}\label{nextstep}
%%%%%%%%%%%%%%%%%%%%%%%%%%%%%%%%%%
\item A largest cut, $\delchisq_{\rm max}$, must now be selected. For example, we might start the process with $\delchisq_{\rm max}=9$. If any of the points have $\Delta \chi^2_i(x_i;\alphabold)>\delchisq_{\rm max}$, reject them---they fell through the ``Sieve''. The choice of $\delchisq_{\rm max}$ is an attempt to pick  the largest ``Sieve'' size (largest $\delchisq_{\rm max}$) that rejects all of the outliers, while minimizing the number of signal points  rejected. \label{redo}
%%%%%%%%%%%%%%%%%%%%%%%%%%%%%%%% 
%%%%%%%%%%%%%%%%%%%%%%%%%%%%%%%%%%
\item Next, make a conventional $\chi^2$ fit to the sifted set---these data points are the ones that have been retained in the ``Sieve''. This  fit is used to estimate   $\chi^2_{\rm min}$.    Since the data set has been truncated by eliminating the points with $\delchisq>\delchisq_{\rm max}$, we must slightly renormalize the $\chi^2_{\rm min}$ found to take this into account, by the factor $\cal R$. This effect is discussed later in detail in Section \ref{section:lessons}.

If the renormalized $\chi^2_{\rm min}$,  i.e., ${\cal R}\times \chi^2_{\rm min}$ is acceptable---in the {\em conventional} sense, using the $\chi^2$ distribution probability function for $\nu$ degrees of freedom---we consider the fit of the data to the  model to be satisfactory  and proceed to the next step. If the renormalized $\chi^2_{\rm min}$ is not acceptable and $\delchisq_{\rm max}$ is not too small, we pick a smaller $\delchisq_{\rm max}$ and go back to step \ref{redo}. The smallest value of $\delchisq_{\rm max}$ that makes much sense, in our opinion, is $\delchisq_{\rm max}>2$.  After all, one of our primary assumptions is that the noise doesn't swamp the signal. If it does, then we must discard the model---we can do nothing further with this model and data set!

%%%%%%%%%%%%%%%%%%%%%%%%%%% 
\item
From the  $\chi^2$ fit that was made to the `sifted' data in the preceding step, evaluate  the parameters $\alphabold$.
Next, evaluate the $M\times M$ covariance (squared error) matrix of the parameter space which was found in the $\chi^2$ fit. We find the new squared error matrix for the $\Lambda^2$  fit by multiplying the covariance matrix by the square of the factor $r_{\chi^2}$ (for example, as shown later in Section \ref{2}, $r_{\chi^2}\sim 1.02,1.05$, 1.11 and 1.14 for $\delchisq_{\rm max}=9$, 6, 4 and 2,  respectively ). The values of $r_{\chi^2}>1$ reflect the fact that a $\chi^2$ fit to the {\em truncated} Gaussian distribution that we obtain---after first making  a robust fit---has a rms (root mean square) width which is somewhat greater than the  rms width of the $\chi^2$ fit to the same untruncated distribution. Extensive computer simulations, summarized in Section \ref{section:lessons}, demonstrate that this {\em robust} method of error estimation yields accurate error estimates and error correlations, even in the presence of large backgrounds. You now have the parameters need to construct the error hyperellipse.
\end{enumerate}

You are now finished.  The initial robust $\Lambda^2_0$ fit has been used  to allow the phenomenologist to find a    sifted data set. The subsequent application of a $\chi^2$ fit to the {\em sifted set} gives stable estimates of the model parameters $\alphabold$, as well as a goodness-of-fit of the data to the model when $\chi^2_{\rm min}$ is renormalized for the effect of truncation due to using the cut $\delchisq_{\rm max}.$   Model parameter errors are found when the covariance (squared error) matrix of the $\chi^2$ fit is multiplied by the appropriate factor $(r_{\chi^2})^2$ for the cut $\delchisq_{\rm max}$. 

It is the {\em combination} of using both  $\Lambda^2_0$ (robust) fitting  and  $\chi^2$ fitting techniques on the sifted set that gives the Sieve algorithm its power to make both a robust estimate of the parameters $\alphabold$ as well as a robust estimate of their errors, along with an estimate of the goodness-of-fit.

Using this same sifted data set, you might then try to fit  to a {\em different} theoretical model and find  $\chi^2_{\rm min}$ for this second model.  Now one can compare the probability of each model in a meaningful way, by  using the standard $\chi^2$ probability distribution function for  the numbers of degrees of freedom $\nu$ for each of the models. If the second model had a very unlikely $\chi^2_{\rm min}$, it could now be eliminated.  In any event, the model maker would now have an objective comparison of the probabilities of the two models.  
%%%%%%%%%%%%%%%%%%%%%%%%%%%%%%%%%%%%%%%%%%%%%%%%%%%5  
\subsubsection{Evaluating the Sieve algorithm}
We will give two separate types of examples  which illustrate the Sieve algorithm.  In the first type, we used computer-generated data, normally distributed about
\begin{itemize}
\item a straight line, along with random  noise to provide outliers,
\item a constant, along with random  noise to provide outliers,
\end{itemize}
the details of which are described below.  The advantage here, of course, is that we know which points are signal and which points are noise.
 
For our second type, a real world example, we took four types of experimental data for elementary particle scattering from the archives of the Particle Data Group\cite{pdg}. For all energies above 6 GeV, we took total cross sections and $\rho$-values and made a fit to these data\cite{bhfroissartnew}. These were all published data points and the entire sample was used in our fit. We then made separate fits to 
\begin{itemize}
\item  $\pi^-p$ and $\pi^+p$ total cross sections $\sigma$ and $\rho$-values, discussed later in detail in Section \ref{section:froissartpip},
\item  $\bar pp$ and $pp$ total cross sections and $\rho$-values, discussed later in detail in Section \ref{sec:lnsqpp},
\end{itemize}
using Eqns. (\ref{sigmapm}), (\ref{rhopm}) and (\ref{derivpm}).
%%%%%%%%%%%%%%%%%%%%%%%%%%%% 
\subsubsection{Studies using large computer-generated data sets}\label{section:computer}
Extensive  computer simulations\cite{sieve} were made using the straight line model $ y_i=1-2x_i$ and the constant model $ y_i=10$. Over 500,000 events were computer-generated, with normal distributions of 100 signal points per event, some with no noise and others with 20\% and 40\% noise added, in order to investigate the accuracy and stability of the ``Sieve'' algorithm.  The cuts $\delchi>9$, 6, 4 and 2 were investigated in detail.  
%%%%%%%%%%%%%%%%%%%%%%%%%%%%%%
\subsubsection{A straight line model}\label{section:line}
An event consisted of generating  100 signal points plus either  20 or 40 background points, for a total of 120 or 140 points, depending on the background level desired.  Let RND be a random number, uniformly distributed from 0 to 1. 
Using random number generators, the first 100 points used $x_i=10\times \rnd$, where $i$ is the point number. This gives a signal randomly distributed between $x= 0$ and $x=10$. For each point $x_i$, a theoretical value $\bar y_i$ was found using $\bar y_i=1-2x_i$. Next, the value of $\sigma_i$, the ``experimental error'',  i.e, the  error bar assigned to point $i$, was generated as $\sigma_i=a_i+\alpha_i\times {\rnd}$. Using these $\sigma_i$, the $y_i$'s were generated, normally distributed about the value of $\bar y_i$ For $i=1$ to 50, $a_i=0.2,\ \alpha_i=1.5$, and for $i=51$ to 100, $a_i=0.2,\ \alpha_i=3$. This sample of 100 points  made up the signal.

The 40 noise points, $i=101$ to 140 were generated as follows. Each point was assigned an ``experimental error'' $\sigma_i=a_i+\alpha_i\times\rnd$. The $x_i$ were generated as $x_i=d_i+\delta_i\times \rnd$. In order to provide outliers, the value of $y_i$ was {\em fixed} at $y_i=1-2x_i+f_{\rm cut}\times{\rm Sign_i}\times(b_i+\beta_i)\times\sigma_i$ and the points were then placed at this fixed value of $ y_i$ and given the ``experimental error'' $\sigma_i$.  The parameter $f_{\rm cut}$ depended only on the value of $\delchisqmax$ that was chosen, being 1.9, 2.8, 3.4 or 4, for $\delchisqmax=2$, 4, 6 or 9, respectively, and was independent of $i$. These choices of $f_{\rm cut}$ made outliers that only existed for values of $\delchisq>\delchisqmax$.

For $i=101$ to 116, $d_i=0,\ \delta_i=10,\ a_i=0.75,\ \alpha_i=0.5,\ b_i=1.0, \ \beta_i=0.6$. To make ``doubles'' at the same $x_i$ as a signal point, if $y_{i-100}>1-2x_{i-100}$ we pick  Sign$_i=+1$; otherwise Sign$_i=-1$, so that the outlier is on the same side of the reference line $1-2x_i$ as is the signal point.

For $i=117$ to 128, $d_i=0,\  \delta_i=10,\ a_i=0.5,\ \alpha_i=0.5,\ b_i=1.0, \ \beta_i=0.6$; Sign$_i$ was randomly chosen as +1 or -1. This generates outliers randomly distributed above and below the reference line, with $x_i$ randomly distributed from 0 to 10.

For $i=129$ to 140, $d_i=8,\ \delta_i=2,\ a_i=0.5,\ \alpha_i=0.5,\ b_i=1.0, \ \beta_i=0.6$; Sign$_i$ = +1. This makes points in a ``corner'' of the plot, since $x_i$ is now randomly distributed at the ``edge'' of the plot, between 8 and 10.  Further, all of this points are above the line, since Sign$_i$ is fixed at +1, giving these points a large lever arm in the fit.  

For the events generated with 20 noise points, the above recipes for background were simply halved.  An example of such an event containing 120 points, for which $\delchisqmax=6$,  is shown in Fig. \ref{noisy}a), with the 100 squares being the normally distributed data and the 20 circles being the noise data. 

After a robust fit to the entire 120 points, the sifted data set retained 100 points after the $\delchi>6$ condition was applied.  This fit had $\chi^2_{\rm min}=88.69$, with an expected $\chi^2=\nu=98$, giving $\chi^2_{\rm min}/\nu=0.905$. Using a renormalization factor  ${\cal R}=1/0.901$, we get a renormalized  $\chi^2_{\rm min}/\nu=1.01$---see Section \ref{section:lessons}  for details of the renormalization factor. After using the Sieve algorithm, by minimizing $\chi^2$ for the sifted set, we found that the best-fit straight line, $y=<a>+<b>x$, had $<a>=0.998\pm0.12$ and $<b>=-2.014\pm0.020$.   The parameter errors given above come from multiplying the errors found in a conventional $\chi^2$ fit to the sifted data by the  factor $r_{\chi^2}=1.05$---for details see Section \ref{section:lessons}.  This turns out to be a  high probability fit ( see footnote \ref{footnote:probability}) with a  probability of 0.48 (since the renormalized $\chi^2_{\rm min}/\nu=1.01$, whereas we expect $<\chi^2/\nu>=1.0\pm0.14$).  

\begin{figure}[h,t,b,p] %Fig. 6
\begin{center}
\mbox{\epsfig{file=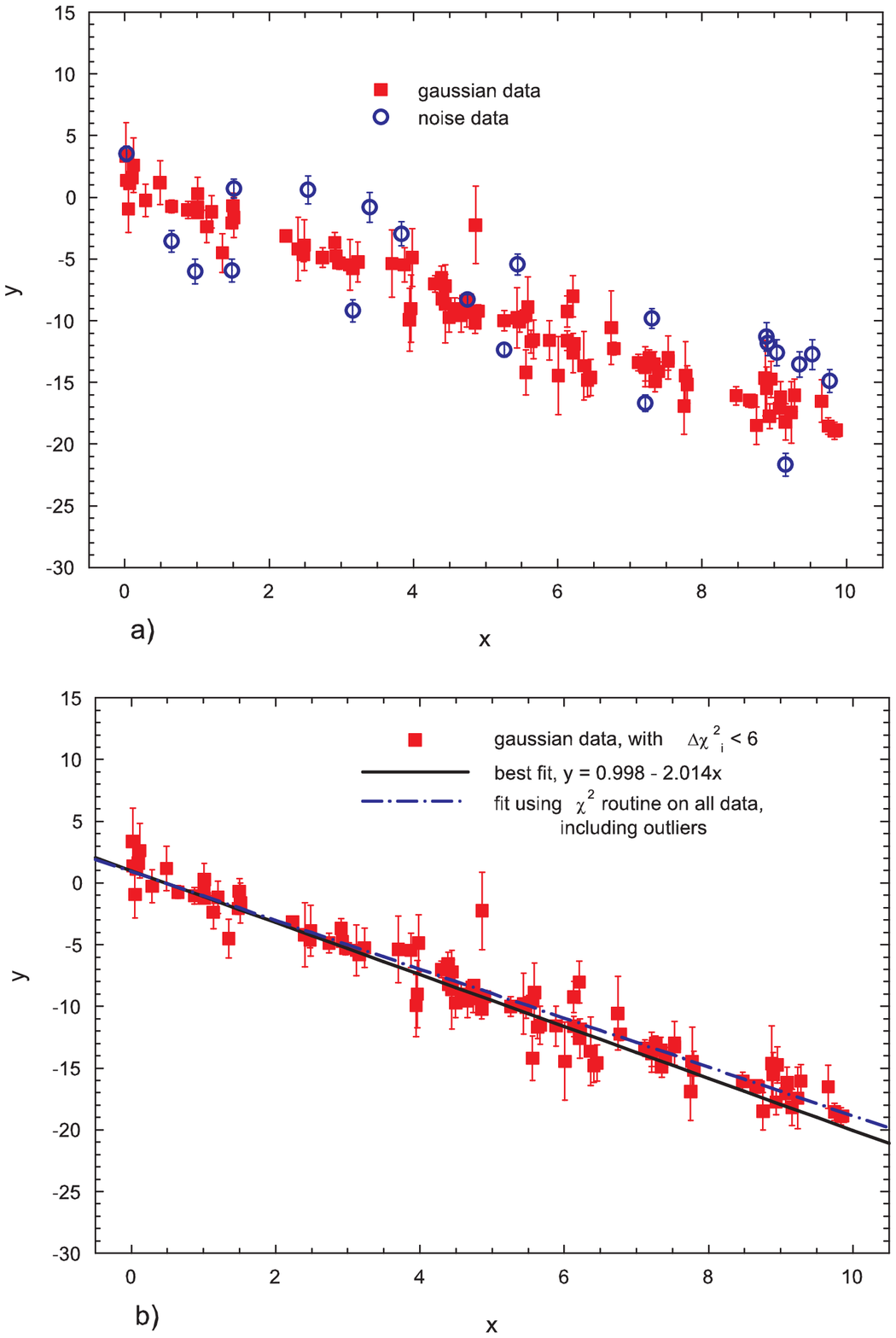,width=5in%
,bbllx=0pt,bblly=0pt,bburx=425pt,bbury=618pt,clip=%
}}
\caption[Computer-generated Gaussianly distributed data about the straight line $y=1-2x$]{\footnotesize
Computer-generated Gaussianly distributed data about the straight line $y=1-2x$.\newline
a) The 100 squares are a computer-generated Gaussianly distributed data set. The 20 open circles are randomly distributed noise data. See Section \ref{section:line} for details.
\newline b) The 100 data points shown  are the result of screening all 120 data points  for those points having $\Delta \chi^2_i<6.$ 
There were no noise points (open circles)  retained in the Sieve and the 100 squares are the Gaussian data retained in the Sieve.
The best fit curve to all points with $\delchi <6$, $y=a+bx$, is the solid curve, where $a=0.998\pm0.12$, $b=-2.014\pm0.020$, and $\chi^2_{\rm min}/\nu=0.91$, yielding a renormalized value ${\cal R}\times\chi^2_{\rm min}/\nu=1.01$ compared to the expected $<\chi^2_{\rm min}>/\nu=1.0\pm0.14$. The dashed-dot  curve  is a $\chi^2$ fit to the totality of data---100 signal plus 20 noise points---which has  $\chi^2_{\rm min}/\nu=4.8$.\label{noisy}
}
\end{center}
\end{figure}
%%%%%%%%%%%%%%%%%

\begin{figure}[h,t,b,p] %Fig. 7
\begin{center}
\mbox{\epsfig{file=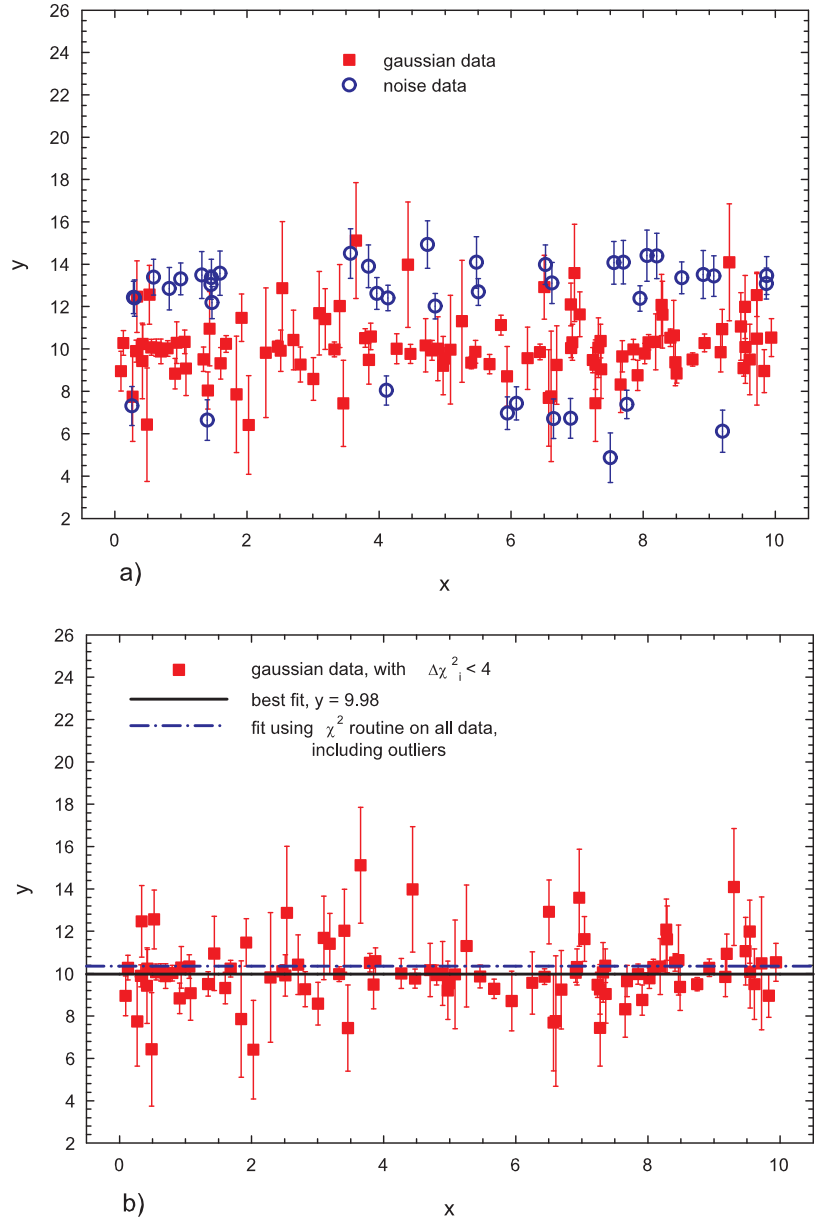,width=5in%
%,bbllx=0pt,bblly=0pt,bburx=500pt,bbury=700pt,clip=%
,bbllx=0pt,bblly=0pt,bburx=475pt,bbury=639pt,clip=%
}}
\end{center}
\caption[Computer-generated Gaussianly distributed data   about the constant $y=10$ ($\Delta \chi^2_i<4.$)]{ \footnotesize
Computer-generated Gaussianly distributed data   about the constant $y=10$ ($\Delta \chi^2_i<4.$).\newline
a) The 100 squares are a computer-generated Gaussianly distributed data set  about the constant $y=10$. The 40 open circles are randomly distributed noise data. See Section \ref{section:constant} for details.
\newline b) The 98 data points shown  are the result of screening all 140 data points  for those points having $\Delta \chi^2_i<4.$ 
There were no noise points (open circles)  retained in the Sieve and the 98 squares are the Gaussian data retained in the Sieve.
The best fit curve to all points with $\delchi <4$, $y=c$, is the solid curve, where $c=9.98\pm0.074$,  and $\chi^2_{\rm min}/\nu=0.84$, yielding a renormalized value ${\cal R}\times\chi^2_{\rm min}/\nu=1.09$  compared to the expected $<\chi^2_{\rm min}>/\nu=1.0\pm0.14$. The dashed-dot  curve  is a $\chi^2$ fit to the totality of data---100 signal plus 40 noise points---which has  $\chi^2_{\rm min}/\nu=4.39$. 
 }
\label{constant140_4}
\end{figure}

%%%%%%%%%%%%%%%%%

\begin{figure}[h,t,b,p] %Fig. 8
\begin{center}
\mbox{\epsfig{file=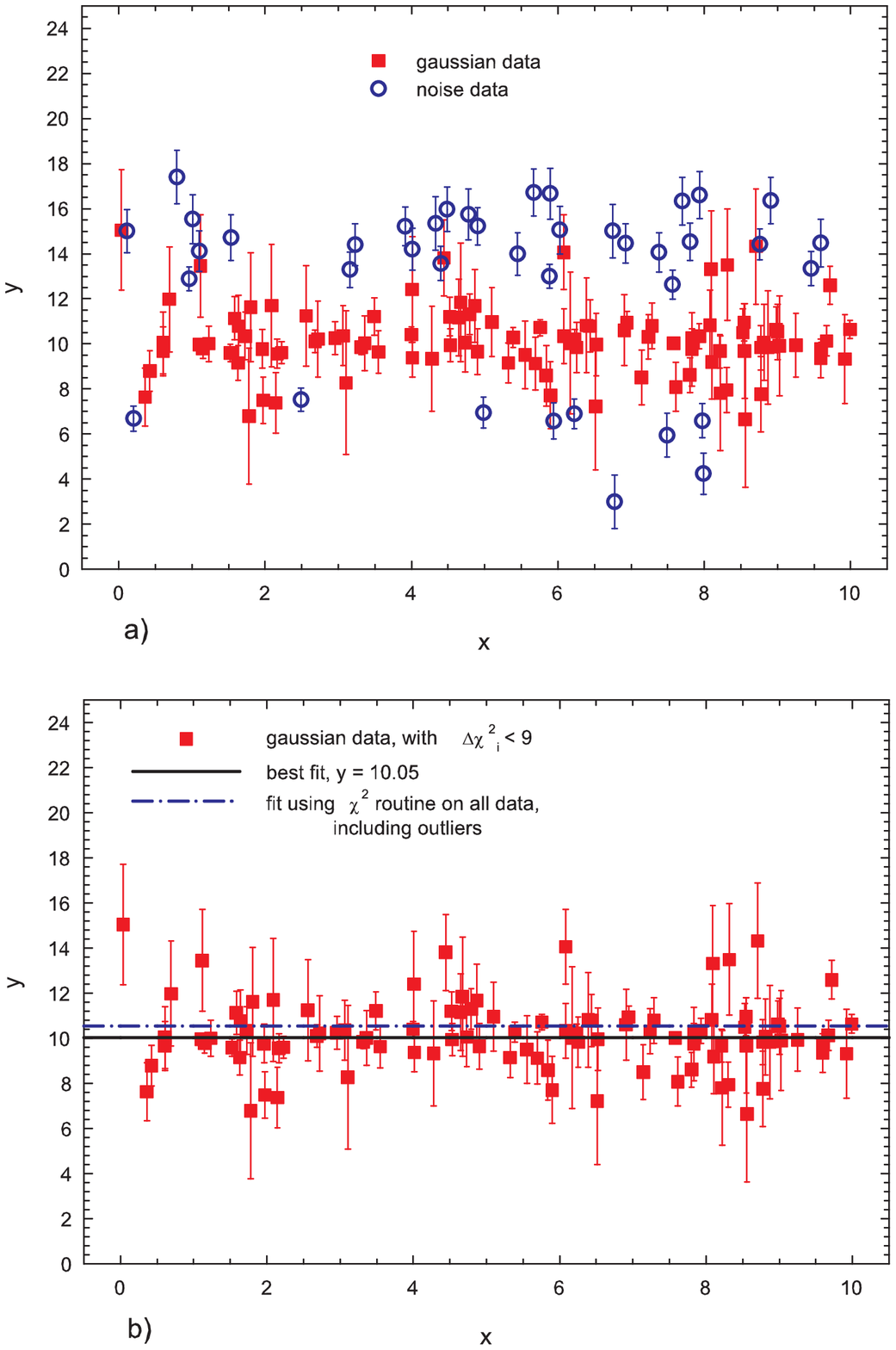,width=5in%
,bbllx=0pt,bblly=0pt,bburx=464pt,bbury=613pt,clip=%
}}
\end{center}
\caption[Computer-generated Gaussianly distributed data   about the constant $y=10$ ($\Delta \chi^2_i<9.$)]
{ \footnotesize
Computer-generated Gaussianly distributed data   about the constant $y=10$ ($\Delta \chi^2_i<9.$).\newline
a) The 100 squares are a computer-generated Gaussianly distributed data set  about the constant $y=10$. The 40 open circles are randomly distributed noise data. See Section \ref{section:constant} for details.
\newline b) The 99 data points shown  are the result of screening all 140 data points  for those points having $\Delta \chi^2_i<9.$ 
There were no noise points (open circles)  retained in the Sieve and the 98 squares are the Gaussian data retained in the Sieve.
The best fit curve to all points with $\delchi <9$, $y=c$, is the solid curve, where $c=10.05\pm0.074$,  and $\chi^2_{\rm min}/\nu=1.08$, yielding a renormalized value ${\cal R}\times\chi^2_{\rm min}/\nu=1.11$ compared to the expected $<\chi^2>/\nu=1.0\pm0.14$. The dashed-dot  curve  is a $\chi^2$ fit to the totality of data---100 signal plus 40 noise points---which has  $\chi^2_{\rm min}/\nu=8.10$. 
 }
\label{constant140}
\end{figure}

Of the original 120 points, all 100 of the signal points were retained (squares), while no noise points (diamonds) were retained. The solid line is the best $\chi^2$ fit,  $y=0.998-2.014x$.   

Had we applied a $\chi^2$ minimization to original 120 point data set, we would have found $\chi^2=570$, which has infinitesimal statistical probability. The straight line resulting from that fit, $y=0.925-1.98x$,  is also shown in Fig. \ref{noisy}b) as the dot-dashed curve. For large x, it tends to overestimate the true values.

To investigate the stability of our procedure with respect to our choice of  $\delchi$, we reanalyzed the full data set for the cut-off, $\delchimax=4$. The evaluation of the parameters $a$ and $b$ was completely stable, essentially independent of the choice of $\delchi$.   The robustness of this procedure on this particular data set is  evident.
%%%%%%%%%%%%%%%%%%%%%%%%%%%%%%%%%%%%%%%%%%%%%%%%%%%%%%%%%%%55
\subsubsection{Distributional widths  for the straight line model}\label{section:width}
We now generate extensive computer simulations of data sets resulting from the straight line $y_i=1-2x_i$ using the recipe of Section \ref{section:line}, with and without outliers, in order to test the Sieve algorithm.
We have generated 50,000 events with 20\% background and 50,000 events with 40\% background, for each cut $\delchimax=9$, 6, 4 and 2. We also generated 100,000 Gaussianly distributed events with no noise. 

%%%%%%%%%%%%%%%%%%%%%%%%%%%%%%%%
\subsubsection{Case 1}\label{1} We  generated 100,000 Gaussianly distributed events with {\em no } noise. Let $a$ and $b$ be the intercept and slope of the straight line $y=1-2x$ and define $<a>$ as the average $a$, $<b>$ as the average $b$ found for the 100,000 straight-line events, each generated with 100 data points, using both a $\Lambda^2_0$ (robust) fit and a $\chi^2$ fit. The purpose of this exercise was to find ${r}(\Lambda^2_0)$,   the ratio of the $\Lambda^2_0$ rms  parameter width $\sigma(\Lambda^2_0)$ divided by $\Sigma$, the parameter error from the $\chi^2$ fit, i.e.,
\[
r_a(\Lambda^2_0)\equiv {\sigma_a(\Lambda^2_0)\over\Sigma_a}, \quad r_b(\Lambda^2_0)\equiv {\sigma_b(\Lambda^2_0)\over\Sigma_b},
\]
as well as demonstrate that there were no biases (offsets) in  parameter determinations found in   $\Lambda_0^2$ and $\chi^2$ fits.

The measured offsets $1-<a_{\chi^2}>$, $1-<a_{\Lambda^2}>$, $-2-<b_{\chi^2}>$ and $-2-<b_{\Lambda^2}>$ were all numerically compatible with zero, as expected, indicating that the parameter expectations were not biased.

%\begin{eqnarray*}
%1-<a_{\chi^2}>&=&7.0\times 10^{-4}\pm 4.41\times10^{-4}\\
%1-<a_{\Lambda^2}>&=&4.3\times 10^{-4}\pm 5.59\times10^{-4}\\
% -2-<b_{\chi^2}>&=&1.72\times10^{-4}\pm 7.62\times10^{-5},\\
%-2-<b_{\Lambda^2}>&=&1.73\times 10^{-4}\pm 9.77\times10^{-5}.
%\end{eqnarray*}
%  All of the offsets are essentially numerically compatible with zero, as %expected, indicating that the parameter expectations were not biased.
  
Let $\sigma$ be the rms width of a parameter distribution and $\Sigma$ the error from the $\chi^2$ covariant matrix. We found:
\begin{eqnarray*}
\sigma_a(\chi^2)&=&0.139\pm0.002\ \quad{\rm and \ }\Sigma_a=0.138\\
\sigma_b(\chi^2)&=&0.0261\pm0.003\quad {\rm and\ }\Sigma_b=0.0241, 
\end{eqnarray*}
showing that the rms widths $\sigma$ and parameter errors $\Sigma$ were the same for the $\chi^2$ fit, as expected.  
% rhoA=1.034+-.0095,  rhoB=1.029+-.011,, AVERAGE: rhoAB=1.032+-.003
Further, the  width ratios $r$ for the $\Lambda^2_0$ fit are given by 
\begin{eqnarray*}
r_a({\Lambda^2_0})&=& 1.034\pm0.010\\
r_b(\Lambda^2_0)&=& 1.029\pm0.011,
\end{eqnarray*}
demonstrating that:
\begin{itemize}
\item the $r$'s of the $\Lambda^2_0$ are almost as good as that of the $\chi^2$ distribution, $r(\chi^2)=1$.
\item the  ratios   of the rms $\Lambda^2$ width to the rms $\chi^2$ width for both  parameters $a$ and $b$ are the {\em same}, i.e., we can now simply write 
\be
r_{\Lambda^2}={\sigma_{\Lambda^2}\over{\Sigma}}\sim 1.03.
\ee
\end{itemize}
 Finally, we find that $1-<\chi^2/\nu>=0.00034\pm0.00044$, which  is approximately  zero, as expected.
%%%%%%%%%%%%%%%%%%%%%%%%%%%%%%%%55
\subsubsection{Case 2}\label{2}
For Case 2, we investigate data generated with 20\% and 40\% noise that  have been subjected to  the  adaptive Sieve algorithm, i.e., the sifted data  after cuts of $\delchisqmax=9$, 6, 4 and 2. We investigated this truncated sample to measure  possible biases and  to obtain numerical values for $r$'s.

We generated 50,000 events, each with 100 points normally distributed and with either 20 or 40 outliers, for each cut. A robust fit was made to the entire sample (either 120 or 140 points) and we 
sifted the data, rejecting all points with either $\delchisq>9,$ 6, 4 and 2, according to how the data were generated. A conventional $\chi^2$ analysis was then made to the sifted data. The results are summarized in Table \ref{table:rchisq}.% 
%%%%%%%%%%%%%%%%%%%
%%Table #1
\begin{table}[h,t]                   % Use "table" environment, but also
				 % use  "tabular" environment below.
%
               \caption[Sieve algorithm results for $r_{\chi^2}=\sigma/\Sigma$;  $<\chi^2_{\rm min}>/\nu$, for both the straight line case  and the constant  case;  $\sigma/\sigma_0$, the total loss of accuracy due to truncation, as  functions of the cut $\delchimax$]
{\protect\small Results for $r_{\chi^2}=\sigma/\Sigma$, the ratio of the rms (root mean square) width to $\Sigma$, the error for the $\chi^2$ fit;  $<\chi^2_{\rm min}>/\nu$, for both the straight line case  and the constant  case;  $\sigma/\sigma_0$, the ratio of the rms width (error) of the parameter relative to what the error would be if the sample were not truncated,  i.e., the total loss of accuracy due to truncation, as  functions of the cut $\delchimax$.   The average results for $r_{\chi^2}$ and $<\chi^2_{\rm min}>/\nu$, where $\nu$ is the number of degrees of freedom, are graphically shown in Fig. \ref{renorm}.
See Sections \ref{section:width}, \ref{section:constant} and \ref{section:lessons} for details. The theoretical values for the renormalization factor ${\cal R}^{-1}$ are from \eq{Rminus1} and the survival fractions $S.F.$ are from \eq{SF}.  See Section \ref{section:lessons} for a discussion of the error-broadening factor $\sigma/\sigma_0$.
\label{table:rchisq}
}

\vspace{.1in}
\def\arraystretch{1.5}            % Make the space between rows in the Table,
				  % 1.5 x bigger than the default spacing.
\begin{center}
\begin{tabular}[b]{ccccc}
    %\cline{2-4}
%\cline{2-5}
	\multicolumn{1}{c}{}
	&\multicolumn{1}{c}{$\delchimax=9$ }
	&\multicolumn{1}{c}{$\delchimax=6$ }
	&\multicolumn{1}{c}{$\delchimax=4$ }
	&\multicolumn{1}{c}{$\delchimax=2$ }\\

\hline
$r_{\chi^2, {\rm str.\ line}}$&1.034&1.054&1.098&1.162\\
$r_{\chi^2,{\rm  constant}}$&1.00&1.05&1.088&1.108\\
\hline
average&1.018&1.052&1.093&1.148\\
\hline\hline
$<\chi^2_{\rm min}>/\nu$&&&&\\
str. line&0.974&0.901&0.774&0.508\\
constant&0.973&0.902&0.774&0.507\\
\hline
average&0.973&0.901&0.774&0.507\\
${\cal R}^{-1}$&0.9733&0.9013&0.7737&0.5074\\
\hline\hline
$S.F.$&0.9973&0.9857&0.9545&0.8427\\
\hline\hline
$\sigma/\sigma_0$&1.02&1.06&1.19&1.25\\
\hline\hline
\end{tabular}
     %\vspace{1in} \\
\end{center}
\end{table}
\def\arraystretch{1}  %Restore the default row spacing in the Table.
%%%%%%%%%%%%%%%%%%%%%%%%%%%%%%%%%%%%%%%%%%%%%%%%%%%%%
As before, we found that the widths from the $\chi^2$ fit were slightly smaller than the widths from a robust fit, so we adopted only the results for the $\chi^2$ fit.

There were negligible offsets $1-<a>$ and $-2-<b>$, being $\sim 1$ to $ 5\%$ of the relevant rms widths, $\sigma_a$ and $\sigma_b$, for both the robust and $\chi^2$ fits. 

In any individual $\chi^2$ fit to the $j$th data set, one measures $a_j,b_i,\Sigma_{a_j},\Sigma_{b_j}$ and $(\chi^2_{\rm min}/\nu)_j$. Thus, we  characterize all of our computer simulations in terms of these 7 observables. 

We again find that the $r_{\chi^2}$ values---defined as $\sigma/\Sigma$---are the same, whether we are measuring $a$ or $b$. They are given by 
$r_{\chi^2}={\sigma}/{\Sigma}=1.034$, 1.054, 1.098 and 1.162  
for the cuts $\delchisqmax=9,$ 6, 4 and 2, respectively\footnote{The fact that $r_{\chi^2}$ is greater than 1 is counter-intuitive.  Consider the case of generating a Gaussian distribution with unit variance about the value $y=0$. If we were to define $\delchi\equiv  (y_i-0)^2=y_i^2$, with $\Delta$ being the cut $\delchimax$,  then the truncated differential probability distribution would be $P(x)=\frac{1}{\sqrt{2\pi}}\exp({-x^2/2})$ for $-\sqrt{\Delta}\le x \le +\sqrt{\Delta}$, whose rms value clearly is {\em less than} 1---after all, this distribution is truncated compared to its parent Gaussian distribution.  However, that is not what we are doing.  What we do is to first make a robust fit  to each  untruncated event that was Gaussianly generated  with unit variance about  the mean value zero.  For every  event we then find the value $y_0$, its best fit parameter, which, although close to zero with a mean of zero, is {\em non-zero}.  In order to obtain the truncated event whose width we sample with the {\em next} $\chi^2$ fit, we use $\delchi\equiv(y_i-y_0)^2$.  It is the jitter in $y_0$ about zero that is responsible for the rms width becoming greater than 1. This result is true even if  the first fit to the untruncated data were a $\chi^2$ fit. 
\label{footnote:subtle}}. 
Further, they are the same for 20\% noise and 40\% noise, since the cuts rejected all of the noise points.  In addition, the $r$ values  were found to be the same as the $r$ values for the  case of truncated pure signal, using  the same $\delchisqmax$ cuts.  The signal retained was 99.7, 98.57, 95.5 and 84.3 \% for the cuts  $\delchisqmax =9,$ 6, 4 and 2, respectively---see Section \ref{section:lessons} and \eq{SF} for theoretical values of the amount of signal retained.

We experimentally determine the rms (root mean square) widths $\sigma$ (the errors of the parameter) by multiplying the $r$ value,  a known  quantity {\em independent} of the particular event, by the appropriate $\Sigma$ which is measured for {\em that} event, i.e.,
\begin{eqnarray*}
{\sigma_a}&=&\Sigma_a\times r_{\chi^2}\\
{\sigma_b}&=&\Sigma_b\times r_{\chi^2}.
\end{eqnarray*}
The  rms widths are now determined for {\em any} particular data set by multiplying the known factors $r_{\chi^2}$ by the appropriate $\Sigma$ found (measured) from the covariant matrix of the $\chi^2$ fit of that data set.

Also shown in Table \ref{table:rchisq} are the values of $\chi^2_{\rm min}/\nu$ found for the various cuts. We will compare these results later with those for the constant case, in   Section \ref{section:constant}

We again see that a sensible approach for data analysis---even where there are large backgrounds of $\sim 40\%$---is to use the parameter estimates for $a$ and $b$ from the truncated $\chi^2$\  fit and assign  their errors as 
\begin{eqnarray}
\sigma_a&=r_{\chi^2}\Sigma_a \nonumber\\
\sigma_b&=r_{\chi^2}\Sigma_b,
\end{eqnarray}
where $r_{\chi^2}$ is a function of the $\delchimax$ cut utilized.
Before estimating the goodness-of-fit, we must renormalize the observed $\chi^2_{\rm min}/\nu$ by the appropriate numerical factor for the $\delchimax$ cut used.

This strategy of using  an adaptive $\delchisqmax$ cut minimizes the error assignments,  guarantees  robust fit parameters with no significant bias and also returns a goodness-of-fit estimate.

%%%%%%%%%%%%%%%%%%
\subsubsection{The constant model, $y_i=10$}\label{section:constant}
For this case, we investigate a different theoretical model ($y_i=10$) with a different background distribution, to measure the values of $r_{\chi^2}$ and $<\chi^2_{\rm min}/\nu>$. 

An event consisted of generating  100 signal points plus either  20 or 40 background points, for a total of 120 or 140 points, depending on the background level desired.  Again, let RND be a random number, uniformly distributed from 0 to 1. 
Using random number generators, for the  first 100 points $i$,  a theoretical value $ \bar y_i=10 $ was chosen. Next, the value of $\sigma_i$, the ``experimental error'',  i.e., the  error bar assigned to point $i$, was generated as $\sigma_i=a_i+\alpha_i\times {\rnd}$. Using these  $\sigma_i$, the $y_i$'s were generated, normally distributed about the value of $\bar y_i=10$ . For $i=1$ to 50, $a_i=0.2,\ \alpha_i=1.5$, and for $i=51$ to 100, $a_i=0.2,\ \alpha_i=3$. This sample of 100 points  made up the signal.

The 40 noise points, $i=101$ to 140 were generated as follows. Each point was assigned an ``experimental error'' $\sigma=a_i+\alpha_i\times\rnd$. In order to provide outliers, the value of $y_i$ was fixed at $y_i=10+f_{\rm cut}\times{\rm sign_i}\times(b_i+\beta_i)\times\sigma_i$ and the points were then placed at this fixed value of $ y_i$ and given the ``experimental error'' $\sigma_i$.  The parameter $f_{\rm cut}$ depended only on the value of $\delchisqmax$ that was chosen, being 1.9, 2.8, 3.4 or 4, for $\delchisqmax=2$, 4, 6 or 9, respectively, and was independent of $i$.

For $i=101$ to 116, $a_i=0.75,\ \alpha_i=0.5,\ b_i=1.0, \ \beta_i=0.6$; Sign$_i$ was randomly chosen at +1 or -1. 

For $i=117$ to 128, $a_i=0.5,\ \alpha_i=0.5,\ b_i=1.0, \ \beta_i=0.6$; This generates outliers randomly distributed above and below the reference line, with $x_i$ randomly distributed from 0 to 10.

For $i=129$ to 140, $a_i=0.5,\ \alpha_i=0.5,\ b_i=1.0, \ \beta_i=0.6$; Sign$_i$ = +1. This forces 12 points to be greater than 10, since Sign$_i$ is fixed at +1.  For the events generated with 20 noise points, the above recipes for background were simply halved. 

Two examples of events with 40 background points are shown in Figures \ref{constant140_4}a) and \ref{constant140}a), with the 100 squares being the normally distributed data and the 40 circles being the noise data. 

In Fig. \ref{constant140_4}b) we show the results after using the cut $\delchisqmax=4$.  No noise points (diamonds)  were retained, and 98 signal  points (circles) are shown. The best fit, $y=9.98\pm0.074$, is the solid line, whereas the dashed-dot curve is the fit to all 140 points.  The observed $\chi^2_{\rm min}/\nu=0.84$ yields a renormalized value ${\cal R}\times \chi^2_{\rm min}/\nu=1.09$, in good agreement with the expected value $\chi^2_{\rm min}/\nu=1\pm0.14$. If we had fit to the entire 140 points, we would find $\chi^2_{\rm min}/\nu=4.39$, with the fit being the dashed-dot curve.  

In Fig. \ref{constant140}b) we show the results after using the cut $\delchisqmax=9$.  No noise points (diamonds)  were retained, and 98 signal  points (circles) are shown. The best fit, $y=10.05\pm0.074$, is the solid line, whereas the dashed-dot curve is the fit to all 140 points.  The observed $\chi^2_{\rm min}/\nu=1.08$ yields a renormalized value ${\cal R}\times \chi^2_{\rm min}/\nu=1.11$, in good agreement with the expected value $\chi^2_{\rm min}/\nu=1\pm0.14$. If we had fit to the entire 140 points, we would find $\chi^2_{\rm min}/\nu=8.10$, with the fit being the dashed-dot curve.  The details of the renormalization of $\chi^2_{\rm min}/\nu$ and the assignment of the errors are given in Section \ref{section:lessons}

We computer-generated a total of 500,000 events,  50,000 events with  20\% noise and an additional 50,000 events with 40\% noise, for each of the cuts $\delchi>9$, 6, 4 and 2, and 100,000 events with no noise. 

For the sample with no cut and no noise, we found $r_{\Lambda^2_0}=1.03\pm0.02$,  equal to the value  $r_{\Lambda^2_0}=1.03$ that was found for the straight line case.

Again, we found that our results for $r_{\chi^2}$ were independent of background, as well as model, and only  depended on the cut.  We also found that the biases (offsets) for the constant case, $(10-<a_{\chi2}>)$, although non-zero for the noise cases,  were small in comparison to   $\sigma$, the rms width.

The results for cuts $\delchimax=9$, 6, 4 and 2  are detailed in Table \ref{table:rchisq}.  
We see in Table \ref{table:rchisq}, compared with the straight line results of Section \ref{2}, that the $r_{\chi^2}$ values for the constant case are essentially identical, as expected. Further, we find the same results for the values of $\chi^2_{\rm min}/\nu$ as a function of the cut $\delchisqmax$.

%%%%%%%%%%%%%%%%%%%%%%%%%%%%%%%%%
%%%%%%%%%%%%%%%%%%%%%%%%%%%%%%%%%%%%%%%
\subsubsection {Lessons learned from computer studies  of  a straight line model and  a constant model}\label{section:lessons}
\begin{itemize}
\item
As found in Sections \ref{2} and \ref{section:constant} and  detailed in Table \ref{table:rchisq}, we have universal values of $r_{\chi^2}$ and $<\chi^2_{\rm min}>/\nu$,  as a function of the cut $\delchisqmax$, independent of both background and model.
\item Use the parameter estimates from the $\chi^2$ fit to the sifted data and assign  the parameter errors to the fitted robust parameters to be 
\begin{eqnarray}
\sigma(\chi^2)&=&r_{\chi^2}\times\Sigma,\nonumber
\end{eqnarray}
where $r_{\chi^2}$ is a function of the cut $\delchisqmax$, given by  the average of the straight line and constant cases of Table \ref{table:rchisq}.
This strategy gives us a minimum parameter error, with only very small  biases to the parameter estimates, working well even for large backgrounds  (less than or the order 40\%).  
\item  Next,  renormalize the value found for $\chi^2_{\rm min}/\nu$ by the appropriate averaged value of $<\chi^2_{\rm min}>/\nu$  for the straight line and  constant case, again as a function of the cut $\delchisqmax$. 
\item Defining  $\Delta$ as the $\delchimax$ cut and $\cal R$ as the renormalization factor that multiplies $\chi^2_{\rm min}/\nu$, we find from inspection of Cases 1 to 2 for the straight line and  of  Section \ref{section:constant} for the case of the constant fit that  a best fit parameterization of $r_{\chi^2}$, valid for $ \Delta\ge2$\ is given by 
\be
r_{\chi^2}=1+0.246e^{-0.263\Delta}.\label{rho}
\ee
 
We note that ${\cal R}^{-1}$, for large $\nu$,  is given analytically by
\begin{eqnarray}
{\cal R}^{-1}&\equiv&{\int^{+\sqrt{\Delta}}_{-\sqrt\Delta} x^2e^{-x^2/2}\,dx}/{\int^{+\sqrt{\Delta}}_{-\sqrt\Delta} e^{-x^2/2}\,dx}\nonumber\\
&=&1-\frac{2}{\sqrt{\pi}}\frac{e^{-\Delta/2}}{{\rm erf}(\sqrt{\Delta/2})}.\label{Rminus1}
\end{eqnarray}
Graphical representations of $r_{\chi^2}$ and ${\cal R}^{-1}$  are shown in Figures  \ref{renorm}a) and \ref{renorm}b), respectively. Some numerical values are given in Table \ref{table:rchisq} and are compared to the computer-generated values found numerically for the straight line and constant cases. The agreement is excellent.
\item We define $\sigma_0$ as the rms parameter width  that we would have had for a $\chi^2$ fit to the uncut sample where the sample had  had no background, and define $\Sigma_0$ the error found from the covariant matrix. They are, of course, equal to each other, as well as being the smallest error possible. We note that the ratio $\sigma/\sigma_0=r_{\chi^2}\times\Sigma/\Sigma_0$. This ratio is a function of the cut $\Delta$  through both $r_{\chi^2}$ {\em and} $\Sigma$, since for a truncated distribution $\Sigma/\Sigma_0$ depends inversely on the square root  of the fraction of signal points that survive the cut $\Delta$. In particular, the survival fraction $S. F.$ is given by
\be
S.F.=\int^{+\sqrt{\Delta}}_{-\sqrt\Delta}\frac{1}{\sqrt{2\pi}} e^{-x^2/2}\,dx= {\rm erf}(\sqrt{\Delta/2})\label{SF}
\ee
and is  99.73, 98.57, 95.45 and 84.27 \% for the cuts  $\Delta =9,$ 6, 4 and 2, respectively. The survival fraction $S.F.$ is shown in Table \ref{table:rchisq} as a function of the cut $\delchimax$, as well as is the ratio $\sigma/\sigma_0$.  We note that the true cost of truncating a Gaussian distribution, i.e., the enlargement of the error due to truncation,  is not $r_{\chi^2}$, but rather $r_{\chi^2}/\sqrt{S.F.}$, which  ranges from $\sim 1.02$ to 1.25 when the cut $\delchimax$ goes from 9 to 2. This rapid loss of accuracy is why the errors become intolerable for cuts $\delchimax$ smaller than 2.  
\end{itemize}
%%%%%%%%%%%%%%%
\begin{figure}[ht] %Fig. 9
\begin{center}
\mbox{\epsfig{file=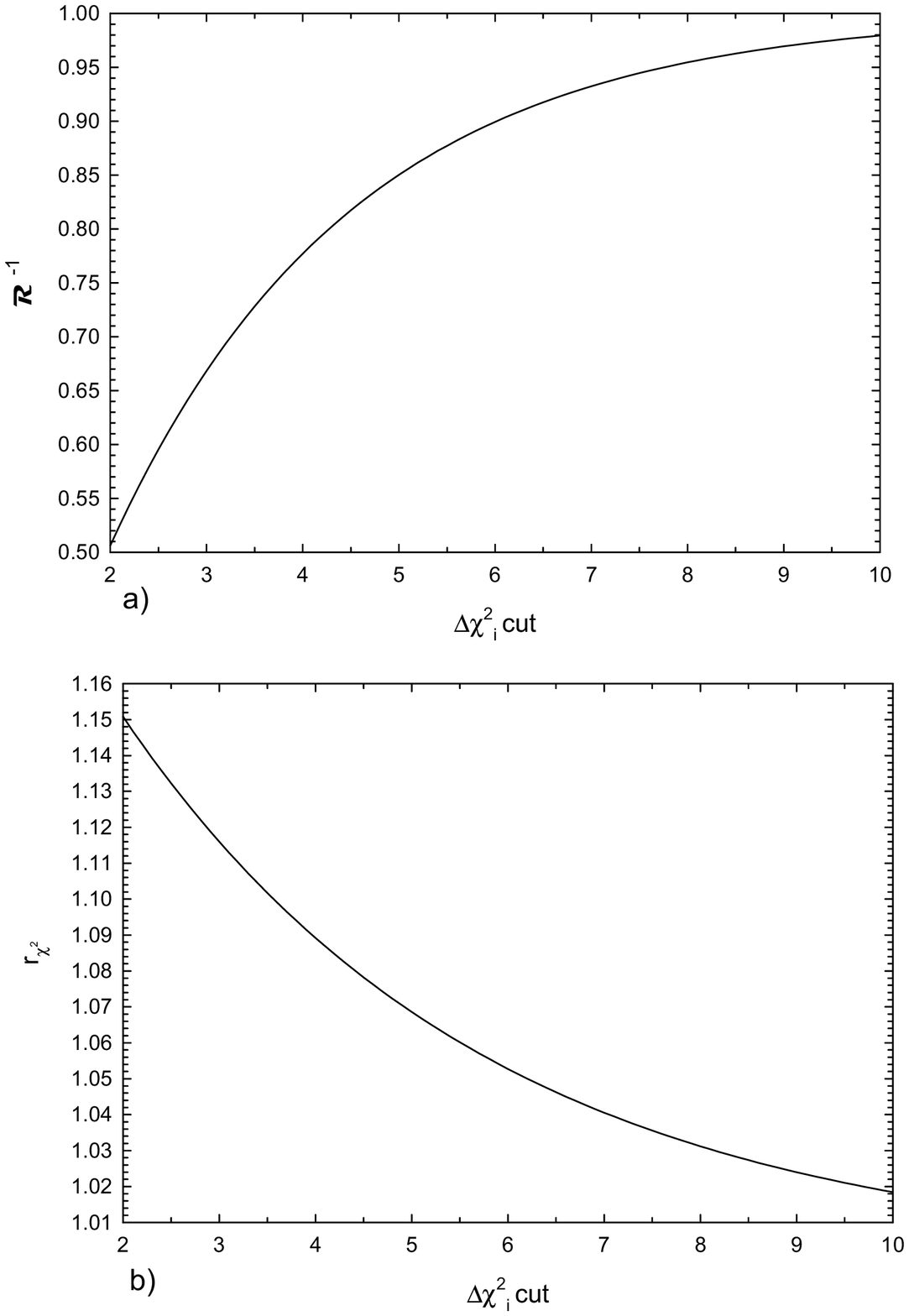,width=3.6in%
,bbllx=46pt,bblly=139pt,bburx=497pt,bbury=758pt,clip=%
}}
\end{center}
\caption[Sieve algorithm factors, ${\cal R}^{-1}$ and $r_{\chi^2}$]{ \footnotesize 
The Sieve algorithm factors, ${\cal R}^{-1}$ and $r_{\chi^2}$ .\newline
a) ${\cal R}^{-1}$ of \eq{Rminus1}, the reciprocal of the factor  that multiplies $\chi^2_{\rm min}/\nu$ found in the $\chi^2$ fit to the sifted data set   vs. $\delchi$ cut, the $\delchimax$ cut.
\newline 
b) $r_{\chi^2}$ of \eq{rho}, the  factor whose square multiplies the covariant matrix found in the $\chi^2$ fit to the sifted data set   vs. $\delchi$ cut, the $\delchimax$  cut.
\newline  
See Sections \ref{section:width}, \ref{section:constant} and \ref{section:lessons} for details. In  \eq{rho} and \eq{Rminus1}, the $\delchi$ cut is called $\Delta$.
  }
\label{renorm}
\end{figure}
%%%%%%%%%%%%%%%%%%%%%%%%%%%%%%
\subsubsection {Fitting strategy}\label{section:strategy}

We conclude that an effective  strategy for both eliminating noise and making robust parameter estimates together  with {\em robust}   error assignments is:
\begin{enumerate}
\item Make an initial $\Lambda^2_0$ fit to the entire data sample. If $\chi^2_{\rm min}/\nu$ is satisfactory, then make a standard $\chi^2$ fit to the data and you are finished. If not, then proceed to the next step.
\item Pick a large value of $\delchisq_{\rm max}$,  e.g., $\delchisq_{\rm max}=9.$ 
\item Obtain a sifted sample by throwing away all points with $\delchisq>\delchisq_{\rm max}$.\label{sifted} 
\item  Make  a conventional $\chi^2$  fit to the sifted sample. Having selected $\delchisq_{\rm max}$, find ${\cal R}^{-1}$ from \eq{Rminus1}. If the renormalized value ${\cal R}\times \chi^2_{\rm min}/\nu$ is sufficiently near 1, i.e., the goodness-of-fit is satisfactory, then go to the next step. If, on the  other hand,
${\cal R}\times \chi^2_{\rm min}/\nu$ is too large, pick a smaller value of $\delchisq_{\rm max}$ and go to step \ref{sifted}. For example, if you had used a cut of 9, now pick $\delchisq_{\rm max}=6$ and start again.  Finally, if you reach $\delchisq_{\rm max}=2$ and you still don't have success, quit---the background has penetrated too much into the signal for the ``Sieve'' algorithm to work properly.  
\item   Use the parameter estimates found from the $\delchisq_{\rm max}$ fit in the previous step.
\item  Find a  new squared error matrix  by multiplying the  covariant matrix $C$ found in the $\chi^2$ fit by $(r_{\chi^2})^2$. Use the value of $r_{\chi^2}$ found in \eq{rho} for the chosen value of the cut  $\delchisq_{\rm max}$  to obtain a robust error estimate essentially independent of background distribution.
\end{enumerate}
You are now finished, having made a robust determination of the parameters, their errors and the goodness-of-fit.  
%%%%%%

The renormalization factors $\cal R$ are only used in  estimating the value of the goodness-of-fit, where small changes in this value are not very important.   Indeed, it hardly matters if the estimated renormalized $\chi^2_{\rm min}/\nu$ is  between 1.00 and 1.01---the possible variation of the expected renormalized $\chi^2_{\rm min}/\nu$ due to the two different background distributions.  After all,  it is a subjective judgment call on the part of the phenomenologist as to whether the goodness-of-fit is satisfactory.  For large $\nu$, only when $\chi^2_{\rm min}/\nu$ starts approaching 1.5  does one really begin to start worrying  about the model. For $\nu\sim100$, the error expected in $\chi^2_{\rm min}/\nu$ is $\sim 0.14$, so uncertainties in the renormalized $\chi^2_{\rm min}/\nu$ of the order of several percent really play no role. The accuracy of the renormalized values is perfectly adequate for the  purpose of judging whether  to keep or discard a model.  

In summary, extensive computer simulations using sifted data sets  show that by  combining the  $\chi^2$ parameter determinations  with the corrected covariance matrix from the $\chi^2$ fit,  a ``robust'' estimate of the errors, basically independent of both the background distribution and the model, is obtained. Further, the renormalized $\chi^2_{\rm min}/\nu$ is a good predictor of the goodness-of-fit.  Having to make  a $\Lambda^2_0$ fit to sift the data and then a $\chi^2$ fit to the sifted data is a small computing cost to pay  compared to the ability to make accurate predictions. Clearly, if the data are not badly contaminated with outliers,  e.g., if a $\delchisq_{\rm max}=6$ fit is satisfactory, the additional penalty paid is that the errors are enlarged by a factor of $\sim1.06$ (see Table \ref{table:rchisq}), which is not unreasonable to rescue a data set. Finally, if you are not happy about the error determinations, you can use the parameter estimates you have found to make Monte Carlo simulations of your model\cite{nr}.  By repeating a $\Lambda_0^2$ fit to the simulated distributions and then sifting them to the same value of $\delchisq_{\rm max}$ as was used in the initial determination of the parameters, and finally, by making a $\chi^2$ fit to the simulated sifted set  you can make an error determination based on the spread in the parameters found from the simulated data sets.  However, this latter method, although essentially fool-proof,  suffers in practice from being very costly in computer time.
%%%%%%%%%%%%%%%%%%%%%%%%%%%%%%%%%%%%%%%%%%%%%%%%%%%%%%%%%%%
\section{Fitting of high energy experimental cross sections $\sigtot$, $\rho$-values and nuclear slope parameters $B$ from accelerators} 
In this Section, we discuss various methods of comparing high energy experimental cross sections and $\rho$-values to theoretical models.   Unfortunately, there are no good models from first principles that we can invoke, and thus we  are forced to use  phenomenological model fits to data. We will discuss two very different methods:
\begin{itemize} 
\item Eikonal fits that are inspired by QCD, sometimes called the QCD-inspired model and often referred to in the literature as the ``Aspen'' model,
\item  Fits using real analytical amplitudes,  
\end{itemize}
with each of these approaches having its own advantages and disadvantages.

One of the  Aspen model's  advantages is that  along with $\sigtot$ and $\rho$ for $pp$ and $p\bar p$ data,  it also uses data from the nuclear slope parameter $B$ in its fit. Further advantages are that, in addition to predicting $\sigtot$ and $\rho$, it also can  calculate $\sigma_{\rm inel}$, $\sigma_{\rm el}$,  $B$ and $d\sigma_{\rm el}/dt$, where $\sigma_{\rm el}$ is the total elastic scattering cross section, $d\sigma_{\rm el}/dt$ is the differential elastic scattering cross section as a function of the squared 4-momentum transfer $t$ and $\sigma_{\rm inel}$ is the total inelastic scattering cross section. Thus, it has more predictive power and therefore many more possible comparisons with different types of experimental data. It has the disadvantage of requiring at least 11 parameters (12 parameters if we also fit a power law to the odd amplitude) and is  model-dependent in trying to emulate QCD in its cross section structure and in assuming that the matter distribution of the quarks and gluons in the nucleon is the same as the electric charge distribution of the proton, albeit each with its own scale.  Further, the model requires extensive numerical integrations over the impact parameter space $b$ each time a parameter is varied in the fitting procedure.

The advantages of fits using real analytical amplitudes are that they are fundamentally model-independent and consequently  are more transparent to interpretation, in terms of Regge poles, Froissart bounds, etc.  Further, they require many fewer parameters---typically, only 6 parameters are required to fit the cross sections.They use simultaneous fits to $\sigtot$ and $\rho$ for both $pp$ and $p\bar p$ data, but in turn, predict only $\sigtot$ and $\rho$.  Since they require no numerical integrations, they are not very computer-intensive are easy to fit. 

Given the two very different approaches, it will be very interesting to compare the results of the two methods, which we will do later.
%%%%%%%%%%%%%%%%%%%%%%%%%%%%%%%%%%%%%%%%%%%%%%%%%%%%%55
\subsection{The Aspen model, a QCD-inspired eikonal model}\label{section:aspen}
The QCD-inspired eikonal was first introduced in its present form in 1990 by Margolis and collaborators\cite{blockfletcherhalzenmargolisvalin}.  For detailed references on its evolvement and applications, see ref. \cite{blockfletcherhalzenmargolisvalin,blockhalzenmargolis,bghp}. As it evolved, it gained the name ``Aspen model''.  

All  cross sections will be computed in an eikonal formalism,  guaranteeing unitarity throughout. 
In Section \ref{sec:eikonal}, we used the conventional definition of the complex eikonal $\chi(b,s)=-2i\delta(b,s)$, in terms of the complex phase shift $\delta$.  

For the Aspen model, an unconventional definition, 
\ba
\chi_R&=&2\delta_R,\nonumber\\
\chi_I&=&-2\delta_I
\ea
has been used, corresponding to the amplitude in impact parameter space being
\be
a(b,s)=\frac{i}{2}\left( 1-e^{-\chi_I(b,s)+i\chi_R(b,s)}\right).
\ee
Rewriting the formulas of Section \ref{sec:eikonal} in terms of the Aspen eikonal, we find
\ba
\sigtot(s)&=&2\int \left[1-e^{-\chi_I(b,s)}\cos\left(\chi_R(b,s)\right)\right]\, d^2\vec b,\label{sigelofb20}\\
\rho(s)&=&\frac{\int e^{-\chi_I(b,s)}\sin(\chi_R(b,s))\,d^2\vec b}{\int\left[ 1-e^{-\chi_I(b,s)}\cos(\chi_R(b,s))\right]\,d^2\vec b}\quad, \label{rhoofb0}\\
B(s)&=&\frac{1}{2}\frac{\int | e^{-\chi_I(b,s)+i\chi_R(b,s)}-1|b^2\,d^2\vec b}{\int | e^{-\chi_I(b,s)+i\chi_R(b,s)}-1|\,d^2\vec b},\\
\frac{d\sigma_{\rm el}}{dt}&=&\pi\left|\int J_0(qb)\left[ e^{-\chi_I(b,s)+i\chi_R(b,s)}-1\right]b\,db\right|^2,\\
\sigma_{\rm el}(s)&=&\int\left| e^{-\chi_I(b,s)+i\chi_R(b,s)}-1\right|^2\,d^2\vec b,\\
\sigma_{\rm inel}(s)&\equiv&\sigtot(s) -\sigma_{\rm el}(s)=\int \left( 1-e^{-2\chi_I(b,s)}\right)\,d^2\vec b,\label{siginel}
\ea 
where $\sigma_{\rm inel}(s)$ is the total inelastic cross section.

The even eikonal profile function $\chi^{ even}$ receives contributions 
from
quark-quark, quark-gluon and gluon-gluon interactions, and therefore
\begin{eqnarray}
\chi^{ even}(s,b) &=& \chi_{qq}(s,b)+\chi_{qg}(s,b)+\chi_{gg}(s,b)
\nonumber \\
&=& i\left [ \sigma_{qq}(s)W(b;\mu_{qq})
+ \sigma_{qg}(s)W(b;\sqrt{\mu_{qq}\mu_{gg}})
+ \sigma_{gg}(s)W(b;\mu_{gg})\right ]\, ,\label{chiintro}
\end{eqnarray}
where $\sigma_{ij}$ is the cross sections of the colliding partons, and
$W(b;\mu)$ is the overlap function in impact parameter space,
parameterized as the Fourier transform of a dipole form factor.

In this model, hadrons asymptotically evolve into black disks of
partons. The rising cross section,
asymptotically associated with gluon-gluon interactions, is simply
parameterized by a normalization and energy scale, and two parameters:
$\mu_{gg}$ which describes the ``area" occupied by gluons in the
colliding hadrons, and $J\,(\,=1+\epsilon)$. Here, $J$ is defined via
the gluonic structure function of the proton, which is assumed to
behave as $1/x^J$ for small x. It therefore controls the soft gluon
content of the proton, and it is meaningful that its value ($\epsilon
\simeq 0.05$) is consistent with the one observed in deep inelastic
scattering. The introduction of the quark-quark and quark-gluon
terms allows one to adequately parameterize the data at all energies,
since the ``size'' of quarks and gluons in the proton can be
different. Values of $\mu_{qq}=0.89$ GeV and $\mu_{gg}=0.73$
GeV.\ were obtained\cite{bghp},  indicating that the gluons occupy a larger area of the proton than do the quarks.

In Appendix \ref{app:qcdeikonal} we give the details of the eikonal appearing in \eq{chiintro}. 
\begin{itemize}
\item In Appendix \ref{app:siggg}, the glue-glue contribution, which is responsible for the increasing cross section at high energies, is given in complete detail. At high energies, it is shown that $\sigma_{gg}\rightarrow s^\epsilon$, and as a consequence, in Appendix \ref{app:highenergysigmagg} we see that the Aspen model satisfies the Froissart bound. Finally, in Appendix \ref{app:sigggevaluation}, the complete evaluation of $\sigma_{gg}$ is made.

We note that  it takes 3 constants to specify
$\sigma_{gg}$;
	the normalization constant $C_{gg}$, which is fitted by the data,
the threshold mass $m_0$, taken as 0.6 GeV and 
 $\epsilon=J-1$, the parameter in the gluon structure function which 
determines 
the behavior at low x ($\propto 1/x ^{1+\epsilon}$),
 taken as 0.05.   
\item In Appendix \ref{app:sigqq}, a toy model is used to get the following approximation to  the quark-quark term :
\begin{equation}
\sigma_{qq}(s)\equiv\Sigma_{gg} \left( C + C_{Regge}^{even}
\frac {m_0}{\sqrt s}\right ), \label{sigmaqq1} \end{equation} 
where
$
\Sigma_{gg}\equiv9\pi \alpha_s^2/m_0^2$ is the cross section scale 
and $C$ and $C_{Regge}^{even}$ are constants to be fitted by data,    
Thus, $\sigma_{qq}(s)$ 
simulates quark-quark interactions with a constant cross
section plus a Regge-even falling cross section.
\item In Appendix \ref{app:sigqg}, a toy model is used to suggest that the
quark-gluon contribution can be approximated by
\begin{equation}
 \sigma_{qg}(s)=\Sigma_{gg} C_{qg}^{log}\log\frac{s}{s_0}, 
\end{equation} 
where $C_{qg}^{log}$ is a constant.
 Here, we attempt to simulate diffraction with the logarithmic term.
We must fit 2 constants, the normalization constant $C_{qg}^{log}$ and  $s_0$, the square of the energy scale in the log term.
\item Using the theorems derived in  Section  \ref{phragmen}, we make the even (under crossing) amplitudes analytic by the substitutions $s\rightarrow se^{-i\pi/2}$ for the even amplitude in Appendix \ref{app:evencontribution}.
\item
It can be shown that a high energy analytic odd amplitude (for
its structure in $s$, see Eq. (5.5b) of reference \cite{bc}, with
$\alpha =0.5$) that fits the data is given by\cite{bghp} 
\begin{eqnarray}
\chii^{odd}(b,s)&=&-\sigma_{odd}\,W(b;\mu_{odd})\nonumber\\
&=&-C_{odd}\Sigma_{gg}\frac{m_0}{\sqrt{s}}e^{i\pi/4}\,W(b;\mu_{odd}),
\end{eqnarray}
with
\be
W(b,\mu_{odd})=\frac{\mu_{odd}^2}{96\pi}(\mu_{odd}
b)^3\,K_3(\mu_{odd} b)
\ee
normalized so that
$\int W(b\,;\mu_{odd})d^2 \vec{b}=1.$

In order to determine the cross section $\sigma_{odd}$, the normalization constant $C_{odd}$ must be fitted to the data. To determine the impact
parameter profile in $b$ space, we also must fit the mass parameter
$\mu_{odd}$ to the data. A mass $\mu_{odd}\approx
0.53$ GeV was found.

We again reiterate that the odd eikonal, which  vanishes like $1/\sqrt s$, accounts for the
difference between $pp$ and $\bar p p$. Thus, at high energies,
the odd term vanishes, and we can neglect the difference between $pp$ and $\ppbar$.
\item In its present incarnation, the Aspen model requires 11 parameters: $C$, $C^{log}_{qg}$,  $C'_{gg}$, $C^{even}_{Regge}$, $C_{odd}$, $s_0$, $m_0$, $ \epsilon$, $\mu_{qq}$, $\mu_{gg},$ and $\mu_{odd}$. Had we left the energy dependence of the odd amplitude as a power law to be fitted by data, as is done in the real analytic amplitude analysis,  there would have been 12 parameters.
\end{itemize}
%%%%%%%%%%%%%%%%%
\subsubsection{Fitting the Aspen model using analyticity constraints: $\sigtot$, $\rho$ and $B$}\label{sec:Bfit}
As discussed earlier in detail in Section \ref{sec:analyticityextensions}, analyticity, in the guise of requiring the high energy data to fit smoothly onto low energy cross sections,  provides powerful constraints on fits to high energy cross section data. Using these constraints with the Aspen model, we will now  simultaneously fit high energy total cross sections $\sigtot$, $\rho$ and nuclear slope parameters $B$, for both $pp$ and $p\bar p$.
In Table \ref{table:qcdparam} we show the results of the new fit made here, using the anchoring of the cross sections,  $\sigtot(pp)=40.2$ mb and $\sigtot(p\bar p)=57.0$ mb at $\sqrt s= 4$ GeV,  i.e., imposing  the analyticity constraint of \eq{sig+=sighigh2} on the Aspen model (QCD-inspired) fit.   This allows us to make a much better fit than in an earlier work\cite{bghp}, since the fit is now severely constrained by low energy cross sections, as well as the high energy cross sections $\sigtot$, $\rho$  and  $B$. 

In Fig. \ref{fig:sigqcd}, using the parameters of Table \ref{table:qcdparam}, we plot the total cross section $\sigtot$, in mb, against $\sqrt s$, the c.m. energy,  in GeV, for both $pp$ and $\bar p p$  scattering, comparing the predictions to the available experimental data with energies greater than 15 GeV.   The solid line and squares are for $pp$ and the dotted line and open circles are for $\bar p p$. 

In Fig. \ref{fig:rhoqcd}, again using the parameters of Table \ref{table:qcdparam}, we plot  $\rho$ against $\sqrt s$, the c.m. energy,  in GeV, for both $pp$ and $\bar p p$  scattering, comparing the predictions to the available experimental data with energies greater than 15 GeV.   The solid line and squares are for $pp$ and the dotted line and open circles are for $\bar p p$. 

In Fig. \ref{fig:Bqcd}, we plot the nuclear slope parameter $B$, in (GeV/c)$^{-2}$ against $\sqrt s$, the c.m. energy,  in GeV, for both $pp$ and $\bar p p$  scattering, comparing the predictions to the available experimental data with energies greater than 15 GeV.   The solid line and squares are for $pp$ and the dotted line and open circles are for $\bar p p$. 

Note that the 6 experimental quantities used in obtaining Table \ref{table:qcdparam} are $\sigtot$, $\rho$ and $B$, for both $pp$ and $\bar p p$ collisions.

In Table \ref{table:QCDpredictions} we show some high energy predictions from the Aspen model   for $\sigma_{\bar pp},\sigma_{pp}, \rho_{\bar pp}, \rho_{pp}$,
$ B_{\bar pp}$ and $B_{pp}$.
%%%%%%%%%%%%%%%%%%%%%%%%%%%%%%
\subsubsection{Aspen model predictions: $\sigma_{\rm el}$, $d\sigma_{\rm el}/dt$ and $\sigma_{\rm el}/\Sigma_{\rm el}$}\label{section:seloverSel}
We now turn our attention to new predictions from our fit, the elastic cross section $\sigma_{\rm el}$ and the differential elastic scattering cross section  $d\sigma_{\rm el}/dt$  as a function of the squared 4-momentum transfer, $t$, using the parameters found in Table  \ref{table:qcdparam}. 

Figure \ref{fig:sigel} is a plot of  the total elastic cross section $\sigma_{\rm el}$, in mb vs. $\sqrt s$, the c.m. energy,  in GeV, for both $pp$ and $\bar p p$  scattering, comparing the predictions to the available experimental data with energies greater than 15 GeV.   The solid line and squares are for $pp$ and the dotted line and open circles are for $\bar p p$. We note that over the entire energy interval, $\sigma_{\rm el}(pp)$ is effectively indistinguishable from $\sigma_{\rm el}(\bar pp)$.  It was alluded to earlier that the model should evolve into a black disk at very high energies, with the ratio of $\sigma_{\rm el}/\sigtot\rightarrow0.5$. Indeed, the energy at which it will  happen is enormous: the ratio rises very slowly from  $\sim 0.18$ at ISR energies(20 to 60 GeV) to $\sim 0.23$ at the S$\bar {\rm p}$pS (550 GeV) to $\sim 0.25$ at the Tevatron (1800 GeV) to $\sim 0.29$ at the LHC (14 TeV).  Indeed, $\sigma_{\rm el}/\sigtot$ only has risen to $\sim 0.32$ at 100 TeV. Clearly, it is rising very slowly and is nowhere near the black disk ratio of 0.5---``true asymptopia'' is still far away!

In Figure \ref{fig:dsdt}, the elastic differential scattering cross section 
$d\sigma_{\rm el}/dt$ , in mb/(GeV/c)$^2$, is plotted against $|t|$, in (GeV/c)$^2$. 
 The solid curve is the prediction for the reaction $ pp\rightarrow pp$ at the LHC, at $\sqrt s=14$ TeV.
The dashed curve is the prediction for the reaction $\bar pp\rightarrow\bar pp$ at $\sqrt{s}=1.8$ TeV, at the Tevatron Collider; the data 
points are from the E710 experiment. 

The data from the E710 experiment\cite{E710,Amos} are compared with our prediction. Unfortunately, because of a Lambertson magnet that was in the way, the maximum $|t|$-value that could be explored in E710 was very near  the predicted first minimum and thus could not  confirm its existence.  However, the observed data are in excellent agreement with the theoretical curve, confirming the prediction of Block and Cahn\cite{bc} that the curvature $C$ (of  $\ln(d\sigma_{\rm el}/dt ))$ should go through zero at the Tevatron energy and should then become positive, as seen in Fig. \ref{fig:dsdt}. 

As expected from diffractive shrinkage, the first minimum of the 14 TeV curve moves to lower $|t|$ than the first minimum of the 1.8 TeV plot. Our new prediction at 14 TeV for the first sharp  minimum at $|t|\sim 0.4$ (GeV/c)$^2$ and a second shallow minimum at $|t|\sim 2$  (GeV/c)$^2$ should be readily verified when the LHC becomes operative. 

In Fig. \ref{fig:sigoverSIG}, the ratio $\sigma_{\rm el}/\Sigma_{\rm el}$ is plotted against the c.m. energy $\sqrt s$, in GeV, where $\sigma_{\rm el}=\int_{-\infty} ^0(d\sigma_{\rm el}/dt)\,dt$ is the true total  elastic scattering cross section,   while $\Sigma_{\rm el}=\sigtot^2/(16\pi B)$, which was defined by \eq{Siggreek2} in Section \ref{sec:measurements}. We recall to your attention that what is typically measured by experimenters is $\Sigma_{\rm el}$ and {\em not} the real $\sigma_{\rm el}$. From Fig. \ref{fig:sigoverSIG} we see that the error made is $\sim 5$--10 \% for energies less than 100 GeV, being $\sim 5$ \% at the S$\bar {\rm p}$pS, $\sim 4$\% at the Tevatron and  less than 1\% at the LHC, and 
hence,  the MacDowell-Martin bound\cite{macdowell}, which  states that 
$\sigma_{\rm el}/\Sigma_{\rm el}\ge 8/9$, is clearly satisfied. 
%%%%%%%%%%%%%%%%%%%%%%%%%%%%%%%%%%%%%%%%%%%%
\subsubsection{Rapidity gap survival probabilities}
We now turn to some interesting properties of the Aspen eikonal, concerning the  validity of the factorization theorem  for nucleon-nucleon, $\gamma p$ and $\gamma\gamma$ collisions. It was shown was that  the survival probabilities of large rapidity gaps in high energy $\pbar p$ and $pp$ collisions are identical (at the {\em same} energy) for $\gamma p$ and $\gamma \gamma$ collisions, as well as for nucleon-nucleon collisions\cite{me}. We will show that neither the factorization theorem nor the reaction-independence of the survival probabilities depends on the assumption of an additive quark model, but, more generally, depends on the {\em opacity} of the eikonal being {\em independent} of whether the reaction is n-n, $\gamma p$ or $\gamma \gamma$.  

Rapidity gaps are an important tool in new-signature physics for ultra-high energy $\pbar p$ collisions. Block and Halzen\cite{gapsurvival} used the Aspen model (QCD-inspired eikonal model) to make a reliable calculation of the survival probability of rapidity gaps in nucleon-nucleon collisions.  We sketch below their arguments. 

From Section \ref{section:aspen}, using \eq{siginel}, we write
the inelastic cross section, $\sigma_{\rm inel}(s)$, as
\be
\sigma_{\rm inel}(s)=\int\,\left [1-e^{-2\chii(b,s)}\right ]\,d^2\vec{b}.\label{sigin}
\ee
It is readily shown, from unitarity and \eq{sigin}, that the differential probability in impact parameter space $b$, for {\em not} having an inelastic interaction, is given by 
\be
\frac{d^2P_{\rm no\  inelastic}}{\,d^2\vec{b}\qquad\qquad}{}=e^{-2\chii(b,s)}.\label{noinelastic}
\ee
Because the parameterization is
 both unitary and analytic, its high energy predictions are
 effectively model--independent, if you require that the proton is  
asymptotically a black disk. 

As an example of a large rapidity gap process, consider the production cross section for Higgs-boson production through W fusion.  The inclusive differential cross section in impact parameter space $b$ is given by % 
$d\sigma/d^2\vec{b}=\sigma_{{\rm WW}\rightarrow {\rm H}}\,W(b\,;\mu_{\rm qq}),$ %\label{eq:xsection}
where it is assumed that $W(b\,;\mu_{\rm qq})$ (the differential impact parameter space {\em quark} distribution in the proton) is the same as that of the W bosons.  

The cross section for producing the Higgs boson {\em and} having a large rapidity gap (no secondary particles) is given by
\be
\frac{d\sigma_{\rm gap}}{d^2\vec{b}}=\sigma_{{\rm WW}\rightarrow {\rm H}}\,W(b\,;\mu_{\rm qq})e^{-2\chii(s,b)}=\sigma_{{\rm WW}\rightarrow {\rm H}}\,\frac{d(|S|^2)}{d\vec{b}^2}. \label{eq:nosecondaries}
\ee
Equation (\ref{noinelastic}) was used in \eq{eq:nosecondaries}  to get the exponential suppression factor, using the normalized impact parameter space distribution $W(b\,;\mu_{\rm qq})=\frac{\mu_{\rm qq}^2}{96\pi}(\mu_{\rm qq} b)^2K_3(\mu_{\rm qq} b)$, with $\mu_{\rm qq}=0.901\pm 0.005$ GeV. 

To generalize, define  $<|S|^2>$, the survival probability of {\em any} large rapidity gap, as
\be 
<|S|^2>=\int W(b\,;\mu_{\rm qq})e^{-2\chii(s,b)}d^2\,\vec{b},\label{eq:survival}
\ee 
which is the integral over the differential probability density in impact parameter space $b$ for {\em no} subsequent interaction (the exponential suppression factor of \eq{noinelastic}) multiplied by the quark probability distribution in $b$ space. It should perhaps be emphasized that \eq{eq:survival} is the probability of {\em survival} of a large rapidity gap and {\em not} the probability for the production and survival of large rapidity gaps, which is the quantity observed experimentally. 
We note that the energy dependence of the survival probability $<|S|^2>$ is through the energy dependence of $\chii$, the imaginary portion of the eikonal. 
For illustration, we show in Fig. \ref{fig:1800}  a plot of ${\rm Im}\, \chi_{\bar pp}$ and the exponential damping factor of \eq{eq:survival}, as a function of the impact parameter $b$, at $\sqrt{s}=1.8$ TeV, .  
The results of numerical integration of \eq{eq:survival} for the survival probability $<S^2>$ at various c.m. energies   are summarized in Table \ref{table:survival}.%  

Further, Block and Halzen\cite{gapsurvival} find for the quark component that the mean squared radius of the quarks in the nucleons, $<R_{\rm nn}^2>$, is given by
$<R_{\rm nn}^2>=\int b^2 W(b;\mu_{\rm qq})\,d^2\vec b=16/\mu_{\rm qq}^2=19.70\ {\rm GeV}^{-2}$. Thus, $b_{\rm rms}$, the rms impact parameter radius is given by $b_{\rm rms}=4/\mu_{\rm qq}=4.44$ GeV$^{-1}$. Inspection of Fig. \ref{fig:1800} (1.8 TeV) at $b_{\rm rms}$ shows a sizeable probability for no interaction  ($e^{-2\chii}$) at that typical impact parameter value.

In Ref. \cite{blockhalzenpancheri}, using the additive quark model, it is shown that the eikonal $\chi^{\gamma p}$ for $\gamma p$ reactions is found by substituting $\sigma\rightarrow \frac{2}{3}\sigma$, $\mu\rightarrow\sqrt{\frac{3}{2}}\mu$ into $\chi^{\rm even}(s,b)$, the even nucleon-nucleon eikonal, found in  Appendix \ref{app:qcdeikonal}, in \eq{chieven}, i.e., the even QCD-inspired eikonal.  

$\chi_{even}$ is given by the sum of
three contributions, gluon-gluon, quark-gluon, and quark-quark, which are
individually factorizable into a product of a cross section $\sigma
(s)$ times an impact parameter space distribution function
$W(b\,;\mu)$, i.e.,
\begin{eqnarray}
\chi^{even}(s,b)&=&\chi_{gg}(s,b)+\chi_{qg}(s,b)+\chi_{qq}(s,b)
\nonumber\\ 
&=&i\left[\sigma_{gg}(s)W(b\,;\mu_{gg}) 
+ \sigma_{qg}(s)W(b\,;\sqrt{\mu_{qq}\mu_{gg}})
+ \sigma_{qq}(s)W(b\,;\mu_{qq})\right]\; ,
\label{chieven2}
\end{eqnarray}
where the impact parameter space distribution functions
$
W(b\,;\mu)=\frac{\mu^2}{96\pi}(\mu b)^3K_3(\mu b)
$
are normalized so that $\int W(b\,;\mu)d^2 \vec{b}=1.$
 
In turn, $\chi^{\gamma \gamma}$ for $\gamma\gamma$ reactions is found  by substituting $\sigma\rightarrow \frac{2}{3}\sigma$, $\mu\rightarrow\sqrt{\frac{3}{2}}\mu$ into $\chi^{\gamma p}(s,b)$. 
Making these quark model substitutions in first into $\chi^{even}(s,b)$ and then into $\chi^{\gamma p}(s,b)$, it was found that 
\ba
\chi^{\gamma p}(s,b)&=& i\left[\frac{2}{3}\sigma_{\rm gg}(s)W\left(b\,;\sqrt{\frac{3}{2}}\mu_{\rm gg}\right)+\frac{2}{3}\sigma_{\rm qg}(s)W\left(b\,;\sqrt{\frac{3}{2}}\sqrt{\mu_{\rm qq}\mu_{\rm gg}}\right)\right.\nonumber\\
&& \qquad\qquad\left.+\frac{2}{3}\sigma_{\rm qq}(s)W\left(b\,;\sqrt{\frac{3}{2}}\mu_{\rm qq}\right)\right], \label{eq:chigp}
\ea
and
\ba
\chi^{\gamma \gamma}(s,b)&=& i\left[\frac{4}{9}\sigma_{\rm gg}(s)W\left(b\,;\frac{3}{2}\mu_{\rm gg}\right)+\frac{4}{9}\sigma_{\rm qg}(s)W\left(b\,;\frac{3}{2}\sqrt{\mu_{\rm qq}\mu_{\rm gg}}\right)\right.\nonumber\\
&&\left.\qquad\qquad+\frac{4}{9}\sigma_{\rm qq}(s)W\left(b\,;\frac{3}{2}\mu_{\rm qq}\right)\right]. \label{eq:chigg}
\ea
%%%%%%%%%%%%%%%%%%%%%

 Let us require that the ratio of elastic to total scattering be process-independent,  i.e.,
\be
\left(\frac{\sigma_{\rm el}}{\sigma_{\rm tot}}\right)^{\rm nn}=
\left(\frac{\sigma_{\rm el}}{\sigma_{\rm tot}}\right)^{\gamma p}=
\left(\frac{\sigma_{\rm el}}{\sigma_{\rm tot}}\right)^{\rm \gamma\gamma}\label{eq:ratio}
\ee
at {\em all} energies, a condition that insures that each process becomes equally black disk-like as we go to high energy.  
For simplicity, we will evaluate $\left(\sigma_{\rm el}/\sigma_{\rm tot}\right)^{\rm nn}$ in the small eikonal limit, utilizing \eq{sigelofb2} and \eq{sigtotofb2}, using for our eikonal the toy version $\chi^{\rm nn}(s,b) =i(\sigma_{gg}W(b;\mu_{gg}))$.  Thus, 
\be
\left(\frac{\sigma_{\rm el}}{\sigma_{\rm tot}}\right)^{\rm nn}= \sigma_{gg}\mu_{gg}^2\times(\frac{1}{96\pi})^2\int y^6(K_3(y))^2d^2\vec y,\quad {\rm where\ } y=\mu_{gg} b.
\ee
Therefore, for the ratio to be process-independent, 
\begin{equation}
(\mu_{gg})^{\gamma \gamma}=\sqrt{\frac{3}{2}}(\mu_{gg})^{\gamma p}=\frac{3}{2}(\mu_{gg})^{\rm nn},\quad 
{\rm since}\label{eq:mus}\end{equation}
\be
(\sigma_{gg})^{\gamma \gamma}=\frac{2}{3}(\sigma_{gg})^{\gamma p}=\frac{4}{9}(\sigma_{gg})^{\rm nn}\label{eq:sigmas}.
\ee
This argument is readily generalized to {\em all} $\mu$, leading to {\em each} $\sigma \mu^2$ being {\em process-independent}. 

Indeed, the consequences of \eq{eq:ratio} that each $\sigma \mu^2$ is  process-independent can be restated more simply in the following language:
\begin{itemize}
\item Require that the eikonal of \eq{chieven2} have the {\em same opacity} for n-n, $\gamma p$ and $\gamma \gamma$ scattering,\end{itemize} 
where the opacity is the value of the eikonal at $b=0$. 

For specificity, however, we will use the eikonal of \eq{chieven2}, with the conditions of \eq{eq:mus} and \eq{eq:sigmas}, hereafter.

Thus, 
\ba
\chi^{\gamma p}(s,b)&=& i\left[\frac{2}{3}\sigma_{\rm gg}(s)W(b\,;\sqrt{\frac{3}{2}}\mu_{\rm gg})+\frac{2}{3}\sigma_{\rm qg}(s)W(b\,;\sqrt{\frac{3}{2}}\mu_{\rm qg})\right.\nonumber\\
&&\qquad\qquad\left.+\frac{2}{3}\sigma_{\rm qq}(s)W(b\,;\sqrt{\frac{3}{2}}\mu_{\rm qq})\right], \label{eq:chigp2}
\ea
and
\be
\chi^{\gamma \gamma}(s,b)= i\left[\frac{4}{9}\sigma_{\rm gg}(s)W(b\,;\frac{3}{2}\mu_{\rm gg})+\frac{4}{9}\sigma_{\rm qg}(s)W(b\,;\frac{3}{2}\mu_{\rm qg})+\frac{4}{9}\sigma_{\rm qq}(s)W(b\,;\frac{3}{2}\mu_{\rm qq})\right]. \label{eq:chigg2}
\ee
Since the normalization of each $W(b;\mu)$ above is proportional to $\mu^2$, it is easy to see, using the new dimensionless variable $x_q=\sqrt{\frac{3}{2}}\mu_{\rm qq}b$
 that
\begin{eqnarray}
\chi^{\gamma p}(s,b)&=&\frac{i}{96\pi}\left[\sigma_{\rm gg}\mu_{\rm gg}^2\left(\frac{\mu_{\rm gg}}{\mu_{\rm qq}}x_q\right)^3K_3(\frac{\mu_{\rm gg}}{\mu_{\rm qq}}x_q)+\sigma_{\rm qg}\mu_{\rm qg}^2\left(\frac{\mu_{\rm qg}}{\mu_{\rm qq}}x_q\right)^3K_3(\frac{\mu_{\rm qg}}{\mu_{\rm qq}}x_q)\right.\nonumber\\
&&\ \ \quad \quad \quad +\left. \vphantom{\frac{\sqrt{ \mu_{\rm gg}\mu_{\rm qq}}}{\mu_{\rm qq}}}\sigma_{\rm qq}\mu_{\rm qq}^2(x_q)^3K_3(x_q)\right].\label{eq:newchigp}
\end{eqnarray}
Thus,
\begin{eqnarray}
\!\!\!\!\!\!\!\!\!<|S^{\gamma p}|^2>&=&\frac{1}{96\pi}\int x_q^3K_3(x_q)\times \nonumber\\
&&{\rm exp}-\frac{1}{48\pi}\left[\sigma_{\rm gg}\mu_{\rm gg}^2\left(\frac{\mu_{\rm gg}}{\mu_{\rm qq}}x_q\right)^3K_3(\frac{\mu_{\rm gg}}{\mu_{\rm qq}}x_q)+\sigma_{\rm qg} \mu_{\rm qg}^2\left(\frac{\mu_{\rm qg}}{\mu_{\rm qq}}x_q\right)^3K_3(\frac{\mu_{\rm qg}}{\mu_{\rm qq}}x_q)\right.\nonumber\\
&&\ \ \quad \quad \quad +\left. \vphantom{\frac{\sqrt{ \mu_{\rm gg}\mu_{\rm qq}}}{\mu_{\rm qq}}}\sigma_{\rm qq}\mu_{\rm qq}^2(x_q)^3K_3(x_q)\right]\,d^2\vec{x_q}
\label{eq:Sgp}
\end{eqnarray} 
and 
\begin{eqnarray}
\!\!\!\!\!\!\!\!\!<|S^{\gamma \gamma}|^2>&=&\frac{1}{96\pi}\int x_g^3K_3(x_g)\times \nonumber\\
&&{\rm exp}-\frac{1}{48\pi}\left[\sigma_{\rm gg}\mu_{\rm gg}^2\left(\frac{\mu_{\rm gg}}{\mu_{\rm qq}}x_g\right)^3K_3(\frac{\mu_{\rm gg}}{\mu_{\rm qq}}x_g)+\sigma_{\rm qg} \mu_{\rm qg}^2\left(\frac{\mu_{\rm qg}}{\mu_{\rm qq}}x_g\right)^3K_3(\frac{\mu_{\rm qg}}{\mu_{\rm qq}}x_g)\right.\nonumber\\
&&\ \ \quad \quad \quad +\left. \vphantom{\frac{\sqrt{ \mu_{\rm gg}\mu_{\rm qq}}}{\mu_{\rm qq}}}\sigma_{\rm qq}\mu_{\rm qq}^2(x_g)^3K_3(x_g)\right]\,d^2\vec{x_g}
\label{eq:Sgg}
\end{eqnarray} 
where we used the variable substitution $x_g=\frac{3}{2}\mu_{\rm qq}b$.
Finally, we have, using the variable substitution $x_n=\mu_{\rm qq}b$,
\begin{eqnarray}
\!\!\!\!\!\!\!\!\!<|S^{\rm even}|^2>&=&\frac{1}{96\pi}\int x_n^3K_3(x_n)\times \nonumber\\
&&{\rm exp}-\frac{1}{48\pi}\left[\sigma_{\rm gg}\mu_{\rm gg}^2\left(\frac{\mu_{\rm gg}}{\mu_{\rm qq}}x_n\right)^3K_3(\frac{\mu_{\rm gg}}{\mu_{\rm qq}}x_n)+\sigma_{\rm qg}\mu_{\rm qg}^2\left(\frac{\mu_{\rm qg}}{\mu_{\rm qq}}x_n\right)^3K_3(\frac{\mu_{\rm qg}}{\mu_{\rm qq}}x_n)\right.\nonumber\\
&&\ \ \quad \quad \quad +\left. \vphantom{\frac{\sqrt{ \mu_{\rm gg}\mu_{\rm qq}}}{\mu_{\rm qq}}}\sigma_{\rm qq}\mu_{\rm qq}^2(x_n)^3K_3(x_n)\right]\,d^2\vec{x_n}.
\label{eq:Seven}
\end{eqnarray} 
Thus, comparing \eq{eq:Sgp}, \eq{eq:Sgg} and \eq{eq:Seven}, we find 
\be 
<|S^{\gamma p}|^2>=<|S^{\gamma \gamma}|^2>=<|S^{\rm even}|^2>.\label{eq:allequal}
\ee
We see from \eq{eq:allequal} that $<|S|^2>$, the survival probability for nucleon-nucleon, $\gamma p$ and $\gamma \gamma$ collisions, is {\em reaction-independent}, depending {\em only} on $\sqrt{s}$, the c.m. energy of the collision.  It should be emphasized that this result is much more general, being true for {\em any} eikonal whose opacity is process-independent---not only for the additive quark model that we have employed.  
%

%%%%%%%%%%
%%%%%%%%%%%%%%% Delete what follow!!!
 The energy dependence of the large rapidity gap survival probability $<|S|^2>$ calculated from \eq{eq:Seven} by Block and Halzen\cite{gapsurvival} is given in Fig. \ref{fig:survival}.

The survival probability $<|S|^2>$ calculated in ref. \cite{gapsurvival} used an eikonal that had been found by fitting accelerator and cosmic ray data over an enormous energy range.  These numerical results are considerably larger than other  calculations\cite{Maor,Fletcher,Bjorken}.  In the case of ref. \cite{Maor} and ref. \cite{Bjorken}, it is probably due to their using a Gaussian probability distribution in impact parameter space, whereas the distribution used here,
$W(b,\mu_{qq})=\frac{\mu_{qq}^2}{96\pi}(\mu_{qq} b)^3\,K_3(\mu_{qq} b)$ which is the Fourier transform of the square of a dipole distribution, has a long exponential tail $e^{-\mu b}$, significantly increasing the probability of survival.  In the case of Ref. \cite{Fletcher}, the authors determine the parameters for their minijet model using only the Tevatron results. The large values of Table \ref{table:survival} are more in line with the earlier predictions of Gotsman et al.\cite{Gotsman} for what they called Regge and Lipatov1 and Lipatov2 models, although with somewhat different energy dependences than that shown in Table \ref{table:survival}. The color evaporation model of \`Eboli  et al.\cite{Eboli} gives somewhat larger values, but again with a different energy dependence. Most recently, Khoze  et al.\cite{Martin}, using a two-channel eikonal, have calculated the survival probabilities for rapidity gaps in single, central and double diffractive processes at several energies, as a function of  the slope of the Pomeron-proton vertex, which they called $b$. For double diffraction, they have a large range of possible parameters. Choosing  $2b=5.5$ GeV$^{-2}$ (corresponding to the slope of the electromagnetic proton form factor),  they obtain $<|S|^2>=0.26$, 0.21 and 0.15 at $\sqrt {s}=0.54$, 1.8 and 14 TeV, respectively. These survival probabilities are in excellent agreement with the values given in Table  \ref{table:survival}.  However, their  calculations for other choices of $2b$ and for single and central diffractive processes do not agree with Table  \ref{table:survival}, being extremely model-dependent, with their results varying considerably with their choice of parameters and model.    

We see that there is a serious model dependence, both in the size of the survival probabilities and in their energy dependence. Further, until now, there has been no estimates for gap survival probabilities for $\gamma p$ and $\gamma\gamma$ reactions. It is hoped that the  quasi model-independent fit to experimental data on $\bar p p$ and $pp$ total cross sections, $\rho$ values and nuclear slopes $B$, over an enormous energy range, $\sqrt s$ = 15 GeV  to 30,000 GeV, of Ref. \cite{gapsurvival} provides a reliable quantitative estimate of the survival probability $<|S|^2>$ as a function of energy, for both $\pbar p$, $pp$, $\gamma p$ and $\gamma \gamma$ collisions.  The fact that these estimates of large rapidity gap survival probabilities are independent of reaction, thus being equal for nucleon-nucleon, $\gamma p$ and $\gamma \gamma$ processes---the equality surviving any particular factorization scheme---has many possible interesting experimental consequences.   
%%%%%%%%%%%%%%%%%%55
%%%%%%%%%%%%%%%%%%%%%%%%%%%%%%%%Table #5

%****************
\begin{table}[thp]
\def\arraystretch{1.2}   
\begin{center}
\caption{\protect\small Values of the parameters used in the constrained Aspen model fit. 
}\label{table:qcdparam}
\begin{tabular}[b]{ll}
\hline\hline
Fixed&Fitted\\
\hline   
$m_0=0.6$ GeV & $C=5.36\pm0.13$ GeV\\ 
$\epsilon=0.05$&$C^{log}_{qg}=0.166\pm0.030$ GeV \\
$\mu_{qq}=0.89$  GeV & $C'_{gg}=0.00103\pm0.00006$ GeV\\
$\mu_{gg}=0.73$ GeV &$C^{even}_{Regge}=29.7\pm0.91$ GeV \\
$\mu_{odd}=0.53$ GeV &$C_{odd}=10.3\pm0.043$ GeV\\
$\alpha_s=0.5$&$s_0=9.53\pm0.63$ GeV$^2$ \\
\hline\hline
\end{tabular}
\end{center}
\end{table}
\def\arraystretch{1}
%****************
%%%%%%%%%%%Table # 6
\begin{table}[tbp]                   % Use "table" environment, but also
				 % use  "tabular" environment below.
%
\def\arraystretch{1.2}            % Make the space between rows in the Table,
				  % 1.5 x bigger than the default spacing.
\begin{center}
     \caption{\protect\small  Predictions of high energy $\bar pp$ and $pp$ total  cross sections and $\rho$-values  from Table \ref{table:qcdparam}, using the  constrained Aspen model.\label{table:QCDpredictions}
}

\begin{tabular}[b]{|l||c|c||c|c||c|c|}
    \cline{1-7}
      \multicolumn{1}{|l||}{ $\sqrt s$}
      &\multicolumn{1}{c|}{$\sigma_{\bar pp}$}
      &\multicolumn{1}{c||}{$\rho_{\bar p p}$}&\multicolumn{1}{c|}{$\sigma_{ pp}$ }&\multicolumn{1}{c||}{$\rho_{pp}$}&\multicolumn{1}{c|}{$B_{\bar pp}$}&\multicolumn{1}{c|}{$B_{pp}$}\\
(TeV)&(mb)&& (mb)&&(GeV/c)$^{-2}$&(GeV/c)$^{-2}$\\
\hline\hline
	0.540&$62.40\pm.38$&$.136\pm.002$&$62.27\pm.38$&$.135\pm.002$&$15.36\pm.05$ &$15.33\pm.05$\\
\hline
1.800&$76.76\pm.67$&$76.72\pm.67$&$.133\pm.002$&$.132pm.002$&$16.76\pm.07$&$16.76\pm.08$\\
\hline
14.00&$106.2\pm1.27$&$106.2\pm1.27$&$.114\pm.001$&$.114\pm.001$&$19.39\pm.13$&$19.39\pm.13$\\
\hline
40.00&$122.8\pm1.6$&$122.8\pm1.6$&$.103\pm.001$&$.103\pm.001$&$20.84\pm.16$&$20.84\pm.16$\\
\hline
100.0&$137.9\pm1.8$&$137.9\pm1.8$&$.095\pm.001$&$.095\pm.001$&$23.19\pm.19$&$23.19\pm.19$\\
\hline
\end{tabular}
\end{center}
     %\vspace{1in} \\
%     \\
\end{table}

\def\arraystretch{1}  %Restore the default row spacing in the Table.

%%%%%%%%%%%%%%%%%%%%%%%%%%%%%%%%%%%%%%%%%%%%%%%%%%%
% table # 7
\begin{table}[ht]                   % Use "table" environment, but also
                                 % use  "tabular" environment below.
%
 \caption[The survival probability, $<|S|^2>$, in \%, for $\bar pp$ and $pp$ collisions, as a function of c.m. energy]{\footnotesize
The survival probability, $<|S|^2>$, in \%, for $\bar pp$ and $pp$ collisions, as a function of c.m. energy. Taken from Ref. \cite{gapsurvival}.}\label{table:survival} 

\vspace{.15in} 
\begin{center}
\begin{tabular}[b]{|l||c|c|}
      \hline
      C.M. Energy (GeV)&Survival Probability($\bar pp$), in \% &Survival Probability, in \%\\ \hline
     63&$37.0\pm 0.9$&$37.5\pm 0.9$\\
     546&$26.7\pm 0.5$&$26.8\pm 0.5$\\
     630&$26.0\pm 0.5$&$26.0\pm 0.5$\\
     1800&$20.8\pm 0.3$&$20.8\pm 0.3$\\
     14000&$\ \  12.6\pm 0.06$&\ \ $12.6\pm 0.06$\\
     40000&$\ \ \ 9.7\pm 0.07$&$\ \ \ 9.7\pm 0.07$\\
     \hline
\end{tabular}
     \vspace{-.2in} \\
\end{center}
\end{table}
%

%%%%%%%%%%%%%%%%%%%%%%%%%%%%%%%%%%%%%%%%%%%%%
\begin{figure}[tbp] %Fig. 10
\begin{center}
\mbox{\epsfig{file=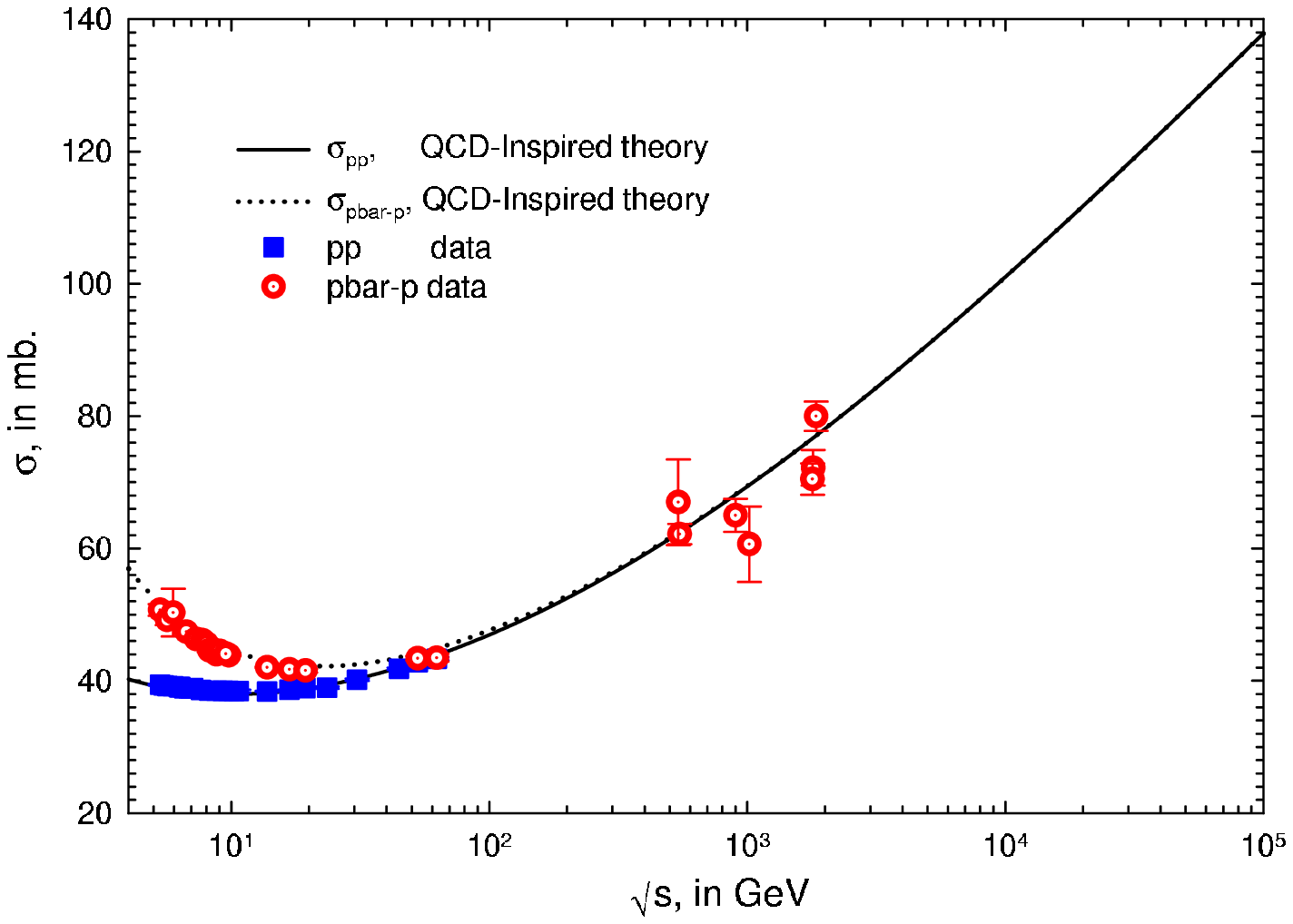,%
		width=4in,bbllx=75pt,bblly=250pt,bburx=495pt,bbury=550pt,clip=%
}}
\end{center}
\caption[ $\sigtot$, for $pp$ and $\bar p p$, using a constrained Aspen model fit] 
{ \footnotesize
The total cross section $\sigtot$, in 
mb,  vs.  the c.m. energy $\sqrt s$, in GeV, for $pp$ and $\bar p p$  scattering, using a constrained Aspen model fit (QCD-inspired theory). 
The solid line and squares are for $pp$ and the dotted line and open 
circles are for $\bar p p$.}
\label{fig:sigqcd}
\end{figure}
%%%%%%%%%%%%%%%%%%%%%%%%%%%%%%%%%%%%%%%%%%%%%%%%%%%

%%%%%%%%%%%%%%%%%%%%%%%%%%%%%%%%%%%%%%%%%%%%%
\begin{figure}[tbp] %Fig.11
\begin{center}
\mbox{\epsfig{file=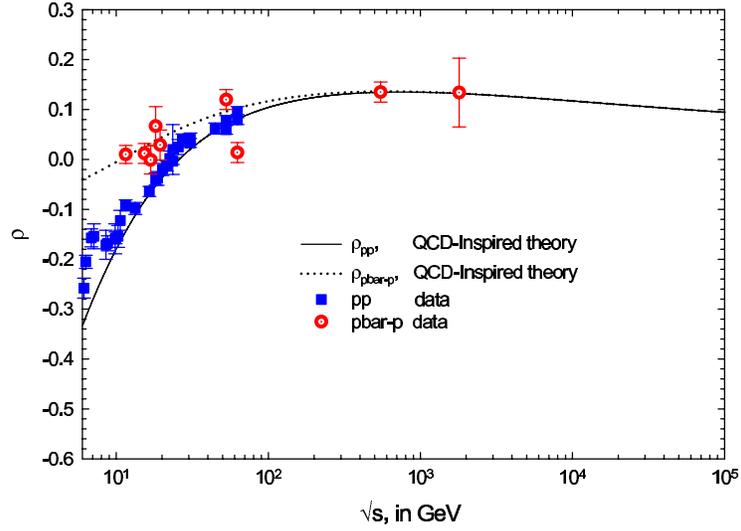,%
		width=4in,bbllx=80pt,bblly=255pt,bburx=510pt,bbury=560pt,clip=%
}}
\end{center}
\caption[$\rho$, for $pp$ and $\bar p p$, using a constrained Aspen model fit]
{ \footnotesize
The ratio of the real to imaginary part of the 
forward scattering amplitude, $\rho$
 vs.  the c.m. energy $\sqrt s$, in GeV, for $pp$ and $\bar p p$  scattering, using a constrained Aspen model fit (QCD-inspired theory).   
The solid line and squares are for $pp$ and the dotted line and open 
circles are for $\bar p p$.}
\label{fig:rhoqcd}
\end{figure}
%%%%%%%%%%%%%%%%%%%%%%%%%%%%%%%%%%%%%%%%%%%%%%%%%%%
%%%%%%%%%%%%%%%%%%%%%%%%%%%%%%%%%%%%%%%%%%%%%
\begin{figure}[tbp] %Fig.12
\begin{center}
\mbox{\epsfig{file=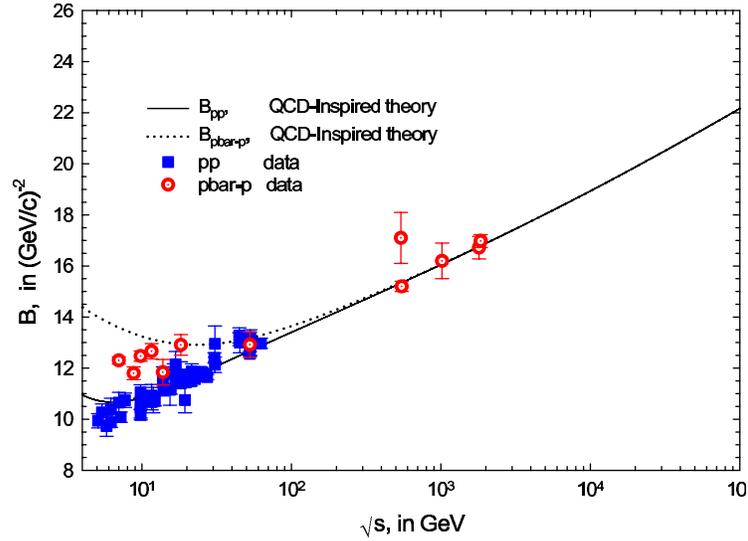,%
		width=4in,bbllx=75pt,bblly=250pt,bburx=495pt,bbury=550pt,clip=%
}}
\end{center}
\caption[The nuclear slope parameter $B$ for elastic $pp$ and $\bar p p$ scattering, using a constrained Aspen model fit]
{ \footnotesize
The nuclear slope parameter $B$, in (GeV/c)$^{-2}$ 
vs. $\sqrt s$, in GeV, for elastic $pp$ and $\bar p p$ scattering,  using a constrained Aspen model fit (QCD-inspired theory). 
The solid line and squares are for $pp$ and the dotted line and open 
circles are for $ \bar p p$.}
\label{fig:Bqcd}
\end{figure}
%%%%%%%%%%%%%%%%%%%%%%%%%%%%%%%%%%%%%%%%%%%%%%%%%%%
%%%%%%%%%%%%%%%%%%%%%%%%%%%%%%%%%%%%%%%%%%%%%
\begin{figure}[tbp] %Fig.13
\begin{center}
\mbox{\epsfig{file=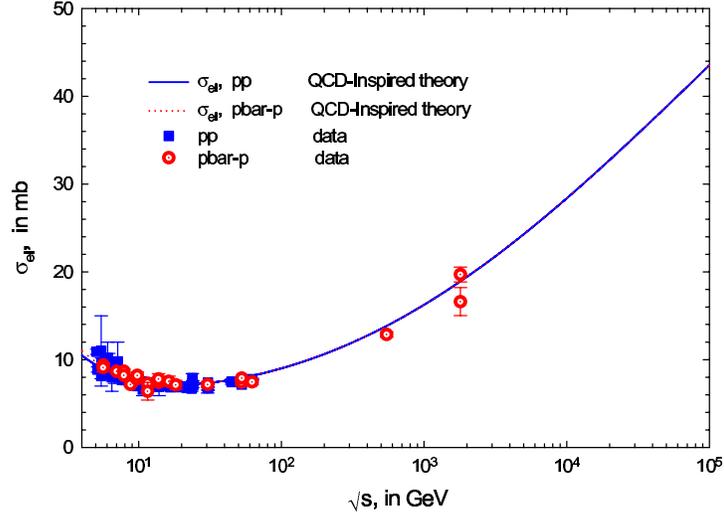,%
		width=4in,bbllx=60pt,bblly=255pt,bburx=500pt,bbury=560pt,clip=%
}}
\end{center}
\caption[$\sigma_{\rm el}$, for $pp$ and $p\bar p$ scattering, using a constrained Aspen model fit ]
{ \footnotesize
Elastic scattering cross sections, 
$\sigma_{\rm el}$, in mb vs. $\sqrt s$, in GeV, 
for $pp$ and $p\bar p$ scattering, using a constrained Aspen model fit (QCD-inspired theory). The solid line and squares 
are for $pp$  and the dotted 
line and open circles are for $\bar p p$.}
\label{fig:sigel}
\end{figure}
%%%%%%%%%%%%%%%%%%%%%%%%%%%%%%%%%%%%%%%%%%%%%%%%%%%

%%%%%%%%%%%%%%%%%%%%%%%%%%%%%%%%%%%%%%%%%%%%%
\begin{figure}[tbp] %Fig.14
\begin{center}
\mbox{\epsfig{file=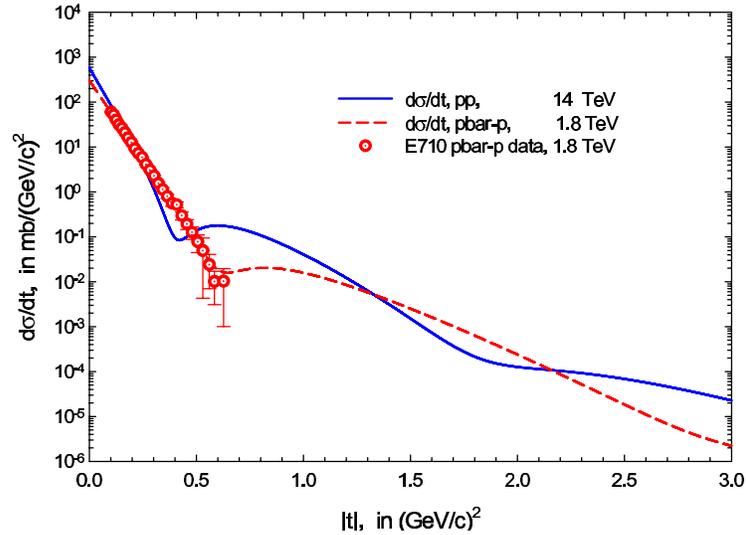,%
		width=4in,bbllx=70pt,bblly=245pt,bburx=500pt,bbury=545pt,clip=%
}}
\end{center}
\caption[$d\sigma/dt$ at the Tevatron and the LHC, using a constrained Aspen model fit]
{ \footnotesize
The elastic differential scattering cross section 
$d\sigma_{\rm el}/dt$, in mb/(GeV/c)$^2$  vs. $|t|$, in (GeV/c)$^2$, using a constrained Aspen model fit (QCD-inspired theory). 
 The solid curve is the prediction for the reaction $ pp\rightarrow pp$ at the LHC, at $\sqrt s=14$ TeV.
The dashed curve is the prediction for the reaction $\bar pp\rightarrow\bar pp$ at $\sqrt{s}=1.8$ TeV, at the Tevatron Collider; the data 
points are from the E710 experiment. }
\label{fig:dsdt}
\end{figure}
%%%%%%%%%%%%%%%%%%%%%%%%%%%%%%%%%%%%%%%%%%%%%%%%%%%
%%%%%%%%%%%%%%%%%%%%%%%%%%%%%%%%%%%%%%%%%%%%%
\begin{figure}[tb] %Fig.15
\begin{center}
\mbox{\epsfig{file=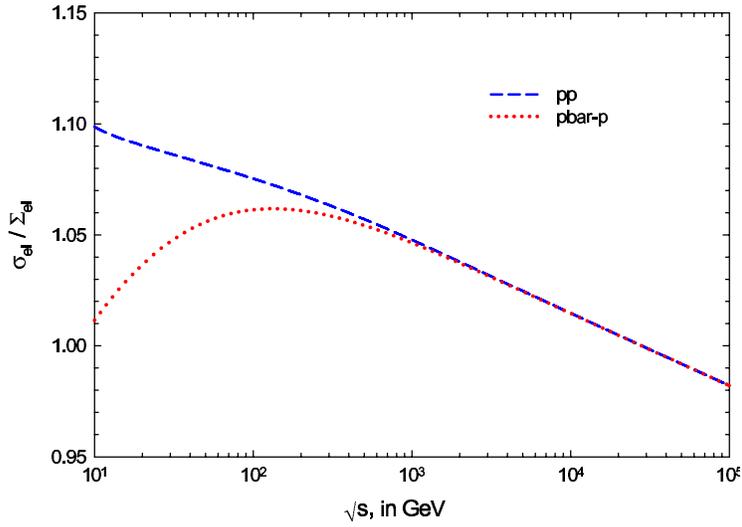,%
		width=4in,bbllx=65pt,bblly=265pt,bburx=500pt,bbury=560pt,clip=%
}}
\end{center}
\caption[$\sigma_{\rm el}$/$\Sigma_{\rm el}$, using a constrained Aspen model fit]
{\protect{\footnotesize
The ratio $\sigma_{\rm el}/\Sigma_{\rm el}$ vs. the c.m. energy $\sqrt s$, in GeV, using a constrained Aspen model fit (QCD-inspired theory).  The elastic cross section is $\sigma_{\rm el}$ and  $\Sigma_{\rm el}\equiv \sigtot^2/16\pi B$. The dashed curve is for $pp$ and the dotted curve is for $\bar pp$.\label{fig:sigoverSIG}}}
\end{figure}
%%%%%%%%%%%%%%%%%%%%%%%%%%%%%%%%%%%%%%%%%%%%%%%%%%%
\begin{figure} %Fig. 16
\begin{center}
\mbox{\epsfig{file=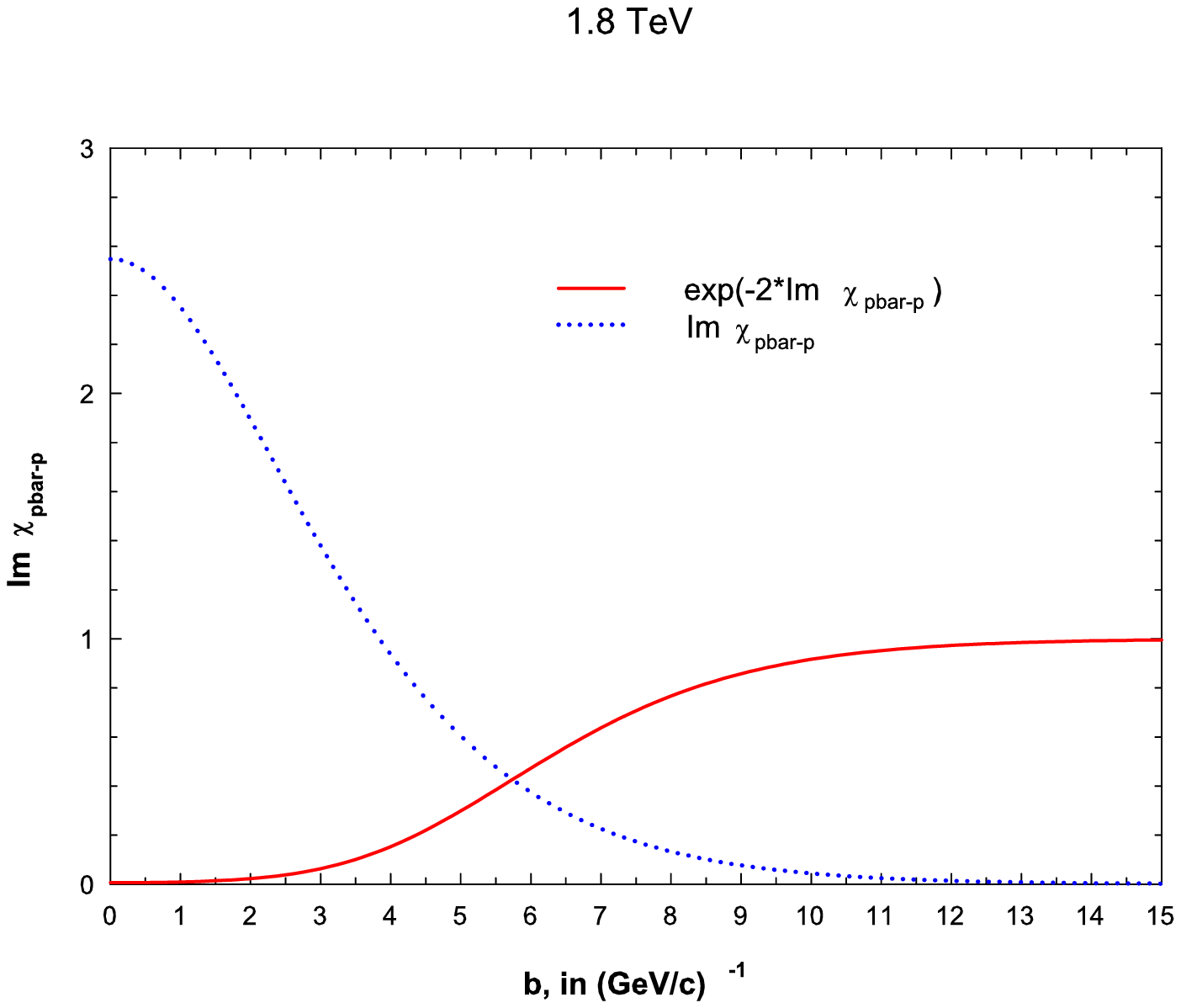%
            ,width=4in,bbllx=0pt,bblly=0pt,bburx=420pt,bbury=310pt,clip=%
}}
\end{center}
\caption[The eikonal ${\rm Im} \chi$ and the exponential damping factor $e^{-2\,{\rm Im} \chi}$ for $\pbar p$ collisions]
{ \footnotesize
The eikonal ${\rm Im}\, \chi$ and the exponential damping factor $e^{-2\,{\rm Im} \chi}$ for $\pbar p$ collisions, at $\sqrt s=1.8$ TeV  vs. the impact parameter $b$, in (GeV/c)$^{-1}$. Taken from Ref. \cite{gapsurvival}.}
\label{fig:1800}
\end{figure}
\begin{figure} % Fig. 17
\begin{center}
\mbox{\epsfig{file=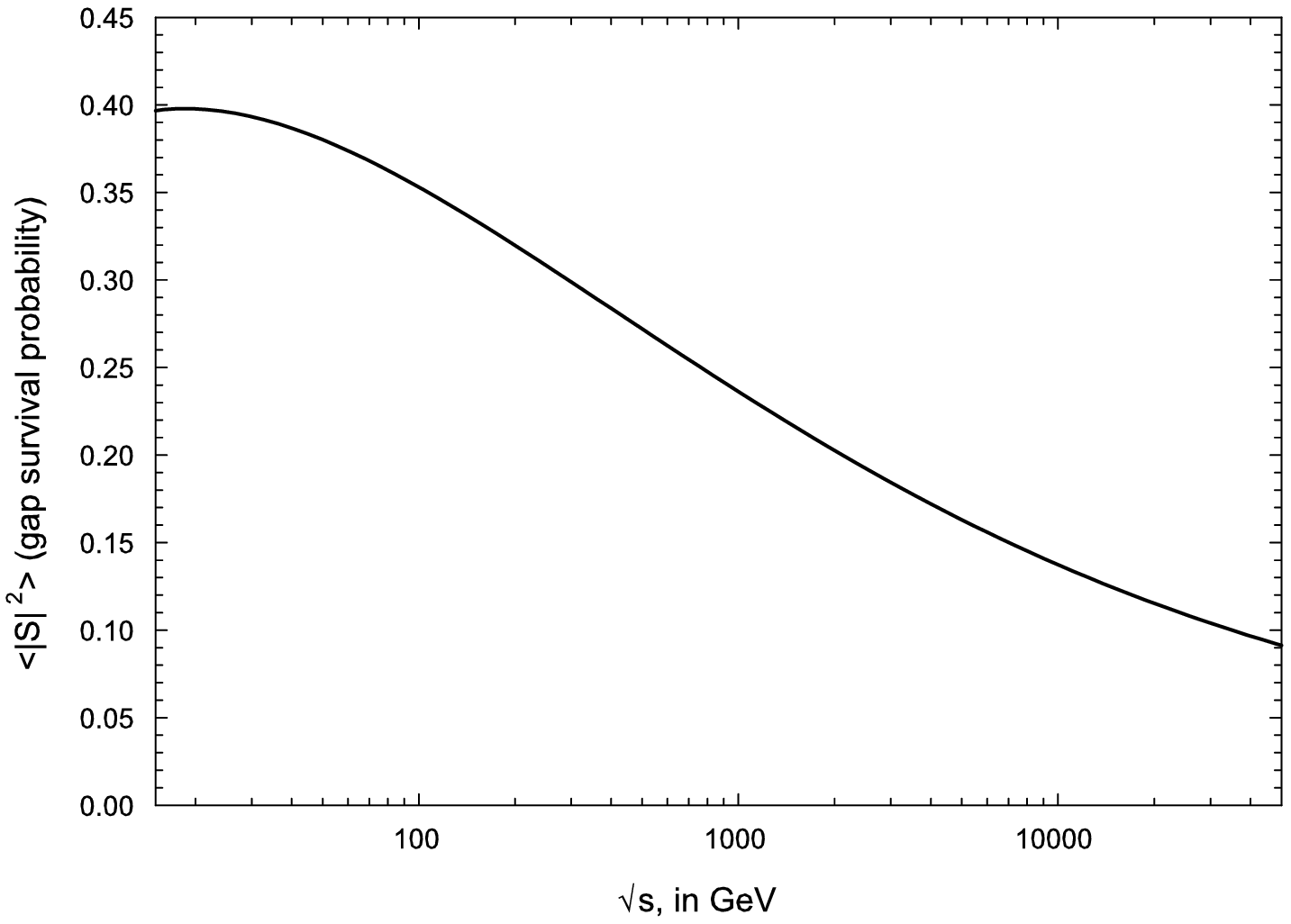%
%            ,width=4in,bbllx=53pt,bblly=250pt,bburx=525pt,bbury=580pt,clip=%
           ,width=4in,bbllx=0pt,bblly=0pt,bburx=420pt,bbury=330pt,clip=%
}}
\end{center}
\caption[The energy dependence of $<|S|^2>$, the large rapidity gap survival probability]
{ \footnotesize
 The energy dependence of $<|S|^2>$, the large rapidity gap survival probability   vs. $\sqrt s$, in GeV. Taken from Ref. \cite{gapsurvival}.}
\label{fig:survival}
\end{figure}
%
%%%%%%%%%%%%%%%%%%%%%%%%%%%%%%%%
\subsubsection{Factorization properties of the eikonal}
Using the additive quark model and meson vector dominance  as an example, it was shown by Block and Kaidalov\cite{me}that for all energies and values of the Aspen eikonal, the factorization theorem $\sigma_{\rm nn}/\sigma_{\gamma p}=\sigma_{\gamma p}/\sigma_{\gamma \gamma}$ holds. In order to calculate the total nucleon-nucleon cross section, they used the variable substitution $x_n=\mu_{\rm qq}b$ and rewrote $\chi^{\rm even}$ of \eq{chieven2} as  
\begin{eqnarray}
\chi^{\rm even}(s,b)&=&\frac{i}{96\pi}\left[\sigma_{\rm gg}\mu_{\rm gg}^2(\frac{\mu_{\rm gg}}{\mu_{\rm qq}}x_n)^3K_3(\frac{\mu_{\rm gg}}{\mu_{\rm qq}}x_n)+\sigma_{\rm qg}\mu_{\rm qg}^2\left(\frac{\mu_{\rm qg}}{\mu_{\rm qq}}x_n\right)^3K_3(\frac{\mu_{\rm qg}}{\mu_{\rm qq}}x_n)\right.\nonumber\\
&&\ \ \quad \quad \quad +\left. \vphantom{\frac{\sqrt{ \mu_{\rm gg}\mu_{\rm qq}}}{\mu_{\rm qq}}}\sigma_{\rm qq}\mu_{\rm qq}^2(x_n)^3K_3(x_n)\right].\label{eq:newchinn}
\end{eqnarray}
Using \eq{sigtot} and approximating $\chi_{\rm even}$ in \eq{eq:newchinn} as pure imaginary, we have 
\begin{eqnarray}
\sigma_{\rm tot}^{\rm nn}(s)&=&2\int \left(1-{\rm exp}-\frac{1}{96\pi}\left[\sigma_{\rm gg}\mu_{\rm gg}^2(\frac{\mu_{\rm gg}}{\mu_{\rm qq}}x_n)^3K_3(\frac{\mu_{\rm gg}}{\mu_{\rm qq}}x_n)\right.\right.\nonumber\\
&&\qquad\left.\left.+\sigma_{\rm qg}\mu_{\rm qg}^2\left(\frac{\mu_{\rm qg}}{\mu_{\rm qq}}x_n\right)^3K_3(\frac{\mu_{\rm qg}}{\mu_{\rm qq}}x_n)\right.\right.\nonumber\\
&&\ \ \quad \quad \quad +\left. \left.\vphantom{\frac{\sqrt{ \mu_{\rm gg}\mu_{\rm qq}}}{\mu_{\rm qq}}}\sigma_{\rm qq}\mu_{\rm qq}^2(x_n)^3K_3(x_n)\right]\right)\frac{1}{\mu_{\rm qq}^2}\,d^2\vec{x_n}.\label{eq:signn}
\end{eqnarray}
Following the arguments of ref. \cite{me}, after
using vector meson dominance and the additive quark model and letting $P_{\rm had}^\gamma$ be the probability that a $\gamma$ ray materialize as a hadron, we find, using $\chi^{\gamma p}$ from \eq{eq:newchigp}, that
\begin{eqnarray}
\sigma_{\rm tot}^{\gamma p}(s)&=&2\int \left(1-{\rm exp}-\frac{1}{96\pi}\left[\sigma_{\rm gg}\mu_{\rm gg}^2(\frac{\mu_{\rm gg}}{\mu_{\rm qq}}x_q)^3K_3(\frac{\mu_{\rm gg}}{\mu_{\rm qq}}x_q)\right.\right.\nonumber\\
&&\qquad\left.\left.+\sigma_{\rm qg}\mu_{\rm qg}^2\left(\frac{\mu_{\rm qg}}{\mu_{\rm qq}}x_q\right)^3K_3(\frac{\mu_{\rm qg}}{\mu_{\rm qq}}x_q)\right.\right.\nonumber\\
&&\ \ \quad \quad \quad +\left. \left.\vphantom{\frac{\sqrt{ \mu_{\rm gg}\mu_{\rm qq}}}{\mu_{\rm qq}}}\sigma_{\rm qq}\mu_{\rm qq}^2(x_q)^3K_3(x_q)\right]\right)\frac{2}{3}P_{\rm had}^\gamma \frac{1}{\mu_{\rm qq}^2}\,d^2\vec{x_q},\label{eq:siggp}
\end{eqnarray}
where $x_q=\sqrt{\frac{3}{2}}\mu_{\rm qq}b$.  Finally, substituting $x_g=\frac{3}{2}\mu_{\rm qq}b$ into \eq{eq:chigp2} and using \eq{sigtotofb2}, we evaluate $\sigma_{\rm tot}^{\gamma \gamma}$ as
\begin{eqnarray}
\sigma_{\rm tot}^{\gamma \gamma}(s)&=&2\int \left(1-{\rm exp}-\frac{1}{96\pi}\left[\sigma_{\rm gg}\mu_{\rm gg}^2(\frac{\mu_{\rm gg}}{\mu_{\rm qq}}x_g)^3K_3(\frac{\mu_{\rm gg}}{\mu_{\rm qq}}x_g)\right.\right.\nonumber\\
&&\left.\left.\qquad+\sigma_{\rm qg}\mu_{\rm qg}^2\left(\frac{\mu_{\rm qg}}{\mu_{\rm qq}}x_g\right)^3K_3(\frac{\mu_{\rm qg}}{\mu_{\rm qq}}x_g)\right.\right.\nonumber\\
&&\ \ \quad \quad \quad +\left. \left.\vphantom{\frac{\sqrt{ \mu_{\rm gg}\mu_{\rm qq}}}{\mu_{\rm qq}}}\sigma_{\rm qq}\mu_{\rm qq}^2(x_g)^3K_3(x_g)\right]\right){\left(\frac{2P_{\rm had}^\gamma }{3}\right)}^2\frac{1}{\mu_{\rm qq}^2}\,d^2\vec{x_g}.\label{eq:siggg}
\end{eqnarray}
 Clearly, from inspection of \eq{eq:signn}, \eq{eq:siggp} and \eq{eq:siggg}, we see that the factorization theorem
\be
\frac{\sigma_{\rm tot}^{\rm nn}(s)}{\sigma_{\rm tot}^{\gamma p}(s)}=\frac{\sigma_{\rm tot}^{\gamma p}(s)}{\sigma_{\rm tot}^{\gamma \gamma}(s)}\label{eq:factorization}
\ee
holds at all energies,  i.e., the factorization theorem survives exponentiation. It should be emphasized  that 
this result is true for {\em any} eikonal which factorizes into sums of $\sigma_i (s)\times W_i(b;\mu)$ having the scaling feature that the product $\sigma_i\mu_i^2$ is reaction-independent---not only for the additive quark model that we have employed here, but for {\em any} eikonal whose opacity is independent of whether the reaction is n-n, $\gamma p$ or $\gamma\gamma$.  It is valid at {\em all} energies, independent of the size of the eikonal and independent of the details of the initial factorization scheme. 

Thus, there are  three high energy factorization theorems:
\begin{equation}
 \frac{\signn(s)}{\siggp(s)}=\frac{\siggp(s)}{\siggg(s)},\label{eq:sig}
\end{equation}
 where the $\sigma$'s are the total cross sections for nucleon-nucleon, $\gamma$p and $\gamma\gamma$ scattering,
\begin{equation}
\frac{\Bnn(s)}{\Bgp(s)}=\frac{\Bgp(s)}{\Bgg(s)},\label{eq:B}
\end{equation}
 where the $B$'s are the nuclear slope parameters for elastic scattering, and 
\begin{equation} \rhonn(s)=\rhogp(s)=\rhogg(s),\label{eq:rho}
\end{equation}
 where the $\rho$'s are the ratio of the real to imaginary portions of the forward scattering amplitudes,  
with the first two  factorization theorems each having their own proportionality constant. By $B_{\rm nn}(s)$, we mean the average nuclear slope, $B_{\rm nn}(s)=(B_{pp}(s)+B_{\bar pp}(s))/2$, whereas  $\Bgp(s)$ is the nuclear slope for  for `elastic' vector meson production,  $\gamma +p\rightarrow V+p$, and  $\Bgg(s)$ is the nuclear slope for `elastic' vector-vector scattering, $\gamma +\gamma\rightarrow V+V$, with $V$ being either $\rho$, $\omega$ or $\phi$.

In the particular scheme of \eq{eq:signn}, \eq{eq:siggp} and \eq{eq:siggg}, i.e., for vector meson dominance and the additive quark model, the proportionality constant is $(2/3)P_{\rm had}^\gamma$.  Clearly, the same proportionality constant holds for both $\sigma_{\rm el}$ and $\sigma_{\rm inel}$, whereas for the nuclear slope parameter $B$ it is easily shown that the proportionality constant is 2/3.  For $\rho$, the proportionality constant is unity. 

%%%%%%%%%%%%%%%%%%%%%%%%%%%%%%
\subsubsection{$d\sigma_{\rm el}/dt$ for vector meson production,  $\gamma +p\rightarrow V+p$ and $\gamma +\gamma\rightarrow V+V$.}

Following the work of Ref. \cite{me}, the elastic differential scattering cross section for nucleon-nucleon scattering can be written as
\be
\frac{d\sigma^{\rm nn}}{dt}(s,t)=\frac{1}{4\pi}\left|\int J_0(qb)
\left(1-e^{i\chi^{\rm even}(b,s)}\right)\,d^2\vec{b}\,\right|^2\; ,
\label{eq:dsdt2}
\ee
where the squared 4-momentum transfer $t=-q^2$. 
It is straightforward, by appropriate variable transformation, to show that the differential cross section for the `elastic' scattering reaction $\gamma +p\rightarrow V+p$ is given by
\be
\frac{d\sigma_{Vp}^{\gamma p}}{dt}(s,t)=\frac{4P_V^\gamma}{9}\frac{d\sigma^{\rm nn}}{dt}\left(s,\frac{2}{3}t\right),\label{eq:dsdtgp}
\ee
where $V$ is the vector meson $\rho$, $\omega$ or $\phi$ and $P_V^\gamma$ is the probability that a photon goes into the vector meson $V$.
For the ``elastic" scattering reaction $\gamma +\gamma\rightarrow V+V$,  we can show that
\be
\frac{d\sigma_{VV}^{\gamma \gamma}}{dt}(s,t)={\left(\frac{4P_V^\gamma}{9}\right)}^2\frac{d\sigma^{\rm nn}}{dt}\left(s,\frac{4}{9}t\right).\label{eq:dsdtgg}
\ee
Thus, a knowledge of $d\sigma^{\rm nn}/dt$ for elastic nucleon-nucleon scattering determines the differential `elastic' scattering cross sections for the reactions $\gamma +p\rightarrow V+p$ and $\gamma +\gamma\rightarrow V+V$.
We now write the factorization theorem for differential elastic scattering as
\be\frac{d\sigma^{\rm nn}}{dt}(s,t){ \  \Bigg / \ }\frac{d\sigma_{Vp}^{\gamma p}}{dt}\left(s,{\frac{3}{2}}t\right)
=\frac{d\sigma_{Vp}^{\gamma p}}{dt}\left(s,{\frac{3}{2}}t\right){ \ \Bigg /\ }\frac{d\sigma_{VV}^{\gamma \gamma}}{dt}\left(s,{\frac{9}{4}}t\right).\label{eq:dsdtfactorize}
\ee
\subsubsection{Experimental evidence for $B$ factorization}\label{sec:Bfactor}
The factorization theorem for the nuclear slopes $B$, \eq{eq:B}, can be rewritten as
\begin{eqnarray}
\Bgg(s)&=\kappa\Bgp(s)&=\kappa^2\Bnn(s)\label{eq:Bfact}.
\end{eqnarray}

For additional evidence involving the equality of the nuclear slopes $B_\rho$, $B_\gamma$ and $B_\phi$ from  differential elastic scattering  data $\frac{d\sigma}{dt}$, see Figures (13,14,15) of ref. \cite{blockhalzenpancheri}.
The additive quark model tells us from quark counting  that $\kappa$ in \eq{eq:Bfact} is given by $\kappa=\frac{2}{3}$.  In this picture, the `elastic scattering' reactions $\gamma +p\rightarrow V +p$, where $V$ is the vector meson $\rho$, $\omega$ or $\phi$, require that  $B_\rho=B_\omega=B_\phi(=\Bgp)$.  To determine the value of $\kappa$ in the relation $\Bgp=\kappa\,\Bnn$, Block, Halzen and Pancheri\cite{blockhalzenpancheri2} made a $\chi^2$ fit to the available $\Bgp$ data.  In Fig. \ref{fig:Bgp} they plotted  $\kappa \Bnn$  vs. the c.m. energy $\sqrt s$, using  the best-fit value of  $\kappa=0.661$, against the experimental values of $\Bgp$, where $B_{\rm nn}$ was obtained from their eikonal model parameters.   The fit gave $\kappa=0.661\pm 0.008$, with a total $\chi^2=16.4$ for 10 degrees of freedom.  Inspection of Fig. \ref{fig:Bgp} shows that the experimental point of $B_\rho$ at $\sqrt s=5.2$ GeV--- which contributes 6.44 to the $\chi^2$---clearly cannot lie on any smooth curve and thus can  safely be ignored.  Neglecting the contribution of this point gives a $\chi^2$/d.f.=0.999, a very satisfactory result.  They emphasized that the experimental $\gamma p$ data thus
\begin{itemize}
\item  require  $\kappa=0.661\pm 0.008$, an $\approx 1\%$ measurement in excellent agreement with the value of 2/3 that is obtained from the additive quark model.
\item clearly verify the nuclear slope factorization theorem of \eq{eq:Bfact} over the available energy range spanned by the data.
\end{itemize}
\begin{figure}[ht] %#18
\begin{center}
\mbox{\epsfig{file=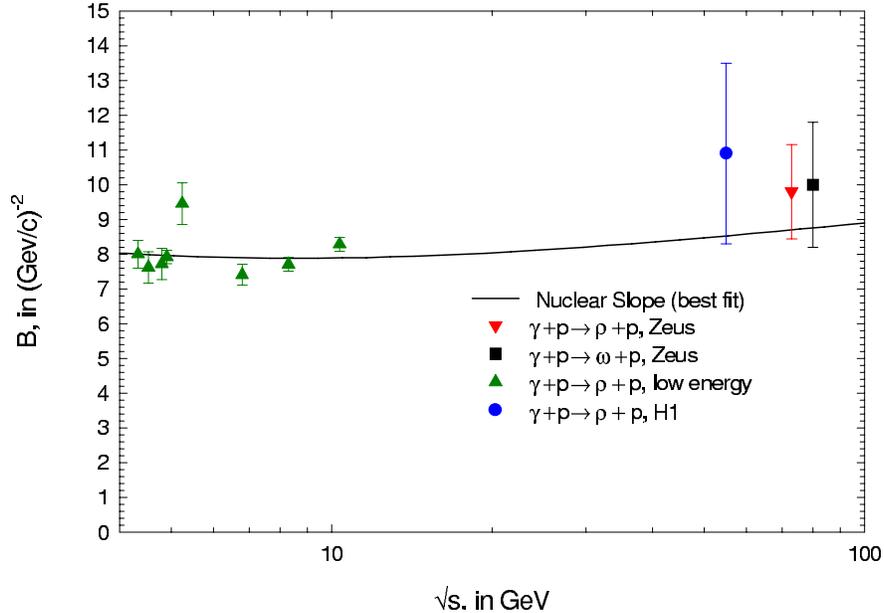,width=4.8in,%
bbllx=50pt,bblly=225pt,bburx=445pt,bbury=495pt,clip=}}
\end{center}
\caption[Evidence for the Additive Quark model: $\Bgp=\kappa \Bnn$] {\footnotesize Evidence for the Additive Quark model. A fit of experimental data for  the nuclear slopes $B$, from the `elastic scattering' reactions $\gamma +p\rightarrow V + p$, to the relation $\Bgp=\kappa \Bnn$ of \eq{eq:Bfact}, where $\kappa=0.661\pm 0.008$. $V$ is $\rho$, $\omega$ or $\phi$. Taken from Ref. \cite{blockhalzenpancheri2}.
}
\label{fig:Bgp}
\end{figure}
%***************
\subsection{Evidence for factorization of nucleon-nucleon, $\gamma p$ and $\gamma\gamma$ total cross sections, using analytic amplitudes}
We next discuss  experimental evidence for the cross section factorization relation 
\be \frac{\sigma_{nn}(s)}{\sigma_{\gamma p}(s)}=\frac{\sigma_{\gamma p}(s)}{\sigma_{\gamma\gamma}(s)},\label{factorizesigma}
\ee
where the $\sigma$'s are the total cross sections and $\sigma_{nn}$, the total nucleon-nucleon cross section, is the {\em even} (under crossing) cross section for $pp$ and $\pbar p$ scattering. These relations were derived by Block and Kaidalov\cite{me}, using eikonals for $\gamma \gamma$, $\gamma p$ and  the even portion of nucleon-nucleon scattering,  and further  assuming that the ratio of elastic scattering to total scattering is process-independent, i.e.,
\be
\left(
\frac{\sigma_{\rm elastic}(s)}{\sigma_{\rm tot}(s)}
\right)_{\gamma \gamma}=
\left(\frac{\sigma_{\rm elastic}(s)}{\sigma_{\rm tot}(s)}
\right)_{\gamma p}
=\left(\frac{\sigma_{\rm elastic}(s)}{\sigma_{\rm tot}(s)}\right)_{nn},\quad {\rm for\  all\ }s,\label{sigratios}
\ee
a result we have previously derived in \eq{eq:factorization} when we discussed eikonal properties in Section \ref{section:aspen}. They have further shown that 
\begin{equation} \rhonn(s)=\rhogp(s)=\rhogg(s),\label{eq:rhofact}
\end{equation}
where $\rho$ is the ratio  of the real to the imaginary portion of the forward scattering amplitude.
These theorems are exact, for {\em all } $s$  (where $\sqrt s$ is the c.m.s. energy), and survive exponentiation of the eikonal (see ref. \cite{me}).
 
Using real analytic amplitudes, Block and Kang\cite{bk} tested factorization ( \eq{factorizesigma}) empirically  by making a {\em global fit} to all of the experimental data
for $pp$, $\bar p p$, $\gamma p$, and $\gamma \gamma$ total cross sections and the  $pp$ and $\pbar p$ $\rho$-values,  i.e., making a simultaneous fit to {\em all} of the available experimental data {\em using} the factorization hypothesis (along with a minimum number of parameters), and seeing if the $\chi^2$ to this global fit gave a satisfactory value. A convenient phenomenological  framework for doing this numerical calculation is to parametrize the data using real analytic amplitudes that give an asymptotic $\ln^2 s$ rise for the total cross sections,  then make the cross sections  satisfy factorization and  finally test the value of the overall $\chi^2$ to see if the factorization hypothesis is satisfied. They showed  that the factorization relation $ \sigma_{nn}(s)/\sigma_{\gamma p}(s)=\sigma_{\gamma p}(s)/\sigma_{\gamma\gamma}(s)$ is satisfied experimentally when they used the PHOJET Monte Carlo analysis of the $\gamma\gamma$ cross section data, rather than the published values\cite{L3, OPAL}. 

The COMPETE collaboration\cite{{compete1},{cudell}} has also done an analysis of these data, using real analytical amplitudes. However, there are  major differences between the Block and Kang (BK) analysis and the one done by the COMPETE group. In order to test factorization, 
\begin{itemize}
\item BK made a {\em simultaneous} fit to $\pbar p$, $pp$, $\gamma p$ and $\gamma\gamma$ data assuming complete factorization using the {\em same} shape parameters, whereas COMPETE fit each reaction separately, using {\em different} shape parameters
\item BK fit {\em individually} the two $\sigma_{\gamma\gamma}$ sets of L3\cite{L3} and OPAL\cite{OPAL} data that are obtained using the PHOJET and PYTHIA Monte Carlos and do not use their average (the {\em published} value quoted in the Particle Data Group\cite{pdg} compilations), since the two sets taken individually have very different shapes and normalizations compared to their experimental errors. They emphasize that this individual fitting of the $\gamma\gamma$ data,  i.e., a detailed understanding of the experimental situation, is key to their analysis. 
\end{itemize}
At the end of their computation, BK investigate whether the overall $\chi^2$ is satisfactory.

Using real analytic amplitudes, they calculated the total cross sections $\sigma_{nn}$, $\sigma_{\gamma p}$ and $\sigma_{\gamma\gamma}$, along with the corresponding $\rho$-values. The cross section $\sigma_{nn}$, referred to in the factorization theorem of \eq{factorizesigma}, is given by
\be
\sigma_{nn}(s)=\frac{4\pi}{p}{\rm Im}f_+(s)=\frac{\sigma_{pp}+\sigma_{\bar pp}}{2},\label{sigmann}
\ee
 i.e., the {\em even} cross section.
The unpolarized (even) total cross sections for $\gamma p$ and $\gamma\gamma$ scattering are, in turn, given by
\be 
\sigma_{\gamma p}(s)=\frac{4\pi}{p}{\rm Im}f_{\gamma p}(s)\quad{\rm and}\quad \sigma_{\gamma \gamma}=\frac{4\pi}{p}{\rm Im}f_{\gamma \gamma}(s)\label{sigmagp}.
\ee

They further assumed that their even (under crossing) amplitude $f_+$ and their odd (under crossing) amplitude $f_-$ are real analytic functions with a simple cut structure\cite{bc},and in the high energy region, and are given by
\begin{equation}
\frac{4\pi}{p}f_+(s)=i\left \{A+\beta[\ln (s/s_0) -i\pi/2]^2+cs^{\mu-1}e^{i\pi(1-\mu)/2}\right\},\label{evenamplitude_nn}
\end{equation}
and
\begin{equation}
\frac{4\pi}{p}f_-(s)=-Ds^{\alpha-1}e^{i\pi(1-\alpha)/2},\label{oddamplitude_nn}
\end{equation}
where $A$, $\beta$, $c$, $s_0$, D, $\mu$ and $\alpha$  are real constants and  they  ignored any real subtraction constants. 
 Using \eq{evenamplitude_nn} and \eq{oddamplitude_nn}, the total cross sections $\sigma_{\pbar p}$, $\sigma_{pp}$ and $\sigma_{nn}$ for high energy scattering are given by
\be
\sigma_{\pbar p}(s)= A+\beta\left[\ln^2 s/s_0-\frac{\pi^2}{4}\right]+c\,\sin(\pi\mu/2)s^{\mu-1}-D\cos(\pi\alpha/2)s^{\alpha-1} , \label{sigmapbarp}
\ee 
\be
\sigma_{p p}(s)= A+\beta\left[\ln^2 s/s_0-\frac{\pi^2}{4}\right]+c\,\sin(\pi\mu/2)s^{\mu-1}+D\cos(\pi\alpha/2)s^{\alpha-1} , \label{sigmapp}
\ee 
\be
\sigma_{nn}(s)= A+\beta\left[\ln^2 s/s_0-\frac{\pi^2}{4}\right]+c\,\sin(\pi\mu/2)s^{\mu-1} , \label{sigmann2}
\ee 
and the $\rho$'s, the ratio of the real to the imaginary portions of the forward scattering amplitudes, are given by
\be
\rho_{\pbar p}(s)=\frac{\beta\,\pi\ln s/s_0-c\,\cos(\pi\mu/2)s^{\mu-1}-D\sin(\pi\alpha/2)s^{\alpha-1}}{\sigma_{\pbar p}}\label{rhopbarp},
\ee  
\be
\rho_{pp}(s)=\frac{\beta\,\pi\ln s/s_0-c\,\cos(\pi\mu/2)s^{\mu-1}+D\sin(\pi\alpha/2)s^{\alpha-1}}{\sigma_{pp}}\label{rhopp},
\ee  
\be
\rho_{nn}(s)=\frac{\beta\,\pi\ln s/s_0-c\,\cos(\pi\mu/2)s^{\mu-1}}{\sigma_{nn}}.\label{rhonn}
\ee  
Assuming  that the  term in $c$ is a Regge descending term, they used  $\mu=0.5$.  

To test the factorization theorem of \eq{factorizesigma}, they wrote the (even) amplitudes $f_{\gamma p}$ and $f_{\gamma\gamma}$ as
\begin{equation}
\frac{4\pi}{p}f_{\gamma p}(s)=iN\left\{A+\beta[\ln (s/s_0) -i\pi/2]^2+cs^{\mu-1}e^{i\pi(1-\mu)/2}\right\},\label{evenamplitude_gp1}
\end{equation}
and 
\begin{equation}
\frac{4\pi}{p}f_{\gamma\gamma}(s)=iN^2\left \{A+\beta[\ln (s/s_0) -i\pi/2]^2+cs^{\mu-1}e^{i\pi(1-\mu)/2}\right\}\label{evenamplitude_gg},
\end{equation}
with  $N$  the proportionality constant in the  relation $\sigma_{nn}(s)/\sigma_{\gamma p}(s)=\sigma_{\gamma p}(s)/\sigma_{\gamma \gamma}(s)=N$. 
We note,  using \eq{evenamplitude_nn}, \eq{evenamplitude_gp1} and \eq{evenamplitude_gg}, that 
\be
\rho_{nn}=\rho_{\gamma p}=\rho_{\gamma\gamma}=\frac{\beta\,\pi\ln s/s_0-c\,\cos(\pi\mu/2)s^{\mu-1}}{A+\beta\left(\ln^2 s/s_0-\frac{\pi^2}{4}\right)+c\,\sin(\pi\mu/2)s^{\mu-1}}\label{3rhos},
\ee
{\em automatically}  satisfying the Block and Kaidalov\cite{me} relation of \eq{eq:rhofact}.

In the additive quark model, using vector dominance, the proportionality constant $N=\frac{2}{3}P_{\rm had}^\gamma$, where $P_{\rm had}^\gamma$ is the probability that a photon turns into a vector hadron. Using (see Table XXXV, p.393 of
Ref.~\cite{bauer}) $\frac{f_{\rho}^2}{4\pi}=2.2$,
$\frac{f_{\omega}^2}{4\pi}=23.6$ and $\frac{f_{\phi}^2}{4\pi}=18.4$,
BK found
\be 
P_{\rm had}^\gamma\approx\Sigma_{V}\frac{4\pi\alpha}{f_V^2}=1/249,\label{Phadestimate}
\ee 
where $V=\rho,\omega,\phi$. In this estimate, they  have neither taken into account the continuum vector channels nor the running of the electromagnetic coupling constant, effects that will tend to increase $P_{\rm had}^\gamma$ by several percent as well as give it a very slow energy dependence, increasing as we go to higher energies.   In the spirit of the additive quark model and vector dominance,  they can now write, using $N=\frac{2}{3}P_{\rm had}^\gamma$ in \eq{evenamplitude_gp} and \eq{evenamplitude_gg}, 
\be
\sigma_{\gamma p}(s)=\frac{2}{3}P_{\rm had}^\gamma \left(A+\beta\left[\ln^2 s/s_0-\frac{\pi^2}{4}\right]+c\,\sin(\pi\mu/2)s^{\mu-1} \right) \label{sigmagp2}
\ee 
and
\be
\sigma_{\gamma \gamma}(s)=\left(\frac{2}{3}P_{\rm had}^\gamma\right)^2 \left(A+\beta\left[\ln^2 s/s_0-\frac{\pi^2}{4}\right]+c\,\sin(\pi\mu/2)s^{\mu-1} \right) \label{sigmagg2}
\ee  
with the real constants $A,\beta,s_0,c,D$ and $P_{\rm had}^\gamma$   being fitted by experiment (assuming $\alpha=\mu=0.5$).  
In  fitting the $\gamma\gamma$ data, BK note that one might be tempted to use the $\gamma\gamma$ cross sections---along with their quoted errors---that are given in the Particle Data Group\cite{pdg} cross section summary. However, on closer inspection of the original  papers, it turns out that the results quoted by the PDG are the {\em averages} of {\em two independent} analyses performed by both the OPAL\cite{OPAL} and L3\cite{L3} groups, using the two different Monte Carlo programs, PHOJET and PYTHIA. The  error quoted by the Particle Data Group was essentially half the  difference between these two very different values, rather than the smaller errors associated with each individual analysis.  

The Monte Carlo simulations used by OPAL and L3 play a critical role in unfolding the $\gamma\gamma$ cross sections from the raw data.  A direct quotation from  the OPAL paper\cite{OPAL} illustrates this:
\begin{itemize}
\item[] ``In most of the distributions, both Monte Carlo models describe the data equally well and there is no reason for preferring one model over the other for the unfolding of the data.  We therefore average the results of the unfolding.  The difference between this cross section and the results obtained by using PYTHIA or PHOJET alone are taken as the systematic error due to the Monte Carlo model dependence of the unfolding.''  
\end{itemize}
For the testing of factorization, there is good reason for possibly preferring one model over another, since the two models give both {\em different normalizations} and {\em shapes}, which are vital to their analysis. BK went  back to the original papers\cite{L3, OPAL} and  deconvoluted the data according to whether PHOJET or PYTHIA was used, with the  results given in Fig. \ref{fig:originaldata}.  Obviously, there are major differences in shape and normalization  due to the different Monte Carlos, with the PYTHIA results being significantly higher and rising much faster for energies above $\approx 15$ GeV.  On the other hand, the OPAL and L3 data agree within errors, for each of the two Monte Carlos, and seem to be quite consistent with each other, as seen in  Fig. \ref{fig:originaldata}. 
%%%%%%%%%%%%%%%%%%%%%%%%
\begin{figure}%[btp] %#19
\begin{center}
\mbox{\epsfig{file=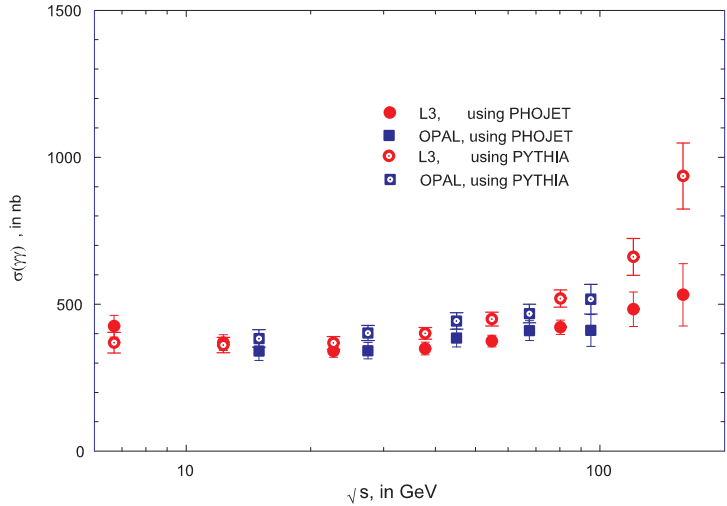,width=4.in,%
bbllx=0pt,bblly=0pt,bburx=438pt,bbury=313pt,clip=}}
\end{center}
\protect\caption[OPAL and L3 total cross sections for $\gamma\gamma$ scattering]
 {\footnotesize OPAL and L3 total cross sections for $\gamma\gamma$ scattering, in nb vs. $\sqrt s$, the c.m.s. energy, in GeV. The data have been unfolded according to the Monte Carlo used. The solid circles are the L3 data, unfolded using PHOJET. The open circles are the L3 data, unfolded using PYTHIA. The solid squares are the OPAL data, unfolded using PHOJET. The open squares are the OPAL data, unfolded using PYTHIA. Taken from Ref.\cite{bk}.}
\label{fig:originaldata}
\end{figure}
%********************************************

For these reasons, they made three different fits, whose results are shown in Table \ref{ta:amp}.%
%%%%%%%%%%%%%%%%%%%%%%%%%%%%%%%%%%%%%%%%%
%**************************************************************
%     Table #8
%
\begin{table}[h,t]                   % Use "table" environment, but also
				 % use  "tabular" environment below.
%
\def\arraystretch{1.2}            % Make the space between rows in the Table,
				  % 1.5 x bigger than the default spacing.
\begin{center}
 \caption[Factorization fit parameters using analytic amplitudes]
{\protect\small Factorization fit parameters using analytic amplitudes. Fit 1 is the result  of a fit to total cross sections and
     $\rho $-values for $\pbar p$ and  $pp$, along with $\sigma_{\gamma p}$. Fit 2 and Fit 3 are the results of fitting total cross sections and
     $\rho $-values for $\pbar p$, $pp$ and $\sigma_{\gamma p}$, as well as including the $\sigma_{\gamma\gamma}$ data from the OPAL and L3 collaborations. Fit 2 uses the results of unfolding  $\sigma_{\gamma\gamma}$ with the PHOJET Monte Carlo, whereas Fit 3 uses the results of unfolding $\sigma_{\gamma\gamma}$ with the PYTHIA Monte Carlo.  The overall renormalization factors $N_{\rm OPAL}$ and $N_{\rm L3}$ are also fitted in both Fit 2 and Fit 3. The  fitted parameters are those that have statistical errors indicated. Taken from Ref.\cite{bk}.\label{ta:amp}}
\vspace{.2in}
\begin{tabular}[b]{|l||l|l|l||}
     %\multicolumn{6}{c}{TABLE \ref{ta:amp}}\\ % Method to get title out of the main
		       % body of the Table, and then space between it and
		       % the Table.
     %\multicolumn{6}{c}{}\\  % Inserted spaces.
     %\multicolumn{6}{c}{}\\
     %\multicolumn{6}{c}{}\\
     %\multicolumn{6}{c}{}\\
     % Start main Table here.
     \cline{2-4}
      \multicolumn{1}{c|}{}
      &\multicolumn{3}{c||}{$\stot \sim \ln^2(s/s_0)$}
      \\
      \hline
      Parameters&\ \ Fit 1: &\ \ \ \ \ \ \ \ Fit 2:  & \ \ \ \ \ \ \ \ Fit 3: \\ %\hline
& no $\sigma_{\gamma\gamma}$&$\sigma_{\gamma\gamma}$ from PHOJET& $\sigma_{\gamma\gamma}$ from PYTHIA\\\hline
     $A$ (mb)&$37.2\pm 0.81$&$37.1\pm$ 0.87&$37.3\pm$ 0.77\\
     $\beta$ (mb)&$0.304\pm 0.023$&$0.302\pm 0.024$&$0.307\pm 0.022$\\
     $s_0$ (${\rm (GeV)}^2$)&$34.3\pm 14$&$32.6\pm 16$&$35.1\pm 14$ \\
     $D$ (mb${\rm (GeV)}^{2(1-\alpha)}$)&$-35.1\pm 0.83$&$-35.1\pm 0.85$
     &$-35.4\pm 0.84$ \\
     $\alpha$&0.5&0.5&0.5 \\
     $c$ (mb${\rm (GeV)}^{2(1-\mu)})$&$55.0\pm 7.5$&$55.9\pm 8.1$
     &$54.6\pm 7.3$ \\
     $\mu$&0.5&0.5&0.5 \\
     $P_{\rm had}^\gamma$&$1/(233.1\pm0.63)$&$1/(233.1\pm0.63)$&$1/(233.0\pm0.63)$ \\
$N_{\rm OPAL}$&\ \ \ \ \ --------&$0.929\pm0.037$&$0.861\pm0.050$\\
$N_{\rm L3}$&\ \ \ \ \ --------&$0.929\pm0.025$&$0.808\pm0.020$\\
     \hline
     degrees of freedom (d.f.)&68&78&78\\
     $\chi^2/$d.f.&1.62&1.49&1.87\\
 total $\chi^2$&110.5&115.9&146.0\\
     \hline
\end{tabular}
\end{center}
     %\vspace{1in} \\
    
%     \\
\end{table}
\def\arraystretch{1}  %Restore the default row spacing in the Table
%%%%%%%%%%%%%%%%%%%%%%%%%%%%%%%%%%%%%%%%%%
%%%%%%%%%%%%%%%%%%%%%%%%%%%%%%%%%%%%%%%%%%%55
Fit 1 is a simultaneous $\chi^2$ fit of \eq{sigmapbarp}, (\ref{sigmapp}), (\ref{rhopbarp}), (\ref{rhopp}) and (\ref{sigmagp2}) 
 to the experimental $\sigma_{\pbar p}$, $\sigma_{pp}$, $\rho_{\pbar p}$, $\rho_{pp}$ and $\sigma_{\gamma p}$ data in the c.m. energy interval $10 {\ \rm GeV}\le \sqrt s\le 1800$ GeV, i.e.,  the $\gamma\gamma$ data are not included.   They next made two different simultaneous $\chi^2$ fits of \eq{sigmapbarp}, (\ref{sigmapp}), (\ref{rhopbarp}), (\ref{rhopp}), (\ref{sigmagp2}) and (\ref{sigmagg2}) to the experimental $\sigma_{\pbar p}$, $\sigma_{pp}$, $\rho_{\pbar p}$, $\rho_{pp}$, $\sigma_{\gamma p}$ and the {\em unfolded} $\sigma_{\gamma \gamma}$, using either PHOJET or PYTHIA results, in the c.m. energy interval $10 {\ \rm GeV}\le \sqrt s\le 1800$ GeV. Fit 2 uses $\sigma_{\gamma \gamma}$ from PHOJET unfolding and Fit 3 uses $\sigma_{\gamma \gamma}$ from PYTHIA unfolding.  To account for possible systematic overall normalization factors in  experimental data, the  cross sections for L3 are multiplied by the overall renormalization factor $N_{\rm L3}$ and those for OPAL are multiplied by an overall renormalization factor $N_{\rm OPAL}$, which are also fit in Fits 2 and 3.

The major fit parameters $A$, $\beta$, $s_0$, $D$, $c$ and $P_{\rm had}^\gamma$ are the same, within errors, as seen from Fits 1, 2 and 3.  The purpose of Fit 1 was to show the robustness of the procedure, independent of the $\gamma\gamma$ data. 

Strikingly, when BK introduced the unfolded $\gamma\gamma$ cross sections in Fits 2 and 3,  the results strongly favor the PHOJET data of Fit 2;  the $\chi^2$/d.f. jumps from 1.49 to 1.87 (the total $\chi^2$ changes from 115.9 to 146.0 for the same number of degrees of freedom). Perhaps more compellingly, the normalizations for both OPAL and L3 are in complete agreement, being $0.929\pm0.037$ and $0.929\pm0.025$, respectively.  The difference  from unity by $\approx 7\pm3$\% is compatible with the experimental systematic normalization error of 5\% quoted by L3 , whereas the PYTHIA results from Fit 3 have normalizations that disagree by $\approx 14$\% and $\approx 19$\% for OPAL and L3, respectively, in sharp disagreement with the 5\% estimate. From here on,  only the PHOJET results of Fit 2 were utilized and these are the parameters given in Table \ref{ta:amp}.
 
BK found that $P_{\rm had}^\gamma=1/(233.1\pm0.63)$, in reasonable agreement  with their preliminary estimate of 1/249, being $\approx 6$\% larger, an effect easily accounted for by continuum vector channels in $\gamma p$ reactions that are not accounted for in the estimate of \eq{Phadestimate}.  

The fitted total cross sections $\sigma_{\pbar p}$ and $\sigma_{pp}$ from \eq{sigmapbarp} and \eq{sigmapp} are shown in Fig. \ref{fig:sigmanucleon}, along with the experimental data.% 
%***************
\begin{figure}[tbp] %#20
\begin{center}
\mbox{\epsfig{file=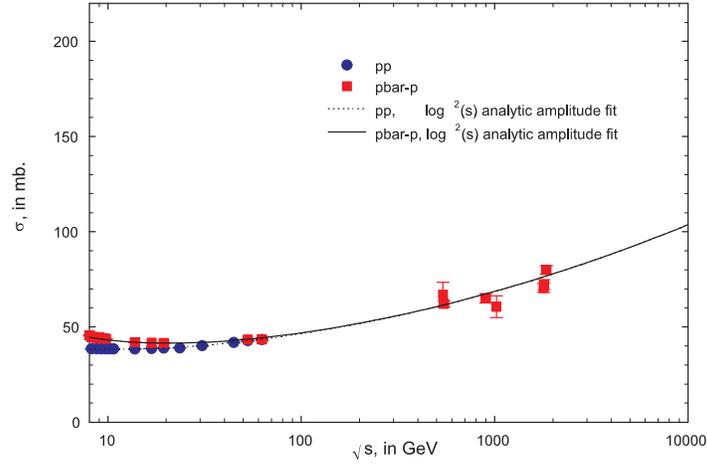,width=3.8in,%
bbllx=0pt,bblly=0pt,bburx=438pt,bbury=313pt,clip=}}
\end{center}
\protect\caption[Total cross sections $\sigma_{pp}$ and $\sigma_{\pbar p}$, using factorization parameters of Table \ref{ta:amp}]
 {\footnotesize Total cross sections $\sigma_{pp}$ and $\sigma_{\pbar p}$, using factorization parameters of Table \ref{ta:amp}. The dotted curve is $\sigma_{pp}$, in mb,  and the solid curve is $\sigma_{\pbar p}$,  in mb  vs. $\sqrt s$, the c.m. energy, in GeV, predictions from Fit 2. The circles are the experimental data for $pp$ reactions  and the squares are the experimental $\pbar p$ data. Taken from Ref.\cite{bk}.
\label{fig:sigmanucleon}}
\end{figure}
%********************************************
%
\begin{figure}[tbp] %#21
\begin{center}
\mbox{\epsfig{file=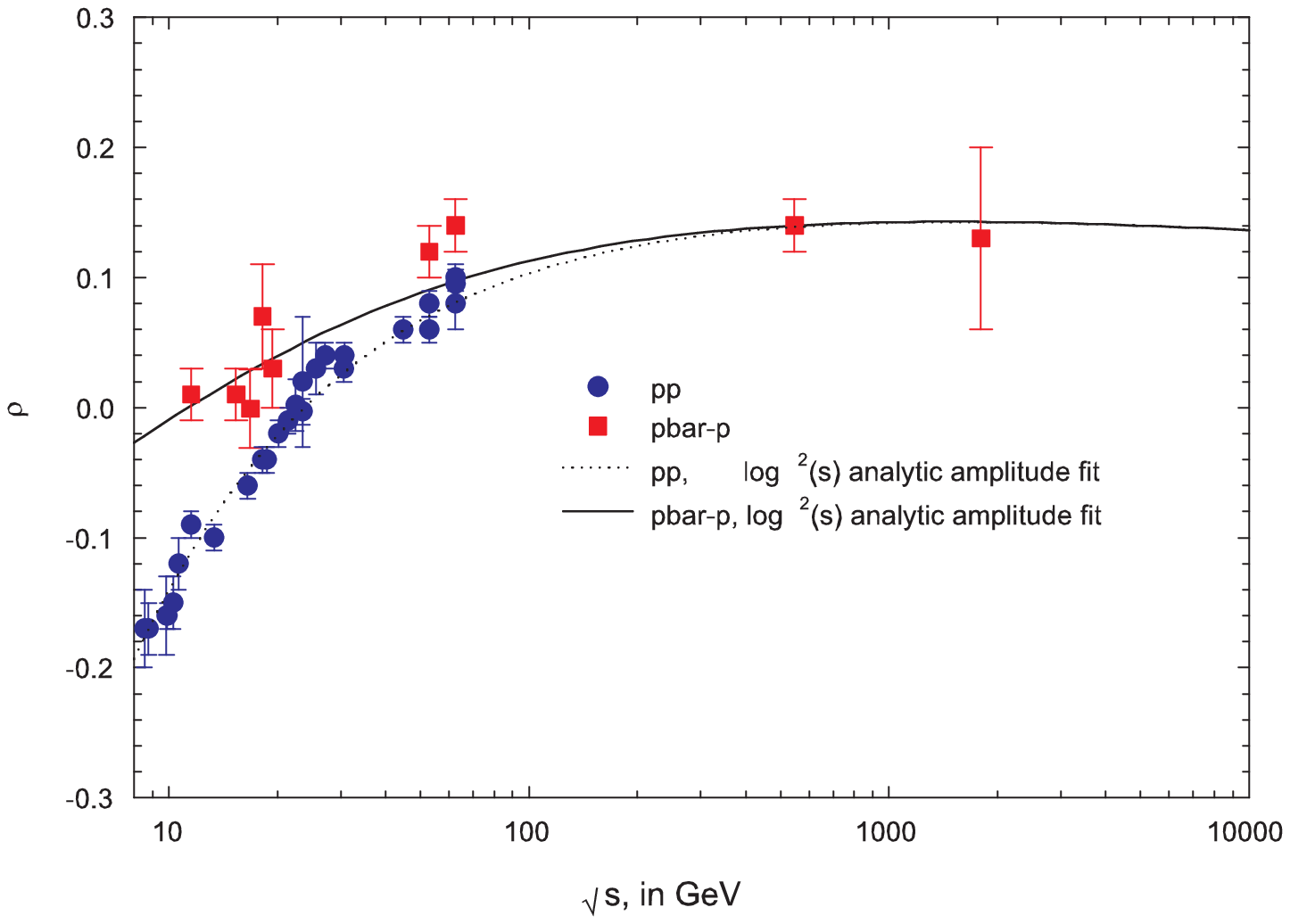,width=3.8in,%
bbllx=0pt,bblly=0pt,bburx=438pt,bbury=313pt,clip=}}
\end{center}
\protect\caption[$\rho_{pp}$ and $\rho_{\pbar p}$, using factorization parameters of Table \ref{ta:amp}]
 {\footnotesize  $\rho_{pp}$ and $\rho_{\pbar p}$, using factorization parameters of Table \ref{ta:amp}. The dotted curve is $\rho_{pp}$ and the solid curve is $\rho_{\pbar p}$ vs. $\sqrt s$, the c.m. energy, in GeV, predictions from Fit 2. The circles  are the experimental data for $p p$ reactions  and the squares are the experimental $\pbar p$ data. Taken from Ref.\cite{bk}.
\label{fig:rhonucleon}}
\end{figure}
%%%%%%%%%%%%%%%
%********************************************
\begin{figure}[tbp] %#22
\begin{center}
\mbox{\epsfig{file=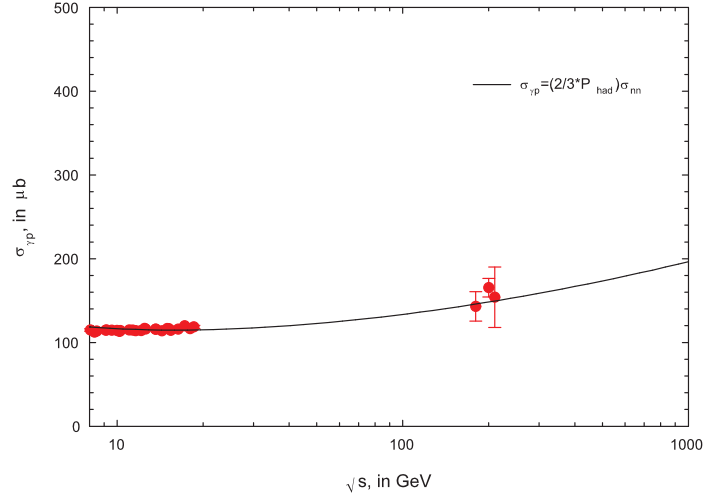,width=3.8in,%
bbllx=0pt,bblly=0pt,bburx=438pt,bbury=313pt,clip=}}
\end{center}
\protect\caption[The total cross section $\sigma_{\gamma p}=\frac{2}{3}{\rm P}_{\rm had}^\gamma \sigma_{\rm nn}$, using factorization parameters of Table \ref{ta:amp}]
 {\footnotesize The total cross section $\sigma_{\gamma p}=\frac{2}{3}{\rm P}_{\rm had}^\gamma \sigma_{\rm nn}$, using factorization parameters of Table \ref{ta:amp}. The solid curve is $\sigma_{\gamma p}$, in $\mu$b  vs. $\sqrt s$, the c.m. energy, in GeV,  a prediction from Fit 2. The circles are the experimental data. Taken from Ref.\cite{bk}. 
\label{fig:sigmagp}}
\end{figure}
%************************************************************************
\begin{figure}[tbp] %#23
\begin{center}
\mbox{\epsfig{file=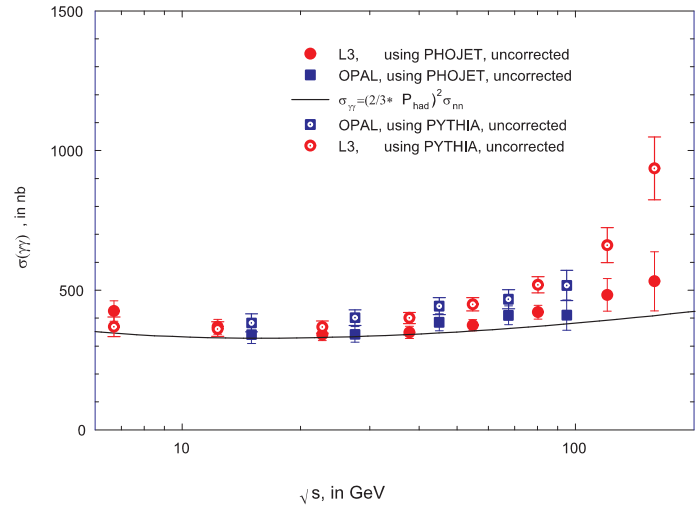,width=3.8in,%
bbllx=0pt,bblly=0pt,bburx=438pt,bbury=313pt,clip=}}
\end{center}
\protect\caption[$\sigma_{\gamma \gamma}=(\frac{2}{3}{\rm P}_{\rm had}^\gamma)^2 \sigma_{\rm nn}$, using factorization parameters of Table \ref{ta:amp}]
 {\footnotesize $\sigma_{\gamma \gamma}=(\frac{2}{3}{\rm P}_{\rm had}^\gamma)^2 \sigma_{\rm nn}$, using factorization parameters of Table \ref{ta:amp}. The solid curve is the total $\gamma \gamma$ cross section, $\sigma_{\gamma \gamma}$, in nb,  vs. $\sqrt s$, the c.m. energy, in GeV, from Fit 2. The open squares and circles are the experimental total cross sections for OPAL and L3, respectively, unfolded using the PYTHIA Monte Carlo. The solid squares and circles are the experimental total cross sections for OPAL and L3, respectively, unfolded using the PHOJET Monte Carlo. Taken from Ref.\cite{bk}.
\label{fig:sigmagguncorrected}}
\end{figure}
%
%********************************************
%%%%%%%%%%%%%%%%%%%%%%%%%%%%%%%%%%

The fitted $\rho$-values, $\rho_{\pbar p}$ and $\rho_{pp}$ from \eq{rhopbarp} and \eq{rhopp} are shown in Fig. \ref{fig:rhonucleon}, along with the experimental data. The fitted total cross section $\sigma_{\gamma p}=\frac{2}{3}P_{\rm had}^\gamma\sigma_{nn}$ from \eq{sigmagp2} is compared to the experimental data in Fig. \ref{fig:sigmagp}, using $P_{\rm had}^\gamma=1/233$. The overall agreement of the $\bar p p$, $pp$ and $\gamma p$ data with the fitted curves is quite satisfactory. 

The fitted total cross section $\sigma_{\gamma \gamma}=(\frac{2}{3}P_{\rm had}^\gamma)^2\sigma_{nn}$ from \eq{sigmagg2} is compared to the experimental data in Fig. \ref{fig:sigmagguncorrected}, again using $P_{\rm had}^\gamma=1/233$. The experimental data plotted in Fig. \ref{fig:sigmagguncorrected} are {\em not} renormalized, but are the results of unfolding the original experimental results,  i.e., use $N_{\rm OPAL}=N_{\rm L3}=1$. We see from Fig. \ref{fig:sigmagguncorrected} that within errors, both the {\em shape} and {\em normalization} of the PHOJET cross sections from both OPAL and L3 are in reasonable agreement with the factorization theorem of \eq{factorizesigma}, whereas the PYTHIA cross sections are in distinct disagreement. This conclusion is born out by the $\chi^2$'s of Fit 2 and Fit 3 in Table \ref{ta:amp}.

The fitted results for $\sigma_{\gamma\gamma}$, using the parameters of Fit 2, are compared to the renormalized OPAL and L3 (PHOJET only) data in Fig. \ref{fig:sigmaggcorrected}.  The agreement in shape and magnitude is quite satisfactory, indicating strong experimental support  for factorization. 

For completeness, we show in Fig. \ref{fig:rho} the expected $\rho$-value for the even amplitude, from \eq{3rhos}. %
%%%%%%%%%%%%%%%%%%%%%%%%%%%%%%%%%%%%%
\begin{figure}[tbp] %#24
\begin{center}
\mbox{\epsfig{file=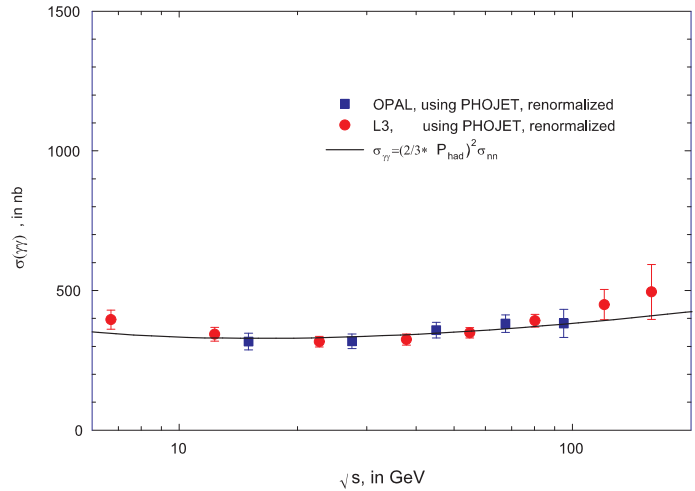,width=3.8in,%
bbllx=0pt,bblly=0pt,bburx=438pt,bbury=313pt,clip=}}
\end{center}
\protect\caption[$\sigma_{\gamma \gamma}=(\frac{2}{3}{\rm P}_{\rm had}^\gamma)^2 \sigma_{\rm nn}$ with renormalized data]
 {\footnotesize  $\sigma_{\gamma \gamma}=(\frac{2}{3}{\rm P}_{\rm had}^\gamma)^2 \sigma_{\rm nn}$,  with renormalized data. The solid curve is the total $\gamma \gamma$ cross section, $\sigma_{\gamma \gamma}$, in nb,  vs. $\sqrt s$, the c.m. energy, in GeV, from Fit 2.   The squares and circles are the  total cross sections for OPAL and L3, respectively, unfolded using PHOJET, {\em after} they have been renormalized by the factors $N_{\rm OPAL}=0.929$ and $N_{\rm L3}=0.929$, found in Fit 2 of Table \ref{ta:amp}. Taken from Ref.\cite{bk}.
\label{fig:sigmaggcorrected}}
\end{figure}
%
%%%%%%%%%%%%%%%%%%%%%%%%%%%%%%%%%%%%%%%
\begin{figure}[tbp] %#25
\begin{center}
\mbox{\epsfig{file=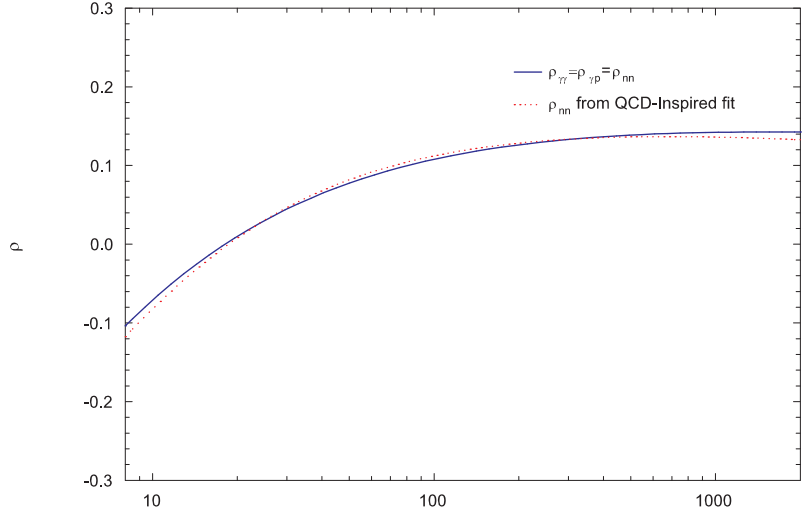,width=4.4in,%
bbllx=0pt,bblly=0pt,bburx=450pt,bbury=319pt,clip=}}
\end{center}
\protect\caption[$\rho_{\gamma\gamma}=\rho_{\gamma p}=\rho_{nn}$, using factorization parameters of Table \ref{ta:amp}]
 {\footnotesize $\rho_{\gamma\gamma}=\rho_{\gamma p}=\rho_{nn}$, using factorization parameters of Table \ref{ta:amp}. The solid curve is the $\rho$-value for the even amplitude  vs. $\sqrt s$, the c.m.s. energy, in GeV. The dotted curve, shown for comparison, is $\rho_{nn}$, the result of a QCD-inspired eikonal fit\cite{BHS} to $\pbar p$ and $pp$ data that included cosmic ray $p$-air data. Taken from Ref.\cite{bk}
\label{fig:rho}}
\end{figure}
%%%%%%%%%%%%%%%%%%%%%%%%%%%
Also shown in this graph is the predicted value for $\rho_{nn}$ found from a QCD-inspired eikonal fit (Aspen model) by Block\cite{BHS}  et al. to  $\pbar p$ and $pp$ total cross sections and $\rho$-values from accelerators plus $p$-air cross sections from cosmic rays. The agreement between these two independent analyses, using very different approaches, with one using  
real analytic amplitudes with a $\ln s^2$ behavior and the other using
a QCD-inspired eikonal model in impact parameter space, giving rise to a cross section also eventually rising as $\ln s^2$,
is most striking.  In both cases, these two analyses give  $\rho_{\rm nn}=\rho_{\gamma p}=\rho_{\gamma\gamma}$, another factorization theorem of Block and Kaidalov\cite{me}.  

BK concluded that the cross section factorization hypothesis of \cite{me}, $\sigma_{nn}(s)/\sigma_{\gamma p}(s)=\sigma_{\gamma p}(s)/\sigma_{\gamma\gamma}(s)$,  is satisfied for nn, $\gamma p$ and $\gamma\gamma$ scattering, if one uses the PHOJET Monte Carlo program to analyze $\sigma_{\gamma\gamma}$. Further, the experimental data also satisfied the additive quark model using vector meson dominance, since 
\begin{eqnarray}
\sigma_{\gamma p}&=&\frac{2}{3}P_{\rm had}^\gamma \sigma_{nn}
\nonumber\\
\sigma_{\gamma \gamma}&=&\left(\frac{2}{3}P_{\rm had}^\gamma\right)^2 \sigma_{nn}\label{vectordominance},
\end{eqnarray}
with $\kappa=2/3$ and $P_{\rm had}^\gamma=1/233$. This result is in excellent agreement with the factorization theorem 
\be 
\frac{B_{nn}(s)}{B_{\gamma p}(s)}=\frac{B_{\gamma p}(s)}{B_{\gamma \gamma}(s)}\label{Bfactorize},
\ee
where the $B$'s are the nuclear slopes for elastic scattering. For $\gamma p$ processes, using vector dominance, the $B$'s are the slopes of the `elastic' scattering reactions 
\be
\gamma +p\rightarrow V+p,\label{elasticgp}
\ee
where the vector meson $V$ is either a $\rho$, $\omega$ or $\phi$ meson. We had earlier seen in Section \ref{sec:Bfactor} that a $\chi^2$ fit to the available $\gamma p$ data (see Ref. \cite{blockhalzenpancheri2}) gave
\be
\kappa =0.661\pm0.008, \label{kappavalue}
\ee
in excellent agreement with the 2/3 value predicted by the additive quark model, again justifying the use of 2/3 in the BK fits.

The authors concluded that if they determined
$\sigma_{nn}(s)$, $\rho_{nn}$ and $B_{nn}$ from experimental $\pbar p$ and $pp$ data for $\sqrt s\ge 8$ GeV, they could then predict rather accurately $\sigma_{\gamma p}(s)$, $\rho_{\gamma p}$, $B_{\gamma p}$ and $\sigma_{\gamma\gamma}(s)$, $\rho_{\gamma \gamma}$, $B_{\gamma \gamma}$,  in essentially a {\em parameter-free} way, by using factorization and the additive quark model with vector dominance.  Certainly, these conclusion would be greatly strengthened by precision cross section measurements of  both $\gamma p$ and $\gamma \gamma$ reactions at high energies. 

%%%%%%%%%%%%%%%%%%%%%%%%%%%%%%%%%%%%%%%%%%%%%%
%%%%%%%%%%%%%%%%%%%%%%%%%%%%%%%%%%%%%%%%%%%%%
\subsection{Testing the saturation of the Froissart bound for $\gamma p,\ \pi^\pm p$, $pp$ and $\bar pp$ collisions, using real analytic amplitudes}
The Froissart bound\cite{froissart} states that the high energy cross section for the scattering of hadrons is bounded by $\sigma \sim \ln^2s$,
where $s$ is the square of the cms energy.   This fundamental result was derived from unitarity and analyticity by Froissart\cite{froissart}, who stated: 
\begin{itemize}
\item[] ``At forward or backward angles, the modulus of the amplitude behaves at most like $s\ln^2s$, as $s$ goes to infinity.  We can use the optical theorem to derive that the total cross sections behave at most like $\ln^2s$, as $s$ goes to infinity".
\end{itemize}
In our context, saturation of the Froissart bound refers to an energy dependence of the total  cross section rising asymptotically as $\ln^2s$.

The question as to whether any of the present day  high energy data for  $\bar pp$, $pp$, $\pi^+ p$, $\pi^-p$, $\gamma p$ and $\gamma \gamma$ cross sections saturate the Froissart bound has not been settled; one can not discriminate between asymptotic fits of $\ln s$ and $\ln^2 s$  using high energy data only\cite{{bkw},{compete1}}.  Some  preference for $\ln^2s$ is found, but a $\ln s$ energy dependence can not be statistically ruled out\cite{compete1}.   We here point out that this ambiguity can be resolved rather elegantly by requiring that real analytic amplitude fits to the high energy data smoothly join the cross section and its derivative  at a transition (low) energy $\nu_0$ just above the resonance region, i.e., by using the analyticity constraints  summarized in \eq{allderiv0}, with $n=0$ and 1. Real analytic amplitudes have previously been introduced in Section \ref{sec:analyticamplitudes}. 

In this Section, we will often use  the notation $\ln^2s$ and $\ln^2(\nu/m)$ interchangeably when referring to the behavior of  an asymptotic high energy cross section that saturates the Froissart bound, where $\nu$ is the projectile laboratory energy and $m$ is the proton (pion) mass.
%%%%%%%%%%%%%%%%%%%%%%%%%%%%%
\subsubsection{Saturating the Froissart bound in  $\gamma p$ scattering}

The new analyticity constraints of \eq{allderiv0} demand that the high energy parametrizations smoothly join on to the low energy experimental $\gamma p$ total cross sections just above the resonance region, at a transition energy $\nu_0$.  Block and Halzen\cite{bhfroissart} (BH) have  shown that only fits to the high energy data behaving as $\ln^2 s$ can  adequately describe the highest energy points. 

For the low energy cross section and its derivative at $\nu_0$, BH used a convenient parametrization by Damashek and Gilman\cite{gilman} of the forward Compton scattering amplitudes yielding a very accurate description of the low energy resonant data that was used. It provided a best fit in the energy region $2m\nu_0+m^2\le \sqrt s\le 2.01$ GeV using five Breit-Wigner resonances and a 6th order polynomial in $(\sqrt s -\sqrt s_{\rm threshold})$.  Here $\nu_{\rm threshold}=m_\pi + m_\pi^2/m$ is the threshold energy and $m$ the proton mass. Their result is shown in Fig.\,\ref{fig:siggpresonances}.

Using \eq{sig0} and \eq{rho0} as their starting point, BH wrote
\begin{eqnarray}
\sigma_{\gamma p}&=&c_0+c_1\ln\left(\frac{\nu}{m}\right)+c_2\ln^2\left(\frac{\nu}{m}\right)+\beta_{\cal P'}\left(\frac{\nu}{m}\right)^{\mu -1},\label{sigma}\\
\rho_{\gamma p}&=& {1\over\sigma} \left\{{\pi}{2}c_1+\pi c_2\ln\left(\frac{\nu}{m}\right)-\cot(\pi\mu/2)\beta_{\cal P'}\left(\frac{\nu}{m}\right)^{\mu -1}+\frac{4\pi}{\nu}f_+(0) \right\}\label{rho1},
\end{eqnarray}
where they introduced the additional real constant $f_+(0)$, the subtraction constant needed  in the singly-subtracted dispersion relation\cite{gilman} for the reaction $\gamma +p\rightarrow\gamma + p$, fixed in the Thompson scattering limit as  $f_+(0)=-\alpha/m=-3.03\  \mu {\rm b\  GeV}$.

Their strategy was to constrain the high energy fit with the precise low energy fit at $\sqrt s\le 2.01$ GeV, which is the energy where Damashek and Gilman\cite{gilman}  join the energy region dominated by resonances to a Regge fit, $a +b/\sqrt{\nu/m}$. They found that the cross section at $\sqrt s_0=2.01$ is 151 $\mu$b and the slope $d\sigma_{\gamma p}/d(\nu/m)$ is $-b/(\nu/m)^{1.5}$, or $-15.66$ in $\mu$b units. Using the asymptotic expression of \eq{sigma}, BH obtained the  two constraints
\begin{eqnarray}
\beta_{\cal P'}&=&73.0+2.68c_1 +3.14c_2\label{deriv},\\
 c_0&=&151-0.586c_1-0.343c_2-0.746\beta_{\cal P'}\label{intercept}
\end{eqnarray}
by matching the values of the cross section derivative and the cross section, respectively. Unless stated otherwise, both analyticity constraints were used in their $\chi^2$ fitting procedure.  

They then fit  \eq{sigma} to the high energy $\sigma_{\gamma p}$ data in the energy range $4\le\sqrt s\le 210$~GeV.  The lower energy data are from the Particle Data Group\cite{pdg}; the high energy points at $\sqrt s=200$ and $\sqrt s=209$~GeV are from the H1 collaboration\cite{h1} and Zeus\cite{zeus} collaboration, respectively. Table~\ref{table:fits} is a summary of their results.  In Fit~1,  the data have been fitted with the $\ln^2 s$ energy dependence of \eq{sigma}, imposing constraints \eq{deriv} and \eq{intercept}; the fitted values for $c_1$ and $c_2$ then determined $c_0$ and $\beta_{\cal P'}$. 

Their fit is excellent, yielding a total $\chi^2$ of 50.34 for 61 degrees of freedom, with a goodness-of-fit probability of 0.83. It is shown as the solid line in Fig.\,\ref{fig:siggplogsq&log}. 

In order to verify that the data discriminate between a $\ln^2s$ fit and a $\ln s$ fit, they made Fit~3, which  assumes a $\ln s$ asymptotic energy dependence,  i.e.,\  $c_2=0$ in \eq{sigma}. After fitting $c_1$, they determined $c_0$ and $\beta_{\cal P'}$ from the constraint equations. The $\ln s$  fit is poor, with a total $\chi^2$ of 102.8 for 62 degrees of freedom, corresponding to a chance probability of $8.76\times10^{-4}$. It is plotted as the dotted line in Fig.\,\ref{fig:siggplogsq&log} and clearly underestimates the high energy cross measurements. 

To test the stability of the $\ln^2 s$ fit, they relaxed the condition that the slopes of the low energy fit and the asymptotic fit are the same at $\sqrt s_0=2.01$ GeV, only imposing the cross section constraint of \eq{intercept}.  Thus, in Fit~2, they fit $c_1,\ c_2$, and $\beta_{\cal P'}$, which then determined $c_0$. This also yielded a good fit, with a total $\chi^2$ of 47.48, for 60 degrees of freedom, corresponding to a chance probability of 0.88. Fit~2 is shown as the dashed-dotted line in Fig.\,\ref{fig:siggplogsq&log}. It fits the data well, indicating stability of the procedure. Clearly, the constraints imposed by the low energy data strongly restrict the asymptotic behavior. 

In conclusion, they  demonstrated that the requirement that high energy cross sections smoothly interpolate into the resonance region, being constrained by the cross section and its derivative at $\sqrt s_0=2.01$ GeV, strongly favors a $\ln^2s$ behavior of the asymptotic cross section---a behavior that saturates the Froissart bound.  Using vector meson dominance (in the spirit of the quark model), they  showed that the shape of  their calculated $\sigma_{\gamma p}(s)$ curve is compatible with the analysis by Igi and Ishida\cite{igiandishidapip} of $\pi^+p$ and $\pi^-p$ data.

%Table #9
\begin{table}[h,t]                   % Use "table" environment, but also
				 % use  "tabular" environment below.
%
\def\arraystretch{1.2}            % Make the space between rows in the Table,
				  % 1.5 x bigger than the default spacing.
\caption[The fitted results for $\gamma p$ scattering]
{\protect\small The fitted results for $\gamma p$ scattering. Fits 1 and 2 have $\sigma\sim\ln^2 s$ and Fit 3 has $\sigma\sim\ln s$. The proton mass is $m$. Taken from Ref. \cite{bhfroissart}.\label{table:fits}}
\begin{center}
\begin{tabular}[b]{|l||c|c||c||}
     \cline{2-4}
      \multicolumn{1}{c|}{}
      &\multicolumn{2}{c||}{$\sigma \sim \ln^2s$}
      &\multicolumn{1}{c|}{$\sigma \sim \ln s$}\\
      \hline
      Parameters&Fit 1 &Fit 2&Fit 3 \\ 
	&$c_0$ and $\beta_{\cal P'}$ constrained&$c_0$ constrained&$c_0$ and 		$\beta_{\cal P'}$ constrained\\
	\hline
     $c_0$$\ $ $(\mu$b)&105.64&92.5&84.22 \\
      $c_1$$\ $  $(\mu$b)&$-4.74\pm1.17$&$-0.46\pm 2.88$&$4.76\pm0.11$  \\ 
	$c_2$$\ $  $(\mu$b)&$1.17\pm0.16$&$0.803\pm 0.273$&-----   \\
      $\beta_{\cal P'}$ $(\mu$b)&$64.0$&$78.4\pm9.1$&$85.8$   \\ 
	$\mu$ &0.5&0.5&0.5\\\hline
     	$\chi^2$&50.34&47.48&102.8\\
	d.f.&61&60&62\\
	$\chi^2/$d.f.&0.825&0.791&1.657\\
Probability&0.83&0.88&$8.76\times10^{-4}$\\
     \hline
\end{tabular}
     %\vspace{1in} \\
\end{center}    
%     \\
\end{table}
\def\arraystretch{1}  %Restore the default row spacing in the Table.

%  Begin Figures
\begin{figure}[tbp] %Fig. 26
\begin{center}
\mbox{\epsfig{file=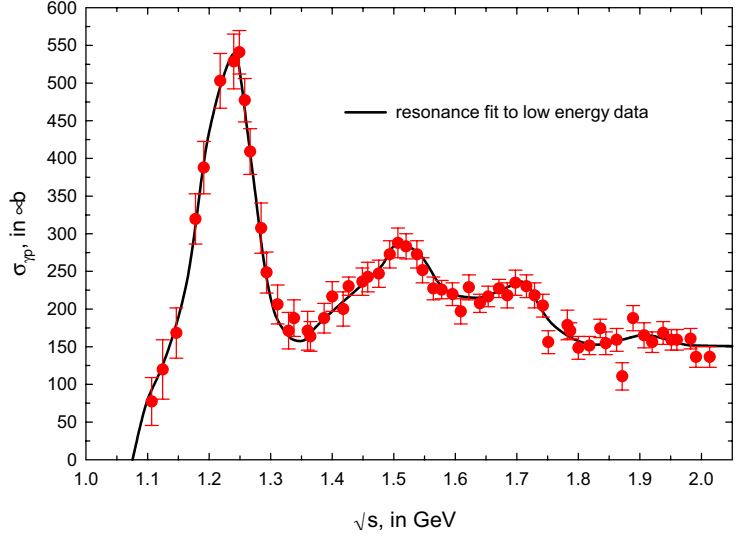
,width=4.4in%
            ,%width=6in,bbllx=122pt,bblly=403pt,bburx=494pt,bbury=660pt,clip=%
		bbllx=60pt,bblly=250pt,bburx=530pt,bbury=585pt,clip=%
}}
\end{center}
\caption[A resonance fit of low energy $\sigma_{\gamma p}$ data]
{ \footnotesize
A resonance fit of low energy $\sigma_{\gamma p}$ data. The heavy line is a fit, by Damashek and Gilman\cite{gilman}, of the low energy $\sigma_{\gamma p}$ data to a sum of five Breit-Wigner resonances plus a sixth-order polynomial background. The fitted value of $\sigma_{\gamma p}$ at $\sqrt s=2.01$ GeV is 151 $\mu$b.  Taken from Ref. \cite{bhfroissart}.
}
\label{fig:siggpresonances}
\end{figure}

\begin{figure}[tbp] %Fig. 27
\begin{center}
\mbox{\epsfig{file=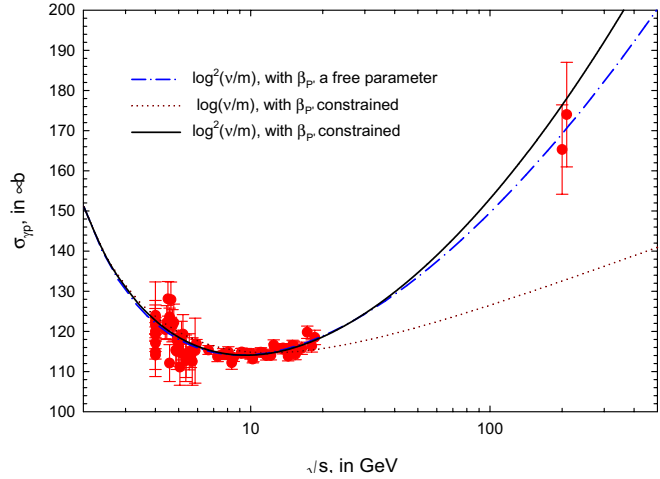,width=3.8in%
            ,%width=6in,bbllx=122pt,bblly=403pt,bburx=494pt,bbury=660pt,clip=%
%		bbllx=60pt,bblly=170pt,bburx=550pt,bbury=585pt,clip=%
bbllx=108pt,bblly=196pt,bburx=565pt,bbury=497pt,clip=%
}}
\end{center}
\caption[$\sigma_{\gamma p}$, from fit parameters of Table \ref{table:fits}]
{ \footnotesize
The $\gamma p$ total cross section, $\sigma_{\gamma p}$, from fit parameters of Table \ref{table:fits}. The 3 curves are  $\sigma_{\gamma p}$, in $\mu$b, vs. $\sqrt s$, in GeV, for various fits. 
The solid curve (Fit~1)  is  of the form : 
$\sigma_{\gamma p}=c_0 +c_1{\ln }(\nu/m)+c_2{\ln }^2(\nu/m)+\beta_{\cal P'}/\sqrt{\nu/m}$,
 with both $c_0$ and $\beta_{\cal P'}$ constrained by \eq{deriv} and \eq{intercept}.  The dot-dashed line is a ${\rm log}^2(\nu/m)$ fit (Fit~2) that constrains $c_0$ only, allowing $\beta_{\cal P'}$ to be a free parameter in the fit.  
The dotted line (Fit~3), uses: $\sigma_{\gamma p}=c_0 +c_1{\ln }(\nu/m)+\beta_{\cal P'}/\sqrt{\nu/m}$, with both $c_0$ and $\beta_{\cal P'}$ constrained by \eq{deriv} and \eq{intercept}.  The laboratory energy of the photon is $\nu$ and $m$ is the proton mass.  The data used in all fits are the cross sections with $\sqrt s \ge 4$ GeV. All fits pass through the low energy anchor point at $\sqrt s_0=2.01$ GeV, where $\sigma_{\gamma p}=151\mu$b. Fits 1 and 3 are further constrained to have the same slope as the low energy fit, at $\sqrt s_0=2.01$ GeV.  Details of the 3 fits are given in Table \ref{table:fits}. Taken from Ref. \cite{bhfroissart}.
}
\label{fig:siggplogsq&log}
\end{figure}
%%%%%%%%%%%%%%%%%
%
%
%
\subsubsection{Saturating the Froissart bound in $\pi^\pm p$ scattering}\label{section:froissartpip}
Two groups have now shown that the Froissart bound is saturated in $\pi^\pm p$ scattering, i.e., that the high energy scattering asymptotically goes as $\ln^2 s$ and not as $\ln s$. Both anchored their fits to low energy data.  Igi and Ishida\cite{igiandishidapip} used 2 FESR constraints on the  even amplitude; Block and Halzen\cite{bhfroissartnew} used the 4 analyticity constraints on both even and odd amplitudes---\eq{allderiv} derived in Section \ref{sec:4constraints}---by fixing both $\pi^+p$ and $\pi^-p$ cross sections and their derivatives at the c.m. transition energy $\sqrt s_0$=2.6 GeV.  For the first time,  the rich amount of accurate low energy cross section data was used by both groups to constrain high energy fits.

Igi and Ishida\cite{igiandishidapip}  fit only  the {\em even} cross section, i.e., $\sigma^0(\nu)=[\sigma_{\pi^+ p}(\nu)+\sigma_{\pi^-p}(\nu)]/2$, for laboratory energies $70\le\nu\le340$ GeV (corresponding to $11.5\le\sqrt s\le 25.3$ GeV) using the form 
 \begin{eqnarray}
\sigma^0(\nu)&=&c_0+c_1\ln\left(\frac{\nu}{m}\right)+c_2\ln^2\left(\frac{\nu}{m}\right)+\beta_{\cal P'}\left(\frac{\nu}{m}\right)^{\mu -1},\label{sig0pi}
\ea
with $\mu=0.5$. For a detailed discussion of the real analytic amplitudes they used, see \eq{sig0} in Section \ref{sec:analyticamplitudes}.
Using  12 experimental points in a $\chi^2$ fit to \eq{sig0pi}, they fit the two free parameters $c_1$ and $c_2$, i.e., they had 10 degrees of freedom (d.f.)  for their fit.  They obtained  $\chi^2$/d.f.=0.075, indicating a good fit and a  high probability  of the hypothesis that asymptotically, the total $\pi p$ cross section goes as $\ln^2s$, , i.e., the Froissart bound is saturated.  In contrast, when they tested the hypothesis of $\ln s$ by setting $c_2=0$ in \eq{sig0pi}, i.e., using 11 degrees of freedom, they found $\chi^2$/d.f.=2.6, indicating a very bad fit to this hypothesis, thus ruling out an asymptotic behavior of $\ln s$.

Block and Halzen\cite{bhfroissartnew} fit the 4 experimental quantities $\sigma_{\pi^+ p}(\nu),\  \sigma_{\pi^- p}(\nu),\  \rho_{\pi^+ p}(\nu)$ and $\rho_{\pi^- p}(\nu)$ using the high energy parametrization (see \eq{sigmapm} and  \eq{rhopm} of Section \ref{sec:analyticamplitudes})
\ba
\sigma^\pm(\nu)&=&\sigma^0(\nu)\pm\  \delta\left({\nu\over m}\right)^{\alpha -1},\label{sigmapmpi}\\
\rho^\pm(\nu)&=&{1\over\sigma^\pm(\nu)}\left\{\frac{\pi}{2}c_1+c_2\pi \ln\left(\frac{\nu}{m}\right)-\beta_{\cal P'}\cot({\pi\mu\over 2})\left(\frac{\nu}{m}\right)^{\mu -1}+\frac{4\pi}{\nu}f_+(0)\right.\nonumber\\
&&\left.\qquad\qquad\qquad\pm \delta\tan({\pi\alpha\over 2})\left({\nu\over m}\right)^{\alpha -1} \right\}\label{rhopmpi},
\ea
where the upper sign is for $\pi^+p$ and the lower sign is for  $\pi^-p$ scattering and $m$ is the pion mass, also with $\mu=0.5$.
 
In the earliest known application of the ``Sieve'' algorithm\cite{sieve}, which was described in detail in Section \ref{section:sievealgorithm},  Block and Halzen\cite{bhfroissartnew} (BH) first  formed a sieved data set, starting with all of the cross sections $\sigma_{\pi^\pm p}$ and $\rho$-values $\rho_{\pi^\pm p}$ in the Particle Data Group\cite{pdg} archive  in the laboratory energy interval $18.7\le\nu\le610$ GeV, which corresponds to $6\le\sqrt s\le 33.8$ GeV.  This is a much larger energy interval and with many more datum points (130) than were used in the fit of Igi and Ishida\cite{igiandishidapip}, since here both $\sigma_{\pi^\pm p}$ and $\rho_{\pi^\pm p}$  points were used. A fit of this type is of course also much more constrained because $\rho$-values are also {\em simultaneously} fit with the cross sections.

For analytic constraints, BH\cite{bhfroissartnew} evaluated experimental values of $\sigma^\pm$ and its derivatives at $\nu_0=2.6$ GeV, the laboratory transition energy, so that the 4 analyticity constraints they used (see \eq{allderiv} in Section \ref{sec:4constraints}) were:
\ba
\sigma^\pm(\nu_0)&=&\sigma^0(\nu_0)\pm\  \delta\left({\nu_0\over m}\right)^{\alpha -1},\label{sigmapmpi0}\\
\frac{d\sigma^{\pm}(\nu_0)}{d(\x}&=&c_1\left\{\frac{1}{(\y}\right\} +c_2\left\{ \frac{2\ln(\y}{(\y}\right\}+\beta_{\cal P'}\left\{(\mu-1)(\y^{\mu-2}\right\}\nonumber\\
&&\ \ \ \ \ \ \ \ \ \ \ \ \ \ \ \ \ \ \ \ \  \pm \ \delta\left\{(\alpha -1)(\y^{\alpha - 2}\right\},\label{derivpmpi0}
\end{eqnarray} 
where the left-hand sides were the experimental values of the cross sections and their derivatives at $\nu_0$. In order to fix these cross sections and their derivatives, they made local fits to the experimental cross sections around the transition energy $\nu_0$.  

At the transition energy $\nu_0$ they introduced the quantities 
\begin{eqnarray}
\sigma_{\rm av}&=&\frac{\sigma^{+}\y+\sigma^-\y}{2}\nonumber\\
&=&c_0+c_1\ln\y+c_2\ln^2\y+\beta_{\cal P'}\y^{\mu-1},\\
\Delta\sigma&=&\frac{\sigma^{+}\y-\sigma^-\y}{2}\nonumber\\
&=&\delta\y^{\alpha -1},\\
m_{\rm av}&=&\frac{1}{2}\left(\frac{d\sigma^{+}}{d\x}+\frac{d\sigma^{-}}{d\x}\right)_{\nu =\nu_0}\nonumber\\
&=&c_1\left\{\frac{1}{\y}\right\}+c_2\left\{ \frac{2\ln\y}{\y}\right\}+\beta_{\cal P'}\left\{(\mu-1)\y^{\mu-2}\right\},\\
\Delta m&=&\frac{1}{2}\left(\frac{d\sigma^{+}}{d\x}-\frac{d\sigma^{-}}{d\x}\right)_{\nu =\nu_0}\nonumber\\
&=&\delta\left\{(\alpha -1)\y^{\alpha - 2}\right\}.
\end{eqnarray}
From these definitions, they  found the  four constraint conditions corresponding to the analyticity constraints summarized in \eq{allderiv} of Section \ref{sec:4constraints}: 
\begin{eqnarray}
\beta_{\cal P'}&=&\frac{\y^{2-\mu}}{\mu -1}\left[m_{\rm av}-c_1\left\{\frac{1}{\y}\right\} -c_2\left\{\frac{2\ln\y}{\y}
\right\}\right],\label{deriveven}\\
c_0&=& \sigma_{\rm av}-c_1\ln\y-c_2\ln^2\y-\beta_{\cal P'}\y^{\mu-1},\label{intercepteven}\\
\alpha&=&1+\frac{\Delta m}{\Delta \sigma}\y,\label{derivodd}\\
\delta&=&\Delta \sigma\y^{1-\alpha}\label{interceptodd},
\end{eqnarray}
 utilizing the two experimental slopes $d\sigma^{\pm}/d\x$  and the two experimental  cross sections $\sigma^{\pm}\x$ at the transition energy $\nu_0$, where they join on to the asymptotic fit. They used $\nu_0=3.12$ GeV, corresponding to $\sqrt s_0=2.6 $ GeV,   as the (very low) energy just after which resonance behavior finishes.  Thus, $\nu_0$  is much below the energy at which they  start their high energy fit, which is at $\nu=18.7$ GeV; however,  $\nu_0$ is at an energy safely above the resonance regions.
%%%%%%%%%%%%%%%%%%%%%%%%%%%%%%%%%%
%%%%%%%%%%%%%%%%%%%%%%%%%%%%%%%%%%%%

The authors\cite{bhfroissartnew} stress that the odd amplitude parameters $\alpha$ and $\delta$---hence the entire odd amplitude---are {\em completely determined} by the experimental values $\Delta m$ and $\Delta \sigma$ at the transition energy $\nu_0$. Thus, at {\em all} energies, the {\em differences} of the cross sections $\Delta\sigma=\sigma^- -\sigma^+$  and the {\em differences} of the real portion of the scattering amplitude are completely fixed {\em before} the fit is made.  

For a $\ln^2s$ $(\ln s$) fit, the even amplitude parameters $c_0$ and $\beta_{\cal P}'$ are determined by $c_1$ and $c_2$ ($c_1$ only) along with the experimental values of $\sigma_{\rm av}$ and $m_{\rm av}$ at the transition energy $\nu_0$. 

Thus, for a $\ln^2s$ ($\ln s$) fit,  only 3 (2) parameters $c_1$, $c_2$, and $f_+(0)$ ($c_1$ and $f_+(0))$ are fit, since the subtraction constant $f_+(0)$  enters only into the $\rho$-value.  Therefore,  only the 2  parameters  $c_1$ and $c_2$  of the original 7 are required for a $\ln^2s$  fit to the cross sections $\sigma^{\pm}$, giving the phenomenologist exceedingly little freedom in making this fit; it is very tightly constrained, with little latitude for adjustment. 

The results of the fits made are summarized in Table \ref{table:pipfitnew}. The authors selected the $\ln^2s$ fit  that corresponded to the  Sieve algorithm cut $\delchimax=6$. Also shown are the results for the cut $\delchimax=9$.  Inspection of Table \ref{table:pipfitnew} shows that the actual parameters of the fit were completely stable, an important property of the Sieve algorithm. Further,  the renormalized $\chi^2$ per degree of freedom, ${\cal R}\times\chi^2_{\rm min}$/d.f., had gone down from 1.555, an unacceptably high value for 135 degrees of freedom to 1.294, a satisfactory value. The uncertainties on the fitted parameters are very small because only 3 parameters are fit, due to the use of 4 analyticity constraints.  

The fourth column of Table \ref{table:pipfitnew} shows the results of a $\ln s$ fit (setting $c_2=0$ in Eqs. (\ref{sig0pi})--(\ref{rhopmpi})) to the {\em same} data set, i.e., the sieved set with the cut $\delchimax=6$  used for the successful fit of the $\ln^2s$. The renormalized $\chi^2$ per degree of freedom, ${\cal R}\times\chi^2_{\rm min}$/d.f.=8.163, gives negligible probability for the goodness-of-fit, completely ruling out an asymptotic cross section behaving as $\ln s$, in complete agreement with Igi and Ishida\cite{igiandishidapip}.

In their $\ln^2s$ fit, Igi and Ishida\cite{igiandishidapip} used two even constraints, the FESRs, and fit 12 datum points  (even cross sections) with two parameters, for a total of 10 degrees of freedom. In contrast, in the Block and Halzen $\ln^2s$ fit\cite{bhfroissart}, they used 2 even and 2 odd constraints, fitting a total of  130 points with 3 parameters, for a total of 127 degrees of freedom. 

In Fig. \ref{fig:sigpip}, we show $\sigma_{\pi^{+} p}$ and $\sigma_{\pi^{-} p}$ as a function of  the c.m. energy $\sqrt s$, for both the $\ln^2s$ fit of Table \ref{table:pipfitnew} and the $\ln s$ fit. For the $\ln^2 s$ fit, the solid curve is $\sigma_{\pi^{-} p}$ and the dash-dotted curve is $\sigma_{\pi^{+} p}$. For the $\ln s$ fit, the long dashed curve is $\sigma_{\pi^{-} p}$ and the short dashed curve is $\sigma_{\pi^{+} p}$.

In Fig \ref{fig:rhopip}, we show  $\rho_{\pi^{+} p}$ and $\rho_{\pi^{-} p}$ as a function of  the c.m. energy $\sqrt s$, for both the $\ln^2s$ fit of Table \ref{table:pipfitnew} and the $\ln s$ fit. For the $\ln^2 s$ fit, the solid curve is $\rho_{\pi^{-} p}$ and the dash-dotted curve is $\rho_{\pi^{+} p}$. For the $\ln s$ fit, the long dashed curve is $\rho_{\pi^{-} p}$ and the short dashed curve is $\rho_{\pi^{+} p}$.

It is clear from examination of Figures \ref{fig:sigpip} and \ref{fig:rhopip} that the curves for $\ln^2s$ are a good representation of the data, whereas the $\ln s$ fit is completely ruled out, going completely below  the experimental data for {\em both} $\sigma$ and $\rho$ at high energies.

We now examine the effect of the Sieve algorithm in cleaning up the data sample by eliminating the outliers. Using a $\ln^2 s$ fit {\em before} imposing the Sieve algorithm,  a value of $\chi^2$/d.f.=3.472 for 152 degrees of freedom was found. When using the $\delchimax=6$ cut,   $\chi^2$/d.f.=1.294 for 127 degrees of freedom was found. The use of the Sieve algorithm eliminated 25 points with energies $\sqrt s\ge6$ GeV (2 $\sigma_{\pi^+p}$, 19 $\sigma_{\pi^-p}$, 4 $\rho_{\pi^+p}$), while changing the total renormalized $\chi^2$ from 527.8 to 164.3. These 25 points that were `sieved'  out had a $\chi^2$ contribution of 363.5, an average value of 14.5. For a Gaussian distribution with no outliers, one would have expected about 2 points having $\delchi>6$, giving a total $\chi^2$ contribution slightly larger than 12, compared to the observed value of 363.5. Clearly, the Sieve algorithm functioned very well on these 4 different data sets, $\sigma_{\pi^{+} p}$, $\sigma_{\pi^{-} p}$, $\rho_{\pi^{+} p}$ and $\rho_{\pi^{-} p}$,  from the PDG\cite{pdg} archives.

Figure \ref{fig:sigpipallenergies} is a plot of {\em all} known experimental $\pi^\pm p$ total cross sections, taken directly from the Particle Data Group\cite{pdg} compendium, as a function of $\sqrt s$, the c.m. energy. The circles are $\sigma_{\pi^-p}$ and the squares are $\sigma_{\pi^+p}$. The dashed curve is the $\ln^2s$ fit taken from Table \ref{table:pipfitnew}, with $\delchimax=6$. The solid  curve is  $210\times \sigma_{\gamma p}$, from a fit of $\gamma p$ cross sections by Block and Halzen\cite{bhfroissart}  of the form: $\sigma_{\gamma p}=c_0 +c_1{\ln }(\nu/m_p)+c_2{\ln }^2(\nu/m_p)+\beta_{\cal P'}/\sqrt{\nu/m_p}$, where $m_p$ is the proton mass.  The $\gamma p$ cross sections were fit for  c.m. energies $\sqrt s\ge2.01$ GeV, whereas the $\pi p$ data (cross sections and $\rho$-values) were fit for c.m. energies $\sqrt s\ge6$ GeV. The two fitted curves are almost numerically identical over the entire energy region shown, $2\le \sqrt s\le 300$ GeV, again giving experimental support for the vector meson dominance model.

In conclusion, it is clear that both the Igi and Ishida\cite{igiandishidapip} and the Block and Halzen\cite{bhfroissart} analyses agree that 
\begin{itemize} 
\item a $\ln s$ behavior for $\pi^\pm p$ cross sections is ruled out.
 \item  the  $\pi^\pm p$ scattering cross sections asymptotically grows as $\ln^2 s$.
\end{itemize}
%%%%%%%%%%%%%%%%%%%%%%%%%%%%%%%%%%%%%%%
%  Table 10 Pi P
\begin{table}[thp]                   % Use "table" environment, but also
				 % use  "tabular" environment below.
%
\def\arraystretch{1.05}            % Make the space between rows in the Table,
				  % 1.5 x bigger than the default spacing.
\begin{center}
     \caption[The fitted results,  $\sigma_{\pi^\pm p}$ and $\rho_{\pi^\pm p}$, for a 3-parameter fit with $\sigma\sim\ln^2 s$ and a 2-parameter fit with $\sigma\sim\ln s$, using 4 analyticity constraints]
{\protect\small The fitted results,  $\sigma_{\pi^\pm p}$ and $\rho_{\pi^\pm p}$, for a 3-parameter fit with $\sigma\sim\ln^2 s$ and a 2-parameter fit with $\sigma\sim\ln s$, using 4 analyticity constraints. The renormalized $\chi^2_{\rm min}$/d.f.,  taking into account the effects of the $\delchimax$ cut, is given in the row  labeled ${\cal R}\times\chi^2_{\rm min}/$d.f. The errors in the fitted parameters have been multiplied by the appropriate $r_{\chi2}$. For a discussion of the Sieve algorithm used in this fit, see Section \ref{section:sievealgorithm}; for details of the renormalization factors ${\cal R}$ and $r_{\chi2}$, see Figures \ref{renorm}a) and \ref{renorm}b) in Section \ref{section:lessons}. Taken from Ref. \cite{bhfroissartnew}. \label{table:pipfitnew}}
\vspace{.1in}
\begin{tabular}[b]{|l||c|c||c||}
    %\cline{2-4}
%\hline
\cline{2-4}
\multicolumn{1}{c|}{}&\multicolumn{2}{c||}{$\sigma\sim \ln^2 s$}&\multicolumn{1}{c||}{$\sigma\sim \ln s$}\\
\cline{1-1}
\multicolumn{1}{|c||}{Parameters }
      &\multicolumn{2}{|c||}{$\delchimax$}&\multicolumn{1}{|c||}{$\delchimax$}\\ 
%\hline 
\cline{2-4}
	\multicolumn{1}{|c||}{}
      &\multicolumn{1}{c|}{6}&\multicolumn{1}{c||}{9} &\multicolumn{1}{c||}{6}\\
      \hline
	\multicolumn{4}{|c||}{\ \ \ \ \  Even Amplitude}\\
	\cline{1-4}
      $c_0$\ \ \   (mb)&$20.11$ &$20.32$&12.75\\ 
      $c_1$\ \ \   (mb)&$-0.921\pm0.110$ &$-0.981\pm0.100$&$1.286\pm 0.0056$\\ 
	$c_2$\ \ \ \   (mb)&$0.1767\pm0.0085$&$0.1815\pm0.0077$&------\\
      $\beta_{\cal P'}$\ \   (mb)&$54.40$ &$54.10$&64.87\\ 
      $\mu$&$0.5$ &$0.5$&0.5\\ 
	$f_+(0)$ (mb GeV)&$-2.33\pm0.36$&$-2.31\pm 0.35$&$0.34\pm 0.36$\\
      \hline
	\multicolumn{4}{|c||}{\ \ \ \ \  Odd Amplitude}\\
	\hline
      $\delta$\ \ \   (mb)&$-4.51$ &$-4.51$&-4.51\\
      $\alpha$&$0.660$ &$0.660$&0.660\\ 
	\cline{1-4}
     	\hline
	\hline
	$\chi^2_{\rm min}$&148.1&204.4&941.8\\
	${\cal R}\times\chi^2_{\rm min}$&164.3&210.0&1044.9\\ 
	degrees of freedom (d.f.)&127&135&128\\
\hline
	${\cal R}\times\chi^2_{\rm min}$/d.f.&1.294&1.555&8.163\\
\hline
\end{tabular}
\end{center}
     %\vspace{1in} \\
\end{table}%
\def\arraystretch{1}  %Restore the default row spacing in the Table.
%%%%%%%%%%%%%%%%%%%%%%%%%%%%%%%%%%

%  
\begin{figure}[tbp] %Fig. 28
\begin{center}
\mbox{\epsfig{file=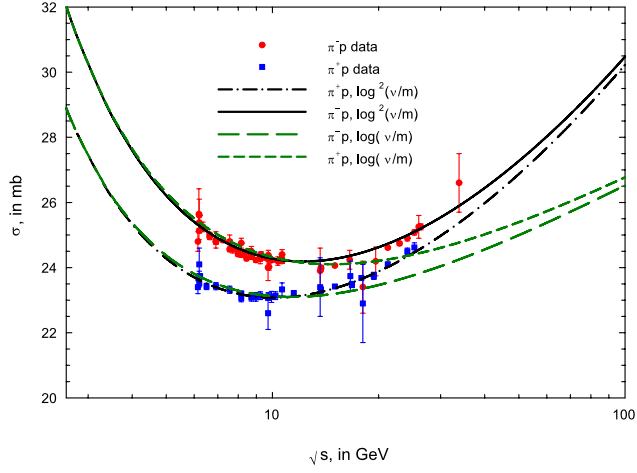
,width=3.4in,%,
bbllx=0pt,bblly=0pt,bburx=420pt,bbury=308pt,clip=%
}}
\end{center}
\caption[$\sigma_{\pi^\pm p}$,  using the fit parameters of Table \ref{table:pipfitnew}]
{ \footnotesize
Total cross sections $\sigma_{\pi^\pm p}$,  using the fit parameters of Table \ref{table:pipfitnew}. 
The 4 curves are the fitted  $\sigma_{\pi^+p}$ and $\sigma_{\pi^-p}$ in mb,  vs. $\sqrt s$, in GeV, using the 4 analyticity constraints of Equations (\ref{deriveven}), (\ref{intercepteven}), (\ref{derivodd}) and (\ref{interceptodd}).  The circles are the sieved  data  for $\pi^-p$ scattering and the squares are the sieved data for $\pi^+p$ scattering for $\sqrt s\ge 6$ GeV. The dash-dotted curve ($\pi^+ p$)  and the solid curve ($\pi^- p$) are $\chi^2$ fits (Table \ref{table:pipfitnew}, $\sigma\sim\ln^2 s$, $\delchimax=6$) of  the high energy data  of the form~: $\sigma_{\pi^{\pm}p}=c_0 +c_1{\ln }\left({\nu\over m}\right)+c_2{\ln }^2\left({\nu\over m}\right)+\beta_{\cal P'}\left({\nu\over m}\right)^{\mu -1}\pm \delta\left({\nu\over m}\right)^{\alpha -1}$. The upper (lower) sign is for $\pi^+p$($\pi^-p$) scattering.  The short dashed curve ($\pi^+ p$) and the long dashed curve ($\pi^- p$) are $\chi^2$ fits (Table  \ref{table:pipfitnew}, $\sigma\sim\ln s$, $\delchimax=6$ ) of  the high energy data  of the form~: $\sigma_{\pi^{\pm}p}=c_0 +c_1{\ln }\left({\nu\over m}\right)+\beta_{\cal P'}\left({\nu\over m}\right)^{\mu -1}\pm\delta\left({\nu\over m}\right)^{\alpha -1}$. The upper sign is for $\pi^+p$ and the lower sign is for $\pi^-p$ scattering. The laboratory energy of the pion is  $\nu$ and $m$ is the pion mass.  Taken from Ref. \cite{bhfroissartnew}.
}
\label{fig:sigpip}
\end{figure}
%%%%%%%%%%%%%%%%%%%%%%%%%%%%%%%%%%%%%%%%
%%%%%%%%%%%%%%%%%%%%%%%%%%%%%%%%%%%%%%%%
\begin{figure}[tbp] %Fig. 29
\begin{center}
\mbox{\epsfig{file=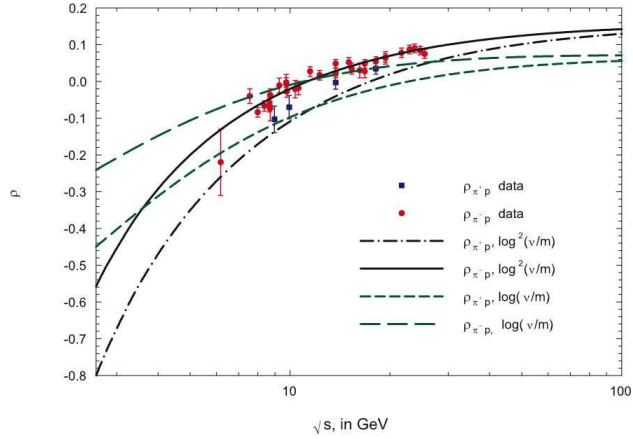,width=3.5in%
,bbllx=0pt,bblly=0pt,bburx=460pt,bbury=308pt,clip=%
}}
\end{center}
\caption[$\rho_{\pi^\pm p}$,  using new analyticity constraints  and the fit parameters of Table \ref{table:pipfitnew}]
{ \footnotesize
 $\rho_{\pi^\pm p}$, using new analyticity constraints  and the fit parameters of Table \ref{table:pipfitnew}. The 4 curves are the  fitted  $\rho_{\pi^+p}$ and $\rho_{\pi^-p}$, vs. $\sqrt s$, in GeV, using the 4 analyticity constraints of Equations (\ref{deriveven}), (\ref{intercepteven}), (\ref{derivodd}) and (\ref{interceptodd}).  The circles are the sieved data  for $\pi^-p$ and the squares for $\pi^+p$ scattering for $\sqrt s\ge 6$ GeV. The  dash-dotted curve ($\pi^+ p$) and the solid curve ($\pi^- p$) are  fits (Table \ref{table:pipfitnew}, $\sigma\sim{\rm ln}^2 s$, $\delchimax=6$) of  the high energy data  of the form~: $\rho^\pm= {1\over\sigma^\pm} \left\{\frac{\pi}{2}c_1+ c_2\pi\ln\left(\frac{\nu}{m}\right)-\beta_{\cal P'}\cot(\pi\mu/2)\left(\frac{\nu}{m}\right)^{\mu -1}+\frac{4\pi}{\nu}f_+(0)\ \pm \ \delta\tan(\pi\alpha/ 2)\left({\nu\over m}\right)^{\alpha -1}\right\}$. The upper (lower) sign is for $\pi^+p$($\pi^-p$) scattering.   The short dashed curve ($\pi^+ p$) and the long dashed curve ($\pi^- p$)  are fits (Table  \ref{table:pipfitnew}, $\sigma\sim\ln(\nu/m_\pi)$, $\delchimax=6$ )  of the form~: $\rho^\pm= {1\over\sigma^\pm} \left\{\frac{\pi}{2}c_1-\beta_{\cal P'}\cot(\pi\mu/2)\left(\frac{\nu}{m}\right)^{\mu -1}+\frac{4\pi}{\nu}f_+(0)\ \pm \ \delta\tan(\pi\alpha/ 2)\left({\nu\over m}\right)^{\alpha -1}\right\}$. The upper sign is for $\pi^+p$ and the lower  is for $\pi^-p$.  The laboratory energy of the pion is  $\nu$ and $m$ is the pion mass.  Taken from Ref. \cite{bhfroissartnew}.
  }
\label{fig:rhopip}
\end{figure}
%%%%%%%%%%%%%%%%%%%%%%%%%%
%%%%%%%%%%%%%%%%%
\begin{figure}[tbp] %Fig. 30
\begin{center}
\mbox{\epsfig{file=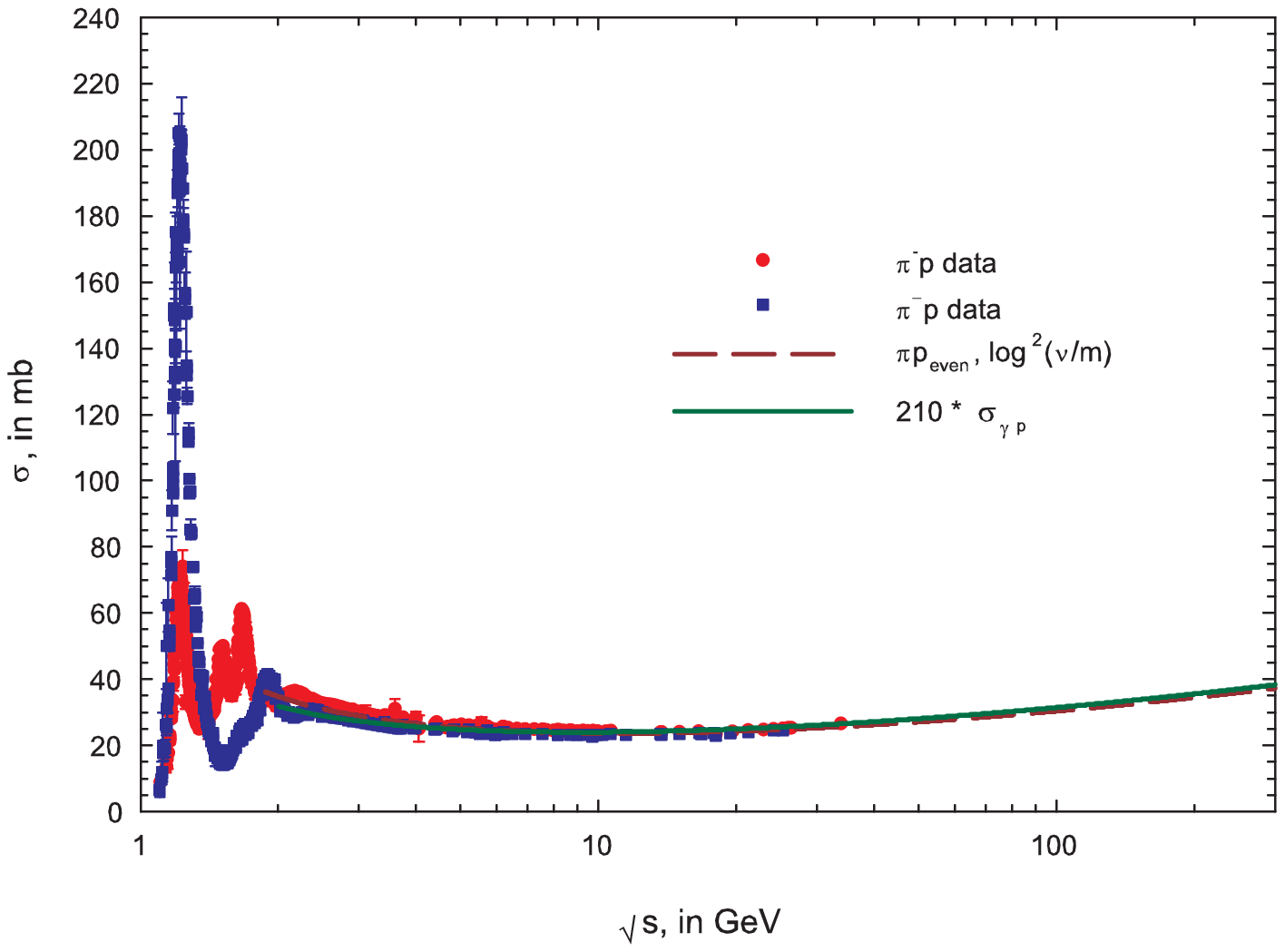,width=3in%
,bbllx=0pt,bblly=0pt,bburx=430pt,bbury=340pt,clip=%
}}
\end{center}
\caption[All known $\sigma_{\pi^\pm p}$ as a function of energy]
{ \footnotesize
All known $\sigma_{\pi^\pm p}$ as a function of energy. The circles are the cross section data  for $\pi^-p$ and the squares are the cross section data for $\pi^+p$, in mb, vs. $\sqrt s$, in GeV. The dashed curve is the  fit (Table \ref{table:pipfitnew}, $\sigma\sim\ln^2(\nu/m_\pi)$, $\delchimax=6$)  of the even amplitude cross section, of the form~: $\sigma_{\pi p^{\rm even}}=c_0 +c_1{\ln }\left({\nu\over m}\right)+c_2{\ln }^2\left({\nu\over m}\right)+\beta_{\cal P'}\left({\nu\over m}\right)^{\mu -1}$, with  $c_0$ and $\beta_{\cal P'}$ constrained by \eq{deriveven} and \eq{intercepteven}.  
The laboratory energy of the pion is  $\nu$ and $m$ is the pion mass. The dashed curve is  $210\times \sigma_{\gamma p}$, from a fit of $\gamma p$ cross sections by Block and Halzen\cite{bhfroissart}  of the form: $\sigma_{\gamma p}=c_0 +c_1{\ln }(\nu/m_p)+c_2{\ln }^2(\nu/m_p)+\beta_{\cal P'}/\sqrt{\nu/m_p}$, where $m_p$ is the proton mass.  The $\gamma p$ cross sections were fit for  c.m. energies $\sqrt s\ge2.01$ GeV, whereas the $\pi p$ data (cross sections and $\rho$-values) were fit for c.m. energies $\sqrt s\ge6$ GeV. The two fitted curves are virtually indistinguishable in the energy region $2\le \sqrt s\le 300$ GeV. Taken from Ref. \cite{bhfroissartnew}.
}
\label{fig:sigpipallenergies}
\end{figure}
%
%%%%%%%%%%%%%%%%% 

\subsubsection{Saturating the Froissart bound in  $pp$ and $\bar pp$ scattering}\label{sec:lnsqpp}
Two groups have now shown that the Froissart bound is saturated in $pp$ and $\bar pp$ scattering, i.e., that the high energy scattering asymptotically goes as $\ln^2 s$ and not as $\ln s$. Both groups anchored their fits to low energy experimental cross section data. Igi and Ishida\cite{{igiandishidapp},{iginew}} used a single FESR constraint on the  even amplitude; Block and Halzen\cite{bhfroissartnew} again used 4 analyticity constraints on both even and odd amplitudes---\eq{allderiv} derived in Section \ref{sec:4constraints}---fixing $pp$ and $\bar pp$ cross sections and their derivatives at the c.m. transition energy $\sqrt s_0$=4.0 GeV,  again exploiting the accurate low energy cross section data. 

Igi and Ishida\cite{{igiandishidapp},{iginew}}, using one FESR as a constraint,   fit {\em only} the  even cross section, $\sigma^0(\nu)=[\sigma_{\pi+ p}(\nu)+\sigma_{\pi^-p}(\nu)]/2$, for laboratory energies $70\le\nu\la 2\times 10^6$ GeV (corresponding to $11.5\le\sqrt s\la 2000$ GeV). This time,however, they also fit the {\em even} $\rho$-value,   simultaneously fitting the forms (see \eq{sig0} and \eq{rho0} in Section \ref{sec:analyticamplitudes}): 
 \ba
\sigma^0(\nu)&=&c_0+c_1\ln\left(\frac{\nu}{m}\right)+c_2\ln^2\left(\frac{\nu}{m}\right)+\beta_{\cal P'}\left(\frac{\nu}{m}\right)^{\mu -1},\label{sig0pp}\\
\rho^0(\nu)&=&{1\over\sigma^0}\left\{\frac{\pi}{2}c_1+c_2\pi \ln\left(\frac{\nu}{m}\right)-\beta_{\cal P'}\cot({\pi\mu\over 2})\left(\frac{\nu}{m}\right)^{\mu -1}+\frac{4\pi}{\nu}f_+(0)\right\},\label{rho0pp}
\ea
 with $\mu=0.5$. Here $m$ is the proton mass.  The {\em even} FESR constraint that they used is \eq{final2},   derived in Section \ref{section:igi} and   reproduced below:
\be
c_0+2.04 c_1+4.26c_2+0.367\beta_{\cal P'}=49.3 {\rm \ mb}.\label{final2pp}
\ee
In ref. \cite{igiandishidapp}, they did not use the subtraction constant $f_+(0)$, but they did include it their fit in ref. \cite{iginew}. They  used 27 experimental points in a $\chi^2$ fit to \eq{sig0pp} and \eq{rho0pp}, the {even} cross sections and $\rho$-values,  fitting the four free parameters $c_0,c_1,c_2$ and $f_+(0)$, i.e., they had 23 degrees of freedom (d.f.)  for their fit.  They obtained  $\chi^2$/d.f.$\approx0.5$, indicating a good fit, with a correspondingly  high probability  of the hypothesis that asymptotically, the nucleon-nucleon total  cross section goes as $\ln^2s$, , i.e., the Froissart bound is saturated.  Unlike  their $\pi^\pm p$ analysis\cite{igiandishidapip},  they did not make a fit for the $\ln s$ hypothesis in order to rule it out for nucleon-nucleon scattering.

Block and Halzen\cite{bhfroissartnew} again fit 4 experimental quantities, $\sigma_{\bar pp}(\nu), \sigma_{p p}(\nu), \rho_{\bar pp}(\nu)$ and $\rho_{p p}(\nu)$, using the high energy parametrization (see \eq{sigmapm} and  \eq{rhopm} of Section \ref{sec:analyticamplitudes})
\ba
\sigma^\pm(\nu)&=&\sigma^0(\nu)\pm\  \delta\left({\nu\over m}\right)^{\alpha -1},\label{sigmapmpp}\\
\rho^\pm(\nu)&=&{1\over\sigma^\pm(\nu)}\left\{\frac{\pi}{2}c_1+c_2\pi \ln\left(\frac{\nu}{m}\right)-\beta_{\cal P'}\cot({\pi\mu\over 2})\left(\frac{\nu}{m}\right)^{\mu -1}+\frac{4\pi}{\nu}f_+(0)\right.\nonumber\\
&&\left.\qquad\qquad\qquad\pm \delta\tan({\pi\alpha\over 2})\left({\nu\over m}\right)^{\alpha -1} \right\}\label{rhopmpp},
\ea
where the upper sign is for $pp$ and the lower sign is for  $\bar pp$ scattering, with $\mu=0.5$ and $m$ the proton mass.

Again the Sieve algorithm\cite{sieve}, described in detail in Section \ref{section:sievealgorithm}, was a critical factor in the nucleon-nucleon analysis.   Block and Halzen\cite{bhfroissartnew}  formed a sieved data set using all of the cross sections, $\sigma_{pp}$ and $\sigma_{\bar pp}$, along with all of the $\rho$-values,  $\rho_{\bar p p}$ and $\rho_{ p p}$,  in the Particle Data Group\cite{pdg} archive that were in the laboratory energy interval $18.3\le\nu\le 1.73\times 10^6$ GeV, i.e., $6\le\sqrt s\le 1800$ GeV.  Just as in their $\pi^\pm p$ analysis, this is a much larger energy interval and with many more datum points (a total of 187 data points were retained in the sieved set) than were used in the fit of Igi and Ishida\cite{{igiandishidapp},{iginew}} (who used 27 data points), since here both $\sigma_{\bar pp }$ and $\sigma_{pp}$ , along with $\rho_{\bar p p}$ and $\rho_{ p p}$  points were used---both even and odd amplitudes were used, not just the even amplitudes used by Igi and Ishida.

As in their $\pi^\pm p$ analysis, Block and Halzen\cite{bhfroissartnew}  used the  four analyticity constraint conditions corresponding to the analyticity constraints summarized in \eq{allderiv} of Section \ref{sec:4constraints}: 
\begin{eqnarray}
\beta_{\cal P'}&=&\frac{\y^{2-\mu}}{\mu -1}\left[m_{\rm av}-c_1\left\{\frac{1}{\y}\right\} -c_2\left\{\frac{2\ln\y}{\y}
\right\}\right],\label{derivevenpp}\\
c_0&=& \sigma_{\rm av}-c_1\ln\y-c_2\ln^2\y-\beta_{\cal P'}\y^{\mu-1},\label{interceptevenpp}\\
\alpha&=&1+\frac{\Delta m}{\Delta \sigma}\y,\label{derivoddpp}\\
\delta&=&\Delta \sigma\y^{1-\alpha}\label{interceptoddpp},
\end{eqnarray}
 utilizing the two experimental slopes $d\sigma^{\pm}/d\x$  and the two experimental  cross sections $\sigma^{\pm}\x$ at the transition energy $\nu_0$, where they join on to the asymptotic fit. They used $\nu_0=7.59$ GeV, corresponding to $\sqrt s_0=4 $ GeV,   as the (very low) energy just after which resonance behavior finishes.  Thus, $\nu_0$  is much below the minimum energy ($\nu_{\rm min} =18.3$ GeV) at which they  start their high energy fit, but is safely above the resonance regions.

Table \ref{table:ppfitnew} summarizes the results of  simultaneous fits to the available accelerator data  from the Particle Data Group\cite{pdg} for  $\sigma_{pp}$, $\sigma_{\bar pp}$, $\rho_{pp}$ and $\rho_{\bar pp}$, using the 4 constraint equations with a transition energy $\sqrt s_0=4$ GeV and a minimum fitting energy of 6 GeV, again using the Sieve algorithm. Two $\delchimax$ cuts,  6 and 9, were made for $\ln^2 s$ fits. The probability of the fit for the cut $\delchimax=6$ was $\sim 0.2$, a most satisfactory probability for this  many degrees of freedom. Block and Halzen chose this data set rather than the data set corresponding to the $\delchimax=9$ cut. As seen in Table \ref{table:ppfitnew}, the fit parameters are very insensitive to this choice.
 The same data set ($\delchimax=6$ cut) was also used for the $\ln s$ fit. The probability of the $\ln s$ fit is $<<10^{-16}$ and is clearly ruled out. This is illustrated most graphically in \fig{fig:sigmapp} and \fig{fig:rhopp}, where the fitted values are always well below the high energy experimental points. 

Again, the Sieve algorithm worked exceedingly well. When using a $\ln^2 s$ fit {\em before} imposing the algorithm,  a value of $\chi^2$/d.f.=5.657 for 209 degrees of freedom was found. This is to be  contrasted to the sieved set's value of $\chi^2$/d.f.=1.095, for  184 degrees of freedom,  when using the $\delchimax=6$ cut. After sifting the data, 25 points with energies $\sqrt s\ge6$ GeV (5 $\sigma_{pp}$, 5 $\sigma_{\pbar p}$, 15 $\rho_{pp}$) were eliminated, while the total renormalized $\chi^2$ changed from 1182.3 to 201.4. Those 25 points that were screened out had a $\chi^2$ contribution of $\sim981$, an average value of $\sim 39$. For a Gaussian distribution, about 3 points with $\delchi>6$ are expected, having a total $\chi^2$ contribution of slightly more than 18 and {\em not} 981. We once more see how the Sieve algorithm had rid the  data sample of outliers.

Using the fit parameters from Table \ref{table:ppfitnew}, Figure \ref{fig:sigmapp} shows the individual fitted cross sections (in mb) for $ pp$ and $\bar pp$ for both $\ln^2 s$ and $\ln s$ for the cut $\delchimax=6$,  plotted against the c.m. energy $\sqrt s$, in GeV. The data shown are the sieved set with $\sqrt s \ge 6$ GeV. The $\ln^2 s$ fits, corresponding to the solid curve for $\bar pp$ and the dash-dotted curve for $pp$,  are excellent. On the other hand, the $\ln s$ fits to the same data sample---the long dashed curve for $\bar p p$ and the short dashed curve for $pp$---are very bad fits. In essence, the $\ln s$ fit clearly undershoots {\em all} of the high energy cross sections. The ability of nucleon-nucleon scattering to distinguish cleanly  between an  energy dependence of $\ln^2 s$ and  an energy dependence of $\ln s$ is even more dramatic than the pion results of Figures \ref{fig:sigpip} and \ref{fig:rhopip}.

Again using the fit parameters from Table \ref{table:ppfitnew}, Figure \ref{fig:rhopp} shows the individual fitted $\rho$-values for  $pp$ and $\bar pp$ for both $\ln^2 s$ and $\ln s$ for the cut $\delchimax=6$, plotted against the c.m. energy $\sqrt s$,  in GeV. The data shown are the sieved data with $\sqrt s \ge 6$ GeV. The $\ln^2 s$ fits,
 corresponding to the solid curve for $\bar pp$ and the dash-dotted curve for $pp$,  fit the data reasonably well. On the other hand, the $\ln s$ fits, the long dashed curve for $\bar pp$ and the short dashed curve for $pp$, are very poor fits, missing completely the precise $\rho_{\bar pp}$ at 546 GeV, as well as  $\rho_{\bar pp}$ at 1800 GeV. These results again strongly support  the $\ln^2 s$ fits that saturate the Froissart bound and once again rule out $\ln s$ fits for the $\bar pp$ and $p p$ system.

A few remarks on the Block and Halzen\cite{bhfroissartnew} (BH) $\ln^2 s$ asymptotic energy analysis for $pp$ and $\bar pp$ are appropriate. It should be stressed that they used {\em both} the CDF and E710/E811 high energy experimental cross sections at $\sqrt s=1800$ GeV in the $\ln^2 s$ analysis---summarized in Table \ref{table:ppfitnew}, $\delchimax=6$, and graphically shown in Figures \ref{fig:sigmapp} and \ref{fig:rhopp}.  Inspection of Fig. \ref{fig:sigmapp} shows that at $\sqrt s=1800$ GeV, their  fit   passes somewhat   below the  cross section point of $\sim$ 80 mb (CDF collaboration).  In particular, to test the sensitivity of their  fit to the differences between the highest energy accelerator $\bar pp$ cross sections from the Tevatron, BH\cite{bhfroissartnew} made an analysis {\em  completely omitting} the CDF ($\sim$ 80 mb) point and refitted the data without it.  This fit, also using $\delchimax=6$,  had a renormalized $\chi^2$/d.f.=1.055, compared to 1.095 with the CDF point included.  Since you only expect, on average, a $\Delta\chi^2$ of $\sim 1$ for the removal of one point, the removal of the CDF point slightly improved  the goodness-of-fit. Moreover, the new parameters of the fit were only {\em very minimally} changed. As an example, the predicted value from  the new fit for the cross section at $\sqrt s=1800$ GeV---{\em without} the CDF point---was $\sigma_{\bar pp}=75.1\pm0.6$ mb, where the error is the statistical error due to the errors in the  fitted parameters.   On the other hand, the predicted value from 
Table \ref{table:predictions}---which used {\em both} the CDF and the E710/E811 point---was $\sigma_{\bar pp}=75.2\pm0.6$ mb, virtually identical. Further, at the LHC energy,  the fit {\em without} the  CDF point had $\sigma_{\bar pp}=107.2\pm1.2$ mb, whereas {\em including} the CDF measurement, they found  $\sigma_{\bar pp}=107.3\pm1.2$ mb, i.e., there was practically no effect of either including or excluding the CDF point, compared to the rather small statistical errors of the fit. The fit was determined almost exclusively by  the E710/E811 cross section---presumably because the asymptotic cross section fit was locked into the cross section and its derivative at the low energy transition energy $\nu_0$, thus utilizing the very accurate low energy experimental cross section data.    

We quote directly from Block and Halzen\cite{bhfroissartnew}:
\begin{itemize}
\item[]
``Our result concerning the (un)importance of the CDF point relative to E710/E811 result is to be contrasted with the statement from the COMPETE Collaboration\cite{{compete1},{cudell}} which emphasized that there is: `the systematic uncertainty coming from the discrepancy between different FNAL measurements of $\sigma_{\rm tot}$',  which contribute large differences to their fit predictions at high energy, depending on which data set they use.  In marked contrast to our results, they conclude that their fitting techniques favor the CDF point. Our results  indicate that {\em both} the cross section and $\rho$-value of the E710/E811 groups are slightly favored. More importantly, we find virtually {\em no sensitivity} to high energy predictions when we do not use the CDF point and only use the E710/E811 measurements. Our method of fitting the data---by anchoring the asymptotic fit at the low transition energy $\nu_0$---shows that our high energy predictions are quasi-independent of the FNAL `discrepancy', leading us to believe that our high energy cross section predictions at both the LHC and at cosmic ray energies  are both robust and accurate.'' 
\end{itemize}

In Table \ref{table:predictions}, we give  predictions from their  $\ln^2 s$ fit for values of $\sigma_{\bar pp}$ and $\rho_{\bar pp}$ at high energies. The  quoted  errors are due to the statistical errors of the fitted parameters $c_1$,  $c_2$ and $f_+(0)$ given in the $\delchimax=6$, $\ln^2 s$ fit  of Table \ref{table:ppfitnew}.

In Fig. \ref{fig:sigppallenergies}, we show an extended energy scale, from threshold up to cosmic ray energies ($1.876\le \sqrt s\le 10^5 $ GeV), 
plotting all available $\bar pp$ and $pp$ cross sections, including cosmic ray  $pp$ cross sections inferred from  cosmic ray $p$-air experiments by Block, Halzen and Stanev\cite{BHS}. We will discuss these cosmic ray points later in Section \ref{section:crpair}. The solid curve is their result from Table \ref{table:ppfitnew} of the {\em even} cross section from  $\ln^2 s$, $\delchimax=6$. The dashed-dot-dot curve is from the Aspen model (an independent QCD-inspired eikonal analysis\cite{BHS}) of the nucleon-nucleon system.  The agreement is quite remarkable---the two independent curves are virtually indistinguishable over almost 5 decades of c.m. energy, from $\sim 3$ GeV to 100 TeV.  Figure \ref{fig:sigppallenergies} clearly indicates that the $p p$ and $\bar pp$ cross section data  greater than $\sim3$ GeV can  be explained by a fit of the form 
$\sigma^\pm=c_0+c_1\ln\left(\frac{\nu}{m_p}\right)+c_2\ln^2\left(\frac{\nu}{m_p}\right)+\beta_{\cal P'}\left(\frac{\nu}{m_p}\right)^{\mu -1}\pm\  \delta\left({\nu\over m_p}\right)^{\alpha -1}$
 over an enormous energy range,  i.e., by a $\ln^2s$ saturation of the Froissart bound.

In Table \ref{table:predictions}, predictions are made of  total cross sections and $\rho$-values for $\bar pp$ and $pp$ scattering---in  the low energy regions covered by RHIC, together with the energies of the Tevatron and LHC, as well as the high energy regions appropriate to cosmic ray air shower experiments.

We now conclude that the three hadronic systems, $\gamma p$, $\pi p$ and  nucleon-nucleon, {\em all} have an asymptotic $\ln^2s$ behavior, thus saturating the Froissart bound.

The predicted values at 14 TeV for the Large Hadron Collider are  $\sigma_{\bar pp}=107.3\pm1.2$ mb and $\rho_{\bar pp}=0.132\pm 0.001$ ---robust predictions that rely critically  on the saturation of the Froissart bound.
%  Table 11 NEW PP
\begin{table}[tbp]                   % Use "table" environment, but also
				 % use  "tabular" environment below.
%
\def\arraystretch{1.2}            % Make the space between rows in the Table,
				  % 1.5 x bigger than the default spacing.
\begin{center}
     \caption[The fitted results for $\sigma_{pp}$, $\sigma_{\bar p p}$, $\rho_{pp}$ and $\rho_{\bar pp}$ for a 3-parameter $\chi^2$ fit with $\sigma\sim\ln^2 s$ and a 2-parameter fit with $\sigma\sim\ln s$, using 4 analyticity constraints]
{\protect\small The fitted results for $\sigma_{pp}$, $\sigma_{\bar p p}$, $\rho_{pp}$ and $\rho_{\bar pp}$ for a 3-parameter $\chi^2$ fit with $\sigma\sim\ln^2 s$ and a 2-parameter fit with $\sigma\sim\ln s$, using 4 analyticity constraints.  The renormalized $\chi^2$/d.f.,  taking into account the effects of the $\delchimax$ cut, is given in the row  labeled ${\cal R}\times\chi^2_{\rm min}/\nu$, where here $\nu$ is the number of degrees of freedom.  The errors in the fitted parameters have been multiplied by the appropriate $r_{\chi2}$.  For a discussion of the Sieve algorithm used in this fit, see Section \ref{section:sievealgorithm}; for details of the renormalization factors ${\cal R}$ and $r_{\chi2}$, see Figures \ref{renorm}a) and \ref{renorm}b) in Section \ref{section:lessons}.  Taken from Ref. \cite{bhfroissartnew}.\label{table:ppfitnew}}
\vspace{.2in}
\begin{tabular}[b]{|l||c|c||c||}
    %\cline{2-4}
%\hline
\cline{2-4}
\multicolumn{1}{c|}{}&\multicolumn{2}{c||}{$\sigma\sim \ln^2 s$}&\multicolumn{1}{c||}{$\sigma\sim \ln s$}\\
\cline{1-1}
\multicolumn{1}{|c||}{Parameters }
      &\multicolumn{2}{|c||}{$\delchimax$}&\multicolumn{1}{|c||}{$\delchimax$}\\ 
%\hline 
\cline{2-4}
	\multicolumn{1}{|c||}{}
      &\multicolumn{1}{c|}{6}&\multicolumn{1}{c||}{9} &\multicolumn{1}{c||}{6}\\
      \hline
	\multicolumn{4}{|c||}{\ \ \ \ \  Even Amplitude}\\
	\cline{1-4}
      $c_0$\ \ \   (mb)&$37.32$ &$37.25$&28.26\\ 
      $c_1$\ \ \   (mb)&$-1.440\pm0.070$ &$-1.416\pm0.066$&$2.651\pm 0.0070$\\ 
	$c_2$\ \ \ \   (mb)&$0.2817\pm0.0064$&$0.2792\pm0.0059$&------\\
      $\beta_{\cal P'}$\ \   (mb)&$37.10$ &$37.17$&47.98\\ 
      $\mu$&$0.5$ &$0.5$&0.5\\ 
	$f(0)$ (mb GeV)&$-0.075\pm0.59$&$-0.069\pm 0.57$&$4.28\pm 0.59$\\
      \hline
	\multicolumn{4}{|c||}{\ \ \ \ \  Odd Amplitude}\\
	\hline
      $\delta$\ \ \   (mb)&$-28.56$ &$-28.56$&-28.56\\
      $\alpha$&$0.415$ &$0.415$&0.415\\ 
	\cline{1-4}
     	\hline
	\hline
	$\chi^2_{\rm min}$&181.6&216.6&2355.7\\
	${\cal R}\times\chi^2_{\rm min}$&201.5&222.5&2613.7\\ 
	$\nu$ (d.f).&184&189&185\\
\hline
	${\cal R}\times\chi^2_{\rm min}/\nu$&1.095&1.178&14.13\\
\hline
\end{tabular}
\end{center}
     %\vspace{1in} \\
\end{table}
\def\arraystretch{1}  %Restore the default row spacing in the Table.
%%%%%%%%%%%%%%%%%%%%%%%%%%%%%
%%%%%%%%%%%%%%%%%%%%%%%%%%%%%%%%
%%%%%%%%%%%%%table #12
\begin{table}[tbp]                   % Use "table" environment, but also
				 % use  "tabular" environment below.
%
\def\arraystretch{1.1}            % Make the space between rows in the Table,
				  % 1.5 x bigger than the default spacing.
\begin{center}
    \caption[Predictions of high energy $\bar pp$ and $pp$ total  cross sections and $\rho$-values,  from Table \ref{table:ppfitnew}, $\sigma\sim\ln^2 s$, $\delchimax=6$]
{\protect\small Predictions of high energy $\bar pp$ and $pp$ total  cross sections and $\rho$-values,  from Table \ref{table:ppfitnew}, $\sigma\sim\ln^2 s$, $\delchimax=6$. Taken from Ref. \cite{bhfroissartnew}.\label{table:predictions}
}
\vspace{.2in}
\begin{tabular}[b]{|l||c|c||c|c||}
    \cline{1-5}
      \multicolumn{1}{|l||}{ $\sqrt s$, in GeV}
      &\multicolumn{1}{c|}{$\sigma_{\bar pp}$, in mb}
      &\multicolumn{1}{c||}{$\rho_{\bar p p}$}&\multicolumn{1}{c|}{$\sigma_{ pp}$, in mb}&\multicolumn{1}{c||}{$\rho_{pp}$}\\

      \hline\hline
	6&$48.97\pm0.01$&$-0.087\pm0.008$&$38.91\pm0.01$&$-.307\pm0.001$\\\hline60&$43.86\pm0.04$&$0.089\pm0.001$&$43.20\pm0.04$&$0.079\pm0.001$\\\hline
	100&$46.59\pm0.08$&$0.108\pm0.001$&$46.23\pm0.08$&$0.103\pm 0.001$\\\hline
	300&$55.03\pm0.21$&$0.131\pm0.001$&$54.93\pm0.21$&$0.130\pm 0.002$\\\hline	
	400&$57.76\pm0.25$&$0.134\pm0.002$&$57.68\pm0.25$&$0.133\pm 0.002$\\\hline
	540&$60.81\pm0.29$&$0.137\pm0.002$&$60.76\pm0.29$&$0.136\pm0.002$\\\hline
 	1,800&$75.19\pm0.55$&$0.139\pm0.001$&$75.18\pm0.55$&$0.139\pm0.001$\\\hline    
 	14,000&$107.3\pm1.2$&$0.132\pm0.001$&$107.3\pm1.2$&$0.132\pm0.001$\\\hline    
 	16,000&$109.8\pm1.3$&$0.131\pm0.001$&$109.8\pm1.3$&$0.131\pm0.001$\\\hline   	
 	50,000&$132.1\pm1.7$&$0.124\pm0.001$&$132.1\pm1.7$&$0.124\pm0.001$\\\hline
 	100,000&$147.1\pm2.0$&$0.120\pm0.001$&$147.1\pm2.0$&$0.120\pm0.001$\\\hline
\end{tabular}
\end{center}
     %\vspace{1in} \\
 
%     \\
\end{table}

\def\arraystretch{1}  %Restore the default row spacing in the Table.
\def\arraystretch{1}  %Restore the default row spacing in the Table.

\begin{figure}[tbp] %Fig.31
\begin{center}
\mbox{\epsfig{file=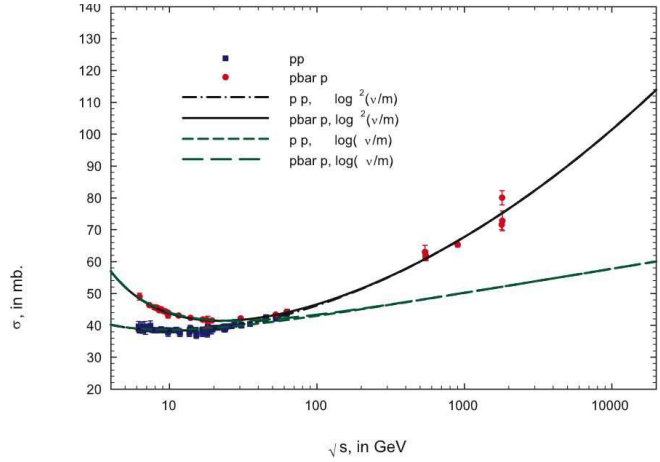,width=3.4in%
,bbllx=0pt,bblly=0pt,bburx=430pt,bbury=300pt,clip=%
}}
\end{center}
\caption[$\sigma_{p p}$ and $\sigma_{\pbar p}$, using  4 analyticity constraints]
{ \footnotesize
The fitted total cross sections $\sigma_{p p}$ and $\sigma_{\pbar p}$ in mb, vs. $\sqrt s$, in GeV, using the 4 constraints of Equations (\ref{derivevenpp}), (\ref{interceptevenpp}), (\ref{derivoddpp}) and (\ref{interceptoddpp}).  The circles are the sieved data  for $\pbar p$ scattering and the squares are the sieved data for $p p$ scattering for $\sqrt s\ge 6$ GeV. The dash-dotted curve ($pp$)  and the solid curve ($\pbar p$) are fits (Table \ref{table:ppfitnew}, $\sigma\sim\ln^2 s$, $\delchimax=6$)  of the form~: $\sigma^\pm=c_0 +c_1{\ln }\left({\nu\over m}\right)+c_2{\ln }^2\left({\nu\over m}\right)+\beta_{\cal P'}\left({\nu\over m}\right)^{\mu -1}\pm \delta\left({\nu\over m}\right)^{\alpha -1}$. The upper (lower) sign is for $p p$ ($\pbar p$) scattering.  The short dashed curve ($p p$) and the long dashed curve ($\pbar p$) are fits (Table  \ref{table:ppfitnew}, $\sigma\sim\ln s$, $\delchimax=6$ ) of  the high energy data  of the form~: $\sigma^\pm=c_0 +c_1{\ln }\left({\nu\over m}\right)+\beta_{\cal P'}\left({\nu\over m}\right)^{\mu -1}\pm\delta\left({\nu\over m}\right)^{\alpha -1}$. The upper (lower) sign is for $p p$ ($\pbar p$) scattering. The laboratory energy of the nucleon is  $\nu$ and $m$ is the nucleon mass. Taken from Ref. \cite{bhfroissartnew}. 
}
\label{fig:sigmapp}
\end{figure}
%
%%%%%%%%%%%%%%%%%%%%%%%%%%
%%%%%%%%%%%%%%%%%%%%%%%%%%%
\begin{figure}[tbp] %Fig. 32
\begin{center}
\mbox{\epsfig{file=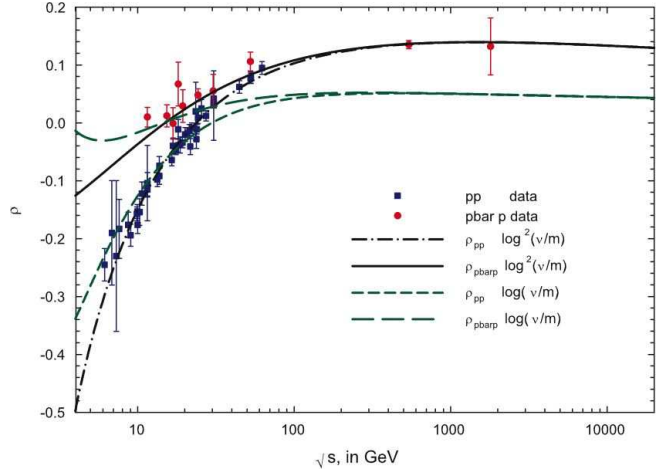,width=3.4in%
,bbllx=0pt,bblly=0pt,bburx=405pt,bbury=292pt,clip=%
}}
\end{center}
\caption[$\rho$-values, $\rho_{p p}$ and $\rho_{\pbar p}$, using 4 analyticity constraints]
{ \footnotesize
The fitted $\rho$-values, $\rho_{p p}$ and $\rho_{\pbar p}$,  vs. $\sqrt s$, in GeV, using the 4 constraints of Equations (\ref{derivevenpp}), (\ref{interceptevenpp}), (\ref{derivoddpp}) and (\ref{interceptoddpp}).  The circles are the sieved data  for $\pbar p$ scattering and the squares are the sieved data for $p p$ scattering for $\sqrt s\ge 6$ GeV. The  dash-dotted curve ($p p$) and the solid curve ($\pbar p$) are  fits (Table \ref{table:ppfitnew}, $\sigma\sim\ln^2 s$, $\delchimax=6$)  of the form~: $\rho^\pm= {1\over\sigma^\pm} \left\{\frac{\pi}{2}c_1+ c_2\pi\ln\left(\frac{\nu}{m}\right)-\beta_{\cal P'}\cot(\pi\mu/2)\left(\frac{\nu}{m}\right)^{\mu -1}+\frac{4\pi}{\nu}f_+(0)\ \pm \ \delta\tan(\pi\alpha/ 2)\left({\nu\over m}\right)^{\alpha -1}\right\} $. The upper (lower) sign is for $p p$ ($\pbar p$) scattering.  The short dashed curve ($p p$) and the long dashed curve ($\pbar p$)  are  fits (Table  \ref{table:ppfitnew}, $\sigma\sim\ln s$, $\delchimax=6$ )  of the form~: $\rho^\pm= {1\over\sigma^\pm} \left\{\frac{\pi}{2}c_1-\beta_{\cal P'}\cot(\pi\mu/2)\left(\frac{\nu}{m}\right)^{\mu -1}+\frac{4\pi}{\nu}f_+(0)\ \pm \ \delta\tan(\pi\alpha/ 2)\left({\nu\over m}\right)^{\alpha -1}\right\}$. The upper (lower) sign is for $p p$ ($\pbar p$) scattering.  The laboratory energy of the nucleon is  $\nu$ and $m$  is the nucleon mass. Taken from Ref. \cite{bhfroissartnew}.
  }
\label{fig:rhopp}
\end{figure}
%%%%%%%%%%%%%%%%%
%
\begin{figure}[h,t,b] %Fig. 33
\begin{center}
\mbox{\epsfig{file=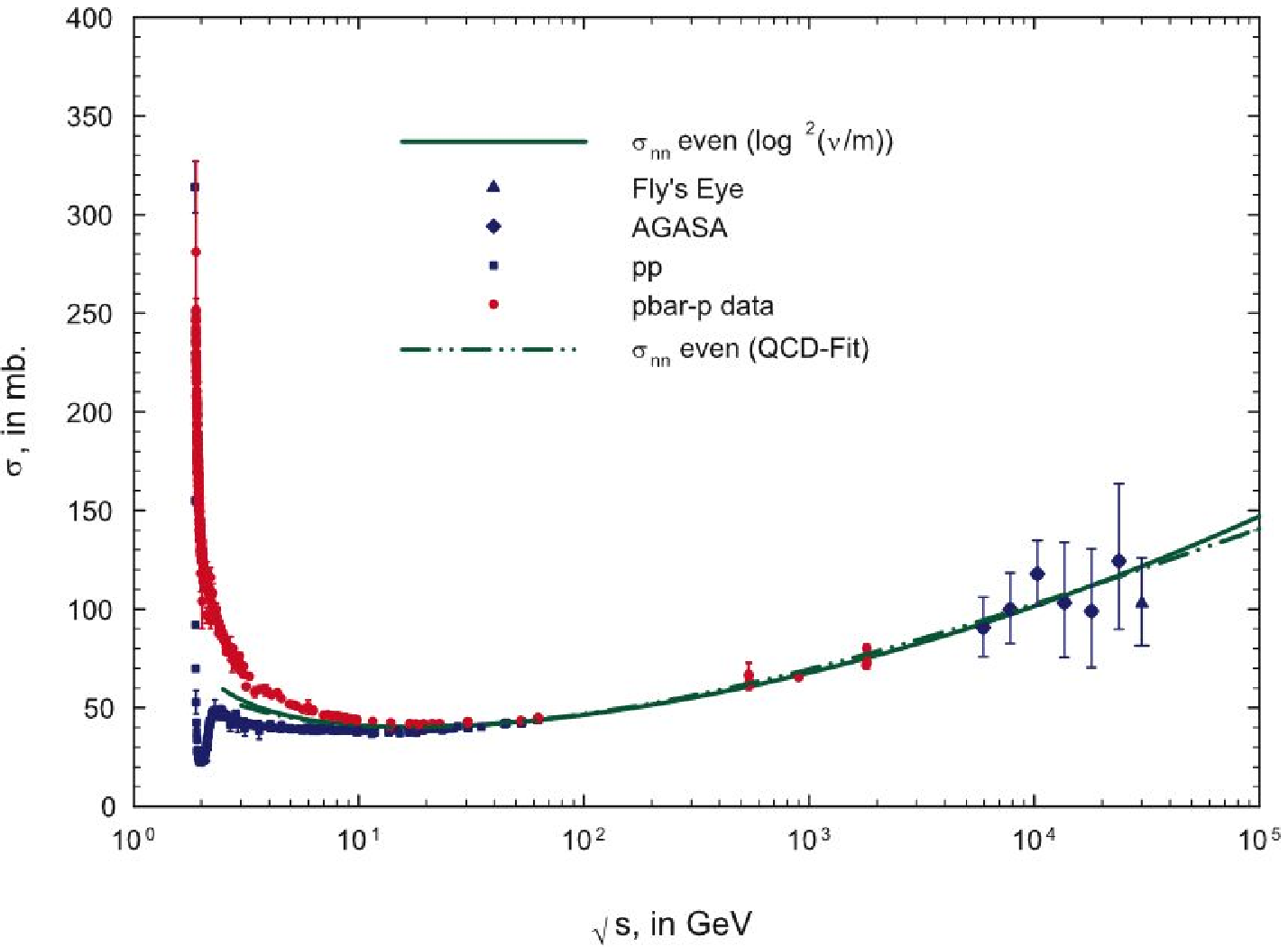,width=3.5in%
,bbllx=0pt,bblly=0pt,bburx=410pt,bbury=300pt,clip=%
}}
\end{center}
\caption[$\sigma_{pp}$ and $\sigma_{\bar pp}$, for all  known accelerator data]
{ \footnotesize
Total cross sections $\sigma_{pp}$ and $\sigma_{\bar pp}$, for all  known accelerator data. The circles are the cross section data  for $\pbar p$  and the squares are the cross section data for $p p$, in mb,  vs. $\sqrt s$, in GeV.  The solid  curve is the fit ((Table \ref{table:ppfitnew}, $\sigma\sim\ln^2 s)$, $\delchimax=6$)  of the even amplitude cross section: $\sigma_{nn}(\nu)=c_0 +c_1{\ln }\left({\nu\over m}\right)+c_2{\ln }^2\left({\nu\over m_p}\right)+\beta_{\cal P'}\left({\nu\over m}\right)^{\mu -1}$, with  $c_0$ and $\beta_{\cal P'}$ constrained by \eq{deriveven} and \eq{intercepteven}.    The dot-dot-dashed curve is the even amplitude cross section $\sigma_{nn}(\nu)$ from a QCD-inspired fit that fit not only  the accelerator $\bar pp$ and $pp$ cross sections and $\rho$-values, but also fit the AGASA and Fly's Eye cosmic ray pp cross sections---work done several years ago by the Block, Halzen and Stanev (BHS) group\cite{BHS}. 
The laboratory energy of the proton is  $\nu$ and $m$ is the proton mass. The two fitted curves for the even cross section, $\sigma_{nn}(\nu)$, using  the $\ln^2 s$ model of this work and  the QCD-inspired model of the BHS group\cite{BHS}, are virtually indistinguishable over 5 decades of c.m. energy,  i.e., in the energy region $3\le \sqrt s\le 10^5$ GeV. Taken from Ref. \cite{bhfroissartnew}.
}
\label{fig:sigppallenergies}
\end{figure}
%%%%%%%%%%%%%%%%%
\subsubsection{Comparison of a FESR to an analytic constraint}\label{section:compare}
In this Section we will show that the constraints of Block and Halzen\cite{bhfroissartnew} which are derived from analyticity and the FESR(2) of Igi and Ishida\cite{igiandishidapp} are, in fact, equivalent, as confirmed by fitting the two apparently very different methods to a common data set\cite{{sieve},{bhfroissartnew}} of $pp$ and $\bar pp$ cross sections.

We again fit the 4 experimental quantities $\sigma_{\bar pp}(\nu), \sigma_{p p}(\nu), \rho_{\bar pp}(\nu)$ and $\rho_{p p}(\nu)$, using the high energy parametrization (\eq{sigmapmpp} and \eq{rhopmpp} of the previous Section)
\ba
\sigma^\pm(\nu)&=&\sigma^0(\nu)\pm\  \delta\left({\nu\over m}\right)^{\alpha -1},\label{sigmapmpp1}\\
\rho^\pm(\nu)&=&{1\over\sigma^\pm(\nu)}\left\{\frac{\pi}{2}c_1+c_2\pi \ln\left(\frac{\nu}{m}\right)-\beta_{\cal P'}\cot({\pi\mu\over 2})\left(\frac{\nu}{m}\right)^{\mu -1}+\frac{4\pi}{\nu}f_+(0)\right.\nonumber\\
&&\left.\qquad\qquad\qquad\pm \delta\tan({\pi\alpha\over 2})\left({\nu\over m}\right)^{\alpha -1} \right\}\label{rhopmpp1},
\ea
where the upper sign is for $pp$ and the lower sign is for  $\bar pp$ scattering, $\mu=0.5$ and $m$ is the proton mass.

At the transition energy $\nu_0$ where we will match the high energy fits to the low energy data,  we define
\begin{eqnarray}
\sigma^0(\nu_0)&=&\frac{\sigma^{+}(\nu_0)+\sigma^-(\nu_0)}{2}\nonumber\\
&=&c_0+c_1\ln\y+c_2\ln^2\y+\beta_{\cal P'}\y^{\mu-1},\label{nu0}
\end{eqnarray}
where $\sigma^+$ ($\sigma^-$) are the total cross sections for $pp$ ($\bar p p$) scattering. From $\sigma^0=48.58$ mb at $\nu_0=7.59$ GeV, we obtain the constraint
\begin{eqnarray}
c_0&=& \sigma^0(\nu_0)-c_1\ln\y-c_2\ln^2\y-\beta_{\cal P'}\y^{\mu-1}\nonumber\\
&=&48.58 -2.091c_1-4.371c_2 -0.3516\beta_{\cal P'}\label{intercepteven1}.
\end{eqnarray}
In brief, we have used the $\bar pp$ and $pp$ cross sections at the transition energy $\nu_0$ to anchor the asymptotic fit to the low energy data. The precise choice of $\nu_0$ is not critical, as we will see further on.  We have previously shown that \eq{intercepteven1} is imposed by analyticity.

To summarize, our strategy is to exploit the rich sample of low energy data just  above the resonance region, but well  below the energies where data are used in  our  high energy fit. At the transition energy $\nu_0$,  the experimental cross sections $\sigma_{\bar pp}(\nu_0)$ and $\sigma_{pp}(\nu_0)$ are used to determine the even cross section $\sigma^0(\nu_0)$ of \eq{intercepteven}. In turn, this constrains the asymptotic high energy fit so that it {\em exactly} matches the low energy data at the transition energy $\nu_0$, constraining the value of $c_0$  in \eq{intercepteven}. Local fits are made to data in the vicinity of $\nu_0$ in order to evaluate the cross sections that are introduced in the above constraint equation, \eq{intercepteven}. We next impose the constraint \eq{intercepteven} on a $\chi^2$ fit to Equations ({\ref{sigmapm}) and (\ref{rhopm}). For safety, we start the  data fitting at much higher  energy,  $\nu_{\rm min}=18.72$ GeV ($\sqrt s_{\rm min}=6$ GeV), well above $\nu_0$.
 
We only consider an asymptotic $\ln^2s$ fit; the even amplitude parameter $c_0$ is  constrained  by \eq{intercepteven},  i.e., by  $c_1$,  $c_2$ and $\beta_{\cal P'}$ and the experimental value of $\sigma_{\rm even}(\nu_0)$. We then perform a simultaneous fit to the experimental high energy values of $\sigma_{\bar pp},\sigma_{pp},\rho_{\bar pp}$ and $\rho_{pp}$ using six parameters: the even parameters $c_1$, $c_2$, $\beta_{\cal P'}$ and $f_+(0)$ and the odd parameters $\delta$ and $\alpha$. Only the first 3 parameters are needed to describe the cross section.

Igi and Ishida\cite{igiandishidapp} derived  the constraint they called FESR(2) (see \eq{final2} of Section \ref{section:igi}), 
\be
c_0= 8.87 -2.04c_1-4.26c_2-0.367\beta_{\cal P'},\label{fesroriginal}
\ee
or, putting their coefficients into in units of mb, their constraint is
\be
c_0= 49.28 -2.04c_1-4.26c_2-0.367\beta_{\cal P'},\label{fesr}
\ee
which will be used in an alternative fit to the high energy data.

In our analysis we will use a  the sieved data set described in detail in ref. \cite{sieve} and  used in ref. \cite{bhfroissartnew}. Table \ref{table:ppfitnew1} shows the results of a 6 parameter $\chi^2$ fit constrained by FESR(2) and, alternatively, by the analyticity constraint that matches $\sigma^0(\nu)$ at $\nu_0$. The resulting $\chi^2$ have been renormalized\cite{sieve} for the cut $\delchisq=6$. Both fits are excellent, each with a $\chi^2$ per degree of freedom slightly less than 1.
 
%  Table 13  PP for FESR and analyticity
\begin{table}[h,t]                   % Use "table" environment, but also
				 % use  "tabular" environment below.
%
\def\arraystretch{1.1}            % Make the space between rows in the Table,
\begin{center}				  % 1.5 x bigger than the default spacing.
     \caption[ $\sigma_{pp}$, $\sigma_{\bar p p}$, $\rho_{pp}$ and $\rho_{\bar pp}$ for a 6-parameter  fit with $\sigma\sim\ln^2(s)$ and the cut $\delchimax=6$, for a FESR constraint and an analyticity constraint]
{\protect\small The fitted results for $\sigma_{pp}$, $\sigma_{\bar p p}$, $\rho_{pp}$ and $\rho_{\bar pp}$ for a 6-parameter  fit with $\sigma\sim\ln^2(s)$ and the cut $\delchimax=6$, for the FESR(2)  constraint $c_0= 49.28 -2.04c_1-4.26c_2-0.367\beta_{\cal P'}
$ and the analyticity constraint $c_0=48.58 -2.091c_1-4.371c_2 -0.3516\beta_{\cal P'}$. The renormalized\cite{sieve} $\chi^2_{\rm min}$/d.f.,  taking into account the effects of the $\delchimax$ cut, is given in the row  labeled ${\cal R}\times\chi^2_{\rm min}$/d.f. The errors in the fitted parameters have been multiplied by the appropriate $r_{\chi2}$ (see ref. \cite{sieve}). \label{table:ppfitnew1}}
\vspace{.1in}
\begin{tabular}[b]{|l||c|c||}
    %\cline{2-4}
%\hline
%\cline{2-3}
%\multicolumn{1}{c|}{}&\multicolumn{2}{c||}{$\sigma\sim \ln^2(\nu/m)$}\\

%\cline{1-1}
\hline
\multicolumn{1}{|c||}{Parameters }
      &\multicolumn{2}{|c||}{$\sigma\sim \ln^2s$, $\delchimax=6$}\\ 
%\hline 
\cline{2-3}
	\multicolumn{1}{|c||}{}
      &\multicolumn{1}{c|}{FESR2 Fit}&\multicolumn{1}{c||}{Analyticity  Fit} \\
      \hline
	\multicolumn{3}{|c||}{\ \ \ \ \  Even Amplitude}\\
	\cline{1-3}
      $c_0$\ \ \   (mb)&$36.68$ &$36.95$\\ 
      $c_1$\ \ \   (mb)&$-1.293\pm0.151$ &$-1.350\pm0.152$\\ 
	$c_2$\ \ \ \   (mb)&$0.2751\pm0.0105$&$0.2782\pm0.105$\\
      $\beta_{\cal P'}$\ \   (mb)&$37.10$ &$37.17$\\ 
      $\mu$&$0.5$ &$0.5$\\ 
	$f(0)$ (mb GeV)&$-0.075\pm0.67$&$-0.073\pm 0.67$\\
      \hline
	\multicolumn{3}{|c||}{\ \ \ \ \  Odd Amplitude}\\
	\hline
      $\delta$\ \ \   (mb)&$-24.67\pm 0.97$ &$-24.42\pm 0.96$\\
      $\alpha$&$0.451\pm 0.0097$ &$0.453\pm 0.0097$\\ 
	\cline{1-3}
     	\hline
	\hline
	$\chi^2_{\rm min}$&158.2&157.4\\
	${\cal R}\times\chi^2_{\rm min}$&180.3&179.4\\ 
	degrees of freedom(d.f).&181&181\\
\hline
	${\cal R}\times\chi^2_{\rm min}$/d.f.&0.996&0.992\\
\hline
\end{tabular}
     %\vspace{1in} \\
\end{center}
\end{table}
\def\arraystretch{1}  %Restore the default row spacing in the Table.
%%%%%%%%%%%%%%%%%%%%%%%%%%%%%%%%%%%%%
%%%%%%%%%%%%%%%%%%%%%%%%%%%%%%%%%%%%%%%

The $\bar pp$ and  $pp$ cross sections, in mb,  derived from the parameters of  Table \ref{table:ppfitnew} are shown in Fig.  \ref{fig:sigmapp1}a) as a function of the c.m. energy $\sqrt s$, in GeV, for both methods. The $\bar p p$ (circles) and $pp$ (squares) data shown are the sieved set.  The short dashed and dot-dashed curves are the analyticity constraint fits to the $\bar p p$ and $pp$ data, respectively. The solid curve and dotted curves are the $\bar p p$ and $pp$ fits for the FESR(2) constraint. The difference between the two fits is completely negligible over the energy interval $4\le\sqrt s\le 20000$ GeV; they agree to an accuracy of about 2 parts in 1000.  It should be emphasized that the FESR(2) fit uses the experimental resonance data below $\sqrt s_0=4$ GeV for evaluating the constraint of \eq{fesr}, whereas the analyticity constraint fit uses the even cross section {\em at} $\sqrt s_0=4$ GeV for the evaluation of its constraint, \eq{intercepteven},  i.e., {\em the alternative fits do not share any data}. Both techniques strongly support  $\ln^2 s$ fits that saturate the Froissart bound. 

Figure  \ref{fig:sigmapp1}b)  shows all of the $\bar pp$ and $pp$  cross section data\cite{pdg}  in the c.m. energy interval $4$ to $6$ GeV, {\em none} of which was used in our high energy fit.  Inspection of Fig.  \ref{fig:sigmapp1}b) reveals that we could have imposed the analyticity constraint at {\em any}  $\nu_0$  from 4 GeV to 6 GeV,  without modifying the result. As required, our conclusions do not depend on the particular choice of $\nu_0$, the transition energy used in \eq{nu0}. 

Figure \ref{fig:rhopp1} shows the fits for $\rho_{\bar pp}$ and $\rho_{pp}$ as a function of the c.m. energy $\sqrt s$; the sieved  experimental data are shown for $\sqrt s \ge 6$ GeV. We conclude that the results are effectively the same for both fits and are in very good agreement with the experimental data. Accommodating $\rho$-values at lower energies allows one to constrain the cross section at higher energies by derivative dispersion relations, giving us additional confidence in our extrapolations.

Summarizing, the  FESR(2) method and the new analyticity constraint introduced here yield fits to $\bar pp$ and $pp$ cross sections and $\rho$-values that agree to 2 parts in 1000 over the large energy interval $4{\rm \ GeV}\le\sqrt s\le 20000$ GeV. In particular, at the LHC energy of 14 TeV, the FESR(2) fit predicts $\sigma_{pp}=107.2\pm 1.4$ GeV and $\rho_{pp}=0.130\pm0.002$, whereas the analyticity fit predicts $\sigma_{pp}=107.4\pm1.5$ GeV and $\rho_{pp}=0.131\pm0.002$. We showed that this agreement was expected; it is numerical confirmation that analyticity, in its two guises---either as a FESR or as a cross section constraint---gives identical numerical results. Further, the fact that the renormalized $\chi^2$ per degree of freedom in Table \ref{table:ppfitnew} is excellent, giving a high probability fit, means that the choice of the high energy even asymptotic amplitude  (giving the cross section of \eq{sigmapmpp1}) {\em satisfies} the analyticity constraint. It did not have to---had we used a poor representation for the even asymptotic amplitude, forcing the fit to go through the even cross section data at $\sqrt s_0=4$ GeV would have resulted in a very high $\chi^2$. This was clearly demonstrated in references \cite{bhfroissartnew} and \cite{igiandishidapp}, where an asymptotic $\ln s$ parametrization was decisively rejected.

The fit of Block and Halzen\cite{bhfroissartnew} using 4 analyticity constraints---given in Table \ref{table:ppfitnew}---which additionally fixes the derivatives of the cross sections at $\sqrt s_0=4$ GeV for both $pp$ and $p\bar p$ as well as the odd cross section, yields essentially the same cross section and $\rho$-value,  but with {\em smaller} errors.  Clearly, from analyticity considerations, this  technique is equivalent to evaluating additional FESRs, but is much more transparent, as well as  more tractable numerically. 

Thus, we have shown that our new tool of analyticity constraints yields  both robust and more precise values for the total cross section at the LHC energy of 14 TeV, as well as at cosmic ray energies.  From Table \ref{table:predictions} which uses all 4 constraints, we see that at the LHC, $\rho_{pp}(14 {\rm \ TeV})=0.132\pm0.001$ and $\sigma_{pp}(14 {\rm \ TeV})=107.3\pm1.2$ mb.
 
%%%%%%%%%%%%%%%%%%%%%%%%%%%%%%%%%%%%%%%%%%%%%%%%
%%%%%%%%%%%%%%%%%%%%%%%%%%%%%%%%%%%%%%%%%%%%%%%%%%%%%
\begin{figure}[tbp] %Fig. 34
\begin{center}
\mbox{\epsfig{file=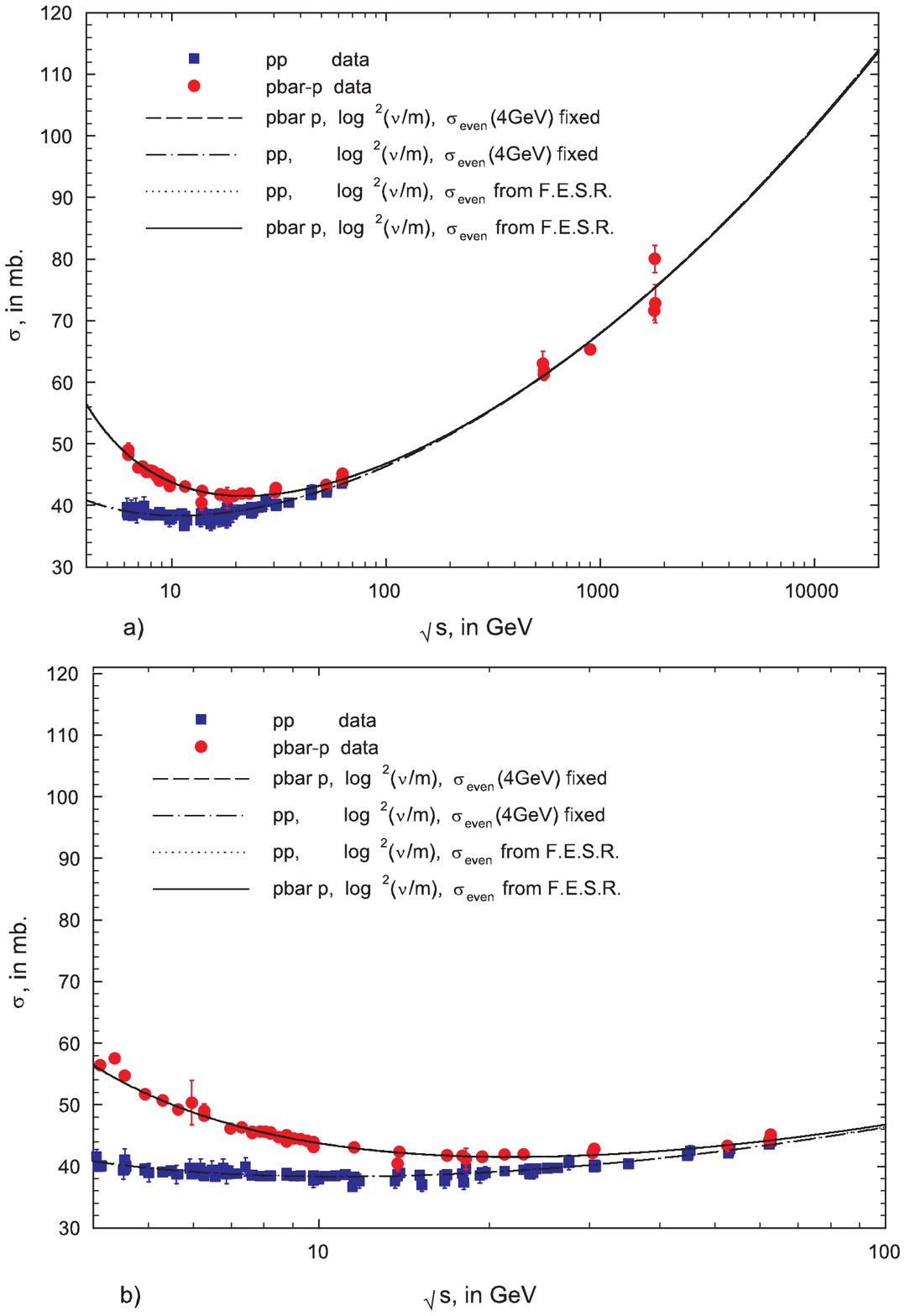,width=4.6in%
,bbllx=0pt,bblly=0pt,bburx=410pt,bbury=605pt,clip=%
}}
\end{center}
\caption[Comparison of cross sections fitted using an analyticity constraint and a FESR]
{ \footnotesize Comparison of cross sections fitted using an analyticity constraint and a FESR. \newline
a) The fitted total cross sections $\sigma_{p p}$ and $\sigma_{\pbar p}$ in mb,  vs. $\sqrt s$, in GeV, using the single constraint of Equations (\ref{intercepteven}) for the analyticity fit and (\ref{fesr}) for the FESR fit of Table \ref{table:ppfitnew}.  The circles are the sieved data  for $\pbar p$ scattering and the squares are the sieved data for $p p$ scattering for $\sqrt s\ge 6$ GeV.  The short dashed curve and dot-dashed curves are the analyticity fits---the {\em even} cross section at 4 GeV was fixed---to the $\bar p p$ and $pp$ data, respectively. The solid curve and dotted curves are the FESR fits to the $\bar p p$ and $pp$ data, respectively. It should be pointed out that the FESR and analyticity curves are essentially indistinguishable numerically for energies between 4 and 20000 GeV.\hspace{.1in} 
\newline b) An expanded energy scale that additionally shows the cross section data\cite{pdg} that  exist between 4 GeV, where $\sigma_{\rm even}$ was fixed, and 6 GeV, the beginning of the fitted data. We emphasize that {\em none} of the data between 4 and 6 GeV were used in the fits. We note that  that the fits go through all of the unused points, with the exception of the $\bar pp$ point at 4.2 GeV, which would have been excluded by the Sieve algorithm\cite{sieve} because of its large $\delchi$, had it been used. 
  }
\label{fig:sigmapp1}
\end{figure}
%%%%%%%%%%%%%%%%%%%%%%%%%%%%%%%%%%%%%%%%
%%%%%%%%%%%%%%%%%%%%%%%%%%%%%%%%%%%%%%%%%%%%%%%%

%%%%%%%%%%%%%%%%%%%%%%%%%%%%%%%%%%%%%%%%%%%%%
%%%%%%%%%%%%%%%%%%%%%%%%%%%%%%%%%%%%%%%%%%%%%%%%%%%%%
\begin{figure}[tbp] %Fig. 35
\begin{center}
\mbox{\epsfig{file=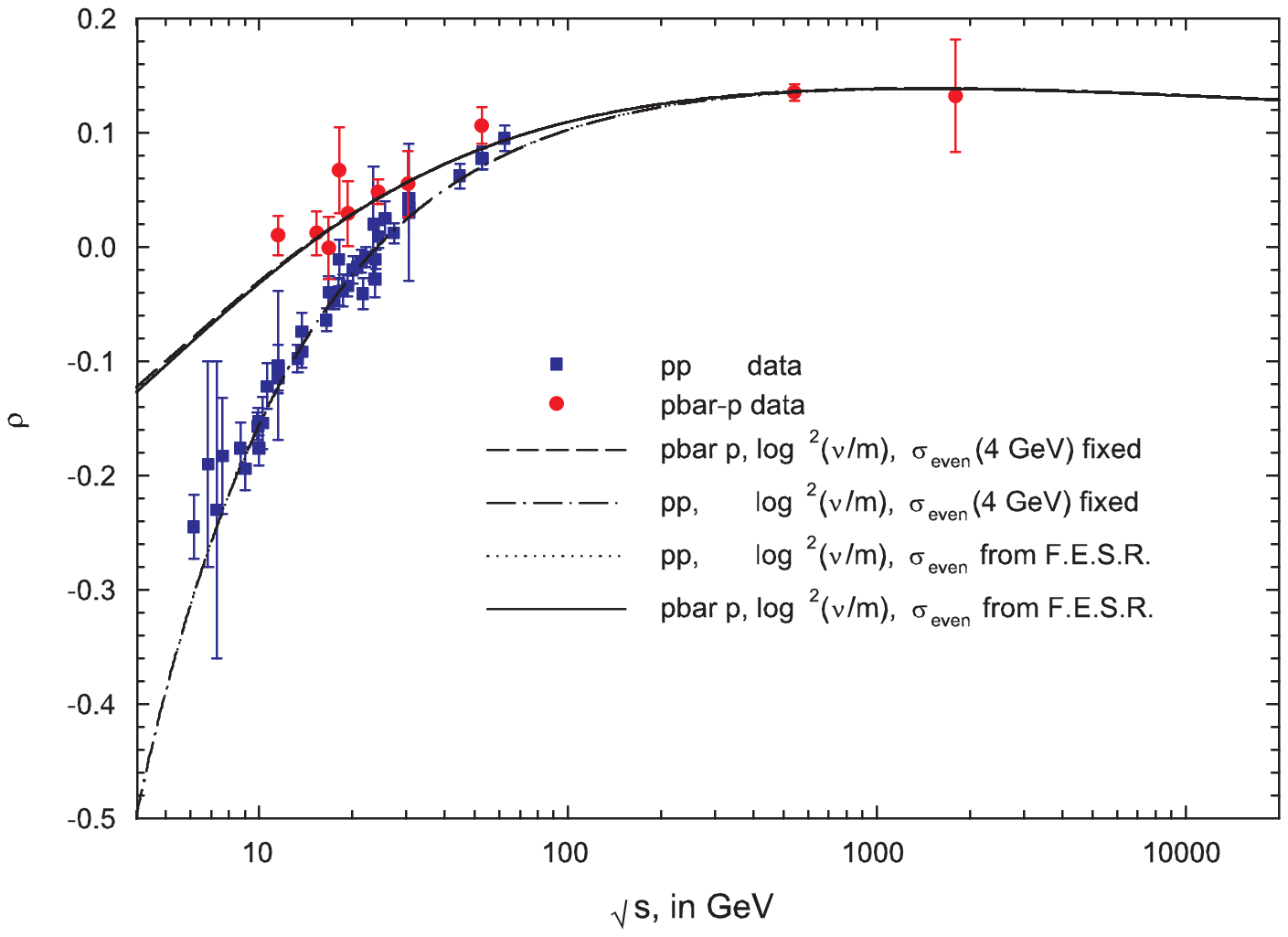,width=4.6in%
,bbllx=0pt,bblly=0pt,bburx=412pt,bbury=300pt,clip=%
}}
\end{center}
\caption[Comparison of $\rho$-values fitted using an analyticity constraint and a FESR]
{ \footnotesize Comparison of $\rho$-values fitted using an analyticity constraint and a FESR. \newline
 The fitted $\rho$-values, $\rho_{p p}$ and $\rho_{\pbar p}$,  vs. $\sqrt s$, in GeV, using the single constraint of Equations (\ref{intercepteven}) for the analyticity fit and (\ref{fesr}) for the FESR fit of Table \ref{table:ppfitnew}.  The circles are the sieved data  for $\pbar p$ scattering and the squares are the sieved data for $p p$ scattering for $\sqrt s\ge 6$ GeV.  The short dashed curve and dot-dashed curves are the analyticity fits---the {\em even} cross section at 4 GeV was fixed---to the $\bar p p$ and $pp$ data, respectively. The solid curve and dotted curves are the FESR fits to the $\bar p p$ and $pp$ data, respectively. It should be pointed out that the FESR and analyticity curves are essentially indistinguishable numerically for energies between 4 and 20000 GeV.
  }
\label{fig:rhopp1}
\end{figure}
%%%%%%%%%%%%%%%%%%%%%%%%%%%%%%%%%%%%%%%%%%%%%%%%%%%%%%%%%%%%%%%%%
%%%%%%%%%%%%%%%%%%%%%%%%%%%%%%%%%%%%%%%%%%%%%%%%%%%%%%%%%%%%%%%%%%%
\subsubsection{New limits on odderon amplitudes}\label{section:oddlimits}
In studies of high energy $pp$ and $\bar pp$ scattering, the odd (under crossing) amplitude accounts for the difference between the $pp$ and $\bar pp$ cross sections.  Conventionally, it is taken as $\frac{4\pi}{p}f_-=-Ds^{\alpha-1}e^{i\pi(1-\alpha)/2}$, which has $\Delta\sigma, \Delta\rho\rightarrow0$ as $s\rightarrow\infty$, where $\Delta \sigma=\sigtot^{pp}-\sigtot^{\bar p p}$; this is the odd amplitude  we have used up to now in our data analyses.  Nicolescu et al.\cite{nicolescu1,nicolescu2,nicolescu3} introduced odd amplitudes, called ``odderons'',  with the interesting properties ranging from having  $\Delta\sigma\rightarrow$non-zero constant to  having $\Delta\sigma \rightarrow \ln s/s_0$ as $s\rightarrow\infty$. 
 We will now reanalyze the high energy $pp$ and $\bar pp$ scattering data, using 4 analyticity constraints, to put new and much more precise limits on the magnitude of odderon amplitudes, using the odderon amplitudes of \eq{odd0}, \eq{odd1} and \eq{odd2} discussed earlier. 

 This Section is very recent work done by Block and Kang\cite{blockandkang2}. They  rewrite the 3 types of odderon amplitudes, $f_-^{(j)}$, where $j=0,1,$ or 2, of \eq{odd0}, \eq{odd1} and \eq{odd2}, again using $\nu$ as the laboratory energy,  as 
\ba
4\pi f_-^{(0)}&=&-\epsilon^{(0)}\nu\label{odd01},\\
4\pi f_-^{(1)}&=&-\epsilon^{(1)}\nu\left[\ln (s/s_0)-i\frac{\pi}{2}\right]\label{odd11},\\
4\pi f_-^{(2)}&=&-\epsilon^{(2)}\nu\left[\ln (s/s_0)-i\frac{\pi}{2}\right]^2\label{odd21},
\ea 
where the $\epsilon^{(j)}$ are all real coefficients. They then combine these odderon amplitudes, called odderon 0, odderon 1 and odderon 2,  individually with the conventional odd amplitude 
\ba
\frac{4\pi}{\nu}\,{\rm Im}f_-&=& \delta\left({\nu\over m}\right)^{\alpha -1}\nonumber\\
\frac{4\pi}{\nu}\,{\rm Re}f_-&=& \delta\tan\left({\pi\alpha\over 2}\right)\left({\nu\over m}\right)^{\alpha -1}\label{conventionalodd}
\ea
 to form a new total odd amplitude.  In the high energy limit, where the momentum $p\rightarrow\nu$, the amplitude $\frac{4\pi}{\nu}f_-^{(0)}=\frac{4\pi}{p}f_-^{(0)}$ is pure real; it only causes a small splitting in the $\rho$-values at high energy; the amplitude $\frac{4\pi}{\nu}f_-^{(1)}=\frac{4\pi}{p}f_-^{(1)}$ has a constant imaginary part, so that it leads to a constant non-zero $\Delta\sigma$, while its real part  causes the $\rho$-values to split at high energy ; finally, the amplitude $\frac{4\pi}{\nu}f_-^{(2)}=\frac{4\pi}{p}f_-^{(2)}$  has an imaginary part  that  leads to $\Delta\sigma\rightarrow\ln(s/s_0)$ as $s \rightarrow\infty$, along with a real part that causes a substantial splitting of the $\rho$-values at high energies; it is often called the ``maximal'' odderon.

Again, in the high energy limit $s\rightarrow2m\nu$, the cross sections $\sigma^\pm_{(j)}$, the $\rho$-values  $\rho^\pm_{(j)}$, along with the cross section derivatives $d\sigma^\pm_{(j)}/d\x,\ j=0,1,2$, can now be written as
\begin{eqnarray}
\sigma^\pm_{(0)}(\nu)&{\!\!\! =\!\!\! }&c_0+c_1\ln\left(\frac{\nu}{m}\right)+c_2\ln^2\left(\frac{\nu}{m}\right)+\beta_{\cal P'}\left(\frac{\nu}{m}\right)^{\mu -1}\pm\  \delta\left({\nu\over m}\right)^{\alpha -1},\label{sigmapm0}\\
\rho^\pm_{(0)}(\nu)&{\!\!\! =\!\!\! }&{1\over\sigma_\pm^{(0)}}\left\{\frac{\pi}{2}c_1+c_2\pi \ln\left(\frac{\nu}{m}\right)-\beta_{\cal P'}\cot\left({\pi\mu\over 2}\right)\left(\frac{\nu}{m}\right)^{\mu -1}+\frac{4\pi}{\nu}f_+(0)\right.\nonumber\\
&&\qquad\qquad\qquad\qquad\qquad\qquad\qquad\qquad\left.\pm \delta\tan\left({\pi\alpha\over 2}\right)\left({\nu\over m}\right)^{\alpha -1} \pm\epsilon^{(0))}\right\}\!\!,\label{rhopm0}\\
\frac{d\sigma^\pm_{(0)}(\nu)}{d\x}&{\!\!\! =\!\!\! }&c_1\left\{\frac{1}{\x}\right\} +c_2\left\{ \frac{2\ln\x}{\x}\right\}+\beta_{\cal P'}\left\{(\mu-1)\x^{\mu-2}\right\}\label{derivpm0} \nonumber\\
&&\ \ \ \ \ \ \ \ \ \ \ \ \ \ \ \ \ \ \ \ \  
\pm \ \delta\left\{(\alpha -1)\x^{\alpha - 2}\right\}
\end{eqnarray}
or
\begin{eqnarray}
\sigma^\pm_{(1)}(\nu)&{\!\!\! =\!\!\! }&c_0+c_1\ln\left(\frac{\nu}{m}\right)+c_2\ln^2\left(\frac{\nu}{m}\right)+\beta_{\cal P'}\left(\frac{\nu}{m}\right)^{\mu -1}\pm\  \delta\left({\nu\over m}\right)^{\alpha -1}\mp\epsilon^{(1)}\frac{\pi}{2},\label{sigmapm1}\\
\rho^\pm_{(1)}(\nu)&{\!\!\! =\!\!\! }&{1\over\sigma_\pm^{(1)}}\left\{\frac{\pi}{2}c_1+c_2\pi \ln\left(\frac{\nu}{m}\right)-\beta_{\cal P'}\cot\left({\pi\mu\over 2}\right)\left(\frac{\nu}{m}\right)^{\mu -1}+\frac{4\pi}{\nu}f_+(0)\right.\nonumber\\
&&\qquad\qquad\qquad\left.\pm \delta\tan\left({\pi\alpha\over 2}\right)\left({\nu\over m}\right)^{\alpha -1} \pm\epsilon^{(1)}\ln(s/s_0)\right\}\!\!,\label{rhopm1}\\
\frac{d\sigma^\pm_{(1)}(\nu)}{d\x}&{\!\!\! =\!\!\! }&c_1\left\{\frac{1}{\x}\right\} +c_2\left\{ \frac{2\ln\x}{\x}\right\}+\beta_{\cal P'}\left\{(\mu-1)\x^{\mu-2}\right\}\label{derivpm1} \nonumber\\
&&\qquad\qquad\qquad
\pm \ \delta\left\{(\alpha -1)\x^{\alpha - 2}\right\}
\end{eqnarray}
or
\begin{eqnarray}
\sigma^\pm_{(2)}(\nu)&{\!\!\! =\!\!\! }&c_0+c_1\ln\left(\frac{\nu}{m}\right)+c_2\ln^2\left(\frac{\nu}{m}\right)+\beta_{\cal P'}\left(\frac{\nu}{m}\right)^{\mu -1}\pm\  \delta\left({\nu\over m}\right)^{\alpha -1}\nonumber\\
&&\qquad\qquad\qquad\mp\epsilon^{(2)}\pi\ln(s/s_0),\label{sigmapm2}\\
\rho^\pm_{(2)}(\nu)&{\!\!\! =\!\!\! }&{1\over\sigma_\pm^{(2)}}\left\{\frac{\pi}{2}c_1+c_2\pi \ln\left(\frac{\nu}{m}\right)-\beta_{\cal P'}\cot\left({\pi\mu\over 2}\right)\left(\frac{\nu}{m}\right)^{\mu -1}+\frac{4\pi}{\nu}f_+(0)\right.\nonumber\\
&&\qquad\qquad\qquad\left.\pm \delta\tan\left({\pi\alpha\over 2}\right)\left({\nu\over m}\right)^{\alpha -1} \pm\epsilon^{(2)}\left(\ln^2(s/s_0)-\frac{\pi^2}{4}\right)\right\}\!\!,\label{rhopm2}\\
\frac{d\sigma^\pm_{(2)}(\nu)}{d\x}&{\!\!\! =\!\!\! }&c_1\left\{\frac{1}{\x}\right\} +c_2\left\{ \frac{2\ln\x}{\x}\right\}+\beta_{\cal P'}\left\{(\mu-1)\x^{\mu-2}\right\}\nonumber\\
&&\qquad\qquad\qquad\mp\epsilon^{(2)}\left\{\frac{\pi}{\x}\right\} 
\pm \ \delta\left\{(\alpha -1)\x^{\alpha - 2}\right\}\label{derivpm2} ,
\end{eqnarray}
where the upper sign is for $pp$ and the lower sign is for $\bar p p$ scattering, $\nu$ is the laboratory energy and $m$ is the proton mass.

At $\sqrt s_0 =4$ GeV, Block and Halzen\cite{bhfroissartnew} found  that
\ba
\sigma^+(\nu_0)&=&40.18 \quad\quad{\rm mb,}
\qquad \sigma^-(\nu_ 0)\quad=\quad\quad 56.99  \quad\quad{\rm mb,}\label{sigs}\\
\left.\frac{d\sigma^+\ \ }{d\x}\right|_{\nu =\nu_0}&=&-0.2305 \quad{\rm mb,}\quad
\left.\frac{d\sigma^-\ \ }{d\x}\right|_{\nu=\nu_0}=-1.4456\quad{\rm mb,}
\ea
using a local fit in the neighborhood of $\nu_0$. 
For $\nu_0=7.59$ GeV, these values yield the 4 experimental analyticity constraints 
\begin{eqnarray}
\sigma_{\rm av}&=&48.59 \quad\quad\,{\rm mb},\qquad
\Delta\sigma=\quad -8.405 \quad{\rm\  mb},\label{delsig&sigav0}\\
m_{\rm av}&=&-0.8381\quad{\rm mb},\qquad
\Delta m=\quad 1.215\quad\quad{\rm \ mb}\label{Delm&mav0}.
\end{eqnarray}

Again introducing the 2 even (under crossing) experimental constraints,  $\sigma_{\rm av}$ and $m_{\rm av}$,  they\cite{blockandkang2} can write 2 of the 4 analyticity constraint equations for the cross sections as 
\begin{eqnarray}
\beta_{\cal P'}&=&\frac{\y^{2-\mu}}{\mu -1}\left[m_{\rm av}-c_1\left\{\frac{1}{\y}\right\} -c_2\left\{\frac{2\ln\y}{\y}
\right\}\right],\label{deriveven1}\\
c_0&=& \sigma_{\rm av}-c_1\ln\y-c_2\ln^2\y-\beta_{\cal P'}\y^{\mu-1}.\label{intercepteven2}
\end{eqnarray}

The situation is a little more complicated for the odd (under crossing) experimental constraints, $\Delta\sigma$  and $\Delta m$. For odderon 0,
\begin{eqnarray}\qquad
\alpha&=&1+\frac{\Delta m}{\Delta \sigma}\times\frac{\nu_0}{m},\ \qquad\qquad\quad\qquad\qquad\qquad j=0,\label{derivodd0}\\
\delta&=&\Delta \sigma\times\left(\frac{\nu_0}{m}\right)^{1-\alpha}\label{interceptodd0},
\end{eqnarray}
whereas for odderon 1,
\begin{eqnarray}
\alpha&=&1+\frac{\Delta m}{\Delta \sigma-\epsilon^{(1)}(\frac{\pi}{2})}\,\times\frac{\nu_0}{m},\qquad\qquad\quad\qquad\qquad j=1,\label{derivodd1}\\
\delta&=&\Delta \sigma\times\left(\frac{\nu_0}{m}\right)^{1-\alpha}\label{interceptodd1},
\end{eqnarray}
and for odderon 2,
\begin{eqnarray}
\alpha&=&1+\frac{\Delta m-\epsilon^{(2)}\left\{\pi\nu_0/m\right\}}{\Delta \sigma-\epsilon^{(2)}\{\pi\ln(2m\nu_0 /s_0)\}}\times\frac{\nu_0}{m},\qquad\qquad j=2,\label{derivodd2}\\
\delta&=&\Delta \sigma\times\left(\frac{\nu_0}{m}\right)^{1-\alpha}\label{interceptodd2},
\end{eqnarray}
where $s_0=22.9$ GeV$^2$, which is the approximate value of $s_0$ found in the fit in Table \ref{table:ppoddfit}.

%%%%%%%%%%%%%%%%%%%%%%%%%%%%%%%%%
%%%%%%%%%%%%%%%%%%%%%%%%%%%%%%%%%%%%
They now  impose the 4 analyticity constraint equations:
\begin{description}
 \item [Odderon 0:]  Eqns. (\ref{deriveven1}), (\ref{intercepteven2}), (\ref{derivodd0}) and (\ref{interceptodd0}) are used in  our $\chi^2$ fit to Equations (\ref{sigmapm0}) and (\ref{rhopm0}).
\item 
[Odderon 1:]Eqns. (\ref{deriveven1}), (\ref{intercepteven2}), (\ref{derivodd1}) and (\ref{interceptodd1})  are used in  our $\chi^2$ fit to Equations (\ref{sigmapm1}) and (\ref{rhopm1}).
\item
[Odderon 2:] Eqns. (\ref{deriveven1}), (\ref{intercepteven2}), (\ref{derivodd2}) and (\ref{interceptodd2})  are used in  our $\chi^2$ fit to Equations (\ref{sigmapm2}) and (\ref{rhopm2}).
\end{description}

 We stress that the odd amplitude parameters $\alpha$ and $\delta$ and thus the odd amplitude itself is {\em completely determined} by the experimental values $\Delta m$ and $\Delta \sigma$ at the transition energy $\nu_0$, together with the value of $\epsilon^{(j)},\  j=0,1,2$.    Further,  the even amplitude parameters $c_0$ and $\beta_{\cal P}'$ are determined by $c_1$ and $c_2$, using the experimental values of $\sigma_{\rm av}$ and $m_{\rm av}$ at the transition energy $\nu_0$. Hence, the only 4  parameters that need to be fitted are $c_1$, $c_2$, $f_+(0)$ and $\epsilon^{(j)}, j=0,1,2$. Since the subtraction constant $f_+(0)$ only enters into the $\rho$-value determinations, only the 3  parameters  $c_1$, $c_2$ and $\epsilon^{(j)},\  j=0,1,2$,  of the original 8 are required for a $\ln^2s$  fit to the cross sections $\sigma^{\pm}(\nu)$, again leaving exceedingly little freedom in this fit---it is indeed very tightly constrained, with little latitude for adjustment. The Block and Halzen\cite{bhfroissartnew} sieved $pp$ and $\bar pp$ data set, used earlier, was again used here for 3 $\chi^2$ fits to $\sigma_{pp},\sigma_{\bar pp},\rho_{pp}$ and $\rho_{\bar p p}$ for $
\sqrt s\ge 6$ GeV, one for each of the three odderon amplitudes.

Table \ref{table:ppoddfit} shows the results of  simultaneous fits to the available accelerator data  from the Particle Data Group\cite{pdg} for  $\sigma_{pp}$, $\sigma_{\bar pp}$, $\rho_{pp}$ and $\rho_{\bar pp}$, using the 4 constraint equations with a transition energy $\sqrt s=4$ GeV and a minimum fitting energy of 6 GeV, again using the Sieve algorithm, for odderons 0, 1 and 2, for the  cut $\delchimax=6$. Very satisfactory probabilities ($\sim  0.2$) for 183 degrees of freedom were found for all 3 odderon choices.

We summarize their\cite{blockandkang2} results below:
\begin{description}
\item[Odderon 0:] Figure \ref{fig:sigmaodd0} shows the individual fitted cross sections (in mb) for $ pp$ and $\bar pp$ for  odderon 0 in Table \ref{table:ppoddfit}, plotted against $\sqrt s$, the  c.m.  energy in GeV. The data shown are the sieved data with $\sqrt s \ge 6$ GeV. The  fits to the data sample with $\delchimax=6$, corresponding to the dotted curve for $\bar pp$ and the solid curve for $pp$,  are excellent, yielding a total renormalized $\chi^2=201.2$, for 183 degrees of freedom, corresponding to a fit probability of $\sim0.2$. Figure \ref{fig:rhoodd0} shows the simultaneously  fitted $\rho$-values for $pp$ and $\bar pp$  for odderon 0 from Table \ref{table:ppoddfit}, plotted against $\sqrt s$, the c.m.  energy in GeV. The data shown are the sieved data with $\sqrt s \ge 6$ GeV. The dotted curve for $\bar pp$ and the solid curve for $pp$ fit the data  well. It should be noted from Table \ref{table:ppoddfit} that the magnitude of odderon 0 is $\epsilon^{(0)}=-0.034\pm0.073$ mb, a very small coefficient which is indeed compatible with 0.
%%%%%%%%%%%%%%%%%%%%%%%%%%%%%%5
\item [Odderon 1:] Figure \ref{fig:sigmaodd1} shows the individual fitted cross sections (in mb) for $ pp$ and $\bar pp$ for  odderon 1 in Table \ref{table:ppoddfit}, plotted against $\sqrt s$, the  c.m.  energy in GeV. The data shown are the sieved data with $\sqrt s \ge 6$ GeV. The  fits to the data sample with $\delchimax=6$, corresponding to the dotted curve for $\bar pp$ and the solid curve for $pp$,  are very good, yielding a total renormalized $\chi^2=200.9$, for 183 degrees of freedom, corresponding to a fit probability of $\sim0.2$. Figure \ref{fig:rhoodd1} shows the simultaneously  fitted $\rho$-values for $pp$ and $\bar pp$  for odderon 1 from Table \ref{table:ppoddfit}, plotted against $\sqrt s$, the c.m.  energy in GeV. The data shown are the sieved data with $\sqrt s \ge 6$ GeV. The dotted curve for $\bar pp$ and the solid curve for $pp$ fit the data reasonably well. It should be noted from Table \ref{table:ppoddfit} that the magnitude of odderon 1 is $\epsilon^{(1)}=-0.0051\pm0.0077$ mb, a very tiny coefficient which is also compatible with 0.
%%%%%%%%%%%%%%%%%%%%%%%%%%%%%%%
\item [Odderon 2:] Figure \ref{fig:sigmaodd2} shows the individual fitted cross sections (in mb) for $ pp$ and $\bar pp$ for  odderon 2 (the `maximal' odderon) in Table \ref{table:ppoddfit}, plotted against $\sqrt s$, the  c.m.  energy in GeV. The data shown are the sieved data with $\sqrt s \ge 6$ GeV. The  fits to the data sample with $\delchimax=6$, corresponding to the dotted curve for $\bar pp$ and the solid curve for $pp$,  are excellent, yielding a total renormalized $\chi^2=196.1$, for 183 degrees of freedom, corresponding to a fit probability of $\sim0.2$. Figure \ref{fig:rhoodd2} shows the simultaneously  fitted $\rho$-values for $pp$ and $\bar pp$  for odderon 2 from Table \ref{table:ppoddfit}, plotted against  $\sqrt s$, the  c.m.  energy in GeV. The data shown are the sieved data with $\sqrt s \ge 6$ GeV. The dotted curve for $\bar pp$ and the solid curve for $pp$ fit are a good fit to the data. It should be noted from Table \ref{table:ppoddfit} that the magnitude of odderon 2 is $\epsilon^{(2)}=0.0042\pm0.0019$ mb, a very tiny coefficient which is only about two standard deviations from 0.
\end{description}

In Table \ref{table:predictionsodd} are the predictions of  total cross sections and $\rho$-values for $\bar pp$ and $pp$ scattering for odderon 2 of Table \ref{table:ppoddfit}. Only at {\em very} high energies, above $\sqrt s=14$ TeV, is there any appreciable difference between $\rho_{\bar pp}$ and $\rho_{pp}$, as seen in Fig. \ref{fig:rhoodd2}.

In conclusion, all three odderon amplitudes, $\epsilon^{(j)},\ j=0,\ 1,$ or 2, are very small in comparison to all of the other amplitudes found in the fit---typically of the order of 1.5 to 40 mb---and indeed, all  are compatible with zero.  These new limits are to be contrasted to the analysis made in 1985 by Block and Cahn\cite{bc}, where they found $\epsilon^{(0)}=-0.25\pm0.13$ mb, $\epsilon^{(1)}=-0.11\pm0.04$ mb and $\epsilon^{(2)}=-0.04\pm0.02$ mb, which were about two standard deviations from zero, but with an  error of almost 2 to 10 times larger than the limits found here.  The marked increase in accuracy is attributable to the use of the 4 analyticity constraints employed in the present analysis, as well as the improved (sieved) data set which  also uses higher energy points than were available in 1985.  
%%%%%%%%%%%%%%%%%%%%%%%%
%%%%%%%%%%%%%%%%%%%%%%%%%%%%
%  Table 14 NEW PPodderon
\begin{table}[tbp]                   % Use "table" environment, but also
				 % use  "tabular" environment below.
%
\def\arraystretch{1.1}            % Make the space between rows in the Table,
\begin{center}				  % 1.5 x bigger than the default spacing.
     \caption[The fitted results for a 4-parameter  fit using odderons 0, 1 and 2, with $\sigma\sim\ln^2 s$, to the total cross sections and $\rho$-values for $pp$ and $\bar pp$ scattering]
{\protect\small The fitted results for a 4-parameter  fit using odderons 0, 1 and 2, with $\sigma\sim\ln^2 s$, to the total cross sections and $\rho$-values for $pp$ and $\bar pp$ scattering, taken from Block and Kang\cite{blockandkang2}. The renormalized $\chi^2_{\rm min}$ per degree of freedom,   taking into account the effects of the $\delchimax$ cut, is given in the row  labeled ${\cal R}\times\chi^2_{\rm min}$/d.f. The errors in the fitted parameters have been multiplied by the appropriate $r_{\chi2}$. For details on the renormalization of the errors by $r_{\chi2}$ and the renormalization of $\chi^2_{\rm min}$ by ${\cal R}$, see ref. \cite{sieve}. \label{table:ppoddfit}}
\vspace{.1in}
\begin{tabular}[b]{|l||c|c|c|}
\hline
Parameters&odderon 0&odderon 1&odderon 2\\
      \hline
	{}&\multicolumn{3}{|c|}{ Even Amplitude}\\
	\cline{1-4}
      $c_0$\ \ \   (mb)&$37.38$ &$37.24$&37.09\\ 
      $c_1$\ \ \   (mb)&$-1.460\pm0.065$ &$-1.415\pm0.073$&$-1.370\pm 0.0074$\\ 
	$c_2$\ \ \ \   (mb)&$0.2833\pm0.0060$&$0.2798\pm0.0064$&$0.2771\pm0.0064$\\
      $\beta_{\cal P'}$\ \   (mb)&$37.02$ &$37.20$&37.39\\ 
      $\mu$&$0.5$ &$0.5$&0.5\\ 
	$f_+(0)$ (mb GeV)&$-0.075\pm0.75$&$-0.050\pm 0.59$&$-.073\pm 0.58$\\
      \hline
	{}&\multicolumn{3}{|c|}{ Odd Amplitude}\\
	\hline
      $\delta$\ \ \   (mb)&$-28.56$ &$-28.53$&-28.49\\
      $\alpha$&$0.415$ &$0.416$&0.416\\ 
$\epsilon^{(j)}$ (mb)$,\quad j=0,1,2$&$-0.034\pm0.073$&$-0.0051\pm0.0077$&$0.0042\pm0.0019$\\
     	\hline
	\hline
	$\chi^2_{\rm min}$&181.3&181.1&176.7\\
	${\cal R}\times\chi^2_{\rm min}$&201.2&200.9&196.1\\ 
	degrees of freedom (d.f).&183&183&183\\
\hline
	${\cal R}\times\chi^2_{\rm min}$/d.f.&1.099&1.098&1.071\\
\hline
\end{tabular}
\end{center}
     %\vspace{1in} \\
\end{table}
\def\arraystretch{1}  %Restore the default row spacing in the Table.
%%%%%%%%%%%%%%%%%%%%%%%%%%%%%
%%%%%%%%%%%%%%%%%%%%%%%%%%%%%%
%%% Table 15
\begin{table}[tbp]                   % Use "table" environment, but also
				 % use  "tabular" environment below.
%
\def\arraystretch{1.2}            % Make the space between rows in the Table,
				  % 1.5 x bigger than the default spacing.
\begin{center}
     \caption[Predictions of high energy $\bar pp$ and $pp$ total  cross sections and $\rho$-values for odderon 2]
{\protect\small Predictions of high energy $\bar pp$ and $pp$ total  cross sections and $\rho$-values for odderon 2,  from Table \ref{table:ppoddfit}.\label{table:predictionsodd}
}
\vspace{.1in} 
\begin{tabular}[b]{|l||c|c||c|c||}
    \cline{1-5}
      \multicolumn{1}{|l||}{ $\sqrt s$, in GeV}
      &\multicolumn{1}{c|}{$\sigma_{\bar pp}$, in mb}
      &\multicolumn{1}{c||}{$\rho_{\bar p p}$}&\multicolumn{1}{c|}{$\sigma_{ pp}$, in mb}&\multicolumn{1}{c||}{$\rho_{pp}$}\\

      \hline\hline
	300&$55.14\pm0.20$&$0.125\pm0.003$&$54.82\pm0.20$&$0.134\pm 0.003$\\
\hline	
	540&$60.89\pm0.29$&$0.129\pm0.004$&$60.59\pm0.29$&$0.141\pm0.003$\\\hline
 	1,800&$75.19\pm0.50$&$0.130\pm0.001$&$74.87\pm0.52$&$0.146\pm0.004$\\\hline    
 	14,000&$107.1\pm1.1$&$0.121\pm0.005$&$106.6\pm1.1$&$0.141\pm0.005$\\\hline    
 	50,000&$131.55\pm1.5$&$0.112\pm0.006$&$131.1\pm1.6$&$0.134\pm0.005$\\\hline
 	100,000&$146.39\pm1.8$&$0.108\pm0.006$&$145.9\pm1.9$&$0.131\pm0.005$\\\hline
\end{tabular}
\end{center}
     %\vspace{1in} \\
%     \\
\end{table}

\def\arraystretch{1}  %Restore the default row spacing in the Table.
\def\arraystretch{1}  %Restore the default row spacing in the Table.

%%%%%%%%%%%%%%%%%%%%%%%%%%%%%%%%%%%%%%%%%%%
\begin{figure}[tbp] %Fig. 36
\begin{center}
\mbox{\epsfig{file=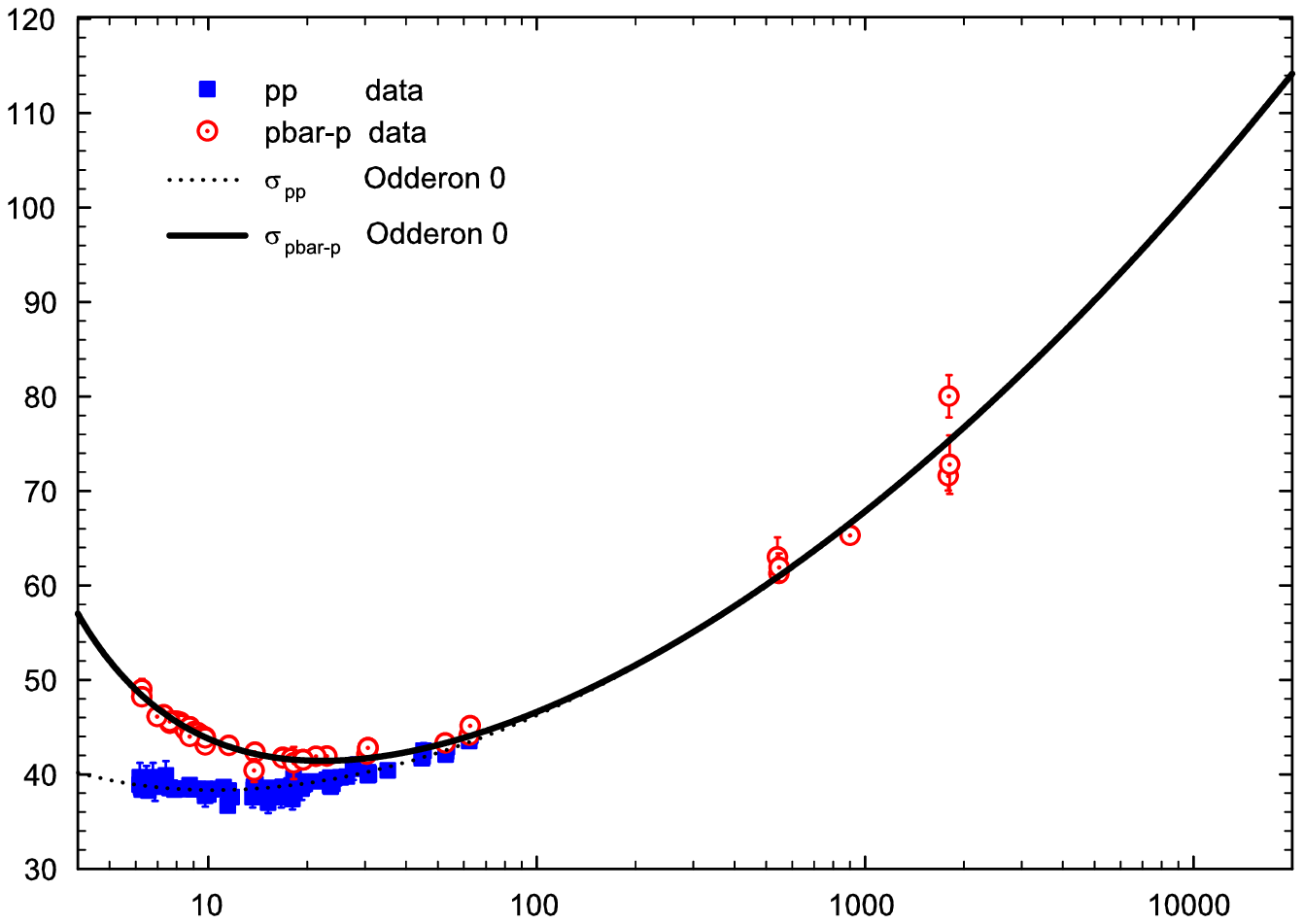,width=4in%
%,bbllx=65pt,bblly=135pt,bburx=450pt,bbury=305pt,clip=%
}}
\end{center}
\caption[Odderon 0: $\sigma_{\bar pp}$ and $\sigma_{pp}$]
{ \footnotesize Odderon 0: $\sigma_{\bar pp}$ and $\sigma_{pp}$. The fitted total cross sections  in mb, vs. $\sqrt s$, in GeV, using the 4 constraints of Equations (\ref{deriveven}), (\ref{intercepteven}), (\ref{derivodd0}) and (\ref{interceptodd0}), for odderon 0 of \eq{odd01}.  The circles are the sieved  data  for $\bar pp$ scattering and the squares are the sieved data for $pp$ scattering for $\sqrt s\ge 6$ GeV. The dotted curve ($pp$)  and the solid curve ($\bar pp$) are $\chi^2$ cross section fits, corresponding to a simultaneous $\ln^2s$ fit to cross sections and $\rho$-values   (Table \ref{table:ppoddfit}, of odderon 0) of  \eq{sigmapm0} and \eq{rhopm0}. }
\label{fig:sigmaodd0}
\end{figure}

%  Begin Figures
\begin{figure}[tbp] %Fig. 37
\begin{center}
\mbox{\epsfig{file=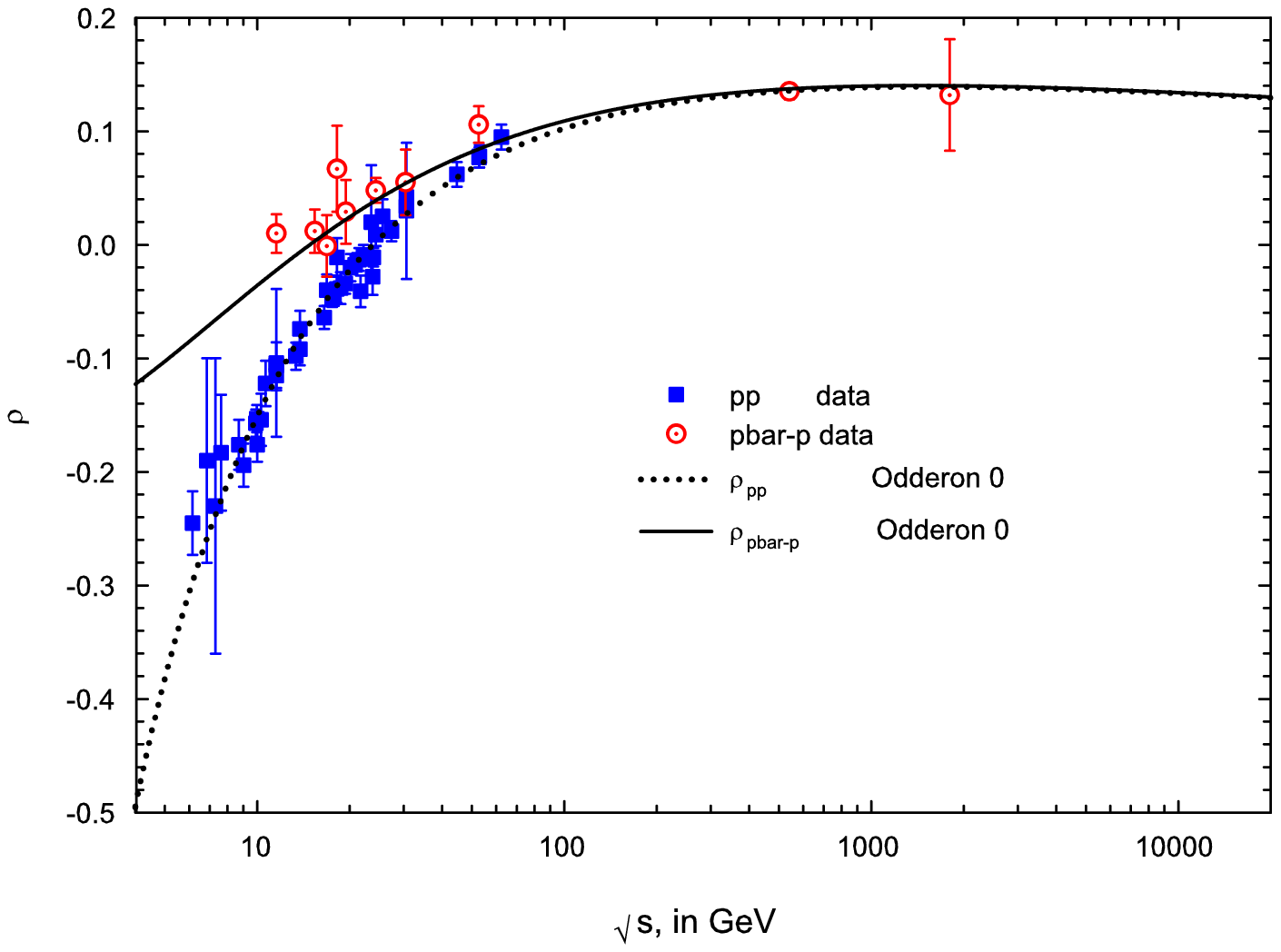
,width=4in,%,
bbllx=0pt,bblly=0pt,bburx=420pt,bbury=305pt,clip=%
}}
\end{center}
\caption[Odderon 0: $\rho_{\bar pp}$ and $\rho_{pp}$]
{ \footnotesize Odderon 0: $\rho_{\bar pp}$ and $\rho_{pp}$. The fitted $\rho$-values  vs. $\sqrt s$, in GeV, using the 4 constraints of Equations (\ref{deriveven}), (\ref{intercepteven}), (\ref{derivodd0}) and (\ref{interceptodd0}), for odderon 0 of \eq{odd01}.  The circles are the sieved  data  for $\bar pp$ scattering and the squares are the sieved data for $pp$ scattering for $\sqrt s\ge 6$ GeV. The dotted curve ($pp$)  and the solid curve ($\bar pp$) are $\chi^2$  $\rho$-value fits, corresponding to a simultaneous $\ln^2s$ fit to cross sections and $\rho$-values  (Table \ref{table:ppoddfit}, odderon 0)  of  \eq{sigmapm0} and \eq{rhopm0}. 
}
\label{fig:rhoodd0}
\end{figure}
%%%%%%%%%%%%%%%%%%%%%%%%%%%%%%%%%%%%%%%%
%%%%%%%%%%%%%%%%%%%%%%%%%%%%%%%%%%%%%%%%
\begin{figure}[tbp] %Fig. 38
\begin{center}
\mbox{\epsfig{file=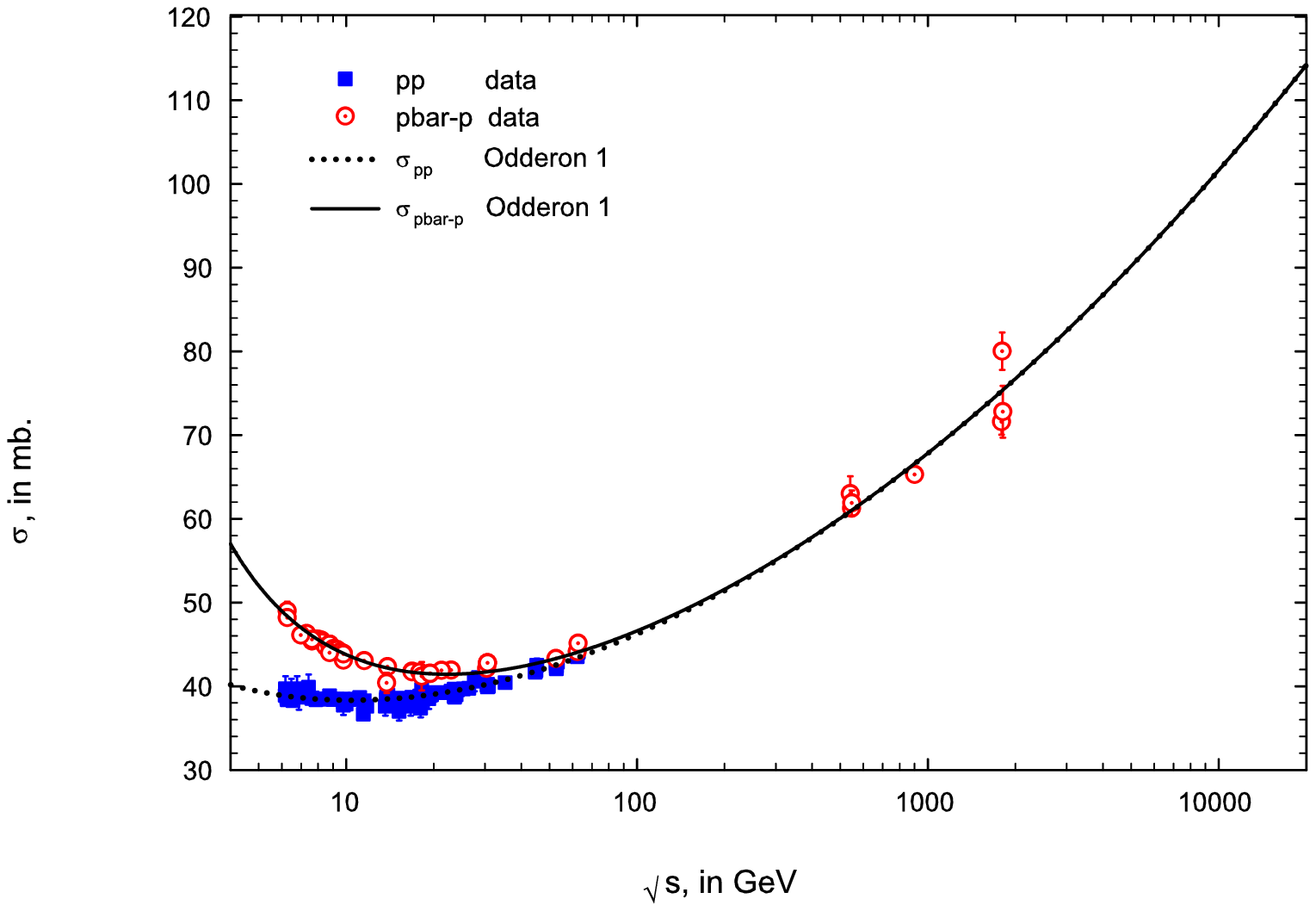,width=4in%
,bbllx=0pt,bblly=0pt,bburx=445pt,bbury=305pt,clip=%
}}
\end{center}
\caption[Odderon 1: $\sigma_{\bar pp}$ and $\sigma_{pp}$]
{ \footnotesize
Odderon 1: $\sigma_{\bar pp}$ and $\sigma_{pp}$. The fitted total cross sections  in mb, vs. $\sqrt s$, in GeV, using the 4 constraints of Equations (\ref{deriveven}), (\ref{intercepteven}), (\ref{derivodd1}) and (\ref{interceptodd1}), for odderon 1 of \eq{odd11}.  The circles are the sieved  data  for $\bar pp$ scattering and the squares are the sieved data for $pp$ scattering for $\sqrt s\ge 6$ GeV. The dotted curve ($pp$)  and the solid curve ($\bar pp$) are $\chi^2$ cross section fits, corresponding to a simultaneous $\ln^2s$ fit to cross sections and $\rho$-values  The dotted curve ($pp$)  and the solid curve ($\bar pp$) are $\chi^2$ cross section fits, corresponding to a simultaneous $\ln^2s$ fit to cross sections and $\rho$-values  (Table \ref{table:ppoddfit}, odderon 1) of  \eq{sigmapm1} and \eq{rhopm1}.  }
\label{fig:sigmaodd1}
\end{figure}
%%%%%%%%%%%%%%%%%
%
%%%%%%%%%%%%%%%%%
\begin{figure}[tbp] %Fig. 39
\begin{center}
\mbox{\epsfig{file=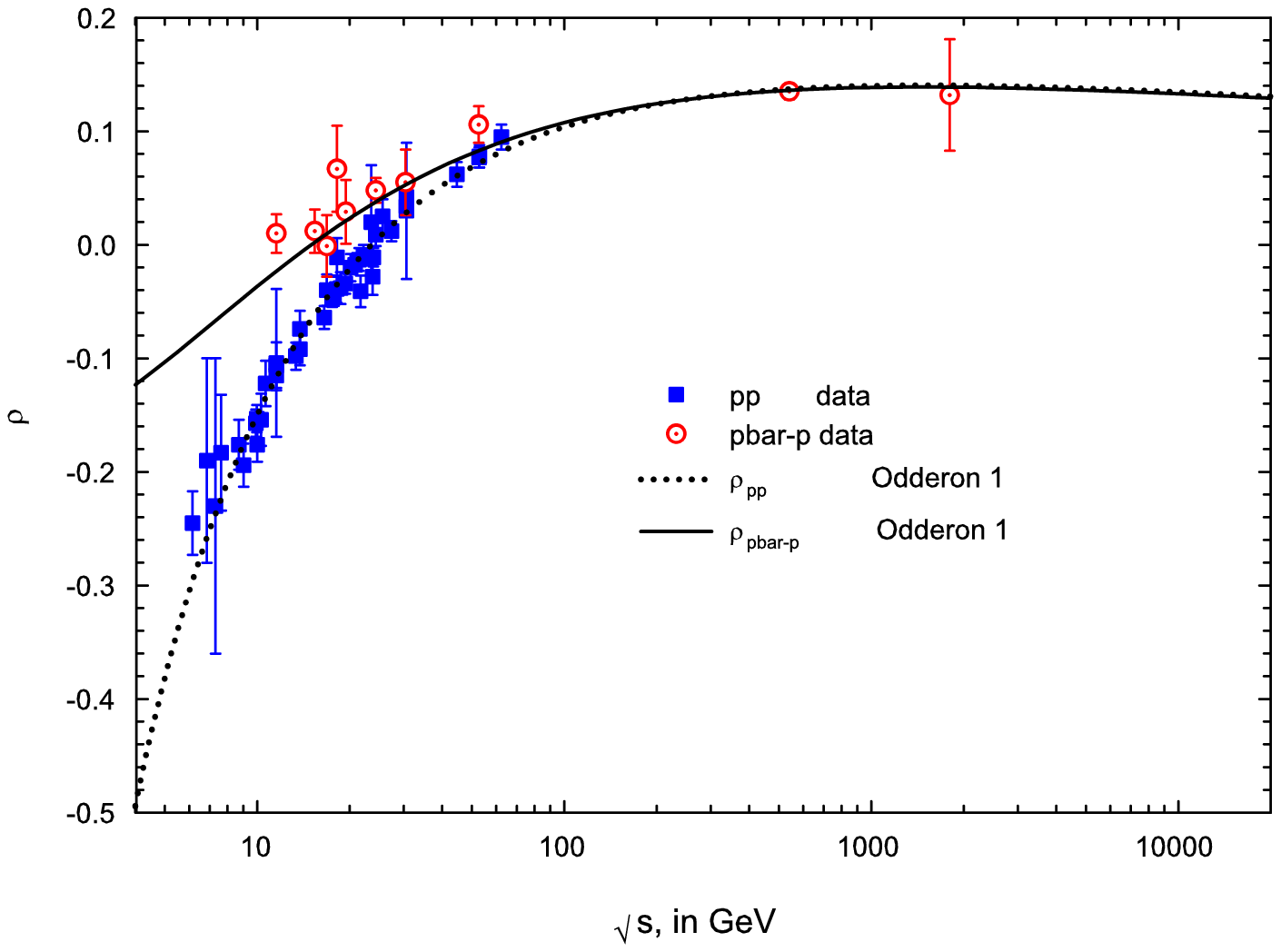,width=4in%
,bbllx=0pt,bblly=0pt,bburx=430pt,bbury=340pt,clip=%
}}
\end{center}
\caption[Odderon 1:  $\rho_{\bar pp}$ and $\rho_{pp}$]
{ \footnotesize
Odderon 1:  $\rho_{\bar pp}$ and $\rho_{pp}$. The fitted $\rho$-values, vs. $\sqrt s$, in GeV, using the 4 constraints of Equations (\ref{deriveven}), (\ref{intercepteven}), (\ref{derivodd1}) and (\ref{interceptodd1}), for odderon 1 of \eq{odd11}.  The circles are the sieved  data  for $\bar pp$ scattering and the squares are the sieved data for $pp$ scattering for $\sqrt s\ge 6$ GeV. The dotted curve ($pp$)  and the solid curve ($\bar pp$) are $\chi^2$  $\rho$-value fits, corresponding to a simultaneous $\ln^2s$ fit to cross sections and $\rho$-values  (Table \ref{table:ppoddfit}, odderon 1)  of  \eq{sigmapm1} and \eq{rhopm1}. 
}
\label{fig:rhoodd1}

\end{figure}
%
%%%%%%%%%%%%%%%%%
\begin{figure}[tbp] %Fig. 40
\begin{center}
\mbox{\epsfig{file=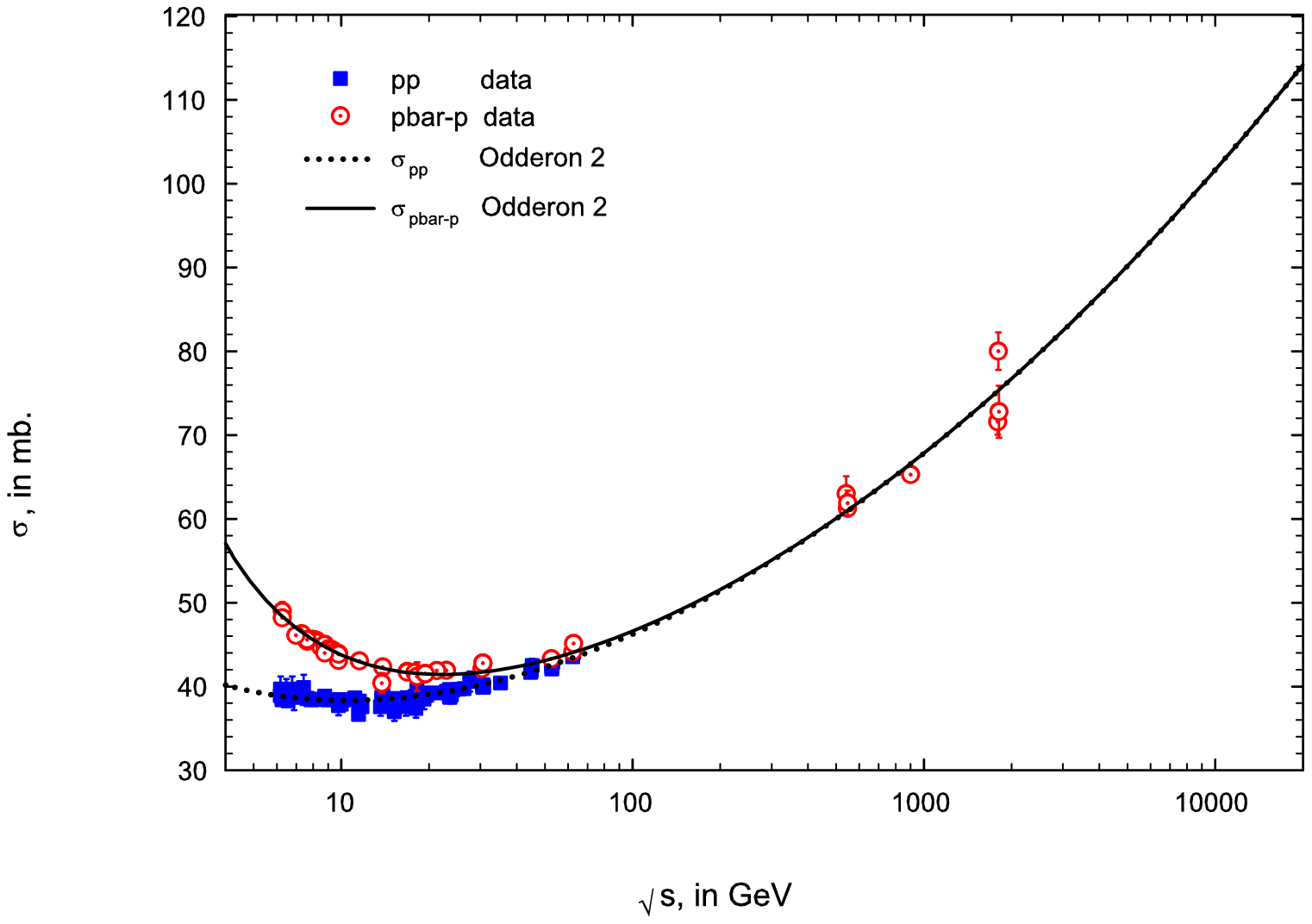,width=4.in%
,bbllx=0pt,bblly=0pt,bburx=440pt,bbury=325pt,clip=%
}}
\end{center}
\caption[Odderon 2: $\sigma_{\bar pp}$ and $\sigma_{pp}$]
{ \footnotesize
Odderon 2: $\sigma_{\bar pp}$ and $\sigma_{pp}$. The fitted total cross sections,  in mb, vs. $\sqrt s$, in GeV, using the 4 constraints of Equations (\ref{deriveven}), (\ref{intercepteven}), (\ref{derivodd2}) and (\ref{interceptodd2}), for odderon 2 of \eq{odd21}.  The circles are the sieved  data  for $\bar pp$ scattering and the squares are the sieved data for $pp$ scattering for $\sqrt s\ge 6$ GeV. The dotted curve ($pp$)  and the solid curve ($\bar pp$) are $\chi^2$ cross section fits, corresponding to a simultaneous $\ln^2s$ fit to cross sections and $\rho$-values  (Table \ref{table:ppoddfit}, odderon 2)  of  \eq{sigmapm2} and \eq{rhopm2}.  }
\label{fig:sigmaodd2}
\end{figure}
\begin{figure}[tbp] %Fig.41
\begin{center}
\mbox{\epsfig{file=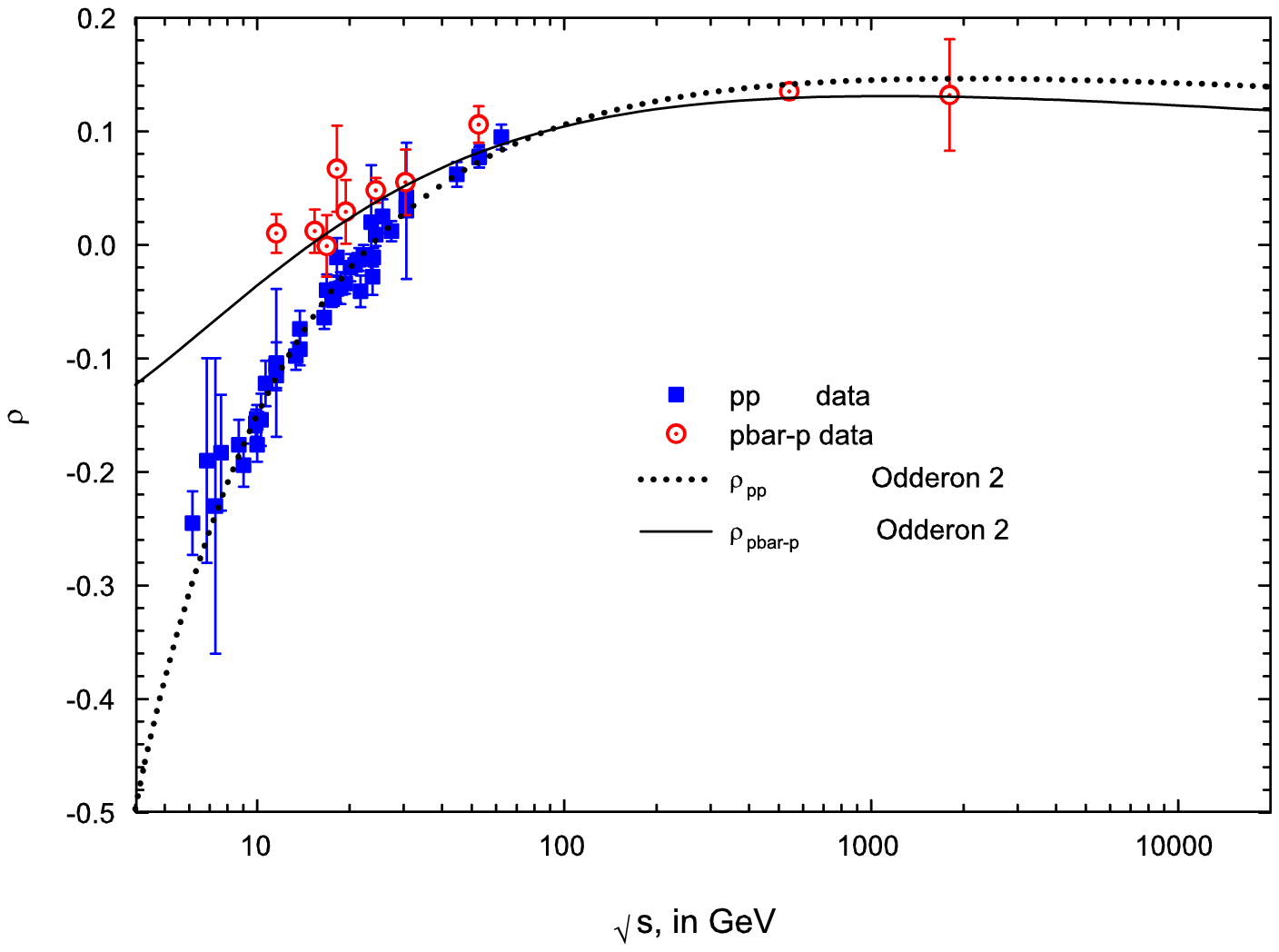,width=4.in%
,bbllx=0pt,bblly=0pt,bburx=445pt,bbury=325pt,clip=%
}}
\end{center}
\caption[Odderon 2: $\rho_{\bar pp}$ and $\rho_{pp}$]
{ \footnotesize
Odderon 2: $\rho_{\bar pp}$ and $\rho_{pp}$. The fitted $\rho$-values vs. $\sqrt s$, in GeV, using the 4 constraints of Equations (\ref{deriveven}), (\ref{intercepteven}), (\ref{derivodd2}) and (\ref{interceptodd2}), for odderon 2 of \eq{odd21}.  The circles are the sieved  data  for $\bar pp$ scattering and the squares are the sieved data for $pp$ scattering for $\sqrt s\ge 6$ GeV. The dotted curve ($pp$)  and the solid curve ($\bar pp$) are $\chi^2$  $\rho$-value fits, corresponding to a simultaneous $\ln^2s$ fit to cross sections and $\rho$-values  (Table \ref{table:ppoddfit}, odderon 2) of  \eq{sigmapm2} and \eq{rhopm2}. 
}
\label{fig:rhoodd2}
\end{figure}
%%%%%%%%%%%%%%%%%%%%%%%%
%%%%%%%%%%%%%%%%%%%%%%%%
\subsubsection{Comparison of a $s^{0.0808}$ fit to a $\ln^2s$ asymptotic fit}
Using an asymptotic term of the form  $s^{0.0808}$, Landshoff and Donnachie\cite{landshoff1,landshoff2} have parameterized the $pp$ and the $\bar pp$ scattering cross section with five parameters, using:
\ba
\sigma^+&=&56.08 s^{-0.4525}+21.70s^{0.0808}\quad {\rm for \ }pp,\label{pp1}\\
\sigma^-&=&98.39 s^{-0.4525}+21.70s^{0.0808}\quad {\rm for \ }p \bar p,\label{ppbar}
\ea
where $s$ is in GeV$^2$. Because of its great simplicity, this type of fit has been very popular, in spite of the fact that it violates unitarity as $s\rightarrow \infty$. To this objection, the authors argue than we have not yet reached high enough energies so that unitarity is important.

Using the our 4 analyticity constraints of \eq{allderiv} and the sieved data set of  of Block and Halzen\cite{bhfroissartnew}, we fit the Landshoff-Donnachie form to the same  set of $pp$ and $p\bar p$ cross section and $\rho$ data that Block and Halzen\cite{bhfroissartnew}  used for an excellent $\ln^2 s$ fit (as we have already shown in Table \ref{table:ppfitnew}).  We will now show that the satisfaction of these new analyticity constraints will require a substantial modification of the Landshoff-Donnachie formulation at lower energies, greatly altering its appeal of simplicity.

In  the high energy limit $s\rightarrow 2 m \nu$, where $\nu$ is the laboratory energy and $m$ is the proton mass, the cross sections of the  Landshoff-Donnachie type, \eq{pp1} and \eq{ppbar}, must satisfy the more general equations of the form
\ba
\sigma^\pm(\nu)&=&A\left(\frac{\nu}{m}\right)^{\alpha-1}    +B\left(\frac{\nu}{m}\right)^{\beta-1}\pm D \left(\frac{\nu}{m}\right)^{\alpha-1}\label{sigma+-},\\
\rho^\pm(\nu)&=&{1\over\sigma^\pm}\left\{-A \cot\left({\pi\alpha\over 2}\right)\left({\nu\over m}\right)^{\alpha -1} - B \cot\left({\pi\beta\over 2}\right)\left({\nu\over m}\right)^{\beta-1}+\frac{4\pi}{\nu}f_+(0)\right.\nonumber\\
&&\qquad\qquad\qquad\qquad\qquad\qquad\left.
\pm \ D\tan\left({\pi\alpha\over 2}\right)\left({\nu\over m}\right)^{\alpha -1} \right\}\label{rho+-},\ea
where the upper sign is for $pp$ and the lower for $p\bar p$ scattering and  where we have now included the $\rho$-values, $\rho^\pm$, in our fit formulas. Taking the derivative of \eq{sigma+-}, we find
\ba
\frac{d\sigma^\pm(\nu) }{d\x}&=&A(\alpha -1)\left(\frac{\nu}{m}\right)^{\alpha-2}+B(\beta -1)\left(\frac{\nu}{m}\right)^{\beta-2}\pm D(\alpha -1)\left(\frac{\nu}{m}\right)^{\alpha-2}\label{deriv+-}.
\ea

There are 6 real parameters in \eq{sigma+-} and \eq{rho+-}. For $\sigma^\pm(\nu)$, the  5 real parameters required are the 3 Regge coefficients, $A, B$ and $D $ in mb, and the 2 Regge powers, $\alpha$ and $\beta$, which are  dimensionless.  The additional real constant $f_+(0)$ introduced in \eq{rho+-} is again the subtraction constant needed for a singly-subtracted dispersion relation\cite{bc}.

Using \eq{pp1} and \eq{ppbar}, along with \eq{sigma+-},  we find  $\alpha=0.5475$ and $\beta=1.0808$, with $A=59.8$ mb, $B=22.71$ and $D=-16.38$ mb.

We again consider a transition energy $\nu_0$ which is a little higher than the energy where the resonances die out, i.e., an energy where the cross sections already have a smooth behavior (we again will use $\nu_0=7.59$ GeV, corresponding to $\sqrt s_0= 4 $ GeV).  At the transition energy $\nu_0$, we now introduce the 4 constraint conditions
\begin{eqnarray}
\sigma_{\rm av}&=&\frac{\sigma^{+}(\nu_0)+\sigma^-(\nu_0)}{2}\qquad\quad
=\quad A\y^{\alpha -1}+B\y^{\beta -1},\\
\Delta\sigma&=&\frac{\sigma^{+}(\nu_0)-\sigma^-(\nu_0)}{2}\qquad\quad
=\quad D\y^{\alpha -1},\\
m_{\rm av}&=&\frac{1}{2}\left[\frac{d\sigma^{+}}{d\left(\frac{\nu}{m}\right)}+\frac{d\sigma^{-}}{d\left(\frac{\nu}{m}\right)}\right]_{\nu =\nu_0}
= \quad A(\alpha -1)\y^{\alpha - 2}+B(\beta -1)\y^{\beta - 2},\\
\Delta m&=&\frac{1}{2}\left[\frac{d\sigma^{+}}{d\left(\frac{\nu}{m}\right)}-\frac{d\sigma^{-}}{d\left(\frac{\nu}{m}\right)}\right]_{\nu =\nu_0}
=\quad D(\alpha -1)\y^{\alpha - 2}.
\end{eqnarray}
At $\sqrt s_0 =4$ GeV, Block and Halzen\cite{bhfroissartnew} found  that
\ba
\sigma^+(\nu_0)&=&40.18 \quad\quad{\rm mb,}
\qquad \sigma^-(\nu_ 0)\quad=\quad\quad 56.99  \quad\quad{\rm mb,}\label{sigs}\\
\left.\frac{d\sigma^+\ \ }{d\x}\right|_{\nu =\nu_0}&=&-0.2305 \quad{\rm mb,}\quad
\left.\frac{d\sigma^-\ \ }{d\x}\right|_{\nu=\nu_0}=-1.4456\quad{\rm mb,}
\ea
using a local fit in the neighborhood of $\nu_0$. 
For $\nu_0=7.59$ GeV, these values yield the 4 analyticity constraints 
\begin{eqnarray}
\sigma_{\rm av}&=&48.59 \quad\quad\,{\rm mb},\qquad
\Delta\sigma=\quad -8.405 \quad{\rm\  mb},\label{delsig&sigav}\\
m_{\rm av}&=&-0.8381\quad{\rm mb},\qquad
\Delta m=\quad 1.215\quad\quad{\rm \ mb}\label{Delm&mav}.
\end{eqnarray}

Using these experimental values, we can now write the 4 constraints of \eq{allderiv} at energy $\nu_0=7.59 $ GeV.  From  \eq{sigma+-} and \eq{deriv+-},  in terms of the one free parameter $A$, we find
\begin{eqnarray}
\alpha&=&1+\frac{\Delta m}{\Delta \sigma}\y,\label{derivoddlandshoff}\\
D&=&\Delta \sigma\y^{1-\alpha}\label{interceptoddlandshoff},\\
\beta(A)&=&1+\frac{m_{\rm av}\y-A(\alpha -1)\y^{\alpha -1}}{\sigma_{\rm av}-A\y^{\alpha -1}},\label{betaofA}\\
B(A)&=&\sigma_{\rm av}\y^{1-\beta}-A\y^{\alpha-\beta}.\label{BofA}
\end{eqnarray}
In the above constraint equations, we have utilized the two experimental cross sections and their slopes at the transition energy $\nu_0$, the energy at which we join on to the asymptotic fit.  It should be emphasized that analyticity (see \eq{allderiv}) requires us to satisfy these 4 constraints.

We note  that the odd amplitude is {\rm completely} specified. This is true even {\em before} we make a fit to the high energy data. The two odd analyticity conditions constrain the odd parameters to be
\ba
D&=&-28.56\quad{\rm mb},\quad\quad\,\quad D_{\rm LD}\quad=\quad-16.38\quad{\rm mb}, \label{delta}\\
\alpha&=&0.4150,\quad
\qquad\qquad\quad\alpha_{\rm LD}\quad=\quad0.545\label{alpha},
\ea
where we have contrasted  these value with the values found by Landshoff and Donnachie, $D_{\rm LD}$ and $\alpha_{\rm LD}$, which are clearly incompatible with our analyticity requirements.

Now, armed with $D$ and $\alpha$,  we use the two constraint equations, \eq{betaofA} and \eq{BofA},  to simultaneously fit a  sieved data set\cite{{sieve},{bhfroissartnew}} of high energy cross sections and $\rho$-values for $pp$ and $p\bar p$ with energies above $\sqrt s =6$ GeV, derived from the Particle Data Group\cite{pdg} archive.  This data set has already  been successfully employed to make an excellent $\ln^2 s$ fit of the type  used in \eq{lnsqs}, enforcing the {\em same} 4 analyticity constraints\cite{bhfroissartnew} as we use here. 

We now make a $\chi^2$ fit to the two free parameters $A$ and $f_+(0)$. As before, it should be noted that the subtraction constant $f_+(0)$ only enters into   $\rho^\pm$-values and {\em not} into  cross section determinations $\sigma^\pm$. In essence, this cross section fit is a one parameter fit, $A$ {\em only}.

The results of the fit are given  in Table \ref{table:Landshoff}.  The $pp$ and $\bar pp$ cross sections from Table \ref{table:Landshoff} can be rewritten as 
\ba
\sigma^+&=&23.97 s^{-0.5850}+33.02s^{0.0255}\quad {\rm for \ }pp,\\
\sigma^-&=&109.1 s^{-0.4525}+33.02s^{0.0255}\quad {\rm for \ }p \bar p,
\ea
which are in sharp contrast to the Landshoff-Donnachie cross sections given in \eq{pp1} and \eq{ppbar}.

For 185 degrees of freedom, the renormalized\cite{sieve} $\chi^2$ per degree of freedom is 21.45, yielding the incredibly large value of 3576.24 for the total $\chi^2$. Thus, there is an essentially zero probability that a fit of the Landshoff-Donnachie type, of the form given in \eq{pp1} and \eq{ppbar}, is a good representation of the high energy data. Certainly, at the very high energy end it violates unitarity. We now see that it does not have the proper shape to satisfy analyticity  at the lower energy end.  Clearly, this form  requires substantial ad hoc structural changes to join on to the low energy constraints. Thus, its simplicity of form---its primary virtue---requires serious modification. This is graphically seen in Figures \ref{fig:sigmapplandshoff} and \ref{fig:rhopplandshoff}, which plot $\sigma$ in mb  and $\rho$, respectively,  against the c.m. energy $\sqrt s$ in GeV. One sees immediately that the fitted curves go completely under all of the higher energy experimental data.

%%%%%%%%%%%%%%%%%%%%%%%%%%%%%%%%%%
%  Table 16 NEW PP landshoff

\begin{table}[tbp]                   % Use "table" environment, but also
				 % use  "tabular" environment below.
%
\begin{center}
\def\arraystretch{1.}            % Make the space between rows in the Table,
				  % 1.5 x bigger than the default spacing.
     \caption[The fitted results for a 2-parameter $\chi^2$ fit of the Landshoff-Donnachie type ($\sigma^\pm\sim A(\nu/m)^{\alpha-1} +B(\nu/m)^{\beta-1}\pm D(\nu/m)^{\alpha-1}$)]
{\protect\small The fitted results for a 2-parameter $\chi^2$ fit of the Landshoff-Donnachie type ($\sigma^\pm\sim A(\nu/m)^{\alpha-1} +B(\nu/m)^{\beta-1}\pm D(\nu/m)^{\alpha-1}$),  where $\nu$ is the laboratory nucleon energy and $m$ is the proton mass) simultaneously to both the total cross sections and $\rho$-values for $pp$ and $\bar pp$ scattering.  The renormalized $\chi^2_{\rm min}$ per degree of freedom,  taking into account the effects of the $\delchimax$ cut\cite{sieve}, is given in the row  labeled ${\cal R}\times\chi^2_{\rm min}$/d.f. The errors in the fitted parameters have been multiplied by the appropriate $r_{\chi2}$\cite{sieve}.    \label{table:Landshoff}}. 

\begin{tabular}[b]{|l||c||}
\hline
{Parameters}&Even Amplitude\\
      \hline
	%\multicolumn{2}{|c||}{\ \ \ \ \  Even Amplitude}\\
	%\cline{1-2}
      $A$\ \ \   (mb)&$44.65\pm 0.0031$ \\ 
      $B$\ \ \   (mb)&$33.60$ \\ 
	$\beta$&$1.0255$\\
	$f_+(0)$ (mb GeV)&$2.51\pm 0.57$\\
      \hline
	%\multicolumn{2}{|c||}{\ \ \ \ \  Odd Amplitude}\\
{}&Odd Amplitude\\
	\hline
      $D$\ \ \   (mb)&$-28.56$\\
      $\alpha$&$0.415$ \\ 
	\cline{1-2}
     	\hline
	\hline
	$\chi^2_{\rm min}$&3576.2\\
	${\cal R}\times\chi^2_{\rm min}$&3968.1\\ 
	degrees of freedom (d.f.)&185\\
\hline
	${\cal R}\times\chi^2_{\rm min}$/d.f.&21.45\\
\hline

\end{tabular}
     %\vspace{1in} \\
\end{center}
\end{table}

\def\arraystretch{1}  %Restore the default row spacing in the Table.

\begin{figure}[tbp] %Fig. 42
\begin{center}
\mbox{\epsfig{file=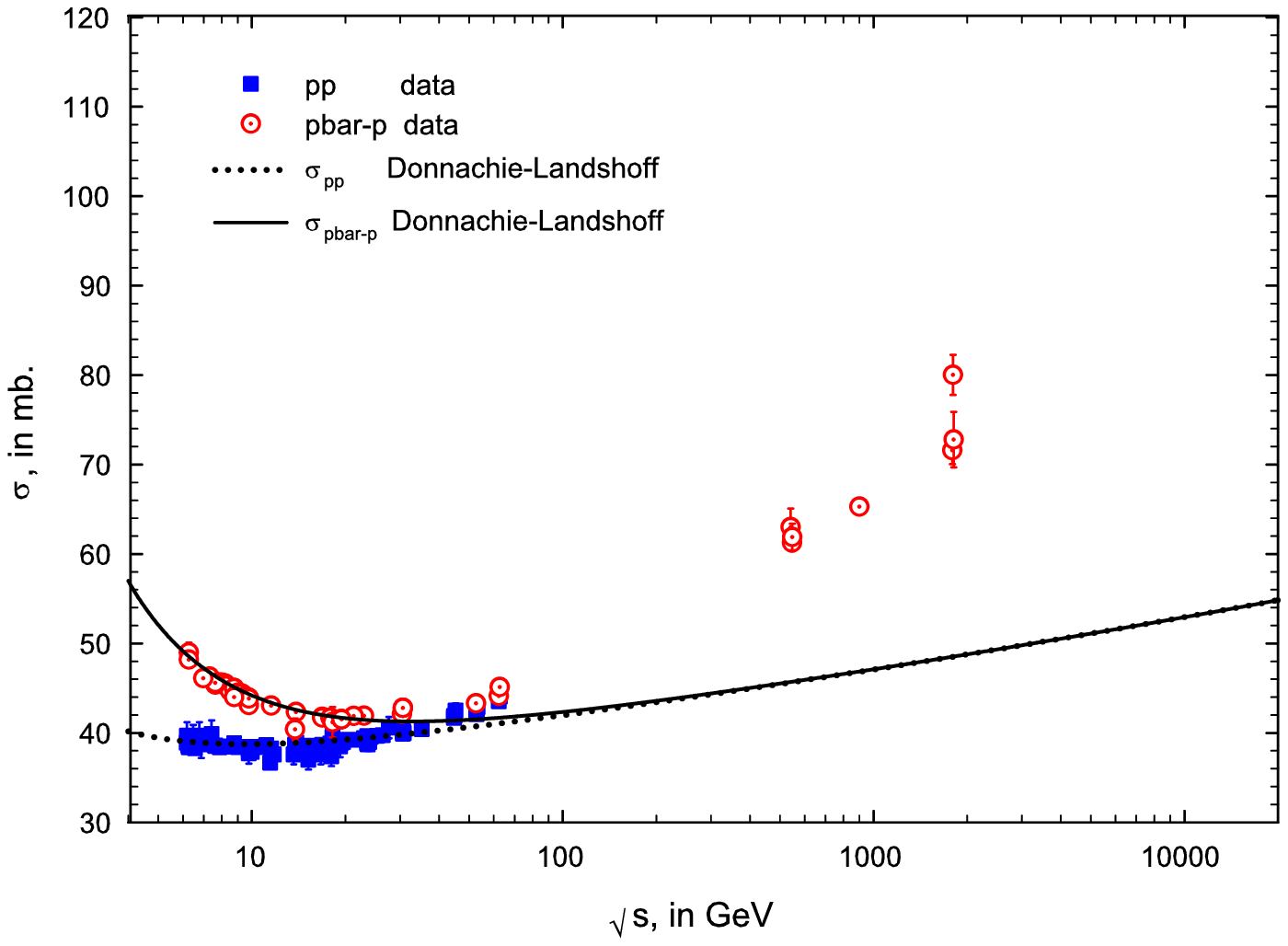,width=3.8in%
,bbllx=0pt,bblly=0pt,bburx=411pt,bbury=325pt,clip=%
}}
\end{center}
\caption[$\sigma_{p p}$ and $\sigma_{\pbar p}$ from a constrained Donnachie-Landshoff fit]
{ \footnotesize $\sigma_{p p}$ and $\sigma_{\pbar p}$ from a constrained Donnachie-Landshoff fit. 
The fitted total cross sections in mb, vs. $\sqrt s$, in GeV, using the 4 constraints of Equations (\ref{delsig&sigav}) and (\ref{Delm&mav}). The circles are the sieved data\cite{{bhfroissartnew},{sieve}}  for $\pbar p$ scattering and the squares are the sieved data\cite{{bhfroissartnew},{sieve}}  for $p p$ scattering for $\sqrt s\ge 6$ GeV. The solid curve ($\pbar p$) and  the dotted curve ($\pbar p$) are the  $\chi^2$ fits from Table \ref{table:Landshoff}.
}
\label{fig:sigmapplandshoff}
\end{figure}
%
%%%%%%%%%%%%%%%%%%%%%%%%%%
%%%%%%%%%%%%%%%%%%%%%%%%%%%
\begin{figure}[tbp] %Fig.43
\begin{center}
\mbox{\epsfig{file=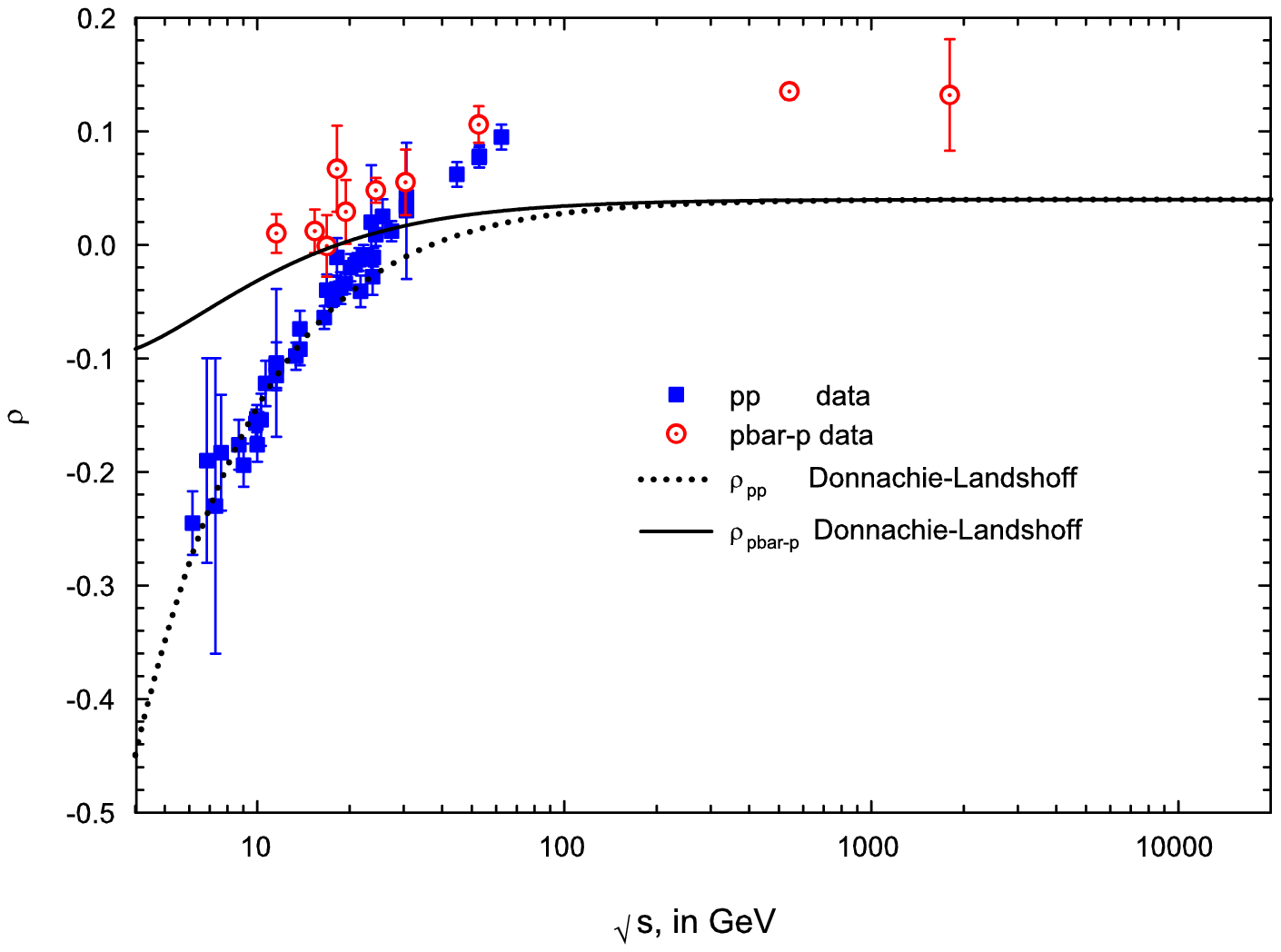,width=3.8in%
,bbllx=0pt,bblly=0pt,bburx=411pt,bbury=325pt,clip=%
}}
\end{center}
\caption[$\rho_{p p}$ and $\rho_{\pbar p}$ from a constrained Donnachie-Landshoff fit]
{ \footnotesize$\rho_{p p}$ and $\rho_{\pbar p}$ from a constrained Donnachie-Landshoff fit. 
The fitted $\rho$-values vs. $\sqrt s$, in GeV, using the 4 constraints of Equations (\ref{delsig&sigav}) and (\ref{Delm&mav}).  The circles are the sieved data\cite{{bhfroissartnew},{sieve}}  for $\pbar p$ scattering and the squares are the sieved data\cite{{bhfroissartnew},{sieve}}  for $p p$ scattering for $\sqrt s\ge 6$ GeV. The solid curve ($\pbar p$) and  the dotted curve ($\pbar p$) are the  $\chi^2$ fits from Table \ref{table:Landshoff}.
  }
\label{fig:rhopplandshoff}
\end{figure}
%%%%%%%%%%%%%%%%%%%%%%%%%%%%%%%%%%%%%%%
%%%%%%%%%%%%%%%%%%%%%%%%%%%%%%%%%%%%%%%%%%%
We bring to the reader's attention that a $\ln^2s$ fit of the form
\be
\sigma^\pm(\nu)=c_0 +c_1\ln(\nu/m)+c_2\ln^2(\nu/m)+\beta_{\cal P'}(nu/m)^{-.5}\pm \delta (\nu/m)^{\alpha -1},\label{lnsqs}
\ee 
was carried out on the {\em same} sieved sample  in ref. \cite{bhfroissartnew}, using the {\em same} 4 analyticity constraints,  where it gave a renormalized $\chi^2$ per degree of freedom of 1.095 for 184 degrees of freedom, an excellent fit (for details of this fit, see Table \ref{table:ppfitnew} that we discussed earlier).

In conclusion,  a functional form of the type
\ba
\sigma(pp)&=&A's^{\alpha-1}+B's^{\beta-1},\\
\sigma(p\bar p)&=&C's^{\alpha-1}+B's^{\beta-1},
\ea 
with $\beta \sim 1.08$, although conceptually very elegant in its simplicity, can not be used to fit high energy scattering using a transition energy   $\sqrt s_0>4$ GeV (and indeed, it also can be shown to be fail for transition energies\footnote{Since the data that are used in the fit {\em start} at 6 GeV, we stop at 6 GeV.} even up to $\sqrt s_0\la$6 GeV), since it can not satisfy the 4 analyticity requirements at these energies. In addition, as already noted, the term in $s^{\beta -1}$ (for $\beta>1$) violates unitarity at the highest energies. Therefore this type of parametrization, without {\em major} modification, is effectively excluded, whereas the $\ln^2s$ fit of Block and Halzen\cite{bhfroissartnew} satisfies  unitarity in a natural way, as well as satisfying the 4 analyticity constraints. 
%%%%%%%%%%%%%%%%%%%%%%%%%%%%%%%%
\subsubsection{Summary of  real analytic amplitude fits to high energy accelerator data}
We have shown that when we use the combination of:
\begin{enumerate}
\item a sifted data set for $\pi^\pm p$ and $pp$ and $\bar pp$ cross sections and $\rho$-values, 
\item  
the 2 new analyticity constraints for $\gamma p$ that anchor its cross and its derivative to their experimental values at a low energy just above resonances, or  the 4 new analyticity constraints that anchor $\pi^\pm p$ and $pp$ and $\bar pp$ the cross sections and their derivatives by the experimental values at low energies just above the resonances, and
\item $\chi^2$ fits to even cross sections  which go as 
\ba
\sigma^0(\nu)&=&c_0+c_1\ln\left(\frac{\nu}{m}\right)+c_2\ln^2\left(\frac{\nu}{m}\right)+\beta_{\cal P'}\left(\frac{\nu}{m}\right)^{\mu -1},\label{s00}
\ea
\end{enumerate}
we find that
\begin{itemize}
\item All give very satisfactory goodness-of-fit to the hypothesis that the cross section goes as $\ln^2s$, as $s\rightarrow\infty$. The new  renormalized goodness-of-fit (${\cal R}\chi^2_{\rm min}$/d.f.$\sim 1$) is due to using a sifted data set.
\item The nucleon-nucleon fits are completely insensitive to whether you use the E710/E811 {\em and} the CDF cross sections at the Tevatron Collider or whether you omit the CDF point, i.e., higher energy predictions are completely stable. This is due to employing the low energy analyticity constraints, as well as using a sifted data set.
\item When fits where made to a $\ln s$ asymptotic behavior (setting the coefficient $c_1=0$ in \eq{s00}), this hypothesis was completely ruled out statistically. Because of the 4 constraints at low energy, the effect on the fit of choosing  a somewhat poorer parametrization of the high energy amplitude is greatly magnified.
\item Odderon amplitudes were shown to be negligibly small, so that an odd amplitude of the form $\frac{4\pi}{p}f_-=-Ds^{\alpha-1}e^{i\pi(1-\alpha)/2}$, which has $\Delta\sigma, \Delta\rho\rightarrow0$ as $s\rightarrow\infty$, is all that required for $\pi^\pm p$ and nucleon-nucleon scattering.
\item Cross section fits of the Landshoff-Donnachie type, 
\ba
\sigma^+&=&56.08 s^{-0.4525}+21.70s^{0.0808}\quad {\rm for \ }pp,\label{pp2}\\
\sigma^-&=&98.39 s^{-0.4525}+21.70s^{0.0808}\quad {\rm for \ }p \bar p,
\ea
going asymptotically as $s^{0.808}$, have been conclusively ruled out---again because of the 4 analyticity constraints at low energy magnifying the poorness of the overall fit.
\item  For the $\ln^2 s$ cross section fit of \eq{s0}, only 2 parameters, $c_1$ and $c_2$, have to be fitted  because of the 4 analyticity constraints.  Therefore, the statistical uncertainties of the fit parameters for nucleon-nucleon scattering are now {\em very small}, thus giving a very accurate cross section prediction ($\sim 1$\%) at the LHC, as well as an accurate $\rho$-value.
\item The use of analyticity constraints allows us to  conclude that the Froissart bound is already saturated at present day energies, thus giving us confidence that very accurate $pp$ cross sections can now be predicted at the high cosmic ray energies where extensive air shower experiments have measured  $p$-air cross sections.
\end{itemize}

%%%%%%%%%%%%%%%%%%%%%%%%%%%%

\section{Cosmic ray $p$-air cross sections}\label{section:crpair}
The energy range of cosmic ray  protons covers not only the energy of the
Large Hadron Collider, but extends well beyond it. Currently, $p$-air cross sections have been measured up to $\sqrt s\sim 80$ TeV. Using extensive air showers, cosmic ray experiments can measure the penetration in the atmosphere of these very high energy protons---however, extracting proton--proton cross sections from cosmic ray observations is far from straightforward\cite{gaisser}. By a variety of experimental techniques, extensive air shower experiments map the atmospheric depth at which cosmic ray initiated showers develop. 
%%%%%%%%%%%%%%%%%%%%%%%%%%%%%%%%%%%%%%%%%%
\subsection{Brief description of  extensive air shower measurements}
We now give a brief description of  extensive air showers development. A high energy particle (proton, alpha particle, iron nucleus, etc.) enters the earth's atmosphere and undergoes a primary (first) interaction with an air atom. This encounter initiates a cascade of secondary particles---called an extensive air shower---which keeps growing in size until ionization losses exceed bremmstrahlung losses. The depth in the atmosphere at which this turnover in size takes place is called  $X_{\rm max}$, the shower maximum. From here on down into the atmosphere, the shower size diminishes until secondary particle energies are below particle production thresholds. Various experimental signals from the shower, such as radio waves, \u Cerenkov and scintillation light are emitted, allowing  shower development to be measured in terrestrial laboratories. In principle, a measurement of  the distribution of $X_1$, the first interaction point  of $p$-air collisions, would allow us to find the mean free path $\lpa$ in a straight-forward way. 

Unfortunately, no existing cosmic ray experiment is capable of detecting the $X_1$ distribution, and hence, $\lpa$ {\em cannot} be directly measured.  %
Instead, what is measured is the $X_{\rm max}$ distribution of extensive air showers. Fluorescence detectors used in the Fly's Eye\cite{fly} and HiRes\cite{hires} experiments measure the $X_{\rm max}$ distributions directly, whereas  ground array detectors such as the AGASA array\cite{akeno} must first convert their direct observations into a $X_{\rm max}$ distribution. In order to find $X_{\rm max}$ from a given extensive air shower, a shower profile function\cite{hillas} function is fitted to the number of electrons as a function of the atmospheric depth. A typical example taken from the Fly's Eye experiment\cite{fly} is shown in Fig. \ref{fig:gaisser}. % 

%%%%%%%%%%%%%%%%%%%%%%%%%%%
\begin{figure}[tbp] %Fig.45
\begin{center}
\mbox{\epsfig{file=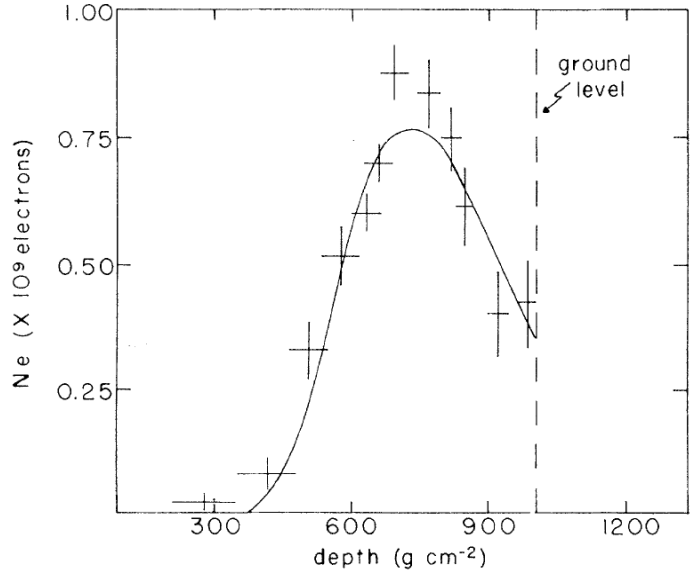,width=3.8in%
,bbllx=0pt,bblly=165pt,bburx=585pt,bbury=645pt,clip=%
}}
\end{center}
\caption[A Fly's Eye extensive air shower]
{ \footnotesize
A Fly's Eye extensive air shower that survives all cuts. The curve is a Gaisser-Hillas shower-development function\cite{hillas}:  Shower parameters $E=1.3$ EeV and $X_{\rm max}=727\pm 33$ g cm$^{-2}$ give the best fit. Figure taken from Ref. \cite{fly}. 
  }
\label{fig:gaisser}
\end{figure}
%%%%%%%%%%%%%%%%%%%%%%%%%%%%%%%%%%%%%%%%
%%%%%%%%%%%%%%%%%%%%%%%%%%%%%%%%%%%%%%%%%%%%%
From these measurements of  $X_{\rm max}$, they plot the logarithm of the  number of  $X_{\rm max}$ per unit atmospheric depth against the atmospheric depth, in g cm$^{-2}$, to measure $\Lambda_m$,   the slope of the exponential tail. The result for the Fly's Eye collaboration\cite{fly} is shown in Fig. \ref{fig:Xmax}, where they measured $\Lambda_m=73\pm9$ g cm$^{-2}$.  Pryke\cite{pryke} has made  extensive Monte Carlo studies of shower profiles and their distribution in the atmosphere, as a function of various shower development models. An example of his computer-generated $X_{\rm max}$ distribution is shown in Fig. \ref{fig:Xmaxpryke}.

%%%%%%%%%%%%%%%%%%%%%%
%%%%%%%%%%%%%%%%%%%%%%%%%%%%%%%%%%%%%%%%%%%%%
\begin{figure}[tbp] %Fig.46
\begin{center}
\mbox{\epsfig{file=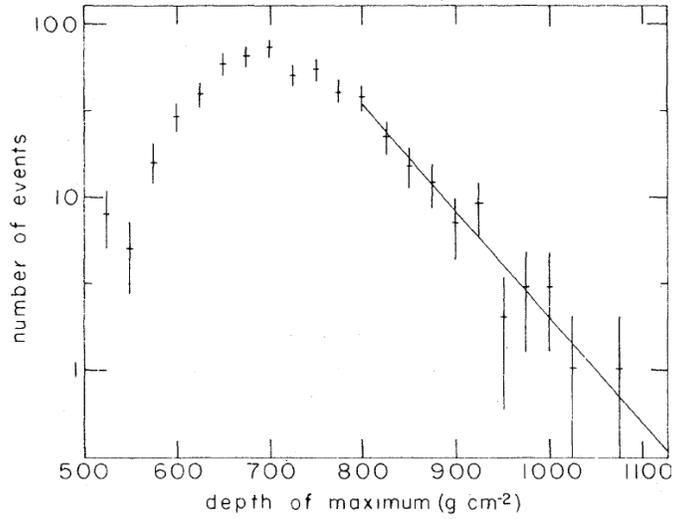,width=3.5in%
,bbllx=0pt,bblly=165pt,bburx=590pt,bbury=645pt,clip=%
}}
\end{center}
\caption[Fly's Eye $X_{\rm max}$ distribution]
{ \footnotesize Fly's Eye $X_{\rm max}$ distribution. Distribution of depth of  shower maxima $X_{\rm max}$ for data whose fitting errors are estimated to be $\delta x<125$ g cm$^{-2}$.  The slope of the exponential tail is $\Lambda_m=73\pm9$ g cm$^{-2}$. Figure taken from the Fly's Eye collaboration, Ref. \cite{fly}. 
  }
\label{fig:Xmax}
\end{figure}
%%%%%%%%%%%%%%%%%%%%%%%%
\begin{figure}[h] %Fig.47
\begin{center}
\mbox{\epsfig{file=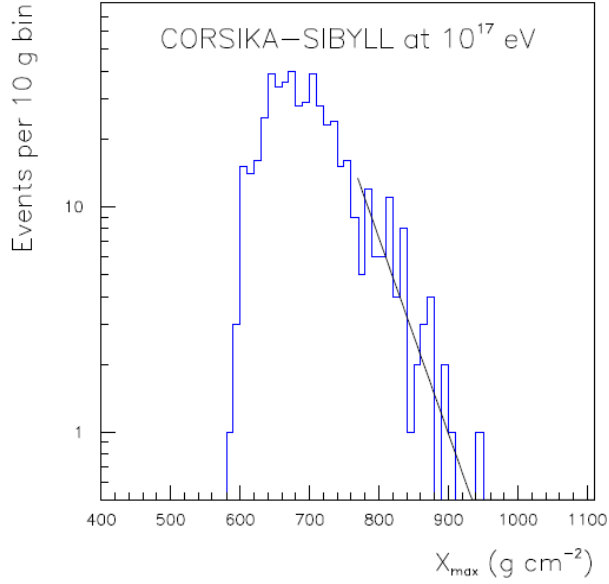,width=3.5in%
,bbllx=50pt,bblly=190pt,bburx=520pt,bbury=600pt,clip=%
}}
\end{center}
\caption[A computer-generated $X_{\rm max}$ distribution]
{ \footnotesize A computer-generated $X_{\rm max}$ distribution.
An example $X_{\rm max}$ distribution with exponential trailing edge fit. Figure taken from Ref. \cite{pryke}. 
  }
\label{fig:Xmaxpryke}
\end{figure}
%%%%%%%%%%%%%%%%%%%%%%%%%%%%%%%%%%%%%%%%

These extensive air shower experiments measure the shower attenuation length of the protons in the atmosphere
%($\lpa\propto \frac{1}{\spai}$), with
with
\begin{equation}
\Lambda_m =k \lpa=k\frac{14.4m}{\spai} =k\frac{24100}{\spai}\,,  \label{eq:Lambda_m}
\end{equation}
(with $\Lambda_m$ in g cm$^{-2}$, the proton mass $m$ in g and $\spai$ in mb; the factor 14.4$m$ being the mean atomic weight of air in g), %}, %
but also depends critically on the proton inelasticity and the properties of the pion interactions, which determines the rate at  
which the energy of the primary proton
 is dissipated into electromagnetic shower energy observed in the
 experiment. The latter effect is taken into account in Eq.\,(\ref{eq:Lambda_m})
 by the parameter $k$. The departure of $k$ from unity depends on the inclusive particle production cross sections for nucleon and meson interactions in the light nuclear targets of the atmosphere and their energy dependences. An a priori  knowledge of $k$, which is model dependent and, in principle, energy dependent, is essential to the extraction of the $p$-air cross section $\spai$. We emphasize that the $p$-air production cross section $\spai$ that is deduced from  \eq{eq:Lambda_m} is the cross section for  {\em particle emission} in the primary interaction, and not the inelastic total cross section, which includes diffaction.

\subsection{Extraction of $\sigma_{pp}$, the total $pp$ cross section}

 The extraction of the pp cross section from the cosmic ray data is a two-stage process. First, one calculates the $p$-air total cross section from
 the  production cross section inferred in Eq.\,(\ref{eq:Lambda_m}),  where
\begin{equation}
\spai = \spa - \spae - \spaqe \,.  \label{eq:spa}
\end{equation}
 Next, the Glauber method\,\cite{yodh} transforms the
 value of $\spai$ into a proton-proton total cross section $\sigma_{pp}$; all the necessary steps are calculable in the theory. In \eq{eq:spa} the measured cross section $\spai$ for particle production is supplemented with $\sigma_{p-{\rm air}}^{\rm el}$ and $\sigma_{p-{\rm air}}^{\rm q-el}$, the elastic and quasi-elastic cross sections---such as the reaction $p+p\rightarrow N^*_{1238}+p$---respectively, as calculated by the Glauber theory, to obtain the total cross section $\spa$. The subsequent relation between $\spai$ and $\sigma_{pp}$ critically involves $B_{pp}$, the slope of the forward scattering amplitude for elastic $pp$ scattering, which from \eq{Bof0}, is given by
\be
B_{pp}=\left.\frac{d}{dt}\left(\ln\frac{d\sigma^{\rm el}_{pp}}{dt}\right)\right|_{t=0}.%\label{Bof0}
\ee
An example Glauber calculation by Engel et al.\cite{gaisser} is shown in Fig. \ref{fig:p-air},
which  plots  $B$ in (GeV/c)$^{-2}$ vs. $\sigma_{pp}$ in mb, with the 5 curves having different values of $\spai$.
 This summarizes the reduction procedure
 from the measured quantity $\Lambda_m$ (of Eq.\,\ref{eq:Lambda_m}) to  
$\sigma_{pp}$\cite{gaisser}.
 Also plotted in Fig.\,\ref{fig:p-air} is a dashed curve, a plot of $B$ against $\sigma_{pp}$ which will be discussed later. 

%%%%%%%%%%%%%%%%%%%%%%%%%%%%%%%%%
%%%%%%%%%%%%%%%%%%%%%%%%%%%%%%%%%%%%%%
\begin{figure}[h]%Fig. 48
\begin{center}
\mbox{\epsfig{file=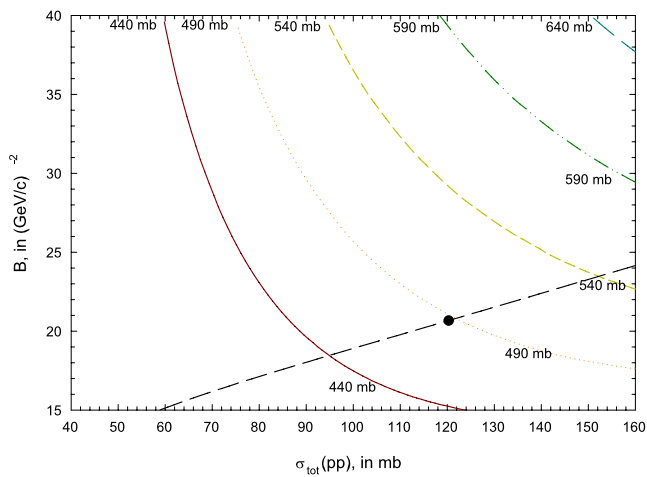%
              ,width=3.8in,bbllx=55pt,bblly=220pt,bburx=520pt,bbury=560pt,clip=%
}}
\end{center}
\caption[The $B$ dependence of the $pp$ total cross section $\sigma_{pp}$]
{ \footnotesize
 The $B$ dependence of the $pp$ total cross section $\sigma_{pp}$, with the nuclear slope $B$ in (GeV/c)$^{-2}$ and $\sigma_{pp}$ in mb.  The
 five curves are lines  of constant  $\spai$,  of 440, 490, 540, 590 and
 640 mb---the central value is the {\em published} Fly's Eye value, and the others
 are $\pm 1\sigma$ and $\pm 2\sigma$. The dashed curve is a plot of an Aspen model (QCD-inspired eikonal) fit of $B$ against $\sigma_{pp}$.  The dot is the fitted value for
 $\sqrt s=30$ TeV, the Fly's Eye energy. Taken from Ref. \cite{BHS}.}
\label{fig:p-air}
\end{figure}%
%%%%%%%%%%%%%%%%%%%%%%%%%%%%%%%%%%%%%%%%%%%%
%%%%%%%%%%%%%%%%%%%%%%%%%%%%%%%%%%%%%%%%%%%

 A significant drawback of this extraction method is that one depends on a detailed model of proton-air interactions to complete the loop between the measured interaction length $\Lambda_m$ and the cross section $\spai$, i.e., to determine the value of $k$ in \eq{eq:Lambda_m}. 
%%%%%%%%%%%%%%%%%%%%%%%%%%%%%%%%%%%%%%%
%%%%%%%%%%%%%%%%%%
\subsection{Original analysis of Fly's Eye and AGASA  experiments} 
In this Section, we will discuss only the {\em published} results of the Fly's Eye\cite{fly} and AGASA\cite{akeno}, as they were originally deduced by their authors some 30 years ago. It is at this point important to recall Eq.\,(\ref{eq:Lambda_m}) and remind ourselves that the measured experimental quantity is $\Lambda_m$ and {\em not} $\lambda_{p-\rm air}$, the $p$-air mean free path. We emphasize that  the extraction of $\spai$ from the measurement of $\Lambda_m$ requires {\em knowledge} of the parameter $k$. The measured depth $X_{\rm max}$ at which a shower reaches maximum development in the atmosphere, which is the basis of the cross section measurement in Ref.~\cite{fly}, is a combined measure of the depth of the first interaction, which is determined by
 the inelastic production cross section, and of the subsequent shower development,
 which has to be corrected for. 
 The rate of shower development and its fluctuation are the origin of the deviation of $k$ from unity in Eq.\,(\ref{eq:Lambda_m}). Its predicted values range from 1.5 for a model where the inclusive cross section exhibits Feynman scaling, to 1.1 for models with large scaling violations\,\cite{gaisser}. The comparison between prediction and experiment in Fig.\,\ref{fig:sigtodorpp} is further confused by the fact that the Akeno Giant Air Shower Array (AGASA)\cite{akeno} and Fly's Eye\cite{fly} experiments used different values of $k$ in the analysis of their data, i.e., AGASA used $k=1.5$ and Fly's Eye used $k=1.6$. Using $k=1.6$, the Fly's Eye group measured $\Lambda_m=45\pm5$ g/cm$^{2}$ and deduced that $\spai=540\pm50$ mb.  The $\spai$ values used in Fig. \ref{fig:p-air} are 5 values, which are the mean value of their measurement, $\pm 1\sigma$ and $\pm 2\sigma$. Block, Halzen and Stanev(BHS)\cite{BHS} converted these original $\spai$ values into total $pp$ cross section $\sigtot(pp)$, by using a QCD-inspired eikonal model (the Aspen model) connecting the nuclear slope $B$ with $\sigtot(pp)$, which is the dashed curve in Fig. \ref{fig:p-air}. The large dot on this curve is the c.m. energy of the Fly's Eye experiment, $\sqrt s=30$ TeV.  They also converted the AGASA published values\cite{akeno} of $\spai$ into $\sigtot(pp)$.  These converted cosmic ray values are shown in Fig. \ref{fig:sigtodorpp}, along with accelerator data and the Aspen model fit that BHS made to the accelerator data.  Clearly, {\em all} of the cosmic ray data are too high, compared to accelerator predictions.
%%%%%%%%%%%%%%%%%%%%%%%%%%%%%%%%5
%%%%%%%%%%%%%%%%%%%%%%%%%%%%%%%%%%%%%%%%%%%%%%%%%%
\begin{figure}[h] %Fig. 49
\begin{center}
\mbox{\epsfig{file=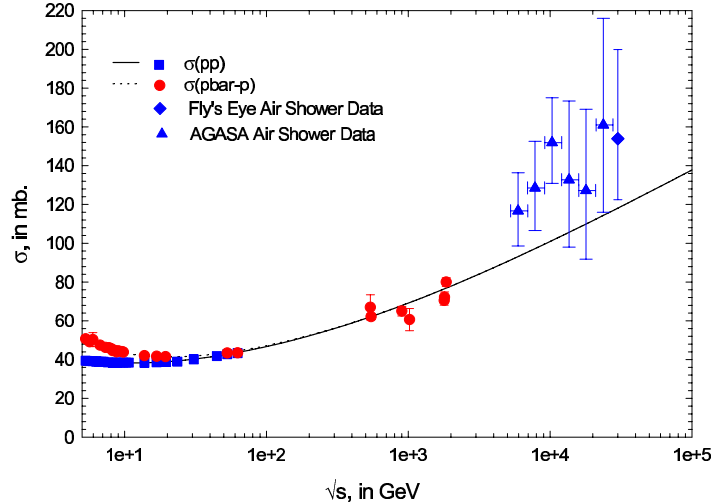%
            ,width=3.8in,bbllx=40pt,bblly=200pt,bburx=580pt,bbury=580pt,clip=%
}}
\end{center}
\caption[Published cosmic ray $\sigma_{pp}$ values from AGASA and Fly's Eye, appended to accelerator data]{ 
\footnotesize Published cosmic ray $\sigma_{pp}$ values from AGASA and Fly's Eye, appended to accelerator data.  A plot of the Aspen model (QCD-inspired eikonal) fit of the total $pp$ and $\bar pp$ cross sections, in mb vs $\sqrt s$, the c.m. energy in GeV.  Using the calculations from Fig. \ref{fig:p-air}, the cosmic ray values that are shown have been converted from their {\em published} values of $\spai$ to $\sigma_{pp}$ and appended to the plot.  The accelerator $pp$  and $\bar p p$ points that were used in the fit are the squares and circles, respectively; the  AGASA points are the triangles and the Fly's Eye point is the diamond.  Taken from Ref. \cite{BHS}.
\label{fig:sigtodorpp}}
\end{figure}
%%%%%%%%%%%%%%%%%%%%%%%%%%%%%%%%%%%%%%%%%%
%%%%%%%%%%%%%%%%%%%%%%%%%%%%%%%%%%%%
%%%%%%%%%%%%%%%%%%%%%%%%%%%%%%%%%%%%%%%%%%%%%%%% 
\subsection{Reanalysis of Fly's Eye and AGASA experiments}\label{flyseyeBHS}
Since the conversion from $\Lambda_m$ to $\lambda_{p-\rm air}$ is highly model-dependent, BHS proposed to minimize the impact of theory on this conversion. They constructed a QCD-inspired parametrization (the Aspen model of Section \ref{section:aspen}) of the forward proton-proton and proton-antiproton scattering amplitudes\,\cite{blockhalzenmargolis} which fits all accelerator data of $\sigma_{\rm tot}$, $\rho$ and $B$, plots of which are shown in Fig. \ref{fig:ppcurves}.%
%%%%%%%%%%%%%%%%%%%%%%%%%%%%%%%%%%%%%%%%
%%%%%%%%%%%%%%%%%%%%%%%%%%%%%%%%%%%%%%%%%%%%%%%%%%%%
\begin{figure}[h] %Fig.49
\begin{center}
\mbox{\epsfig{file=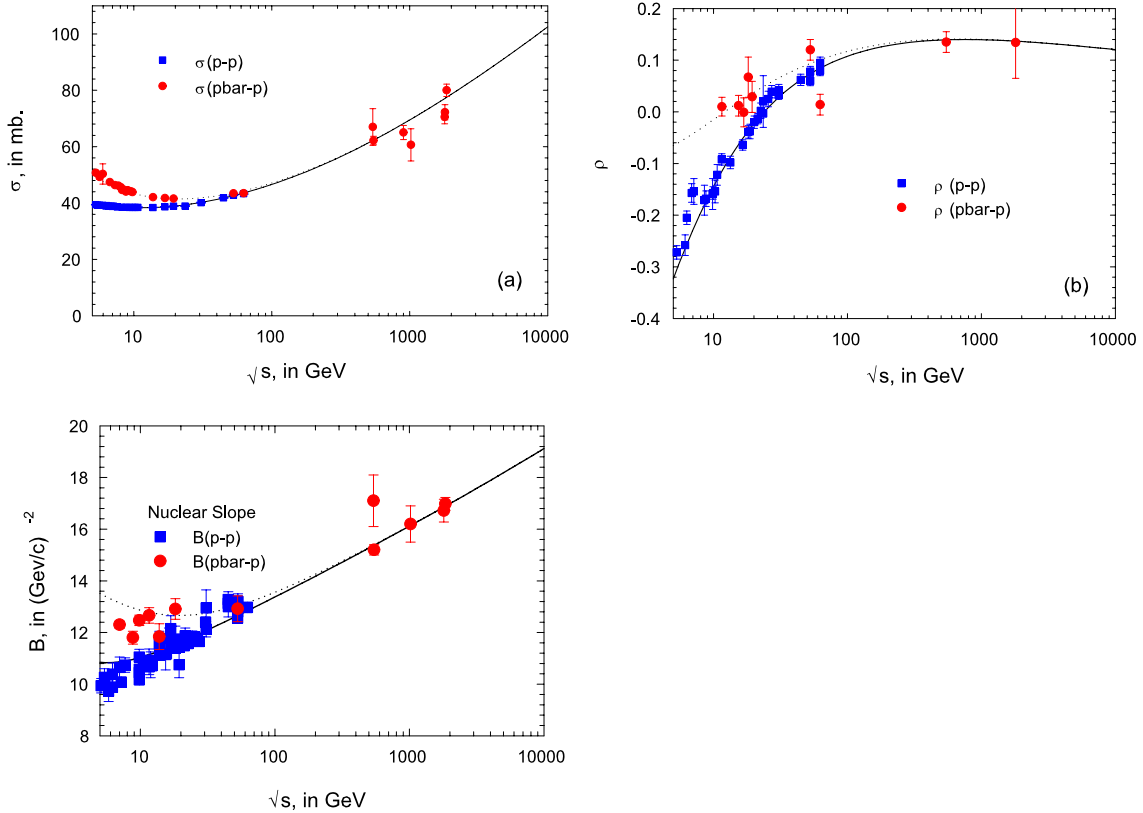%
            ,width=6in,bbllx=0pt,bblly=0pt,bburx=597pt,bbury=430pt,clip=%
}}
\end{center}
\caption[The simultaneous Aspen model (QCD-inspired eikonal) fit of  total nucleon-nucleon cross section $\sigtot$, $\rho$ and $B$]
{ \footnotesize
 The simultaneous Aspen model (QCD-inspired eikonal) fit of the total cross section $\sigtot$, $\rho$ and the nuclear slope parameter $B$ vs. $\sqrt s$, in GeV, for pp
 (squares) and $\bar {\rm   p}$p (circles) accelerator data:
 (a) $\sigtot$, in mb,\ \  (b) $\rho$,\ \  (c) Nuclear slope $B$,
 in (GeV/c)$^{-2}$. The solid line is $pp$ and the dotted line is $\bar pp$. Taken from Ref. \cite{BHS}.}
\label{fig:ppcurves}
\end{figure}
%
%%%%%%%%%%%%%%%%%%%%%%%%%%%%%%%%%%%%%%%%%%%%%%%%%%%%%%%%%%555
In addition, the {\em measured}   high energy cosmic ray $\Lambda_m$ values of the Fly's Eye\,\cite{fly} and  
AGASSA\,\cite{akeno} experiments are also simultaneously used,  i.e., $k$ from  Eq. (\ref{eq:Lambda_m}) is also a fitted quantity. They refer to this fit as a {\em global} fit, emphasizing that in the global fit, all 4 quantities, $\sigma_{\rm tot}$, $\rho$, $B$ {\em and} $k$, are {\em simultaneously} fitted. Because the parametrization is both unitary and analytic, its high energy predictions are effectively model-independent, if one requires that the proton  asymptotically becomes a black disk. Using vector meson dominance and the additive quark models, there is further support for their QCD fit---it accommodates a wealth of data on photon-proton and photon-photon interactions without the introduction of new parameters\cite{bghp}.  In particular, it also {\em simultaneously} fits $\sigma_{pp}$ and $B$, forcing a relationship between the two. Specifically, the $B$ vs. $\sigma_{pp}$ prediction of the Aspen model fit shown in Fig. \ref{fig:p-air} completes the relation needed (using the Glauber model) between $\sigma_{pp}$  and $\spai$. The percentage error in the prediction of $\sigma_{pp}$ at $\sqrt s=30$ TeV is $\approx 1.2$\%, due to the statistical error in the fitting parameters (see  
Refs. \cite{blockhalzenmargolis,bghp}). In Fig.\,\ref{fig:sigpp_p-air},%
\begin{figure}[tbp] %Fig.51
\begin{center}
\mbox{\epsfig{file=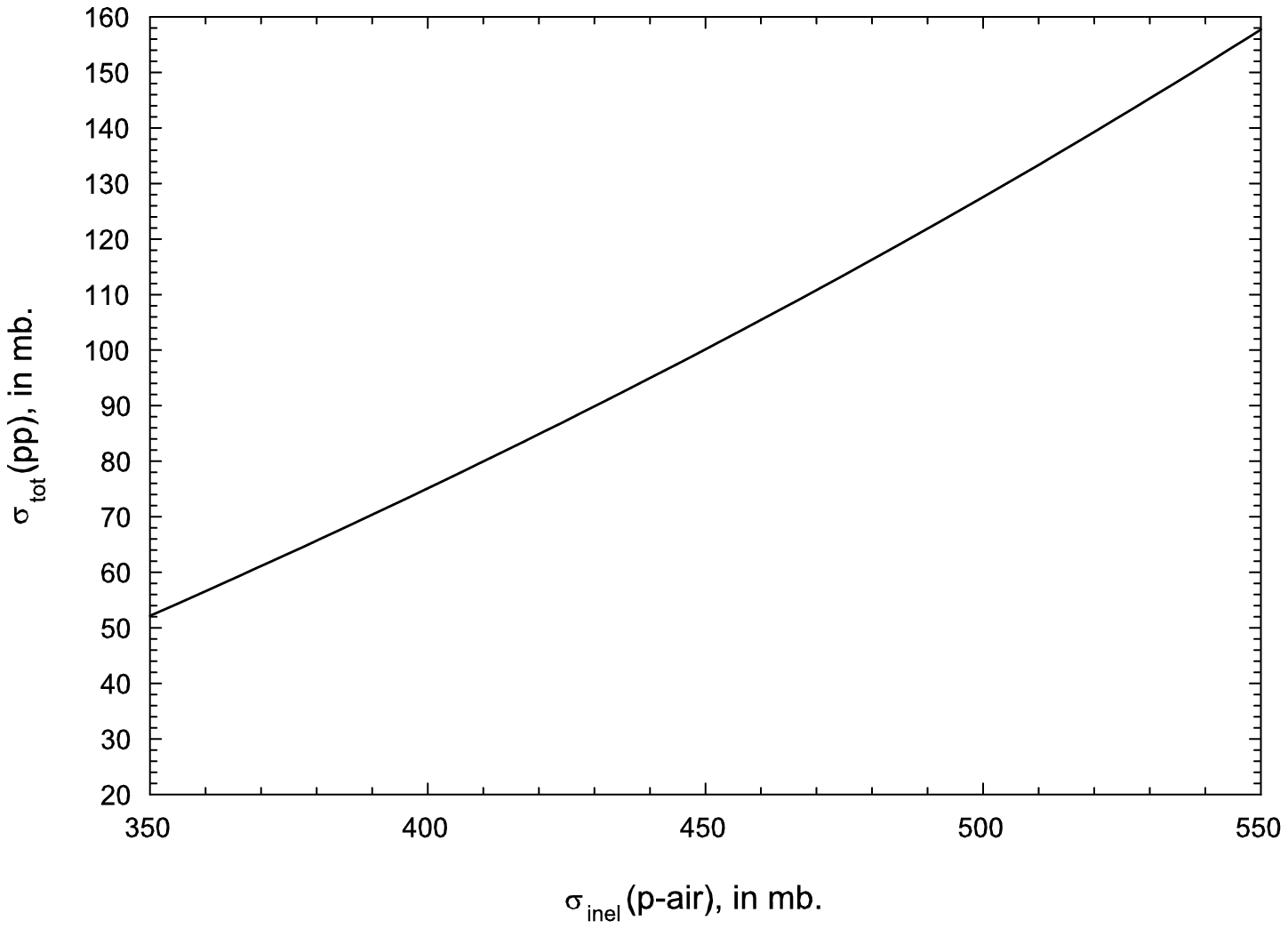%
              ,width=3.8in,bbllx=0pt,bblly=0pt,bburx=420pt,bbury=310pt,clip=%
}}
\end{center}
\caption[$\sigma_{pp}$ as a function of $\spai$]
{\footnotesize $\sigma_{pp}$ as a function of $\spai$. 
A plot of the predicted total $pp$ cross section , in mb
vs. the measured $p$-air cross section,  in mb, using the calculation of Fig. \ref{fig:p-air}. Taken from Ref. \cite{BHS}.
}
\label{fig:sigpp_p-air}
\end{figure}
%%%%%%%%%%%%%%%%%%%%%%%%%%%%%%%%%%%%%%%%
we have plotted the values of $\sigma_{pp}$ vs. $\spai$ that are deduced from the intersections of their $B$-$\sigma_{pp}$ curve  with the $\spai$ curves of Fig.\,1. Figure~\,\ref{fig:sigpp_p-air} allows the conversion of measured $\spai$ cross sections to $\sigma_{pp}$ total cross sections. %
The percentage error in $\spai$ is $\approx 0.8$ \% near $\spai = 450 $mb, due  
to the errors
in $\sigma_{pp}$ and $B$ resulting from the errors in the fitting parameters.  
Again, the global fit gives an error of a factor of about 2.5 smaller than our  
earlier result, a {\em distinct} improvement.

Confronting their predictions of the p-air cross sections $\spai$ as a function of energy with published cross section measurements of the Fly's Eye\,\cite{fly}---see Fig. \ref{fig:p-air}---and AGASSA\,\cite{akeno} groups, they find that the predictions systematically are about one standard deviation below the {\em published} cosmic ray values. Letting $k$ be a free parameter in a global fit  
and neglecting the possibility that $k$ might show a weak energy dependence over the range measured\footnote{ Recently,  Monte Carlo model simulations made by Pryke\cite{pryke} indicate that $k$ is compatible with being energy-independent in this energy region.}%
, they refit simultaneously both the accelerator and the  cosmic ray data.  Using an energy-independent $k$, BHS\cite{BHS} find that $k=1.349\pm 0.045$, where the error is the statistical error of the global fit.
By combining the results of Fig.\,\ref{fig:ppcurves}\,(a) and  
Fig.\,\ref{fig:sigpp_p-air}, they predict the variation of $\spai$ with the c.m. energy $\sqrt s$. In Fig.\,\ref{fig:p-aircorrected2}%
%%%%%%%%%%%%%%%%%%%%%%%%%%%%%%%%%%%%%%%%%5
\begin{figure}%Fig. 51
\begin{center}
\mbox{\epsfig{file=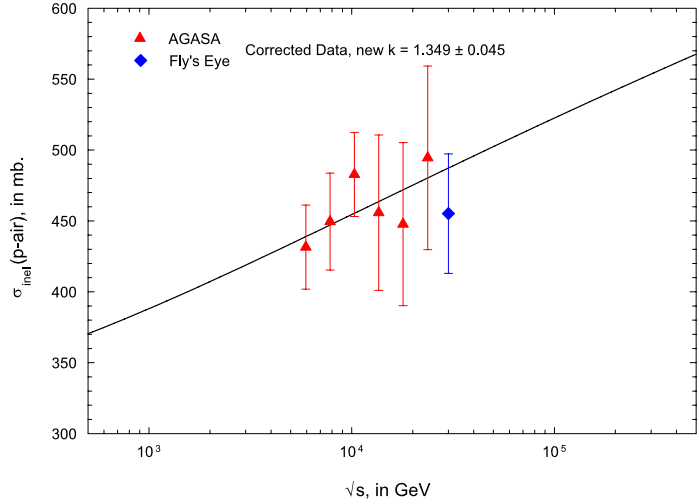%
              ,width=3.7in,bbllx=0pt,bblly=0pt,bburx=420pt,bbury=303pt,clip=%
}}
\end{center}
\caption[Renormalized values of $\spai$ for AGASA and Fly's Eye]{\footnotesize
\protect
{Renormalized values of $\spai$ for AGASA and Fly's Eye. The AGASA and Fly's Eye data for $\spai$, in mb, as a function of the energy, $\sqrt s$, in GeV, as found in a global fit that used the common value of $k=1.349$. Taken from Ref. \cite{BHS}.}
}
\label{fig:p-aircorrected2}
\end{figure}
%%%%%%%%%%%%%%%%%%%%%%%%%%%%%%%%%%%%%%%%%%%%%%%%%
they have {\em rescaled} the published high energy data for $\spai$ (using the common value of $k=1.349$), and plotted the  
revised data against their prediction of $\spai$  vs. $\sqrt s$.%
%%%%%%%%%%%%%%%%%%%%%%%%%%%%%%
%%%%%%%%%%%%%%%%%%%%%%%%%%%%%%%%
\begin{figure}%Fig. 52
\begin{center}
\mbox{\epsfig{file=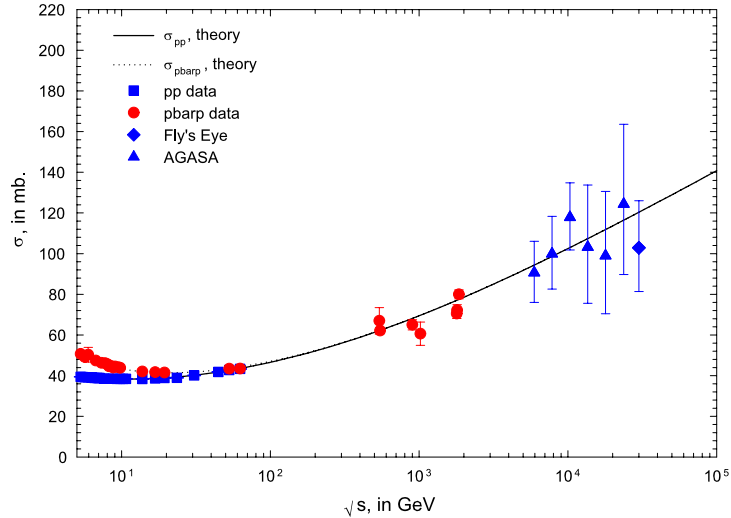%
              ,width=3.8in,bbllx=0pt,bblly=0pt,bburx=410pt,bbury=297pt,clip=%
}}
\end{center}
\caption[AGASA and Fly's Eye  $p$-air cross sections converted to $\sigma_{pp}$]
{\footnotesize
\protect
{AGASA and Fly's Eye  $p$-air cross sections converted to $\sigma_{pp}$. A plot of the Aspen model (QCD-inspired eikonal) fit of the total nucleon-nucleon cross section $\sigma$, in mb  vs. $\sqrt s$, in Gev. The solid line is for $pp$ and the dotted line is for $\bar pp$. The cosmic ray data that are shown have been converted from $\spai$ to $\sigma_{pp}$ using the results of Fig.~\ref{fig:sigpp_p-air} and the common value of $k=1.349$, found from a global fit. Taken from Ref. \cite{BHS}.
}
}
\label{fig:4sigtotcr}
\end{figure}
%%%%%%%%%%%%%%%%%%%%%%%%%%%%%%
%%%%%%%%%%%%%%%%%%%%%%%%%%%%%%%%%%%%
%%%%%%%%%%%%%%%%%%%%%%%%%%%%%%%%%%%%
The plot of $\sigma_{pp}$ vs. $\sqrt s$, including the  rescaled cosmic  
ray data, is shown in Fig.\,\ref{fig:4sigtotcr}. %
Clearly, the fit is excellent, with good agreement between AGASA and Fly's Eye. Since the cosmic ray spectrum varies very rapidly with energy, there must be allowance for systematic errors in $k$ due to possible energy misassignments.  At the quoted experimental energy resolutions, $\Delta{\rm Log}_{10}(E_{\rm lab}({\rm eV}))=0.12$ for AGASSA\cite{akeno} and $\Delta{\rm Log}_{10}(E_{\rm lab}({\rm eV}))=0.4$ for Fly's Eye\cite{fly}, where  $E_{\rm lab}$ is the proton laboratory energy,
BHS find from the curve in Fig.\,\ref{fig:p-aircorrected2} that $\Delta  
k/k=0.0084$ for AGASSA\cite{akeno} and $\Delta k/k=0.0279$ for Fly's  
Eye\cite{fly}. They estimated conservatively that experimental energy resolution  
introduces a systematic error in $k$ such that $\Delta k_{\rm systematic}=\sqrt  
{(\Delta k_{\rm AGASSA}^2+\Delta k_{\rm FLYSEYE}^2)/2}= 0.028$.  Thus, the final BHS result\cite{BHS} was $k=1.349\pm 0.045\pm 0.028$, where the first error is statistical and the last error is systematic.

Pryke\cite{pryke} has published a comparative study of  high  
statistics simulated air showers for proton primaries, using four combinations  
of the MOCCA\cite{mocca} and
CORSIKA\cite{corsika} program frameworks, and SIBYLL\cite{sibyll} and  
QGSjet\cite{qgsjet} high energy hadronic interaction models. He finds $k=1.30 \pm 0.04$ and $k=1.32 \pm 0.03$ for the CORSIKA-QGSjet and MOCCA-Internal models, respectively, which are in excellent agreement with the BHS measured result, $k=1.349\pm 0.045\pm 0.028$.

Further, Pryke\cite{pryke} obtains  $k=1.15 \pm 0.03$ and $k=1.16 \pm 0.03$ for the CORSIKA-SIBYLL and MOCCA-SIBYLL models, respectively, whereas the SYBILL\cite{gaisser} group finds $k=1.2$, which is not very different from the Pryke value.  However, the SYBILL-based models, with $k=$1.15---1.20, are significantly different from the BHS measurement of $k=1.349\pm 0.045\pm 0.028$. At first glance, this appears somewhat strange, since their model for forward scattering amplitudes and SIBYLL share the same underlying physics. The increase of the total cross section with energy to a black disk of soft partons is the shadow of increased particle production which is modeled in SYBILL by the production of (mini)-jets in QCD. The difference between the $k$ values of 1.15---1.20 and 1.349 results from the very rapid rise of the $pp$ cross section in SIBYLL at the highest energies. This is not a natural consequence of the physics in the model---it's an artifact of a fixed transverse momentum cut-off. In most other codes (QGSjet\cite{qgsjet}), it is remedied by the use of an energy-dependent transverse momentum cut-off in the computation of the mini-jet production cross section. 

As demonstrated in Fig.\,\ref{fig:4sigtotcr}, the overall agreement between the accelerator and the cosmic ray $pp$ cross sections with the Aspen model (QCD-inspired eikonal) fit  is striking. The accelerator and cosmic ray $pp$ cross sections  are readily reconcilable using  a  value of $k=1.349\pm 0.045\pm 0.028$, which is both model independent and energy independent. Using the Aspen model fit, this determination of $k$ severely constrains any model of high energy hadronic interactions. 
%%%%%%%%%%%%%%%%%%%%%%%%%%%%%%%%%%%%%%%%
\subsection{The HiRes experiment}
Using two fluorescence detector stations, HiRes1 and HiRes2, located in the Dugway Proving Ground and separated by 12.6 km, the HiRes group\cite{hires} has measured the total inelastic production cross section at a mean laboratory energy of $10^{18.5}$ eV ($\sqrt s=77.0$ TeV) as
\be
\spai=456\pm17({\rm statistical})+39({\rm systematic})-11({\rm systematic})\quad {\rm mb}\label{sigpair}.
\ee

The maximum of an extended air shower distribution is given by 
\be
X_{\rm max}=X'+X_1, 
\ee
where $X'$ is the distance (in g cm$^{-2}$) of the shower maximum relative to the first interaction point $X_1$.  The distribution of $X'$ depends on the extensive air shower model employed as well as the energy of the primary proton;  $X'$ is subject to large fluctuations. An example of the $X'$ distribution for $E=10^{18.5}$ eV, using the Monte Carlo air shower framework CORSIKA\cite{corsika}, in conjunction with the particle production model  QGSJet\cite{qgsjet}, is shown in Fig. \ref{fig:xprimedistribution}. A computer-simulated $X_1$ distribution whose logarithmic slope is $\lpa$, taken from the HiRes\cite{hires} experiment, is shown in Fig. \ref{fig:X1}. The distribution in $X_{\rm max}$ is the result of the  convolution of the $X_1$ distribution into the $X'$ distribution. 
%%%%%%%%%%%%%%%%%%%%%%%%%%
%%%%%%%%%%%%%%%%%%%%%%%%
\begin{figure}[]%Fig. 53
\begin{center}
\mbox{\epsfig{file=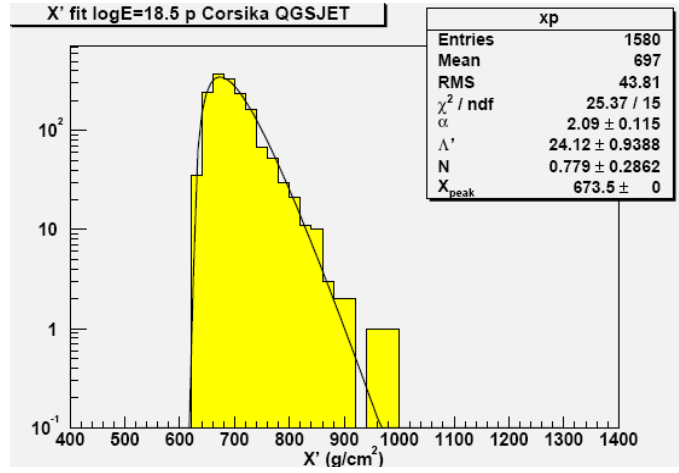%
              ,width=3.5in,bbllx=80pt,bblly=240pt,bburx=537pt,bbury=557pt,clip=%
}}
\end{center}
\caption[HiRes $X'$ distribution]
{\footnotesize
\protect
{HiRes $X'$ distribution. The $X'$ distribution, using Monte Carlo data.   Taken from Ref. \cite{hires}.
}
}
\label{fig:xprimedistribution}
\end{figure}
%%%%%%%%%%%%%%%%%%%%%%%%%%%%%%
%%%%%%%%%%%%%%%%%%%%%%%%%%%%%%

\begin{figure}[tbp] %Fig. 54
\begin{center}
\mbox{\epsfig{file=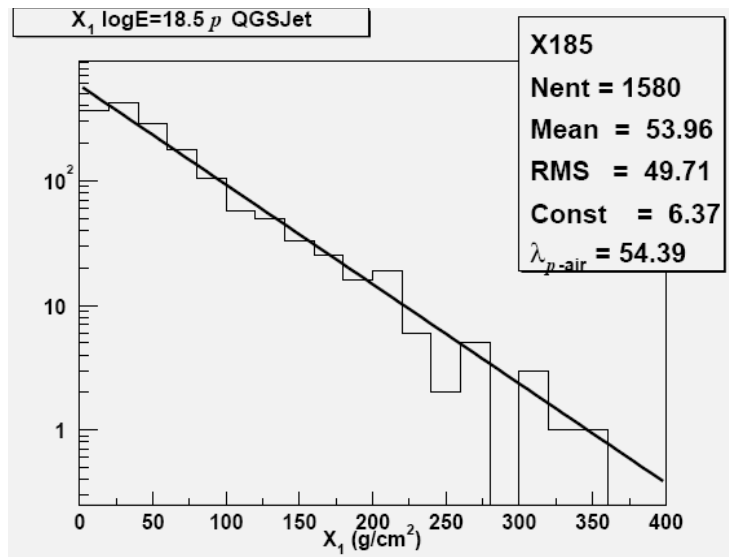,width=3.8in%
,bbllx=72pt,bblly=220pt,bburx=537pt,bbury=574pt,clip=%
}}
\end{center}
\caption[A computer-simulated $X_{1}$ distribution]
{ \footnotesize
A computer-simulated $X_{1}$ distribution, taken from Ref. \cite{hires}. 
  }
\label{fig:X1}
\end{figure}
%%%%%%%%%%%%%%%%%%%%%%%%%%%%%%

As emphasized  earlier, neither the $X'$ nor the $X_1$ distribution can be measured in air shower experiments. However, the $X'$ distribution can be simulated using  Monte Carlo data, utilizing  various air shower and particle production models (for some details about generating Monte Carlo extensive air showers, see Pryke\cite{pryke}).  Convolution of the initial interaction point $X_1$, which  depends {\em only} on the single parameter $\lambda_{\rm p-air}$, into a $X'$ distribution  for  a fixed production model\footnote{A shower library of 12000 Monte Carlo events using a an $E^{-3}$ energy spectrum was created for this purpose, which was run through the HiRes standard reconstruction routines and cuts.} %
%%%%%%%%%%%%%%%%%%%%%%%%%%%%%%%%
gives the potentially experimentally observable $X_{\rm max}$ distribution, as a function of the single variable $\lambda_{\rm p-air}$. A best fit is then made  to the actual experimental distribution, in order to find the mean free path $\lambda_{\rm p-air}$. This technique has the advantage of potentially using the entire experimental $X_{\rm max}$  distribution, and not just the exponential-like tail of the distribution---the part of the distribution that was used by Fly's Eye\cite{fly} and AGASA\cite{akeno} to obtain $\Lambda_m$---which has sensitive production model and cosmic ray particle composition dependencies.  

It was found by HiRes that this new analysis technique, fitting the entire $X_{\rm max}$ distribution,  was rather insensitive to the production models, being quite stable over the energy range $10^{17}\ {\rm eV}\le E\le10^{20.5}\ {\rm eV}$. Shown in Fig. \ref{fig:xmaxproductionmodel} is the fractional deviation 
$(\lambda_{\rm p-air}^{\rm input}-\lambda_{\rm p-air}^{\rm deconvoluted})/\lambda_{\rm p-air}^{\rm input}$ plotted as a function of the logarithm of the laboratory energy in eV, for the two particle production models QGSJet\cite{qgsjet} (small diamonds) and SIBYLL2.1\cite{sibyll} (large open circles). The Monte Carlo data used for this exercise was taken from the shower library. The input mean-free-path to the Monte Carlo is called $\lambda_{\rm p-air}^{\rm input}$ and the best fit to a production model $X_{\rm max}$ distribution is called $\lambda_{\rm p-air}^{\rm deconvoluted}$. The model sensitivity is seen to be quite small, unlike the methods used in Fly's Eye and AGASA.

The effect of cosmic ray composition, assuming 20\% CNO and 20\% Fe, compared to 100\% protons, is shown in Fig. \ref{fig:crcomposition}. The authors claim that a safe cut that eliminates Fe contamination is to use only values of  $X_{\rm max}>700 $ gm cm$^{-2}$. 

Figure \ref{fig:crenergydistribution} shows the energy distribution of the  observed cosmic ray air showers.  The number of showers in a bin is plotted against the logarithm of proton primary energy, in eV. The mean primary energy used, after all cuts, was $E=10^{18.52\pm0.39}$ eV. 

Finally, Fig. \ref{fig:xmaxdata} shows a fit to the HiRes observed extensive air shower $X_{\rm max}$ distribution, which has an excellent $\chi^2$/d.f.  They conclude that their deconvolution of the $X_{\rm max}$ distribution leads to the measured value
\be
\lambda_{\rm p-air}=52.88\pm1.98\  {\rm g\  cm}^{-2},
\ee
corresponding to a $p$-air cross section  at a mean laboratory energy of $E=10^{18.52\pm0.39}$ eV (a c.m. energy of $\sqrt s=78.82 ^{+45.0}_{-28.5}$ TeV ) of
\be
\spai=456\pm17({\rm statistical})+39({\rm systematic})-11({\rm systematic})\ {\rm mb}\label{sigpair2},
\ee
where the asymmetric systematic errors are due to a possible 5\% gamma ray flux contamination. 

Clearly, in addition to being the highest energy $\spai$ measurement made, the HiRes measurement is the first to attempt a relatively model-free measurement of the $p$-air cross section.  A more detailed analysis of the gamma ray flux, promised by the authors, should improve the systematic uncertainty of their cross section measurement.  

Using a Glauber calculation, we will convert $\spai$ to $\sigtot(pp)$ in the next Section.
%%%%%%%%%%%%%%%%%%%%%%%%%%%%%%%%%%%%%%%
%%%%%%%%%%%%%%%%%%%%%%%%%%%%%%%%%%%%%%%%
%%%%%%%%%%%%%%%%%%%%%%%%%%%%%%%%
%%%%%%%%%%%%%%%%%%%%%%%%%%%%%%%%
\begin{figure}[]%Fig. 55
\begin{center}
\mbox{\epsfig{file=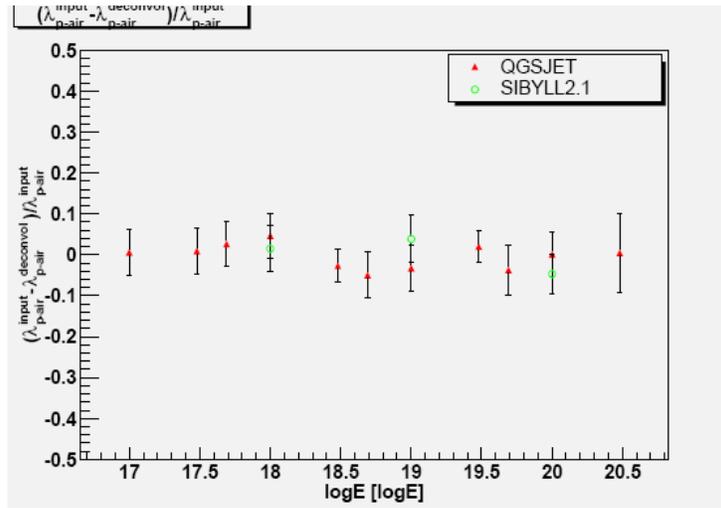%
              ,width=3.8in,bbllx=75pt,bblly=235pt,bburx=535pt,bbury=555pt,clip=%
}}
\end{center}
\caption[HiRes stability plot of $\lambda_{\rm p-air}$]
{\footnotesize
\protect
{ HiRes stability plot of $\lambda_{p-\rm air}$. A plot of the fractional deviation, $(\lambda_{p-\rm air}^{\rm input}-\lambda_{p-\rm air}^{\rm deconvoluted}/\lambda_{p-\rm air}^{\rm input})$ vs  $\log_{10}E$, where $E$ is the  laboratory energy in eV, for the two particle production models QGSJet\cite{qgsjet} (small diamonds) and SIBYLL2.1\cite{sibyll} (large open circles). Taken from Ref. \cite{hires}.
}\label{fig:xmaxproductionmodel}
}
\end{figure}
%%%%%%%%%%%%%%%%%%%%%%%%%%%%%%
%%%%%%%%%%%%%%%%%%%%%%%%%%%%%%
%%%%%%%%%%%%%%%%%%%%%%%%%%%%%%%%
\begin{figure}[]%Fig. 56
\begin{center}
\mbox{\epsfig{file=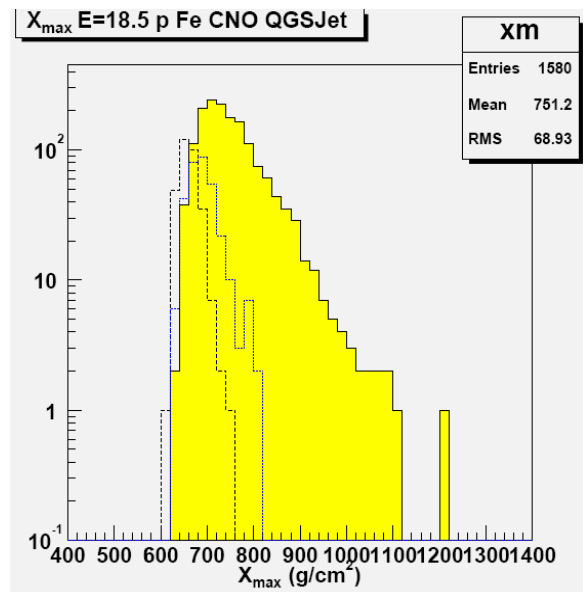%
              ,width=3in,bbllx=75pt,bblly=160pt,bburx=535pt,bbury=632pt,clip=%
}}
\end{center}
\caption[The $X_{\rm max}$ distribution for $p$, CNO and Fe]
{\footnotesize
\protect
{ The $X_{\rm max}$ distribution for $p$, CNO and Fe. The dashed curve is for 20\% CNO and the dotted curve is for 20\% Fe, shown on the same scale as 100\% protons. Taken from Ref. \cite{hires}.
}
}
\label{fig:crcomposition}
\end{figure}
%%%%%%%%%%%%%%%%%%%%%%%%%%%%%%
%%%%%%%%%%%%%%%%%%%%%%%%%%%%%%
\begin{figure}[tbp]%Fig. 57
\begin{center}
\mbox{\epsfig{file=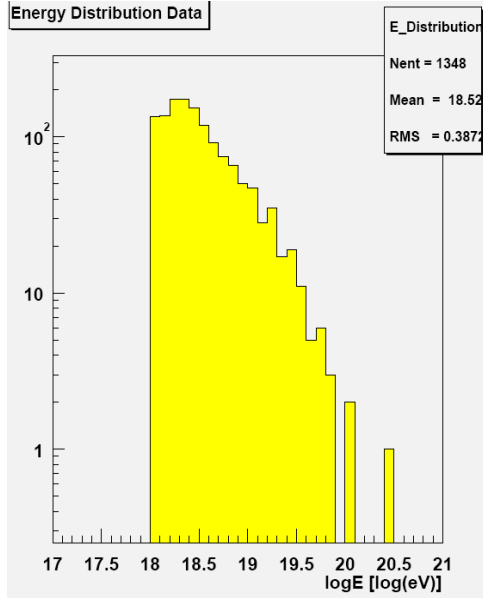%
              ,width=2.5in,bbllx=155pt,bblly=180pt,bburx=509pt,bbury=623pt,clip=%
}}
\end{center}
\caption[The energy distribution of the experimental extensive air  showers in HiRes]
{\footnotesize
\protect
{ The energy distribution of the experimental extensive air showers in HiRes.  Shown are the number of showers per bin vs. $\log_{10}E$, where $E$ is the laboratory energy of the proton, in eV. The mean energy is $E=10^{18.52\pm0.39}$ eV. Taken from Ref. \cite{hires}.
}
}
\label{fig:crenergydistribution}
\end{figure}

%%%%%%%%%%%%%%%%%%%%%%%%%%%%%%%%%%%%%%%%%%%%%%
%%%%%%%%%%%%%%%%%%%%%%%%%%%%%%
\begin{figure}[tbp]%Fig. 58
\begin{center}
\mbox{\epsfig{file=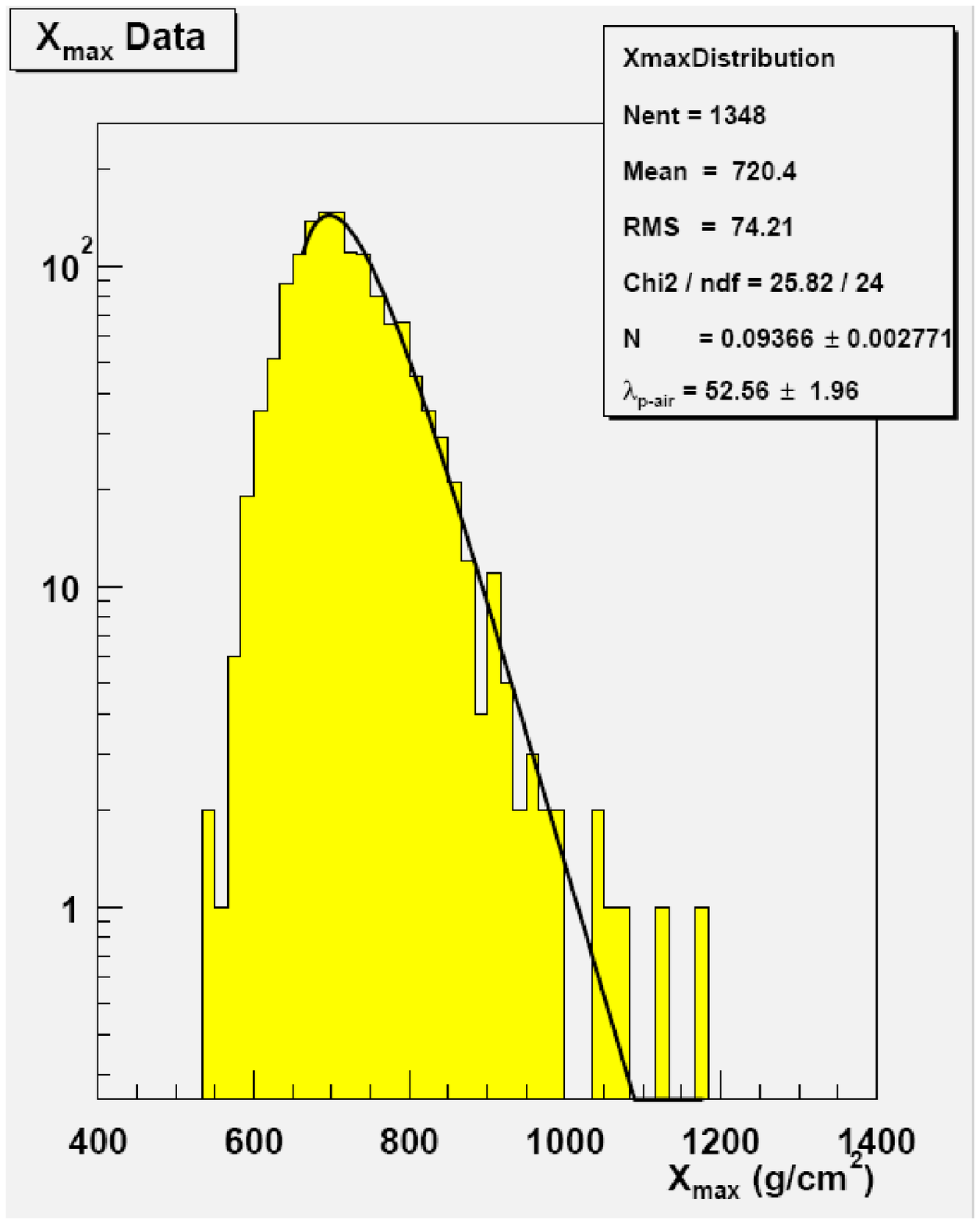%
              ,width=3in,bbllx=115pt,bblly=155pt,bburx=490pt,bbury=625pt,clip=%
}}
\end{center}
\caption[The $X_{\rm max}$ distribution for the  experimental extensive air showers in HiRes]
{\footnotesize
\protect
{ The $X_{\rm max}$ distribution for the  experimental extensive air showers in HiRes, fitted for $X_{\rm max}>667$ gm cm$^{-2}$.  Taken from Ref. \cite{hires}.
}
}
\label{fig:xmaxdata}
\end{figure}
%%%%%%%%%%%%%%%%%%%%%%%%%%%%%%%%%%%%%%%%%%%%%%
\subsection{Converting $\spai$ to $\sigma_{pp}$}
Figure \ref{fig:newBvssigpp} shows 5 contours  of various values of $\spai$ in the $B$-$\sigma_{pp}$ space, where $B$ is the  nuclear slope parameter and $\sigma_{pp}$ is the total $pp$ cross section, derived from a Glauber calculation that used a 2-channel shadowing model\cite{engelprivate}. $B$ is in (GeV/c)$^{-2}$ and $\sigma_{pp}$ is in mb. The five contours shown  are lines  of constant $\spai$,  of 414, 435, 456, 499 and 542 mb, where the central value is the HiRes\cite{hires} measurement, and the others are $\pm 1\sigma$ and $\pm 2\sigma$. The total $pp$ cross section $\sigma_{pp}$ is called $\sigtot(pp)$ in Fig. 
\ref{fig:newBvssigpp}.

The solid line is a hybrid curve which requires some explanation. Both $B$ and $\sigma_{pp}$ are both implicit functions of $s$. The choice for $B(s)$ was the analyticity-constrained Aspen model eikonal fit of $B_{pp}$  to $\sigtot$, $\rho$ and $B$ for $pp$ and $\bar pp$ data, which was shown in Fig. \ref{fig:Bqcd} in Section \ref{sec:Bfit}. The choice for $\sigma_{pp}( s)$ was the 4 constraint $\ln^2 s$ fit  to $\sigtot$ and $\rho$  for $pp$ and $\bar pp$ data, made  from the parameters of Table \ref{table:ppfitnew} of Section \ref{sec:lnsqpp}, and shown as the dot-dashed line in Fig. \ref{fig:sigmapp}. The plots $B(s) $ and $\sigma_{pp}( s)$ are then combined to make the curve $B(\sigma_{pp})$, the solid line of Fig. \ref{fig:newBvssigpp}. The large open dot on this curve is our $B$-$\sigma_{pp}$ prediction for the c.m. energy $\sqrt s=77.02$ TeV, the mean HiRes c.m. energy. We see that our HiRes prediction is safely within the +1 $\sigma$ limit of the HiRes experiment.  
%%%%%%%%%%%%%%%%%%%%%%%%%%%%%%
\begin{figure}[tbp]%Fig. 59
\begin{center}
\mbox{\epsfig{file=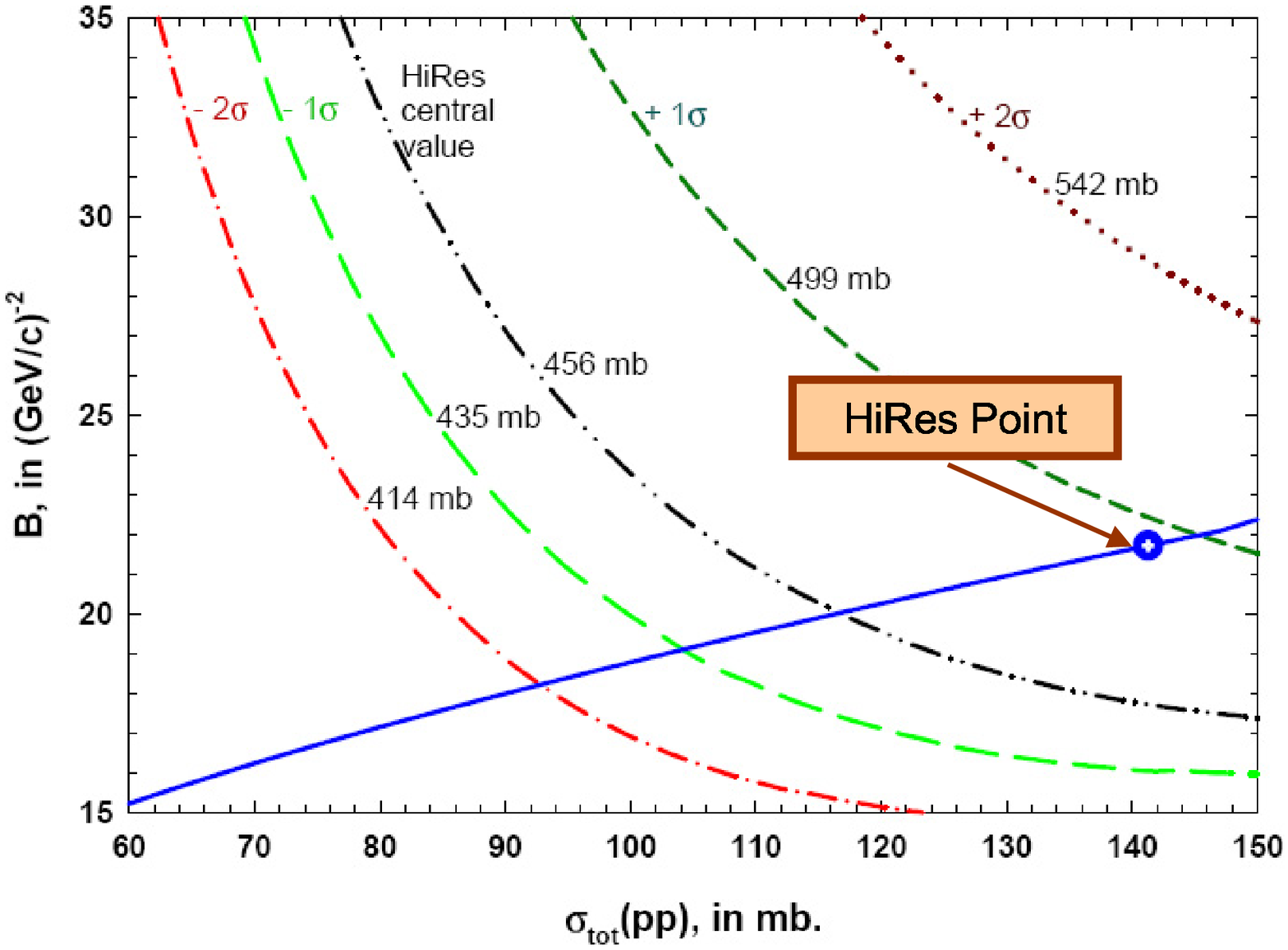%
              ,width=3.8in,bbllx=15pt,bblly=198pt,bburx=579pt,bbury=595pt,clip=%
}}
\end{center}
\caption[The $B$ dependence of the $pp$ total cross section $\sigma_{pp}$, showing our HiRes prediction for $\spai$]
{\footnotesize
\protect
{ The $B$ dependence of the $pp$ total cross section $\sigtot(pp)$, showing our HiRes prediction for $\spai$. $B$ is in (GeV/c)$^{-2}$ and $\sigtot(pp)$ is in mb. The five curves are lines of constant $\spai$,  of 414, 435, 456, 499 and
 542 mb---the central value is the HiRes\cite{hires} value, and the others
 are $\pm 1\sigma$ and $\pm 2\sigma$. The solid curve is a fit from accelerator data of
 $B$ vs. $\sigtot(pp)$. The large point on this curve corresponds to the HiRes energy of $\sqrt s=77.02$ TeV. The $B$ values, at a common energy,  are from an analytically constrained Aspen model (eikonal) fit and  the  $\sigtot(pp)$ values,  at the same energy, are from  an analytically  constrained $\ln^2 s$ fit. See the text for details.
}
}
\label{fig:newBvssigpp}
\end{figure}

%%%%%%%%%%%%%%%%%%%%%%%%%%%%%%%%%%%%%
\subsubsection{The predicted energy dependence of $\spai$}
We project out the energy dependence of the $\spai$ prediction in Fig. \ref{fig:newsigpairvsE}, in which we plot $\spai$, in mb, against the c.m. energy $\sqrt s$, in GeV. The published HiRes point\cite{hires} at $\sim 80$ TeV, along with its error, is the open diamond point.   The  experimental cross sections for AGASA and Fly's Eye are the triangles and the diamond, respectively. Their values were obtained by first making an analyticity-constrained {\em global} $\ln^2 s$ fit to {\em both} accelerator data ($\sigma_{pp},\ \sigma_{\bar pp},\ \rho_{pp}$ and $\rho_{\bar pp}$), {\em and} cosmic ray data ($\sigma_{pp}$ obtained  from $\spai$ using the 2-channel Glauber calculation shown in Fig. \ref{fig:newBvssigpp}), and next, rescaling the original published values of $\spai$ for AGASA and Fly's Eye using the best-fit value of $k=1.287$.  We estimate that our uncertainty in our $\spai$ prediction is $\sim6$ mb, due to projecting the statistical fitting errors in $B$ and $\sigtot(pp)$ into $\spai$ errors. An additional theoretical error of several mb, due to uncertainties in the Glauber calculation that are hard  to evaluate, lead us to estimate an overall prediction error of $\Delta\spai =\pm6 \mbox{\rm \ mb (statistical)}\pm 8 {\mbox{\rm \ mb ( systematic)}}.$  We see from Fig. \ref{fig:newsigpairvsE} that the HiRes measurement fits our prediction quite nicely, as do all of the other cosmic ray points, using $k=1.287\pm 0.06$. 

In Section \ref{flyseyeBHS} we showed that Block, Halzen and Stanev's analysis\cite{BHS} of the AGASA and Fly's Eye data---using $B$ and $\sigtot(pp)$ fits made with an unconstrained Aspen model and the Glauber calculation of Ref. \cite{gaisser}---led to a value of $k=1.348\pm0.053$, which is not inconsistent with the present determination of $k=1.287\pm 0.06$, a value in accord with all of the experimental cosmic ray data over the enormous energy range $6\la\sqrt s\la 80$ TeV.
%%%%%%%%%%%%%%%%%%%%%%%%%%%%%%
\begin{figure}[tbp]%Fig. 60
\begin{center}
\mbox{\epsfig{file=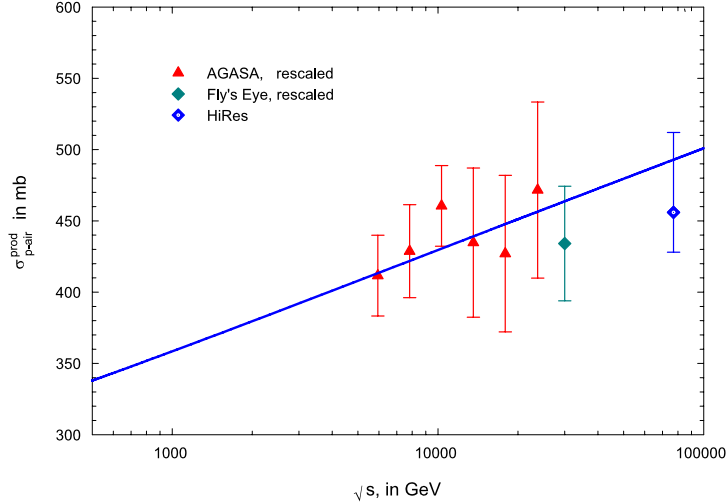%
              ,width=3.8in,bbllx=0pt,bblly=48pt,bburx=429pt,bbury=350pt,clip=%
}}
\end{center}
\caption[Predictions for $\spai$, showing the HiRes point and renormalized AGASA and Fly's Eye data]{\footnotesize
\protect
{ The AGASA, Fly's Eye and HiRes data for $\spai$, in mb,
 as a function of the c.m. energy $\sqrt s$, in GeV. The AGASA and Fly's  Eye values were found from an analytically constrained  global $\ln^2 s$ fit to both accelerator and cosmic ray data, rescaled using the best-fit  value of $k=1.287$. The  HiRes measurement is taken directly from Ref. \cite{hires}.
}
}
\label{fig:newsigpairvsE}
\end{figure}
%%%%%%%%%%%%%%%%%%%%%%%%%%%%%%%%%%%%%%%%%%%%%%
%%%%%%%%%%%%%%%%%%%%%%%%%%%%%%%%%%%%%%%%%%%%%%
%%%%%%%%%%%%%%%%%%%%%%%%%%%%%%
\begin{figure}[tbp]%Fig. 61
\begin{center}
\mbox{\epsfig{file=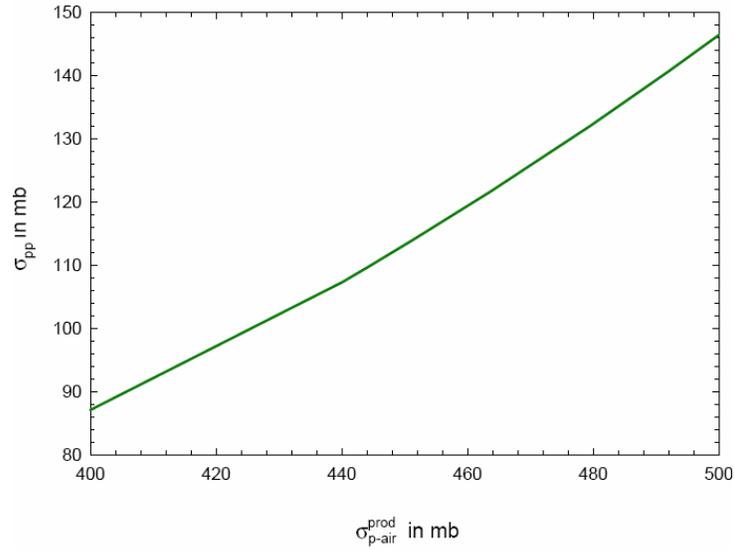%
              ,width=3.8in,bbllx=20pt,bblly=180pt,bburx=575pt,bbury=605pt,clip=%
}}
\end{center}
\caption[ $\sigtot(pp)$ as a function of
 $\spai$]
{\footnotesize
\protect
{ The predicted total $pp$ cross section $\sigma_{pp}$ , in mb, 
vs. the measured $p$-air cross section, $\spai$, in mb, using the calculation of Fig. \ref{fig:newBvssigpp}. 
}
}
\label{fig:newsigppvssigpair}
\end{figure}
%%%%%%%%%%%%%%%%%%%%%%%%%%%%%%%%%%%%%%%%%%%%%%
%%%%%%%%%%%%%%%%%%%%%%%%%%%%%%
\begin{figure}[tbp]%Fig. 62
\begin{center}
\mbox{\epsfig{file=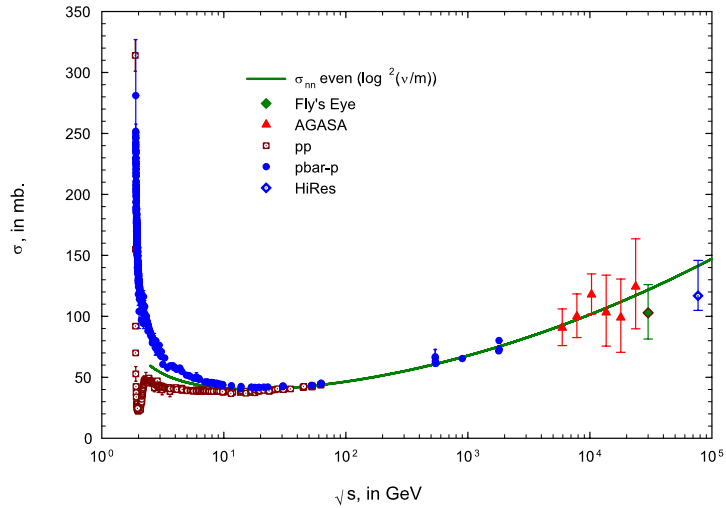%
              ,width=3.8in,bbllx=0pt,bblly=0pt,bburx=430pt,bbury=300pt,clip=%
}}
\end{center}
\caption[All known $\sigma_{pp}$ and $\sigma_{\bar pp}$ accelerator total cross sections, together with $\sigma_{pp}$ deduced from the AGASA, Fly's Eye and HiRes cosmic ray experiments]
{\footnotesize
\protect
{ All known $\sigma_{pp}$ and $\sigma_{\bar pp}$ accelerator total cross sections\cite{pdg},  together with $\sigma_{pp}$ deduced from the AGASA, Fly's Eye and HiRes cosmic ray experiments. The $pp$ and $\bar pp$  accelerator total cross sections, in mb, are plotted against the c.m. energy $\sqrt s$, in GeV. The circles are $\bar pp$ and the open squares are $pp$ data. The solid curve is a plot of the even (under crossing) nucleon-nucleon cross section $\sigma^0=(\sigma_{pp}+\sigma_{\bar pp})/2$, taken  from an analytically constrained $\ln^2 s$ global fit which included the cosmic ray data.  The AGASA data are the triangles, the Fly's Eye point is the diamond and the HiRes point is the open diamond. A value of k=1.287 was used. See the text for details.
}
}
\label{fig:hirespp}
\end{figure}
%%%%%%%%%%%%%%%%%%%%%%%%%%%%%%%%%%%%%%%%%%%%%%

%%%%%%%%%%%%%%%%%%%%%%%%%%%%%%%%%%%%%%%%%
\subsubsection{Final experimental $k$ value}
We consider $k=1.287\pm 0.06$ to be the final and best value, since it is derived from our new analysis that  combines  the 4-constraint $\ln^2 s$ fit of Section \ref{sec:lnsqpp} that used a sieved data set in its fit, together with a 2-constraint $B$ fit from the Aspen model of Section \ref{sec:Bfit}. We next compare our best-fit value $k=1.287\pm 0.06$---which is an experimental value---with Pryke's\cite{pryke} Monte Carlo values of $k=1.30\pm0.04$ and $k=1.32\pm0.03$ for CORSIKA-QGSjet and MOCCA-Internal models, respectively.  Their close agreement strongly suggests that we understand the development of air showers, both in structure and in particle production model, when CORSIKA and QGSjet are used. 

The good agreement between the revised values of $\spai$ from  the AGASA and Fly's Eye experiments and the published HiRes experiment is most gratifying. It finally gives a  strong foundation and consistency to the analysis of cosmic ray extensive air showers whose understanding  require  shower development and particle production models, thus putting the interpretation of these experiments on a much sounder footing. 
%%%%%%%%%%%%%%%%%%%%%%%%%%%%%%%%
%%%%%%%%%%%%%%%%%%%%%%%%%%%%%%%
\subsubsection{Predicting $\sigma_{pp}$ from a measurement of $\spai$}
From Fig. \ref{fig:newBvssigpp}, we can project out the values of $\sigma_{pp}$ as a function  of $\spai$. A plot of the predicted total  cross section $\sigma_{pp}$ , in mb, 
vs. the measured $p$-air cross section $\spai$, in mb, is shown in Fig. \ref{fig:newsigppvssigpair}, enabling  cosmic ray experimenters to convert        
their measured $\spai$ cross sections into $pp$ total cross sections. 

Shown in Fig. \ref{fig:hirespp} is a plot of all known accelerator cross section measurements of $\sigma_{pp}$  and $\sigma_{\bar pp}$, from threshold to $\sqrt s=1800$ GeV, taken from the Particle Data Group\cite{pdg} archives.
Using the results of Fig. \ref{fig:newsigpairvsE} together with Fig. \ref{fig:newsigppvssigpair}, we have converted the cosmic ray $p$-air measurements into $pp$ total cross sections.  The accelerator $pp$ and $\bar pp$ points are the open squares and circles, respectively; the cosmic ray points use triangles for AGASA, a diamond for Fly's Eye and an open diamond for the HiRes experiment. The theoretical curve, the solid line, is the even (under crossing) nucleon-nucleon cross section, 
$\sigma^0= (\sigma_{pp}+\sigma_{\bar pp})/2$. Because of the very  high density of low energy accelerator points, for the sake of visibility we show the prediction of the even cross section $\sigma^0$, rather than for the predicitions for $\sigma_{pp}$ and $\sigma_{\bar pp}$,   for $\sqrt s\ge 3$ GeV, which splits the difference of the $\sigma_{pp}$ and $\sigma_{\bar pp}$ curves.  
 
The excellent agreement between the $\ln^2 s$ theoretical curve  that saturates the Froissart bound and the  experimental accelerator and cosmic ray cross sections extends over 5 decades of c.m. energy; indeed, it is a triumph of phenomenology.
\section{C2CR: Colliders to Cosmic Rays}
We have finally reached our goal.  This long energy tale of accelerator experiments, extending over some 55 years, starting with Van der Graafs, next with cyclotrons, then  synchrotrons  and  finally,  colliders, has now  been unified with those ultra-high energy experiments that use high energy cosmic rays as their beams.  The accelerator experiments always had large fluxes and accurate energy measurements, allowing for precision measurements; typically, the lower the energy the more the precision. On the other hand, the cosmic ray experiments have the highest attainable energy, but have always suffered from low fluxes of particles and poor energy determinations of their events. 

The ability to clean up accelerator cross section and $\rho$-value data by the Sieve algorithm (see Section \ref{section:sievealgorithm}), along with new fitting techniques using analyticity constraints (see Section \ref{sec:4constraints}) in the form of anchoring high energy cross section fits to the value of low energy $pp$ and $\bar pp$ experimental cross sections and their derivatives, have allowed us to make constrained fits, using the a $\ln^2 s$ form that saturates the Froissart bound, i.e.,
\ba
\sigma^\pm(\nu)&=&c_0+c_1\ln\left(\frac{\nu}{m}\right)+c_2\ln^2\left(\frac{\nu}{m}\right)+\beta_{\cal P'}\left(\frac{\nu}{m}\right)^{\mu -1}\pm\  \delta\left({\nu\over m}\right)^{\alpha -1},\\
\rho^\pm(\nu)&=&{1\over\sigma^\pm(\nu)}\left\{\frac{\pi}{2}c_1+c_2\pi \ln\left(\frac{\nu}{m}\right)-\beta_{\cal P'}\cot({\pi\mu\over 2})\left(\frac{\nu}{m}\right)^{\mu -1}+\frac{4\pi}{\nu}f_+(0)\right.\nonumber\\
&&\left.\qquad\qquad\qquad\pm \delta\tan({\pi\alpha\over 2})\left({\nu\over m}\right)^{\alpha -1} \right\}.
\ea

We can now make  precision determinations of the coefficients $c_0,c_1,c_2$, $\beta_{\cal P'}$, $\delta$ and $\alpha$, allowing the phenomenologist to make accurate cross section and $\rho$-value extrapolations into the LHC and cosmic ray energy regions, extrapolations guided by the principles of analyticity and unitarity embodied in the Froissart bound.

%%%%%%%%%%%%%%%%%%%%%%%%%%%
%%%%%%%%%%%%%%%%%%%%%%%%%%%%%%%%%%%%%%%%%%%%%%
\section{Acknowledgements}
I would like to acknowledge valuable discussions with my great friend and often colleague, Prof. Francis Halzen of University of Wisconsin, who has helped me immensely; with Prof. Steven M. Block of Stanford University, who has been of great editorial aid;  to Dr. Robert N. Cahn, of Lawrence Berkeley Laboratory, with whom I commenced my work in the phenomenology of nucleon-nucleon scattering and with whom I was co-author of our 1985 Review of Modern Physics paper, ``High-energy $p\bar p$ and $pp$ forward elastic scattering and total cross sections'' and who taught me so much; to the Aspen Center for Physics for its hospitality and use of its resources; and finally, to my wife Beate Block, for her patience and forbearance during the long writing of this monograph. 

%\appendix  is contained in file {APP_ABC_2.TEX}
%\input{APP_ABC.TEX}
\appendix
%%%%%%%%%%%%%%%%%%%%%%%%
\section{QCD-inspired eikonal: the Aspen Model}\label{app:qcdeikonal}

The even QCD-inspired eikonal $\chi_{even}$ is given by the sum of
three contributions, gluon-gluon, quark-gluon, and quark-quark, which are
individually factorizable into a product of a cross section $\sigma
(s)$ times an impact parameter space distribution function
$W(b\,;\mu)$, i.e.,:
\begin{eqnarray}
\chi^{even}(s,b)&=&\chi_{gg}(s,b)+\chi_{qg}(s,b)+\chi_{qq}(s,b)
\nonumber\\ 
&=&i\left[\sigma_{gg}(s)W(b\,;\mu_{gg}) 
+ \sigma_{qg}(s)W(b\,;\sqrt{\mu_{qq}\mu_{gg}})
+ \sigma_{qq}(s)W(b\,;\mu_{qq})\right]\; ,
\label{chieven}
\end{eqnarray}
where the impact parameter space distribution function $W(b\,;\mu)=\frac{\mu^2}{96\pi}(\mu b)^3K_3(\mu b)$is normalized so that $
\int W(b\,;\mu)d^2 \vec{b}=1$, where $b$ is the 2-dimensional impact parameter. Hence, the $\sigma$'s in \eq{chieven} have the dimensions of a cross section.

The factor $i$ is inserted in \eq{chieven} since the high energy
eikonal is largely imaginary (the $\rho$-value for nucleon-nucleon
scattering is rather small).

As a consequence of both factorization and the normalization chosen
for the $W(b\,;\mu)$, it should be noted that $\int \chi^{even}(s,b)\, d^2\vec b=i\left[\sigma_{gg}(s)+
\sigma_{qg}(s)+\sigma_{qq}(s)\right], \label{integrateeven}$ so that, after using \eq{sigtot} for small $\chi$, 
$
\sigma_{\rm tot}^{even}(s)=2\,{\rm Im}\left\{i\left[\sigma_{gg}(s)+\sigma_{qg}(s)+\sigma_{qq}(s)\right]\right\}$
%\label{smallevensigma}.

In \eq{chieven}, the inverse sizes (in impact parameter space)
$\mu_{gg}$ and $\mu_{gg}$ are to be fit by experiment, whereas
the quark-gluon inverse size is taken as $\sqrt{\mu_{qq}\mu_{\rm
gg}}$.

\subsection{The $\sigma_{gg}$ contribution}
\label{app:siggg}

Modeling the gluon-gluon interaction after QCD, we write the cross
section $\sigma_{gg}(s)$ in \eq {chieven} as
\begin{equation}
\sigma_{gg}(s)=C_{gg}N_{g}^2\int\Sigma_{gg}
\theta(\hat s-m_0^2)F_{gg}\left ( x_1x_2=
\frac{\hat s}{s}\right )\,d\left (\frac{\hat s}{s}\right )\label{sigggqcd},
\end{equation}
where
\begin{equation}
\Sigma_{gg}=\frac{9\pi \alpha_s^2}{m_0^2}.
\end{equation}
The normalization constant $C_{gg}$ and the threshold $m_0$ are to
be fitted by experiment (in practice, the threshold is taken as $m_0 =
0.6$~GeV, and the strong coupling constant $\alpha_s$ is fixed at
0.5). The constant $N_g$ in \eq{sigggqcd} is given by 
$N_{g}=\frac{3}{2}\frac{(
5-\epsilon)(4-\epsilon)(3-\epsilon)(2-\epsilon)(1-\epsilon)}{5!}$.
Using the gluon structure function
\begin{equation}
f_{g}(x)=3\frac{(1-x)^5}{x^{1+\epsilon}},\label{fgepsilon}
\end{equation} 
we can now write the function $F_{gg}$ in \eq{sigggqcd} as
\begin{equation}
F_{gg}=\int\int f_{g}(x_1)f_{g}(x_2)\delta
(x_1x_2=\tau)\,dx_1\,dx_2.
\end{equation}
After carrying out the integrations, we can now explicitly express
$\sigma_{gg}(s)$ as a function of $s$. The parameter $\epsilon$ in
\eq{fgepsilon} is to be fitted by experiment (in practice, we fix it
at 0.05).

\subsubsection{High energy behavior of $\sigma_{gg}(s)$:  the Froissart bound}
\label{app:highenergysigmagg}

We note that the high energy
behavior of $\sigma_{gg}(s)$ is controlled by
\begin{eqnarray}
\lim_{s \rightarrow\infty}\mbox{ }\int^1_{m^2_0/s} d\tau F_{gg}(\tau)&\sim&
\int^1_{m^2_0/s} d\tau\frac{{}-\log\tau}{\tau^{1+\epsilon}}\nonumber\\
&&\mbox{}\nonumber \\
&\sim&\left(\frac{s}{m^2_0}\right)^{\epsilon}\log
\left(\frac{s}{m^2_0}\right),
\end{eqnarray}
where $\epsilon >0$.  The cut-off impact parameter $b_{\rm max}$ is given by
\begin{equation}
cW_{gg}(b_{\rm max};\mu_{gg})s^{\epsilon}\log(s)\sim 1,\label{about1}
\end{equation}
where $c$ is a constant.  For large values of $\mu b$, we can now
write \eq{about1} as
\begin{equation}
c'(\mu_{gg}b_{\rm max})^{3/2}e^{-\mu_{gg}b_{\rm max}}s^{\epsilon}\log(s)\sim 1
\end{equation}
with $c'$ another constant, and therefore,
\begin{equation}
b_c=\frac{\epsilon}{\mu_{gg}} \log \frac{s}{s_0}+O\left (\log\log
\frac{s}{s'_0}\right),
\end{equation}
where $s_0$ and $s_0'$ are scale constants.
As in Section \ref{sec:froissart}, we again reproduce the Froissart bound, this time  from QCD arguments, i.e.,
\begin{equation}
\sigtot=2\pi \left(\frac{\epsilon}{\mu_{gg}}\right)^2\log^2
\frac{s}{s_0},\label{Froissart}
\end{equation}
when we go to very high energies, as long as $\epsilon>0.$ The usual
Froissart bound coefficient of the $\log^2\frac{s}{s_0}$ term,
$1/m^2_{\pi}=20$ mb, is now replaced by
$\left(\epsilon/\mu_{gg}\right)^2\sim 0.002$ mb. Note that $\mu_{gg}$
controls the size of the area occupied by the gluons inside the
nucleon.

\subsubsection{Evaluation of the $\sigma_{gg}$ contribution:} 
\label{app:sigggevaluation}

In the following, we set the
matrices $a(0)=-a(5)=-411/10$, $a(1)=-a(4)=-975/2,$ $a(2)=-a(3)=-600$
and $b(0)=b(5)=-9,$ $b(1)=b(4)=-225,$ $b(2)=b(3)=-900$.  The result is
\begin{eqnarray}
\sigma_{gg}(s)&=&C_{gg}\Sigma_{gg}N_{g}^2 
\int_{\tau_0}^{1}{F_{gg}\,d\tau}\nonumber \\
&=&C_{gg}\Sigma_{gg}N_{g}^2 \times \nonumber\\
&&\sum_{i=0}^{5}
\left\{
\frac{a(i)-\frac{b(i)}{i-\epsilon}}{i-\epsilon}
-\tau_0^{i-\epsilon}
\left(
\frac{a(i)-\frac{b(i)}{i-\epsilon}}{i-\epsilon}+\frac{b(i)}{i-\epsilon}
\log (\tau_0)
\right)
\right\}
\nonumber\\	
&=&C_{gg}\Sigma_{gg}N_{g}^2 \times\nonumber\\ &
&\!\!\!\!\!\!\!\!\left\{\quad \frac{
\frac{411}{10}+\frac{9}{\epsilon}}{\epsilon}
\ \quad -\tau_0^{-\epsilon}
\left(\frac{ \frac{411}{10}+\frac{9}{\epsilon}}{\epsilon}+
\frac {9\log(\tau_0)}{\epsilon}\right)\right.\nonumber \\
 &+ & \frac{ \frac{-975}{2}+\frac{225}{1-\epsilon}}{1-\epsilon}
 -\tau_0^{1-\epsilon}
\left(\frac{ \frac{-975}{2}+\frac{225}{1-\epsilon}}{1-\epsilon}-
\frac {225\log(\tau_0)}{1-\epsilon}\right)\nonumber  \\
 & + & \frac{-600+\frac{900}{2-\epsilon}}{2-\epsilon}
 -\tau_0^{2-\epsilon}
\left(\frac{ -600+\frac{900}{2-\epsilon}}{2-\epsilon}-
\frac {900\log(\tau_0)}{2-\epsilon}\right)\nonumber  \\
 & + & \frac{600+\frac{900}{3-\epsilon}}{3-\epsilon}
 -\tau_0^{3-\epsilon}
\left(\frac{600+\frac{900}{3-\epsilon}}{3-\epsilon}-
\frac {900\log(\tau_0)}{3-\epsilon}\right)\nonumber  \\
& + & \frac{ \frac{975}{2}+\frac{225}{4-\epsilon}}{4-\epsilon}
-\tau_0^{4-\epsilon}
\left(\frac{ \frac{975}{2}+\frac{225}{4-\epsilon}}{4-\epsilon}-
\frac {225\log(\tau_0)}{4-\epsilon}\right)\nonumber \\
 &+& \left.\frac{ \frac{411}{10}+\frac{9}{5-\epsilon}}{5-\epsilon}
 -\tau_0^{5-\epsilon}
\left(\frac{ \frac{411}{10}+\frac{9}{5-\epsilon}}{5-\epsilon}-
\frac {9\log(\tau_0)}{5-\epsilon}\right)\right\}
,\qquad {\rm where\ }\tau_0= \frac{{m_0}^2}{s}.
\label{Fggintegrated}
\end{eqnarray}

We note that we must fit the following 3 constants in order to specify
$\sigma_{gg}$:
\begin{enumerate}
\item the normalization constant $C_{gg}$.
\item the threshold mass $m_0$.
\item $\epsilon$, the parameter in the gluon structure function which 
determines 
the behavior at low x  ($\propto 1/x ^{1+\epsilon}$).
\end{enumerate}

\subsection{The $\sigma_{qq}$ contribution}
\label{app:sigqq}

If we use the toy structure function
\begin{equation}
f_{q}(x)=\frac{(1-x)^3}{\sqrt x},\label{fq}
\end{equation}
we can write
\begin{eqnarray}
 \sigma_{qq}(s)& \propto &
 \frac{m_0}{\sqrt{s}}\log\frac{s}{s_0}+{\cal P}\left(\frac{m_0}{\sqrt
 s}\right) \nonumber\\ & \approx & {\rm constant}+ \frac{m_0}{\sqrt
 s},
\end{eqnarray}
where $\cal P$ is a polynomial in $m_0/{\sqrt s}$.

Thus, we approximate the quark-quark term by
\begin{equation}
\sigma_{qq}(s)=\Sigma_{gg} \left( C + C_{Regge}^{even}
\frac {m_0}{\sqrt s}\right ), \label{sigmaqq} \end{equation} where
$C$ and $C_{Regge}^{even}$ are constants. Thus, $\sigma_{qq}(s)$ 
simulates quark-quark interactions with a constant cross
section plus a Regge-even falling cross section.

We must fit the following 2 constants in order to specify $\sigma_{\rm
qq}$:
\begin{enumerate}
\item the normalization constant $C$.
\item the normalization constant $C_{Regge}^{even}$.
\end{enumerate}

\subsection{The $\sigma_{qg}$ contribution}
\label{app:sigqg}

If we use the toy structure function
\begin{equation}
f_{g}(x)=\frac{(1-x)^5}{x},\label{fg}
\end{equation}
and the toy structure function $f_{q}(x)$ of \eq{fq} we can write
\begin{eqnarray}
 \sigma_{qg}(s)& \propto & C''\log \frac{s}{s_0}+C'{\cal
 P'}\left(\frac{m_0}{\sqrt s}\right) \nonumber\\ & \approx & C''\log
 \frac{s}{s_0}+ C',
\end{eqnarray}
 where $C'$ and $C''$ are constants and $\cal P'$ is a polynomial in
 $m_0/{\sqrt s}$.

Thus, if we absorb the constant piece $C'$ into the quark-quark term,
we can approximate the quark-gluon term by
\begin{equation}
 \sigma_{qg}(s)=\Sigma_{gg} C_{qg}^{log}\log\frac{s}{s_0},
 \label{sigmaqg} \end{equation} where $C_{qg}^{log}$ is a constant.
 Hence, we attempt to simulate diffraction with the logarithmic term
 $\sigma_{qg}(s)$.

We must fit the following 2 constants in order to specify $\sigma_{\rm
qg}$:
\begin{enumerate}
\item the normalization constant $C_{qg}^{log}$.
\item $s_0$, the square of the energy scale in the log term of \eq{sigmaqg}.
\end{enumerate}

\subsubsection{Making the even contribution analytic}
\label{app:evencontribution}

The total even contribution, which is not yet analytic, can be written
as the sum of equations \ref{Fggintegrated}, \ref{sigmaqq} and
\ref{sigmaqg},  i.e.,
\begin{eqnarray}
\chi_{even}&=&i\left\{
\vphantom{
\left.\Sigma_{gg}\left [ \left( C + C_{Regge\  even} 
\frac {m_0}{\sqrt s}\right )W(b\,;\mu_{qq})+ C_{qg\ log}
\log\frac{s}{s_0}W(b\,;\sqrt{\mu_{qq}\mu_{gg}})\right]\right\}}
\sigma_{gg}(s)W(b\,;\mu_{gg})\right.\nonumber\\
&&\quad+\Sigma_{gg} \left( C + C_{Regge\ even} \frac
{m_0}{\sqrt s}\right )W(b\,;\mu_{qq})\nonumber\\ &&\,\,\,\quad +
\left. \Sigma_{gg} C_{qg\
log}\log\frac{s}{s_0}W(b\,;\sqrt{\mu_{qq}\mu_{gg}})\right\}.\label{finaleven}
\end{eqnarray}
For large $s$, the even amplitude in \eq{finaleven} can be made
analytic by the substitution 
\[s\rightarrow se^{-i\pi/2}.\] 
Thus, we finally rewrite the even contribution of \eq{finaleven},
which is now analytic, as
\begin{eqnarray}
\chi_{even}&=&i\left\{
\vphantom{
\left.\Sigma_{gg}\left [ \left( C + C_{Regge\  even} \frac{m_0}
{\sqrt s}
\right )W(b\,;\mu_{qq})+ C_{qg\ log}\log\frac{s}{s_0}W(b\,;
\sqrt{\mu_{qq}
\mu_{gg}})\right]\right\}}
\sigma_{gg}(se^{-i\pi/2})W(b\,;\mu_{gg})\right.\nonumber\\
&&\quad+\Sigma_{gg}\left( C + C_{Regge\ even} \frac
{m_0}{\sqrt s}e^{i\pi/4}\right )W(b\,;\mu_{qq})\nonumber\\
&&\,\,\,\quad + \left. \Sigma_{gg}C_{qg\log}
\left(\log\frac{s}{s_0}-i\frac{\pi}{2}\right)W(b\,;\sqrt{\mu_{qq}\mu_{gg}})
\right\}.\label{evenanalytic}
\end{eqnarray}

To determine the impact parameter profiles in $b$ space, we also must
fit the mass parameters $\mu_{gg}$ and $\mu_{qq}$ to the
data. We find masses $\mu_{gg}\approx 0.73$ GeV and 
$\mu_{qq}\approx 0.89$ GeV.

\subsection{The Odd eikonal}
\label{app:oddeikonal}

It can be shown that a high energy analytic odd amplitude (for
its structure in $s$, see Eq. (5.5b) of reference \cite{bc}, with
$\alpha =0.5$) that fits the data is given by
\begin{eqnarray}
\chii^{odd}(b,s)&=&-\sigma_{odd}\,W(b;\mu_{odd})\nonumber\\
&=&-C_{odd}\Sigma_{gg}\frac{m_0}{\sqrt{s}}e^{i\pi/4}\,W(b;\mu_{odd}),
\label{oddanalytic}
\end{eqnarray}
with
\be
W(b,\mu_{odd})=\frac{\mu_{odd}^2}{96\pi}(\mu_{odd}
b)^3\,K_3(\mu_{odd} b)\label{Woddnormalization},
\ee
normalized so that
\begin{equation}
\int W(b\,;\mu)d^2 \vec{b}=1. \label{oddWintegral}
\end{equation}
Hence, the $\sigma_{odd}$ in \eq{oddanalytic} has the dimensions
of a cross section.

In order for $C_{odd}$ to be positive, a minus sign has been
inserted in \eq{oddanalytic}.

With the normalization (\eq{Woddnormalization} and \eq{oddWintegral})
chosen for $W(b,\mu_{odd})$, we see that
\be
\int \chi^{odd}(s,b)\, d^2\vec b=\sigma_{odd}(s)\label{integrateodd},
\ee
so that, using \eq{sigtot} for small $\chi$,
\be
\sigma_{tot}^{odd}(s)=2\,{\rm Im}\,\sigma_{odd}(s)
\label{smalloddsigma}.
\ee

In order to determine the cross section $\sigma_{odd}$, we must
fit the normalization constant $C_{odd}$. To determine the impact
parameter profile in $b$ space, we also must fit the mass parameter
$\mu_{odd}$ to the data. We find a mass $\mu_{odd}\approx
0.53$ GeV.

We again reiterate that the odd eikonal, which we see (from
\eq{oddanalytic}) vanishes like $1/\sqrt s$, accounts for the
difference between $pp$ and $\bar p p$. Thus, at high energies,
the odd term vanishes, and we can neglect the difference between $pp$
and $\bar p p$ interactions .

\section{2-body, 3-body and n-body phase space}\label{sec:phspace} 
In  Appendix \ref{sec:phspace} we derive the Lorentz-invariant phase space for 2-body, 3-body, up to n-body phase space, for reactions such as 
$$M_{n}\rightarrow m_0+m_1+\cdots+m_{n-1},\qquad i=0,1,\cdots,n-1$$ or 
$$a+b\rightarrow m_0+m_1+\cdots+m_{n-1},\qquad  i=0,1,\cdots,n-1,$$
 where we have a decay into $n$ particles or an inelastic reaction with $n$ particles in the final state.  A decay into 5 particles, illustrating the notation we will employ,  is shown in Figure \ref{phspd24}. 
\subsection {2-body kinematics}\label{ap:2bodykinematics}
Consider a system of two particles, of rest masses $m_1$ and $m_2$. We
will work in  the  center-of-mass (c.m. or *)
frame, defined as the frame where $\vec p^{\,*}_1=-\vec p_2^{\,*}$ (it would
perhaps be better named as the center-of-momentum frame). In general, 
$P=P_1+P_2$, and thus $P^2=(P_1+P_2)^2$, is a Lorentz invariant, normally called $s$. Defining 
$\vec k=\vec p^{\,*}_1$ and $E^*=E_1^*+E_2^*$ in the c.m. frame,  we have
$P^*_1=(E^*_1,\vec k)$ and $P^*_2=(E^*_2,-\vec k)$. We find that 
$P^2=(E^*_1+E^*_2)^2={E^*}^2$. Using ${E^*_1}^2={k}^2+m_1^2$ and
${E^*_2}^2={k}^2+m_2^2$ to get
${E^*}^2-2E^*E^*_1+{k}^2+m_1^2={k}^2+m_2^2$ by squaring $E^*-E^*_1$,
we find the 
well-known
two-body c.m. frame kinematics relation,
\begin{equation}
E^*_1=\frac{{E^*}^2+m_1^2-m_2^2}{2E^*}\qquad \mbox{\ and}
\qquad E^*_2=\frac{{E^*}^2+m_2^2-m_1^2}{2E^*}.\label{twobody}
\end{equation}
Using Eq.~(\ref{twobody}), we deduce that 
\begin{eqnarray}
\frac{k}{E^*}&=&\frac{1}{E^*}\left({E^*_1}^2-m_1^2\right)^{1/2}
\nonumber\\
&=&\frac{1}{2{E^*}^2}\left[{E^*}^4-2{E^*}^2(m_1^2+m_2^2)+m_1^4+m_2^4-
2m_1^2m_2^2\right]^{1/2}
\nonumber\\
&=&\frac{1}{2}\left[1-2\left(\frac{m_1^2}{{E^*}^2}+\frac{m_2^2}{{E^*}^2}
\right)+\left(\frac{m_1^2}{{E^*}^2}-\frac{m_2^2}{{E^*}^2}
\right)^2\right]^{1/2}.\label{PoverE1}
\end{eqnarray}
At this point it becomes convenient to introduce a new function 
$\F(x,y),$ defined as
\begin{equation}
\F (x,y)\equiv \sqrt{1-2(x+y)+(x-y)^{2}},
\label{eq:f1}
\end{equation}
having the  properties 
\begin{eqnarray}
\F(x,y)&=&{\cal F}_{1}(y,x),\nonumber\\
\F(x,0)&=&{\cal F}_{1}(0,x)=1-x,\nonumber\\
\F(0,0)&=&1.%\label{Fproperties}
\end{eqnarray}
Using Eq.~(\ref{eq:f1}), we then rewrite Eq.~(\ref{PoverE1}) as
\begin{equation}
k=\frac{E^*}{2}\F \left(\frac{m_1^2}{{E^*}^2},
\frac{m_2^2}{{E^*}^2}\right).%\label{PoverE}
\end{equation}
\subsection{2-body phase space}
Consider the decay sequence $M_2\rightarrow m_0+m_1$, where
$M_{2}^{2}=(P_{0}+P_{1})^{2}$. Let us define  $I_{2}$, which is
proportional to the integrated 
two-body phase space, as 
\begin{eqnarray}
I_{2} & = & \int_{\vec{p}_{1_{min}}}^{\vec{p}_{1_{max}}}\int_{\vec{p}_{0_{min}}}^
{\vec{p}_{0_{max}}} \frac {d^{3}\vec{p}_{1}}{2 E_{1}}
\frac {d^{3}\vec{p}_{0}}{2 E_{0}}
\delta^{4}(P_{2}-p_{1}-p_{0}).\label{I2}
\end{eqnarray}
Since it  is Lorentz-invariant, 
we can take advantage of its invariance and we can evaluate it 
(most simply) in the c.m. 
system, where we denote the four vectors by $P^*=(E^*,{\vec p}^{\,*})$. 
Here we have 
$\vec{p}_{0}^{\,*}+\vec{p}_{1}^{\,*}=0$, the case where  $M_{2}$ 
decays {\em at rest} into $m_{1}$ and  $m_{0}$.  
\subsubsection{Method 1}\label{method1}
We rewrite it as
\begin{eqnarray}
 I_{2} & = & \int_{ { {\vec{p}}_{1_{min}} }^{\,*} }
^{ { {\vec{p}}_{1_{max}} }^{\,*} }
\int_{{\vec{p}_{0_{min}}^{\,*}}}^{{\vec{p}_{0_{max}}^{\,*}}}
\frac {d^{3}\vec{p}_{1}^{*}}
{2 E_{1}^{*}} \frac {d^{3}\vec{p}_{0}^{*}}{2 E_{0}^{*}}\, \delta^{3}
(\vec{p}_{0}^{\,*}+\vec{p}_{1}^{\,*})\, \delta\!\left(
M_{2}- E_{1}^{*}- E_{0}^{*}\right) 
 \nonumber \\ 
& = &
\int_{ { {\vec{p}}_{1_{min}} }^{\,*} }^{ { {\vec{p}}_{1_{max}} }^{\,*} }
\int_{{\vec{p}_{0_{min}}^{\,*}}}^{{\vec{p}_{0_{max}}^{\,*}}}
\frac {d^{3}\vec{p}_{1}^{*}}
{2 E_{1}^{*}} \frac {d^{3}\vec{p}_{0}^{*}}{2 E_{0}^{*}}\, \delta^{3}
(\vec{p}_{0}^{*}+\vec{p}_{1}^{*}) 
\,\delta\!\left(M_{2}- \sqrt{{p_1^*}^2+m_1^2}
- \sqrt{{p_1^*}^2+m_0^2}\right) 
 \nonumber \\  
       & = & \int_{\Omega_{1}^{*}} \frac {p_{1}^{* 2} d\Omega_{1}^{*}}
{2 E_{1}^{*} 2 E_{0}^{*}} \frac {E_{1}^{*} E_{0}^{*}}{p_{1}^{*}M_{2}}
\nonumber  \\
       & = & 4\pi \frac{p_{1}^{* 2}}{2 E_{1}^{*} 2 E_{0}^{*}}
\frac {E_{1}^{*} E_{0}^{*}}{p_{1}^{*}M_{2}}.
\label{eq:idue}
\end{eqnarray}
In Eq.~(\ref{eq:idue}), we will use the Dirac delta function relation, 
\begin{equation}
\int_{x_{\rm min}}^{x_{\rm max}} g(x)\delta[f(x)]\,dx=
\sum_{i=1}^n\frac{g(x)
_{x={x_0}_i}}{
\left|\frac{\partial f(x)}
{\partial x}\right|_{x={x_0}_i}}
\label{Diraclemma}\,,
\end{equation}
where ${x_0}_i$  is defined as the $i$-th root of $f(x)=0$ in
the interval $x_{\rm min} \le x\le x_{\rm max}$.
After integrating over the Dirac delta function 
$\delta\!\left(M_{2}- \sqrt{{p_1^*}^2+m_1^2}
- \sqrt{{p_1^*}^2+m_0^2}\right)$, using  the energy conservation 
relation
$M_2=E_1^*+E_0^*$,
we evaluate $I_2$ as
\begin{equation}
I_{2}=\frac {4\pi}{2^2} \frac{p_{1}^{*}}{M_{2}}\,.
%\label{eq:11bis}
\end{equation}
Defining 
$ U_{0} \equiv\left(\frac{m_0}{M_2}\right)^2$  and 
$U_{1}\equiv\left(\frac{m_1}{M_2}\right)^2$, we find
\begin{equation}
\frac {p_{1}^{* }}{M_{2}}=\frac {\F(U_{1},U_{0})}{2}
%\label{eq:psue}
\end{equation}
and now rewrite $I_2$ in its final form as
\begin{eqnarray}
I_{2}  & = & \frac {4 \pi }{2^{3}}\F(U_{1},U_{0}).\label{trueI2}
\end{eqnarray}
\subsubsection{Method 2}
We here show that 
\begin{equation}
\frac{1}{2E}=\int ^{+\infty}_{-\infty}\delta(P^2-m^2)
\theta (E)\,dE, \label{delfn}
\end{equation}
where $P^2=E^2-{\vec p}^{\,\, 2}$, and the theta function is defined as
\begin{eqnarray}
\theta (x)&\equiv &1\quad\mbox{if } x\ge 0,\nonumber\\
\theta (x)&\equiv &0\quad\mbox{if } x < 0.%\label{theta}
\end{eqnarray}

The proof of \eq{delfn} goes as follows:
\be
\int ^{+\infty}_{0}\delta(P^2-m^2)\,dE=
\int ^{+\infty}_{0}\delta(E^2-{\vec p}^{\,\,2}-m^2)\,dE=\frac{1}{2E},
\label{proof1}
\ee
using the results of the Dirac lemma of Eq.~(\ref{Diraclemma}), with
$g(E)=1$ and $f(E)=E^2-{\vec p}^{\,\,2}-m^2$, and $E=\sqrt{{\vec p}^{\,\,2}+m^2}$.  
We can rewrite Eq.~(\ref{proof1}), using the $\theta$ function, as 
\be
\frac{1}{2E}=\int ^{+\infty}_{-\infty}\delta(P^2-m^2)
\theta (E)\,dE, %\label{delfn2}
\ee
which completes our proof.

Hence,
\be
I_{2} = \int_{\vec{p}_{1_{min}}}^{\vec{p}_{1_{max}}}\int_{\vec{p}_{0_{min}}}^
{\vec{p}_{0_{max}}} \frac {d^{3}\vec{p}_{1}}{2 E_{1}}
\frac {d^{3}\vec{p}_{0}}{2 E_{0}}
\delta^{4}(P_{2}-p_{1}-p_{0})%\label{I2new}
\ee
can be rewritten as
\begin{eqnarray}
I_2&=&\int_{\vec{p}_{1_{min}}}^{\vec{p}_{1_{max}}}
 \frac {d^{3}\vec{p}_{1}}{2 E_1}\int\theta(E_0)
\delta(P_0^2-m_0^2)\delta^4(P-p_1-p_0)\,d^4P_0 \nonumber\\
&=&\int_{\vec{p}_{1_{min}}}^{\vec{p}_{1_{max}}}
 \frac {d^{3}\vec{p}_{1}}{2 E_1}\delta\left((P-p_1)^2-m_0^2)\right)
\theta (E-E_1).\label{i2} 
\end{eqnarray}
To evaluate the second line in \eq{i2}, we integrate over the 4-dimensional
delta function and set $P-p_1-p_0=0$.
Since $\vec P=0$ in the c.m. frame, we can rewrite $I_2$ as
\begin{eqnarray}
I_2&=&4\pi \int_0^{p_{1max}} p_1^2\frac{dp_1}{2E_1}
\delta\left(P^2+m_1^2-m_0^2-2EE_1\right)
\theta (E-E_1)\nonumber\\
&=&\frac{4\pi}{2^2} \frac{p_1^*}{M_2},
\end{eqnarray}
because $p_1=p_1^*$ and $E=E^*=M_2$ in the c.m. system .
%%%%%%%%%%%%%%%%%%%%%%%%%%%%%%
\subsubsection{Method 3}\label{method3}
We can also write $I_2$ as
\begin{eqnarray}
 I_{2} & = & \int_{ { {\vec{p}}_{1_{min}} }^{\,*} }
^{ { {\vec{p}}_{1_{max}} }^{\,*} }
\int_{{\vec{p}_{0_{min}}^{\,*}}}^{{\vec{p}_{0_{max}}^{\,*}}}
\frac {d^{3}\vec{p}_{1}^{*}}
{2 E_{1}^{*}} \frac {d^{3}\vec{p}_{0}^{*}}{2 E_{0}^{*}} \delta^{3}
(\vec{p}_{0}^{\,*}+\vec{p}_{1}^{\,*}) \delta(M_{2}- E_{1}^{*}- E_{0}^{*}) 
 \nonumber \\  
& = & \int_{ { {\vec{p}}_{1_{min}} }^{\,*} }
^{ { {\vec{p}}_{1_{max}} }^{\,*} }
\frac {d^{3}\vec{p}_{1}^{*}}
{2 E_{1}^{*}}  \delta\left(M_{2}- \sqrt{{p^*_1}^2+m_1^2}-
\sqrt{{p^*_1}^2+m_0^2}\,\right) 
\nonumber  \\
       & = &\int_{\Omega_{1}^{*}} \frac {p_{1}^{* 2} d\Omega_{1}^{*}}
{2 E_{1}^{*} 2 E_{0}^{*}} \frac {dp_{1}^{*}}{dM_{2}}=
4\pi \frac{p_{1}^{* 2}}{2 E_{1}^{*} 2 E_{0}^{*}}
\frac {dp_{1}^{*}}{dM_{2}}. 
%\label{eq:idue2}
\end{eqnarray}
Using energy conservation,
\begin{equation}
M_{2}=E_{1}^{*}+E_{0}^{*}=\sqrt{p_{1}^{* 2}+m_{0}^{2}}+
\sqrt{p_{1}^{* 2}+m_{1}^{2}},
\end{equation}
after differentiating with respect to 
$dp_{1}^{*}$  and inverting the relation, we get 
\begin{equation}
\frac {dp_{1}^{*}}{dM_{2}}=\frac {E_{1}^{*} E_{0}^{*}}{p_{1}^{*}M_{2}},
\end{equation}
and hence, 
\begin{equation}
I_{2}=\frac {4\pi}{4} \frac{p_{1}^{*}}{M_{2}}=
\frac {4 \pi }{2^{3}}\F(U_{1},U_{0})\,,
\end{equation}
as before. This method of  differentiating with respect to $dM$ 
will be exploited later.
\subsection{Energy and momentum conservation for the many-body problem}
In Figure \ref{phspd24}, we show a decay of a particle $M_5$ into 5 
particles, i.e., 
 \begin{equation}
M_5\rightarrow m_4+m_3+m_2+m_1+m_0.
\end{equation}
In order to easily conserve momentum and energy  simultaneously,
we visualize the decay as occurring {\em sequentially}, as follows.  
\begin{figure}[htb] %#64
\centerline{\psfig{figure=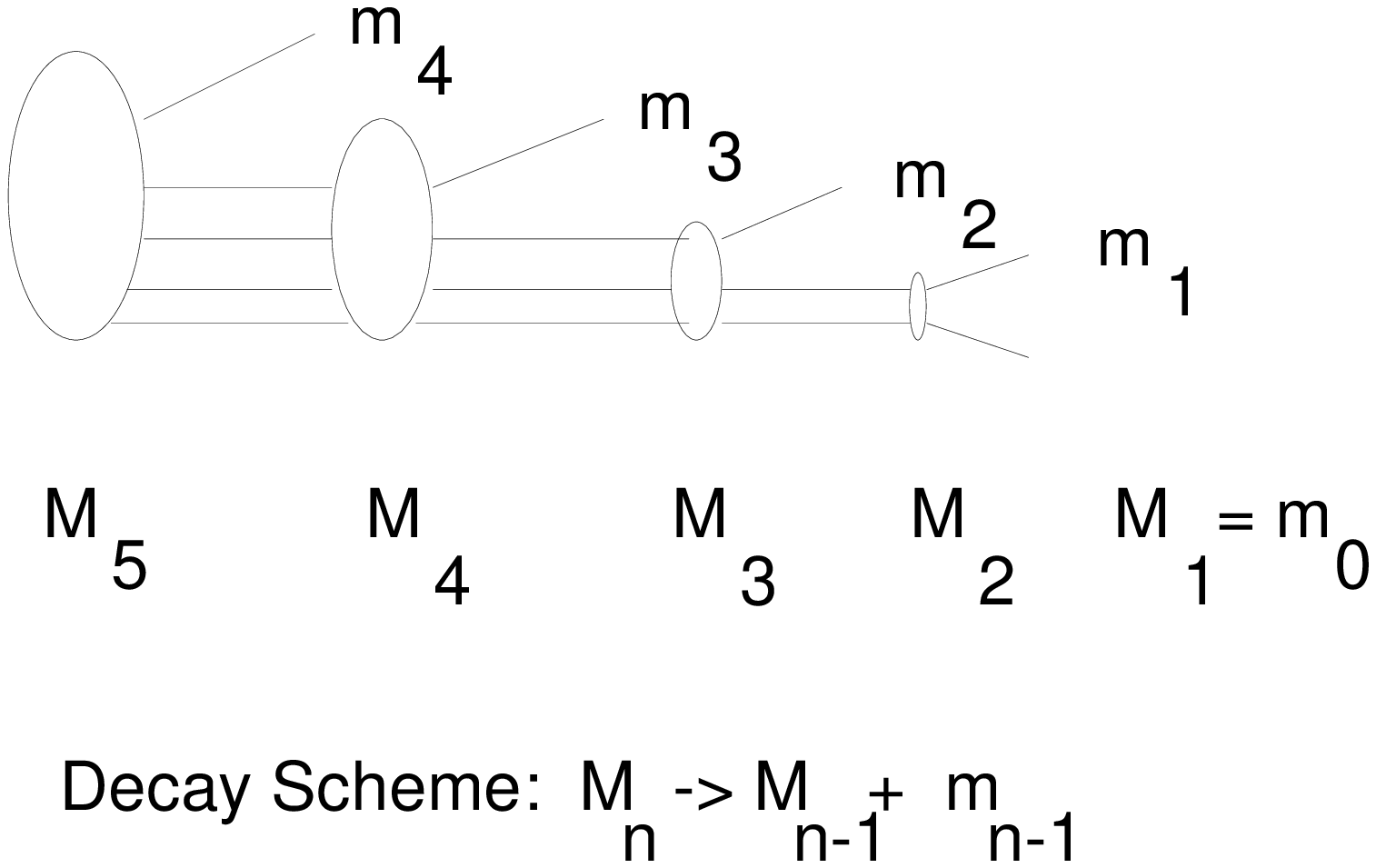,width=4.25in}}%True width =4.25in
%\centerwmf{4in}{3in}{c:/files/pctex/mytex/latex/phspd24.wmf}
\caption[The `decay' scheme for 5 particles ($m_4,m_3,\ldots,m_0$) in the 
final-state]
{\footnotesize The `decay' scheme for 5 particles ($m_4,m_3,\ldots,m_0$) in the 
final-state, where the `decays' are: $\qquad $
$M_n\rightarrow M_{n-1}+m_{n-1}$, with $M_1\equiv m_0$.}\label{phspd24}
\end{figure}

First,
$M_5$ decays into the physical particle $m_4$ and the 4-particle system of 
effective mass $M_4$. We conserve energy and momentum trivially in the 
 rest  frame of $M_5$, since in that frame we are dealing with 
two-body kinematics.

We next Lorentz transform to the rest frame
of $M_4$.  There, the particle $M_4$ decays at rest into the physical 
particle $m_3$ and the 3-particle system of effective mass $M_3$.  
Energy and momentum are  simple to conserve here, since again we have 
reduced the problem to two-body kinematics. We continue in this manner until
we get to the final decay, $M_2\rightarrow m_1+M_1$.  If we define 
$M_1\equiv m_0$, we now have the decay of $M_5$ into the 5 physical particles $m_4,m_3,\ldots,m_0$, 
with energy and momentum conserved at every step of the way.

To insure momentum and energy conservation at every vertex, we must work from {\em left to right } in Figure \ref{phspd24},  as
follows. We assume that $M_5$ is given. The value of $M_4$ can lie between
$\sum_{i=0}^{3}m_i$ and $M_5-m_4$.  The lower limit occurs when all 4 
particles that constitute $M_4$ go off with zero relative velocity, recoiling
against $m_4$. The upper limit corresponds to the decay of $M_5$ into   $m_4$ and  $M_4$ when both $m_4$ and  $M_4$ are at rest.  Thus, if
\begin{equation}
\sum_{j=0}^{i-1}m_j\le M_i\le M_{i+1}-m_i,\qquad i=2,3,\ldots,n, 
\label{Mconserve}
\end{equation} 
we automatically will have both energy conservation and momentum 
conservation.  The implementation of this idea is straightforward---
at each stage, we Lorentz transform to the appropriate rest frame and then
apply two-body kinematics.  In terms of the masses $m_i$ and $M_i$, the 
energies at this point in the decay chain are:
\begin{eqnarray}
E_{m_i}&=&\frac{M_{i+1}^2+m_i^2-M_i^2}{2M_{i+1}},\nonumber \\
E_{M_i}&=&\frac{M_{i+1}^2+M_i^2-m_i^2}{2M_{i+1}},%\label{Econserve}
\end{eqnarray}
where $E^*$ of Eq.~(\ref{twobody}) is replaced by the effective mass 
$M_{i+1}$. Later, we will discuss the Lorentz transforms needed to go to
the rest frames of the `decaying' particles.
%
%%%%%%%%%%%%%%%%%%%%%
\subsection{3-body phase space}
Consider the 3-body decay, $M_3\rightarrow m_2+m_1+m_0$.
Here,  
$P_{3}^{2}=M_{3}^{2}$.  We write $I_3$, similar to the two-body result of the 
preceding section, as
\begin{eqnarray}
I_{3}&=&
\int_{ { {\vec{p}}_{2_{min}} } }^{ { {\vec{p}}_{2_{max}} } }
\int_{ { {\vec{p}}_{1_{min}} } }^{ { {\vec{p}}_{1_{max}} } }
\int_{{\vec{p}_{0_{min}}}}^{{\vec{p}_{0_{max}}}}
\frac {d^{3}\vec{p}_{2}}{2 E_{2}}
\frac {d^{3}\vec{p}_{1}}{2 E_{1}}
\frac {d^{3}\vec{p}_{0}}{2 E_{0}}
\delta^{4}(P_{3}- \sum_{i=0}^{2} p_{i})  \\
     & = & \int_{ { {\vec{p}}_{2_{min}} }^{\,*} }
^{ { {\vec{p}}_{2_{max}} }^{\,*} }
\frac {d^3\vec{p_{2}^*}}{2 E_{2}^*}I_{2} \nonumber \\
     & = & \int_{ { {\vec{p}}_{2_{min}} }^{\,*} }
^{ { {\vec{p}}_{2_{max}} }^{\,*} }
\frac {d^3\vec{p_{2}^*}}{2 E_{2}^*}
\left[\frac {4 \pi }{2^{3}}\F (U_{1},U_{0})\right]
,\qquad \mbox{where }P_3=P_2+p_2\label{threebody}.
\end{eqnarray}
Again, we used the facts that both $I_2$ and $I_3$ are Lorentz invariants 
(they are both Lorentz scalars), and once 
again, we pick  special frames  where their evaluation is the simplest. In Eq.~(\ref{threebody}), we have used the value of $I_2$ found in Eq.~(\ref{trueI2}). We now choose to evaluate $p_2$ in the frame where $M_{3} \rightarrow M_{2}+m_{2}$, with $M_3$ at rest, where its value is $p^*_2$. In this frame, 
\begin{equation}
p^*_2=\frac{M_3}{2}\,\F \!\left(\frac{m_2^2}{M_3^2},
\frac{M_2^2}{M_3^2}\right),\qquad \mbox{where }M^2_2=M_3^2+m_2^2-2M_3E^*_2.
\end{equation}
Further,
\begin{eqnarray} 
\frac {d^{3}\vec{p}_{2}^{*}}{2 E_{2}^{*}} = 
\frac {p_{2}^{* 2} dp_{2}^{*} d\Omega_{2}^{*}}{2 E_{2}^{*}} =  
\frac{1}{2}p_{2}^{*} dE_{2}^{*}d\Omega_{2}^{*},
%\label{eq:cemasfra}
\end{eqnarray}
where
\begin{equation}
m_{2}\leq E_{2}^{*} \leq
\frac{M_{3}^{2}+m_{2}^{2}-(m_{0}+m_{1})^{2}}{2 M_{3}},
\end{equation}
which corresponds to 
\begin{equation}
m_0+m_1\le M_2\le M_3-m_2.
\end{equation}

We simplify the above by introducing a new {\em dimensionless}
variable
\begin{equation}
\xi_2=\left(\frac{M_2}{M_3}\right)^2 \label{xi},
\end{equation}
the ratio of the {\em square} of the effective masses. Since 
\begin{equation}
E_{2}^{*}=\frac{M_{3}^{2}+m_{2}^{2}-M_{2}^{2}}{2 M_{3}}
\label{eq:energiam2},
\end{equation}
we immediately find that
\begin{equation}
|dE_{2}^{*}|=\frac{dM_{2}^{2}}{2M_{3}^{2}}=\frac{M_3}{2}\,d\xi_2.
\end{equation} 
In terms of the new variable $\xi_2$, the integration limits are
\begin{eqnarray}
{\xi_{2_{min}}}&=&\left(\frac{m_{0}+m_{1}}{M_{3}}\right)^{2}\nonumber \\
{\xi_{2_{max}}}&=&\left(1-\frac {m_{2}}{M_{3}}\right)^{2}.%\label{xilimits}
\end{eqnarray}

Putting it all together after defining 
$U_2\equiv \left(\frac{m_2}{M_3}\right)^2$, we get
\begin{eqnarray}
\frac{1}{2}p_{2}^{*} dE_{2}^{*}d\Omega_{2}^{*} & = & \frac{1}{2^{3}}M_{3}^{2}
\F(\xi_{2},U_{2}) d\xi_{2}d\Omega_{2}^{*}.
\end{eqnarray}
The angular integration over $d\Omega_{2}^{*}$
trivial, giving $4\pi$.
Finally, we rewrite the 3-body phase space $I_3$ in its final form as
\begin{eqnarray}
I_{3} & = & (\frac {4 \pi }{2^{3}})^{2}M_{3}^{2} 
\int_{\xi_{2_{min}}}^{\xi_{2_{max}}} \F (\xi_{2},U_{2})d\xi_{2}
\times \F(U_{1},U_{0}).%\label{I3}
\end{eqnarray}
\subsubsection{Analytic energy and momentum spectra}
We can easily analytically calculate the energy spectrum of particle of mass $m_2$ for {\em arbitrary} masses $m_2,m_1,m_0$. This three-body 
case is the only
\Em{general} case in which an analytic solution is possible.
We note that 
\be
{E_2^*}=\frac{M_3^2+m_2^2-M_2^2}{2M_3}\quad {\rm or }\quad 
M_2^2=M_3^2+m_2^2-2M_3{E_2^*}.
%\label{mofe}
\ee
Thus, since ${E_2^*}\,d{E_2^*}={p_2^*}\,d{p_2^*}$,
we can write the differential energy spectrum $d\Gamma/d{E_2^*}$
(in arbitrary units) of \eq{threebody} as 
\be
\frac{d\Gamma}{d{E_2^*}}={p_2^*}\,
\F\!\left(\frac{m_1^2}{M_3^2+m_2^2-2M_3{E_2^*}},\frac{m_0^2}
{M_3^2+m_2^2-2M_3{E_2^*}}
\right),\label{espectrum}
\ee
where ${E_2^*}$ is the energy of particle 2 and ${p_2^*}$ is its momentum 
in the rest frame of the 3-particle system.

Since ${E_2^*}\,d{E_2^*}={p_2^*}\,d{p_2^*}$, we can rewrite \eq{espectrum} as the momentum
spectrum $d\Gamma/d{p_2^*}$,
\be
\frac{d\Gamma}{d{p_2^*}}=\frac{{p_2^*}^2}{{E_2^*}}\,
\F\!\left(\frac{m_1^2}{M_3^2+m_2^2-2M_3{E_2^*}},
\frac{m_0^2}{M_3^2+m_2^2-2M_3{E_2^*}}
\right),%\label{pspectrum}
\ee
where ${E_2^*}$ is the energy of particle 2 and ${p_2^*}$ is its momentum,
in the rest frame of the 3-particle
system.
\subsection{4-body phase space}
We now consider the 4-body system.
We write $I_4$,  after defining $P_4=p_3+p_2+p_1+p_0$ and  noting   that $P_4^2=M_4^2$, as
\begin{eqnarray}
I_4&=&\int_{ { {\vec{p}}_{3_{min}} }^{\,*} }^{ { {\vec{p}}_{3_{max}} }^{\,*} }
\int_{ { {\vec{p}}_{2_{min}} }^{\,*} }^{ { {\vec{p}}_{2_{max}} }^{\,*} }
\int_{ { {\vec{p}}_{1_{min}} }^{\,*} }^{ { {\vec{p}}_{1_{max}} }^{\,*} }
\int_{{\vec{p}_{0_{min}}^{\,*}}}^{{\vec{p}_{0_{max}}^{\,*}}}
\frac{d^3{\vec{p}_3}}{2E_3}\frac{d^3{\vec{p}_2}}{2E_2}
\frac{d^3{\vec{p}_1}}{2E_1}\frac{d^3{\vec{p}_0}}{2E_0}\,
\delta^4\!\!\left(\!P_4-\sum_{i=0}^{3}p_i\!\right) \nonumber \\
&=&\int_{ { {\vec{p}}_{3_{min}} }^{\,*} }^{ { {\vec{p}}_{3_{max}} }^{\,*} }
\frac{d^3{\vec{p}_3}}{2E_3}I_3.
\end{eqnarray}
Since $I_4$ is a Lorentz scalar,  we choose to evaluate it in the rest
frame of the ``decay'' $M_4\rightarrow M_3 + m_3$, the
$*$-frame.  Proceeding as before, after introducing the dimensionless variable
$\xi_3$, defined as
\begin{equation}
\xi_3\equiv\left(\frac{M_3}{M_4}\right)^2, \end{equation}
it is straightforward to show that
\begin{eqnarray}
\lefteqn{ I_4=\left(\frac{4\pi}{2^3}\right)^3 
\left[
\left( M_4^2 \right)^{4-2}
\!\!\int_{{\xi}_{3_{min}}}^{{\xi}_{3_{max}}}\,
\!\!\!\!\left({\xi_3}\right)^{3-2}{\cal F}_1\left(\xi_3,{u_3}\right)
\,d\xi_3\,
\!\!\int_{{\xi}_{2_{min}}}^{{\xi}_{2_{max}}}\,
\!\!\!\!\left({\xi_2}\right)^{2-2}{\cal F}_1\left(\xi_2,\frac{u_2}
{\xi_3}\right)\,d\xi_2\,
\right]\times} \nonumber\\
&&\,{\cal F}_1\left(\frac{u_1}{\xi_2\xi_3},
\frac{u_0}{\xi_2\xi_3}\right),
\end{eqnarray}
with $u_i$ defined by
\begin{equation}
u_i\equiv \left(\frac{m_i}{M_n}\right)^2, \ \ i=0 {\rm \ to\ }3.
%\label{eq:ui}
\end{equation}
%%%%%%%%%%%%%%%%%%%%%%%%%%%5
\subsection{$n$-body phase space}
The phase space for the $n$-body system is defined as
\begin{equation}
\Phi_n\left(M_n^2;m_{n-1}^2,m_{n-2}^2,\ldots ,m_1^2,m_0^2\right) =
\prod_{i=0}^{n-1} \int_{{\vec{p}_{i_{min}}}}^{{\vec{p}_{i_{max}}}}\,\frac{d^3
{\vec{p}_i}}{(2\pi )^3 2E_i}\, \delta^4\!\!\left(
\!P_n-\sum_{i=0}^{n-1}p_i\!\right),
\label{eq:phspace}
\end{equation}
where $p_i$ is the 4-momentum $\left(E_i,{\vec p}_i\right)$  of the
$i$th particle, $P_n$ is the 4-momentum of the whole system, and, in our
metric, $M^2_n=P_n^2 $.
It is easy to show that $\Phi_n\left(M_n^2;0\right)$, the phase 
space integral for
{\em{massless}} particles\cite{BlockJackson} whose total effective mass is 
$M_n$,
is given by
\begin{equation}
\Phi_n\left(M_n^2;0\right)=\frac {8\,\left(M^2_n\right)^{n-2}
}{{(4\pi)}^{2n+1}
\,(n-1)!\,(n-2)!}.  \label{eq:Phi0}
\end{equation}

For ease of writing, we again consider the
somewhat simpler quantity, $I_n$, as before.
It is clear that the generalization of $I_n$ for  $n$ particles is
\begin{eqnarray}
\lefteqn{I_{n}=\left(\frac{4\pi}{2^3}\right)^{n-1}
\times }\nonumber\\
&&\left[\left(M_n^2\right)^{n-2}
\prod_{i=2}^{n-1}\left\{
\!\!\int_{{\xi}_{i_{min}}}^{{\xi}_{i_{max}}}\,
\!\!\!\!\left({\xi_i}\right)^{i-2}{\cal F}_1\!\left(\xi_i,\frac{u_i}
{{\cal P}_i}\right)
\,d\xi_i\,\right\}
\right]\,{\cal F}_1\!\left(\frac{u_1}{{\cal P}_1},
\frac{u_0}{{\cal P}_1}\right),    \label{eq:In}
\end{eqnarray}
for 
\begin{equation}
u_k\equiv \left(\frac{m_k}{M_n}\right)^2,\qquad k=0,1,\ldots,i,
\end{equation}
where we now introduce the  dimensionless variable $\xi_i$,
defined as 
\begin{equation}
\xi_i\equiv\left(\frac{M_i}{M_{i+1}}\right)^2,\qquad i=2,3,\ldots,n-1,
\end{equation}
and express the kinematics limits as
\begin{eqnarray}
\xi_{i_{min}}&=&\left( \sum_{j=0}^{i-1}
\,\sqrt{u_j}\right)^2\!\! / \,{{\cal P}_i},\nonumber\\
\xi_{i_{max}}&=&\left(1-\sqrt{{u_i}/{{\cal P}_i}}\right)^2, 
\qquad i=2,\,3,\,\dots ,\,n-1,\label{eq:kinematics}
\end{eqnarray}
where  ${\cal P}_i$ is defined as
\begin{equation}
{\cal P}_i \equiv \prod_{j=i+1}^n \xi_j,\\
i=1,\,2,\,\dots ,\,n-1\\
{\rm \ and\ } \xi_n\equiv 1.  %\label{eq:Psubi}
\end{equation}
By inserting factors of $i(i-1)$ in {\em each} integrand in Eq.~(\ref{eq:In}),
and noting the difference between $I_n$ of Eq.~({\ref{eq:In}) and
$\Phi_n\left(M_n^2;m_{n-1}^2,m_{n-2}^2,\ldots ,m_1^2,m_0^2\right)$
of Eq.~(\ref{eq:phspace}), we can now write the $n$-dimensional phase space
as 
\begin{eqnarray}
\lefteqn{\Phi_n\left(M_n^2;m_{n-1}^2,m_{n-2}^2,\ldots ,m_1^2,
m_0^2\right) =\frac {8\,\left( M^2_n \right)^{n-2}}{{(4\pi)}^{2n+1}
\,(n-1)!\,(n-2)!} \times }\nonumber \\
&&\left[\prod_{i=2}^{n-1}\left\{
\int_{{\xi}_{i_{min}}}^{{\xi}_{i_{max}}}\,
\!\!\!\!i(i-1)\left({\xi_i}\right)^{i-2}{\cal F}_1\!\left(\xi_i,\frac{u_i}
{{\cal P}_i}\right)
\,d\xi_i\,\right\}
\right] \times {\cal F}_1\!\left(\frac{u_1}{{\cal P}_1},
\frac{u_0}{{\cal P}_1}\right)   \nonumber \\
&\quad=& \Phi_n\left(M_n^2;0\right) \nonumber\\&&\times
\left[\,
\prod_{i=2}^{n-1} \,\left\{ \int^{{\xi_i}_{max}}_{{\xi_i}_{min}}
i(i-1) \left( \xi_i \right)^{i-2}
{\cal F}_1 \!\left( \xi_i,\frac {u_i}{{\cal P}_i} \right) \,d\xi_i\,
\right\}
\right]\times
{\cal F}_1 \!\left( \frac{u_0}{{\cal P}_1},\frac{u_1}{{\cal P}_1} 
\right)\!. \label{eq:qed}
\end{eqnarray}
} % end of same page
To obtain the last transformation of Eq.~(\ref{eq:qed}),
we note from Eq.~(\ref{eq:Phi0}) that
$\Phi_n\left(M_n^2;0\right)$, the phase space
integral\cite{BlockJackson} for
{\em{massless}} particles whose total effective mass is $M_n$, 
now appears in Eq.~(\ref{eq:qed}) as a multiplying factor,
by virtue of our having multiplied each integrand by $i(i-1)$.  This also
has the effect of making each of the integrals {\em unity}  for massless 
particles,
because  ${\cal F}_1 \!\left( \xi_i,0 \right)=1-\xi_i$, as well as making  the integration 
limits of Eq.~(\ref{eq:qed}) become ${\xi_i}_{min}=0$ and ${\xi_i}_{max}=1$. 
We note that since $\F (0,0)=1$, 
Eq.~(\ref{eq:qed}) simplifies and becomes 
$\Phi_n\left(M_n^2;0\right)$,
which is, of course, the total phase space for $n$ massless particles. 

To understand better the kinematic limits of Eq.~(\ref{eq:qed}), as well 
as the
method that we will use to integrate it numerically (using Monte
Carlo techniques!), let us consider explicitly the example of the 
4-particle state,
$M_4\rightarrow m_3+m_2+m_1+m_0$.  For $n=4$, 
${\cal P}_3=\xi_4=1$, ${\cal P}_2=\xi_3 \xi_4=\xi_3$,
${\cal P}_1=\xi_2 \xi_3 \xi_4=\xi_2\xi_3$, and 
the kinematic limits for $\xi_3$ and 
$\xi_2$
are given by 
\begin{eqnarray}
{\xi_3}_{ min}&=&\frac{\left(\sqrt {u_0} +\sqrt{u_1}+\sqrt{u_2}\right)^2}
{\xi_4}
=\left(\frac{m_0+m_1+m_2}{M_4}\right)^2\nonumber\\
{\xi_3}_{max}&=&\left(1-\frac{m_3}{M_4}\right)^2,\label{xi3}\\
{\xi_2}_{ min}&=&\left(\sum_{j=0}^1\sqrt{u_j}\right)^2=
\frac{
\left(
\frac{m_0+m_1}{M_4}
\right)^2}{\xi_3} 
=\left(\frac{m_0+m_1}{M_3}\right)^2\nonumber\\
{\xi_2}_{max}&=&\left(1-\frac{m_2}{M_4\sqrt{\xi_3}}\right)^2
=\left(1-\frac{m_2}{M_3}\right)^2.  \label{xi2}
\end{eqnarray}
Thus, if we integrate Eq.~(\ref{eq:qed}) from {\em left} to {\em right}, 
rather
than the conventional right to left technique, we see from 
Eq.~(\ref{eq:qed}) that the limits of 
$\xi_3$  only depend on the c.m. energy $M_4$ and the
masses $m_0,m_1$ and $m_2$, all fixed quantities. We can then pick a value of $\xi_3$ from Eq.~(\ref{xi3}) and thus pick a value of the 
effective mass $M_3=\xi_3M_4$. Having picked $M_3$ (or its equivalent, $\xi_3$), we can then use  the limits of Eq.~(\ref{xi2}) to pick a value of $\xi_2$, since 
we already know $\xi_3$. This of course simultaneously chooses for us the
effective mass $M_2$ which subsequently decays into $m_1$ and $m_0$. 

We see that we can start our integration of Eq.~(\ref{eq:qed})
by going from {\em left} to {\em right}, each time picking an effective
mass within appropriate limits that conserve both energy and momentum. To
get the laboratory quantities of the momentum of the individual 
physical particles 
$m_0, m_1,\ldots,m_{n-1}$, we must take these effective masses that we just
chose, pick `decay' angles in the various c.m.systems, and finally, inverse 
Lorentz transform the particles' vector momenta into the laboratory system.

\section{Monte Carlo techniques}\label{sec:MCtechniques}
In this Section, we will discuss various Monte Carlo techniques, with the eventual goal of formulating a fast computer program to both evaluate $n$-body phase space and simulate  experimental `events' in Appendix \ref{sec:MCphasespace}.

Monte Carlo techniques are often used for two distinctly different goals.  

Goal 
number one is to find the integral of a function, $f(x_1,x_2,x_3,\ldots,x_n)$
in $n$-dimensional space, where the boundaries are often very complex. The
Monte Carlo method estimates the integral of the function over the 
$n$ dimensional volume V as
\be
\int f\,dV\approx V\overline{f}\pm V
\sqrt{\frac{\overline{f^2}-(\,\overline{f}\,)^2}{N}},\label{mc} 
\ee
where $N$ is the number of points sampled.
The bars in \eq{mc} are arithmetic means over the $N$ points,
\be
\overline{f}\equiv \frac{1}{N}\sum_{i=1}^N f(x_i)\quad
\overline{f^2}\equiv \frac{1}{N}\sum_{i=1}^N f^2(x_i). \label{means}
\ee
We note that the $\pm$ term in \eq{mc}, which is meant to be an estimate 
of a one standard
deviation error, should not be taken too literally since 
the error does not have to be  distributed as a Gaussian.

Goal number two is to build an `experimental' distribution of the function
\linebreak 
$f(x_1,x_2,x_3,\ldots,x_n)$, with the equivalent of `experimental events' which in the limit of large $N$,
reproduce the theoretical distribution $f$.  In other words, we would like 
to have a distribution that would have been achieved experimentally for 
$N$ events, consisting of the events, 
${x_1}_i,{x_2}_i,{x_3}_i,\ldots,{x_n}_i$, for $i=1,2,\ldots,n$, if the 
physical world had been governed by the distribution function $f$ and the
quantities $x_1,x_2,x_3,\ldots,x_n$ had been physical observables. Particle
physics uses this technique {\em very} often, depending on Monte Carlo
calculations to simulate an experiment in order to plan the experiment,
simulate the apparatus, calculate experimental
efficiencies, etc.
%%%%%%%%%%%%%%%%%%%%%%%%%%%%%%5
\subsection{`Crude' Monte Carlo}\label{section:crude}
To understand `crude' Monte Carlo, consider a one-dimensional function,
$g(z)$, where $g_{\rm max}$ and $g_{\rm min}$ are both \Em{finite}, and where
$z_{\rm max}$ and $z_{\rm min}$ are also both \Em{finite}.  Thus both
$g(z)$ and $z$ are bounded.  
Let us define $\Delta z\equiv z_{\rm max}-z_{\rm min}$ and 
$\Delta g\equiv g_{\rm max}-g_{\rm min}$. We then renormalize the function
to lie in a unit square, by defining
\be x=(z-z_{\rm min})/\Delta z,\qquad y=f(x)=(g-g_{\rm min})/\Delta g, 
\label{insquare}
\ee
so that $0\le x \le 1$, and $0\le y=f(x) \le 1$, as illustrated in 
Figure \ref{fofx}. We then pick a random number $r_1$ in the interval 
$0\le r_1 \le 1$, calling it $x_1$. We next calculate the function weight
$w_1\equiv f(x_1)$. We continue this process, obtaining the $N$ weights $w_i,\ i=1,2,\ldots,N$.  The integral of $g$ 
is given by
\be
\int g\,dz\approx \Delta  g\Delta z\,\overline{w}\pm \Delta g\Delta z
\sqrt{\frac{\overline{w^2}-\overline{w}^2}{N}},%\label{mc2} 
\ee
where 
\be
\overline{w}\equiv \frac{1}{N}\sum_{i=1}^N w_i\quad
\overline{w^2}\equiv \frac{1}{N}\sum_{i=1}^N w_i^2. %\label{means2}
\ee
The problem with this technique, aside from the fact that it converges
slowly ($\approx \frac{1}{\sqrt{N}}$), is that we have generated \Em{fractional
events}, some with very tiny weights.  For example, if we trace these 
low weight events through an
experimental apparatus, perhaps to see if it hits a counter in our apparatus,
we take as much computer time for a process that might be of negligible 
weight (and thus unimportant) as for
an event of high weight which has important experimental consequences.  For
these reasons, `crude' Monte Carlo, although easy to implement (after all, it
basically requires little thinking and essentially no analysis) should often \Em{not} be the method of choice.  We will use `hit-or-miss' Monte Carlo, described  in Section \ref{hitormiss} or `importance sampling', described in Section \ref{importance}, in order to generate our `experimental events'.
%
%
%%%%%%%%%%%%%%%%%%%%%%%%%%%%%%%%%%%%%%%5
\subsection{`Hit-or-Miss' Monte Carlo}\label{hitormiss}
`Hit-or-Miss' Monte Carlo can be likened to throwing darts at a dart
board.  Pretend that the function $y=f(x)$ that you wish to reproduce
is plotted in such a way that $0\le x \le 1$, and that $0\le y \le 1$,
( see \eq{insquare}) so that a square of area unity is generated for the 
boundary of the
function, as shown in Fig.~\ref{fofx}.
If you first pick a random number $r_1$ ({\em all} random numbers
$r_i$ will be assumed to be in the interval $0\le r_i\le 1$), and choose
$x_1=r_1$, you {\em keep} the choice of $x_1$ if, upon picking a random
number $r_2$, the value of $r_2$ is such that $r_2\le f(x_1)$.  If not,
you discard the value $x_1$. This way, you build up a distribution of 
$x_i$, which
in the large number limit, goes into $f(y)$.  The assumptions here are that
you can renormalize {\em both} $x$ and $f(x)$ to be in the interval 0 to
1. 
%%%%%%%%%%%%%%%  
\begin{figure}[ht]%#65
\begin{center}
\mbox{\epsfig{file=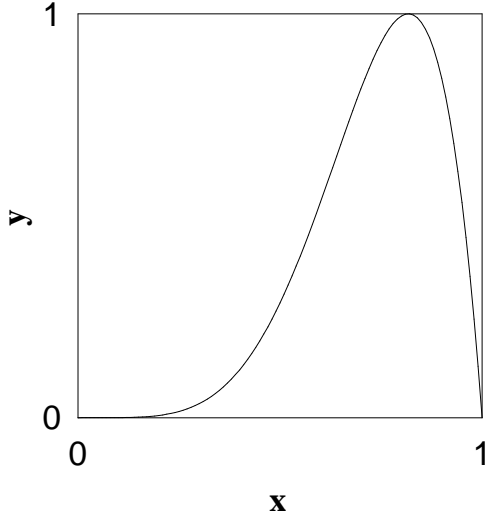
,width=3in,%,
bbllx=90pt,bblly=310pt,bburx=485pt,bbury=670pt,clip=%
}}
\end{center}
%\centerline{\psfig{figure=montecar.eps,width=4.25in}}%True width =4.25in
\caption[The renormalized probability distribution
$y=x^5(1-x)\cos (x^{1/2.3})/y_{\rm max}$]
{\footnotesize The renormalized probability distribution
$y=x^5(1-x)\cos (x^{1/2.3})/y_{\rm max}$, where 
$0\le x \le 1$ and $0\le y \le 1$ , i.e., $x$ and $y$ lie in the unit square.}\label{fofx}
\end{figure}
%%%%%%%%%%%%%%%%%%%%%%%%%%%%%%%%%

If the above assumptions are satisfied, this provides a conceptually simple
way of generating the distribution $f(x)$.  However, one pays for this, in
that the {\em efficiency} of generation (defined as the fraction of the number of the $x$ values  that you keep over the total number of times that you throw the two random numbers $r_1$ and $r_2$) is obviously
\begin{equation}
{\rm efficiency}\equiv e=\int_0^1 f(x)\,dx, 
\end{equation}
since the area of the circumscribed square is 1.  Namely, you are throwing
`darts' at the board, and any `dart' that `hits' inside the desired function 
is a winner, whereas any that `miss' are
losers. One pays a substantial price if the function is very narrow 
(if, for example, it is approximately a delta function), because the efficiency $e$ then
becomes vanishingly small---basically, the width of the function, which  by definition is 
$\ll 1$. 

It is clear from the above that the estimate of the area under the function 
$g(z)$ described in Section \ref{section:crude} is 
given by 
\be
\int g(z)\,dz=e\times \Delta g\Delta z,
\ee
where the error is given by a binomial variance.
%%%%%%%%%%%%%%%%%%%%%%%%%%%%%%%%%%%%%%%%%%%
\subsubsection{Example 1---$dP/dx=2x$}
Let us try to reproduce the {\em normalized} probability distribution
$dP/dx=2x$, for $0\le x \le 1$. We note that 
$\int_0^1 dP/dx\,dx=1$. 
We define $h(x)=f(x)/2$, such that $0\le h(x) \le 1$, which is 
the triangle shown in  Figure \ref{triangle}
and then use the following algorithm:
\begin{enumerate}
\item Pick a random number $0\le r_1 \le 1$, which we use to pick an x value.
\item  Calculate $h(r_1)$
\item Pick a random number
$r_2$. If $r_2\le h(r_1)$, {\em keep} the x-value, otherwise start 
again by going back to step (1). 
\end{enumerate}

%%%%
\begin{figure}[htb] %#66
\centerline{\psfig{figure=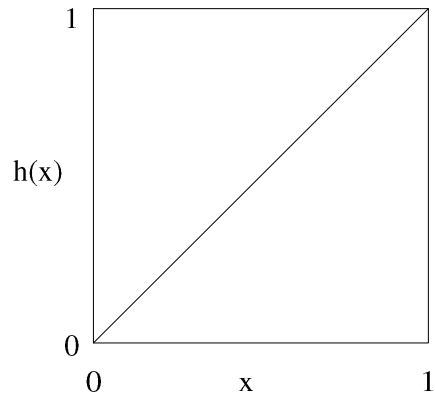,width=3in}}%True width =4.25in
%\centerwmf{1.5in}{1.5in}{c:/files/pctex/mytex/latex/triangle.wmf}
\caption[$h(x)=x$]
{\footnotesize $h(x)=x$, with $0\le x \le 1$ and $0\le h(x) \le 1$ .}
\label{triangle}
\end{figure}
%%%%%%%%%%%%%%%%%%%%%%%%%%%%%%%%%
Clearly the efficiency in this case is given by $\int_0^1x\,dx=1/2$.
%%%%%%%%%%%%%%%%%%%%%%%%%%%%%%%%%%%%%%%%%%%%%%%%%%%%%%%%%
%%%%%%%%%%%%%%%%%%%%%%%%%%%%%%%%%%%%%%%%%%%%%%%%%%%%%%%%%
\subsection
{Importance sampling}\label{importance}
\subsubsection{Example 1---$dP/dx=2x$, revisited}
We can improve this efficiency to 100\% by the following method, called 
importance sampling.  Let 
\be
g(x)\equiv\int_0^x \frac{dP}{dz}\,dz=\int_0^x2x\,dx=x^2,
\ee
where now $0\le g(x) \le 1$.  We note that the probability distribution in $g$ is
now {\em flat}, since 
\be
\frac{dP}{dg}=\frac{dP}{dx}\frac{dx}{dg}=\frac{dP}{dx}\raisebox{-.22em}{ \Huge /}\frac{dg}{dx}=1.
\ee
This is illustrated by the square shown in Figure \ref{square}, where we plot
$dP/dg$  vs. $g$.  Clearly, {\em every} choice of $g$ chosen
randomly is a success.  We can now replace the preceding algorithm with the 
new algorithm:
\begin{enumerate}
\item Pick a random number $r_1$ between 0 and 1 and calculate $g_i$, where 
$g_i=r_i\Delta g+g_0$. Let $\Delta g=g_{1}-g_{0}$.  For 
this case, $g_0=g(x=0)=0$ and
$g_1=g(x=1)=1$, so $\Delta g=1$.
\item Since $g=x^2$, the corresponding x-value {\em automatically accepted}
for this distribution is $x_1=\sqrt{g_1}$.
\item Go to step (1) again.
\end{enumerate}
\begin{figure}[htb] %#67
  \textheight 800pt \textwidth 450pt
\begin{picture}(20000,20000)
\drawline\fermion[\W\REG](28500,10000)[10000]
\put(18200,8800){\footnotesize 0}
\put(28200,8800){\footnotesize 1}
\put(23200,8300){$g$}
\drawline\fermion[\N\REG](\fermionbackx,\fermionbacky)[10000]
\put(17200,9700){\footnotesize 0}
\put(17200,19700){\footnotesize 1}
\put(16500,14700){$\frac{dP}{dg}$}
\THICKLINES
\drawline \fermion[\E\REG](\fermionbackx,\fermionbacky)[10000]
\drawline \fermion[\E\REG](18500,19970)[10000]
\drawline \fermion[\E\REG](18500,19940)[10000]
\THINLINES
\drawline \fermion[\S\REG](28500,20000)[10000]
\end{picture}
\vspace{-1in}
\caption[$dP/dg=1$ vs. $g$]
{\footnotesize $dP/dg=1$ vs. $g$, for $0\le g(x) \le 1$.}
\label{square}
\end{figure}
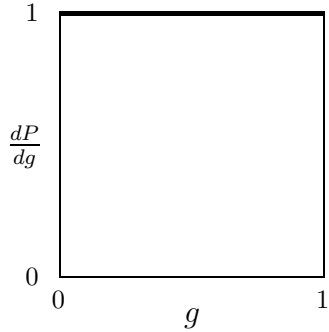 %
%%%%%%%%%%%%%%%%%%%%%%%%%%%%%%
In addition to the 100\% efficiency achieved using this technique, 
we only have to pick one random number per accepted event instead of the
two numbers needed for the `hit-or-miss' technique. 
\subsubsection{Example 2---$dP/dx=xe^{-x}$}
Consider the distribution $dP/dx=xe^{-x}$. For this distribution,
$0\le x \le +\infty$, and the $x$-range {\em can not} be rescaled to the 
interval from 0 to 1.  In order to use `hit-or-miss', we must 
{\em truncate} the $x$-scale. If we truncate at a maximum length $L$, it is 
easy to show that the efficiency is $e/L$, which becomes very small if we
must sample large $x$ values.  We do no have this problem if we use 
`importance sampling'. We introduce
\be
g(x)\equiv-\int_0^{+\infty}xe^{-x}=e^{-x}(x+1).\label{exp}
\ee
Thus, 
\begin{eqnarray}
x=\infty&&\mbox{\rm for }g(x)=0,\nonumber\\
x=0&&\mbox{\rm for }g(x)=1,
\end{eqnarray}
and, again, 
\be
\frac{dP}{dg}=\frac{dP}{dx}\left |\frac{dx}{dg}\right |=1,
\ee
giving us again 100\% efficiency, and {\em no} truncation. The price 
we pay for this is that the function $g(x)=e^{-x}(x+1)$ in \eq{exp} is not 
invertible,
\ie we  \Em{can not} solve for $x_i=x(g_i)$ by known functions.  We must 
invert
\Em{numerically}, perhaps by the Newton-Raphson method of 
Section \ref{newtonmethod}, to solve for $x_i$.
\subsection{Digression---Newton-Raphson method}\label{newtonmethod}
Sometimes called Newton's Rule, this method requires us to be able to 
evaluate both the
function $y=f(x)$ and its derivative $y'=f'(x)$, at arbitrary values of $x$.
Let us expand the function by a Taylor's series around the point $x+\Delta$ 
as
\be
f(x+\Delta)\approx f(x)+f'(x)\Delta +\frac{f''(x)}{2!}\Delta^2 +\ldots
\label{taylor}
\ee
For a sensibly behaved function and $\Delta$ small enough, we can ignore 
terms
beyond the linear.  Hence, to solve for $x$ from $f(x+\Delta)=0$, we can
write
\be
\Delta=-\frac{f(x)}{f'(x)}.\label{newton}
\ee
The problem with \eq{newton} is that, if $x$ is \Em{far} from a root, the 
higher order 
terms of \eq{taylor} \Em{are} important and the method can yield terribly
inaccurate and hopelessly wrong corrections.  For example, if by chance, 
your initial 
guess for $x$
happens to be near a local extremum, the reciprocal of the 
derivative $f'(x)$ can be \Em{huge} and lead to ridiculous results, 
when used in
\eq{newton}.  On the other hand, if the initial guess of $x$ is reasonably 
good, the method converges \Em{very rapidly}, a convergence called 
`quadratic' (see the book ``Numerical Recipes''\cite{NumRec}). The ``Function RNEWT'', a concise
Fortran program  for finding the roots using Newton's method is found in
``Numerical Recipes''\cite{NumRec}.
\subsection{Generating the Gaussian distribution}
%{$\frac{dP}{dx}=\frac{e^{-x^2/2}}{\sqrt{2\pi}}$}}
To generate the Gaussian distribution 
\be
\frac{dP}{dx}=\frac{e^{-x^2/2}}{\sqrt{2\pi}}
\ee
using importance sampling \Em{without truncation}, we will use a trick. We note that we can write
\begin{eqnarray}
d^2P&\equiv&\frac{dP}{dx}dx\times\frac{dP}{dy}dy, \quad -\infty\le x,y\le +\infty\nonumber\\
&=&\frac{1}{2\pi}e^{-x^2/2}e^{-y^2/2}\,dx\,dy 
=\frac{1}{2\pi}e^{-(x^2+y^2)/2}\,dx\,dy\nonumber\\
&=&\frac{1}{2\pi}e^{-r^2/2}r\,dr\,d\theta,\qquad 0\le \theta \le 2\pi,0\le r
\le +\infty\nonumber\\
&=& \frac{1}{2\pi}\left[\frac{d\left(e^{-r^2/2}\right)}{d(r^2/2)}
d(r^2/2)\,
d\theta\right],
%\label{gauss}
\end{eqnarray}
where
\begin{eqnarray}
r&=&\sqrt{x^2+y^2},\quad \theta = \arctan \frac{y}{x}\nonumber\\
x&=&r\cos\theta,\quad y=r\sin\theta . %\label{transform}
\end{eqnarray}
Since $z=e^{-r^2/2}$ is distributed \Em{uniformly} , with  $0 \le z \le 1$, 
we can reproduce this distribution by picking a random 
number $z_1$. We then solve for the  
variable $r_1$, which is given by $r_1=\sqrt{-2\ln z_1}$. Next, we pick a 
second random number $z_2$.  The 
variable $\theta$ is given by $\theta_1=2\pi z_2$ (we simply rescale 
$z_2$ to the
interval $0\le z_2\le 2\pi$). Finally, we find \Em{two} entries for our 
Gaussian
distribution, called $x_1$ and $y_1$, by:
\begin{eqnarray}
x_1&=&r_1\cos \theta_1,\quad y_1=r_1\sin \theta_1 .
\end{eqnarray}
\subsection{Practical problems in importance sampling}
What do we do if we \Em{can not} integrate the function
$dP/dx=f(x)$ by quadratures (a situation which happens only too
often)? 

We attempt to make a change of variable,
$ 
z=\int g(x)\,dx, 
$
where
$g(x)$ is, first of all, integrable, and second, is the most rapidly varying
portion of the original function $f(x)$. Then, after this variable change,
we find that we now want the distribution
\be
\frac{dP}{dz}=\frac{dP}{dx}\frac{dx}{dz}=\frac{f(x)}{g(x)}\label{smooth}.
\ee
In essence, by a judicious choice of $g(x)$, we have managed (hopefully!) to 
have changed a rapidly varying function $f(x)$ into a gently varying 
function $h(x)=f(x)/g(x)$. We then do `hit-or-miss' Monte Carlo
on the new (and slowly-varying) function $h(z)$. The problem here is to 
find a suitable
function $g(x)$, which is a delicate choice depending on 
both insight and luck, as well as on the skill of the reader.
\subsubsection{Example---$dP/dx=x^5(1-x)\cos (x^{1/2.3})$}
As an example of importance sampling, consider the problem of 
reproducing the (somewhat unlikely and artificial) probability distribution
shown in Figure \ref{fofx},
\be
\frac{dP}{dx}=x^5(1-x)\cos (x^{1/2.3}),\quad 0\le x \le 1. \label{artificial}
\ee
Let us define $f(x)\equiv x^5(1-x)\cos (x^{1/2.3})$. The function $f(x)$ is 
not integrable, so our techniques used up until now are not applicable.
The \Em{rapidly}
varying portion of $f(x)$ in the interval 0 to 1 is clearly the term
$x^5(1-x)$, so we choose to define $g(x)=x^5(1-x)$. Hence, we introduce the new variable $z$ as
\be
z\equiv\int_0^x \xi^5(1-\xi)\,d\xi=\frac{x^6}{6}-\frac{x^7}{7},
\ee 
with $z_{1}=1/6-1/7=1/42$ and $z_{0}=0$.  In order to use a random
number $r_i$ with $0\le r_i \le 1$, we must rescale $z$ such that
$z_i=r_i\Delta z+z_0$, where $\Delta z=z_{1}-z_{0}=z(x=1)-z(x=0)$.
The probability distribution in $z$ is given by
\be
\frac{dP}{dz}=\frac{dP}{dx}\left|\frac{dx}{dz}\right|
=h(x)=\cos (x^{1/2.3}). \label{better}
\ee
We will next evaluate $\frac{dP}{dz}$ by `hit-or-miss' Monte Carlo.
The net result is that we have replaced a the rapidly varying distribution 
of \eq{artificial} with the smooth distribution $h(x)$ of \eq{better}, 
since the function $h(x)=\cos (x^{1/2.3})$ is 
slowly-varying 
in the 
interval  0 to 1, as shown in Figure \ref{cos}.
\begin{figure}[htb]%#68
%\centerline{\psfig{figure=cos23.eps,width=4.25in}}%True width =4.25in
%\centerwmf{2in}{2in}{c:/files/pctex/mytex/latex/cos23.wmf}
\begin{center}
\mbox{\epsfig{file=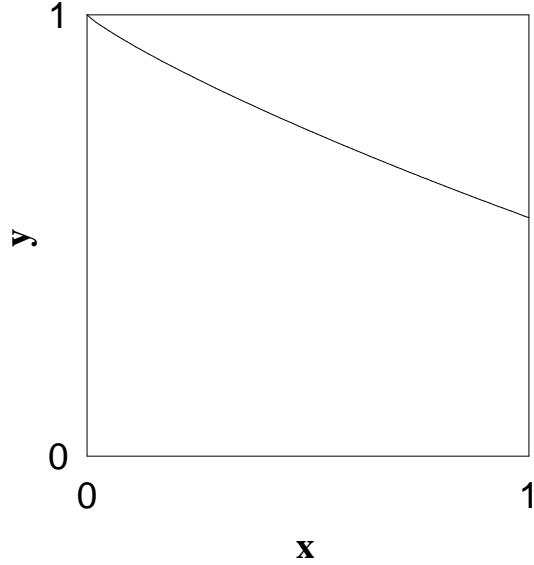
,width=3in,%,
bbllx=105pt,bblly=285pt,bburx=470pt,bbury=660pt,clip=%
}}
\end{center}

\caption[The probability distribution
$h(x)=\cos (x^{1/2.3})$]
{\footnotesize The probability distribution
$h(x)=\cos (x^{1/2.3})$, 
where 
$0\le x \le 1$ and $0\le y \le 1$ .}\label{cos}
\end{figure}
By using 
`hit-or-miss' to evaluate the distribution $h(x)$ of Figure \ref{cos}, 
we now have 
a \Em{high} efficiency calculation 
replacing the \Em{low} efficiency method that we would have had 
using `hit-or-miss' on the 
\Em{original} distribution of \eq{artificial} shown in Figure \ref{fofx}.
The good news is that we 
gain about a factor of $\approx 20$
in efficiency.  The bad news is that we need to invert the equation 
$z=\frac{x^6}{6}-\frac{x^7}{7}$ to solve for $x$ (possibly using Newton's
method, described in Section \ref{newtonmethod}). The good news strongly 
outweighs the bad news in this case.

We will use this important technique later for
get a phase space distribution for massive particles.
\subsection{Distribution generators}
We will derive here some distribution generators, to be later used with Monte Carlo
techniques (see ref. \cite{blcomputer} for a full discussion) to evaluate $n$-body phase space. 
\subsubsection{Mellin transformation} \label{sec-Mellin}
If we have a normalized probability distribution $dP(y)/dy$,
the $n^{\it th}$ moment of this distribution $Q_n$ is given by
\begin{equation}
Q_n=<y^n>=\int^1_0\,y^n\,\frac{dP(y)}{dy} \,dy.\label{moments}
\end{equation}
The inverse Mellin transformation allows one to determine
the distribution $dP(y)/dy$, if we know the analytic continuation $Q(t)$  of all of the moments $Q_n$. The appropriate transformation is given by
\begin{equation}
\frac{dP(y)}{dy} = \frac{1}{2\pi i}\,\int_{c-i\infty}^{c+i\infty}
\frac {Q(t)}{y^{\,t+1}}\,dt.
\label{eq:Mellin}
\end{equation}
\subsubsection{Massless particle generators}
We now derive the generator for the $i^{\rm th}$ massless
particle distribution, which can be written as 
\begin{equation}
\frac{dz}{d\xi_i}=i(i-1){\left(\xi_i\right)}^{i-2}(1-\xi_i), 
\label{eq:massless2}
\end{equation}
for $0\leq \xi_i \leq 1$, using the Mellin transformation technique of
Section {\ref{sec-Mellin}. We will later 
derive the generator of the distribution for {\em massive} particles, 
\begin{equation}
\frac{dz_i}{d\eta_i}=i(i-1){\left(\eta_i\right)}^{i-2}
  (1+\frac {u_i}{{\cal P}_i} - \xi_{i_{min}}-\eta_i).
\end{equation}
\subsubsection{Applications of the Mellin transformation}
We would now like to use the Mellin transformation to 
find the probability distribution $dP(y)/dy$
that is determined by setting $y$ in Eq.~(\ref{moments}) to be
\begin{equation}
y=\prod_{i=1}^k \left( r_i \right)^\frac{1}{m_i},
\end{equation}
in the domain $0\leq y\leq 1$, where the $m$'s are subject to the conditions
$m_1>m_2>m_3\cdots >m_k>0$, and the $r_i$'s are random numbers
between 0 and 1,  i.e., $y$ is determined by using the $k$ random numbers $r_1,r_2\ldots,r_k$. We  observe that it is simple to obtain the
$n^{\it th}$ moment of $y$. Elementary integration, using the
independence of the random numbers $r_i$ in the above equation, yields
\begin{equation}
Q_n=\prod_{i=1}^k \frac{m_i}{n+m_i}. \label{eq:moments}
\end{equation}
By analytically continuing the moments in Eq.~(\ref{eq:moments}),
and deforming the contour to
close with an infinite semicircle in the negative half-plane (for $c>0$),
we use the
Cauchy theorem to evaluate Eq.~(\ref{eq:Mellin}) as
\begin{equation}
\frac{dP(y)}{dy} = \left(m_1\times m_2\times m_3 \cdots \times m_k\right)
\,\sum_{j=1}^k\left[
\,\prod_{i=1}^k\! \raisebox{1.2ex}{$\prime$} \,
\frac{1}{m_i-m_j}\,y^{m_j-1}\right],   \label{eq:Cauchy}
\end{equation}
where $\prod '$ means $i\neq j$.  If we take $k=2$ and $m_1=m_2+1=i$  in
Eq.~(\ref{eq:Cauchy}), we get
\begin{equation}
\frac{dP(y)}{dy}= i(i-1){\left(y\right)}^{i-2}(1-y), %\label{eq:dist2}
\end{equation}
for
\begin{equation}
y=r_1^{\,\frac{1}{i}}\times r_2^{\,\frac{1}{i-1}}, \ \  0\leq y\leq1,
\label{eq:tworan}
\end{equation}
which is the same as Eq.~(\ref{eq:massless2}) for {\em massless} particles,
by substituting $\xi_i$ for
$y$, and $z$ for $P$.

For later use in Section {\ref{sec:addthem}, we note that for $k=1$ and
$m_1=i$, we get
\begin{equation}
\frac{dP(y)}{dy}= i{\left(y\right)}^{i-1}, %\label{eq:dist1}
\end{equation}
for
\begin{equation}
y=r_3^{\,\,\frac{1}{i}}, \ \  0\leq y\leq1.
\label{eq:oneran}
\end{equation}

The case where $k=3$ is rather interesting.  It yields
\begin{eqnarray}
\lefteqn{\frac{dP(y)}{dy} = \frac{m_1m_2m_3}{(m_1-m_2)(m_1-m_3)(m_2-m_3)}
\times} \nonumber \\
&& \left\{(m_2-m_3)y^{m_1-1} -(m_1-m_3)y^{m_2-1} +(m_1-m_2)y^{m_3-1}\right\},
\label{eq:3}
\end{eqnarray}
corresponding to the generator
\begin{equation}
y=r_1^{\,\frac{1}{m_1}}\times r_2^{\,\frac{1}{m_2}} \times r_3^{\,\frac{1}{m_3}}.
\end{equation}
A particularly interesting special case of Eq.~(\ref{eq:3}) is to pick
$m_1=2i+1$, $m_2=i+1$, and $m_3=1$, resulting in the generator
\begin{equation}
y=r_1^{\,\frac{1}{2i+1}}\times r_2^{\,\frac{1}{i+1}} \times r_3,
\end{equation}
and the distribution
\begin{equation}
\frac{dP(y)}{dy} = \frac{(i+1)(2i+1)}{2i^2}\left(y^i-1\right)^2,
\end{equation}
for $i>0$.

Other useful distributions are easily obtained from the generator
\begin{equation}
y=r_1\times r_2\times r_3\cdots\times r_k,
\end{equation}
for
\begin{equation}
0\leq y\leq 1 ,  \nonumber
\end{equation}
where the $r_i$ are, again,
independent random numbers between 0 and 1.  Clearly,
$<y^n>=\left(\frac{1}{n+1}\right)^k$. Using the Mellin transformation
of Eq.~(\ref{eq:Mellin}),  we obtain a $k$th-order pole, whose residue,
after some
manipulation, yields
\begin{equation}
\frac{dP(y)}{dy} = \frac{1}{(k-1)!}\left(-\log y\right)^{k-1}.
\end{equation}
If we make the variable substitution $x=-\log y$ in the above distribution,
we find, using
\begin{equation}
x=-\log \left( r_1\times r_2\times r_3\cdots\times r_k\right),
\end{equation}
that
\begin{equation}
\frac{dP(x)}{dx} = \frac{1}{(k-1)!} x^{k-1} e^{-x},
\end{equation}
where the new variable's domain is $0\leq x \leq \infty$.

Our technique of using the Mellin transformation obviously
can be extended to many
other distributions which have not been considered in this work. 
\subsubsection{Massive particles---fractional addition of distributions}
\label{sec:addthem}
The distribution that we want to generate for massive particles 
(see \eq{eq:dzi})
is given by
\begin{equation}
\frac{dP(\eta)}{d\eta}\propto
  i(i-1){\left(\eta_i\right)}^{i-2}
  \left(
{1-\frac{\eta_i}
{
1+\frac {u_i}{{\cal P}_i}-\xi_{i_{min}}
}
}\right),  \label{eq:dPdeta}
\end{equation}
%as can be seen from Eq.~(\ref{eq:dzi}).  
To this end, we consider a
probability $p$, where $0\leq p \leq 1$, calling $dP_2/dy$ the
probability distribution of Eq.~(\ref{eq:tworan}) and $dP_1/dy$ the
probability distribution of Eq.~(\ref{eq:oneran}), and then form a new,
normalized probability distribution
\begin{equation}
\frac{dP(\eta)}{dy}\equiv p\times \frac{dP_2(y)}{dy} + (1-p)\times \frac{dP_1(y)}{dy}.
\end{equation}
After some algebraic manipulation, we can rewrite it as
\begin{equation}
\frac{dP(\eta)}{dy}=i(i-1)py^{i-2}\left[
1-\left(1-\frac{1-p}{(i-1)p}\right)y\right].
\end{equation}
Substituting $y=\eta_i/\Delta\xi_i$ in the preceding equation, we get
\begin{equation}
\frac{dP(\eta)}{d\eta}\propto
  i(i-1){\left(\eta_i\right)}^{i-2}
  \left(
{1-\frac{\eta_i}
{
1+u_i/{\cal P}_i-\xi_{i_{min}}
}
}\right),
\end{equation}
the desired distribution of Eq.~(\ref{eq:dPdeta}), if we set
\begin{equation}
1-\frac{1-p}{(1-i)p}=\frac{\Delta\xi_i}{1+u_i/{\cal P}_i-\xi_{i_{min}}},
\end{equation}
 i.e.,
\begin{equation}
p=\frac{
1+u_i/{\cal P}_i-{\xi_i}_{min}
}
{
i\left({1+u_i/{\cal P}_i-{\xi_i}_{min}}\right) - (i-1)\Delta\xi_i
}.         \label{eq:p}
\end{equation}

Thus, if we pick
\begin{equation}
{\bar \xi}_i={\xi_i}_{min}+\Delta\xi_i\times
r_2^{\,\,\frac{1}{i}} \times r_3^{\,\,\frac{1}{i-1}}, \label{eq:tworr}
\end{equation}
with the probability $p$ of Eq.~({\ref{eq:p}), and
pick
\begin{equation}
{\bar \xi}_i={\xi_i}_{min}+\Delta\xi_i\times
r_4^{\,\,\frac{1}{i}},\label{eq:onerr}                 
\end{equation}
with the probability $1-p$, we automatically generate the
distribution we need, if $r_2,r_3,r_4$ are random numbers between 0 and 1.
We insure the proper distribution of chosen $\xi$'s by first
picking a random number $r_1$ between 0 and 1, and if $r_1$ is less than the
probability $p$, using the ${\bar \xi}_i$ determined by
Eq.~(\ref{eq:tworr}), and, if
not, using the ${\bar \xi}_i$ determined by Eq.~(\ref{eq:onerr}).
In this way, we now have at our disposal a fast algorithm
for generating our desired
distribution, by the technique of {\em fractional addition} of two
independent distributions.
%%%%%%%%%%%%%%%%%%%%%%%%
\section{Monte Carlo formulation of $n$-body phase space}\label{sec:MCphasespace}
\subsection{Generation of effective masses}
For high efficiency in doing `hit-or-miss'
Monte Carlo\cite{James}\cite{Kleiss}\cite{Byckling}\cite{NumRec}, it is
important to be able to introduce a sampling function,  i.e.,
a function integrable by quadratures which is similar in form to our desired
function.  This technique, described in Section \ref{importance}, is known as importance sampling.
This variance-reducing method is particularly important for
Eq.~(\ref{eq:qed}), which, for large numbers of particles $n$
at very high energies,
has terms in it (the integrals)  which are {\em{very}} narrow, and
hence are exceedingly poorly sampled otherwise.

Fortunately, the massless particle case is integrable by quadratures, and
a variant of it furnishes us with a valuable sampling 
function\cite{blcomputer}, which
we define as
\begin{equation}
dz_i\equiv i(i-1)\left(\xi_i-\xi_{i_{min}}\right)^{i-2}
(1+\frac {u_i}{{\cal P}_i}-\xi_i)\,d\xi_i . \label{eq:dzi}
\end{equation}
Introducing the new variable
\begin{equation}
\eta_i = \xi_i-\xi_{i_{min}},   \label{eq:eta}
\end{equation}
with
\begin{eqnarray}
\Delta\xi_i&\equiv &\xi_{i_{max}}-\xi_{i_{min}},\\ 
\eta_{i_{max}}&=&\Delta\xi_i, \nonumber\\
\eta_{i_{min}}&=&0, \nonumber
\end{eqnarray}
we rewrite
Eq.~(\ref{eq:dzi})  to get
\begin{equation}
\frac{dz_i}{d\eta_i}=i(i-1){\left(\eta_i\right)}^{i-2}(1+\frac {u_i}{{\cal P}_i}
  -\xi_{i_{min}}-\eta_i),
\label{eq:dzdeta}
\end{equation}
which we integrate by elementary means to get
\begin{equation}
z_i=\left[i\left(1+\frac {u_i}{{\cal P}_i}-\xi_{i_{min}}\right)-(i-1)\eta_i\right]
\eta_i^{i-1}.  \label{eq:z}
\end{equation}
Noting that
\begin{eqnarray}
z_{i_{min}}&=&z_i(\xi_{i_{min}})=0 \nonumber\\
z_{i_{max}}&=&z_i(\xi_{i_{max}}) = \left[i\left(1+\frac {u_i}{{\cal P}_i}-
\xi_{i_{min}}\right)-(i-1){\Delta\xi_i}\right]{\Delta\xi_i}^{i-1},
\label{eq:zmax}
\end{eqnarray}
we get, defining $\Delta z_i=z_{i_{max}}\!- z_{i_{min}}$, and using the
kinematics of Eq.~(\ref{eq:kinematics}) in Eq.~(\ref{eq:zmax}),
\begin{equation}
\Delta z_i=z_{i_{max}}. \label{eq:delz}
\end{equation}
In terms of our new variable $z_i$, we rewrite Eq.~(\ref{eq:qed})  as
\begin{eqnarray}
\lefteqn{\Phi_n\left(M_n^2;m_{n-1}^2,m_{n-2}^2,\ldots ,
m_1^2,m_0^2\right)=\Phi_n\left(M_n^2;0\right) \times} \nonumber \\
& & 
\left[\,
\prod_{i=2}^{n-1} \,
\left\{
\int_0^{z_{i_{max}}}
\left( \frac{\xi_i}{\xi_i-\xi_{i_{min}}} \right)^{i-2}
\frac{{\cal F}_1(\xi_i,u_i/{\cal P}_i)}{1+u_i/{\cal P}_i-\xi_i}\,\,dz_i
\right\}\right] \times
{\cal F}_1 \left( \frac{u_0}{{\cal P}_1},\frac{u_1}{{\cal P}_1} \right).
\label{eq:sampleit}
\end{eqnarray}
We now sample Eq.~(\ref{eq:sampleit}), running the product {\em backwards},
starting with $i=n-1$ and going down to $i=2$. We next pick a value
${\bar \xi}_i$  using the following novel algorithm, which was proved in 
Section \ref{sec:addthem}:

Step 1---Calculate a probability $p_i$, where $0\leq p_i\leq 1$, from
\begin{equation}
p_i=\frac{
1+u_i/{\cal P}_i-{\xi_i}_{min}
}
{
i\left({1+u_i/{\cal P}_i-{\xi_i}_{min}}\right) - (i-1)\Delta\xi_i
}.
\end{equation}

Step 2---Pick a random number $r_1$ , with $0<r_1<1$.  If $r_1<p_i$,
go to Step 3; else, go to Step 4.

Step 3---Choose random numbers $r_2,\ r_3$ such that
$0 \leq r_2 \leq 1$
and $0 \leq r_3 \leq 1$. Pick the value
\begin{equation}
{\bar \xi}_i={\xi_i}_{min}+\Delta\xi_i\times
r_2^{\,\,\frac{1}{i}} \times r_3^{\,\,\frac{1}{i-1}}, \label{eq:twor}
\end{equation}
and exit.

Step 4---Choose a random numbers $r_4$ such that
$0 \leq r_4 \leq 1$ . Pick the value
\begin{equation}
{\bar \xi}_i={\xi_i}_{min}+\Delta\xi_i\times
r_4^{\,\,\frac{1}{i}},                  \label{eq:oner}
\end{equation}
and exit.

Thus, we pick ${\bar \xi}_i$ from Eq.~(\ref{eq:twor}) with a probability $p_i$
and from Eq.~(\ref{eq:oner}) with a probability $1-p_i$, in order to
reproduce the importance sampling distribution given in Eq.~(\ref{eq:dzi}),
or, alternatively, in Eq.~(\ref{eq:dzdeta}).  This technique gives us a very
fast generator. In  contrast,
in the traditional method for generating such a distribution, one first picks a
random number
$r_i$, such that $0\le r_i\le 1$, then calculates
\begin{equation}
{\bar z}_i=r_i \times {z_i}_{max},
\end{equation}
and finally solves for ${\bar \xi}_i=\xi ({\bar z}_i)$, from
Eq.~(\ref{eq:eta}) and Eq.~(\ref{eq:z}), using an iterative numerical 
technique
for the solution  ${\bar \xi}_i=\xi ({\bar z}_i)$.
Since our generating technique does not require time-consuming iterative
numerical
inversion routines, we gain tremendously in computer speed, particularly
when high accuracy in the numerical inversion is required.

Having picked a value of ${\bar \xi}_i$,
we then evaluate the weight factor $w_i$, which is
\begin{equation}
w_i=\Delta z_i \left(\frac{{\bar \xi}_i}{{\bar \xi}_i-
\xi_{i_{min}}}\right)^{i-2}
\frac{{\cal F}_1({\bar \xi}_i, u_i/\bar{{\cal P}_i})}
{1+u_i/\bar{{\cal P}_i}-
{\bar \xi}_i}
,\label{eq:wi}
\end{equation}
where
\begin{equation}
\bar{{\cal P}_i}= \prod_{j=i+1}^n {\bar \xi}_j,\\
i=1,\,2,\,\dots ,\,n-1\\
{\rm and\ } {\bar \xi}_n=1.
\end{equation}
The reason we choose to {\em descend} in $i$ is that, at every stage of
the calculation,
$\bar{{\cal P}_i}$ contains only values
of $\bar{\xi_i}$ for $i$'s that have {\em already} been calculated.
We continue the process for $i=n-2$, {\it etc.}, decreasing $i$ until we 
have finished with $i=2$.
From inspection of Eq.~(\ref{eq:sampleit}), using Eq.~(\ref{eq:wi}), we 
arrive at the
final result
\begin{eqnarray}
\lefteqn{\Phi_n\left(M_n^2;m_{n-1}^2,m_{n-2}^2,\ldots ,
m_1^2,m_0^2\right)=} \nonumber \\
& & \Phi_n\left({\cal M}_n^2;0\right) \times
\left[\,
\prod_{i=2}^{n-1} \,w_i   \right]\times
{\cal F}_1\left(u_0/\bar{{\cal P}_1},u_1/\bar{{\cal P}_1}\right)=\\
&& \Phi_n\left(M_n^2;0\right) \times W,
\label{eq:answer}
\end{eqnarray}
where $W$ is the {\em total} weight.
The above procedure, which automatically conserves energy and momentum,
allows us to generate an event (let us call it the $k$th event)
with effective masses $M_2,M_3,\dots ,M_{n-1},$ having
a total weight $W_k$ which is large (comparable to unity).  To generate
individual events which have unit weight, and therefore are the equivalent
of experimental data, we must find (empirically) $W_{max}$, the maximum
possible weight for the given kinematics. We throw another random number,
$r_k$, where $0\le r_k\le 1$, after the $k$th event is generated. We
accept the $k$th event if $W_k\ge r_k \times W_{max}$, and reject it,
otherwise.  Because of the ``importance sampling'' we employ, most of the
generated events have weights $W_k$ near
$W_{max}$, and we thus efficiently generate individual events of unit weight.
\subsection{Generation of `decays'}\label{decays}
If the event is accepted, we then generate the `decays' of the
individual particles $m_i$, in the rest frame $i+1$,  i.e., in the rest
frame of $M_{i+1}\rightarrow M_{i}+m_i$, which are then
Lorentz-transformed into the appropriate laboratory reference frame.
For simplicity, we will limit our arguments to
isotropic decays.  We use the notation: ${\vec P}_{i+1}$ is the 3-momentum
vector of the mass $M_{i+1}$, and ${\vec p}_i^{\,*}$ is the 3-momentum
vector and $E_i^*=\sqrt{(p^*_i)^2+m_i^2}$ is the energy of the particle of mass $m_i$ in the $*$-frame (which is the rest frame
of $M_{i+1}$) and ${\vec p}_i$ is its 3-momentum in the laboratory
frame.  Using the Lorentz transformation
along the direction of ${\vec P}_{i+1}$,
\begin{equation}
{\vec p}_i={\vec p}_i^{\,*} +\frac{{\vec P}_{i+1}}{M_{i+1}}\left(
E_i^* + \frac{{\vec P}_{i+1} \cdot {\vec p}_i^{\,*}}{M_{i+1}
+ \sqrt{M_{i+1}^2+{\vec P}_{i+1}^2}} \right)   \label{eq:Lorentz}
\end{equation}
and
\begin{equation}
{\vec P}_{i}={\vec P}_{i+1} - {\vec p}_i.
\end{equation}
To calculate $p_i^*$, we first compute $E_i^*$, using elementary kinematics, as
\begin{equation}
E_i^*  = \frac{M_{i+1}^2+m_i^2-M_{i}^2}{2M_{i+1}},
\end{equation}
and then $p_i^*=\sqrt{(E_i^*)^2-m_i^2}$.
We next generate a random direction $(\theta^*,\,\phi^*)$ for
${\vec p}_i^{\,*}$ by
using $\cos \theta^*_i=2r_i-1$ and $\phi^*_i=2\pi r_j$, where $r_i,\,r_j$ are
random numbers between 0 and 1.  The vector components of
${\vec p}_i^{\,*}$ are obtained from
\begin{eqnarray}
p^*_{i_x}&=&p_i^*\sin \theta_i^*\cos\phi^*_i  \nonumber\\
p^*_{i_y}&=&p_i^*\sin \theta_i^*\sin\phi^*_i  \nonumber\\
p^*_{i_z}&=&p_i^*\cos \theta_i^*.   \label{eq:momenta}
\end{eqnarray}
These components of ${\vec p}_i^{\,*}$  are then used in
Eq.~(\ref{eq:Lorentz}).  By sequentially applying Eq.~(\ref{eq:Lorentz}),
starting with $i=n-1$, and the {\em known} ${\vec P}_n,\ M_n$, we get
${\vec p}_{n-1}$, the laboratory 3-momentum of particle of mass $n-1$, and
${\vec P}_{n-1}$, the laboratory 3-momentum of the cluster $n-1$.  With
successive application of Eq.~(\ref{eq:Lorentz}), we eventually get down to
$i=1$, and obtain ${\vec p}_{1}$.  Since particle $m_0$ earlier had been labeled
as $M_1$, we see that ${\vec P}_{1}$ is the laboratory momentum of
$m_0$, so our goal of transforming all particles to a common laboratory frame
of reference has been completed.
%%%%%%%%%%%%%%%%%%%%%%%%%%%%%%%%%%
\subsection{NUPHAZ, a computer implementation of $n$-body 
phase space}
The program NUPHAZ is a computer implementation of $n$-body phase space described above. 
A Fortran version  of  NUPHAZ
is described in reference \cite{blcomputer}.

As input, we require the matrix of the
$n$ masses, $m_{n-1},m_{n-2},\dots ,m_1,m_0$, along with the 4-vector of the system, $P_n$, together with the desired number of events and the name of the data file to be made, along with the settings of the NUPHAZ\cite{blcomputer} switches. You set a flag to decide whether or not to Lorentz transform an event after it has been generated in order to make a data file containing the three momentum components $p_x,p_y,p_z$ and energy $E$ of each particle in an `event'. To simulate experimental data and make individual `events' which all have the same (unit) weight, it is necessary to know $W_{max}$, the maximum possible weight for the given kinematics. The strategy adopted is to turn off both the Lorentz transformation generator flag and the unit weight event selection flag and then make a relatively short run in order to determine empirically the maximum value of $W$. This value is then inserted for $W_{\rm max}$, the event generator flag is turned back on (with Lorentz transformations, if desired) and the program is rerun for the desired number of unit weight events.  These events are then
selected using the input value of $W_{\rm max}$. However, if during a long run, a value of $W$ is found that is greater than the input value of $W_{max}$, we update $W_{max}$ with this new number. We note that the upper bound on $W_{max}$ is 1, which is its value when the particles are all massless.

This method of evaluating $W_{max}$ by means of a preliminary run is
rather efficient at {\em all} energies. At very high energies where the
particles become ultra-relativistic, the values of $W_{max}$ are {\em very} close to unity and, thus, the number of events required to be
generated for a given accuracy is consequently small (since the efficiency of the calculation is approximately 100\%). In contradistinction, the calculation time of each $n$-bodied  event is correspondingly long because of the large number $n$ of particles. In contrast, at very low
energies  where $W_{max}$ is quite small---since the sums of rest masses of the particles are a very large fraction of the total c.m.  energy---the number of events required for an accurate determination is considerably larger. However, the time to generate a single event is very small since there are very few particles in the event, and the overall computer time required is not very different for high and low energies.
%
%

%%%%%%%%%%%%%%%%%%%%%%%%%%%%%%%%%%%%%%%%%%%%%%%%%%%%%%
%%%%%%%%%%%%%%%%%%%%%%%%%%%%%%%%%%%%%%%%%%%%%%%%%%%%
%%%%%%%%%%%%%%%%%%%%%%%%%%%%%%%%%%


\begin{thebibliography}{00}
\bibitem{pdg} %#1
Particle Data Group, K. Hagiwara S. Eidelman et al., Phys. Lett. B {\bf  592}, 1 (2004).
%
\bibitem{froissart} %#2
M. Froissart, Phys. Rev. {\bf 123}, 1053 (1961).%; A. Martin, Nuovo Cimento A {\bf 42}, 930 (1965).
%%%%%%%%%%%%%%%%%%%%%%%%%%%%%%%%
\bibitem{bc} % #3
M.~M.~Block and R.~N.~Cahn, 
      Rev. Mod. Phys. {\bf 57}, 563 (1985). 
%%%%%%%%%%%%%%%%%%%%%%%%%%%%%
\bibitem{nulouvain} %#4
Amos et al., Phys. Lett. B {\bf 120}, 460 (1983); Phys. Lett. B {\bf  128}, 343 (1983).
\bibitem{ua42}%#5
 UA4 Collaboration, C.~Augier  et al., 
     CERN-PPE-93-115 July (1993);\\ 
     International Conference on Elastic and Diffractive Scattering,\\ 
     Proceedings of the Vth Blois Workshop, Brown University, \\ 
     World Scientific, Editors, H.~Fried, K.~Kang and C.-I.~Tan, 
     p. 7, June (1993);\\ 
     Phys. Lett. {\bf B316} , 448 (1993). 
\bibitem{ua4old}%#6
UA4 Collaboration, M.~Bozzo  et al.,
      Phys. Lett. B {\bf 147} , 392 (1984). 
\bibitem{bethephase} %#7
H. A. Bethe, Ann. Phys. (N. Y.) {\bf 3}, 190 (1958). 
\bibitem{westyennie}%#8
 G. B. West and D. Yennie, Phys. Rev. {\bf 172}, 1413 (1968).
\bibitem{cahnphase} %#9
R.~N.~Cahn, 
     Zeitschr. f\"{u}r Phys. C {\bf 15}, 253 (1982). 
\bibitem{vandermeer} %#10
S. van der Meer, CERN Report ISR-PO/68-31, unpublished (1968).
\bibitem{amos2} %#11
N. Amos et al, Phys. Lett. B {\bf 243}, 158 (1990).
\bibitem{goldenrule} %#12
Editors, J. Orear, A. H. Rosenfeld and R. A. Schluter ``Nuclear Physics, Lectures of Enrico Fermi'', p. 142, The University of Chicago Press, Chicago (1950).
\bibitem{Fermi}%#13
 E. Fermi, Prog. Theor. Phys. (Japan) {\bf 5}, 570 (1950).
\bibitem{Block} %#14
M. M. Block, Phys. Rev. {\bf 101}, 796 (1956).
\bibitem{Kretzschmar}%#15
 M. Kretzschmar, Ann. Rev. Nucl. Sc. {\bf 11}, 1 (1961).
\bibitem{Hagedorn}%#16
F. Cerulus and R. Hagedorn, Nuovo Cimento Suppl. {\bf 9} (2), 646
(1958).
\bibitem{sudarshan} %#17
Prem Prakash Srivastava and George Sudarshan, Phys. Rev. {\bf 110}, 765 (1958).
\bibitem{BlockJackson} %#18
M. M. Block and J. D. Jackson, Z. f\"{u}r Physik {\bf C},
Particles and Fields {\bf 3}, 255 (1980).
\bibitem{erdelyi} % #19
A. Erd\`{e}lyi, ``Higher Transcendental Functions'', Vol. II, p. 58, McGrawHill, New York (1953).
\bibitem{Stegun}%#20
 M. Abramowitz and I. A. Stegun, Editors, ``Handbook of Mathematical Physics'', Natl. Bur. Stand., US GPO, Washington, D. C. (1964).
\bibitem {macdowell} %#21
S. W. MacDowell and A. Martin, Phys. Rev. B {\bf 135}, 960 (1964).
\bibitem{durand} %#22
L. Durand and R. Lipes, Phys Rev. Lett. {\bf 20}, 637 (1968).
%{jackson1, eden, martinandcheung,jackson2}
\bibitem{jackson1} %#23
J. D. Jackson, in ``Dispersion Relations, Scottish Universities' Summer School'', Editor, G. R. Screaton, Interscience, Edinburgh (1960).
\bibitem{eden} %24
R. J. Eden, ``High Energy Collisions of Elementary Particles'', Cambridge University Press, Cambridge (1967).
\bibitem{martinandcheung} %#25
A. Martin and F. Cheung, ``Analytic Properties and Bounds of Scattering Amplitudes'', Gordon and Breach, New York (1970).
\bibitem{jackson2} %#26
J. D. Jackson, ``Phenomenology of Particles at High Energies: Proceedings of the 14th Scottish Universities' Summer School'', Editors, R. L. Crawford and R. Jennings, Academic, London (1974).
\bibitem{titchmarsh} %#27
E. C. Titchmarsh, ``The Theory of Functions'', Oxford University Press, Oxford (1939).
\bibitem{soding} %#28
P. S\"oding, Phys. Rev. Lett {\bf 8}, 285 (1964).
\bibitem{amaldi1} %#29
U. Amaldi et al., Phys. Lett.  B {\bf 66}, 390 (1977).
\bibitem{delprete} %#30
T. Del Prete, ``Antiproton Proton Physics and the W Discovery, Proceedings of the 18th Recontre de Moriond on Elementary Particle Physics'', Editor,  J. Tran Thanh Van, Editions Fronti\`eres, Gif-sur-Yvette, Vol. III, p. 49 (1983).
\bibitem{dolen-horn-schmid} %#31
R. Dolen, D. Horn and C. Schmid, Phys. Rev. {\bf 166}, 178 (1968).
\bibitem{bargerandphillips} %#32 Inserted june 5, 2006
V. Barger and R. J. N. Phillips, Phys. Rev. {\bf 187}, 2210 (1969).
\bibitem{igiandishidapip} %#33
K. Igi and M. Ishida, Phys. Rev. D {\bf{ 66}}, 034023 (2002).
\bibitem{igiandishidapp} %#34
K. Igi and M. Ishida, Phys. Lett.  B {\bf 622}, 286 (2005).
\bibitem{blockalone} %# 35 Added June 5, 2006
Martin M. Block, hep-ph/0601210 (2006)---accepted for publication in Eur. Phys. J. C (2006).
\bibitem{iginew} %#36
 K. Igi and M. Ishida, Prog. Theor. Phys. {\bf 115}, 601 (2006).
\bibitem{bhfroissart} %#37
M. M. Block and F. Halzen, Phys. Rev. D {\bf 70}, 091901 (2004).
\bibitem{bhfroissartnew} %#38
M. M. Block and F. Halzen, Phys. Rev. D {\bf 72}, 036006 (2005).
\bibitem{eden2} %#39
R. J. Eden, Phys. Rev. Lett. {\bf 16}, 39 (1966).
\bibitem{kinoshita} %#40
T. Kinoshita, in ``Perspectives in Modern Physics'' Editor, R. E. Marshak,  Wiley, New York (1966).
\bibitem{sieve} %#41
M. M. Block,  Nucl. Inst. and Meth. A. {\bf 556}, 308 (2006).
%%%%%%%%%%%%%%%%%%%%%%%%%%%%%%%%%%%%%
\bibitem{nr}%#42
``Numerical Recipes, The Art of Scientific Computing", W. H. Press, B. P. Flannery, S. A. Teukolsky and W. T. Vettering, Cambridge University Press, p. 289-293 (1986). There is also an excellent discussion of modeling of data, including a section on confidence limits by Monte Carlo simulation,  in Chapter 14.
\bibitem{huber}%#43
``Robust Statistics'', P. J. Huber, John Wiley (1981).
\bibitem{hampel}%#44
``Robust Statistics: The Approach Based on Influence Functions'', F. Hampel, John Wiley (1986).
\bibitem{regression} %#45
``Robust Regression and Outlier Detection'', P. J. Rousseeuw and A. M. Leroy, John Wiley (1987). Robust regression is also included in the R and S languages for statistical analysis.
%%%%%%%%%%%%%%%%%%%%%%%%%%
\bibitem{blockfletcherhalzenmargolisvalin} % #46
M. M. Block, R. Fletcher, F. Halzen, B. Margolis and P. Valin, Phys. Rev. D {\bf 41}, 978 (1990).
%%%%%%%%%%%%%%%%%%%%
\bibitem{blockhalzenmargolis} %#47
M. M. Block, F. Halzen and B. Margolis, Phys. Rev. D {\bf 45}, 839 (1992).
%%%%%%%%%%%%%%%%%%%%%%%%%%%%%%%%%%%%%%%%%%%%%%%%%%%%%%%%%%%% 
\bibitem{bghp}%#48
 M. M. Block, E. M. Gregores, F. Halzen and G. Pancheri, Phys.  Rev. D {\bf 60}, 054024 (1999).
%%%%%%%%%%%%%%%%%%%%%%%
\bibitem{E710} %#49
E710 Collaboration, N. A.  Amos et. al., Phys. Rev. Lett. {\bf 63}, 2784 (1989).
%%%%%%%%%%%%%%%%%%%%%%%%%%%%%%%%%%%%%%
\bibitem{Amos}%#50
E710 Collaboration, N.~A.~Amos  et al.,  
     Phys. Rev. Lett. {\bf 68}, 2433 (1992).
%%%%%%%%%%%%%%%%%%%%%%%%%%%%%%%%%%%%
\bibitem{gapsurvival} %#51
M. M. Block and F. Halzen, Phys. Rev. D {\bf 63}, 114004 (2001).
%%%%%%%%%%%%%%%%%%%%%%%%%%%%%%%%%%%%%%%%
\bibitem{me}%52
M. M. Block and A. B. Kaidalov, Phys. Rev. D {\bf 64}, 076002 (2001).
%%%%%%%%%%%%%%%%%%%%%%%%%%%%%%%%%%%%%%%%%%%%%
\bibitem{blockhalzenpancheri}%53
M. M. Block, F. Halzen and G. Pancheri, 
 Phys. Rev. D {\bf 60}, 054024 (1999). %changed MMB 9/28/99
%%%%%%%%%%%%%%%%%%%%%%%%%%%%%%%%%%%%
\bibitem{Maor}%54
E.~Gotsman, E.~Levin and U.~Maor, Phys. Lett. B {\bf 438}, 229 (1998).
%
\bibitem{Fletcher}%55
R.~S.~Fletcher and T.~Stelzer, Phys. Rev. D {\bf 48}, 5162 (1993).
%
\bibitem{Bjorken}%56
J.~D.~Bjorken, Phys. Rev. D{\bf 47}, 101 (1993).
\bibitem{Gotsman}%57
E.~Gotsman, E.~Levin and U.~Maor, Phys. Lett. B {\bf 309}, 199 (1993).
%
\bibitem{Eboli}%58
O. J. P. \`Eboli, E. M. Gregores and F. Halzen, Phys. Rev. D {\bf 61}, 034003 (2000).
%
\bibitem{Martin}%59
V.~A.~Khoze, A.~D.~Martin and M.~G.~Ryskin, Eur. Phys. J. C {\bf 18}, 167 (2000).
%%%%%%%%%%%%%%%%%%%%%%%%%%%%%
\bibitem{blockhalzenpancheri2}%60
M. M. Block, F. Halzen and G. Pancheri, 
Eur. Phys. J. C {\bf 23}, 329 (2002).
%%%%%%%%%%%%%%%%%%%%%%%%%%%%%%%%%%%%
\bibitem{bk} % #61
M. M. Block and K. Kang, Int. J. Mod. Phys. A {\bf 20}, 2781 (2005).
%%%%%%%%%%%%%%%%%%%%%%%%%%%%%%%%%%%
\bibitem{L3} % #62
L3 Collaboration, M.~Acciarri  et al., 
Phys.\ Lett. B {\bf 519}, 33 (2001).
%
\bibitem{OPAL} %#64
 OPAL Collaboration, G. Abbiendi  
et al., Eur. Phys. J. C {\bf 14}, 199 (2000).
%%%%%%%%%%%%%%%%%%%%%
\bibitem{bauer} %#65
T.~H.~Bauer  et al., 
Rev.\ Mod.\ Phys.\ {\bf 50}, 261 (1978).
%%%%%%%%%%%%%%%%%%%%%%%%%%%%%%%%
\bibitem{bkw} %#65
M. M. Block, K. Kang and A. R. White, Int. J. Mod. Phys. A{\bf 7}, 4449 (1992).
\bibitem{compete1} %# 66
 The COMPETE Collaboration, J. R. Cudell {\em et al.}, Phys. Rev. D {\bf 65}, 074024 (2002). 
%%%%%%%%%%%%%%%%%%%%%%%%%%%%%%%%%
\bibitem{gilman}%#67
M. Damashek and F. J. Gilman, Phys. Rev. D {\bf 1}, 1319 (1970).
%%%%%%%%%%%%%%%%%%%%%%%%%
\bibitem{compton} %#68
M. M.  Block, Phys. Rev. D {\bf 65}, 116005 (2002).
%%%%%%%%%%%%%%%%%%%%%%%%%%%%%%%%%%%%%%%%%%%%%%%%
\bibitem{h1} %#69
H1 Collaboration, S. Aid  et al., Z. Phys. C {\bf 75},421 (1997).
%%%%%%%%%%%%%%%%%%%%%%%%%%%%%%%%%%%%%%%%%%%%%
\bibitem{zeus} %#70
Zeus Collaboration, S. Chekanov et al., Nucl. Phys. B {\bf 627}, 3 (2002). 
%
\bibitem{BHS} %#71
M. M. Block, F. Halzen and T. Stanev, Phys. Rev.  Lett. {\bf 83}, 4926 (1999); Phys. Rev. D {\bf 62}, 077501 (2000).
\bibitem{cudell} %#72
The COMPETE Collaboration,  J. R. Cudell et al., Phys. Rev. Lett. {\bf 89}, 201801 (2002).
%%%%%%%%%%%%%%%%%%%%%%%%%%%%%%%%%%%%%%%%
\bibitem{nicolescu1} %#73
L.~Lukaszuk and B.~Nicolescu, Lett. Nuovo Cimento {\bf 8}, 405 (1973).
%%%%%%%%%%%%%%%%%%%%%%%%%%%%%%%%%%%%%%%
%%%%%%%%%%%%%%%%%%%%%%%%%%%%%%%%%%%%%%%%%%
\bibitem{nicolescu2}%#74
K. Kang and B.~Nicolescu, Phys. Rev. D {\bf 11}, 2461 (1975).
%%%%%%%%%%%%%%%%%%%%%%%%%%%%%%%%%%%%%%%%
\bibitem{nicolescu3}%#74
D. Joynson et al, Nuovo Cimento A {\bf 30}, 345 (1975).
%%
\bibitem{blockandkang2} %# 76 Added June 5, 2006.
Martin M. Block and Kyungsik Kang, Phys. Rev. D {\bf 73}, 094003 (2006). 
\bibitem{landshoff1}% #77
P. V. Landshoff, Nucl. Phys. B  (Proc. Suppl.) {\bf 99}, 311 (2001).
%%%%%%%%%%%%%%%%%%%%%%%
\bibitem{landshoff2} % #78
P. V. Landshoff, hep/ph 0509240, (2005).
%%%%%%%%%%%%%%%%%%%%%%%%%%%%%%%%%%%%
\bibitem{gaisser} %#79
R. Engel, T. K. Gaisser, P. Lipari and T. Stanev,  Phys. Rev. D {\bf 58}, 014019, 1998.
%%%%%%%%%%%%%%%%%%%%%%%%%%%%%%%%
\bibitem{hires} %#80
K. Belov for the HiRes collaboration,  Nucl. Phys. B (Proc. Suppl.) {\bf 151}, 197 (2006).
%%%%%%%%%%%%%%%%%%%%%%%%%%%%%%%%%%%
\bibitem{fly} % 81
R. M. Baltrusaitis  et al., Phys. Rev. Lett. {\bf 52}, 1380, (1984).
%%%%%%%%%%%%%%%%%%%%%%%%%%%%%%
\bibitem{akeno} %#82
M. Honda et al., Phys. Rev. Lett. {\bf70}, 525, (1993).
%%%%%%%%%%%%%%%%%%%%%%%%%%%%
\bibitem{hillas} %#83
T. K. Gaisser and A. M. Hillas, Proc. 15$^{\rm th}$ Int. Cosmic Ray Conf. (Plovdiv) {\bf 8}, 353 (1979).
%%%%%%%%%%%%%%%%%%%%%%%%%%%%%%%%%%
\bibitem{pryke} % #84
C. L. Pryke, (2000), Astropart. Phys. {\bf 14}, 319 (2001).
%%%%%%%%%%%%%%%%%%%%%%%%%%%%%%%%%
\bibitem{yodh} % #85
T. K. Gaisser, U. P. Sukhatme and G. B. Yodh, Phys. Rev. D {\bf 36}, 1350, (1987).
%%%%%%%%%%%%%%%%%%%%3%%%%%%%%%%%%%%
\bibitem{mocca} %#86
A. M. Hillas, Nuc. Phys. B (Proc. Suppl.) {\bf 52}, 29, (1997). 
%%%%%%%%%%%%%
\bibitem{corsika} % #87
J. Knapp  et al., Reports {\bf FZKA 6019} and {\bf FZKA 5828} (1998 and  
1996), Forschungszentrum Karlsruhe. Available from {\tt  
http://www-ik3.fzk.de/\~{ }heck/corsika/}.
%
\bibitem{sibyll}% #88
R. S. Fletcher  et al., Phys. Rev. D {\bf 50}, 5710, (1994).
%
\bibitem{qgsjet}%#89
N. N. Kalmykov  et al., Nuc. Phys. B (Proc. Suppl.) {\bf 52B}, 17, 1997.\
%%%%%%%%%%%%%%%%%%%%%%%%%%%%%%%%%%%%%%
\bibitem{engelprivate} %#90
R. Engel, private communication, Karlsruhe (2005).
%%%%%%%%%%%%%%%%%%%%%%%%%%%%%%%%%%%%%%
%%%%%%%%%%%%%%%%%%%%%%%%%%%%%%%%%%%%%
%%%%%%%%%%%%%%%%%%%%%%%%%%%%%%%%%%%%%
%%%%%%%%%%%%%%%%%%%%%%%%%%%%%%%%%%%%%%%
\bibitem{James} %Appendix 1 #91
 F. James, ``Monte Carlo Phase Space'', Lecture Series, CERN
68-15, May, 1968.  He describes the unpublished Raubold-Lynch generator. 
\bibitem{Kleiss} %#2 appendix #92
 R.~Kleiss, W.J.~Sterling and S.D.~Ellis, Comp. Phys. Comm.
{\bf 40}, 359 (1986).
\bibitem{Byckling} %#3 appendix #93
 E.~Byckling and K.~Kajantie, ``Particle Kinematics'',
Wiley, New York (1973). They generalized the Raubold-Lynch generator to allow exponential importance sampling.
\bibitem{NumRec} %#4 appendix #94
 W.H. Press, B.P. Flannery, S.A.
Teukolsky, and W.T. Vetterling, ``Numerical Recipes'',
Cambridge University Press, Cambridge (1987).
\bibitem{blcomputer} %#5 appendix #95
 M.~M.~Block, Comp. Phys. Comm.
{\bf 69}, 459 (1992).
%%%%%%%%%%%%%%

\end{thebibliography}
\end{document}